\documentclass[twoside,openright]{report}
\usepackage[french,english]{babel}
\usepackage[utf8]{inputenc}
\usepackage[T1]{fontenc}
\usepackage{lmodern}
\usepackage{graphicx}
\usepackage{amsmath}  
\usepackage{amsfonts}  
\usepackage{xcolor}
\usepackage{amssymb}
\usepackage{microtype}
\usepackage[a4paper,
	width=150mm,
	top=25mm,
	bottom=25mm,
	bindingoffset=6mm]{geometry}
\usepackage[many]{tcolorbox}
\usepackage{empheq}
\tcbset{colback=white, colframe=black, highlight math style= {enhanced, colframe=black,colback=white,boxsep=0pt}}
\usepackage[pdftex, 
	pdfauthor={Florent Christian Lucien Marie Michel}, 
	pdftitle={Nonlinear and quantum effects in analogue gravity},
	pdfsubject={Lorentz-violating quantum field theories in curved spacetimes},
	pdfkeywords={Quantum fields, Partial differential equations, Gravity},
	colorlinks=true,
	filecolor=blue,
	linkcolor=blue,
	citecolor=blue,
	urlcolor=blue, 
	runcolor=blue,
	linktoc=all]{hyperref}
\usepackage{tikz}
\usetikzlibrary{calc}
\usetikzlibrary{arrows}
\usepgflibrary{fpu}
\usepackage{fancyhdr}
\newcommand{\chaptertitle}{???}
\newcommand{\newchapter}[1]{\chapter{#1}\renewcommand{\chaptertitle}{#1}}
\fancypagestyle{plain}{%
	\fancyhf{} 
	\fancyfoot[LE,RO]{\hyperref[toc]{\color{black}{\thepage}}}
	\fancyhead[RE,LO]{}

}
\fancypagestyle{thefancy}{%
	\fancyhf{} 
	\fancyhead[RE,LO]{} 
	\fancyhead[LE,RO]{\nouppercase{\chaptername\ \thechapter. \chaptertitle}}
	\fancyfoot{}
	\fancyfoot[LE,RO]{\hyperref[toc]{\color{black}{\thepage}}}
}
\usepackage{enumitem}
\usepackage{csquotes}
\usepackage[style = alphabetic, labelnumber, defernumbers = true, backend=biber,sorting=nyt]{biblatex}
\renewbibmacro{in:}{\ifentrytype{article}{}{\printtext{\bibstring{in}\intitlepunct}}}
\usepackage{import}
\usepackage{titlesec}
\usepackage{background}
\usepackage{xifthen}
\usepackage{totcount}
\usepackage{doi}
\usepackage[font={small}]{caption}
\regtotcounter{chapter}

\backgroundsetup%
{   contents={%
        \begin{tikzpicture}[overlay]
            \pgfmathtruncatemacro{\mytotalchapters}{\totvalue{chapter} > 0 ? \totvalue{chapter} : 20}
            \pgfmathsetmacro{\mypaperheight}{\paperheight/28.453}
            \pgfmathsetmacro{\mytop}{-(\thechapter-1)/(\mytotalchapters-1)*\mypaperheight}
            \pgfmathsetmacro{\mybottom}{-\thechapter/(\mytotalchapters-1)*\mypaperheight}
            \ifcase\thechapter
                \xdef\mycolor{white}
                \or \xdef\mycolor{black!13!blue}
                \or \xdef\mycolor{black!24!blue}
                \or \xdef\mycolor{black!39!blue}
                \or \xdef\mycolor{black!52!blue}
                \or \xdef\mycolor{black!65!blue}
                \or \xdef\mycolor{black!78!blue}
                \else \xdef\mycolor{black!91!blue}
            \fi
            \ifthenelse{\isodd{\value{page}}}
            {\fill[\mycolor] ($(current page.north east)+(0,\mytop)$) rectangle ($(current page.north east)+(-0.5,\mybottom)$);}
            {\fill[\mycolor] ($(current page.north west)+(0,\mytop)$) rectangle ($(current page.north west)+(0.5,\mybottom)$);}
        \end{tikzpicture}
    },
    scale=1,
    opacity=1,
    angle=0
}

\DeclareSourcemap{
  \maps[datatype=bibtex, overwrite]{
    \map{
      \perdatasource{biblio/bibliothesis0.bib}
      \step[fieldset=KEYWORDS, fieldvalue=primary]
    }
    \map{
      \perdatasource{biblio/bibliothesis.bib}
      \step[fieldset=KEYWORDS, fieldvalue=secondary]
    }
  }
}

\DeclareFieldFormat{labelnumberwidth}{\mkbibbrackets{#1}}
\renewbibmacro*{cite}{%
  \printtext[bibhyperref]{%
    \printfield{labelprefix}%
    \ifkeyword{primary}
      {\printfield{labelnumber}}
      {\printfield{labelalpha}%
       \printfield{extraalpha}}}}

\defbibenvironment{bibliographyNUM}
  {\list
     {\printtext[labelnumberwidth]{%
        \printfield{prefixnumber}%
        \printfield{labelnumber}}}
     {\setlength{\labelwidth}{\labelnumberwidth}%
      \setlength{\leftmargin}{\labelwidth}%
      \setlength{\labelsep}{\biblabelsep}%
      \addtolength{\leftmargin}{\labelsep}%
      \setlength{\itemsep}{\bibitemsep}%
      \setlength{\parsep}{\bibparsep}}%
      }
  {\endlist}
  {\item}

\newcommand{\nn}{\nonumber \\}
\newcommand{\eq}[1]{Eq.~\eqref{#1}}
\newcommand{\eqs}[2]{Eqs.~(\ref{#1}) and~(\ref{#2})}
\newcommand{\fig}[1]{Fig.~\ref{#1}}
\newcommand{\ket}[1]{\ensuremath{\left\lvert #1 \right\rangle}}
\newcommand{\bra}[1]{\ensuremath{\left\langle #1 \right\rvert}}
\newcommand{\eg}{ & \hspace*{-0.1 cm}=}
\newcommand{\Fig}[1]{Fig.~\ref{#1}}
\newcommand{\be}{\begin{eqnarray}}
\newcommand{\ee}{\end{eqnarray}}
\newcommand{\om}{\ensuremath{\omega}}

\newcommand{\pd}{\ensuremath{\partial}}
\newcommand{\la}{\ensuremath{\lambda}}
\newcommand{\ep}{\ensuremath{\epsilon}}
\newcommand{\lp}{\ensuremath{\left(}}
\newcommand{\rp}{\ensuremath{\right)}}
\newcommand{\lb}{\ensuremath{\left\lbrace}}
\newcommand{\rb}{\ensuremath{\right\rbrace}}
\definecolor{darkcyan}{rgb}{0,0.7,0.7}
\definecolor{cYan}{rgb}{0.,1.,1.}
\definecolor{lightcyan}{rgb}{0.5,1.,1.}

\newcommand{\abs}[1]{\ensuremath{\left \lvert #1 \right\rvert}} 
\newcommand{\norm}[1]{\left\lVert#1\right\rVert}
\newcommand{\deft}[1]{#1}
\newcommand{\defe}[1]{#1}
\newcommand{\higt}[1]{#1}

\newcommand{\rese}[1]{#1}
\newcommand{\smallres}[1]{#1}

\newcommand{\thisisblue}[1]{\textcolor{blue}{\boxed{\textcolor{black}{#1}}}}

\newcommand{\psiu}[1]{\ensuremath{\psi_#1^{u}}}
\newcommand{\psiv}[1]{\ensuremath{\psi_#1^{v}}}
\newcommand{\phiu}[1]{\ensuremath{\phi_#1^{u}}}

\newcommand{\psil}[1]{\ensuremath{\lp \psi_{#1}^{u, \leftarrow}} \rp^*}

\newcommand{\psir}[1]{\ensuremath{\lp \psi_{#1}^{u, \rightarrow}}\rp^*}

\newcommand{\symbain}{\phi_\om^{\rightarrow,d,{\rm in}}}
\newcommand{\symbAtin}{\phi_\om^{\leftarrow,{\rm in}}}
\newcommand{\symbAin}{\phi_\om^{\rightarrow,{\rm in}}}
\newcommand{\symbbin}{\lp \phi_{-\om}^{\rightarrow,d,{\rm in}} \rp^*} 
\newcommand{\symbaout}{\phi_\om^{\rightarrow,d,{\rm out}}}
\newcommand{\symbAtout}{\phi_\om^{\leftarrow,{\rm out}}}
\newcommand{\symbAout}{\phi_\om^{\rightarrow,{\rm out}}}
\newcommand{\symbbout}{\lp \phi_{-\om}^{\rightarrow,d,{\rm out}} \rp^*} 

\newcommand{\e}{\mathrm{e}}
\newcommand{\dd}{\mathrm{d}}
\newcommand{\ii}{\mathrm{i}}
\newcommand{\capf}{\captionof{figure}}
\newcommand{\s}{\nobreak\hspace{.08em plus .04em}}
\definecolor{skyblue}{RGB}{0,191,255}
\colorlet{dskyblue}{skyblue!85!black}
\newenvironment{myfig} 
    {\vspace*{0.5cm}
    \noindent
    \begin{minipage}{1.0\linewidth}
    \begin{center}
    \captionsetup{type=figure}}
    {\end{center}
    \end{minipage}
    \vspace*{0.5cm}
    }

\titleclass{\subsubsubsection}{straight}[\subsection]
\newcounter{subsubsubsection}[subsubsection]
\renewcommand\thesubsubsubsection{\thesubsubsection.\arabic{subsubsubsection}}
 
\titleformat{\subsubsubsection}
  {\normalfont\normalsize\itshape}{\thesubsubsubsection}{1em}{}
\titlespacing*{\subsubsubsection}{0pt}{3.25ex plus 1ex minus .2ex}{1.5ex plus .2ex}
\makeatletter
\renewcommand\paragraph{\@startsection{paragraph}{5}{\z@}%
  {3.25ex \@plus1ex \@minus.2ex}%
  {-1em}%
  {\normalfont\normalsize\bfseries}}
\renewcommand\subparagraph{\@startsection{subparagraph}{6}{\parindent}%
  {3.25ex \@plus1ex \@minus .2ex}%
  {-1em}%
  {\normalfont\normalsize\bfseries}}
\def\toclevel@subsubsubsection{4}
\def\toclevel@paragraph{5}
\def\toclevel@paragraph{6}
\def\l@subsubsubsection{\@dottedtocline{4}{7em}{4em}}
\def\l@paragraph{\@dottedtocline{5}{10em}{5em}}
\def\l@subparagraph{\@dottedtocline{6}{14em}{6em}}
\makeatother

\title{Nonlinear and quantum effects in analogue gravity}
\date{\today}
\author{Florent Michel}

\usepackage[absolute]{textpos}
\usepackage{multicol}

\newcommand{\PhDTitleFR}{Effets non-lin\'eaires et effets quantiques en gravit\'e analogue} 
\newcommand{\PhDkeywordsFR}{Th\'eorie quantique des champs, \'Equations aux d\'eriv\'ees partielles, Espaces courbes} 
\newcommand{\PhDsumFR}{
Cette th\`ese concerne l'\'etude des propri\'et\'es de champs scalaires classiques et quantiques en pr\'esence d'un environnement inhomog\`ene et/ou d\'ependant du temps. Nous nous concentrerons sur des mod\`eles pouvant être d\'ecrits, fondamentalement ou de mani\`ere effective, par un espace-temps courbe contenant un horizon des \'ev\'enements. Nous verrons en particulier comment une correspondance math\'ematique, provenant d'une sym\'etrie de Lorentz effective à basse \'energie, permet de relier les comportements des ondes dans un cadre non relativiste à la physique des trous noirs, quelles en sont les limites et dans quelle mesure les r\'esultats ainsi obtenus sont \og analogues \fg{} à leurs pendants gravitationnels. Apr\`es un premier chapitre d'introduction rappelant quelques bases de relativit\'e g\'en\'erale puis une d\'erivation de la radiation de Hawking et de la correspondance avec des syst\`emes non relativistes, je pr\'esenterai le d\'etail de quatre travaux effectu\'es durant ma th\`ese. Les autres articles \'ecrits dans ce cadre sont r\'esum\'es dans le dernier chapitre, pr\'ec\'edant une conclusion g\'en\'erale.

Mes collaborateurs et moi nous sommes concentr\'es sur trois aspects du comportement des champs pr\`es de l'analogue d'un horizon des \'ev\'enements dans des mod\`eles avec une sym\'etrie de Lorentz effective à basse \'energie. Le premier concerne les effets non lin\'eaires, cruciaux pour comprendre l'\'evolution de la radiation de Hawking ainsi que pour les r\'ealisations exp\'erimentales mais auparavant peu \'etudi\'es. Nous montrerons comment ceux-ci d\'eterminent les possibles comportements aux temps longs pour des syst\`emes stables ou instables. Le second aspect a trait aux effets lin\'eaires et quantiques, en particulier la radiation de Hawking elle-même, son devenir lorsque l'horizon est continûment effac\'e, ainsi que les diverses instabilit\'es à même de survenir dans diff\'erents mod\`eles. Enfin, nous avons particip\'e à l'\'elaboration, à l'analyse et à l'\'etude d?exp\'eriences dites de \og gravit\'e analogue \fg{} dans des condensats de Bose-Einstein et des syst\`emes hydrodynamiques ou acoustiques, dont je rapporte les principaux r\'esultats.}

\newcommand{\PhDTitleEN}{Nonlinear and quantum effects in analogue gravity} 
\newcommand{\PhDkeywordsEN}{Quantum field theory, Partial differential equations, Curved spaces} 
\newcommand{\PhDsumEN}{The present thesis deals with some properties of classical and quantum scalar fields in an inhomogeneous and/or time-dependent background, focusing on models where the latter can be described as a curved space-time with an event horizon. While naturally formulated in a gravitational context, such models extend to many physical systems with an effective Lorentz invariance at low energy. We shall see how this effective symmetry allows one to relate the behavior of perturbations in these systems to black-hole physics, what are its limitations, and in which sense results thus obtained are ``analogous'' to their general relativistic counterparts. The first chapter serves as a general introduction. A few notions from Einstein's theory of gravity are introduced and a derivation of Hawking radiation is sketched. The correspondence with low-energy systems is then explained through three important examples. The next four chapters each details one of the works completed during this thesis, updated and slightly reorganized to account for new developments which occurred after their publication. The other articles I contributed to are summarized in the last chapter, before the general conclusion. 

My collaborators and I focused on three aspects of the behavior of fields close to the (analogue) event horizon in models with an effective low-energy Lorentz symmetry. The first one concerns nonlinear effects, which had been given little attention in view of their crucial importance for understanding the evolution in time of Hawking radiation as well as for experimental realizations. We showed in particular how they determine the late-time behavior in stable and unstable configurations. The second aspect concerns linear and quantum effects. We studied the Hawking radiation itself in several models and what replaces it when continuously erasing the horizon. We also characterized and classified the different types of linear instabilities which can occur. Finally, we contributed to the design and analysis of ?analogue gravity? experiments in Bose-Einstein condensates, hydrodynamic flows, and acoustic setups, of which I report the main results.} 

\label{Personnal}
\newcommand{\PhDname}{M. Florent MICHEL} 
\newcommand{\NNT}{2017SACLS164} 
\newcommand{\ecodocnum}{564} 
\newcommand{\ecodoctitle}{Physique en Île-de-France} 
\newcommand{\PhDspeciality}{Physique}
\newcommand{\PhDworkingplace}{Universit\'e Paris-Sud} 
\newcommand{\defenseplace}{Laboratoire de Physique Th\'eorique} 
\newcommand{\defensedate}{23 juin 2017} 
\newcommand{\logoED}{\includegraphics[width = 0.8\linewidth]{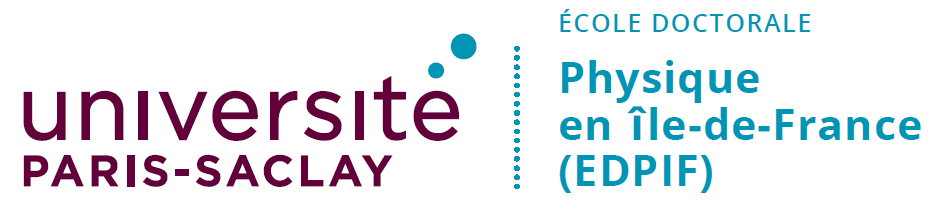}} 
\newcommand{\logoEt}{\includegraphics[scale=1.2]{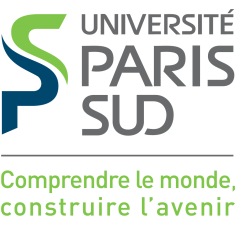}} 
\newcommand{\vpos}{1} 

\label{Jury}
\newcommand{\jurynameA}{Denis Boiron}
\newcommand{\jurygenderA}{M.} 
\newcommand{\juryadressA}{Institut d'optique graduate school}
\newcommand{\jurygradeA}{PR1}
\newcommand{\juryroleA}{Pr\'esident du jury} 
\newcommand{\jurynameB}{Theodore A. Jacobson}
\newcommand{\jurygenderB}{M.}
\newcommand{\juryadressB}{University of Maryland}
\newcommand{\jurygradeB}{Professeur}
\newcommand{\juryroleB}{Rapporteur}
\newcommand{\jurynameC}{Daniele Faccio}
\newcommand{\jurygenderC}{M.}
\newcommand{\juryadressC}{Heriot-Watt University}
\newcommand{\jurygradeC}{Professeur}
\newcommand{\juryroleC}{Rapporteur}
\newcommand{\jurynameD}{Iacopo Carusotto}
\newcommand{\jurygenderD}{M.} 
\newcommand{\juryadressD}{Università degli Studi di Trento}
\newcommand{\jurygradeD}{Professeur}
\newcommand{\juryroleD}{Examinateur}
\newcommand{\jurynameE}{Ruth Gregory}
\newcommand{\jurygenderE}{Mme}
\newcommand{\juryadressE}{Durham University}
\newcommand{\jurygradeE}{Professeur}
\newcommand{\juryroleE}{Examinatrice}
\newcommand{\jurynameF}{Karim Noui}
\newcommand{\jurygenderF}{M.} 
\newcommand{\juryadressF}{Universit\'e de Tours}
\newcommand{\jurygradeF}{PR1}
\newcommand{\juryroleF}{Examinateur}
\newcommand{\jurynameG}{Renaud Parentani}
\newcommand{\jurygenderG}{M.} 
\newcommand{\juryadressG}{LPT Orsay}
\newcommand{\jurygradeG}{PR1}
\newcommand{\juryroleG}{Directeur de th\`ese}

\begin{document}

\hypersetup{pageanchor=false}
\makeatletter
\pagenumbering{gobble}
\label{cotutelle}

\begin{tikzpicture}[remember picture,overlay,color=blue!20!red!45!black!75!]
	\draw[very thick]
		([yshift=-160pt,xshift=45pt]current page.north west)--     
		([yshift=-160pt,xshift=-25pt]current page.north east)--    
		([yshift=35pt,xshift=-25pt]current page.south east)--      
		([yshift=35pt,xshift=45pt]current page.south west)--cycle; 
\end{tikzpicture}

\begin{textblock}{13}(1.15,3.3)
  NNT : \NNT
\end{textblock}

\begin{textblock}{1}(1.15,1)
\includegraphics[height=2.4cm]{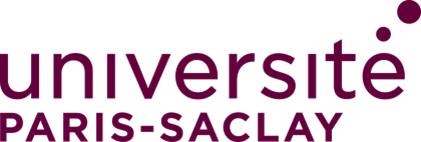} 
\label{Logo Paris Saclay}
\end{textblock}

\begin{textblock}{1}(12,\vpos)
\logoEt 
\label{Logo Etablissement}
\end{textblock}

\vspace{6cm}
\color{blue!20!red!45!black} 
  \begin{center}    
    \LARGE\textsc{Th\`ese de doctorat\\ de l'Universit\'e Paris-Saclay} \\
    \LARGE{\textsc{pr\'epar\'ee \`a l'\PhDworkingplace}} \\ \bigskip
  \color{black} 
	\vfill
    \Large{\'Ecole doctorale n$^{\circ}\ecodocnum$}\\ 
     \Large{\ecodoctitle}  \\

     \Large{Sp\'ecialit\'e de doctorat: \PhDspeciality} 
    \vfill  
   \Large{par}
   \vfill
   \LARGE{\textbf{\textsc{\PhDname}}} 
    \vfill
    \Large{\PhDTitleFR} 
    \vfill
    \bigskip
\end{center}
\color{black}
\begin{flushleft}
Th\`ese pr\'esent\'ee et soutenue au \defenseplace, le \defensedate. \\
\bigskip
Composition du Jury :
\end{flushleft}

\begin{center}
\begin{tabular}{llll}

    \jurygenderA & \textsc{\jurynameA}  & \jurygradeA & (\juryroleA) \\
    \null & \null & \juryadressA &\\   
   
    \jurygenderB & \textsc{\jurynameB}  & \jurygradeB & (\juryroleB) \\
    \null & \null & \juryadressB &\\ 
    
    \jurygenderC & \textsc{\jurynameC}  & \jurygradeC & (\juryroleC) \\
    \null & \null & \juryadressC &\\ 
    
    \jurygenderD & \textsc{\jurynameD}  & \jurygradeD & (\juryroleD) \\
    \null & \null & \juryadressD &\\ 
    
    \jurygenderE & \textsc{\jurynameE}  & \jurygradeE & (\juryroleE) \\
    \null & \null & \juryadressE &\\ 
    
    \jurygenderF & \textsc{\jurynameF}  & \jurygradeF & (\juryroleF) \\
    \null & \null & \juryadressF &\\ 
    
    \jurygenderG & \textsc{\jurynameG}  & \jurygradeG & (\juryroleG) \\
    \null & \null & \juryadressG &\\ 
   
  \end{tabular}    
\end{center}

\cleardoublepage
\makeatother
\hypersetup{pageanchor=true}

\chapter*{}
\pagenumbering{roman}

\vspace*{-4 cm}

\textcolor{white}{test}

\begin{huge}\textbf{Abstract}\end{huge} \\

The present thesis deals with some properties of classical and quantum scalar fields in an inhomogeneous and/or time-dependent background, focusing on models where the latter can be described as a curved space-time with an event horizon. While naturally formulated in a gravitational context, such models extend to many physical systems with an effective Lorentz invariance at low energy. We shall see how this effective symmetry allows one to relate the behavior of perturbations in these systems to black-hole physics, what are its limitations, and in which sense results thus obtained are “analogous” to their general relativistic counterparts. The first chapter serves as a general introduction. A few notions from Einstein's theory of gravity are introduced and a derivation of Hawking radiation is sketched. The correspondence with low-energy systems is then explained through three important examples. The next four chapters each details one of the works completed during this thesis, updated and slightly reorganized to account for new developments which occurred after their publication. The other articles I contributed to are summarized in the last chapter, before the general conclusion. 

My collaborators and I focused on three aspects of the behavior of fields close to the (analogue) event horizon in models with an effective low-energy Lorentz symmetry. The first one concerns nonlinear effects, which had been given little attention in view of their crucial importance for understanding the evolution in time of Hawking radiation as well as for experimental realizations. We showed in particular how they determine the late-time behavior in stable and unstable configurations. The second aspect concerns linear and quantum effects. We studied the Hawking radiation itself in several models and what replaces it when continuously erasing the horizon. We also characterized and classified the different types of linear instabilities which can occur. Finally, we contributed to the design and analysis of “analogue gravity” experiments in Bose-Einstein condensates, hydrodynamic flows, and acoustic setups, of which I report the main results.
\\ \\

\begin{otherlanguage}{french}
\begin{huge}\textbf{R\'esum\'e}\end{huge} \\

Cette th\`ese concerne l'\'etude des propri\'et\'es de champs scalaires classiques et quantiques en pr\'esence d'un environnement inhomog\`ene et/ou d\'ependant du temps. Nous nous concentrerons sur des mod\`eles pouvant être d\'ecrits, fondamentalement ou de mani\`ere effective, par un espace-temps courbe contenant un horizon des \'ev\'enements. Nous verrons en particulier comment une correspondance math\'ematique, provenant d'une sym\'etrie de Lorentz effective à basse \'energie, permet de relier les comportements des ondes dans un cadre non relativiste à la physique des trous noirs, quelles en sont les limites et dans quelle mesure les r\'esultats ainsi obtenus sont \og analogues \fg{} à leurs pendants gravitationnels. Apr\`es un premier chapitre d'introduction rappelant quelques bases de relativit\'e g\'en\'erale puis une d\'erivation de la radiation de Hawking et de la correspondance avec des syst\`emes non relativistes, je pr\'esenterai le d\'etail de quatre travaux effectu\'es durant ma th\`ese. Les autres articles \'ecrits dans ce cadre sont r\'esum\'es dans le dernier chapitre, pr\'ec\'edant une conclusion g\'en\'erale.

Mes collaborateurs et moi nous sommes concentr\'es sur trois aspects du comportement des champs pr\`es de l'analogue d'un horizon des \'ev\'enements dans des mod\`eles avec une sym\'etrie de Lorentz effective à basse \'energie. Le premier concerne les effets non lin\'eaires, cruciaux pour comprendre l'\'evolution de la radiation de Hawking ainsi que pour les r\'ealisations exp\'erimentales mais auparavant peu \'etudi\'es. Nous montrerons comment ceux-ci d\'eterminent les possibles comportements aux temps longs pour des syst\`emes stables ou instables. Le second aspect a trait aux effets lin\'eaires et quantiques, en particulier la radiation de Hawking elle-même, son devenir lorsque l'horizon est continûment effac\'e, ainsi que les diverses instabilit\'es à même de survenir dans diff\'erents mod\`eles. Enfin, nous avons particip\'e à l'\'elaboration, à l'analyse et à l'\'etude d’exp\'eriences dites de \og gravit\'e analogue \fg{} dans des condensats de Bose-Einstein et des syst\`emes hydrodynamiques ou acoustiques, dont je rapporte les principaux r\'esultats.

\end{otherlanguage}

\chapter*{Acknowledgements}

First and foremost I would like to warmly thank my advisor, Renaud Parentani, for his guidance, encouragements, and support during the years spent at LPT Orsay. 
His enthusiasm, foresight, broad culture, and endeavor to understand the physics at play beyond the mathematical description of physical systems made working with him extremely pleasant and rewarding. 
I will long remember the time spent discussing with him about topics as varied as the interpretation of quantum mechanics, current  trends in science, history, or topology changes in alternative models of gravity, his perception of links between seemingly disconnected question always providing deep and original insight into the underlying concepts. 

These years have been the opportunity to start fruitful collaborations, which was, each time, both a pleasure and a great opportunity to learn from my collaborators' experience and different points of view. 
I thank, in no particular order, Xavier Busch for having introduced me to some of the concepts of analogue gravity as well as the hurdles of a PhD, Antonin Coutant for interesting discussions on water waves and the structure of linear equations, Robin Zegers for his hindsight on mathematical aspects of our works, Scott Robertson for discussions about various topics on and around analogue gravity, Germain Rousseaux and L\'eo-Paul Euv\'e for their clear vision of the experimental aspects of water waves, Betti Hartmann for enlightening exchanges about cosmic strings, solitons, and science in general, as well as Tom Philbin, Yves Aur\'egan, Vincent Pagneux, Patrick Peter, Pierre Fromholz, and Jean-François Coupechoux. 
I hope working together was as rewarding for them as it has been for me, and that these collaborations will continue long after the end of my PhD.

I am very grateful to Theodore Jacobson and Daniele Faccio for accepting to be my thesis referees, as well as to the other members of the jury: Karim Noui, Denis Boiron, Iacopo Carusotto, and Ruth Gregory. 
I know that, working in an ever-evolving field, their time is precious, and I am honored that they accepted to devote some of it to my work. 

I also thank the other members of the cosmology group: Christos Charmousis, Alessandro Fabbri, Bartjan van Tent, Eugeny Babichev, Antoine Lehebel, and Gabriel Jung, with whom I had frequent discussions on cosmology, alternative theories of gravity, and links with other fields, and to whom I owe much of what I know about theoretical physics, as well as Vincent Rivasseau for his support and the enlightening exchanges we had about the links between mathematical physics and cosmology, and William Unruh, Silke Weinfurtner, and Yves Brihaye for interesting discussions. 

Last but not least, I am very grateful to S\'ebastien Descotes-Genon, head of the LPT, and to the administrative staff: Olivier Brand-Foissac, Mireille Calvet, Odile Heckenauer, Yvette Mamilonne, Philippe Molle, Marie-Agn\`es Poulet, and Jocelyne Raux, whose support, warmheartedness, and willingness to help greatly contributed to making my years here a very pleasant experience.

\hypersetup{linkcolor=black!0!blue}
\tableofcontents
\label{toc}
\hypersetup{linkcolor=blue}

\pagestyle{thefancy}

\newchapter{(Analogue) Gravity and Hawking Radiation}
\pagenumbering{arabic}
\label{chap:intro}
\begin{tikzpicture}[overlay]
\newcommand*{\xA}{-0.2}
\newcommand*{\xB}{14.5}
\newcommand*{\yA}{5.5}
\newcommand*{\yB}{1.5}
\newcommand*{\epsil}{0.75}
\draw[overlay] (\xA-\epsil,\yA) -- (\xB-\epsil,\yA);
\draw[overlay] (\xA,\yA+\epsil) -- (\xA,\yB);
\draw[overlay] (\xB,\yB-\epsil) -- (\xB,\yA);
\draw[overlay] (\xB+\epsil,\yB) -- (\xA+\epsil,\yB);
\end{tikzpicture}
\begin{flushright}
``In primisque hominis est propria veri inquisitio atque investigatio.'' \\
Marcus Tullius Cicero (-106 -- -43), \textit{De officiis, I, IV}
\end{flushright}
\newpage

\section{Introduction}

\subsection{A viewpoint on analogue gravity}

General relativity~\cite{Misner1973, wald, weinbergGR} and quantum mechanics~\cite{cohen1991quantum,bellac2011quantum,weinberg1996quantum1} are two of the most fascinating scientific developments of the last century. 
Beyond explaining previous observations, they extended the reach of our understanding to uncharted territories, from the length scales of elementary particles to those of galaxy superclusters and beyond, and predicting new phenomena like Bose-Einstein condensation and gravitational waves which have since then been observed and are in close agreement with theoretical predictions. 
It is, however, difficult to devise a self-consistent theoretical framework which would embrace them both. 
While a lot of effort has been devoted to finding a formulation of general relativity amenable to quantization, so far such quantum gravity theories remain elusive. 
It is thus useful to find setups where some concepts from general relativity and quantum mechanics can be studied without the main difficulties arising when trying to quantize gravity. 

To this end, one should first determine which features of these theories to include, which must be sufficiently related to the conceptual problems raised by quantum gravity to yield nontrivial results. 
On the ``gravity'' side, two ideas one would like to include are the description of the early universe and black holes. 
Indeed, both were fundamentally new aspects brought by general relativity without clear, well-defined counterparts in Newtonian physics\footnote{Although the notion of ``dark star'', not unrelated to black holes, was already proposed in 1783~\cite{Michell}.} and are expected to require quantum effects to be accurately described due to the high energies they involve. 
On the ``quantum'' side, one would like to include the particle pair production and entanglement, two distinctive, generic features from quantum field theories which are at the root of problems encountered when quantizing gravity. 
The ideal setup would thus be one where the main ingredients of a black hole or the early universe would be present alongside the possibility to produce pairs of entangled particles. 

Fortunately, such models exist -- and were studied, for some of them, long before general relativity and quantum mechanics. 
Indeed, as explained in W.~Unruh's seminal work~\cite{Unruh:1980cg}, some classical, non-relativistic fluid mechanics problems have an effective Lorentz invariance at low energies and show a precise correspondence with the behavior of fields in curved space-times. 
In particular, one can devise setups where they have the analogue of a horizon at the outer boundary of a black hole, or similarities with the post-inflation phase of the early universe. 
Although initially formulated in a classical hydrodynamic context, this analysis extends to the quantum realm, both formally at the level of the mathematical description and on concrete examples when applying it to quantum systems. 
It thus provides ``analogue'' models of gravity, where the behavior of (quantum) fields in curved space-times can be studied in a self-consistent mathematical framework and probed experimentally. 

Besides the aesthetic motivation of unifying some of the peculiar features of general relativity and quantum mechanics in a consistent -- and more general -- frame, I think the study of these models, often referred to as ``analogue gravity'', is interesting for three main reasons. 
The first one is, of course, the possibility to realize experimentally the analogue of a black hole, or other gravitational systems, where quantum effects such as Hawking radiation and reheating could be observed. 
The second reason is that the way divergences occurring in a gravitational context are regularized in ``analogue'' models can provide inspiration for aiming at theories of quantum gravity. 
As an example, the dispersion which regularizes the redshift of an incoming wave close to the horizon in these models bears many similarities with the higher-order terms of Ho\v{r}ava gravity and Einstein-\AE ther theory, both proposed as effective theories to include the low-energy corrections due to quantum effects, so that results drawn from condensed-matter ``analogues'' can be extended to these two theories or show possible paths to incorporate similar regularizations. 
The third reason is that analogue gravity sheds new light on the ``analogue'' models themselves, motivating to study them in new setups where the analogy with gravity is clearest and providing fresh understanding of their mathematical structure. 

\subsection{Aims of this thesis}

The main objective of this manuscript is to present the work done before and during my PhD at LPT Orsay, under the supervision of Renaud Parentani.  
To this end, after a brief description of Hawking radiation and analogue gravity, I show in the following chapters updated versions of four articles we published, partially rewritten and reorganized to clarify the aspects on which we have obtained a better understanding since the publication and include relevant material from later works. 
Although this introduces some redundancy, part of the information being repeated in more than one article, the overlap between two chapters is, I hope, small enough not to bore the reader. 
The motivation for including the full articles was to make each chapter essentially self-contained, so that they can be read independently. 
The choice of articles is of course not devoid of arbitrariness, although I believe it gives an accurate view of what has been done during the last three years. 
It was based on two criteria: relevance for the the global project of the thesis and my perception of my personal contribution. 
The other articles written during my PhD are briefly presented in Chapter~\ref{ch:concl}. 

The second objective is to show the links between the different works, as well as with recent advances in analogue gravity, more explicitly than the articles taken independently can do. 
To this end, each of the following four chapters begins by a short summary, akin to an extended abstract, explaining its relevance for the longer-term project of the thesis. 
In each chapter, a section “Additional remarks” contains information which is not required to understand the main idea but important for the technical aspects of the work. 
They are based on the articles' appendices as well as new, so far unpublished, material. 

This thesis was essentially devoted to nonlinear effects, needed to understand the formation and evolution of the flow itself, and linear quantum ones in flows with one or several analogue horizons. 
To be more specific, it can be divided into three main axes:
\begin{itemize}
\item Nonlinear dynamics of the background flows at the classical level;
\item Quantum effects and Hawking radiation in the presence of Lorentz-violating terms;
\item Design and/or interpretation of analogue gravity experiments. 
\end{itemize}
The works presented in the following chapters all deal with one or several of these aspects, to various degrees.  
Taken globally, they aim at providing a better understanding of black- and white-hole-like solutions and their perturbations in realistic models with broken Lorentz invariance and determine what can be learned from their theoretical study or experimental realization. 

\subsection{Notations}

Throughout this manuscript, I adopt mostly standard notations for mathematical symbols and expressions. 
For completeness, let me recall some of them which, sometimes used with different meanings in other contexts, might otherwise be a source of confusion: 
\begin{enumerate}
\item Quantifiers and logical connectives: 
\begin{itemize}
\item $\exists$ is the existential quantifier (``there exists''),
\item $\forall$ is the universal quantifier (``for all''),
\item If $P$ and $Q$ are two statements,
\begin{itemize}
\item $\neg P$ (negation of $P$) is true if and only if $P$ is false,
\item $P \wedge Q$ (``$P$ and $Q$'') is true if and only if $P$ and $Q$ are both true,
\item $P \vee Q$ (``$P$ or $Q$'') is true if and only if at least one of the statements $P$ and $Q$ is true,
\item $P \Rightarrow Q$ (``$P$ implies $Q$'') is true if and only if $Q$ is true or $P$ is wrong (equivalent to $\neg P \vee Q$),
\item $P \Leftrightarrow Q$ (``$P$ is equivalent to $Q$'') is true if and only if $P$ and $Q$ ate either both true or both false.
\end{itemize}
\end{itemize}
\item Sets of numbers: 
\begin{itemize}
\item The symbols $\mathbb{N}$, $\mathbb{Z}$, $\mathbb{Q}$, $\mathbb{R}$, and $\mathbb{C}$ denote, respectively, the sets of natural integers, relative integers, rational numbers, real numbers, and complex numbers.
\item A subscript ``$+$'' on  $\mathbb{Z}$, $\mathbb{Q}$, or $\mathbb{R}$ denotes their restrictions to positive numbers. For instance, $\mathbb{R}_+ = \left[ 0, \infty \right[$. 
\item A superscript ``$*$'' means that $0$ is not included. For instance, $\mathbb{R}_+^* = \left] 0, \infty \right[$ and $\mathbb{R^*} = \left] -\infty, 0 \right[ \cup \left] 0, + \infty \right[$.
\end{itemize}
\item \label{intr:pt2} Let $\lp n_1, n_2 \rp \in \mathbb{N}^{*2}$, $D_1 \subset \mathbb{C}^{n_1}$, and $D_2 \subset \mathbb{C}^{n_2}$. For any $n \in \mathbb{N}$, $C^n \lp D_1, D_2 \rp$ is the set of functions from $D_1$ to $D_2$ which are differentiable $n$ times and have continuous $n$th derivatives. The set of functions from $D_1$ to $D_2$ (without any regularity condition) is sometimes denoted as $D_2^{D_1}$. 
\item \label{intr:pt3} Landau notations: With the same definitions, let $x_1 \in D_1$ and $f$, $g$ two functions from $D_1$ to $D_2$, 
\begin{itemize}
\item $\displaystyle{f(x) \mathop{=}_{x \to x_1} O \lp g(x) \rp}$ is equivalent to
\begin{equation*}
\exists \, A > 0, \; \exists \, \eta > 0, \; \forall \, x \in D_1, \; \abs{x - x_1} < \eta \Rightarrow \abs{f(x)} \leq A \, \abs{g(x)},
\end{equation*}
\item $\displaystyle{f(x) \mathop{=}_{x \to x_1} o \lp g(x) \rp}$ is equivalent to
\begin{equation*}
\forall \, \ep > 0, \; \exists \, \eta > 0, \; \forall \, x \in D_1, \; \abs{x - x_1} < \eta \Rightarrow \abs{f(x)} \leq \ep \, \abs{g(x)},
\end{equation*}
\item $\displaystyle{f(x) \mathop{\sim}_{x \to x_1} g(x)}$ is equivalent to
\begin{equation*}
\forall \, \ep > 0, \; \exists \, \eta > 0, \; \forall \, x \in D_1, \; \abs{x - x_1} < \eta \Rightarrow \abs{f(x) - g(x)} \leq \ep \, \abs{g(x)}.
\end{equation*}
\end{itemize}
Similar definitions hold when choosing $x_1 = \pm \infty$, with ``$\abs{x - x_1} < \eta$'' replaced by ``$\pm x > \eta$''. 
\item Unless otherwise stated, $\theta$ denotes Heaviside's theta function defined as
\begin{equation*}
\theta: \lp 
\begin{aligned}
& \mathbb{R} \to \mathbb{R} \\
& x \mapsto \lb 
\begin{aligned}[2]
& 0 && x < 0 \\
& 1/2 && x = 0 \\
& 1 && x > 0
\end{aligned}
\right.
\end{aligned}
\rp . 
\end{equation*}
$\delta$ denotes Dirac's delta distribution, defined in $d$ dimensions by: \\ for all open domain $D$ of $\mathbb{R}^d$ such that $0 \in D$, for all $f \in C^0 \lp D, \mathbb{C} \rp$,
\begin{equation*}
\int_D \delta (x) \, f(x) \, \dd x^d = f(0).
\end{equation*}
\end{enumerate}
As is usual in the literature, the reduced Planck constant is denoted by $\hbar$ and the Boltzmann constant by $k_B$. 

I will also use the following conventions, maybe less standard:
\begin{itemize}
\item The complex number with unit modulus and phase $\pi / 2$ is denoted by ``$\ii$'' (instead of the usual ``$i$'').
\item The Euler constant is denoted by ``$\e$'' (instead of ``$e$'').
\item Integration measures are denoted by ``$\dd$'' (instead of ``$d$'').
\item The derivative of a function $f$ of the variable $x$ is denoted by ``$f'$'', ``$\pd_x f$'', or ``$\frac{\dd f}{\dd x}$'', the first one being used only if $f$ depends on only one variable. 
\item The symbol $\equiv$ will occasionally be used with the sense of ``defined as''. 
\item With the notations of points \ref{intr:pt2} and \ref{intr:pt3} above, the writing $\displaystyle{f(x)\mathop{\propto}_{x \to x_1} g(x)}$ (with the condition $x \to x_1$ implicit when no confusion is possible) will be used with the meaning of
\begin{equation*}
f(x)\mathop{=}_{x \to x_1} O \lp g(x) \rp \, \wedge \, g(x)\mathop{=}_{x \to x_1} O \lp f(x) \rp.
\end{equation*}
If no point $x_1$ is specified, the notation $f \propto g$ will be used for
\begin{equation*}
\exists \, a \in \mathbb{C}^*, \; f = a \s g.
\end{equation*}
\end{itemize}
The use of roman-style letters for $\ii$, $\e$, and $\dd$ was chosen to avoid possible ambiguities when $i$, $e$, and/or $d$ also represent variables or indices. 
Unless otherwise specified, vectors are denoted by an overarrow. 
To shorten the mathematical expressions, functions will occasionally be written instead of their values at a given point, or conversely, when no confusion is possible. 
For instance, if $f$ and $g$ are two functions from a set $I_1$ to a set $I_2$, and if $x$ is unspecified, the writing
\begin{align*}
f(x) = g(x)
\end{align*}
is understood as~\footnote{Although in this particular case one could save a few characters while remaining rigorous by writing $f = g$.}
\begin{align*}
\forall \, x \in I_1, \; f(x) = g(x). 
\end{align*}
Two functions representing the same physical quantity after a change of variable will also occasionally be denoted by the same symbol when I feel that a more rigorous notation would not be worth introducing a new one and might obscure the reasoning rather than clarifying it. 
Similarly, tensors and coordinates will often be referred to by their components with unspecified indices. 
For instance, if $T_{\mu \nu}$ and $V_\mu$ are the components, respectively, of covariant tensors of rank $2$ and $1$ in a $4$-dimensional spacetime, the expression
\begin{align*}
T_{\mu \nu} \lp x^\lambda \rp = V_\mu \lp x^\lambda \rp \, V_\nu \lp x^\lambda \rp
\end{align*} 
with unspecified $\mu$, $\nu$, $\lambda$, and $\lp x^\eta \rp_{\eta \in [\![ 0, 3 ] \! ]} \in \mathbb{R}^4$ stands for
\begin{align*}
\forall \, \lp \mu, \nu \rp \in [\![ 0, 3 ] \! ]^2, \; \forall \, \lp x^0, x^1, x^2, x^3 \rp \in \mathbb{R}^4, \hspace{4 cm} \\
T_{\mu \nu} \lp x^0, x^1, x^2, x^3 \rp = V_\mu \lp x^0, x^1, x^2, x^3  \rp \, V_\nu \lp x^0, x^1, x^2, x^3  \rp. 
\end{align*}

\section{Hawking radiation in a nutshell}
\label{sec:HR}

The discovery of S.~Hawking~\cite{Hawking:1974sw} (see also the work of W.~Unruh~\cite{Unruh:1976db}) that black holes can amplify vacuum fluctuations to trigger a thermal emission of particles, the so-called Hawking radiation, was an important step forward in the quest to the still elusive theory of quantum gravity. 
Indeed, it is one of the few known results where general relativity and quantum mechanics both play a fundamental role: the former describes the event horizon where the amplification mechanism takes place while the latter provides the vacuum fluctuations which are turned to real particles. 
This mechanism thus sheds light on the behavior of quantum fields in a curved space-time and provides fundamental questions to be addressed by models aimed at quantizing gravity. 

In this section I present some of the main concepts and ideas leading to Hawking radiation. 
I first recall a few elements of general relativity and properties of its black hole solutions in four dimensions. 
I then sketch a derivation of Hawking radiation and mention the (currently unsolved) issues it raises. 
One idea to keep in mind is that, in spite of the technicalities involved by the use of both general relativity and quantum field theory, the root of the Hawking effect is a simple, classical, wave amplification. 
In a sense, the interior of the black hole acts as an anomalous region containing negative-energy waves, so that the positive-energy ones outside the black hole can be amplified while conserving the total energy. 
The role of quantum mechanics is to provide irreducible fluctuations which source the amplified waves. 

\subsection{The Schwarzschild black hole}

In this subsection I introduce the elements needed to obtain the simplest black hole solutions of general relativity and the extension of bosonic field equations to curved space times.  
My aim is to give the main ideas and technical tools while avoiding the use of abstract concepts as far as possible:  
I shall sketch how to obtain solutions and determine some of their properties, but leave aside the question of their interpretation and the deeper mathematical structure of the theory. 
The interested reader will find in the classical textbooks~\cite{Misner1973, weinbergGR, largescale, landau1975classical, wald} a more detailed and rigorous discussion of these aspects as well as many more advanced topics. 

In all this subsection, $x^\mu$ denotes the space-time coordinates, with $\mu$ (or any other Greek index) implicitly running from $0$ to $d$, where $d$ is the number of space dimensions. $\mu = 0$ denotes the time coordinate, and $\mu > 0$ denotes the spatial ones.  
A Latin index, on the other hand, will run from $1$ to $d$, and thus always denotes a space coordinate. 
I shall also not distinguish explicitly vectors, matrices, or higher-order tensors from their components when no confusion seems reasonably possible. 
So, for instance, $g_{\mu \nu}$ will refer to the metric matrix as well as its components in a particular coordinate system.  
Finally, I adopt Einstein's summation convention that repeated lower and upper indices are summed over. For instance, if $T$ is a tensor of rank 2,
\begin{align}
T_\mu \phantom{} ^\mu \equiv \sum_{\mu = 0}^{d} T_\mu \phantom{}  ^\mu \; \text{and} \;  T_i \phantom{}  ^i \equiv \sum_{i = 1}^{d} T_i \, ^i.
\end{align}

\subsubsection{The Einstein equation}

General relativity is based on two fundamental principles. The first one is an extension of the principle of special relativity, which I now recall. 
It states that the fundamental laws of physics must be invariant under a change of reference frame provided the initial and final ones do not accelerate with respect to each other, i.e., under any Lorentz transformation sending one inertial frame into another one. 
These transformations form a noncompact Lie group, the Poincaré group, consisting of 
\begin{itemize}
\item translations (in space and time) $x^\mu \to x^\mu + t^\mu$, where $x^\mu$ denotes the coordinates and $t^\mu$ is a constant vector,
\item time-reversal $T: \, x^0 \to - x^0$ and space reversal $P: \, x^i \to - x^i$ for $i \neq 0$,
\item Lorentz transformations $x^\mu \to \Lambda^\mu_{\, \nu} x^\nu$, where $\Lambda$ belongs to the restricted Lorentz group ${\rm SO^+ (1,d)}$ in $d$ space dimensions. The latter consists of all matrices $\Lambda$ with a unit determinant, preserving the space orientation, and such that $\Lambda^\mu_{\,\alpha} \Lambda^\nu_{\,\beta} \eta^{\alpha \beta} = \eta^{\mu \nu}$, where $\eta^{\mu \nu} \equiv {\rm diag}(c^2, -1, -1, ..., -1)$ and $c$ denotes the celerity of light.
\end{itemize} 
In general relativity, this invariance is extended to all smooth changes of coordinates $x^\mu \to x'^\mu (x^\nu)$. 

The second principle is that space-time behaves like an ``elastic'' structure, deformed by the matter content of the universe. 
This deformation gives rise to gravitational forces through the peculiar coupling between matter and geometry needed to enforce the invariance under changes of coordinates. 
More precisely, space-time is seen as a differentiable Riemanian manifold, that is, roughly speaking, an infinite set of points endowed with a notion of continuity (a topology) and differentiability up to some order $n \in \mathbb{N}^* \cup \left\lbrace \infty \right\rbrace$, as well as a metric. 
The manifold is covered by open sets with a smooth mapping into open sets of $\mathbb{R}^{d+1}$, which defines a coordinate system. 
Smooth coordinate transformations (called diffeomorphisms) are those which preserve continuity and differentiability up to the order $n$. 

The metric $g_{\mu \nu}$, which is a symmetric matrix, gives a notion of distance between two infinitesimally close points on the manifold. It defines the line element $\dd s^2$ between two points whose coordinates differ by $\dd x^\mu$ through
\begin{align}\label{eq:intr:ds}
\dd s^2 = g_{\mu \nu} \, \dd x^\mu \, \dd x^\nu.
\end{align}
From this expression, one sees that $\dd s^2$ will be invariant under a smooth change of coordinates $x^\mu \to x'^\mu$ if and only if the metric transforms as
\begin{align} \label{eq:intro_trans_g}
g'_{\mu \nu} \lp x'^\lambda \rp = \frac{\partial x^\alpha}{\partial x'^\mu} \frac{\partial x^\beta}{\partial x'^\nu} g_{\alpha \beta} \lp x^\lambda \rp, 
\end{align}
where a prime denotes a quantity evaluated in the new coordinate system. 
We denote by $g^{\mu \nu}$ the inverse matrix of $g_{\mu \nu}$. 
By a direct calculation using that $\frac{\partial x^\mu}{\partial x'^\nu}$ is the inverse of $\frac{\partial x'^\mu}{\partial x'^\nu}$ in the matrix sense, one finds that $g^{\mu \nu}$ transforms as
\begin{align} \label{eq:intro_trans_g_inv}
g'^{\mu \nu}\lp x'^\lambda \rp = \frac{\partial x'^\mu}{\partial x^\alpha} \frac{\partial x'^\nu}{\partial x^\beta} g^{\alpha \beta}\lp x^\lambda \rp. 
\end{align}
This shows that $g_{\mu \nu}$ is a tensor of rank $(0,2)$ and $g^{\mu \nu}$ a tensor of rank $(2,0)$. 
More generally, a tensor $T$ of rank $(n,p)$ is an $(n+p)$-dimensional array of real numbers which transforms as
\begin{align} \label{eq:intr_T}
T'_{\rho'_1...\rho'_n} \phantom{}^{\nu'_1...\nu'_p} \lp x'^\mu \rp = 
	\lp \prod_{j=1}^n \frac{\pd x^{\rho_j}}{\pd x'^{\rho'_j}} \rp \lp \prod_{j=1}^p \frac{\pd x'^{\nu'_j}}{\pd x^{\nu_j}} \rp T_{\rho_1...\rho_n} \phantom{}^{\nu_1...\nu_p} \lp x^\mu \rp.
\end{align}
Although this definition does not show the (deep) links with the mathematical properties of the underlying manifold (see for instance~\cite{wald} for a more geometrical one), it will be enough for our purposes. 
From \eqref{eq:intr_T}, one can easily show that the multiplication of two tensors of ranks $(n_1, p_1)$ and $(n_2, p_2)$ gives a tensor of rank $(n_1+n_2, p_1+p_2)$ and that a contraction of a tensor or rank $(n,p)$, obtained by equating a lower index with an upper one and summing over their common value, is a tensor of rank $(n-1,p-1)$. 
The indices of a tensor can be raised and lowered by contracting them with $g^{\mu \nu}$ and $g_{\mu \nu}$, respectively. 

Another ingredient we shall need is the notion of covariant derivative. 
As we did for tensors, we here introduce it in a pedestrian way, focusing on its usefulness for practical calculations rather than its geometric interpretation. 
It can be easily seen from \eqref{eq:intr_T} that the ordinary derivative of a tensor, $\partial_{x^\mu} T_{\rho_1...\rho_n} \,^{\nu_1...\nu_p}$, is not a tensor unless $n = p = 0$. 
Indeed, the derivative $\partial_{x'^\mu}$ acting on the prefactor on the right-hand side gives additional terms in
\begin{align}
\frac{\partial x^{\rho_j}}{\partial x'^\mu \, \partial x'^{\rho'_j}}
\; \; {\rm and} \; \; 
\frac{\partial}{\partial x'^\mu} \frac{\partial x'^{\nu'_j}}{\partial x^{\nu_j}} . 
\end{align}
One can circumvent this by defining the covariant derivative $\nabla_\mu$ as
\begin{align}
\nabla_\mu T_{\rho_1...\rho_n} \phantom{}^{\nu_1...\nu_p} \equiv 
	\pd_\mu T_{\rho_1...\rho_n} \phantom{}^{\nu_1...\nu_p} 
	- \sum_{j=1}^n \Gamma_{\mu {\rho_j}}^{\lambda} T_{\rho_1...\lambda...\rho_n} \phantom{}^{\nu_1...\nu_p}
	+ \sum_{j=1}^p \Gamma_{\mu \lambda}^{\nu_j} T_{\rho_1...\rho_n} \phantom{}^{\nu_1...\lambda...\nu_p}
\end{align}
(where $\lambda$ is at the position $j$ in the second term and $n+j$ in the third one)  
provided each additional term cancels the corresponding term where $\pd_\mu$ acts on the matrix of the change of variable.
We also impose that $\nabla_\mu = \pd_\mu$ if $x^{\mu}$ is a locally Minkowskian coordinate system in which $g_{\mu \nu} = \eta_{\mu \nu}$ and $\pd_\rho g_{\mu \nu} = 0$ at the point $x_\textrm{M}^\alpha$ we are considering.~\footnote{Such a coordinate system can always be found. One simple proof is to first note that, given an arbitrary coordinate system $y^\mu$, $g_{\mu \nu}(y^\rho)$ is a symmetric real matrix, and thus diagonalizable by an orthogonal matrix, and has a signature $(+1, -1,...,-1)$. One can thus always find a real matrix $\lp M^\mu_\alpha \rp$ such that $M^\mu_\alpha M^\nu_\beta g_{\mu \nu} = \eta_{\alpha \beta}$. ($\lp M^\mu_\alpha \rp$ can be chosen as an orthogonal matrix diagonalizing $(g_{\mu \nu})$ multiplied on the left by a diagonal matrix whose diagonal components are the inverse square roots of the absolute values of the eigenvalues of $g$.)
Choosing $\pd y^{\mu} / \pd x^{\alpha} = M^\mu_\alpha$ at $x^\alpha = x_\textrm{M}^\alpha$ thus gives $g'_{\alpha \beta} = \eta_{\alpha \beta}$ at this point, where $g'_{\mu \nu}$ denotes the metric in the coordinate system $x^\alpha$. 
The first derivatives of $g'$ can then be canceled by noting that
\begin{equation*}
\pd_\alpha g'_{\beta \gamma} + \pd_\beta g'_{\gamma \alpha} - \pd_\gamma g'_{\beta \alpha} = 
	2 \frac{\pd^2 y^\mu}{\pd x^\alpha \pd x^\beta} \frac{\pd y^\nu}{\pd x^\gamma} g_{\mu \nu} + A_{\alpha \beta \gamma},
\end{equation*}
where $A_{\alpha \beta \gamma}$ is symmetric in $(\alpha, \beta)$ and involves only the first derivatives of $y^\mu$ and $g_{\mu \nu}$. 
We thus have $\pd_\alpha g'_{\beta \gamma} + \pd_\beta g'_{\gamma \beta} - \pd_\gamma g'_{\beta \alpha} = 0$, and thus $\pd_\alpha g'_{\beta \gamma} = 0$ after symmetrizing over $(\beta, \gamma)$, at the point $x = x_M$ provided 
\begin{equation*}
\frac{\pd^2 y^\mu}{\pd x^\alpha \pd x^\beta} = - \frac{1}{2} A_{\alpha \beta \gamma} \frac{\pd y^\gamma}{\pd x^\nu} g^{\mu \nu} 
\end{equation*} 
at $x = x_M$, which can always be satisfied as both sides of this equation are symmetric in $(\alpha, \beta)$. (Indeed, if $f_\nu^\mu$ denotes the wanted first derivatives if $y^\mu$ and $h^\mu_{\nu \rho}$ the wanted second derivatives, since $h^\mu_{\nu \rho}$ is symmetric in $(\nu,\rho)$ one can choose $y^\mu = f^\mu_\nu \s (x^\nu - x_M^\nu) + h^\mu_{\nu \rho} \s (x^\nu - x_M^\nu) \s (x^\rho - x_M^\rho) / 2$.)} 
Let $y^{\mu}$ be an arbitrary coordinate system, we need
\begin{align} \label{intr:cond_Gamma}
\left\lbrace
\begin{aligned}
\Gamma_{\mu \lambda}^\nu \frac{\pd y^\lambda}{\pd x^\eta} & = - \pd_{y^\mu} \lp \frac{\pd y^\nu}{\pd x^\eta} \rp \\
\Gamma_{\mu \rho}^\lambda \frac{\pd x^\eta}{\pd y^\lambda} & = \pd_{y^\mu} \lp \frac{\pd x^\eta}{\pd y^\rho} \rp
\end{aligned}
\right. ,
\end{align}
where the coefficients $\Gamma^\alpha_{\mu \beta}$ are evaluated in the coordinate system $y^\nu$. 
(By definition, they locally vanish in the coordinate system $x^\nu$.)
The first line is equivalent to
\begin{align}
\Gamma_{\mu \lambda}^\nu \frac{\pd y^\lambda}{\pd x^\eta} = \frac{\pd y^\nu}{\pd x^\alpha} \pd_{y^\mu} \lp \frac{\pd x^\alpha}{\pd y^\beta} \rp \frac{\pd y^\beta}{\pd x^\eta}
\end{align}
\begin{align}
\Gamma_{\mu \beta}^\nu = \frac{\pd y^\nu}{\pd x^\alpha} \pd_{y^\mu} \lp \frac{\pd x^\alpha}{\pd y^\beta} \rp 
\end{align}
\begin{align}
\Gamma_{\mu \rho}^\lambda \frac{\pd x^\eta}{\pd y^\lambda} = \pd_{y^\mu} \lp \frac{\pd x^\eta}{\pd y^\rho} \rp .
\end{align}
The two lines of \eq{intr:cond_Gamma} are thus equivalent and give
\begin{align}
\Gamma_{\mu \nu}^\rho = \frac{\pd y^\rho}{\pd x^\eta} \frac{\pd^2 x^\eta}{\pd y^\mu \pd y^\nu}.
\end{align}
To make link with $g_{\mu \nu}$, notice that from \eq{eq:intro_trans_g},
\begin{align}
g_{\mu \nu} = \frac{\pd x^\alpha}{\pd y^\mu} \frac{\pd x^\beta}{\pd y^\nu} \eta_{\alpha \beta}.
\end{align}
So,
\begin{align}
\pd_\eta g_{\mu \nu}  = \pd_{y^\eta} g_{\mu \nu} = \frac{\pd x^\alpha}{\pd y^\mu} \frac{\pd^2 x^\beta}{\pd y^\nu \pd y^\eta} \eta_{\alpha \beta}
	+ \frac{\pd^2 x^\alpha}{\pd y^\mu \pd y^\eta} \frac{\pd x^\beta}{\pd y^\nu} \eta_{\alpha \beta}
\end{align}
and
\begin{align}
\pd_\mu g_{\eta \nu} + \pd_\nu g_{\mu \eta} - \pd_\eta g_{\mu \nu} = 
	2 \frac{\pd^2 x^\alpha}{\pd y^\mu \pd y^\nu} \frac{\pd x^\beta}{\pd y^\eta} \eta_{\alpha \beta}.
\end{align}
Using that the inverse matrix of $g_{\mu \nu}$ is, from \eq{eq:intro_trans_g_inv},
\begin{align}
g^{\lambda \eta} = \frac{\pd y^\lambda}{\pd x^\alpha} \frac{\pd y^\eta}{\pd x^\beta} \eta^{\alpha \beta},
\end{align}
where $\eta^{\alpha \beta}$ denotes the inverse matrix of $\eta_{\alpha \beta}$, we find
\begin{align}
\frac{1}{2} g^{\rho \eta} \lp \pd_\mu g_{\eta \nu} + \pd_\nu g_{\mu \eta} - \pd_\eta g_{\mu \nu} \rp = 
	\frac{\pd y^\rho}{\pd x^\eta} \frac{\pd^2 x^\eta}{\pd y^\mu \pd y^\nu}.
\end{align}
We thus obtain the relation
\begin{align}
\Gamma_{\mu \nu}^\rho = \frac{1}{2} g^{\rho \eta} \lp \pd_\mu g_{\eta \nu} + \pd_\nu g_{\mu \eta} - \pd_\eta g_{\mu \nu} \rp.
\end{align}
The coefficients $\Gamma_{\mu \nu}^\rho$ are called \textit{Christoffel symbols}. 
From this expression, one easily obtains the expression of the covariant derivative of the metric tensor:
\begin{align}
\nabla_\lambda g_{\mu \nu} &= \pd_\lambda g_{\mu \nu} - \Gamma_{\lambda \mu}^\eta g_{\eta \nu} - \Gamma_{\lambda \nu}^\eta g_{\mu \eta} \nn
& =  \pd_\lambda g_{\mu \nu} - \frac{1}{2} \lp \pd_\lambda g_{\nu \mu} + \pd_\mu g_{\nu \lambda} - \pd_\nu g_{\lambda \mu} \rp - \frac{1}{2} \lp \pd_\lambda g_{\nu \mu} + \pd_\nu g_{\mu \lambda} - \pd_\mu g_{\lambda \nu} \rp \nn
& = 0. 
\end{align}

The covariant derivative may be used to write equations of motion or Lagrangian densities in a form invariant under diffeomorphisms. For instance, the Klein-Gordon equation for a free massive vector field $V^\mu$, which reads in Minkowski space
\begin{align}
\pd_\mu \pd^\mu V^\nu + m^2 V^\nu = 0,
\end{align}
is not invariant under a change of coordinates -- unless the latter belongs to the Poincaré group. 
But it can easily be made invariant by writing it as
\begin{align}\label{eq:intr_V}
\nabla_\mu \nabla^\mu V^\nu + m^2 V^\nu = 0,
\end{align} 
which coincides with the above expression in a locally Minkowski frame where $g_{\mu \nu} = \eta_{\mu \nu}, \, \pd_\eta g_{\mu \nu} = 0$, and is now invariant under any smooth coordinate transformation. 
It can be easily derived from a least-action principle with the action
\begin{align}\label{eq:intr_S}
S_V = \int \dd^{d+1}x \, \sqrt{\abs{g}} \, \lp \lp \nabla_\mu V_\nu \rp \lp \nabla^\mu V^\nu \rp - m^2 V_\nu V^\nu \rp,
\end{align}
where $g$ denotes the determinant of the metric tensor $g_{\mu \nu}$. 
This action is also invariant under a change of coordinates, as the factor ${\rm det} \lp \frac{\partial x}{\partial x'} \rp$ from the variation of $\sqrt{\abs{g}}$ exactly cancels the Jacobian of the transformation. 
In fact, the invariance of \eqref{eq:intr_V} can be seen as a consequence of that of the action \eqref{eq:intr_S}.

Notice that, contrary to the usual derivatives, the covariant ones generally do not commute when applied to a tensor of nonvanishing rank. 
This allows to define the Riemann tensor $R^\mu \phantom{}_{\nu \rho \sigma}$ by
\begin{align}
\nabla_\rho \nabla_\sigma V^\mu - \nabla_\sigma \nabla_\rho V^\mu = R^\mu \phantom{}_{\nu \rho \sigma} V^\nu
\end{align}
for any vector field $V^\mu$. 
Using the explicit expressions of the covariant derivative, one obtains 
\begin{align}
\nabla_\rho \nabla_\sigma V^\mu &= \pd_\rho \nabla_\sigma V^\mu - \Gamma_{\rho \sigma}^\eta \nabla_\eta V^\mu + \Gamma^\mu_{\rho \eta} \nabla_\sigma V^\eta \nn 
&= \pd_\rho \pd_\sigma V^\mu + \pd_\rho \lp \Gamma_{\sigma \lambda}^\mu V^\lambda \rp
- \Gamma_{\rho \sigma}^\eta \pd_\eta V^\mu - \Gamma_{\rho \sigma}^\eta \Gamma_{\eta \lambda}^\mu V^\lambda
+ \Gamma^\mu_{\rho \eta} \pd_\sigma V^\eta + \Gamma^\mu_{\rho \eta} \Gamma^\eta_{\sigma \lambda} V^\lambda \nn
&= \pd_\rho \Gamma_{\sigma \lambda}^\mu V^\lambda 
+ \Gamma^\mu_{\rho \eta} \Gamma^\eta_{\sigma \lambda} V^\lambda+...,
\end{align}
where the other terms are symmetric under the exchange $\rho \leftrightarrow \sigma$. 
Since all terms involving derivatives of $V^\mu$ are symmetric, the Riemann tensor is well defined (in that the above definition is independent of $V^\mu$), and equal to
\begin{align}
R^\mu \phantom{}_{\nu \rho \sigma} = \pd_\rho \Gamma_{\sigma \nu}^\mu - \pd_\sigma \Gamma_{\rho \nu}^\mu
+ \Gamma^\mu_{\rho \eta} \Gamma^\eta_{\sigma \nu} - \Gamma^\mu_{\sigma \eta} \Gamma^\eta_{\rho \nu}. 
\end{align}
Informally speaking, the Riemann tensor gives a measure of the local curvature of space-time. 
One defines from it the Ricci tensor and Ricci scalar by successive contractions: $R_{\mu \nu} \equiv R^\lambda \phantom{}_{\mu \lambda \nu}$ and $R \equiv R^\mu \phantom{}_\mu$. 

We now have all the ingredients to determine the behavior of bosonic fields in a curved space-time. What remains to be discussed is the dynamical part of the theory, namely how these matter fields affect the metric, and thus generate the gravitational force. 
This is given by a least-action principle with the Einstein-Hilbert action:~\footnote{This expression assumes a vanishing cosmological constant $\Lambda = 0$. A non-vanishing one can be included by replacing $R$ with $R - 2 \Lambda$. \label{foo:Lambda}}
\begin{align}\label{eq:SEH}
S_{\rm E-H} = \int \lp \frac{-c^4}{16 \pi G} R + \mathcal{L}_M \rp \sqrt{\abs{g}} \, d^{d+1}x,
\end{align}
where $c$ is the celerity of light, $G$ is Newton's constant, and $\mathcal{L}_M$ is the matter Lagrangian density. 
The latter defines the energy-momentum tensor:
\begin{align}
T_{\mu \nu} \equiv \frac{2}{\sqrt{\abs{g}}} \frac{\delta \lp \sqrt{\abs{g}} \mathcal{L}_M \rp}{\delta g^{\mu \nu}}
 = 2 \frac{\delta \mathcal{L}_M}{\delta g^{\mu \nu}} - g_{\mu \nu} \mathcal{L}_M. 
\end{align}
A long but straightforward calculation shows that the variation of the first term in \eqref{eq:SEH} gives the Einstein tensor $G_{\mu \nu}$:
\begin{align}
\frac{1}{\sqrt{\abs{g}}}\frac{\delta \lp \sqrt{\abs{g}} R \rp}{\delta g^{\mu \nu}} = R_{\mu \nu} - \frac{1}{2} R \, g_{\mu \nu} \equiv G_{\mu \nu}. 
\end{align}
(The second term can be easily obtained by noting that
\begin{align}
\frac{\delta \sqrt{\abs{g}}}{\delta g^{\mu \nu}} = \frac{-1}{2 \sqrt{\abs{g}}} \frac{\delta g}{\delta g^{\mu \nu}} = \frac{1}{2 \sqrt{\abs{g}}} g_{\mu \rho} g_{\nu \sigma} \frac{\delta g}{\delta g_{\rho \sigma}}
= \frac{1}{2 \sqrt{\abs{g}}} g_{\mu \rho} g_{\nu \sigma} g^{\rho \sigma} g
 = - \frac{\sqrt{\abs{g}}}{2} g_{\mu \nu}. 
\end{align}
The first one can be determined by computing explicitly the variations of the Christoffel coefficients.) 
The least-action principle thus gives the Einstein equation:
\begin{align}\label{eq:intr_E}
R_{\mu \nu} - \frac{1}{2} R g_{\mu \nu} = \frac{8 \pi G}{c^4} T_{\mu \nu}. 
\end{align}
Including a cosmological constant $\Lambda$ (see footnote~\ref{foo:Lambda}), it becomes
\begin{align}\label{eq:intr_E2}
R_{\mu \nu} - \frac{1}{2} R g_{\mu \nu} + \Lambda g_{\mu \nu} = \frac{8 \pi G}{c^4} T_{\mu \nu}. 
\end{align}
Equation \eqref{eq:intr_E} or \eqref{eq:intr_E2} determines how ``matter deforms space-time'', i.e., how the energy-momentum of the matter field affects the metric $g_{\mu \nu}$ at the source of the gravitational forces.
Interestingly, in 4 space-time dimensions, the Einstein equations are the only field equations deriving from a Lagrangian density built using $g_{\mu \nu}$ and its two first derivatives to be of order two~\cite{Lovelock1969}. (However, in higher dimension other choices are possible.)

\subsubsection{The Schwarzschild solution}

Let us now look for simple vacuum solutions (i.e. with $T_{\mu \nu} = 0$) in 3+1 dimensions, with a vanishing cosmological constant $\Lambda = 0$. 
It is easily seen that any uniform metric with $\pd_\rho g_{\mu \nu} = 0$ is a solution: then all Christoffel coefficients, and thus the Riemann and Einstein tensors, identically vanish. 
Imposing that the signature of $g_{\mu \nu}$ be $(+1, -1, -1, -1)$, i.e., with one time dimension and 3 space dimensions, there exists a global change of coordinates sending $g_{\mu \nu}$ to $\eta_{\mu \nu}$. 
The line element is then that of special relativity:
\begin{align}
\dd s^2 = c^2 \, \dd t^2 - \dd x^2 - \dd y^2 - \dd z^2:
\end{align}
the space-time is flat and there is no gravitational force. 
To obtain the simplest solutions with gravitational force, let us make the three following assumptions:~\footnote{In fact, the Jebsen-Birkhoff theorem states that the first and third hypotheses are consequences of the second one.}
\begin{itemize}
\item The metric is {\it static}, i.e., there exists a coordinate system in which it is independent of time and irrotational.~\footnote{More precisely, a solution is said static if it has an irrotational, asymptotically timelike Killing vector.} 
It thus takes the form
\begin{align}
\dd s^2 = g_{0 0} \lp x^i \rp \dd t^2 - g_{i j} (x^k) \dd x^i \, \dd x^j.
\end{align}
\item It is spherically symmetric. This means that its isometry group contains ${\rm SO(3)}$ as a subgroup, with spacelike two-surfaces as the group orbits. By a convenient choice of coordinates, the metric can then be written in the form
\begin{align} \label{eq:intr:anszm}
\dd s^2 = g_{0 0} (r) \dd t^2 - g_{r r}(r) \dd r^2 - r^2 \lp \dd  \theta^2 + \sin^2 \theta \, \dd  \varphi^2 \rp,
\end{align}
where $\theta \in \left[ 0, \pi \right[$ and $\varphi \in \left[ 0, 2 \pi \right[$. 
\item It is asymptotically flat, i.e., the metric for $r \to \infty$ must describe a Minkowski space. Up to a rescaling of the coordinate $t$, this is equivalent to $g_{00}(r) \to c^2$ and $g_{rr} \to 1$. 
\end{itemize}
Plugging the ansatz \eq{eq:intr:anszm} into the Einstein equations \eqref{eq:intr_E} and imposing the asymptotic flatness condition, one finds the one-dimensional space of solutions:
\begin{align} \label{eq:Scharzschild}
\dd s^2 = c^2 \lp 1 - \frac{r_S}{r} \rp \dd t^2 - \frac{\dd r^2}{1 - \frac{r_S}{r}} - r^2 \lp \dd  \theta^2 + \sin^2 \theta \, \dd  \varphi^2 \rp,
\end{align}
where $r_S$ is the Schwarzschild radius, given by $r_S = 2 G M / c^2$. 
The metric~\eqref{eq:Scharzschild} is called the Schwarzschild metric. 
The parameter $M$ can be interpreted as the mass of the solution, as in the limit $r \gg r_S$ its gravitational field becomes identical to the Newtonian field of a spherically-symmetric body of total mass $M$. 

\begin{figure}
\centering
\def\svgwidth{0.49 \linewidth}
{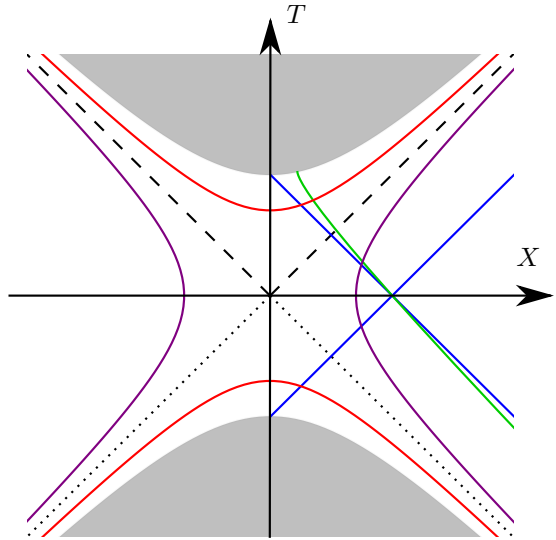}
\caption{Kruskal diagram of the Schwarzschild solution. 
The grey domains correspond to $c^2 T^2 > X^2 + r_S^2$. 
Their boundaries are the black hole (up) and white hole (bottom) singularities. 
The two black, oblique lines give the black hole horizon (dashed) and the white hole horizon (dotted). 
The region described by the Schwarzshild coordinates is the right quadrant $X > c \abs{T}$. 
Red and purple lines are loci of uniform $r$, inside (red) or outside (purple) the horizons. 
The green curve is the trajectory of a massive, freely-falling particle, which is at rest at $r \to - \infty$ for $t \to - \infty$. 
The blue lines show two light rays. 
One of them was emitted from the white hole singularity at $T = - r_S, \, X = 0$, and goes to $r \to \infty$ as $t \to + \infty$. 
The other one comes from infinity $r \to \infty$ at $t \to - \infty$ and reaches the black hole singularity at $T = r_S, \, X = 0$. 
} \label{fig:intro_kruskal}
\end{figure}

The metric \eq{eq:Scharzschild} has two singularities at $r = 0$ and $r = r_S$. 
One can show that the former is a true, geometrical singularity in that the space-time curvature is divergent.~\footnote{For instance, the Kretschmann invariant $R_{\mu \nu \rho \sigma}R^{\mu \nu \rho \sigma}$ diverges like $r^{-6}$ in this limit.} 
However, the surface $r = r_S$ is not: although the coefficients of the metric tensor, expressed in this particular coordinate system, are divergent, the geometry remains perfectly regular. 
To see this, let us consider the trajectories of radial light rays. 
The latter follow null characteristics, obtained by solving $\dd s^2 = 0$. 
In the radial case, this becomes
\begin{align}
\dd t = \pm \frac{\dd r}{c \lp 1 - \frac{r_S}{r} \rp}. 
\end{align}
Integrating over $r$ gives
\begin{align}\label{eq:intre:geo1}
c \, t = \pm \lp r + r_S \ln \lp \frac{r}{r_S} - 1 \rp \rp + c \, t_0,
\end{align}
where $t_0$ is an integration constant. 
Said otherwise, the quantity
\begin{align}
c\, t \mp \lp r + r_S \ln \lp \frac{r}{r_S} - 1 \rp \rp
\end{align}
is a constant along the trajectory. The upper sign corresponds to outgoing light rays, and the lower sign to incoming ones. 
Taking the exponential of these constants, one defines the light-like coordinates (which will also play an important role in the derivation of Hawking radiation)
\begin{align}
U \equiv  \lp \frac{r}{r_S} - 1 \rp^{1/2} \e^{(r - c \, t) / (2 r_S)}
\end{align} 
and
\begin{align}
V  \equiv \lp \frac{r}{r_S} - 1 \rp^{1/2} \e^{(r + c \, t) / (2 r_S)}.
\end{align}
Then, $U$ is constant along outgoing light-rays, while $V$ is constant along incoming ones. 
Their variations are given by
\begin{align} \label{eq:intro_dU}
dU  = \lp \frac{r}{r_S} - 1 \rp^{-1/2} \e^{(r - c \, t)/(2r_S)} \frac{r \, dr}{2r_S^2} - \lp \frac{r}{r_S} - 1 \rp^{1/2} \e^{(r - c \, t)/(2r_S)} \frac{c \, dt}{2 r_S} 
\end{align}
and 
\begin{align} \label{eq:intro_dV}
dV  = \lp \frac{r}{r_S} - 1 \rp^{-1/2} \e^{(r + c \, t)/(2 r_S)} \frac{r \, dr}{2r_S^2} + \lp \frac{r}{r_S} - 1 \rp^{1/2} \e^{(r + c \, t)/(2r_S)} \frac{c \, dt}{2 r_S}. 
\end{align}
Using \eqs{eq:intro_dU}{eq:intro_dV}, one readily obtains 
\begin{align}
\dd s^2 = - 4 \frac{r_S^3}{r} \e^{- r / r_S} \dd U  \, \dd V  - r^2 \lp \dd  \theta^2 + \sin^2 \theta \, \dd  \varphi^2 \rp. 
\end{align}
This may be written in a slightly more transparent form using the Kruskal–Szekeres coordinates $T \equiv r_S \lp V  - U  \rp/(2c)$ and $X \equiv r_S \lp V + U \rp/2$. We obtain
\begin{align}\label{eq:Scw_K}
\dd s^2 = 4 \frac{r_S}{r} \e^{- r / r_S} \lp c^2 \dd T^2 - \dd X^2 \rp - r^2 \lp \dd  \theta^2 + \sin^2 \theta \, \dd  \varphi^2 \rp, 
\end{align}
where 
\begin{align}
T = \frac{r_S}{c} \lp \frac{r}{r_S} - 1 \rp^{1/2} \e^{r/(2 r_S)} \sinh \lp \frac{c \, t}{2 r_S} \rp
\end{align}
and
\begin{align}
X = r_S \lp \frac{r}{r_S} - 1 \rp^{1/2} \e^{r/(2 r_S)} \cosh \lp \frac{c \, t}{2 r_S} \rp. 
\end{align}
This construction works only for $r > r_S$, so that $X$ and $T$ are in the quadrant $X \geq c \abs{T}$. 
However, in this region $r$ is an analytical function of $X$ and $T$ since
\begin{align}\label{eq:defr}
\frac{X^2}{r_S^2} - \frac{c^2 T^2}{r_S^2} = \lp \frac{r}{r_S} - 1 \rp \e^{r / r_S}. 
\end{align}
One can use this relation to define $r$ for arbitrary real values of $X$ and $T$ such that $X^2 - c^2 T^2 > - r_S^2$, giving the maximal extension of the Schwarzshild space-time, with the metric given by \eq{eq:Scw_K}, containing a black hole and a white hole (see Fig.~\ref{fig:intro_kruskal}). 
In particular, it extends beyond $r = r_S$, which gives the locus $X = \pm c T$, up to the singularities $X = \pm \sqrt{c^2 T^2 - r_S^2}$. 
In this coordinate system, the metric is perfectly regular at $r = r_S$, showing that the geometry also is. 
A Kruskal-Szekeres diagram showing the geometry of the solution and a few geodesics is shown in \fig{fig:intro_kruskal}. 

Although the geometry is regular at $r = r_S$, this surface has peculiar properties. 
First, particles following time-like or light-like geodesics~\footnote{I remind that a geodesic between two points in spacetime is a path between them extremizing the quantity $\int \dd s$. They are given by \eq{eq:intr:geo}. A geodesic is said timelike if $\dd s^2 > 0$ along it, spacelike if $\dd s^2 > 0$, and null if $\dd s^2 = 0$.} can cross it in one direction only: particles following them can enter a black hole but never escape from inside it; conversely, they can escape from inside a white hole but never enter it. 
Let us show this for the null radial ones. As these are the geodesics with highest radial velocity (it can be easily shown that adding a transverse velocity and/or a mass term reduces the radial speed), the result will extend to all time-like or null geodesics. 
We first notice that the locus $r = r_S$, corresponding to $X = \pm c \, T$, is itself formed by two light-like geodesics. 
Let us concentrate on the line $X = c \, T$, $X > 0$, which is the black hole horizon described by the Schwarzschild metric~\eq{eq:Scharzschild}.~\footnote{The locus $X = -c \, T$, $X > 0$, corresponding to a white hole horizon in the extended metric, will play no role in the following where a collapsing object forming the black hole is included. } 
Around this region, the geodesics with constant $X + c \, T$ are incoming. Indeed, as $t$ increases, $X / (c \, T)$ increases too. Since $X / (c \, T) = X / (X + c \, T - X)$ is an increasing function of $X$ at fixed $X + c \, T$, we deduce that $X$ increases, and thus that $T$ decreases. 
From the expression of $T$, one finds that $r$ decreases. So $r$ decreases with $t$ and the light ray falls inside the black hole. 
The outgoing geodesics instead correspond to constant $X - c \, T$. As said above, the horizon $r = r_S$ is precisely one of them. 
The upshot of this is that the black hole horizon separates outgoing radial light rays into two classes: those which are inside (respectively outside) at one time remain inside (respectively outside) at all times. 
So, anything occurring inside the surface $r = r_S$ can not have any causal effect on the outside world (assuming no information propagates faster than the speed of light). 
Such a surface is called an \textit{event horizon}.

Another property of the horizon is the redshift experienced by outgoing light rays in its vicinity. 
To see this, let us consider a freely-falling massive particle which emits an electromagnetic plane wave with a fixed frequency in its reference frame, and compute the frequency received by an observer at infinity. 
The trajectory of the particle is a geodesic, given by a least-action principle with Lagrangian
\begin{align} 
L = \frac{1}{2} g_{\mu \nu} (x^\lambda) \, \frac{\dd x^\mu}{\dd  \tau} \frac{\dd x^\nu}{\dd  \tau},
\end{align}
where $\tau$ is its proper time. 
Since the Lagrangian does not depend explicitly on $\tau$, the corresponding Hamiltonian $H$ is conserved. 
It is easily shown that in the present case $H = L$. So, $L$ is a constant of motion. 
For a massless system, it is equal to zero and we recover the equation $\dd s^2 = 0$. 
For a massive one, by convention $L = c^2 / 2$.~\footnote{The only constraint is that $L$ must be strictly positive to describe a massive particle. Its value then depends on the normalization of the  parameter $\tau$. With the choice $L = c^2 / 2$, $\tau$ coincides with the proper time of the particle, since $\dd s^2 = c^2 \, d \tau^2$ along its trajectory.}
The equations of motion read
\begin{align}
\frac{\dd }{\dd  \tau} \lp g_{\mu \nu} \frac{\dd  x^\nu}{\dd  \tau} \rp = \frac{1}{2} \pd_\mu g_{\rho \sigma} \frac{\dd x^\rho}{\dd  \tau} \frac{\dd x^\sigma}{\dd  \tau},
\end{align}
which may be rewritten as
\begin{align} \label{eq:intr:geo}
\frac{\dd^2 x^\mu}{\dd \tau^2} = - \Gamma_{\rho \sigma}^\mu \frac{\dd x^\rho}{\dd \tau} \frac{\dd x^\sigma}{\dd \tau}.
\end{align}
Let us consider the radial component $\mu = r$. 
A straightforward calculation shows that the only non-vanishing relevant Christoffel coefficients are
\begin{align}
\Gamma_{r r}^r = - \frac{1}{2} \lp 1 - \frac{r_S}{r} \rp^{-1} \frac{r_S}{r^2} 
\; \text{and} \; 
\Gamma_{t t}^r = \frac{c^2}{2} \lp 1 - \frac{r_S}{r} \rp \frac{r_S}{r^2}. 
\end{align} 
We thus obtain
\begin{align}\label{eq:intr:int1}
\frac{\dd ^2 r}{\dd  \tau^2} = \frac{1}{2} \lp 1 - \frac{r_S}{r} \rp^{-1} \frac{r_S}{r^2} \lp \frac{\dd r}{\dd \tau} \rp^2
 - \frac{c^2}{2} \lp 1 - \frac{r_S}{r} \rp \frac{r_S}{r^2} \lp \frac{\dd t}{\dd \tau} \rp^2. 
\end{align}
Moreover, the condition $L = c^2/2$ gives
\begin{align} \label{eq:intr:int2}
\lp 1 - \frac{r_S}{r} \rp c^2 \lp \frac{\dd t}{\dd \tau} \rp^2 - \lp 1 - \frac{r_S}{r} \rp^{-1} \lp \frac{\dd r}{\dd \tau} \rp^2 = c^2. 
\end{align}
Plugging this into \eq{eq:intr:int1}, we obtain
\begin{align} \label{eq:intro:d2rdt2au}
\frac{\dd^2 r}{\dd \tau^2} = - \frac{c^2}{2} \frac{r_S}{r^2}. 
\end{align}
This can be easily integrated after multiplication by $\frac{\dd r}{\dd \tau}$. 
Assuming for simplicity that the infalling particle is at rest at infinity for $\tau \to -\infty$, we obtain
\begin{align}\label{eq:intr:drdtau}
\frac{\dd r}{\dd \tau} = - c \sqrt{\frac{r_S}{r}}. 
\end{align}
Using \eq{eq:intr:int2} and \eq{eq:intr:drdtau} gives
\begin{align}\label{eq:intr:dtdtau}
\frac{\dd t}{\dd \tau} = \lp 1 - \frac{r_S}{r} \rp^{-1}. 
\end{align}
Combining \eqref{eq:intro:d2rdt2au} and \eqref{eq:intr:dtdtau}, we obtain
\begin{align}
\frac{\dd r}{\dd t} = - c \sqrt{\frac{r_S}{r}} \lp 1 - \frac{r_S}{r} \rp. 
\end{align}
To go further, it is useful to define the auxiliary variable $x \equiv \sqrt{r / r_S}$. 
Then,
\begin{align}
\frac{c \, t}{r_S} = - 2 \int \frac{x^2}{1 - \frac{1}{x^2}} dx = -2 \int \frac{x^4}{x^2 - 1} dx 
= - 2 \int \lp x^2 + 1 + \frac{1}{x^2 - 1} \rp dx .
\end{align}
This integral can be done straightforwardly, giving (up to an integration constant which we set to $0$)
\begin{align} \label{eq:intr_exactr}
\frac{r^{3/2}}{3 r_S^{3/2}} + \sqrt{\frac{r}{r_S}} + \frac{1}{2} \ln \lp \frac{\sqrt{r/r_S} - 1}{\sqrt{r/r_S} + 1} \rp
 = - \frac{c \, t}{2 r_S}. 
\end{align}
Close to the horizon, for $r \to r_S^+$, this simplifies as
\begin{align}
\frac{r}{r_S} - 1 \propto \exp \lp - \frac{c \, t}{r_S} \rp. 
\end{align}

To see the physical implications of this behavior, let us consider an external, stationary observer at $r \gg r_S$, aligned with the enter of the black hole and the infalling particle. 
We assume the latter emits some electromagnetic radiation with a fixed angular frequency $\om_e$ (in its own reference frame). 
That is, the proper time $d \tau$ between two instants where the wave has a given phase is equal to $2 \pi / \om_e$. 
From \eqref{eq:intr:dtdtau}, the corresponding Schwarzschild time interval $dt_e$ is larger by a factor $\lp 1 - r_S / r_e \rp^{-1}$, where $r_e$ is the value of the $r$ coordinate at the emission (assuming $r / r_S$ does not change significantly during this interval~\footnote{
This approximation is valid provided $\om_e \gg \abs{dr/d\tau}/r_S$. 
It thus holds from $r = \infty$ to $r = r_S$ iff $\om_e \gg c/r_S$.}).  
During the time interval $dt_e$, the infalling particle has moved by $d r_e$. 
The time interval $dt_r$ of reception is thus
\begin{align}
dt_r &= dt_e  - \lp \frac{\dd t}{\dd r} \rp_{\text{outgoing light ray}} dr_e
= dt_e \lp 1 + \frac{1}{c \lp 1 - \frac{r_S}{r_e} \rp} \, c \, \sqrt{\frac{r_S}{r}} \lp 1 - \frac{r_S}{r_e} \rp  \rp \nn
&= \lp 1 - \sqrt{\frac{r_S}{r}} \rp^{-1} d \tau_e .
\end{align}
So, the external observer will receive it with an angular frequency $\om_r \approx \om_e \lp 1 - \sqrt{r_S / r_e} \rp < \om_e$. 
This is the redshift effect. 

\begin{figure}
\centering
\includegraphics[width=0.49 \linewidth]{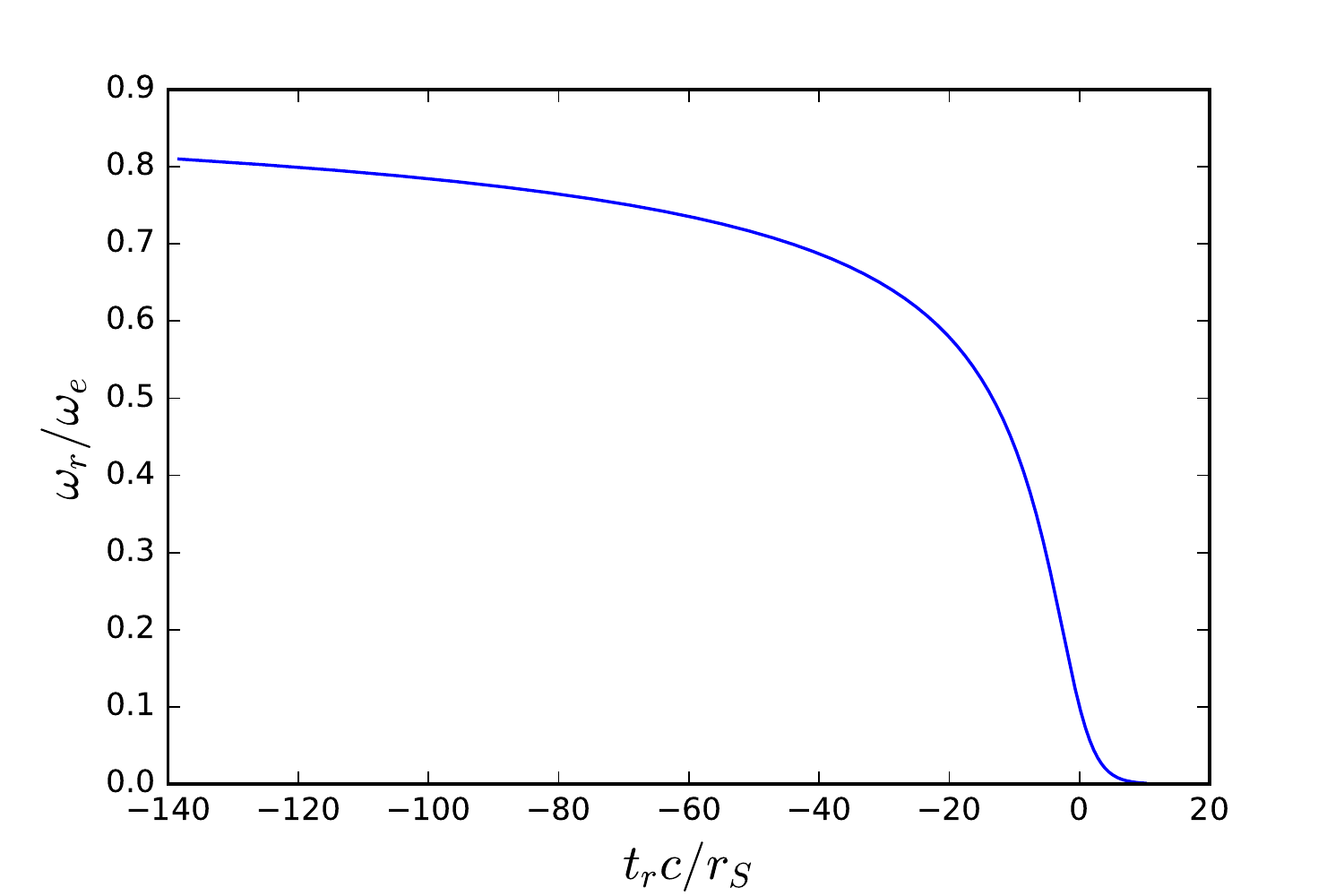}
\includegraphics[width=0.49 \linewidth]{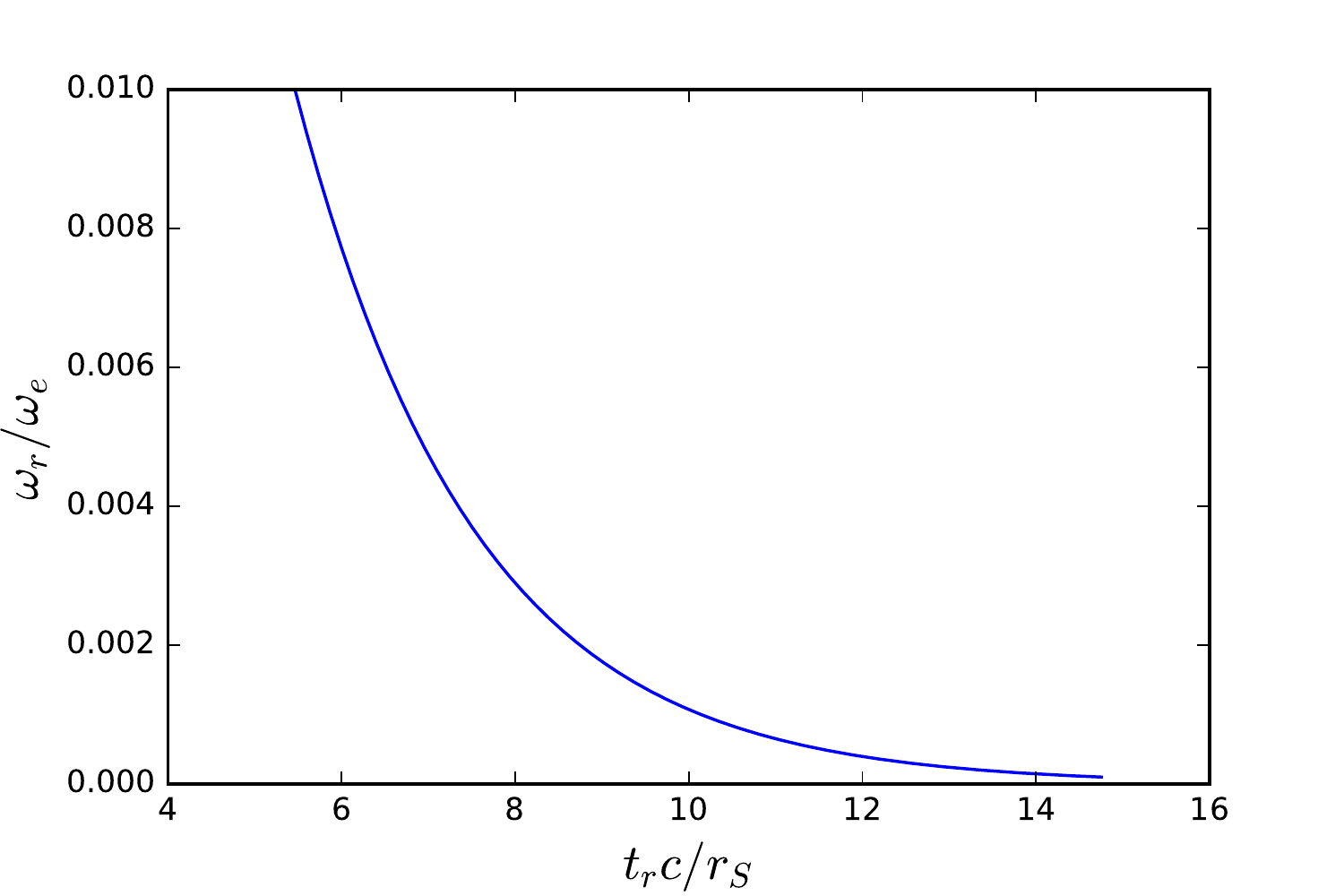}
\caption{Ratio of the received and emitted frequencies for an electromagnetic wave sent by a massive object initially at rest at infinity, falling radially into a Schwarzschild black hole, given by \eqref{eq:intro:int3} and \eqref{eq:intr:dtdtau}. The integration constant is set to $t_{r,0} = 0$.} \label{fig:redshift}
\end{figure}

To relate this to how the observer will perceive the particle falling into the black hole, let us compute the ratio $\om_r / \om_e$ as a function of time. 
The wave emitted at time $t_e$ was sent when the infalling object was at $r = r_e$ given by \eqref{eq:intr_exactr}. 
Integrating the outgoing null geodesic from this point, one obtains that the time it takes to reach the external observer is
\begin{align}
\Delta t = \frac{r_0 - r_e}{c} + \frac{r_S}{c} \ln \lp \frac{r_0 - r_S}{r_e - r_S} \rp. 
\end{align}
The relation between the reception time $t_r$ and the point of emission is thus
\begin{align}\label{eq:intro:int3}
\frac{c \, \lp t_r - t_{r,0} \rp}{r_S} = - 2 \ln \lp \sqrt{\frac{r_e}{r_S}} - 1 \rp 
- \frac{2}{3} \lp \frac{r_e}{r_S}  \rp^{3/2}
- \frac{r_e}{r_S} 
- 2 \lp \frac{r_e}{r_S}  \rp^{1/2},
\end{align}
where $t_{r,0}$ is a constant. 
One can then determine the relation between $t_r$ and the received frequency using that $r_e / r_S = \lp 1 - \om_r / \om_e \rp^{-2}$. 
This relation is shown in \fig{fig:redshift}. 
At early time, the redshift effect is small since gravitational forces are weak an the emitter has a slow velocity in the frame of the receiver. 
However, when the former approaches the black hole, gravitational effects strongly reduce the frequency of photons reaching the observer.

We have thus seen that although the Schwarzschild geometry is regular at $r = r_S$, this surface plays a crucial role in the propagation of outgoing light rays, which can not cross it from inside to outside and are strongly redshifted when escaping to infinity from a region close to it. 
As we shall see below, these properties are at the root of the Hawking effect. 
Importantly, they still appear (with some modifications due to dispersion) in the ``analogue'' models discussed in Section~\ref{sec:intr:AG}, triggering a similar wave amplification.

\subsubsection{Other black hole solutions}

Before moving on to the behavior of matter fields in black hole space-times and the Hawking effect, let us briefly discuss the other black hole solutions. 
First, a generalization of the Schwarzschild metric can be obtained by relaxing the assumptions of staticity and spherical symmetry, assuming instead that the metric be only {\it stationary}. 
This means that it is independent on time but not necessarily irrotational.~\footnote{A metric is said stationary if it has a global, asymptotically timelike Killing vector.} 
One then finds that the general solution, the Kerr metric has one more parameter: the angular momentum $J$. 
An introduction to the Kerr metric can be found in~\cite{Visser:2007fj}. 
Using the same notations as in that reference, it may be written in Boyer-Lindquist coordinates as
\begin{align}
\dd s^2 =& - \lp 1 - \frac{r_S \, r}{r^2 + a^2 \cos^2 \theta} \rp \dd t^2 - \frac{2 r_S \, r \, a \, \sin^2 \theta}{r^2 + a^2 \cos^2 \theta} \dd t \, \dd \varphi + \frac{r^2 + a^2 \cos^2 \theta}{r^2 - r_S \, r + a^2} \dd r^2 \nn
 & + \lp r^2 + a^2 \cos^2 \theta \rp \dd \theta^2 + \lp r^2 + a^2 + \frac{r_S \, r \, a^2 \sin^2 \theta}{r^2 + a^2 \cos^2 \theta} \rp \sin^2 \theta \, \dd \varphi^2,
\end{align}
where $a = J / (M c)$ is a radius associated with the angular momentum of the black hole. 
This metric describes an actual black hole solution, with all curvature singularities located inside an event horizon, provided $a^2 \leq r_S^2 / 4$. 
Notice that the Schwarzschild metric~\eqref{eq:Scharzschild} is recovered in the limit $a \to 0$. 
The Kerr solution has two disconnected event horizons: an outer one at $r = r_S/2 + \sqrt{r_S^2/4 - a^2}$ and an inner one at $r = r_S/2 - \sqrt{r_S^2/4 - a^2}$. As in the case $r \to r_S$ for the Schwarzschild solution, the geometry remains regular at these surfaces although the behavior of geodesics is strongly affected. 
There is also a curvature singularity ar $r = 0, \, \theta = \pi / 2$, whose geometry may be though of as a ring. 

A new concept, which is not present in static solutions, is that of ergoregion. 
The latter may be defined as the region ``above'' the event horizon, i.e., from which it is possible to escape, but inside which nothing can remain stationary from the point of view of an observer at infinity: the rotating motion of the black hole drags the spacetime around it, so that an object sufficiently close to the horizon must rotate in the same direction. 
Interestingly, it is possible to extract energy from this region, by a mechanism not unrelated to the Hawking effect discussed below. 
By definition, one boundary of the ergoregion is the event horizon. The other boundary is called an ergosphere. 
A Kerr black hole has two ergoregions, and thus two ergospheres: one outside the outer horizon, at $r = r_S/2 + \sqrt{r_S^2/4 - a^2 \cos^2 \theta}$, and one inside the inner horizon, at $r = r_S/2 - \sqrt{r_S^2/4 - a^2 \cos^2 \theta}$. 

New solutions can be found when adding an electromagnetic field. 
Looking for static, spherically symmetric, asymptotically flat solutions~\footnote{Once again, by the Jebsen-Birkhoff theorem the second assumption implies the first and third ones.}, the general solution is the Reissner-Nørdström metric
\begin{align}
\dd s^2 = \lp 1 - \frac{r_S}{r} + \frac{r_Q^2}{r^2} \rp c^2 \dd t^2 - \lp 1 - \frac{r_S}{r} + \frac{r_Q^2}{r^2} \rp^{-1} \dd r^2 - r^2 \lp \dd \theta^2 + \sin^2 \theta \, \dd \varphi^2 \rp,
\end{align}
where 
\begin{align}
r_Q \equiv \sqrt{\frac{Q^2 G}{4 \pi \epsilon_0 c^4}}
\end{align}
and $Q$ is the charge of the black hole. Here we denote by $\epsilon_0$ the vacuum permittivity. 
This solution supports a static electric field with vector potential $A_\mu = \lp Q/r, 0, 0, 0 \rp$. 
It has a point-like curvature singularity at $r=0$, shielded by event horizons if $r_Q^2 \leq r_S^2 / 4$. 
When this inequality is satisfied, there are two event horizons at $r = \lp r_S \pm \sqrt{r_S^2 - 4 r_Q^2} \rp / 2$. 

Finally, when including an electric charge and rotation, one obtains the Kerr-Newman metric
\begin{align}
\dd s^2 = \lp c \, \dd t - a \, \sin^2 \theta \, \dd \varphi \rp^2 \frac{\Delta}{\rho^2} - \lp \frac{\dd r^2}{\Delta} + \dd \theta^2 \rp \rho^2 - \lp (r^2 + a^2) \dd \varphi - a \, c \, \dd t \rp^2 \frac{\sin^2 \theta}{\rho^2},
\end{align}
where $\Delta = r^2 - r_S r + a^2 + r_Q^2$. 
It corresponds to an actual black hole provided $a^2 + r_Q^2 \leq r_S^2/4$. 
When restricting to Einstein-Maxwell theory, this is the most general asymptotically flat black hole metric in (3+1) dimensions~\cite{Chrusciel:2012jk} up to coordinate transformations. 
However, other solutions can be found in higher dimensions or when including more exotic matter fields. 
For instance, in (4+1) dimensions, one finds (unstable~\cite{Gregory93}) ``black string'' solutions whose horizons have a cylinder topology. 
Even in (3+1) dimensions, ``hairy'' black hole solutions with additional features have been found, for instance when including a scalar field~\cite{Zloshchastiev:2004ny}. 
The study of these solutions and their stability is a rich and fascinating topic, still very active from both analytical and numerical perspectives. 
However, as far as the Hawking radiation is concerned the essential point is the presence of an event horizon. 
The other features of black holes will introduce technical complications but will not affect the main results. 
For this reason, in the following we shall focus on the simplest black hole solution, namely the Schwarzschild one~\eqref{eq:Scharzschild}. 

\subsection{Real scalar field in a collapsing geometry and Hawking effect}

We now consider a real, massless scalar field $\phi$, with the action
\begin{align}\label{eq:action_phi}
S_\phi = \int \dd^4x \, \sqrt{\abs{g}} \, \lp \pd_\mu \phi \rp \lp \pd^\mu \phi \rp,
\end{align}
in a space-time with a Schwarzschild black hole centered at the origin $r = 0$. 
Notice that $S_\phi$ is manifestly invariant under changes of coordinates. 
The corresponding Euler-Lagrange equation is
\begin{align} \label{eq:intro_phi}
\pd_\mu \lp \sqrt{\abs{g}} g^{\mu \nu} \pd_\nu \phi \rp = 0,
\end{align}
which may be rewritten as
\begin{align}
\nabla_\mu \nabla^\mu \phi = 0. 
\end{align}
Our aim is to solve this equation in the presence of a Schwarzschild black hole and then quantize the field to exhibit the mechanism of Hawking radiation. 
Our presentation closely follows reference~\cite{Brout:1995rd}, in which the interested reader more details and a deeper discussion of the results. 

\subsubsection{Collapsing mass shell}

\begin{figure}
\centering
\def\svgwidth{0.49 \linewidth}
{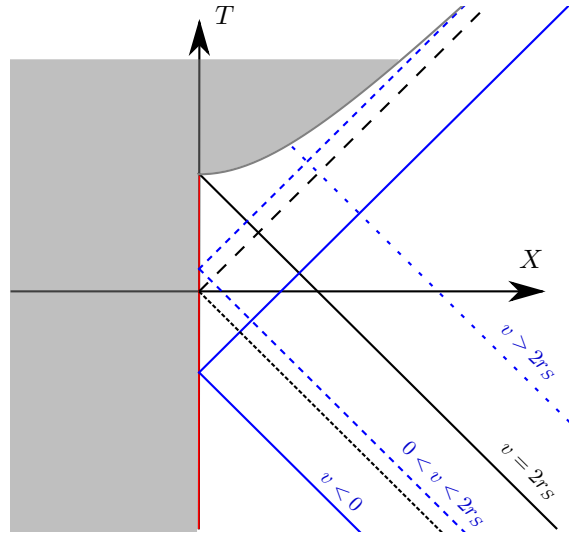}
\begin{tikzpicture}[overlay]
\put(-30,30){\color[rgb]{0,0,0}\rotatebox{-45}{\scalebox{0.8}{\makebox(0,0)[lb]{$v = 2 r_S$}}}}
\put(-30,75){\color[rgb]{0,0,0.8}\rotatebox{-45}{\scalebox{0.8}{\makebox(0,0)[lb]{$v > 2 r_S$}}}}
\put(-67,30){\color[rgb]{0,0,0.8}\rotatebox{-45}{\scalebox{0.8}{\makebox(0,0)[lb]{$0 < v < 2 r_S$}}}}
\put(-97,20){\color[rgb]{0,0,0.8}\rotatebox{-45}{\scalebox{0.8}{\makebox(0,0)[lb]{$v < 0$}}}}
\end{tikzpicture}
\caption{Collapsing-shell geometry. The coordinates $X$ and $T$ are the Kruskal–Szekeres ones for the Schwarzschild metric outside the shell, i.e., $v > 2 r_S$, and Minkowski coordinates inside. 
The grey area is the domain (formally) beyond $r = 0$. 
Its boundary for $v > 2 r_S$, materialized by a darker grey line, is the curvature singularity $r = 0$ of the Schwarzshild metric. 
The oblique, black, continuous line shows the mass shell trajectory. 
The dashed one shows the event horizon, and the dotted one shows the line $v = 0$ separating the incoming geodesics that will escape the black hole (for $v < 0$) or remain trapped (for $v > 0$). 
The vertical red line is the (regular) center of the shell $r = 0$. 
Blue lines show null characteristics (light rays). 
The continuous one, incoming with $v < 0$, is reflected at $r = 0$ without crossing the event horizon. 
It then escapes to infinity. 
The dashed one is incoming with $0 < v < 2 r_S$. It is initially inside the mass shell, but reaches $r = 0$ only after crossing the event horizon. 
It thus reaches the singularity after exiting the mass shell. 
The dotted line shows an incoming light ray with $v > 2 r_S$, falling directly into the black hole. 
} \label{fig:intro_kruskal2}
\end{figure}

To avoid the complications due to the presence of the white hole in the maximal extension of the Schwarzschild solution, we shall work with a collapsing geometry. 
Specifically, we consider a lightlike, spherical, collapsing matter shell. 
The metric outside the shell is the Schwarzschild one, while the inside metric is Minkowski:
\begin{equation}
\dd s^2 = 
\left\lbrace
\begin{alignedat}{2}
& \lp 1 - \frac{r_S}{r} \rp c^2 \dd t^2 - \lp 1 - \frac{r_S}{r} \rp^{-1} \dd r^2 - r^2 \lp \dd \theta^2 + \sin^2 \theta \, \dd \varphi^2 \rp && \; \; \; \text{outside} \\
& \, c^2 \dd t^2 - \dd r^2 - r^2 \lp \dd \theta^2 + \sin^2 \theta \, \dd \varphi^2 \rp && \; \; \; \text{inside}
\end{alignedat}
\right. .
\end{equation}
The geometry is shown in \fig{fig:intro_kruskal2}. 
We also assume the mass shell interacts with $\phi$ only through it gravitational field.~\footnote{As we shall see, the relevant modes must have a very large frequency when crossing the mass shell. It is thus realistic to assume the latter is transparent for these modes.}

\begin{figure}
\centering
\includegraphics[width=0.5 \linewidth]{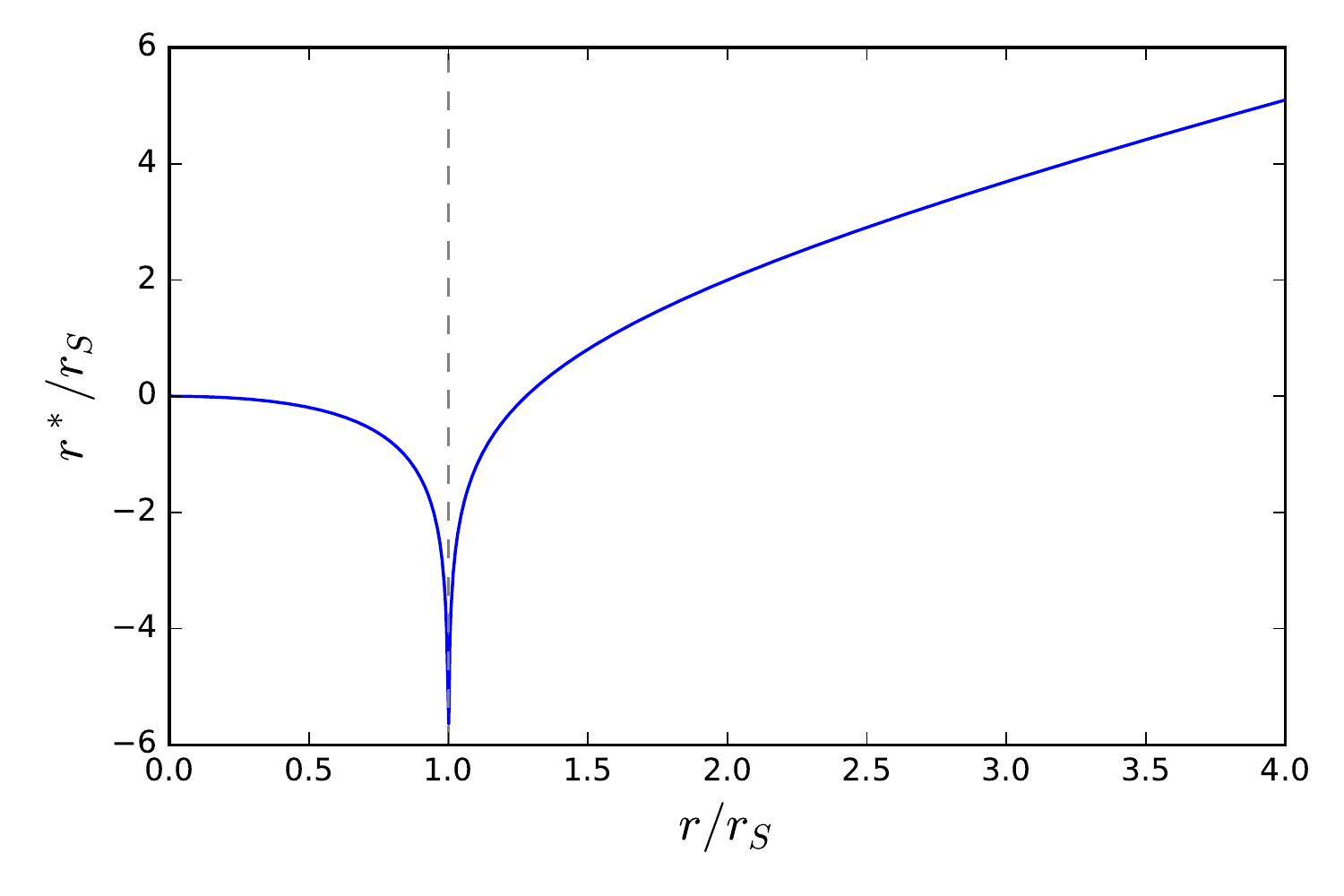}
\caption{The tortoise coordinate $r^*$ as a function of $r$, in units of the Schwarzschild radius $r_S$.
The vertical dashed line shows the event horizon $r = r_S$.} \label{fig:intro_rstar}
\end{figure}

In the following, it will be useful to work with the coordinates $u$ and $v$ constructed as follows. 
We first define the ``tortoise'' coordinate~\cite{Misner1973}~\footnote{This name comes from one of the Greek philosopher Zeno's ``paradoxes'', referred to as ``Achilles and the tortoise''. 
It involves an imaginary race between the hero Achilles and a tortoise, the former supposedly running faster than the latter. 
If the tortoise is allowed a head start, Zeno claims, Achilles can never overtake it.
Indeed, to do so he would first need to reach the point where the tortoise was when he began running. 
During that time, the tortoise has advanced by some distance, which Achilles will then have to cover, and so on. 
This and other paradoxes were used by Zeno to support Parmenides' doctrine that change, and motion in particular, is an illusion. 
They are among the first known examples of tentative proofs by contradictions, whith the aim ``to show that their hypothesis that existences are many, if properly followed up, leads to still more absurd results than the hypothesis that they are one.''~\cite{Parmenides}} 

\begin{align}
r^* \equiv r + r_S \ln \abs{\frac{r}{r_S} - 1}. 
\end{align}
The behavior of $r^*$ with $r$ is shown in \fig{fig:intro_rstar}. 
It goes to infinity linearly in the limit $r \to \infty$ and to $- \infty$ for $r \to r_S$. 
The light-like coordinates $u$ and $v$ are then defined by
\begin{align}
u \equiv c \, t - r^*, \, v \equiv c \, t + r^*. 
\end{align}
They are related to the Kruskal-Szekeres coordinates through
\begin{align} \label{eq:intr:def_U1}
U = \e^{-u / (2 r_S)}, \, V = \e^{v / (2 r_S)}
\end{align}
in the right quadrant $X > \abs{T}$. 
Notice that, contrary to $U$ and $V$, $u$ and $v$ do not cover the maximal extension of the Schwarzshild space-time. 
Indeed, $u$ (respectively $v$) only cover the part $U > 0$ (respectively $V > 0$). 
When including the mass shell, $v$ now covers the whole space-time as the part $V < 0$ is removed. 
However, $u$ only covers the region outside the event horizon. 
We thus define another lightlike coordinate $u_i$ for $r < r_S$ by
\begin{align} \label{eq:intr:def_U2}
U = - \e^{u_i / (2 r_S)}. 
\end{align}
The change of sign in the exponent ensures that $U$ is a monotonously decreasing function of both $u$ and $u_i$. 

The mass shell trajectory is taken to be $v = 2 r_S$. 
It intersects the singularity at $r = 0$, $c \, t = 2 r_S$. 
The event horizon appears at the center $r = 0$ at $c \, t = 0$ and grows linearly, following $r = c \, t$, until it reaches the mass shell at $r = c \, t = r_S$. 
As shown in \fig{fig:intro_kruskal2}, incoming, radial, null geodesics have three qualitatively different behaviors depending on the value of $v$. 
If $v < 0$, the lightray reaches $r = 0$ at the Minkowski time $t = v/ c$ prior to the formation of the horizon. 
It is then reflected and escape to infinity. 
If $0 < v < 2 r_S$, it reaches $r = 0$ after the formation of the horizon. 
The reflected wave escapes the mass shell, but is then trapped inside the event horizon and falls towards the singularity. 
Finally, if $v > 2 r_S$ the light-ray falls directly into the black hole. 

\subsubsection{Mode decomposition}

We now would like to solve \eq{eq:intro_phi} in the collapsing shell geometry. 
To this end, let us first write down explicitly this field equation in Schwarzshild coordinates. 
Using that $\sqrt{\abs{g}} = c \, r^2 \sin \theta$, one easily obtains 
\begin{align}
\frac{1}{c^2}\lp 1 - \frac{r_S}{r} \rp^{-1} \pd_t^2 \phi 
- \frac{1}{r^2} \pd_r \lp \lp 1 - \frac{r_S}{r} \rp r^2 \pd_r \phi \rp
- \frac{1}{r^2 \sin \theta} \pd_\theta \lp \sin \theta \, \pd_\theta \phi \rp
- \frac{1}{r^2 \sin^2 \theta} \pd_\varphi^2 \phi 
= 0.
\end{align}
This equation may be simplified using the tortoise coordinate $r^*$. Indeed, using
\begin{align}
\frac{\dd r^*}{\dd r} = \lp 1 - \frac{r_S}{r} \rp^{-1},
\end{align}
one obtains
\begin{align}
\frac{1}{c^2}\lp 1 - \frac{r_S}{r} \rp^{-1} \lp \pd_t^2 \phi 
- \frac{1}{r^2} \pd_{r^*} \lp r^2 \pd_{r^* }\phi \rp \rp
- \frac{1}{r^2 \sin \theta} \pd_\theta \lp \sin \theta \, \pd_\theta \phi \rp
- \frac{1}{r^2 \sin^2 \theta} \pd_\varphi^2 \phi 
= 0.
\end{align}
Since solutions of the Laplace equation in (3+1) dimensions decay like $1/r$ at infinity, it is useful to work with the field $\psi \equiv r \, \phi$. It obeys the equation 
\begin{align} \label{eq:intr:int4}
\frac{1}{c^2}\lp 1 - \frac{r_S}{r} \rp^{-1} \lp \pd_t^2 \psi 
- \pd_{r^*}^2 \psi 
+ \frac{1}{r} \lp \pd_{r^*}^2r \rp \psi \rp
- \frac{1}{r^2 \sin \theta} \pd_\theta \lp \sin \theta \, \pd_\theta \psi \rp
- \frac{1}{r^2 \sin^2 \theta} \pd_\varphi^2 \psi 
= 0.
\end{align}
In the following we concentrate on s-wave solutions depending only on $t$ and $r$. 
Equation~\eqref{eq:intr:int4} then becomes
\begin{align}
\frac{1}{c^2} \pd_t^2 \psi 
- \pd_{r^*}^2 \psi 
+ \lp 1 - \frac{r_S}{r} \rp \frac{r_S}{r^3} \psi
= 0.
\end{align}
The third term in this equation describes a potential barrier with a maximum at $r = 4 r_S / 3$. 
Notice that it is equal to zero at $r = r_S$, where the emission process we are interested in essentially takes place (see below), and vanishes like $r_S \, r^{-3}$ at infinity. 
Hence it does not affect the near-horizon behavior nor the asymptotic properties of the solutions. 
For this reason, we shall neglect it in the following. 
(Notice however that it must be included in precise estimations of the emitted spectrum, as the potential barrier will reflect waves with frequencies smaller than or of the order of $M \, c^2 / \hbar$.) 

Neglecting the potential barrier, the equation on $\psi$ becomes
\begin{align}
\pd_U \pd_V \psi= 0 \; \text{for} \; v > 2 r_S.
\end{align}
Similarly, defining the light-like coordinates $\mathcal{U} \equiv c \, t - r$ and $\mathcal{V} \equiv c \, t + r$ inside the mass shell, one obtains
\begin{align}
\pd_\mathcal{U} \pd_\mathcal{V} \psi= 0 \; \text{for} \; v < 2 r_S.
\end{align}
The field $\psi$ may thus be written, in each region, as a sum of functions depending of one coordinate only:
\begin{align}
\psi(\mathcal{U},v) = \left\lbrace 
\begin{array}{ll}
X(V) + \Xi (U) & v > 2 r_S \\
\chi(\mathcal{V}) + \xi (\mathcal{U}) & v < 2 r_S
\end{array}
\right. ,
\end{align}
where $X$, $\Xi$, $\chi$, and $\xi$ are differentiable functions. 
In this expression, we used that $v = \mathcal{V}$ along the mass shell, so that we can unify them in the single coordinate $v$, defined to be equal to $\mathcal{V}$ for $v < 2 r_S$, and that the definition of $\mathcal{U}$ can be extended to $v > 2 r_S$ by imposing that it is constant along outgoing light rays. 
Moreover, $\psi$ must vanish at $r = 0$ for $\phi$ to be nonsingular. 
This gives
\begin{align}
\forall \, \mathcal{U}, \, \xi(\mathcal{U}) = - \chi(\mathcal{U}).
\end{align}
Since we assumed the mass shell does not interact with the field $\phi$, the latter must be smooth across $v = 2 r_S$. 
So, $\psi$ may be written everywhere as
\begin{align}
\psi(\mathcal{U},v) = \chi(\mathcal{U}) - \chi(v),
\end{align}
where $\chi$ is an arbitrary differentiable function. 

In the following, we will need the relation between $\mathcal{U}$ and $u$ along the shell trajectory. 
Just inside the shell, we have $\mathcal{U} = v - 2 r = 2 \lp r_S - r \rp$ at $v = 2 r_S$. 
Just outside the shell, we have $u = v - 2 r^* = 2 \lp r_S - r^* \rp$. 
So,
\begin{align} \label{eq:intr:uU}
u = 2 \lp r_S - r  - r_S \ln \abs{\frac{r}{r_S} - 1}\rp = \mathcal{U}  - 2 r_S \ln \abs{\frac{\mathcal{U}}{2 r_S}}. 
\end{align}
Close to but outside the event horizon, this becomes
\begin{align} \label{eq:Uvsu_1}
\mathcal{U} \approx - 2 r_S \e^{- u / (2 r_S)}. 
\end{align}

\subsubsection{Quantization}
\label{subsub:quant}

Before moving on to the determination of the scattering coefficients encoding the Hawking effect, let us briefly review the quantization of the scalar field $\phi$. 
Our aim is not to give a rigorous presentation (a deeper and more precise one can be found, for instance, in~\cite{wald1994quantum}). Rather, we try to motivate, and give the feeling of, the main features which will be needed in deriving the Hawking effect and in the following chapters. 
In this section we work in a system of units where the reduced Planck constant $\hbar$ is equal to $1$.
We first work in flat space-time, and write explicitly only the radial part of $\phi$ to avoid unnecessary technical complications. 
The general solution of the field equation $\pd_\mu \pd^\mu \phi = 0$ may be written as
\begin{align}\label{eq:intr:clsol}
\phi (u, v, \theta, \varphi) = 
\int_0^\infty \dd \om \, \left[ a_\om \lp \frac{\e^{-\ii \om v /c}}{4 \pi \sqrt{\om} r} - \frac{\e^{-\ii \om u /c}}{4 \pi \sqrt{\om} r} \rp + {\rm c.c.} \right] + ...,
\end{align}
where ``${\rm c.c.}$'' denotes the complex conjugate and the unwritten terms have a non-vanishing angular momentum. 
They thus vanish after integration over the unit sphere. 
Differentiating the action with respect to $\pd_t \phi$ gives the conjugate momentum
\begin{align}
\Pi (u,v,\theta,\varphi) = c^{-1} r^2 \sin \, \theta \, \pd_t \phi. 
\end{align}
Leaving aside some technical issues, quantizing the theory amounts to promoting the fields $\phi$ and $\Pi$ to the status of operators in an infinite-dimensional Hilbert space. 
That is, we assume the existence of a Hilbert space $\mathcal{H}$ and two smooth maps $\hat{\phi}$, $\hat{\Pi}$ from Minkowski space to the set of operators in $\mathcal{H}$, such that
\begin{itemize}
\item $\pd_\mu \pd^\mu \hat{\phi} = 0$,
\item $\hat{\Pi} =c^{-1} r^2 \sin \theta \, \pd_t \hat{\phi}$,
\item $\forall \, t \in \mathbb{R}, \, \left[\hat{\phi}(t,r',\theta',\varphi'), \hat{\Pi}(t,r,\theta,\varphi) \right] = i \delta (r - r') \, \delta (\theta - \theta') \, \delta (\varphi - \varphi')$.
\end{itemize}
The general solution to the equation $\pd_\mu \pd^\mu \hat{\phi} = 0$ is 
\begin{align}
\hat{\phi} (u, v, \theta, \varphi) = 
\int_0^\infty \dd \om \left[ \hat{a}_\om  \lp \frac{\e^{-\ii \om v /c}}{4 \pi \sqrt{\om} r} - \frac{\e^{-\ii \om u /c}}{4 \pi \sqrt{\om} r} \rp + {\rm h.c.} \right] + ...,
\end{align}
where ``${\rm h.c.}$'' denotes the hermitian conjugate and, as in \eq{eq:intr:clsol}, the unwritten terms vanish after integration over the unit sphere. 
Here the $\hat{a}_\om$ are constant operators in $\mathcal{H}$, called annihilation operators. 
Their hermitian conjugates are the creation operators.  
They define the vacuum state $\left\lvert 0 \right\rangle$ by the condition 
\begin{align}
\forall \, \om > 0, \, \hat{a}_\om \left\lvert 0 \right\rangle = 0,
\end{align}
and similarly for the annihilation operators of modes with a non-vanishing angular momentum. 
A straightforward calculation gives
\begin{align}
\hat{a}_\om = \frac{\ii}{c} & \int_0^{2 \pi} \dd \varphi \, \int_0^{\pi} \dd \theta \, \int_0^\infty \dd r \nn
&\left[ 
	\lp \frac{\e^{\ii \om v /c}}{4 \pi \sqrt{\om} r} - \frac{\e^{\ii \om u /c}}{4 \pi \sqrt{\om} r} \rp  r^2 \sin \theta \, \pd_t \hat{\phi} 
	- r^2 \sin \theta \, \pd_t \lp \frac{\e^{\ii \om v /c}}{4 \pi \sqrt{\om} r} - \frac{\e^{\ii \om u /c}}{4 \pi \sqrt{\om} r} \rp \hat{\phi} 
\right].
\end{align}
This may be rewritten as
\begin{align} \label{eq:intro:an}
\hat{a}_\om = \ii \int_0^{2 \pi} \dd \varphi \, \int_0^{\pi} \dd \theta \, \int_0^\infty \dd r \,
\left[ 
	\phi_\om^* \hat{\Pi} - \Pi_\om^* \hat{\phi}
\right],
\end{align}
where
\begin{align}
\phi_\om : (t,r) \mapsto \frac{e^{-i \om v /c}}{4 \pi \sqrt{\om} r} - \frac{e^{-i \om u /c}}{4 \pi \sqrt{\om} r} 
\end{align}
and $\Pi_\om : (t,r,\theta) \mapsto r^2 \sin \theta \, \pd_t \phi_\om(t,r)$. 
The commutation relations between creation and annihilation operators can be easily deduced from \eq{eq:intro:an}. 
It gives
\begin{align}\label{eq:commafs}
\left[ \hat{a}_\om, \hat{a}_{\om'} \right] &= 
\int_0^{2 \pi} \dd \varphi \, \int_0^{\pi} \dd \theta \, \int_0^\infty \dd r \, 
\int_0^{2 \pi} \dd \varphi' \, \int_0^{\pi} \dd \theta' \, \int_0^\infty \dd r' \nn
& \hspace*{1 cm}\, \left[ \phi_\om^*(x) \hat{\Pi}(x) - \Pi_\om^*(x) \hat{\phi}(x), \phi_{\om'}^*(x') \hat{\Pi}(x') - \Pi_{\om'}^*(x') \hat{\phi}(x') \right] \nn
& = \ii \int_0^{2 \pi} \dd \varphi \, \int_0^{\pi} \dd \theta \, \int_0^\infty \dd r \, \lp \phi_{\om'}^*(x) \Pi_\om^*(x)  - \phi_\om^*(x') \Pi_{\om'}^*(x')  \rp  \nn
& = 0
\end{align}
and
\begin{align}\label{eq:commabfs}
\left[ \hat{a}_\om, \hat{a}_{\om'}^\dagger \right] &= 
\int_0^{2 \pi} \dd \varphi \, \int_0^{\pi} \dd \theta \, \int_0^\infty \dd r \, 
\int_0^{2 \pi} \dd \varphi' \, \int_0^{\pi} \dd \theta' \, \int_0^\infty \dd r' \nn
& \hspace*{1 cm} \left[ \phi_\om^*(x) \hat{\Pi}(x) - \Pi_\om^*(x) \hat{\phi}(x), \phi_{\om'}(x') \hat{\Pi}(x') - \Pi_{\om'}(x') \hat{\phi}(x') \right] \nn
& = \ii \int_0^{2 \pi} \dd \varphi \, \int_0^{\pi} \dd \theta \, \int_0^\infty \dd r \, \lp \phi_{\om}^*(x) \Pi_{\om'}(x)  - \phi_{\om'}(x) \Pi_{\om}^*(x)  \rp \nn
& = \delta ( \om - \om'). 
\end{align}
The creation operators can be applied on the vacuum to define states with an arbitrary number of particles. The state
\begin{align}
\left\lvert \om_1,n_1 ; \om_2, n_2 ; ... ; \om_l, n_l\right\rangle \equiv
\lp n_1! \, n_2!... n_l! \rp^{-1/2} 
\lp \hat{a}_{\om_1}^\dagger \rp^{n_1} 
\lp \hat{a}_{\om_2}^\dagger \rp^{n_2}
...
\lp \hat{a}_{\om_l}^\dagger \rp^{n_l}  \left\lvert 0 \right\rangle
\end{align}
(where the prefactor ensure its norm is equal to $1$) 
contains $n_1$ particles with energy $\om_1$, $n_2$ particles with energy $\om_2$,..., and $\om_l$ particles with energy $\om_l$. 
The operator $\hat{n}_\om \equiv \hat{a}_{\om}^\dagger \hat{a}_{\om}$ gives the number density of particles with frequency $\om$. 
Indeed, straightforward calculation using the commutation relations~\eqref{eq:commafs} and~\eqref{eq:commabfs} shows that 
\begin{align}
& \hat{n}_\om \left\lvert \om_1,n_1 ; \om_2, n_2 ; ... ; \om_l, n_l\right\rangle = \nn
& \hspace{2 cm} \lp n_1 \delta(\om - \om_1) + n_2 \delta(\om - \om_2) + ... + n_l \delta(\om - \om_l) \rp
\left\lvert \om_1,n_1 ; \om_2, n_2 ; ... ; \om_l, n_l\right\rangle .
\end{align}

Let us now turn to the quantization in curved spacetime. To this end, it is useful to first extend the action \eq{eq:action_phi} to complex scalar fields as
\begin{align}
S_\phi^c = \int \dd^4x \, \sqrt{\abs{g}} \, \lp \pd_\mu \phi^* \rp \lp \pd^\mu \phi \rp.
\end{align}
In fact, the discussion below is more general and can be applied to fields with a potential or dispersive terms. 
The important point is that the action $S_\phi^c$ has a global $\mathrm{U}(1)$ invariance under $\phi \to \e^{\ii \alpha} \phi$, $\alpha \in \mathbb{R}$. 
The corresponding Noether charge is
\begin{align}\label{eq:intr:NC}
\ii \int \dd^3 x \, \lp \Pi^* \, \phi^* - \Pi \, \phi \rp,
\end{align}
where $\Pi = \sqrt{\abs{g}} g^{00} \pd_t \phi^*$. 
(A proof for dispersive fields, maybe less standard than the relativistic case, is given in Section~\ref{sub:proofNoether}.)
Notice that this conserved quantity vanishes for a real solution, but will also be important for the original real field which can (formally) be expanded on a basis of complex solutions. 
Let us define the following inner product between two solutions $\phi_1$ and $\phi_2$:
\begin{align}\label{eq:intro_KGnorm}
\lp \phi_1, \phi_2 \rp \equiv \ii \int \dd^3 x \, \lp \Pi_2^* \, \phi_1^* - \Pi_1 \, \phi_2 \rp,
\end{align}
where $\Pi_i$ is the momentum conjugate to $\phi$, evaluated for $\phi = \phi_i$. 
This inner product is linear in $\phi_2$, antilinear in $\phi_1$, and satisfies $\lp \phi_2, \phi_1 \rp = \lp \phi_1, \phi_2 \rp^* = -\lp \phi_1^*, \phi_2^* \rp$. 
Let us show that it is conserved in time. 
We first note that, for any solution $\phi$ of the field equation, $(\phi, \phi)$ is conserved as it is exactly the Noether charge \eq{eq:intr:NC}. 
Since the field equation is linear, given two solutions $\phi_1$ and $\phi_2$, the four functions $\phi_1 \pm \phi_2$ and $\phi_1 \pm \ii \phi_2$ are also solutions. Moreover,
\begin{align}
\lp \phi_1 \pm \phi_2, \phi_1 \pm \phi_2 \rp = \lp \phi_1, \phi_1 \rp + \lp \phi_2, \phi_2 \rp \pm 2 \Re \lp \phi_1, \phi_2 \rp
\end{align}
and
\begin{align}
\lp \phi_1 \pm \ii \, \phi_2, \phi_1 \pm \ii \, \phi_2 \rp = \lp \phi_1, \phi_1 \rp + \lp \phi_2, \phi_2 \rp \pm 2 \ii \,  \Im \lp \phi_1, \phi_2 \rp.
\end{align}
So, $\lp \phi_1, \phi_2 \rp$ may be written as
\begin{align}
\lp \phi_1, \phi_2 \rp = \frac{1}{4} \lp 
	\lp \phi_1 + \phi_2, \phi_1 + \phi_2 \rp 
	- \lp \phi_1 - \phi_2, \phi_1 - \phi_2 \rp 
	+ \lp \phi_1 + \ii \phi_2, \phi_1 + \ii \phi_2 \rp 
	- \lp \phi_1 - \ii \phi_2, \phi_1 - \ii \phi_2 \rp 
\rp 
\end{align}
As the sum of $4$ conserved quantities, it is itself a conserved quantity. 

The general solution of the field equation can be expanded on an orthonormal basis $\phi_\om^{(i)}, \phi_\om^{(i)*}$ of solutions, where the label $i$ may contain continuous and/or discrete values. 
For definiteness, we here treat it as a discrete parameter, but the generalization to a continuous one, or a mixture of continuous and discrete ones, is straightforward. 
We choose these solutions such that $\lp \phi_\om^{(i)}, \phi_{\om'}^{(j)} \rp =\delta_{i j} \delta (\om - \om')$ and $\lp \phi_\om^{(i)}, \phi_{\om'}^{(j)*} \rp = 0$.  
Then, $\lp \phi_\om^{(i)*}, \phi_{\om'}^{(j)*} \rp = - \delta_{i j} \delta (\om - \om')$. 
The quantum fields $\hat{\phi}$ and $\hat{\Pi}$ may be written as
\begin{align}
\hat{\phi} = \int \dd \om  \, \sum_i \lp \hat{a}_\om^{(i)} \phi_\om^{(i)} + \hat{b}_\om^{(i) \dagger} \phi_\om^{(i)*}  \rp
\end{align} 
and
\begin{align}
\hat{\Pi} = \int \dd \om  \, \sum_i \lp \hat{a}_\om^{(i)} \Pi_\om^{(i)} + \hat{b}_\om^{(i) \dagger} \Pi_\om^{(i)*}  \rp,
\end{align}
where $\hat{a}_\om^{(i)}$ and $\hat{b}_\om^{(i)}$ are constant operators, and where $\hat{a}_\om^{(i)} = \hat{b}_\om^{(i)}$ for real fields. 
As done in the case of flat space, $\hat{a}_\om^{(i)}$ and $\hat{b}_\om^{(i)}$ can be obtained using 
\begin{align}\label{eq:intr:aphi}
\hat{a}_\om^{(i)} = \lp \phi_\om^{(i)}, \hat{\phi} \rp, \, \hat{b}_\om^{(i) \dagger} = - \lp \phi_\om^{(i)*}, \hat{\phi} \rp.
\end{align}
The commutation relations of the creation and annihilation operators can be easily obtained from these expressions. 
For definiteness, let us return to the case of a real field (the case of a complex one is very similar). 
We have
\begin{align}
\left[ \hat{a}_\om^{(i)}, \hat{a}_{\om'}^{(j)} \right] &= 
\left[ \ii \int \dd^3 x \, \lp \hat{\Pi} \, \phi_\om^{(i)*} - \Pi_\om^{(i)} \, \hat{\phi} \rp, \ii \int \dd^3 x' \, \lp \hat{\Pi} \, \phi_{\om'}^{(j)*} - \Pi_{\om'}^{(j)} \, \hat{\phi} \rp \right] \nn 
& = \ii \int \dd^3 x \, \lp \Pi_\om^{(i)} \phi_{\om'}^{(j)*} - \Pi_{\om'}^{(j)} \phi_\om^{(i)*} \rp \nn
& = \lp \phi_{\om'}^{(j)}, \phi_\om^{(i)*} \rp = 0
\end{align}
and 
\begin{align}
\left[ \hat{a}_\om^{(i) }, \hat{a}_{\om'}^{(j)\dagger} \right] &= 
\left[ \ii \int \dd^3 x \, \lp \hat{\Pi} \, \phi_\om^{(i)*} - \Pi_\om^{(i)} \, \hat{\phi} \rp, -\ii \int \dd^3 x' \, \lp \hat{\Pi} \, \phi_{\om'}^{(j)} - \Pi_{\om'}^{(j)*} \, \hat{\phi} \rp \right] \nn 
& = \ii \int \dd^3 x \, \lp \Pi_{\om'}^{(j)*} \phi_{\om}^{(i)*} - \Pi_{\om}^{(i)} \phi_{\om'}^{(j)} \rp \nn
& = \lp \phi_{\om}^{(i)}, \phi_{\om'}^{(j)} \rp = \delta_{i j} \delta (\om - \om').
\end{align}
Like in the case of flat space, the vacuum state $\left\lvert 0 \right\rangle$ is defined by
\begin{align}
\forall \om > 0, \, \forall i, \, \hat{a}_\om^{(i)} \left\lvert 0 \right\rangle = 0.
\end{align}
States with definite numbers of particles can be constructed by acting on $\left\lvert 0 \right\rangle$ with the creation operators $a_\om^{(i) \dagger}$. 

Another important concept is that of change of orthonormal basis. 
Let us assume we have two orthonormal bases $\phi_\om^{(i)}$ and $\varphi_\la^{(j)}$ related by
\begin{align}
\forall \la, \, \forall j, \, 
\varphi_\la^{(j)} = \int \dd \om \sum_i \lp A_{\la, \om}^{j,i} \phi_\om^{(i)} + B_{\la, \om}^{j,i} \phi_\om^{(i)*}\rp. 
\end{align}
The coefficients $A_{\la, \om}^{j,i}$ and $B_{\la, \om}^{j,i}$ are given by
\begin{align}
A_{\la, \om}^{j,i} = \lp \phi_\om^{(i)}, \varphi_\la^{(j)} \rp, \, B_{\la, \om}^{j,i} = -\lp \phi_\om^{(i)*}, \varphi_\la^{(j)} \rp. 
\end{align}
Let us denote by $\hat{c}_\la^{(j)}$ the annihilation operator corresponding to $\varphi_\la^{(j)}$. 
The quantum field $\hat{\phi}$ may be written as
\begin{align}
\hat{\phi} = \int \dd \la \sum_j \lp \hat{c}_\la^{(j)} \varphi_\la^{(j)} + \hat{c}_\la^{(j)\dagger} \varphi_\la^{(j)*} \rp.
\end{align}
Using~\eqref{eq:intr:aphi}, we deduce that
\begin{align} \label{eq:Bogointr1}
\hat{a}_\om^{(i)} = \int d \la \sum_j \lp A_{\la, \om}^{j,i} \hat{c}_\la^{(j)} + B_{\la, \om}^{j,i *}  \hat{c}_\la^{(j)\dagger}\rp.
\end{align}

Such a change of basis is rather innocuous if $B_{\la, \om}^{j,i} = 0$. Indeed, in that case the annihilation operators in one basis are simply linear combinations of those in the other basis, so that the two vacuum states are the same. 
However, this ceases to be true when $B_{\la, \om}^{j,i} \neq 0$. 
In that case, the vacuum state $\left\lvert 0 \right\rangle_\phi$ in the first basis is not annihilated by the operators $\hat{c}_\la^{(j)}$. 
Conversely, the vacuum state $\left\lvert 0 \right\rangle_\varphi$ in the second basis is not annihilated by the operators $\hat{a}_\la^{(j)}$. 
To see the relationship between these two vacua in a simple model, let us assume the two bases are chosen such that $A_{\la, \om}^{j,i} = B_{\la, \om}^{j,i} = 0$ unless $i = j$ and $\om = \la$. 
One can then write $A_{\la, \om}^{j,i} = \delta_{j i} \delta(\om - \om') A_{\om}^{(i)}$ and $B_{\la, \om}^{j,i} = \delta_{j i} \delta(\om - \om') B_{\om}^{(i)}$. 
The commutation relations on $a_\om^{(i)}$ and $a_\om^{(i) \dagger}$ then give
\begin{align}
\forall \om, \, \forall i, \, \abs{A_{\om}^{(i)}}^2 - \abs{B_{\om}^{(i)}}^2 = 1.
\end{align}
Let us write the relationship between the two vacua as
\begin{align} \label{eq:intr_deff}
\ket{0}_\phi = f \lp \left\lbrace \hat{c}_\la^{(j) \dagger} \right\rbrace \rp \ket{0}_\varphi,
\end{align}
where the notation ``$\left\lbrace \hat{c}_\la^{(j) \dagger} \right\rbrace$'' means that $f$ is a function of all the creation operators in the basis $\lp \varphi_\la^{j)} \rp_{\la, j}$. 
(All terms in $f$ containing some powers of the annihilation operators $c_\om^{(i)}$ can be eliminated using the commutation relations and the definition of the vacuum state.)
Applying the annihilation operator $a_\om^{(i)}$ gives
\begin{align}
\lp A_{\om}^{(i)} \hat{c}_\om^{(i)} + B_{\om}^{(i)*} \hat{c}_\om^{(i) \dagger}  \rp f \lp \left\lbrace \hat{c}_\la^{(j) \dagger} \right\rbrace \rp \ket{0}_\varphi = 0.
\end{align}
This may be rewritten using the commutation relations on $\hat{c}_\la^{(j)}$ and $\hat{c}_\la^{(j) \dagger}$ as
\begin{align}
\lp A_{\om}^{(i)} \frac{\delta}{\delta \hat{c}_\om^{(i)\dagger}} f \lp \left\lbrace \hat{c}_\la^{(j) \dagger} \right\rbrace \rp + B_{\om}^{(i)*} {\hat{c}_\om^{(i)\dagger}} f \lp \left\lbrace \hat{c}_\la^{(j) \dagger} \right\rbrace \rp \rp \ket{0}_\varphi = 0.
\end{align}
This is satisfied if and only if
\begin{align}
A_{\om}^{(i)} \frac{\delta}{\delta \hat{c}_\om^{(i)\dagger}} f \lp \left\lbrace \hat{c}_\la^{(j) \dagger} \right\rbrace \rp + B_{\om}^{(i)*} {\hat{c}_\om^{(i)\dagger}} f \lp \left\lbrace \hat{c}_\la^{(j) \dagger} \right\rbrace \rp  = 0.
\end{align}
Solving this differential equation gives $\ket{0}_\phi$ up to a normalization factor:
\begin{align}
\ket{0}_\phi \propto \exp \lp - \frac{1}{2} \int \dd \om \sum_i \frac{B_{\om}^{(i)*}}{A_{\om}^{(i)}} \lp \hat{c}_\om^{(i) \dagger} \rp^2 \rp \ket{0}_\varphi.
\end{align}
This is the simplest case of mixing between two bases of modes, referred to as \textit{one-mode mixing} as the annihilation operator in one basis is a combination of the creation and annihilation operators of a single mode of the other basis. 
However, in the following we shall encounter more the \textit{two-mode mixing}. 
Let us assume that each mode $i$ is associated with a ``partner'' mode $p(i)$, where $p$ is a permutation such that $p(p(i)) = i$. 
We further assume that $B_{\la, \om}^{j,i} = 0$ unless $\om = \la$ and $j = p(j)$, while as before $A_{\la, \om}^{j,i} = 0$ unless $\om = \la$ and $j = i$.
One can then write $A_{\la, \om}^{j,i} = \delta_{j \s i} \delta(\om - \om') \alpha_{\om}^{(i)}$ and $B_{\la, \om}^{j,i} = \delta_{j \s p(i)} \delta(\om - \om') \beta_{\om}^{(i)}$ where, for each value of $\om$ and $i$, $\alpha_\om^{(i)}$ and $\beta_\om^{(i)}$ are two complex numbers. 
Eq.~\eqref{eq:Bogointr1} then becomes
\begin{equation}
\hat{a}_\om^{(i)} = \alpha_{\om}^{(i)} \s \hat{c}_\om^{(i)} + \beta_{\om}^{(i)*} \s \hat{c}_\om^{(p(i))\dagger}.
\end{equation}
The commutation relations give
\begin{equation}
\forall \om, \, \forall i, \, \abs{\alpha_\om^{(i)}}^2 - \abs{\beta_\om^{(i)}}^2 = 1.
\end{equation}
Imposing that the state $\ket{0}_\varphi$ be annihilated by $\hat{a}_\om^{(i)}$ now gives, with the definition \eq{eq:intr_deff}, 
\begin{align}
\alpha_{\om}^{(i)} \frac{\delta}{\delta \hat{c}_\om^{(i)\dagger}} f \lp \left\lbrace \hat{c}_\la^{(j) \dagger} \right\rbrace \rp + \beta_{\om}^{(p(i))*} {\hat{c}_\om^{(i)\dagger}} f \lp \left\lbrace \hat{c}_\la^{(j) \dagger} \right\rbrace \rp  = 0,
\end{align}
leading to
\begin{align}\label{eq:intr:rel0}
\ket{0}_\phi \propto \exp \lp - \int \dd \om \sum_i \frac{B_{\om}^{(i)*}}{A_{\om}^{(i)}}  \hat{c}_\om^{(i) \dagger} \hat{c}_\om^{(p(i)) \dagger}  \rp \ket{0}_\varphi.
\end{align}
This shows that the state $\ket{O}_\phi$ , when viewed in the basis $\lp \varphi_\la^{(j)} \rp_{\la,j}$, contains pairs of entangled particles, related by the permutation $p$, i.e., which are each other's partners. 
This notion will play a crucial role in the Hawking mechanism.  

\subsubsection{Scattering coefficients and Hawking radiation}

From then on until the end of this subsection, we work in natural units where $\hbar = c = G = k_B = 1$ (where $k_B$ is Boltzmann's constant) to shorten notations and to make the link with the presentation of~\cite{Brout:1995rd}, where the approximations done here are discussed more thoroughly, more evident. 
As we will now see, the aforementioned difference between the vacua associated with different bases of modes triggers a spontaneous thermal emission of particles: the Hawking radiation. 
Let us assume that the state of the field is vacuum at $t \to - \infty$, i.e., for the incoming modes
\begin{align}
\varphi_\om^{\rm in}(\mathcal{U},v) = \frac{\e^{- \ii \om v}}{\sqrt{4\pi \om}} - \frac{\e^{- \ii \om \mathcal{U}}}{\sqrt{4\pi \om}}.
\end{align}
(The different normalization with respect \eq{eq:intr:clsol} is due to the fact that we consider only the modes in the $(r,v)$ plane, i.e., we will not integrate over $\theta$ and $\varphi$.) 
On the other hand, the outgoing modes at infinity are
\begin{align}
\varphi_\la^{\rm out}(u) = \frac{\e^{- \ii \la u}}{\sqrt{4 \pi \abs{\la}}}. 
\end{align}
Following~\cite{Brout:1995rd}, let us decompose the incoming modes as
\begin{align} \label{eq:intr_decompphiin}
\varphi_\om^{\rm in} = \int_0^\infty \lp \alpha_{\la \om} \varphi_\la^{\rm out} + \beta_{\la \om} \varphi_\la^{\rm out*} \rp \dd \la. 
\end{align}
The coefficients $\beta_{\la \om}$ encode the difference between the incoming vacuum and that of the observer at infinity. 
They thus characterize the production of particles by amplification of quantum fluctuations, which is directly due to this difference. 

The coefficients $\alpha_{\om \la}$ and $\beta_{\om \la}$ are given by
\begin{align} \label{eq:intr_alpha_rel}
\alpha_{\la \om} = \ii \int_{-\infty}^{+\infty} \dd u \lp \varphi_\la^{\rm out*} \pd_u \varphi_\om^{\rm in} - \varphi_\om^{\rm in} \pd_u \varphi_\la^{\rm out*} \rp
\end{align}
and
\begin{align}
\beta_{\la \om} = - \ii \int_{-\infty}^{+\infty} \dd u \lp \varphi_\la^{\rm out} \pd_u \varphi_\om^{\rm in} - \varphi_\om^{\rm in} \pd_u \varphi_\la^{\rm out} \rp.
\end{align}
Because of the exponential redshift close to the event horizon, the late-time radiation, emitted from a point very close to the horizon, will come from modes with $\om \gg \la$. 
In this limit, the above integrals can be computed explicitly using \eq{eq:Uvsu_1}. 
One finds
\begin{align}
\alpha_{\la \om} \approx \frac{r_S}{\pi} \sqrt{\frac{\la}{\om}} \Gamma \lp - 2 \ii \, r_S \la \rp \lp 2 r_S \om \rp^{2 i \, r_S \la} \e^{\pi \, r_S \la} 
\end{align}
and $\beta_{\la \om} \approx  \e^{-2 \pi \, r_S \la} \alpha_{\la \om}^*$.~\footnote{We remind that Euler's $\Gamma$ function is defined by
\begin{equation*}
\Gamma: z \mapsto \int_0^\infty t^{z-1} \, \e^{-t} \, \dd t . 
\end{equation*}}
In particular, under these approximations,
\begin{align}\label{eq:intr_betaoveral}
\abs{\frac{\beta_{\la \om}}{\alpha_{\la \om}}}^2 = \e^{-4 \pi r_S \la}.
\end{align}
Moreover, using the complement formula $\Gamma(z) \, \Gamma(-z) = - \pi / (z \sin \lp \pi z \rp)$ gives
\begin{align} \label{eq:beta2}
\abs{\beta_{\la \om}}^2 \approx & \, \frac{r_S^2}{\pi^2} \frac{\la}{\om} \abs{\Gamma(2 \ii r_S \la)}^2 \e^{- 2 \pi r_S \la} \nn
\approx & \, - \frac{r_S}{2 \pi \om} \frac{\e^{- 2 \pi r_S \la}}{\ii \sin \lp 2 \pi \ii r_S \la \rp} \nn
\approx & \, \frac{r_S}{\pi \om} \frac{1}{\e^{4 \pi r_S \la} - 1}. 
\end{align}
This is very reminiscent of a thermal law, with a temperature given by $T = \lp 4 \pi r_S \rp^{-1}$. 
To give some weight to this assertion, let us estimate the flux of particles received by an asymptotic observer. 
From \eq{eq:Bogointr1} with $A_{\om,\la} = \alpha_{\la \om}$ and $B_{\om,\la} = \beta_{\la \om}$, a straightforward calculation shows that the mean number of particles with angular frequency $\la$ received by an observer at infinity if the state of the field is the incoming vacuum is~\footnote{Notice that the first term in \eq{eq:Bogointr1} cancels when applied to $\ket{0}_\varphi$, so that only the second one contributes.}
\begin{align}
\left\langle \hat{a}_\la \hat{a}_\la^\dagger \right\rangle = \int \dd \om \abs{\beta_{\la \om}}^2. 
\end{align}
To determine the flux of particles, one should include in the integral only the frequencies $\om$ giving modes that will be detected between $u$ and $u + \Delta u$, and then divide by $\Delta u$ to obtain the number of particles per unit $u$. 
To this end, we notice that the relationship between $\la$ and $\om$ at fixed $u$, given by the continuity of $\phi$ along the mass shell (which gives $\la = \om \, \pd_u \mathcal{U}$) and the relation \eq{eq:Uvsu_1}, is
\begin{align}
\om = \la \e^{u / (2 r_S)}.
\end{align}
The mean flux $f_\la$ of particles detected at infinity is thus
\begin{align}
f_\la \approx  \mathop{\rm lim}_{\Delta u \to \infty} \frac{1}{\Delta u} \int_{\la \e^{u / (2 r_S)}}^{\la \e^{(u + \Delta u) / (2 r_S)}} \abs{\beta_{\la \om}}^2 \dd \om.
\end{align}
Using that, for large values of $u$ and thus of $\om$, $\abs{\beta_{\la \om}}^2$ depends on $\om$ only through a factor $1 / \om$, one sees that the integral becomes independent of $u$ and $\Delta u$ at late times. 
This is a nontrivial result, which implies that the late-time radiation is nonvanishing and stationary.
Using \eqref{eq:beta2}, one finds
\begin{align}
f_\la \approx \frac{1}{2 \pi} \lp \e^{4 \pi r_S \la} - 1 \rp^{-1}. 
\end{align}
This is a thermal spectrum, with the Hawking temperature $T_H = 1 / (4 \pi r_S)$. 

At this point it is useful to pause and consider the implications of \eq{eq:intr_decompphiin}. 
Since the in and out bases are both orthonormal for the inner product of \eq{eq:intr_alpha_rel}, the coefficients $\alpha_{\la \om}$ and $\beta_{\la \om}$ satisfy
\begin{align}
\forall \lp \la, \la' \rp \in \mathbb{R}_+^2, \, \int_0^\infty \lp \alpha_{\la \om}^* \alpha_{\la' \om} - \beta_{\la \om}^* \beta_{\la' \om} \rp \dd \om = \delta \lp \la - \la' \rp.
\end{align}
Because of the minus sign in this expression, the positive-frequency part of the outgoing wave in the classical field theory can be larger (in the sense of the inner product of \eq{eq:intr_alpha_rel}) than the incoming one. 
Moreover, this amplification is encoded in the coefficient $\beta_{\la \om}$, which also gives the flux of particles $f_\la$ emitted through Hawking radiation. We thus see that the (classical) wave amplification and Hawking effect have the same mathematical origin.  
This link will become more evident in the next section. 

While this calculation can be used to obtain the asymptotic flux from a black hole, it fails to capture another crucial ingredient of Hawking radiation: the entanglement between particles emitted at infinity and their ``partners'' falling into the black hole. 
To see this, following~\cite{Massar:1996tx}, let us define the modes $\chi_{\la}^{R}$ for $\la > 0$ by
\begin{equation}
\chi_{\la}^{R}(\mathcal{U}) = \varphi_\la^{\textrm{out}}(u),
\end{equation} 
where $\mathcal{U}$ is defined outside the mass shell by~\eqref{eq:intr:uU}.
A straightforward calculation using that $\varphi_\la^{\textrm{out}}$ is defined only for $\mathcal{U}<0$ gives
\begin{equation} \label{eq:intr:chiR}
\chi_{\la}^{R}(\mathcal{U}) = \frac{\e^{- \ii \la \mathcal{U}}}{\sqrt{4 \pi \abs{\la}}} \lp \frac{-\mathcal{U}}{2 r_S} \rp^{2 \ii r_S \la} \Theta(-\mathcal{U}). 
\end{equation}
These modes are associated with particles emitted at infinity. 
To describe the field inside the black hole, one defines the modes $\chi_{\la}^{L}$ by
\begin{equation}
\chi_{\la}^{L} : \mathcal{U} \mapsto \lp \chi_{\la}^{R}(-\mathcal{U}) \rp^*. 
\end{equation}
One easily checks that they are also normalized, and that wave packets built using superpositions of them for $\la > 0$ have positive inner products. 
Indeed, they satisfy
\begin{equation}
\chi_{\la}^{L}(\mathcal{U}) = \frac{\e^{- \ii \la u_i}}{\sqrt{4 \pi \abs{\la}}}
\end{equation}
for $\mathcal{U} < 0$. 
To relate them to the incoming modes, it is useful to introduce the ``Unruh'' modes~\cite{Unruh:1976db}
\begin{equation}
\psi_{\la,U} \equiv \lim_{\ep \to 0} \frac{1}{\sqrt{8 \pi \abs{\la} \sinh (2 \pi r_S)}} \lp 
	\lp \frac{\ep + \ii \s v}{2 r_S} \rp^{2 \ii r_S \la} - \lp \frac{\ep + \ii \s \mathcal{U}}{2 r_S} \rp^{2 \ii r_S \la}
\rp, \, \la \in \mathbb{R}^*,
\end{equation}
where the limit $\ep \to 0$ should be taken only at the end of the calculations.  
The outgoing (``$\mathcal{U}$'') part of this mode may be written as
\begin{equation}
\frac{-1}{\sqrt{8 \pi \abs{\la} \sinh (2 \pi r_S)}} \lp
	\e^{\pi r_S \la} \lp \frac{-\mathcal{U}}{2 r_S} \rp^{2 \ii r_S \la} \Theta(-\mathcal{U})
	+ \e^{-\pi r_S \la} \lp \frac{\mathcal{U}}{2 r_S} \rp^{2 \ii r_S \la} \Theta(\mathcal{U})
\rp .
\end{equation}
At late times $t \to \infty$ at fixed $r$, $\mathcal{U} \to 0$. 
One can thus approximate the factor $\e^{- \ii \la \mathcal{U}}$ in \eq{eq:intr:chiR} by $1$ and the above expression becomes
\begin{equation}
\psi_{\la,U} \approx - \lp
	\frac{\e^{\pi r_S \la}}{\sqrt{2 \sinh(2 \pi r_S)}} \chi_\la^R
	+ \frac{\e^{- \pi r_S \la}}{\sqrt{2 \sinh(2 \pi r_S)}} \chi_\la^{L*}
\rp.
\end{equation}
Let us define $\alpha_\la \equiv - \e^{\pi r_S \la} / \sqrt{2 \sinh(2 \pi r_S)}$ and $\beta_\la \equiv - \e^{-\pi r_S \la} / \sqrt{2 \sinh(2 \pi r_S)}$. 
We have:
\begin{equation}
\begin{pmatrix}
\psi_{\la,U} \\
\psi_{-\la,U}^*
\end{pmatrix}
=
\begin{pmatrix}
\alpha_\om & \beta_\om \\
\beta_\om & \alpha_\om
\end{pmatrix}
\begin{pmatrix}
\chi_{\la}^R \\
\chi_{\la}^{L*}
\end{pmatrix} .
\end{equation}
Using that $\alpha_\la^2 - \beta_\la^2 = 1$, this relation can be inverted as
\begin{equation}
\begin{pmatrix}
\chi_{\la}^R \\
\chi_{\la}^{L*}
\end{pmatrix}
=
\begin{pmatrix}
\alpha_\om & -\beta_\om \\
-\beta_\om & \alpha_\om
\end{pmatrix}
\begin{pmatrix}
\psi_{\la,U} \\
\psi_{-\la,U}^*
\end{pmatrix} .
\end{equation}
From this, using \eq{eq:intr:rel0}, one deduces that the relationship between the ``Unruh'' vacuum $\ket{0}_U$ for the modes $\psi_{\la,U}$ and the ``Boulware'' vacuum $\ket{0}_B$ for the modes $\chi_\la^R$ and $\chi_\la^L$ is of the form
\begin{equation}
\ket{0}_U = \exp \lp \int \dd \la \frac{\beta_\la}{\alpha_\la} \hat{a}_{\la,R}^\dagger \hat{a}_{\la,L}^\dagger \rp \ket{0}_B,
\end{equation}
where $\hat{a}_{\la,R}^\dagger$ and $\hat{a}_{\la,L}^\dagger$ are respectively the creation operators for quanta of $\chi_{\la}^R$ and $\chi_{\la}^L$. 
The point to retain is that the Hawking emission is accompanied by ``partner'' particles falling into the black hole, which are entangled -- sharing, roughly speaking, the same quantum state.

\subsubsection{The ``transplanckian'' and unitarity problems}

The above calculation raises two important issues, both related to the fact that it relies on a semi-classical approximation where the scalar field $\phi$ is quantized whereas the gravitational field is treated as a classical, fixed background. 
The first one, sometimes referred to as the ``transplanckian problem'', is due to the exponential redshift between the inside and outside frequencies when approaching the horizon:
\begin{align}
\om = \la \s \e^{u / (2 r_S)}.
\end{align}
Considering for instance a moderate frequency $\la$ of order $r_S^{-1}$, much smaller than the Planck mass (equal to $1$ in the system of units where $\hbar = c = G = 1$) for macroscopic black holes with $r_S \gg 1$, the corresponding value of $\om$ will reach the Planck scale after a time close to $2 \, r_S \ln r_S$, much smaller than the expected black-hole lifetime, of order $r_S^3$.~\footnote{To get this estimate, notice that the power radiated by a thermal system with temperature $T$ is proportional to $T^2$. In our case, $T \propto r_S^{-1}$ and the mass of the black hole is $r_S / 2$. The time needed to radiate away all (or a significant fraction of) its energy thus scales as $r_S^3$.} 
(When this condition is not satisfied, the semi-classical approximation should break down and quantum gravity effects are expected to dominate.) 
The energy $\omega$ even reaches that of the black hole after a time close to $4 \, r_S \ln r_S$.
This means that, after a time which is short compared with the timescale on which the black hole evaporates by radiating its energy through the Hawking effect, the particles emitted with moderate energies come from vacuum fluctuations inside the shell with sufficiently high energies for quantum gravity effects and backreaction to become important.~\footnote{A maybe less abstract way to see this problem is that, as explained in the introduction of~\cite{Unruh2014}, the typical particles produced just one second after the formation of a solar-mass black hole originate from vacuum fluctuations with frequencies of order $\e^{10^5}$, larger than virtually any imaginable energy scale.} 
The free quantum field theory used above very probably breaks down in this regime. 

Although the full resolution of this problem would require a complete theory of quantum gravity, it is interesting to note that analogue models of gravity already provide a possible way out. 
Indeed, as explained below, the later have dispersive terms, i.e., higher-order space derivatives in the wave equation, which regularize the redshift close to the horizon. 
(See~\cite{Jacobson:1991gr,Unruh:1994je} for an early discussion of these additional terms.) 
Their effects on the (analogue) Hawking radiation is now well understood, see for instance~\cite{Macher:2009tw}. 
In particular, on very general grounds, it is found that in spite of the important changes they induce in the dynamics of waves close to the (analogue) horizon, they do not affect the emission spectrum at energies small before the dispersive scale. 
If one assumes for a moment that the first correction to classical general relativity from a theory of quantum gravity can be modeled by such dispersive terms, the dispersive scale will presumably be of the order of the Planck mass, above the range currently accessible to observations and experiments. 
The emission spectrum of black holes at observationally accessible energies should thus be nearly indistinguishable from the results of S.~Hawking summarized above.~\footnote{One possible issue with this scenario, however, is that it would break Lorentz invariance at high energies.} 
I shall come back to this point in Section~\ref{sec_intr:WcbblftA}. 

Another issue, sometimes referred to as the ``information loss paradox'', is the apparent breakdown of unitarity during the gravitational collapse and subsequent evaporation. 
Indeed, from the uniqueness theorem, the metric of a black hole formed by, say, the gravitational collapse of a star, does not depend on the latter's precise structure. 
If the black hole then evaporates through a purely thermal radiation, most of the information about the star seems unavoidably lost. 
In a quantum language, a pure quantum state before the collapse would then become a mixed, thermal state after the evaporation, violating unitarity. 
Actually, during the evaporation, each Hawking quanta emitted at infinity comes with a ``partner'' which falls into the black hole, and to which it is correlated. 
This is very reminiscent of a liquid / gas phase transition, where each evaporating molecule has a small back-reaction on its neighbors, producing correlations between the liquid phase and evaporated cloud as well as between the molecules comprising the latter. 
In that case, these correlations contain the the information which seems to be lost in a thermodynamic, coarse-grained description involving an increase of entropy. 
Importantly, it can in principle be recovered -- although this is practically impossible for a macroscopic system with more than $10^{23}$ molecules -- by a determination of the position and velocity of each particle. 
The apparent loss of information is thus a practical problem, coming from our limited ability to measure the trajectories of many particles, more than a fundamental one.
However, in the case of a black hole, assuming the semi-classical theory is valid, the ``partners'' remain at all times inside the horizon, inaccessible to an external observer. 
It is thus far from certain that the information they encode can possibly be recovered by an external observer, even if assuming unlimited computational and technical means. 
As mentioned in~\cite{Brout:1995rd,Preskill:1992tc}, there seem to be three options:
\begin{itemize}
\item Unitarity may indeed be broken from the point of view of an external observer, part of the information being either fundamentally lost or inaccessible (for instance hidden in a disconnected part of space-time if quantum gravity allows for a dynamical change of topology
), as originally claimed by S.~Hawking~\cite{Hawking:1974sw}.
\item The semiclassical treatment may fail when the black hole radius reaches the Planck scale, for instance leaving a stable remnant, correlated with the emitted radiation, which contains all the missing information. 
\item The semiclassical theory presented above may be too rough to capture the correlations between emitted quanta, which would carry the missing information~\cite{Page,Hawking:2016msc}. 
\end{itemize}

\section{Analogue Gravity}
\label{sec:intr:AG}

In spite of its importance for the physics of black holes in the presence of quantum fields, the notion of Hawking radiation in astrophysical black holes does not straightforwardly lead to physically relevant predictions. 
Indeed, apart from the aforementioned theoretical issues, it faces a problem of order of magnitude: assuming for definiteness that the mass of the black hole is of the order of that of the Sun, one finds that the Hawking temperature is about $10^{-7} \textrm{K}$. 
This is significantly smaller than the temperature of the cosmic microwave background (CMB), at about $2.7 \textrm{K}$. 
Detecting the Hawking radiation would thus require a knowledge of the CMB, as well as all other sources of radiations contributing to the received signal, to an unprecedented accuracy. 

Fortunately, W.~Unruh showed in 1981 that astrophysical black holes are not the only objects able to trigger Hawking radiation~\cite{Unruh:1980cg}. 
Indeed, under some conditions (to be detailed below), the physics of linear waves propagating in nonrelativistic fluids bears a mathematical correspondence with that of fields in a black hole space-time. 
They may thus be used to realize experimentally the analogue of a black hole, in which the Hawking process could be observed. 
Moreover, these analogue models generically come with a natural cutoff due to the breakdown of the fluid description at wavelengths close to or below some physical scale, for instance the typical intermolecular distance in a gas.  
They offer a self-consistent framework in which regularizing effects such as those expected to arise, maybe in an effective way, in a quantized theory of gravity can be studied, allowing one to determine how they change the properties of the particle emission -- or to what extent they leave it unaffected.

In this subsection I briefly present a few analogue models. 
I first focus on sound waves in a classical, nonrelativistic fluid used in the original derivation of~\cite{Unruh:1980cg} to introduce the main concepts and ideas. 
I then turn to two more widely used systems, namely sound waves in Bose-Einstein condensates and surface gravity waves in classical fluids. 
The interested reader will find a more extensive discussion in the reviews~\cite{Visser:2001fe,Barcelo:2005fc}

\subsection{Sound waves in a nonrelativistic fluid}

Let us consider a classical, irrotational gas of nonrelativistic particles, described by the local density $\rho$, velocity $\vec{v}$, and pressure $p$, in an infinite space.~\footnote{A similar model with boundaries used to reduce the effective sound velocity was studied in~\cite{2015ASAJ..138..605A,Auregan:2015uva}, see also Chapter~\ref{ch:concl}, section~\ref{sub:slowsound}.} 
Conservation of the number of atoms gives
\begin{align}
\partial_t \rho + \vec{\nabla} \cdot \lp \rho \, \vec{v} \rp = 0.
\end{align}
We assume that $p$ depends only on $\rho$, that $\dd p / \dd \rho > 0$,  and that all the external forces acting on the gas are conservative. 
We call $\Phi$ the sum of their potentials. 
Newton's second law applied to an infinitesimal volume of gas gives
\begin{align} \label{eq:intr_2N}
\rho \, \lp \pd_t \vec{v} + \lp \vec{v} \cdot \vec{\nabla} \rp \, \vec{v} \rp = - \vec{\nabla}p - \rho \, \vec{\nabla} \Phi.
\end{align}
Since $\vec{v}$ is irrotational, one can define a velocity potential $\phi$ such that $\vec{v} = \vec{\nabla} \phi$.  
\eqref{eq:intr_2N} becomes
\begin{align} \label{eq:intr_2N2}
\pd_t \vec{\nabla} \phi + \frac{1}{2} \vec{\nabla} \lp \vec{\nabla} \phi \rp^2 = - \frac{\vec{\nabla} p(\rho)}{\rho} - \vec{\nabla} \Phi. 
\end{align}
This may be simplified by defining 
\begin{align}
\gamma : \rho \mapsto \int^\rho \frac{1}{\tilde{\rho}}\frac{\dd p(\tilde{\rho})}{\dd \tilde{\rho}} \dd \tilde{\rho}:
\end{align}
then, \eqref{eq:intr_2N2} becomes
\begin{align}
\vec{\nabla} \, \lp 
	\pd_t \phi
	+ \frac{1}{2} \lp \vec{\nabla} \phi \rp^2 
	+ \gamma(\rho)
	+ \Phi
\rp = \vec{0}.
\end{align}
This implies that $\pd_t \phi
	+ \frac{1}{2} \lp \vec{\nabla} \phi \rp^2 
	+ \gamma(\rho)
	+ \Phi \equiv F$ is a function of time only. 
Notice that $\phi$ is defined only up to a function of time, which can be chosen so that $F = 0$. 
Doing so, one obtains 
\begin{align}\label{eq:intr_s_int1}
\pd_t \phi
	+ \frac{1}{2} \lp \vec{\nabla} \phi \rp^2 
	+ \gamma(\rho)
	+ \Phi
	= 0.
\end{align}
Let us assume that we know a solution $(\phi_0, \rho_0)$. 
We look for perturbations of the form $\phi = \phi_0 + \delta \phi$, $\rho = \rho_0 + \delta \rho$. 
To first order in the perturbations, \eq{eq:intr_s_int1} becomes
\begin{align}
\pd_t \delta \phi + \vec{v}_0 \cdot \vec{\nabla} \delta \phi + \gamma'(\rho_0) \, \delta \rho = 0,
\end{align}
where $\vec{v}_0 \equiv \vec{\nabla} \phi_0$ is the velocity field of the unperturbed solution. 
The continuity equation gives to the same order
\begin{align}
\pd_t \delta \rho + \vec{\nabla} \cdot \lp \vec{v}_0 \,  \delta \rho \rp + \vec{\nabla} \cdot \lp \rho_0 \, \vec{\nabla} \delta \phi \rp = 0. 
\end{align}
Combining these two equations, one obtains
\begin{align}\label{eq:intr_soundwaves}
\pd_t \lp \frac{1}{\gamma'(\rho_0)} \pd_t \delta \phi \rp
+ \pd_t \lp \frac{\vec{v}_0}{\gamma'(\rho_0)} \cdot \vec{\nabla} \delta \phi \rp 
+ \vec{\nabla} \cdot \lp \frac{\vec{v}_0}{\gamma'(\rho_0)} \, \pd_t \delta \phi \rp 
\hspace{3 cm} \nn
+ \vec{\nabla} \cdot \lp \lp \vec{v}_0 \frac{1}{\gamma'(\rho_0)} \vec{v}_0 \cdot - \rho_0 \rp \vec{\nabla} \delta \phi \rp
= 0. 
\end{align}
The local sound velocity in the frame of the fluid, $c_0$, is obtained by setting $\vec{v}_0 = \vec{0}$ and neglecting $\vec{\nabla} \rho_0$ as well as $\pd_t \rho_0$.  
One obtains $c_0^2 = \rho_0 \, \gamma'(\rho_0)$. 

Equation~\eqref{eq:intr_soundwaves} may be rewritten as $\pd_\mu \lp F^{\mu \nu} \pd_\nu \delta \phi \rp = 0$ with $F^{00} = 1/\gamma'(\rho_0)$, $F^{0i} = F^{i0} = v_0^i/\gamma'(\rho_0)$, and $F^{ij} = v_0^i v_0^j  / \gamma'(\rho_0) - \delta_{i,j} \, \rho_0$,
where $\delta$ denotes the Kronecker delta. 
A straightforward calculation shows that the determinant of this matrix is $(-\rho_0)^d / \gamma'(\rho_0)$, where $d$ is the number of space dimensions. 
To make link with field theory in curved space-time, one must find a metric $g_{\mu \nu}$ such that $\sqrt{\abs{g}} g^{\mu \nu} = F^{\mu \nu}$. 
That is, we need $\abs{g}^{(d+1) / 2 - 1} = \rho_0^d / \abs{\gamma'(\rho_0)}$, i.e., $\abs{g} = \rho_0^{2 d / (d-1)} \, \gamma'(\rho_0)^{2 / (1-d)}$. 
Assuming $d > 1$, one possibility is
\begin{align}
g^{\mu \nu} = \rho_0^{d / (1-d)} \, \gamma'(\rho_0)^{1 / (d-1)} \, F^{\mu \nu}. 
\end{align}
$\delta \phi$ then obeys the d'Alembert equation
\begin{align}
\nabla_\mu \nabla^\mu \phi = 
\frac{1}{\sqrt{\abs{g}}} \pd_\mu \lp \sqrt{\abs{g}} \, g^{\mu \nu} \pd_\nu \delta \phi \rp = 0
\end{align}
in an effective curved spacetime with metric $g_{\mu \nu}$ given by the inverse of $g^{\mu \nu}$. 
A straightforward calculation shows that $\dd s^2 \equiv g_{\mu \nu} \, \dd x^\mu \, \dd x^\nu$ reads
\begin{align}
\dd s^2 = \lp \frac{\rho_0}{\gamma' \lp \rho_0 \rp} \rp^{1 / (d - 1)} \, \lp 
	\lp c_0^2 - \vec{v}_0^2 \rp \, \dd t^2
	+ 2 \, \vec{v}_0 \cdot \dd \vec{x} \, \dd t
	- \dd \vec{x} \cdot \dd \vec{x}
\rp .
\end{align}

For simplicity, let us assume that $\rho_0$ $\vec{v}_0$ depend only on one Cartesian space coordinate $x$, and that $\vec{v}_0$ is parallel to $\pd_x$. 
Denoting collectively by $\vec{x}_\perp$ the coordinates in the orthogonal directions, the metric becomes
\begin{align}
\dd s^2 = \lp \frac{\rho_0}{\gamma' \lp \rho_0 \rp} \rp^{1 / (d - 1)} \, \lp 
	c_0^2 \, \dd t^2
	- \lp v_0 \, \dd t - \dd x \rp^2
	- \dd \vec{x}_\perp \cdot \dd \vec{x}_\perp
\rp .
\end{align}
The characteristic lines in the $(t,x)$ plane are given by
\begin{align}
\frac{\dd x}{\dd t} = \lp v_0 \pm c_0 \rp ,
\end{align}
from which one defines the light-cone coordinates
\begin{align}
U \equiv t + \int^x \frac{\dd x'}{c_0 - v_0}, \, V \equiv t - \int^x \frac{\dd x'}{c_0 + v_0}. 
\end{align}
The metric can then be written as
\begin{align}
\dd s^2 = \lp \frac{\rho_0}{\gamma' \lp \rho_0 \rp} \rp^{1 / (d - 1)} \, \lp 
	\lp c_0^2-v_0^2 \rp \, \dd U \, \dd V
	- \dd \vec{x}_\perp \cdot \dd \vec{x}_\perp
\rp .
\end{align}
To write this metric in a form closer to the Schwarzschild one~\eqref{eq:Scharzschild}, let us define the new time coordinate
\begin{align}
T \equiv t + \int^x \frac{v_0}{c_0^2 - v_0^2} \, \dd x'.
\end{align}
We have: 
\begin{align}
\dd U = \dd T + \frac{c_0}{c_0^2 - v_0^2} \, \dd x , \, 
\dd V = \dd T - \frac{c_0}{c_0^2 - v_0^2} \, \dd x .
\end{align}
The metric thus becomes
\begin{align}
\dd s^2 = \lp \frac{\rho_0}{\gamma' \lp \rho_0 \rp} \rp^{1 / (d - 1)} \, \lp 
	\lp 1 - \frac{v_0^2}{c_0^2} \rp \, c_0^2 \,  \dd T^2
	- \frac{\dd x^2}{1 - \frac{v_0^2}{c_0^2}}
	- \dd \vec{x}_\perp \cdot \dd \vec{x}_\perp
\rp .
\end{align}
Up to a smooth prefactor, the metric in the $(T,x)$ plane has the same structure as the Schwarzschild one, with $c_0$ playing the role of the celerity of light\footnote{The fact that $c_0$ can now depend on $x$ does not affect the subsequent discussion provided it is smooth at the point where $v_0 = c_0$.} and $v_0^2 / c_0^2$ that of $r_S / r$. 
In particular, assuming for definiteness $v_0 > 0$, there is a horizon at any point $x_h$ where $v_0 / c_0$ crosses the value $1$. 
Close to this point, we have
\begin{align}
1 - \frac{v_0^2}{c_0^2} = 2 \frac{v_0'(x_H) - c_0'(x_H)}{c_0(x_H)} \, \lp x - x_H \rp + O \lp \lp \frac{x - x_h}{x_0} \rp^2 \rp,
\end{align}
where $x_0$ is the typical scale of variation of $v_0 - c_0$
By comparison,
\begin{align*}
1 - \frac{r_S}{r} = \frac{1}{r_S} \lp r - r_S \rp + O \lp \lp \frac{r}{r_S} - 1 \rp^2 \rp. 
\end{align*}
Close to the horizon, $\abs{v_0'-c_0'}/c_0$ thus plays the role of $1 / (2 r_S)$. 
As explained in the previous subsection, this leads to a Hawking temperature of (putting back the factor of $c$ using that $T_H$ is a frequency in units where $k_B = \hbar = 1$) $T_H = c / (4 \pi r_S)$. 
The Hawking temperature in the present analogue model is thus
\begin{align}
T_H = \frac{1}{2 \pi} \abs{v_0'(x_h) - c_0'(x_h)}. 
\end{align}

Notice that the horizon at $x = x_H$ corresponds to either a black hole or a white hole depending on the sign of $v_0' - c_0'$. 
To see this, let us consider the characteristics of counter-propagating waves close to $x_h$, represented in \fig{fig:intro_char}. 
They are given by
\begin{align}
\frac{\dd x}{\dd t} = \lp v_0 - c_0 \rp \approx \lp v_0' \lp x_h \rp - c_0' \lp x_h \rp \rp \, \lp x - x_h \rp .
\end{align}
The general solution is
\begin{align}
x(t) \approx x_h + A \, \e^{\lp v_0' \lp x_h \rp - c_0' \lp x_h \rp \rp \, t}, \, A \in \mathbb{R}. 
\end{align}
When increasing $t$, the characteristics move away from the horizon if $v_0' \lp x_h \rp - c_0' \lp x_h \rp > 0$, showing the exponential behavior typical of black holes. 
If $v_0' \lp x_h \rp - c_0' \lp x_h \rp < 0$, they instead approach the horizon, the corresponding wave vector being exponentially blueshifted as occurs near a white hole. 

\begin{figure}
\centering
\includegraphics[width = 0.49 \linewidth]{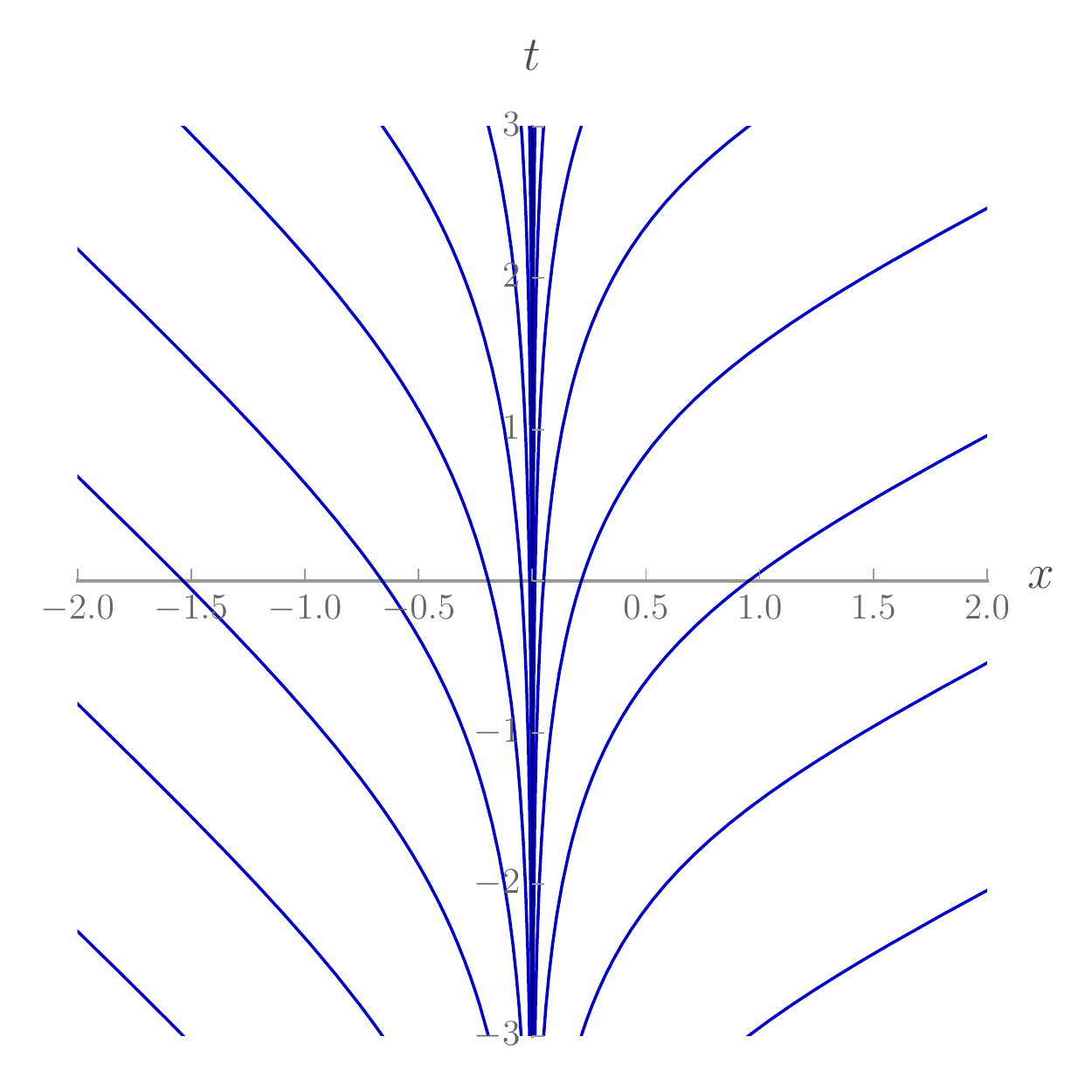}
\includegraphics[width = 0.49 \linewidth]{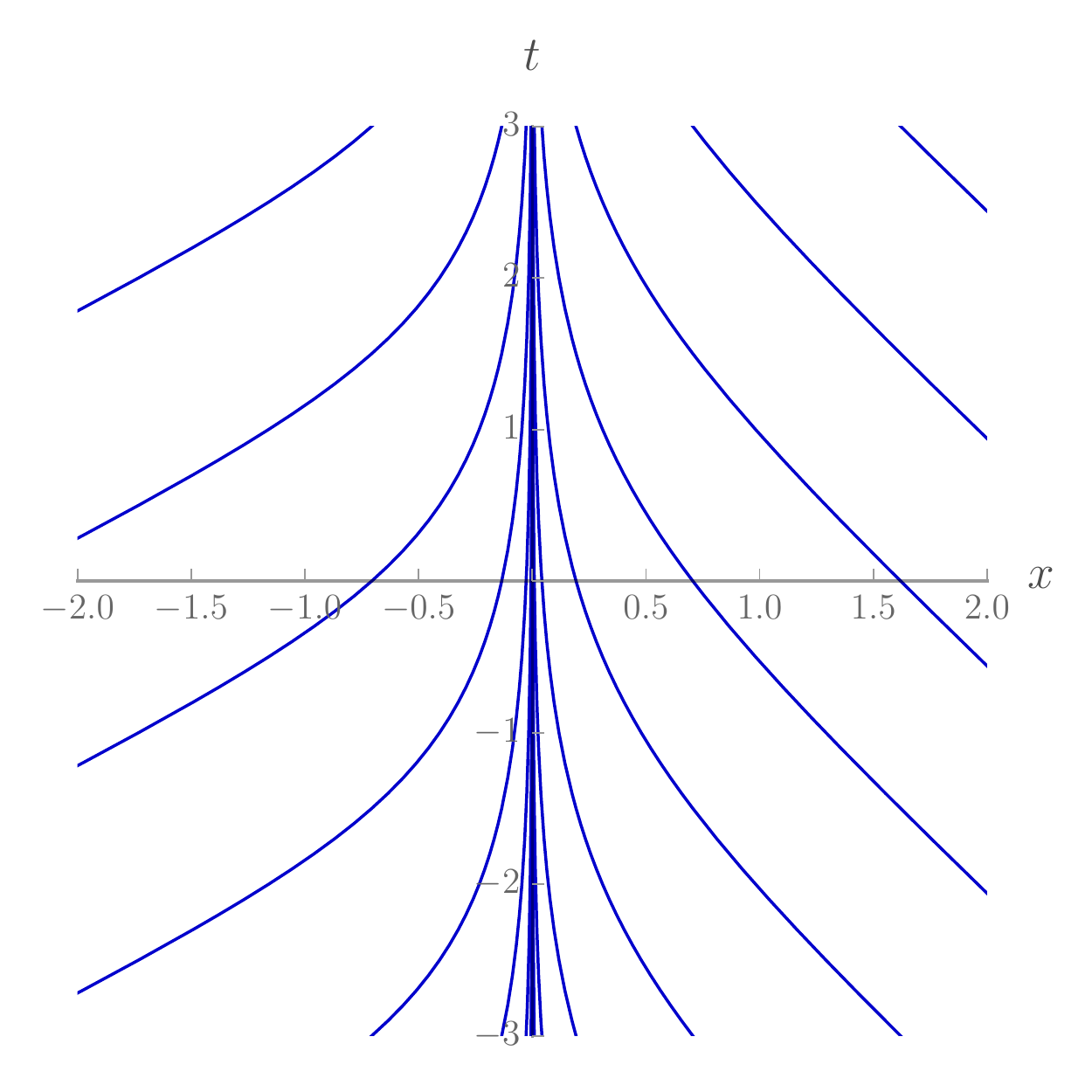}
\includegraphics[width = 0.49 \linewidth]{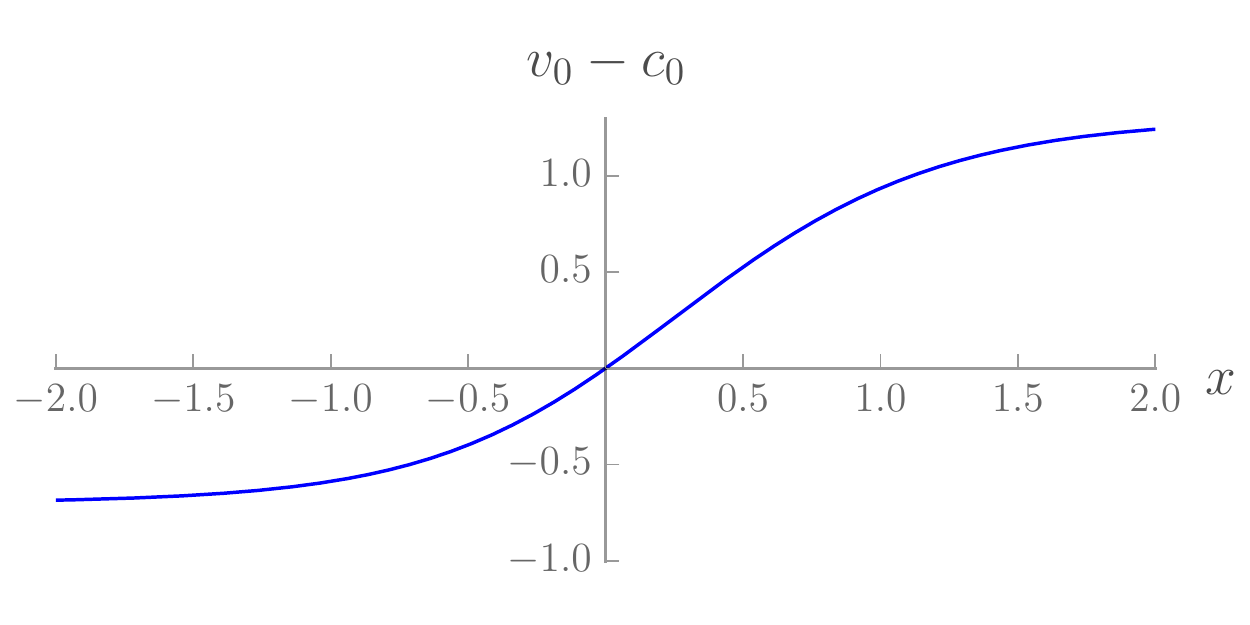}
\includegraphics[width = 0.49 \linewidth]{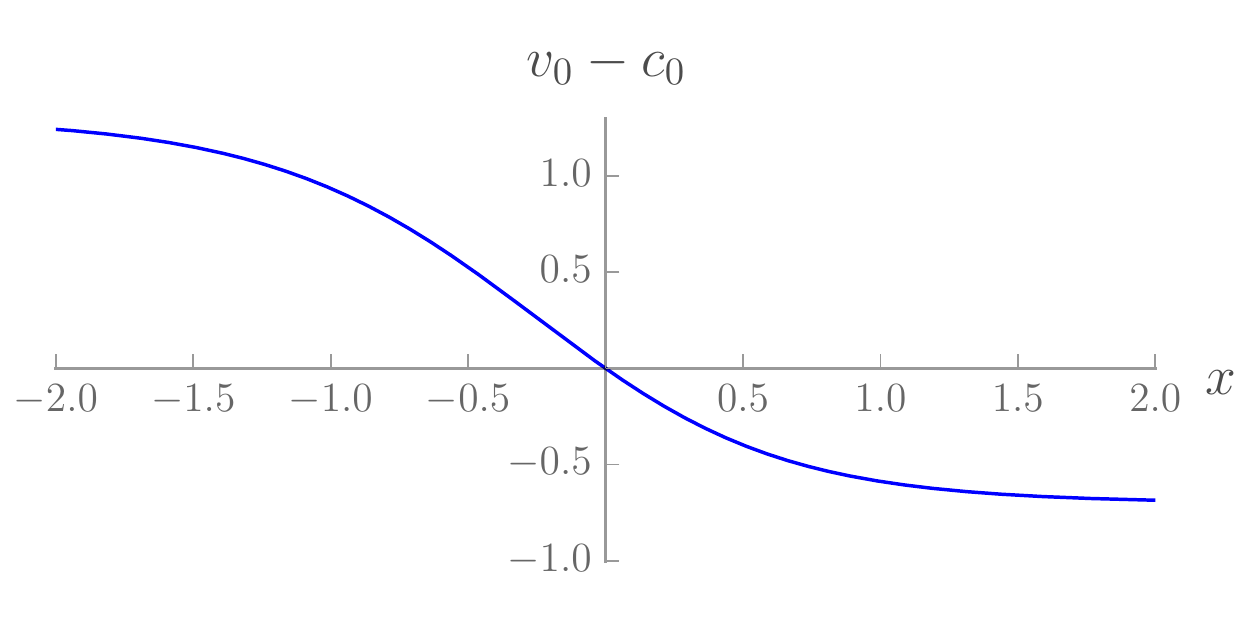}
\caption{Characteristics of \eqref{eq:intr_soundwaves} in the $(t,x)$ plane for $\vec{v}_0$ oriented along the $x$ direction. The left panel corresponds to a “black hole” flow with $v_0(x) - c_0(x) = \tanh(x) + 0.3 \tanh(x)^2$. The right panel corresponds to a “white hole” flow with $v_0(x) - c_0(x) = -\tanh(x) + 0.3 \tanh(x)^2$. In both cases, the analogue horizon is at $x=0$. The bottom panels show the corresponding profiles of $v_0 - c_0$.}\label{fig:intro_char}
\end{figure}

To end this subsection, let us briefly comment on the case $d = 1$. Then, the linear equation on $\delta \phi$ can generally not be written as a d'Alembert equation. 
However, it can be turned to a d'Alembert equation by multiplying $F^{\mu \nu}$ by $\sqrt{\gamma_0' \lp \rho_0 \rp / \rho_0}$, which is smooth and nonvanishing at the horizon(s) and thus does not modify the behavior of the waves in its vicinity. 
The derivation of the Hawking effect of the previous section therefore still applies. 

\subsection{Bose-Einstein condensates}

The phenomenon of Bose-Einstein condensation occurs when a gas of bosons is cooled to sufficiently low temperatures for a large number of particles to be in the same quantum state, (the ground state of the system at equilibrium), forming a condensate.   
In that case, the classical description of the gas fails: 
The condensed particles instead follow a collective quantum behavior exhibiting peculiar properties like superfluidity. 
A detailed account of the theoretical description and experimental relevance of Bose-Einstein condensates (BEC) can be found in the textbook~\cite{pitaevskii2003bose}. 
In the present subsection I only mention a few aspects which will be important for the following chapters. 
For definiteness, I consider here a condensate of nonrelativistic atoms, although condensation also occurs for bosonic molecules~\cite{PhysRevLett.91.250401} and relativistic Bose gas (see for instance~\cite{PhysRevLett.99.200406} and references therein). 

When the density and interatomic interactions are sufficiently small, a BEC can be described by a (quantum) complex scalar field $\hat{\psi}$, with the action
\begin{align} \label{intr:action_GP}
S = \int \dd t \int \dd^d x \, \lp 
	\frac{\ii}{2} \lp \psi^* \, \pd_t \psi - \psi \, \pd_t \psi^* \rp
	- \frac{1}{2} \, \abs{\vec{\nabla} \psi}^2 
	- V \lp \vec{x} \rp \, \abs{\psi}^2
	- \frac{g \lp \vec{x} \rp}{2} \, \abs{\psi}^4
\rp,
\end{align}
where $V$ is the external potential, $g$ characterizes local interactions between two particles, and we use units in which Planck's constant $\hbar$ and the mass of the atoms are equal to $1$. 
Experimentally, $V$ can be tuned by applying a magnetic field and using the polarisability of the atoms. 
The interaction strength $g$ is usually taken to be uniform, but can also be varied by changing the internal state of the atoms \textit{via} a resonant magnetic field~\cite{RevModPhys.82.1225}. 

Using a semi-classical approximation, the gas is described by a set of quantum states on which the action of $\hat{\psi}$ can be approximated by the identity operator times a scalar function $\psi_0$. 
More precisely, we write $\hat{\psi}$ as
\begin{align}
\hat{\psi} \lp t, \vec{x} \rp = \psi_0 \lp t, \vec{x} \rp + \delta \hat{\psi} \lp t, \vec{x} \rp,
\end{align}
where the norm of the restriction of $\delta \hat{\psi}$ to the set of relevant states is sufficiently small to be negligible in a first approximation. 
Physically, $\psi_0$ describes the condensed atoms. 
It thus corresponds to the wave function of the ground state, deformed by the interactions between particles and normalized so that $\int \abs{\psi_0}^2 \, \dd^dx$ gives the total number of condensed atoms. 
$\rho_0 \equiv \abs{\psi_0}^2$ is the local number density of particles in the condensate and $\vec{v}_0 \equiv \vec{\nabla} \arg \psi_0$ their velocity.~\footnote{Notice that $\vec{v}_0$, being defined as the gradient of a scalar function, is locally irrotational.}
$\delta \hat{\psi}$ describes the uncondensed atoms. 

To leading order, i.e., neglecting $\delta \hat{\psi}$, extremizing the action \eqref{intr:action_GP} gives the Gross-Pitaevskii equation (GPE):
\begin{align} \label{intr:eq_psi0}
\ii \pd_t \psi_0 \lp t, \vec{x} \rp = \frac{1}{2} \Delta \psi_0 \lp t, \vec{x} \rp  + V \lp \vec{x} \rp \, \psi_0 \lp t, \vec{x} \rp  + g \lp \vec{x} \rp  \, \abs{\psi_0 \lp t, \vec{x} \rp }^2 \psi_0 \lp t, \vec{x} \rp . 
\end{align}
Neglecting the nonlinear term, one recovers the time-dependent Schrödinger equation. 
The term $g \, \abs{\psi_0}^2 \psi_0$ accounts for pointwise interactions between the atoms in the same quantum state. 
In all this thesis, $g$ is assumed to be positive, corresponding to repulsive interactions. 
Negative values of $g$ are less interesting for analogue gravity experiments because they tend to make solutions with asymptotically homogeneous, nonvanishing values of $\rho$ unstable to collapse.
Indeed, increasing the local density lowers the total energy if $g < 0$. 

It is also interesting to consider the hydrodynamic-like equations on $\rho_0$ and $\vec{v}_0$. 
Writing $\psi_0 = \sqrt{\rho_0} \, \exp \lp \ii \theta \rp$ with $\theta  = \int \vec{v_0} \cdot \dd \vec{x}$, \eqref{intr:eq_psi0} becomes
\begin{equation}
\lb 
\begin{aligned}
& \displaystyle{\pd_t \rho_0 + \vec{\nabla} \cdot \lp \rho_0 \, \vec{v}_0 \rp = 0} \\
& \displaystyle{\pd_t \theta = \frac{1}{2 \, \sqrt{\rho_0}} \Delta \sqrt{\rho_0} - \frac{\vec{v}_0^2}{2} - V - g \, \rho}
\end{aligned}
\right. .
\end{equation}
The first line is simply the conservation law for the number of atoms in the condensate (continuity equation). 
Taking the gradient of the second line, one obtains the Euler equation
\begin{align}
\lp \pd_t + \vec{v}_0 \cdot \vec{\nabla} \rp \vec{v}_0 = - \vec{\nabla} V_\textrm{eff},
\end{align}
in the effective potential $V_\textrm{eff} \equiv V + g \, \rho - \lp \Delta \sqrt{\rho_0} \rp / (2 \sqrt{\rho_0})$. 
In particular, the absence of viscous term shows that the flow is superfluid (although a massive impurity can still see its energy dissipated through the emission of linear waves if moving faster than the sound speed in the local rest frame of the fluid, or through nonlinear effects~\cite{Hakim1997}). 

To study the non-condensed part of the gas, it is convenient to work with the relative perturbation $\hat{\phi} \equiv \psi_0^{-1} \, \delta \hat{\psi}$. 
Expanding the action to second order and extremizing the result, one obtains the Bogoliubov-de Gennes (BdG) equation
\begin{equation}\label{int:eq:BdG}
\ii \lp \pd_t + \vec{v}_0 \cdot \vec{\nabla} \rp \hat{\phi} = - \frac{1}{2 \, \rho_0} \vec{\nabla} \cdot \lp \rho_0 \, \vec{\nabla} \hat{\phi} \rp + g \, \rho_0 \, \lp \hat{\phi} + \hat{\phi}^\dagger \rp + O \lp \hat{\phi}^2 \rp. 
\end{equation}
In the following, we shall drop the terms in $O \lp \hat{\phi}^2 \rp$ to simplify the notations. 
For simplicity, let us now discuss the classical version of \eqref{int:eq:BdG}. 
As usual for bosons in linear quantum field theory, the field $\phi$ can be quantized \textit{a posteriori} once a basis of classical solutions is found by replacing the coefficients in the mode decomposition by an operator obeying suitable commutation relations, along the lines of subsection~\ref{subsub:quant}. 

To obtain the analogue metric, let us first separate $\phi$ into its real and imaginary parts. 
One obtains the system of coupled equations:
\begin{equation}\label{int:BEC:int1}
\lb 
\begin{aligned}
& \lp \pd_t + \vec{v}_0 \cdot \vec{\nabla} \rp \, \Im(\phi) = \frac{1}{2 \, \rho_0} \vec{\nabla} \cdot \lp \rho_0 \, \vec{\nabla} \Re(\phi) \rp - 2 \, g \, \rho_0 \, \Re(\phi) \\
& \lp \pd_t + \vec{v}_0 \cdot \vec{\nabla} \rp \, \Re(\phi) = - \frac{1}{2 \, \rho_0} \vec{\nabla} \cdot \lp \rho_0 \, \vec{\nabla} \Im(\phi) \rp
\end{aligned}
\right. .
\end{equation}
Dividing the first equation by $g \, \rho_0$, applying the operator $\pd_t + \vec{v}_0 \cdot \vec{\nabla}$, and using the second equation gives
\begin{align} \label{int:BEC:int2}
\lp \pd_t + \vec{v}_0 \cdot \vec{\nabla} \rp \lp \frac{1}{g \, \rho_0} \lp \pd_t + \vec{v}_0 \cdot \vec{\nabla} \rp \, \Im(\phi) \rp = 
	\frac{1}{2} \lp \pd_t + \vec{v}_0 \cdot \vec{\nabla} \rp \lp \frac{1}{g \, \rho_0^2} \vec{\nabla} \cdot \lp \rho_0 \, \vec{\nabla} \Re(\phi) \rp \rp
	\nn 
	+ \frac{1}{\rho_0} \vec{\nabla} \cdot \lp \rho_0 \, \vec{\nabla} \Im(\phi) \rp.
\end{align}
The analogy with fields in a curved space-time holds in the long-wavelength approximation. 
In this limit, from the first equation in \eqref{int:BEC:int1} and the dispersion relation (see for instance~\cite{pitaevskii2003bose} and Eq.~\eqref{eq:DRBdG} below), we find $\abs{\Re(\phi)} \ll \abs{\Im(\phi)}$. 
This relation extends to the space- and time-derivatives. 
One can thus neglect the first term in the right-hand side of \eqref{int:BEC:int2}. 
Moreover, for any smooth function $f$, we have
\begin{align}
\lp \pd_t + \vec{v}_0 \cdot \vec{\nabla} \rp \lp \frac{1}{g \, \rho_0} \,  f \rp = 
\frac{1}{g \, \rho_0} \lp \pd_t + \vec{v}_0 \cdot \vec{\nabla} \rp \, f  
	- \frac{1}{g \, \rho_0^2} \lp \pd_t \rho_0 + \vec{v}_0 \cdot \vec{\nabla} \rho_0 \rp \, f.
\end{align}
Using the continuity equation, this becomes
\begin{align}
\lp \pd_t + \vec{v}_0 \cdot \vec{\nabla} \rp \lp \frac{1}{g \, \rho_0} \, f \rp = 
\frac{1}{g \, \rho_0} \lp \pd_t + \vec{v}_0 \cdot \vec{\nabla} \rp \, f  
	+ \frac{1}{g \, \rho_0} \lp \vec{\nabla} \cdot \vec{v}_0 \rp \, f = 
\frac{1}{g \, \rho_0} \lp \pd_t f + \vec{\nabla} \cdot \lp \vec{v}_0 \, f \rp \rp. 
\end{align}
This allows us to rewrite \eqref{int:BEC:int2} as
\begin{align} \label{int:eq:int3}
\pd_t^2 \Im(\phi)
+ \pd_t \lp \vec{v}_0 \cdot \vec{\nabla} \Im(\phi) \rp
+ \vec{\nabla} \cdot \lp \vec{v}_0 \, \pd_t \Im(\phi) \rp
+ \vec{\nabla} \cdot \lp \vec{v}_0 \, \vec{v}_0 \cdot \vec{\nabla} \Im(\phi) \rp
- g \, \vec{\nabla} \cdot \lp \rho_0 \, \vec{\nabla} \Im(\phi) \rp \approx 0.
\end{align}
In a Cartesian coordinate system, this may be written as $\pd_\mu \lp F^{\mu \nu} \pd_\nu \Im(\phi) \rp = 0$, where $F^{00} = 1$, $F^{0i} = F^{i0} = v_{0,i}$ is the component of the velocity in the $i^\textrm{th}$ direction, and $F^{ij} = v_{0,i} \, v_{0,j} - g \, \rho_0 \, \delta_{ij}$. 
One can also read from this equation the sound velocity $c_0$ in the fluid frame, given by $c_0^2 = g \, \rho_0$. 
As done above for sound waves in a classical gas, if $d \neq 1$ one can define a metric $g_{\mu \nu}$ such that \eqref{int:eq:int3} becomes
\begin{align} \label{int:eq:int4}
\nabla_\mu \nabla^\mu \Im \lp \phi \rp = 
\frac{1}{\sqrt{\abs{\det\lp g_{\alpha \beta} \rp}}} \pd_\mu \lp \sqrt{\abs{\det\lp g_{\alpha \beta} \rp}} \, g^{\mu \nu} \pd_\nu \Im (\phi) \rp \approx 0.
\end{align}
In $d$ space dimensions, this gives $\abs{\det\lp g^{\mu \nu} \rp}^{(1 - d)/2} \propto \abs{\det\lp F^{\mu \nu} \rp} = c_0^{2d}$. 
One possible choice for $d \neq 1$ is $g^{\mu \nu} = c_0^{2d/(1-d)} F^{\mu \nu}$. 
As also discussed above, such a choice does not exist for $d = 1$, but the analogue Hawking mechanism is still expected to occur since the solutions of equations \eqref{int:eq:int3} and \eqref{int:eq:int4} have the same behavior when $\abs{\vec{v}_0} \approx c_0$. 

Importantly, \eqref{int:eq:int4} is valid only for very long wavelengths. 
Returning to \eqref{int:eq:BdG}, and looking for solutions of the form
\begin{align}
\phi \lp t,\vec{x} \rp = U \, \e^{\ii \lp \vec{k} \cdot \vec{x} - \om \, t \rp} + V^* \, \e^{-\ii \lp \vec{k}^* \cdot \vec{x} - \om^* \, t \rp},
\end{align}
one finds that nontrivial solutions exist if and only if (see Section~\ref{sub:BdGDR})
\begin{align}\label{eq:DRBdG}
\lp \om - \vec{v}_0 \cdot \vec{k} \rp^2 = c_0^2 \, k^2 + \frac{k^4}{4}. 
\end{align}
This dispersion relation is shown in \fig{fig:DR}. 
Because of the quartic term, the group and phase velocities of small-wavelength perturbations are both larger than $c$ in the fluid frame. 
In the limit $k \to \infty$, one recovers the relation $\abs{\om} \sim k^2 / 2$ for a massive, non-relativistic particle. 
Interestingly, this dispersive term regularizes the red- or blue-shift experienced by a wave approaching an analogue horizon~\cite{Jacobson:1993hn,Robertson:2011xp,Robertson:2012ku}. 

This system has two important advantages with respect to the original one of~\cite{Unruh:1980cg}. 
First, BECs are by essence low-temperature systems, with temperatures $T$ usually smaller than the $\mathrm{\mu K}$. 
They are thus more suitable to observe the quantum Hawking radiation, which will be larger than the thermal noise if $T < T_H$.  
Second, the absence of viscosity and vorticity reduces parasite effects like boundary layers, generation of vortices, and turbulence. 

One remark is in order concerning the case $d = 1$, which will often be considered in the following chapters. 
Strictly speaking, there is no phase transition for systems of particles with only local interactions in infinite space in 1 or 2 space dimensions~\cite{PhysRevLett.17.1133,PhysRev.158.383}, and thus no Bose-Einstein condensation. 
However, quasi-condensates can form provided the temperature is sufficiently low and the density of particles is large enough. 
More precisely, the condition for quasi-condensation in $d=1$ is $\rho_0 \, \la, \rho_0 \, \xi \gg 1$, see~\cite{PhysRevA.67.053615}, where $\la$ is the thermal de Bröglie wavelength and $\xi$ the healing length of the condensate, respectively given by $\la = \sqrt{2\pi \hbar^2 / (m k_BT)}$ and $\xi = \hbar / \sqrt{m \mu}$, where $\mu$ is the chemical potential and $m$ is the atomic mass. 
Locally, the physics of a quasi-condensate is very similar to that of a true condensate, the main difference being that the correlation of the phase $\arg{\psi_0}$ between two points decreases, exponentially in their separation at finite temperature (with a length scale of order $\rho \, \la^2$, see Eq.~(188) in~\cite{PhysRevA.67.053615}) and polynomially at zero temperature, with an exponent in $1 / (\rho \, \xi)$ (see Eq.~(185) in that reference). 
In particular, the Bogoliubov description of perturbations sketched above remains valid over sufficiently short length scales. 
In the following we shall thus mostly ignore these additional complications and assume the system can be approximated as a true condensate.

Before moving on to water waves, I would like to mention two recent experiments realized by J.~Steinhauer. 
In the first one~\cite{BHLaser-Jeff} was reported the formation of a “black hole laser” configuration, where two analogue horizons form a resonant cavity in which the self-amplified Hawking radiation grows exponentially in time, see~\cite{Corley:1998rk,Leonhardt:2008js}, in a condensate of rubidium atoms. 
One instability was observed, which was interpreted as resulting from the self-amplified Hawking radiation. 
This may turn out to be the first observed manifestation of the Hawking mechanism in the quantum regime, although there are some controversies concerning the origin of the observed instability~\cite{Tettamanti:2016ntx,Wang:2016jaj,Steinhauer:2016hfa}. 
The second experiment~\cite{Steinhauer:2015saa} was devoted to the quantum Hawking radiation from an analogue black hole flow. 
J.~Steinhauer measured both the amplitude of the density fluctuations generated from the horizon and their correlations to ascertain their quantum origin. 
Although there remain a few differences between the theoretical predictions and experimental results (see Chapter~\ref{ch:concl}, Section~\ref{sub:JE}), if confirmed, this will be a very important step forward in the analogue gravity program, constituting the first direct observation of the quantum Hawking radiation.

\subsection{Water waves}

The third system we consider is surface gravity waves in water or another liquid. 
One practical advantage over cold atoms is that it is macroscopic and can be realized at room temperature. 
Moreover, the velocity of water waves is usually significantly smaller than the sound speed in room-temperature air or water, allowing for the realization of analogue horizons with velocities of the order of $1 \mathrm{m.s^{-1}}$. 

To obtain a simple description of the fluid and its perturbations, throughout this thesis we work under the following approximations. 
First, we model the system by a \textit{perfect fluid}, i.e., entirely described by its local density $\rho$ and pressure $p$. 
In particular, we neglect heat conduction phenomena, viscosity, and shear stresses.  
Second, we assume the fluid is \textit{incompressible}, i.e., that $\rho$ is a constant. 
Third, we assume the flow is \textit{laminar}, i.e., can be described as the union of an infinite number of (non-intersecting) streamlines, with no gaps between the streamlines. 
Fourth, the velocity field $\vec{v}$ is assumed to point everywhere in a fixed plane containing the vertical direction and the flow is homogeneous in the direction orthogonal to this plane, making it effectively two-dimensional.
Finally, we assume the flow is \textit{irrotational}, i.e., $\vec{\nabla} \times \vec{v} = \vec{0}$.  

These five hypotheses all break down to some extent in experiments. 
For instance, real flows have a non-vanishing viscosity giving rise to boundary layers at the interfaces with stationary surfaces and vortices or turbulence can be generated when an incident laminar, irrotational flow passes over an obstacle to make it inhomogeneous. 
However, this simple model has the important advantage of leading to an explicit wave equation~\cite{Unruh:2012ve,Coutant:2012mf} which can be used to study analytically~\cite{Coutant:2016vsf} or numerically the scattering and analogue Hawking effect. 

We consider a flow of water (or another liquid) in a long, linear flume with parallel planar, vertical sides. 
The system is assumed to be invariant under translations in the transverse horizontal direction. 
The wave equation under the above approximations has been derived in~\cite{Unruh:2012ve,Coutant:2012mf}. 
We give a more pedestrian proof along the same lines in Chapter~\ref{ch:probing}, Section~\ref{sub:waveeqder}. 
For the time being, let us only mention that in the limit where the vertical velocity $v_y$ is small and neglecting dispersive terms, it becomes
\begin{align}
\left[ \lp \pd_t + \pd_x \times v_x \rp \lp \pd_t + v_x \, \pd_x \rp - g \, \pd_x \times h \times \pd_x \right] \delta \phi \approx 0,
\end{align}
where $h$ is the water depth, $v_x$ the horizontal velocity at the free surface, $g$ the gravitational acceleration, $x$ a Cartesian coordinate along the direction of the flume, $\delta \phi$ the perturbation of the velocity potential, and $\times$ denotes multiplication in the sense of operators (so that the last term, for instance, is equal to $g \, \pd_x \lp h \, \pd_x \delta \phi \rp$ and not $g \, h' \, \pd_x \delta \phi$).  
This equation may be written as $\pd_\mu \lp F^{\mu \nu} \pd_\nu \delta \phi \rp = 0$, where $F^{00} = 1$, $F^{01} = F^{10} = v_x$, $F^{11} = v_x^2 - c^2$, and $c^2 \equiv g \, h$ is the velocity of long-wavelength waves in the frame moving with the fluid. 
As discussed above, up to multiplication of $F^{\mu \nu}$ by a smooth function (or, equivalently, up to viewing this equation as a $(1+1)$-dimensional reduction of a higher-dimensional one), this is the d'Alembert equation in an effective curved space-time which has a horizon provided the flow is transcritical, i.e., provided $F \equiv v_x^2 / c^2$ crosses the value $1$.~\footnote{A flow is said \textit{subcritical} if $F$ is everywhere smaller than $1$, \textit{supercritical} if it is everywhere larger than $1$, and \textit{transcritical} if it takes values smaller than and larger than $1$ when varying $x$.}
Like in the case of cold atoms, in practice dispersive terms regularize the infinite redshift (respectively blueshift) close to a black hole (respectively white hole) horizon.~\footnote{One important difference is that the dispersion relation of water waves is subluminal, meaning that high-frequency waves move slower than low-frequency ones, while they move faster in a BEC.}

While thermal fluctuations and other sources of classical noise like the vibrations of the pump producing the flow in practice make it virtually impossible to observe the quantum Hawking effect in this system, the latter may be used to study its classical counterpart. 
Indeed, as explained in subsection~\ref{sec:HR}, the phenomenon of Hawking radiation is directly related to the structure of the (classical) solutions of the field equations. 
In the present context, these also manifest themselves through wave amplification: as explained in more details in Chapter~\ref{ch:probing}, sending a counter-propagating wave in a flow with an analogue horizon can result in the emission of two reflected waves: one with an energy greater than that of the incident wave, and a negative-energy one. 
In fact, the Hawking effect can be seen as amplification of quantum fluctuations through the same mechanism, turning them into pairs of real particles. 

The first observation of the conversion from an incoming positive-energy wave to a negative-energy one in a water flume was reported in~\cite{Rousseaux:2007is}. 
Although the flow had seemingly no analogue horizon, the Froude number $F = v_x / c$ remaining smaller than unity, waves with sufficiently high frequencies (whose velocity is reduced by dispersive effects) were blocked and reflected over an immersed obstacle. 
Negative-energy waves were produced along with the reflected ones and detected by the experimental team.
In~\cite{Weinfurtner:2010nu} the ratio of the two scattering coefficients determining the amplitude of the positive- and negative-energy reflected waves was measured for different frequencies and found to follow a Boltzmann law to a good accuracy. 
This behavior was interpreted as a signature of the thermal character of the spectrum. 
As the flow was apparently subcritical, with $F(x) < 1$ for all $x$, this conclusion raises some questions concerning the relevant definition of temperature, see Chapter~\ref{ch:probing}. 
However, the experimental results mentioned in that reference are in good agreement with numerical estimates. 
Several new experiments have been performed in Poitiers during this thesis, which I mention in Chapter~\ref{ch:concl}, Sections~\ref{sub:exp_paper_1} and~\ref{sub:exp_paper_2}. 
The first one focused on the low-frequency transmission in subcritical flows, showing that the transmission coefficient shows a sharp transition from $0$ to $1$ when decreasing the angular frequency below the minimum one $\om_\textrm{min}$ for which waves are blocked. 
The second one was devoted to an independent measurement of the two scattering coefficients used in the characterization of the analogue Hawking effect, using correlations between the incident and reflected waves to rseparate the contribution of the incident wave from that of the noise. 
This method also allows to extract information on the scattering from the random fluctuations of the height of the free surface when no macroscopic wave is sent.  

\subsection{What can be learned from the gravitational ``analogy''?}
\label{sec_intr:WcbblftA}

Analogue models of black holes cover wide areas of non-relativistic physics. 
Apart from classical or quantum gases and water waves, they have also been proposed in optical fibers~\cite{Philbin1367}, polariton condensates~\cite{PhysRevB.86.144505,PhysRevLett.114.036402}, and magnetohydrodynamics~\cite{Noda:2016pva}. 
This diversity of models shows that the mechanism of Hawking radiation is much more general than could be expected from its original derivation and raises the question of the extent to which the results obtained in one system can be applied to the others -- or to astrophysical black holes. 
As discussed in~\cite{Parentani:2002bd} and emphasized in the recent reference~\cite{Thebault:2016udp}, this is a nontrivial question whose answer, in the latter case, depends on the hypotheses one makes on the (unknown) quantum theory of gravity.

Let me first emphasize that the “analogy” on which these models are based is more a mathematical correspondence than an analogy in the usual sense. 
Indeed, as explained above, their linear perturbations obey, in the long-wavelength limit, the same equation as a scalar field in curved spacetime does. 
Under this approximation, the mechanism at play is thus literally the Hawking effect, not a mere illustration of it. 
For this reason, results drawn from some “analogue” model which depend only on the behavior of the fields in this limit can be directly extended to other models, as well as to the gravitational realm. 
They offer a self-consistent framework in which the redshift of a perturbation escaping from the near-horizon region of a black hole is regularized by dispersive effects, and the equivalent of Hawking's calculation can be performed without invoking ultra-high-energy particles. 
As shown in the two references~\cite{Jacobson:1991gr,Unruh:1994je}, this provides precious insight on the stability of Hawking radiation under modifications of the high-energy physics and the first expected deviations from the thermal result. 
In particular, building on the the description of dispersive effects on the propagation of waves given by T.~Jacobson, W.~Unruh showed that, if one assumes that the first corrections due to quantum gravity or modifications of the standard model of particle physics can be described (either fundamentally or at an effective level) by the addition of a dispersive term, then the results of~\cite{Hawking:1974sw,Unruh:1976db} remain valid at lower energies, in spite of the dramatic change of behavior of high-energy particles. 
This is the case, for instance, in Ho\v{r}ava gravity~\cite{Horavasummary,Sotiriou:2010wn}, which has been put forward to describe high-energy corrections and relies on (effective) breaking of the Lorentz invariance. 
Dispersive terms have the welcome effect of making the theory power-counting renormalizable, offering hope that it may be fully renormalizable and thus potentially serve as a well-defined quantum theory of gravity~\cite{Visser:2009ys,Barvinsky:2015kil}. 
The Einstein-\AE ther theory~\cite{Jacobson:2000gw,Jacobson:2000xp,Eling:2004dk,Jacobson:2005bg}, which contains a unit vector field, the \ae ther, in addition to the metric, also allows dispersive terms.  These can be included by coupling matter fields to the \ae ther, as will be done in Chapte~\ref{ch:universal}.

It must be noted, however, that analogue gravity alone can not settle the transplanckian problem. 
Indeed, the high-energy regulator is by essence beyond the correspondence between relativistic fields in curved spacetimes and perturbations of nonrelativistic fluids. 
Whether results obtained on one side can be applied to the other one depends on the eventual similarities between their high-energy behaviors, which is for the time being unknown for gravity. 
Similarly, resolution of the information loss paradox may depend on the late-time effects of the back-reaction of the emitted quanta on the geometry, and it is at present not\label{key} clear whether the correspondence can be extended to them. 
This is further complicated by two differences. 
First is the fact that analogue black holes usually have no singularity, which may play an important role in the late-time evaporation of astrophysical ones. 
Second, analogue models lack diffeomorphism invariance, which plays an important role in discussions related to holography~\cite{2009JPhCS.171a2009B}.

In fact, some of the soundest lessons one can draw from the correspondence may well concern the “analogue” models themselves rather than gravity. 
Indeed, the notions of curved metric, horizons and Hawking radiation offer a new point of view on these systems, allowing the discovery of new effects or simpler interpretations of already-known ones. 
To give one example, in~\cite{Finazzi:2014tqa} it is shown that a dynamical instability of a superfluid flowing through a penetrable barrier coincides with the “black hole laser effect”, i.e., the spontaneous amplification of Hawking-like radiation bouncing back and forth between a black hole and a white hole horizons. 
Similarly, the link with the Hawking effect provides a physical picture for the negative-energy waves observed in~\cite{Rousseaux:2007is} as well as motivation to study and realize transonic solutions in Bose-Einstein condensates~\cite{PhysRevA.85.013621,Larre:2013tba,2015PhRvL.115b5301B}. 

\section{Additional remarks}

\subsection{Conserved charge for dispersive fields}
\label{sub:proofNoether}

Let us consider a complex scalar field $\phi$ with the action
\begin{align}
S_\phi^d = \int d^{d+1} x \lp A_t (x) \lp \pd_t \phi^*(x) \rp \lp \pd_t \phi(x) \rp + \sum_{l=0}^n \sum_{i_1, i_2, ..., i_l} A_{i_1 i_2 ... i_l}(x) \Re \lp \phi^*(x) \pd_{i_1 i_2 ... i_l} \phi(x) \rp \rp,
\end{align}
where $A_t$, $A_0$, and the $A_{i_1 i_2 ... i_l}$, $l = 1..n$, $i_k = 1..d$ are real-valued functions. 
Without loss of generality, one can assume that only even orders appear in the second term. 
Indeed, if $l$ is odd, performing $l$ integration by parts turns the corresponding term 
\begin{align} \label{intr_app_nt1}
A_{i_1 i_2 ... i_l}(x) \Re \lp \phi^*(x) \pd_{i_1 i_2 ... i_l} \phi(x) \rp  = \frac{A_{i_1 i_2 ... i_l}(x)}{2} 
\lp \phi^*(x) \pd_{i_1 i_2 ... i_l} \phi(x) + \phi(x) \pd_{i_1 i_2 ... i_l} \phi(x)^* \rp
\end{align}
to
\begin{align}
\frac{1}{2}
\lp A_{i_1 i_2 ... i_l}(x) \phi^*(x) \pd_{i_1 i_2 ... i_l} \phi(x) - \phi^*(x) \pd_{i_1 i_2 ... i_l} \lp A_{i_1 i_2 ... i_l}(x) \phi(x) \rp \rp.
\end{align}
Expanding the second term, one finds that the two terms where all the derivatives act on $\phi$ cancel each other. 
The remaining ones all have strictly less than $l$ derivatives acting on $\phi$, and none on $\phi^*$. 
After symetrizing the resulting expression between $\phi$ and $\phi^*$ (which does not change its value since \eq{intr_app_nt1} is real), they can all be absorbed in a redefinition of the functions $A_{i_1 i_2 ... i_k}$ with $k< l$. 
Doing so for all odd values of $l$ from the largest one to the smallest, one can get rid of all odd-order terms. 
The Euler-Lagrange equations read
\begin{align}
\sum_{i=0}^n \pd_{\mu_1 \mu_2 ... \mu_l} \lp \frac{\delta S_\phi^d}{\delta \pd_{\mu_1 \mu_2 ... \mu_l} \phi^*} \rp = 0,
\end{align}
i.e.,
\begin{align}
\pd_t \lp A_t(x) \pd_t \phi(x) \rp &=   
\frac{1}{2}\sum_{l=2}^n \sum_{i_1, i_2, ..., i_l} A_{i_1 i_2 ... i_l}(x) \pd_{i_1 i_2 ... i_l} \phi(x)
+ \frac{1}{2}\sum_{l=2}^n \sum_{i_1, i_2, ..., i_l} \pd_{i_1 i_2 ... i_l}  \lp A_{i_1 i_2 ... i_l}(x) \phi(x) \rp \nn
 &=  \sum_{l=2}^n \sum_{i_1, i_2, ..., i_l} \sqrt{A_{i_1 i_2 ... i_l}(x)} \pd_{i_1 i_2 ... i_l}  \lp \sqrt{A_{i_1 i_2 ... i_l}(x)} \phi(x) \rp. 
\end{align}
Using this, one easily obtains
\begin{align} \label{eq:intr_app_int2}
&\pd_t \lp A_t(x) \lp \phi^*(x) \pd_t \phi(x) - \phi(x) \pd_t \phi^*(x) \rp \rp = \nn 
& \hspace*{1 cm} = \sum_{l=2}^n \sum_{i_1, i_2, ..., i_l}  \lp 
	- \sqrt{A_{i_1 i_2 ... i_l}(x)} \phi^*(x) \pd_{i_1 i_2 ... i_l} \lp \sqrt{A_{i_1 i_2 ... i_l}(x)} \phi(x) \rp \right. \nn 
	& \hspace*{2 cm} \left.\sqrt{A_{i_1 i_2 ... i_l}(x)} \phi(x) \pd_{i_1 i_2 ... i_l} \lp \sqrt{A_{i_1 i_2 ... i_l}(x)} \phi^*(x) \rp 
 \rp.
\end{align}
Moreover, for any smooth function $\Phi$, and any even integer $l$,
\begin{align}
\Phi(x) \pd_{i_1 i_2 ... i_l} \Phi^*(x) = & \pd_{i_1} \lp \Phi(x) \pd_{i_2 i_3 ... i_l} \Phi^*(x) \rp
 - \pd_{i_2} \lp \pd_{i_1}\Phi(x) \pd_{i_3 i_4 ... i_l} \Phi^*(x) \rp
 + ... \nn
 & - \pd_{i_l}  \lp \pd_{i_1 i_2 ... i_l} \Phi(x) \, \Phi^*(x) \rp
 + \Phi^*(x) \pd_{i_1 i_2 ... i_l} \Phi(x). 
\end{align}
Using this relation with $\Phi = \sqrt{A_{i_1 i_2 ... i_l}} \phi$, one finds that the right-hand side of \eq{eq:intr_app_int2} is a sum of derivatives with respect to a space variable of a smooth function quadratic in the field. 
Assuming suitable boundary conditions, for instance that $\phi$ and all its derivatives go to zero at spatial infinity, one thus obtains
\begin{align}
\pd_t \int d^dx \lp A_t(x) \lp \phi^*(x) \pd_t \phi(x) - \phi(x) \pd_t \phi^*(x) \rp \rp = 0.
\end{align}
Finally, noticing that the momentum conjugate to $\phi$ is $\delta S_\phi^d / \delta \pd_t \phi  = A_t \pd_t \phi^*$ gives
\begin{align}
\pd_t \int d^dx \lp \Pi(x) \phi(x) - \Pi^*(x) \phi^*(x) \rp = 0.
\end{align} 

The reason why this calculation works can be traced to the $\mathrm{U}(1)$ global invariance of the action under $\phi \to \e^{i \theta} \phi, \, \theta \in \mathbb{R}$.
A similar reasoning can be done for other global invariance, showing that the corresponding charges are unaffected by the dispersive terms. 

\subsection{Dispersion relation for the Bogoliubov-de Gennes equation}
\label{sub:BdGDR}

For completeness, I here show the calculation leading to~\eqref{eq:DRBdG}. 
This will also allow me to introduce a more convenient form of the BdG equation~\ref{int:eq:BdG}. 
Let us start from the BdG equation for a scalar function $\phi$:
\begin{align}\label{eq:int_BdGphi}
\ii \lp \pd_t + \vec{v}_0 \cdot \vec{\nabla} \rp \phi = - \frac{1}{2 \, \rho_0} \vec{\nabla} \cdot \lp \rho_0 \, \vec{\nabla} \phi \rp + g \, \rho_0 \, \lp \phi + \phi^* \rp.
\end{align}
To proceed, it will be convenient to work with a linear equation, without antilinear term. 
To this end, let us define the vector
\begin{align}
W \equiv \begin{pmatrix}
\phi \\
\phi^*
\end{pmatrix}. 
\end{align}
It obeys the equation
\begin{align}\label{eq:int_BdGW}
\ii \lp \pd_t + \vec{v}_0 \cdot \vec{\nabla} \rp W = 
\begin{pmatrix}
- \frac{1}{2 \, \rho_0} \vec{\nabla} \cdot \rho_0 \, \vec{\nabla} + g \, \rho_0 & g \, \rho_0 \\
- g \, \rho_0 & \frac{1}{2 \, \rho_0} \vec{\nabla} \cdot \rho_0 \, \vec{\nabla} - g \, \rho_0
\end{pmatrix}
W. 
\end{align}
Conversely, if $W = \begin{pmatrix}
\varphi_1 \\
\varphi_2
\end{pmatrix}$ is a solution of~\eqref{eq:int_BdGW}, a straightforward calculation shows that $\phi \equiv \lp \varphi_1 + \varphi_2^* \rp / 2$ satisfies~\eqref{eq:int_BdGphi}. 
There is thus a surjective map from the set of solutions of~\eqref{eq:int_BdGW} to that of~\eqref{eq:int_BdGphi}.~\footnote{This map is not injective. However, this is not a problem here since the solutions we find for different values of $\lp \om,\vec{k} \rp \in \mathbb{R}^+ \times \mathbb{R}^d$ clearly correspond to different functions $\phi$.} 
Let us now assume that $\rho_0$ and $\vec{v}_0$ are constants. 
One can then look for a basis of solutions of the form
\begin{align}
W \lp t, \vec{x} \rp = \begin{pmatrix}
\mathcal{U} \\
\mathcal{V}
\end{pmatrix}
\e^{\ii \lp \vec{k} \cdot \vec{x} - \om \, t \rp},
\end{align}
where $\lp \om,\vec{k} \rp \in \mathbb{R}^+ \times \mathbb{R}^d$. (This domain can be extended to $\lp \om,\vec{k} \rp \in \mathbb{C} \times \mathbb{C}^d$ if $\rho_0 \lp t, \vec{x} \rp$ and $v_0 \lp t, \vec{x} \rp$ are uniform only for $\vec{x}$ in a subdomain of $\mathbb{R}^d$, see Chapter~\ref{ch:saturation}.)  
Plugging this form into~\eqref{eq:int_BdGW} gives
\begin{align}
\begin{pmatrix}
- \lp \om - \vec{v}_0 \cdot \vec{k} \rp + \frac{k^2}{2} + g \, \rho_0 & g \, \rho_0 \\
- g \, \rho_0 & - \lp \om - \vec{v}_0 \cdot \vec{k} \rp - \frac{k^2}{2} - g \, \rho_0
\end{pmatrix}
\begin{pmatrix}
\mathcal{U} \\
\mathcal{V}
\end{pmatrix}
= 0.
\end{align}
Nontrivial solutions exist if and only if the determinant of the 2 by 2 matrix on the left-hand side vanishes. 
A straightforward calculation shows that this condition is equivalent to~\eqref{eq:DRBdG}. 

\newchapter{Saturation of black hole lasers in Bose-Einstein condensates}
\label{ch:saturation}
\begin{tikzpicture}[overlay]
\newcommand*{\xA}{-0.2}
\newcommand*{\xB}{14.35}
\newcommand*{\yA}{5.5}
\newcommand*{\yB}{1.5}
\newcommand*{\epsil}{0.75}
\draw[overlay] (\xA-\epsil,\yA) -- (\xB-\epsil,\yA);
\draw[overlay] (\xA,\yA+\epsil) -- (\xA,\yB);
\draw[overlay] (\xB,\yB-\epsil) -- (\xB,\yA);
\draw[overlay] (\xB+\epsil,\yB) -- (\xA+\epsil,\yB);
\end{tikzpicture}
\begin{small}
This chapter is mostly based on~\cite{Michel:2013wpa}, with some new results from~\cite{Michel:2015pra} and~\cite{Coutant:2016bgk}. 
\cite{Michel:2013wpa} was the first project I completed with Renaud Parentani, and it laid the foundation for many of the following ones. 
Indeed, focusing on ``black hole laser'' configurations with two analogue horizons, we introduced in this work the main ideas behind the three axes of my thesis:
\begin{itemize}
\item First, as will be detailed in this chapter, linear instabilities due to the Hawking effect and dispersion play a crucial role, triggering the ``laser'' effect and controlling the transition between lasing solutions and more stable ones. 
\item Second, nonlinear effects are used to find the set of stationary solutions of the system for given asymptotic conditions. 
We were particularly interested in stable solutions, which provide a natural candidate for the late-time solution once the ``laser'' instability is triggered. 
\item The link with experiments is fainter as we used a simple model of black hole laser which is convenient for analytical calculations but relatively far from those which have been realized experimentally. 
However, many ideas and results from this work were then used in~\cite{Michel:2015pra} and, more recently, in~\cite{Michel:2016tog}, both of which have deeper connections with the experimental realizations of J.~Steinhauer reported in~\cite{BHLaser-Jeff} and~\cite{Steinhauer:2015saa}.
\end{itemize}
While the existence of two horizons for a black hole laser flow at first sight complicates the analysis with respect to a single black- or white-hole flow, it is actually simpler in some respects. 
From an experimental point of view, the emission due to the laser effect is in principle easier to observe than the standard quantum Hawking effect. 
Indeed, while the latter is typically smaller than thermal fluctuations unless the temperature is below the Hawking one, the former grows exponentially in time until it reaches an amplitude where nonlinear effects dominate, making it much easier to observe. 
From a theoretical point of view, the subsonic character of the asymptotic regions gives natural asymptotic conditions for both linear and nonlinear solutions and allows us to define thermodynamic potentials which remain finite for all the relevant solutions. 
As we shall see in this chapter, this property is very useful to characterize the stationary solutions and to understand the interplay between linear and nonlinear effects. 

With the black hole ``information paradox'' in mind, our main goal was to determine the late-time evolution of an (unstable) black hole laser using a mean-field theory. 
As we shall see, in full generality this is a difficult question as the condensate does not always become stationary at late times.  
However, when working with a sufficiently small cavity (or with fine-tuned initial conditions), some stationary solution is eventually reached. 
This end-state can then be characterized analytically. 
Moreover, the situation then bears some similitude with that of astrophysical black holes in the semi-classical theory, in that an infinite set of different initial conditions lead to the same final state. 
More precisely, when working in quantum settings, the phonons emitted by the laser effect will be entangled with the ones inside the supersonic region, leading to an apparent loss of information from the point of view of an external observer. 
But this information loss may be resolved through correlations between phonons emitted at different times once the instability saturates. 
It would then be interesting to follow explicitly the state of phonons during the evolution to see precisely how the information is recovered and the ``information loss'' paradox avoided in this analogue model. 
While the work presented in this chapter does not go this far, it provides a first step by characterizing the background on which entangled phonons will evolve. It also finds more practical uses in describing the possible outcomes of experiments using black hole lasers, such as the one reported in~\cite{BHLaser-Jeff}. 
\end{small}
\newpage

\renewcommand*{\theHsection}{\theHchapter.\the\value{section}}
\renewcommand\thesection{\arabic{section}}

\section{Introduction}
\label{intro}

The stability of inhomogeneous solutions of the Gross-Pitaevskii equation (GPE) 
is a rich and wide topic, even in one dimension. 
Indeed, while the one-dimensional GPE in a uniform potential is integrable by the inverse scattering method, nontrivial phenomena appear in a the presence of a potential inhomogeneity. 
In the uniform case, it is known that dark and gray solitons~\cite{SolitonGPE,pitaevskii2003bose,Shlyapnikovcourse}, which are the only solutions with asymptotically homogeneous densities to be stationary up to a galilean transformation, are stable. 
When imposing asymptotic homogeneity on one side only, one finds another solution ~\cite{belokolos1994algebro,PhysRevA.64.033602}, which is divergent at a finite point but plays an important role in semi-infinite setups.  
As we shall see in this chapter, they are also crucial to understanding the behavior of black hole lasers, where the supersonic region regularizes their divergence. 

In this chapter we shall mostly focus on {\it transonic} flows, in which the velocity $v$ crosses the sound speed $c$. 
The most commonly studied class of such flows is subsonic in one asymptotic region and supersonic in the other one, with a monotonic Mach number $M$. 
These flows support linear excitations with negative energies. 
The mixing of these modes with the usual positive-energy ones induces a super-radiance, i.e., the amplification of incoming positive-energy waves by the production of negative-energy ones. 
In quantum settings, this process is triggered by quantum fluctuations, leading to the spontaneous production or pairs of entangled quanta with opposite energies. 
This is directly related to the Hawking prediction, according to which incipient black holes should spontaneously emit a thermal flux of radiation, through the effective metric~\cite{Unruh:1980cg,Unruh:1994je} (see also Chapter~\ref{chap:intro}) of the flow, which has a Killing horizon at the point where $\abs{v} = c$.

Here we instead consider flows which are asymptotically subsonic on both sides, with a finite supersonic region in between. 
These lead to an even richer phenomenology. 
In particular, it has been understood~\cite{Corley:1998rk,Leonhardt:2008js} that these flows can be dynamically unstable because of a self-amplification of the super-radiance (the Hawking effect) occurring at each subsonic/supersonic transition. 
The intuitive picture is that the two horizons act as mirrors in an optical cavity, reflecting part of the waves which then bounce back and forth between them. 
But, because of the Hawking effect, the reflected wave at each horizon has a {\it larger} amplitude than the incident one provided the horizons are sufficiently well separated to avoid interferences. 
This over-reflection at each horizon gives rise to an exponential growth of the waves, hence the name ``black hole laser'' (although this denomination is slightly misleading as a standard laser operates in the nonlinear regime~\cite{2015arXiv150900795D}).  
More precisely, it was shown that the spectrum of linearized perturbations contains a discrete set of complex-frequency modes which characterizes the dynamical instability~\cite{Coutant:2009cu,Finazzi:2010nc}: the supersonic region acts as an unstable resonant cavity, and the distance between the two sonic horizons governs the number of unstable modes. Below a certain value, there is no unstable mode and no pair production: the flow is stable. 
When increasing the distance, unstable modes appear one by one, and the density fluctuations associated with the $n^{\rm th}$ one have $n$ nodes, as expected from the above analogy with optics. 
For large inter-horizon lengths, the number of unstable modes increases linearly with the distance.
In the present chapter, we complete the analysis in the particular case of piecewise-constant potentials such that the GPE admits a homogeneous solution with two sonic horizons. 
The main interest of this model is that it allows to obtain the main properties of both the linear and nonlinear solutions analytically while exhibiting the same behavior as more realistic ones with smooth potentials. 
Similar configurations with a single horizon were considered in~\cite{Carusotto:2008ep,PhysRevA.85.013621}. 
In addition, as done in~\cite{Finazzi:2012iu} for a single horizon, we show numerically that the results obtained within the steplike approximation apply to smooth profiles when the transition regions are sufficiently narrow.

Using the distance $2 L$ between the horizons as our control parameter, we first study 
the onset of the dynamical instability. We show that for a finite range of $L$, it is first described by an unstable mode with a purely imaginary frequency.\footnote{This was independently found by I. Carusotto, J.R.M. de Nova, and S. Finazzi (private communication).}
For larger distances, we recover the ordinary situation~\cite{Coutant:2009cu,Finazzi:2010nc} of complex-frequency modes with properties directly linked to the Hawking effect. 
More precisely, each unstable degree of freedom originates from a single quasinormal mode (QNM) when the latter's frequency, which is purely imaginary, crosses the real axis. 
This frequency then leaves the imaginary axis when a second QNM merges with it, so as to form a two-dimensional unstable system. 
These steps can be understood from the holomorphic properties of the determinant encoding the junction conditions across the two horizons which define the complex-frequency modes, severely restricting the conditions under which complex-frequency modes can appear~\cite{Fullingbook}. 
Indeed, the spectrum of the system under consideration has two symmetries: $\lambda \to \lambda^*$ and $\lambda \to - \lambda$. 
The number of eigenfrequencies thus changes when one of them leaves $\mathbb{R} \cup i \mathbb{R}$, as a doublet of frequencies $(\lambda, -\lambda)$ becomes a quartet $(\lambda, \lambda^*, - \lambda, - \lambda^*)$. 
As we shall see, these complex frequencies are given by the zeroes of a function which is holomorphic in the relevant domain of the complex plane and depends smoothly on the distance between the horizons. 
As the number of roots of this function is conserved when increasing $L$ by continuity, an eigenfrequency can acquire non-vanishing real and imaginary parts only when two eigenfrequencies merge.  

Following~\cite{PhysRevA.64.033602,Piazza,Baratoff,PhysRevB.53.6693}, we also study the set of stationary nonlinear solutions of the GPE. 
We show that it is closely related to the discrete set of complex-frequency modes which triggers the dynamical instability of the initial flow. 
Indeed, each unstable mode can be associated with a set of nine nonlinear solutions. In each set, the solution with the smallest energy may be conceived as the end point of the instability, in that it is the solution reached at late times provided the dynamics leads to a stationary configuration minimizing the energy.~\footnote{As we shall see, the relevant energy functional is bounded from below when restricting to stationary solutions.} 
This assumption in fact turns out to be too strong: by solving the GPE numerically we motivate that, depending on the initial data and parameters, the solution may never become stationary but instead emits periodic or quasiperiodic, superposed soliton trains. 
(This was also observed in~\cite{2015arXiv150900795D}.) 
However, these numerical simulations also confirm that when a stationary solution is reached it corresponds to the lowest-energy one.
The nine types of stationary solutions can be separated in two categories, as four of them are smoothly connected to the homogeneous one, while the five additional ones contain either one or two solitons. 
When considering the full set of solutions for a given value of $L$, 
we show that the ground state of the system has no node, and contains no soliton. 
Finally, by a perturbative expansion of the Gross-Pitaevskii energy functional similar to that used by Pitaevskii~\cite{Pitaevskii1984} and Baym and Pethick~\cite{Baym} to study the spontaneous appearance of layered structures in flowing superfluids with a roton-maxon spectrum, we directly relate the set formed by the union of QNM and unstable modes to the above-mentioned four-dimensional subset of nonlinear solutions connected to the homogeneous one.

One should bear in mind that the transition from the initial unstable 
solution to the lowest-energy state described in Section~\ref{stat_sol} provides an interesting example of a process which mimics a unitary black hole evaporation. 
We remind the reader that, when considering gravitational black holes, it is still unknown whether the evaporation process is nonunitary, as originally suggested by Hawking, or if it satisfies unitarity, as is the case for standard quantum mechanical processes. We also remind the reader that in order for the emitted Hawking radiation to end up in a pure state at the end of the evaporation (when starting from a pure state), it is necessary to have a nondegenerate final black hole state.
As argued by Page~\cite{Page}, this implies that the Hawking quanta emitted after a certain time must be correlated to the former ones. Using a mean field treatment of the metric, that is, when adopting the so-called semiclassical scenario,
this conclusion is highly nontrivial since the Hawking quanta are entangled with their negative energy partners~\cite{Brout:1995rd,Massar:1996tx} but are uncorrelated with each other.
One can of course hope that when working beyond the mean field approximation, quantum backreaction effects will restore the unitarity. The difficulty one then faces is to find some microscopic description of black holes in which this can be shown to occur. 
(A recent proposal~\cite{Hawking:2016msc} is that soft gravitons may contain enough information to restore unitarity.) 
The main virtue of the present model is that it combines in a nontrivial way two essential elements. First, at early times, using a linearized treatment, the emitted phonons can be shown to be entangled with the negative energy partners which are trapped in central region $I_2$, as is the case for the Hawking process.~\footnote{After a while, as noticed in~\cite{Corley:1998rk}, because the laser effect is taking place, there exist correlations among the emitted quanta. However these correlations are not sufficient to restore unitarity, as the correlations to the partners are still present.} 
Second, the full  Hamiltonian possesses a unique ground state. One can therefore deduce, like Page, that after some time, the emitted phonons should be correlated with the former ones. Another virtue of the model is that these correlations should, in principle, be calculable without encountering the uncontrolled divergences which occur in perturbative treatments of quantum gravity, although the existence of solutions which apparently do not become stationary at late times might introduce important qualitative differences. 
We hope to return to these questions in the future. 

This chapter is organized as follows. In Section~\ref{Slt}, we present the model, linearize the GPE, and find both the modes responsible for the dynamical instability and the QNM from which they originate.
Exact stationary solutions of the GPE are studied in Section~\ref{stat_sol} and their links with the linear solutions are given in Section~\ref{thermo}. 
In Section~\ref{Timeevolution} we report the results of time-dependent numerical simulations. 
Subsection~\ref{App:M-matrix} details the method we used to find complex-frequency modes. Stationary solutions of the GPE with one single discontinuity are discussed in subsection~\ref{App:single-h}.
Explicit formulas used to compute properties of solutions are given in subsection~\ref{App:eqs_G_L}. 
Finally, subsection~\ref{sec:degvsnondeg} sets the main linear results within a more general context and explains how they can be deduced from the symplectic structure of the theory.

\section{Setting and linearized treatment}
\label{Slt}

\subsection{Setting}
\label{Settings}

We consider a one-dimensional flowing condensate with two-body coupling $g$ and external potential $V$. 
We assume that $V$ and $g$ are piecewise-constant with two discontinuities located at $z=-L$ and $z=L$ (where $z$ denotes a Cartesian spatial coordinate).  
We denote as $g_1,V_1$ the parameters in the left region $I_1 : - \infty < z < -L$, $g_2,V_2$ the parameters in the central region $I_2 : - L < z < L$, and $g_3,V_3$ the parameters in the right region $I_3 : L < z < \infty$. 
In each region $I_j$ ($j \in \lbrace 1,2,3 \rbrace$), the Gross-Pitaevskii equation (GPE) reads
\begin{align} \label{GPE1}
\ii \frac{\partial \psi }{\partial t} = -\frac{1}{2}\frac{\partial ^2\psi }{\partial z^2}+V_j \psi +g_j \left|\psi ^2\right|\psi.
\end{align}
We work in units in which the reduced Planck constant $\hbar$ and the atom mass are equal to unity. 
There are two independent residual rescaling of the solutions. 
Indeed, \eq{GPE1} is invariant under the transformation $\psi \to \eta_\psi \psi$, $t \to \eta_t t$, $z \to \eta_t^{1/2} z$, $V \to \eta_t^{-1} t$, $g \to \eta_t^{-1} \eta_\psi^{-2}$ for any $(\eta_t, \eta_\psi) \in \lp \mathbb{R}_+^* \rp ^2$.
These can be used to fix two quantities to unity, for instance $g_2$ and the asymptotic limit of the density. 
We consider stationary solutions, 
\begin{align}\label{eq:psiststanz}
\psi : (t,z) \mapsto \binom{\mathbb{R}^2 \to \mathbb{C}}{\e^{-\ii \mu t} f(z) \,\e^{\ii \theta (z)}},
\end{align}
where $f$ and $\theta$ are two real-valued functions, and $\mu \in \mathbb{R}$. 
The former is related to the local atomic density $\rho$ by $\rho(z) = f(z)^2$. 
Plugging \eq{eq:psiststanz} into \eq{GPE1} and taking the imaginary part of the resulting equation gives $\partial_z J = 0$, where $J \equiv f^2 \pd_z \theta$. 
Using this definition, the real part of the GPE becomes
\begin{align} \label{eq:f}
f''= -2 \mu_j  f + 2 g_j f^3+\frac{J^2}{f^3},
\end{align}
where $\mu_j \equiv \mu - V_j$, and where a prime denotes differentiation with respect to $z$. 

We work with $g_j, \mu_j > 0$, and assume that the current is smaller than the critical value $J_{\rm max}$, 
so that homogeneous solutions exist in each region (see subsection~\ref{App:single-h}, \eq{eq:Jmax}).~\footnote{We here adopt the choice of working at fixed chemical potential $\mu$, which in an experimental setup will be fixed by that of the clouds of uncondensed atoms with which the condensate interacts.} 
We also assume that $g_j, \mu_j$ are chosen such that there is a global homogeneous solution $f(z)=f_0 > 0$ 
with a subsonic flow in $I_1$, $I_3$, and a supersonic one in $I_2$. Hence this flow is a particular case of the black hole laser system studied in Refs.~\cite{Corley:1998rk,Leonhardt:2008js,Coutant:2009cu,Finazzi:2010nc}. 
Notice that the flow velocity $v$ is uniform in our homogeneous solution.   
To characterize the flow, it is convenient to work with 
\be \label{civ}
c_j \equiv \sqrt{g_j f_0^2},\, 
v \equiv \frac{J}{f_0^2},
\ee
where $c_j$ is the sound speed in $I_j$, and $v = \pd_z \theta$ is the constant condensate velocity. 

Setups with similar step-like variations of the external potential $V$ and $g$ leaving invariant the Hartree interaction energy $V + g \s \rho$ have previously been used in~\cite{Carusotto:2008ep,Balbinot:2007de,PhysRevA.85.013621} to study the analogue Hawking radiation from black holes in Bose-Einstein condensates. 
Although such variations seem less easy to achieve experimentally than changes of $V$ only, they may provide a better control in time-dependent setups.  
Indeed, imposing that the Hartree interaction energy be uniform at all times during the formation of the horizons suppresses back-scattering of condensate atoms and the formation of solitons~\cite{2002PhRvA..66a3610P,Carusotto:2008ep}, which would otherwise interfere with the dynamics of the black hole laser itself. 

\subsection{Complex-frequency modes}
\label{cfm}

The main properties of the set of unstable modes have been obtained by algebraic techniques in~\cite{Coutant:2009cu} and numerically in~\cite{Finazzi:2010nc}. 
These two works focused on the evolution of the complex frequencies when varying $L$. However, they were not able to describe the birth of these modes. 
In the present setting, this can be analyzed in detail, revealing an interesting two-step process. In the body of the text we discuss the method and main results. 
The details of the calculation are presented in subsection~\ref{App:M-matrix}.

To obtain the equations for perturbations on the homogeneous solution $(f_0,\theta_0 (z)= z\, J/f_0^2)$,
we write $f(t,z)=f_0 + \delta f(t,z)$ and $\theta(t,z)=\theta_0(z)+\delta \theta(t,z)$, linearize \eq{GPE1}, and look for solutions of the form
\be \label{eq:fandtheta}
 \left\lbrace 
\begin{array}{ll}
 \delta  f_\om(t,z)=\Re \lp \delta F_\om \, \e^{\ii (k_\om z - \omega  t)}  \rp &   \\
 \delta \theta_\om (t,z)=\Re \lp \delta  \Theta_\om \, \e^{\ii (k_\om z - \omega  t)} \rp &.  
\end{array}
\right.
\ee
Doing so, one obtains
\be \label{eq:thetavsf}
\delta  \Theta_\om =2 \ii \lp \frac{v k_\om-\omega }{f_0 \, k_\om^2} \rp \delta  F_\om,
\ee 
and the dispersion relation
\be \label{eq:disprel1}
\Omega^2 = \frac{1}{4}k_\om^4 + c_j^2 k_\om^2,
\ee
where $\Omega \equiv \omega -v k_\om$ is the frequency in the reference frame moving with the condensate. 
The roots of this polynomial equation in $k$ at fixed $\om$ are described in subsection~\ref{App:M-matrix}. 
\eq{eq:thetavsf} and \eq{eq:disprel1} characterize the four linearly independent solutions in each region $I_j$. 
Solutions in different regions are related by matching conditions at $z = \pm L$, which follow from the continuity and differentiability of $f$ and $\theta$.

We are interested in computing the discrete set of complex-frequency modes with eigenfrequencies $\om_a \in \mathbb{C}-\mathbb{R}$, which trigger the laser effect.~\footnote{Notice that \eq{GPE1} is invariant under $\psi(t,z) \to \psi(-t,z)^*$, which sends $\om$ to $\om^*$. The discrete spectrum is thus invariant under complex conjugation. In particular, each complex frequency with a non-vanishing imaginary part indicates the existence of two modes. One of them grows exponentially as $\exp (\abs{\Im(\omega) t})$ while the other decays as $\exp (-\abs{\Im(\omega) t})$.}
To determine them, we consider asymptotically bounded modes (ABM), \textit{i.e.}, keep the waves $e^{i k_\om z}$ which decay exponentially as $z \rightarrow \pm \infty$~\cite{Coutant:2009cu}.
In $I_1$ and $I_3$ there are two such solutions for a given sign of $\Gamma \equiv \Im \om$. For $\Gamma > 0$, they correspond to the analytical continuations in $\om \in \mathbb{C}$  of the outgoing wave and the exponentially decreasing wave.
In the central region $I_2$, the four waves are kept. Eight boundary conditions must be satisfied: continuity and differentiability of $\delta f$ and $\delta \theta$ at $z=\pm L$. 
They impose eight linear relations between the coefficients of the waves, which can be written as an 8-by-8 matrix $M(\om)$. This system has nontrivial solutions if and only if $\det M(\om)=0$, which selects the sought-for discrete set of frequencies. 

To study how these ABM appear as $L$ increases, we will consider a larger (still discrete) set which includes quasinormal modes (QNM) which are not asymptotically bounded.
Using this larger set, we will see that every complex-frequency ABM arises from two QNM in two steps. To understand their origin, one should recall that, in the general case, each dynamical instability is described by a two-dimensional system which corresponds to a complex unstable oscillator; see~\cite{PhysRevA.72.032101} and Appendix C in~\cite{Coutant:2009cu}. 
This system is composed of two {\it complex} eigenmodes of frequencies $\om_a = \Re \, \om_a  \pm  i\Gamma_a $, with $\Re \, \om_a, \Gamma_a  > 0$. Only the mode which grows in time is {\it outgoing}.  
The definition of ``outgoing'' modes for complex values of $\om$ requires some care. 
When $\om \in \mathbb{R}$, the direction of propagation of the energy is given by the group velocity $v_G \equiv \lp \partial_\omega k_\omega \rp^{-1}$. 
Outgoing waves are then defined by the condition that $v_G$ be oriented away from the supersonic region. 
The group velocity acquires a non-vanishing imaginary part when $\om$ leaves the real axis, leading to an ambiguity in the definition of outgoing waves. 
Here we define them by analytical continuation towards $\om \in \mathbb{R}$: a solution with wave vector $k_\om$ is called ``outgoing'' if its analytical continuation to $\om \in \mathbb{R}$ has a group velocity oriented away from the supersonic region. 
This definition  makes sense provided we remain in a domain of the complex plane where the analytical continuations of the wave vectors $k_\om$ are well-defined. 
The latter become ambiguous when two roots cross each other, i.e., when the dispersion relation has a double root. 
A straightforward calculation shows that this happens at the four frequencies $\om \in \left\lbrace 0, +\om_{\rm max}, -\om_{\rm max}, + i \, \Gamma_0, - i \, \Gamma_0 \right\rbrace$, where $\om_{\rm max}$ and $\Gamma_0$ are given in Eqs.~(\ref{eq:omMax},\ref{eq:Gamma0}). 
The double root at $\om = 0$ is innocuous. Indeed, the two roots which cross each other have different, finite group velocities and can thus be analytically continued across $\om = 0$. 
Our definition of outgoing modes is thus unambiguous provided one works in a simply connected domain of $\mathbb{C}$ which does not contain the points $\pm \om_{\rm max}$, $\pm i \, \Gamma_0$. 
Its relevance was established in~\cite{Coutant:2016bgk}, where it was shown that the modes thus defined can be used to build the retarded Green function. 

Besides this ``standard'' scenario where both the real and imaginary parts of $\om$ are generally non-vanishing and the corresponding modes are described by two complex degrees of freedom, there also exists a degenerate case, not considered in~\cite{Coutant:2009cu,Finazzi:2010nc}, described by only two {\it real} modes with imaginary frequencies $\pm i \Gamma_a$~\cite{PhysRevA.72.032101,Fullingbook}. 
In this case too, the ABM which grows in time is outgoing in the above sense. 
Interestingly, the two-step process we found is directly associated with this degenerate case, see subsection~\ref{sec:degvsnondeg}.

When looking for QNM, we should also pay attention to the implementation of the outgoing boundary conditions because there are four roots in \eq{eq:disprel1}, but only two in the standard definition of QNM~\cite{reviewKokotas,QNM_BH_BB}.
We adopt the same definition as the one above determining the ABM: We keep the analytical continuations in the complex lower half-plane of the outgoing wave and the exponentially decreasing wave for $\om \in \mathbb{R}$. 
With this definition, the condition $\det M = 0$ gives all the outgoing modes, that is, the spatially ABM for $\Gamma > 0$, and the QNM for $\Gamma < 0$, both for the standard and the degenerate case with $\Re \om = 0$. 

\subsection{Results}
\label{res_ABM/QNM}

To study the two-step process for increasing values of $L$, we work with $c_3=c_1$, and then briefly discuss the changes when $c_3 \neq c_1$. 
We first find that every new ABM appears in the degenerate sector, at $\om = 0$, and for values of $L$ given by
\be \label{eq:Lm} 
L_m \equiv L_0 + \frac{\lambda_0}{2} m, 
\ee
where 
\be \label{eq:Lcrit}
L_0 &=&  \frac{1}{2 \sqrt{v^2 - c_2^2}} \arctan \lp \sqrt{\frac{c_1^2 - v^2}{v^2 - c_2^2}} \rp, 
\\
\label{eq:lambda}
\lambda_0 &=& \frac{\pi}{\sqrt{v^2 - c_2^2}},
\ee
and where $m \in \mathbb{N}$. This is the first ``step''.
For each $m$, it is followed at $L = L_{m +1/2}$ by 
a merging process when the frequency of a QNM 
crosses the real line and equals that of the degenerate ABM. For $L > L_{m+1/2}$, the $m$th unstable sector is described by the nondegenerate case, i.e. by a complex ABM whose frequency has non-vanishing real and imaginary parts. 

We now show that \eq{eq:Lm} gives the set of values of $L$ at which a frequency can cross $\om = 0$.
Linearizing \eq{eq:f} in $\delta f$ and assuming the solution is static and bounded at infinity, 
i.e., setting $\om = 0$ and discarding the exponentially growing modes (in space) in the regions $I_1$ and $I_3$, one obtains 
\be 
\delta f(z)=
\left\lbrace 
\begin{array}{ll}
 A_L \exp \left(2 \sqrt{c_1^2-v^2} \, z\right), & z<-L ,\\
 A \cos \left(2 \sqrt{v^2-c_2^2} \, z+\varphi \right), & -L<z<L, \\ 
 A_R \exp \left(-2 \sqrt{c_1^2-v^2} \, z\right), & z>L ,
\end{array}
\right. 
\ee
where $A_L$, $A$, $A_R$ and $\varphi$ are real constants. 
Continuity of $\delta f$ and $\delta f'$ at $z = \pm L$ gives 
\be \label{eq:tan}
\tan \left(2 \sqrt{v^2-c_2^2} \, L+\varphi \right)=-\tan \left(-2 \sqrt{v^2-c_2^2} \, L+\varphi \right)=\sqrt{\frac{c_1^2-v^2}{v^2-c_2^2}}.
\ee
This implies that $\varphi =0\, \text{modulo} \, {\pi }/{2}$ 
 and $L$ obeys \eq{eq:Lm} with $m$ 
 integer or half-integer. 
At this level, it might seem that a new ABM is obtained for all these values of $L$. This is not quite correct, as is revealed by studying the solutions of \eq{eq:fandtheta} with $\Re \om \neq 0$, 
 see Fig.~\ref{fig:QNM} and Sec~\ref{thermo}.
\begin{figure} \centering
\includegraphics[scale=0.59]{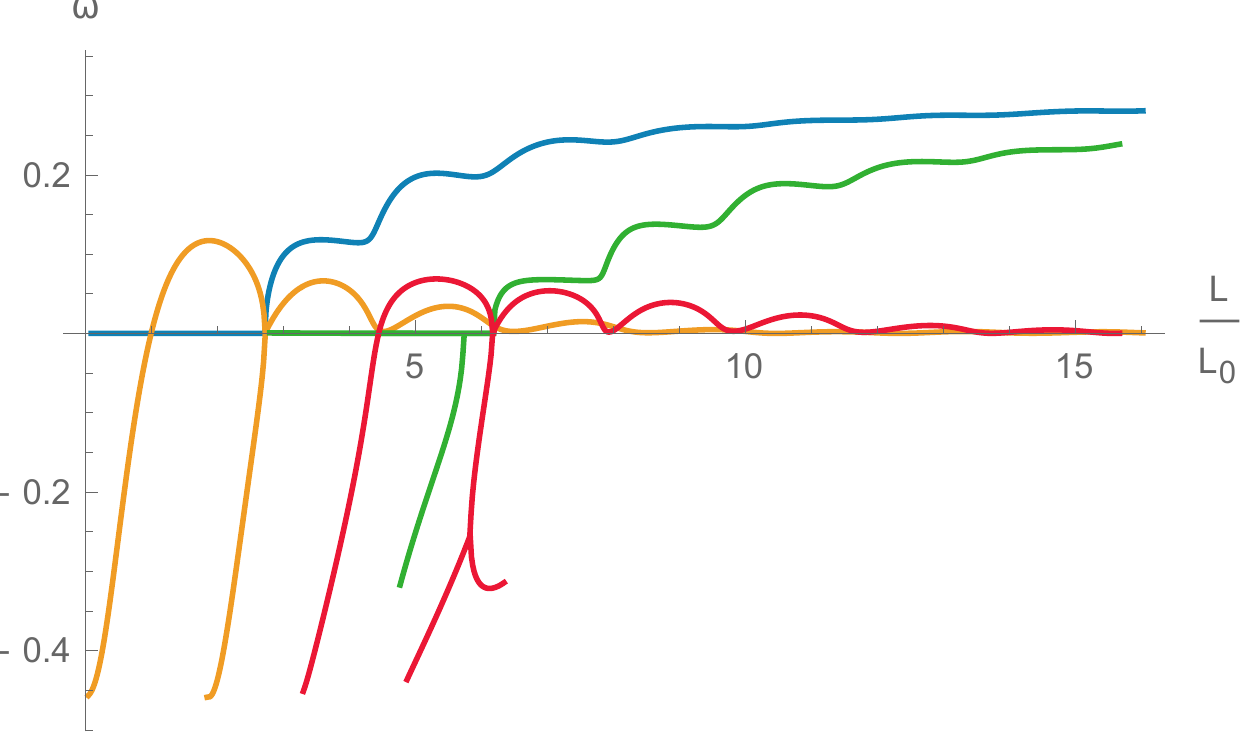} \hspace*{0.2 cm}
\includegraphics[scale=0.4]{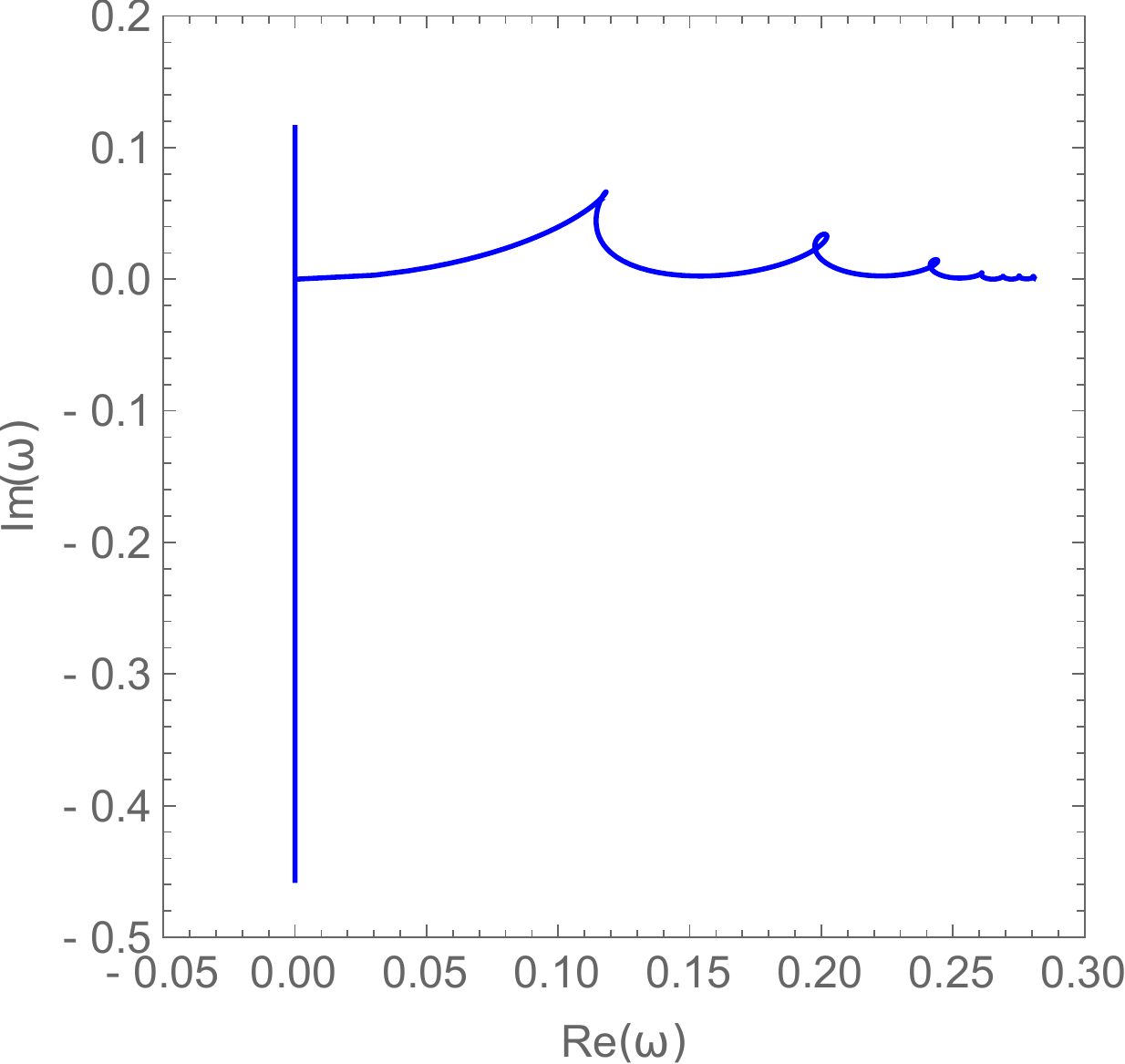}
\caption{Left: Evolution of the two complex frequencies composing the first two sectors $n=0$ and $n=1$ as functions of $L/L_0$.  Orange (blue) line: Imaginary (real) part of the two frequencies associated with $n=0$. Green (red) line: 
Imaginary (real) part for $n=1$.
The red line for negative values gives $\Re \om$
of the second QNM. At $L = L_0$ of \eq{eq:Lcrit}, the first ABM appears, as the QNM frequency crosses the real axis. 
The ABM frequency remains purely imaginary until $L = L_0 + \lambda_0/ 4 \sim 2.8 \, L_0$, where the second QNM of the first sector merges with it. 
For larger values of $L$, the frequency is complex (only the solution with $\Re \om > 0$ is represented). The story is similar for the second sector $n=1$. Note that the second QNM has a complex frequency for small values of $L$. This complex QNM splits into two purely imaginary QNMs for some value $L_i$ close to $L_{3/2}$, as  can be seen in the bottom of the figure. 
The parameters are: $v=1.0$, $c_1=1.5$ and $c_2=0.5$.
Right: Trajectory of the QNM and ABM frequencies with $n = 0$ in the complex plane.
} \label{fig:QNM}
\end{figure}

When including the QNM, the picture gets clearer, as one can see that both steps occur
when a QNM frequency crosses the real axis.
Starting with $L=0$, we obtain the following sequence; see Fig.~\ref{fig:QNM}.
When $L = 0$, there is no ABM, but there is already one QNM.
It is the ``ancestor'' of the first ABM. Indeed, when $L$ increases, its frequency moves along the imaginary axis, and when it crosses the real axis, it becomes the first ABM. 
The onset of instability occurs at $L=L_0$ of \eq{eq:Lcrit}.
As shown in Section~\ref{stat_sol}, $L_0$ is also the value of $L$ at which the thermodynamic potential of a nontrivial nonlinear solution becomes smaller than that of the homogeneous solution. 
As expected~\cite{PhysRevA.72.032101}, the dynamical (linear) instability thus appears together with an energetic instability.   
When $v < c_2$, the homogeneous solution is everywhere subsonic and there is no dynamical instability. There are still QNM, but these never cross the real axis to become ABM. 

When further increasing $L$, the ABM frequency keeps moving along the imaginary axis.
$\Im \om$ reaches a maximum value $\Gamma_M$ and then starts to decrease, still along the imaginary axis. 
Besides this, a second QNM appears on the negative imaginary axis and moves up.
This new QNM merges with the ABM at $\Gamma=0$ for $L = L_{1/2}$.
For higher values of $L$, the ABM eigenfrequency leaves the imaginary axis. The evolution is then similar to what was found in \cite{Finazzi:2010nc}: the imaginary part shows oscillations with a decreasing amplitude, while the real part goes to $\om_{\rm max}$ given by \eq{eq:omMax}.
By a numerical analysis of the equation $\det M = 0$, we found couples of QNM for $n \equiv  \lfloor m \rfloor \in \left[0,8 \right]$.
As we will see in Section~\ref{stat_sol}, nonlinear solutions are classified by the same integer
number $n$~, which labels the harmonics in the central region.
Notice also that $n$ 
coincides with the Bohr-Sommerfeld number $n_{\rm BS}$ used in~\cite{Finazzi:2010nc}; see subsection~\ref{App:M-matrix}.
For each value of $n$, one QNM crosses the real axis with a vanishing real part at $L = L_n$, therefore becoming the new ABM. Then the latter merges with another QNM at $\Gamma=0$ for  $L = L_{n+1/2}$ before leaving the imaginary axis. The subsequent evolution is similar to the case $n=0$. 
We conjecture that this remains true for any $n \in \mathbb{N}$ since the stationary analysis giving 
\eq{eq:tan} applies to all $n$. The cases $n=0$ and $n=1$ are represented in Fig.~\ref{fig:QNM}. 

In all cases, we notice that QNM and ABM frequencies never leave the imaginary axis except when they merge with another one. This is due to the continuity and differentiability of $\det M$ in $\om$, as well as its symmetry under $\om \rightarrow - \om^*$, $k \rightarrow - k^*$. (This property holds if $\abs{\Gamma}$ is smaller than $\Gamma_0$, so that the wave vectors $k_\om$ entering in the coefficients of $M$ are smooth functions of $\om$. 
This is always the case for the modes we describe.) Indeed, a mode leaving the imaginary axis must turn into two modes $\om$ and $-\om^*$. The change in the phase of $\det M$ when turning around them in the complex $\om$ plane is then equal to $4 \pi$ times some integer. But turning around one single ABM (or QNM) frequency gives, in general, a change of phase of $\pm 2 \pi$ since $\det M$ is linear close to it. So, by continuity of the phase of $M$, a frequency cannot leave the imaginary axis, except when two frequencies merge. 

When considering $c_1 \neq c_3$ we find the following. \eq{eq:Lm} remains true, with $\lambda_0$ still given by \eq{eq:lambda} and $L_0$ given by
\be \label{eq:lambda_c3neqc1} 
L^{c_1 \neq c_3}_0 =  
\frac{1}{4 \sqrt{v^2 - c_2^2}} \lp \arctan \lp \sqrt{\frac{c_1^2 - v^2}{v^2 - c_2^2}} \rp + \arctan \lp \sqrt{\frac{c_3^2 - v^2}{v^2 - c_2^2}} \rp \rp.
\ee 
The simplest way to see this is to shift the origin of $z$ so that $\delta f$ takes the form
\begin{align*}
\delta f(z)=
\left\lbrace 
\begin{array}{ll}
 A_L \exp \left(2 \sqrt{c_1^2-v^2} \, z\right), & z<-L_1 ,\\
 A \cos \left(2 \sqrt{v^2-c_2^2} \, z \right), & -L_1<z<L_3, \\ 
 A_R \exp \left(-2 \sqrt{c_1^2-v^2} \, z\right), & z>L_3 ,
\end{array}
\right. ,
\end{align*}
where $L_1 + L_3 = 2 L$. 
The matching conditions at $z = \pm L$ become
\begin{align*}
\tan \lp 2 \sqrt{v^2 - c_2^2} L_1 \rp = \frac{\sqrt{c_1^2 - v^2}}{\sqrt{v^2 - c_2^2}}, \; 
\tan \lp 2 \sqrt{v^2 - c_2^2} L_3 \rp = \frac{\sqrt{c_3^2 - v^2}}{\sqrt{v^2 - c_2^2}}.
\end{align*}
This gives for $j \in \left\lbrace 1,3 \right\rbrace$:
\begin{align*}
L_j \in \frac{1}{2 \sqrt{v^2 - c_2^2}} \arctan \lp \sqrt{\frac{c_j^2 - v^2}{v^2 - c_2^2}} \rp + \frac{\pi}{2 \sqrt{v^2 - c_2^2}} \mathbb{N},
\end{align*}
hence the above result. 
One also finds that ABM have a finite imaginary part when they leave the imaginary axis, and the first QNM appears at a finite value of $L$. 

We end this section by noting that we observe in Fig.~\ref{fig:QNM} a strong parallelism between the curves followed by the QNM frequencies, especially the first two. This indicates that there might be an approximative discrete translation invariance. This is reinforced by the fact that the difference in $L$ between them is $\lambda_0/4$, which corresponds to a symmetry of $\det M$ in the limit $\Gamma \rightarrow 0$. It is currently unclear whether this symmetry alone can explain the observed parallelism.

\subsection{Linear stability of inhomogeneous solutions}
\label{sub:BHLIaddpaper}

To study the dynamical stability of the stationary solutions described below in subsection~\ref{sub:statsoltuned}, we look for the set of ABM whose angular frequencies $\omega = \lambda + i \Gamma$, $\lp \omega,\Gamma \rp \in \mathbb{R}^2$, have positive imaginary parts  $\Gamma >0$. These modes thus grow exponentially in time, triggering a laser effect~\cite{Coutant:2009cu}.
We consider the inhomogeneous solutions which are smoothly connected to the  homogeneous one, i.e., those of types 1, 2, 3, and 4 in the classification of subsection~\ref{sub:statsoltuned}. (See also \fig{fig:profiles}.)
In the present subsection we only present the main results. The details of the analysis can be found in subsection~\ref{ABM}. 

\begin{figure}
\centering
\def\svgwidth{0.6 \linewidth}
{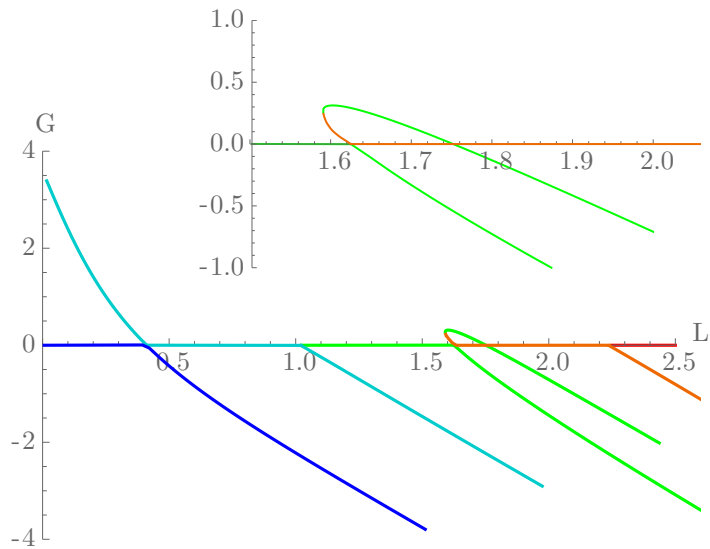}
\caption{Thermodynamic potential $G$ and dynamical stability of the first nonlinear solutions found when increasing the interhorizon distance $2L$. 
The parameters are $g_1=8$, $g_2=1$, $\mu_1=28/3$, $\mu_2=14/6$ and $J^2=8/3$. 
The number of instabilities is indicated by the color of the curve: blue indicates stable, cyan indicates one degenerate dynamical instability, green indicates one nondegenerate dynamical instability, orange indicates one degenerate and one nondegenerate instabilities, and red indicates two nondegenerate dynamical instabilities. 
The inset shows a zoomed-in picture on the point where the second type 3 solution appears. As could have been expected, when a new instability occurs for increasing $L$, the 
number and types of instabilities 
of the homogeneous solution is transmitted to a new inhomogeneous solution with a smaller energy. 
Therefore the only stable solution is the type 1 solution with $n=0$ (or the homogeneous one for $L < L_0$).} \label{fig:Soltun}
\end{figure}
Figure~\ref{fig:Soltun} shows the grand potential $G$, defined by 
\be \label{eq:G_off}
G[\psi]=\int \left(\frac{1}{2}\left\lvert \frac{\partial \psi }{\partial  z}\right\rvert^2 -\mu  \left\lvert \psi \right\rvert^2 +\frac{1}{2}g \left\lvert \psi \right\rvert^4 \right) \, \dd z + C^{te},
\ee 
of the first connected stationary solutions which are homogeneous in both asymptotic regions, along with the number of instabilities. 
The constant is chosen such that $G$ vanishes when evaluated on the homogeneous solution.
The grand potential is the relevant thermodynamic quantity for systems at equilibrium in the grand canonical ensemble, suitable to describe condensates interacting with a cloud of uncondensed atoms. 
Notice that it is bounded from below since $g > 0$. A straightforward calculation using Eq.~\eqref{GPE1} shows that it is conserved in time. 
Its on-shell expression and relation to the boundary conditions are mentioned in Section~\ref{thermo_app}.
We notice that, for each $n \in \mathbb{N}$, there exists a series of solutions for $L > L_n$ which has the same set of ABM (same numbers of degenerate and nondegenerate ones) as the homogeneous solution for $L < L_n$. This solution merges with the homogeneous one for $L=L_n$. So, each time the number of instabilities of the homogeneous solution is increased, a new inhomogeneous solution which preserves the number and type of dynamical instabilities continuously emerges at $L=L_n$. As a result, for any value of $L$ there is only one dynamically stable inhomogeneous solution. 
When $L > L_0$, it corresponds to the type 1 solution with $n=0$. 
This result remains valid when including the stationary solutions which are not connected to the homogeneous one, as all these solutions are dynamically unstable.

\begin{figure}
\centering
\includegraphics[scale=.7]{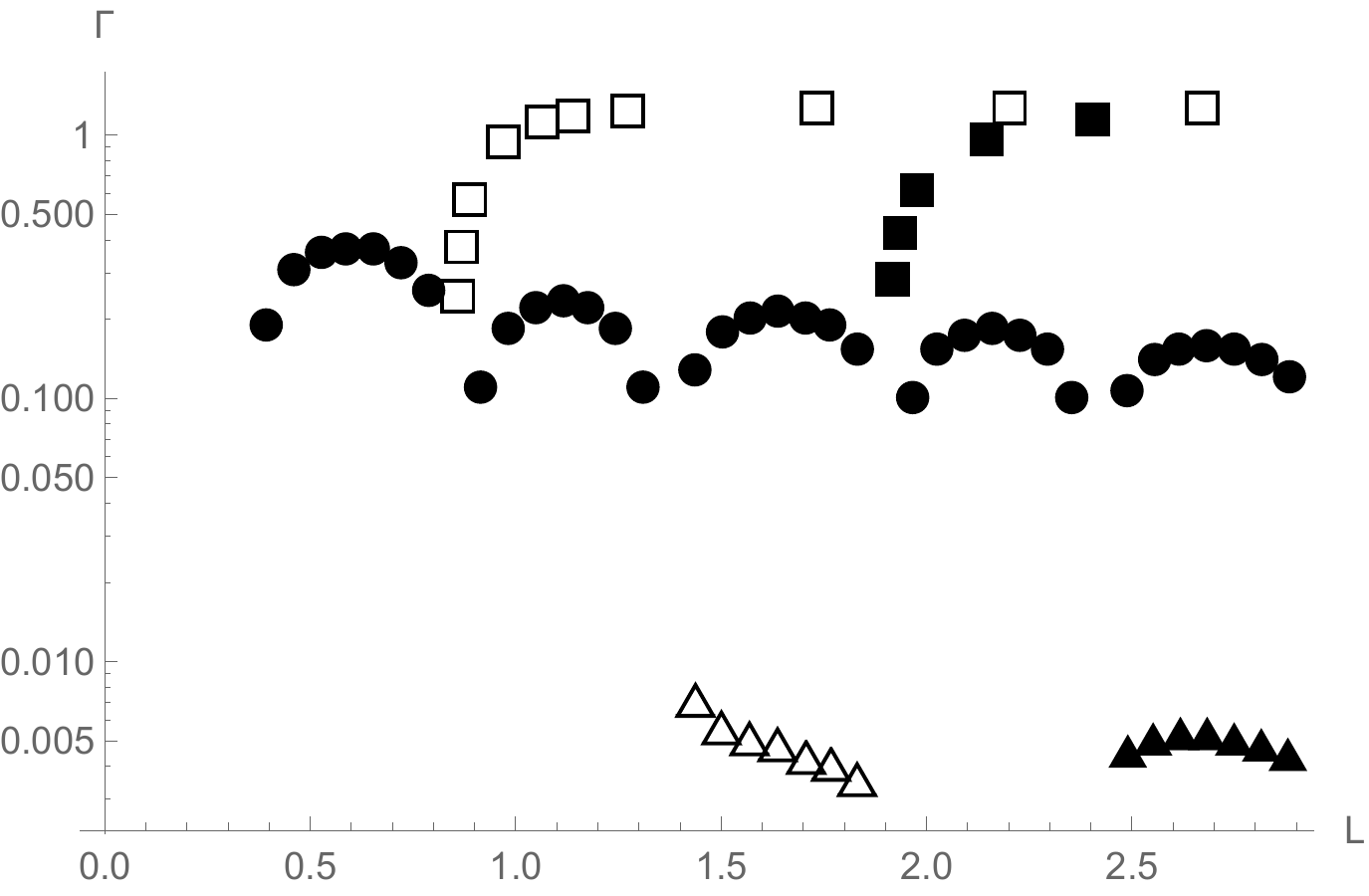}
\caption{As a function of $L$, we represent the imaginary part of the frequency of the most unstable mode on the homogeneous solution (circles), the first type 2 solution with $n= 1$  (empty squares), the second type 2 solution with $n= 2$ (filled squares), the type 1 solution with $n=2$  (empty triangles), and the type 1 solution with $n=3$ (filled triangles). One clearly sees that type 1 solutions are only mildly unstable compared to the other ones. 
The parameters are the same as those of \fig{fig:Soltun} except that $f_{b, \rm int} = 2$.
} \label{fig:Gammamax}
\end{figure}
To complete the analysis, we now study the relative magnitude of the instability of the above flows. Figure~\ref{fig:Gammamax} shows the imaginary parts of the frequencies of the most unstable modes on the homogeneous solution, as well as on type 1, type 2, and type 4 solutions. 
It can be seen that, for a fixed value of $n$, the instabilities on type 1 solutions are much milder than those on the homogeneous solution, while those on type 2 and type 4 solutions are stronger. Consequently, for sufficiently short time scales the type 1 solutions can be seen as nearly stable, while the other solutions are strongly unstable. The interested reader can find more details in subsection~\ref{ABM}. The implications of this hierarchy will become clear when studying time-dependent effects. 

\section{Nonlinear stationary solutions}
\label{stat_sol}

In this section we derive the exact solutions of the time-independent GPE in black hole laser configurations.
Our method is similar to that used in \cite{PhysRevA.64.033602} to describe a propagating Bose-Einstein condensate through a wave guide with an obstacle. Related ideas were also used in \cite{periodic_gradini}. We limit ourselves to solutions for which the density goes to $f_0^2$ at $z \rightarrow \pm \infty$. 
Only those have a finite energy difference with the homogeneous-density solution (see footnote~\ref{ft:DVE}).  
Our aim is to classify the set of solutions and to find the ground state of the system when the homogeneous configuration is unstable, \textit{i.e.}, for $L > L_0$. For simplicity, unless explicitly stated otherwise, we assume the microscopic parameters $g$ and $\mu$ are identical in $I_1$ and $I_3$: $ g_1 = g_3 , \, \mu_1 = \mu_3$. 

\subsection{Stationary solutions in the tuned case}
\label{sub:statsoltuned}

We work at fixed chemical potential $\mu$, temperature (set to zero), and current. 
In an experiment similar to that of~\cite{BHLaser-Jeff}, assuming a quasi-equilibrium is reaches, these quantities are fixed by the chemical potential and temperature of the uncondensed atoms, and the speed at which the potential creating the horizons moves in the laboratory frame.  
Considering the corresponding solutions as a statistical ensemble, the relevant thermodynamic potential is then (see Section~\ref{thermo_app}) 
\begin{equation}
E \equiv G - \int \, J \pd_z \theta \, \dd z.
\end{equation} 
The system is characterized by the parameters $g_1$, $g_2$, $\mu_1$, $\mu_2$, $J$, and the interhorizon length $2 L$. They are not independent: The assumption that a globally uniform solution exists gives a relation between them since the two polynomials 
$2 g_j \, f^6- 2 \mu_j \, f^4 +J^2$ evaluated in regions $1$ and $2$ must have a common root $f_0$;
see \eq{poly1}. When setting $f_0$ to unity by a rescaling of the unit of length, this condition becomes
\be \label{eq:cond_tun}
2 g_1 - 2 \mu_1  +J^2=2 g_2 - 2 \mu_2  +J^2=0. 
\ee 
The system thus depends only on four parameters, for instance $(c_1,c_2,v,L)$; see \eq{civ}. In the black hole laser case, we have $0 < c_2 < |v| < c_1$.
\begin{figure}
\centering
\includegraphics[width = 0.49 \linewidth]{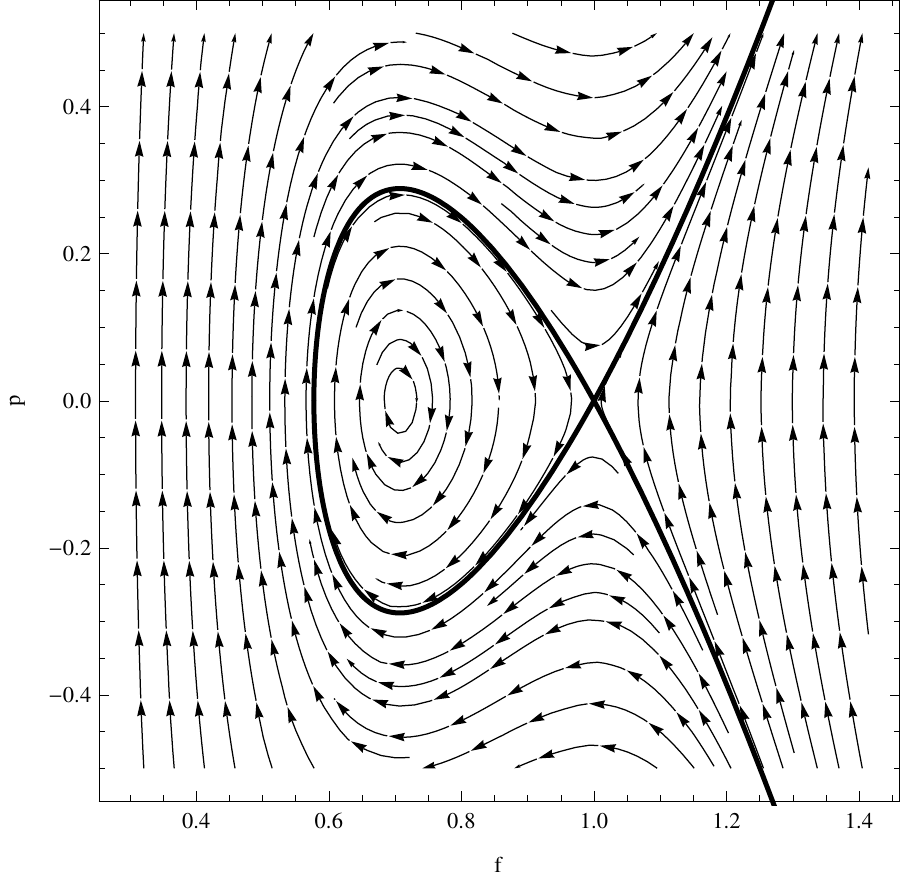}
\includegraphics[width = 0.49 \linewidth]{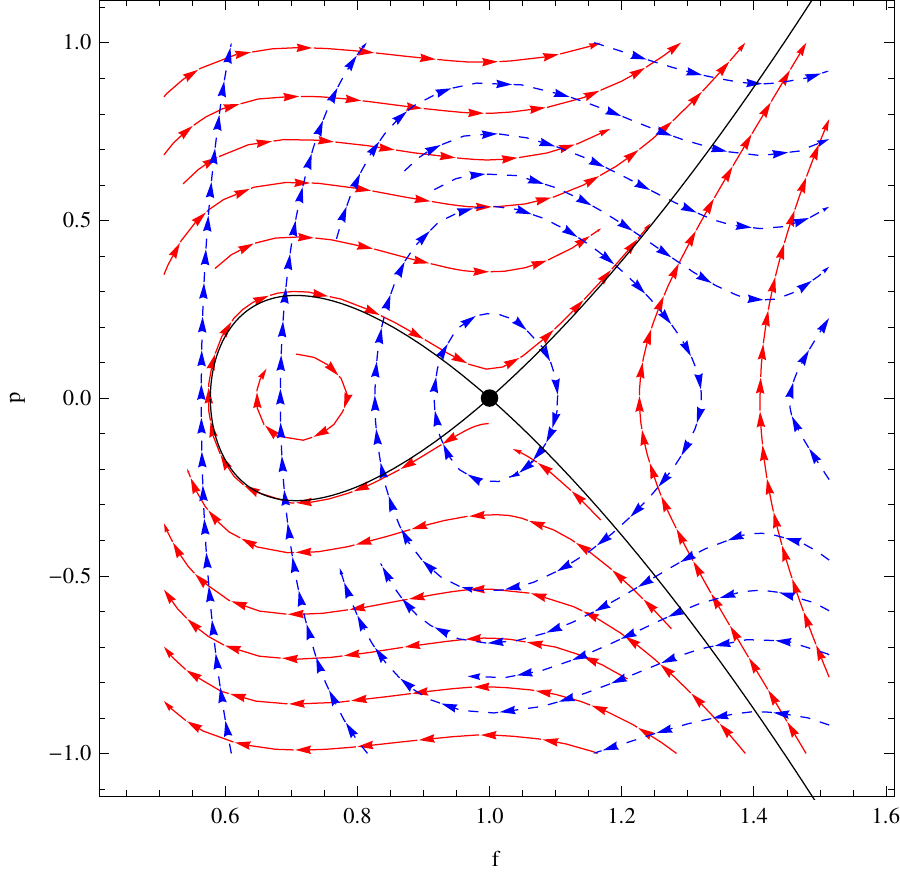}
\caption{Left panel: Phase portrait $p$ versus $f$ of \eq{keyE}.
It contains three qualitatively different regions, separated by the thick lines. 
The latter correspond to solutions which go asymptotically to a finite value. 
The middle domain contains periodic bounded solutions. Solutions in the right and left domains are divergent at finite values of $z$.  
Right panel: Two superimposed phase portraits corresponding to regions $I_1, I_3$ (solid, red) and $I_2$ (dashed, blue). The black dot represents the globally homogeneous solution $f = f_0$, and the black lines are the solutions which reach $f_0$ at infinity. The parameters are $g_1=8$, $g_2=1$, $\mu_1=28/3$, $\mu_2=7/6$ and $J^2=8/3$.}\label{phase_portrait2}
\end{figure}

The main properties of the solutions can be seen on the phase portrait, which shows the trajectories of the solutions in the $(f,p=f')$ plane; see Fig.~\ref{phase_portrait2}, left panel. The key equation in $I_j$ is given by the integral of \eq{eq:f}, namely,
\be \label{keyE}
p^2 = \frac{1}{f^2}\left(g_j f^6-2 \mu_j  f^4+C_j f^2-J^2\right),
\ee
where $C_j$ is the integration constant. 
Figure~\ref{phase_portrait2} right panel shows a superposition of the two phase portraits for the regions $I_1$ and $I_3$ (red,solid) and $I_2$ (blue, dashed).  Its qualitative properties, in particular the ordering of the three stationary points and the behavior of solutions around them, do not depend on the precise values of the parameters. 
(They would change if we allowed $c_1<|v|$ or $c_2>|v|$.) 

\begin{figure}
\centering
\includegraphics[width = 0.24 \linewidth]{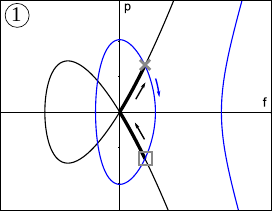}
\includegraphics[width = 0.24 \linewidth]{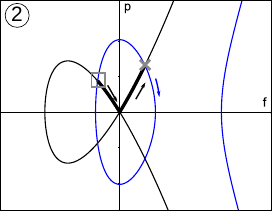}
\includegraphics[width = 0.24 \linewidth]{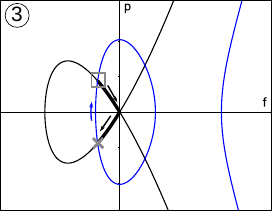}
\includegraphics[width = 0.24 \linewidth]{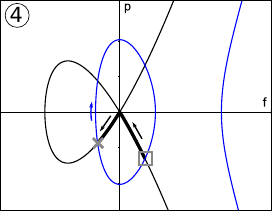}
\includegraphics[width = 0.24 \linewidth]{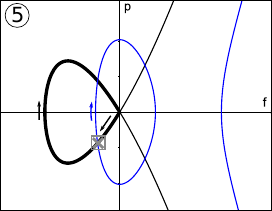}
\includegraphics[width = 0.24 \linewidth]{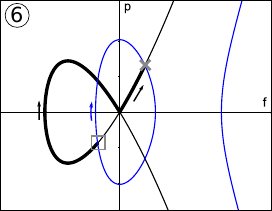}
\includegraphics[width = 0.24 \linewidth]{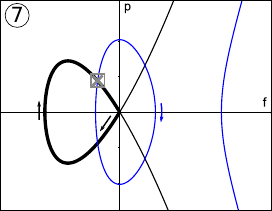}
\includegraphics[width = 0.24 \linewidth]{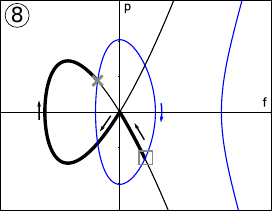}
\includegraphics[width = 0.24 \linewidth]{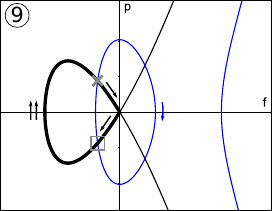}
\caption{The nine different types of trajectories in phase space: 
The first four solutions (top line) continuously connect to the homogeneous one, the next
four solutions have one soliton (middle), and the last solution (bottom) has two solitons.  
On each plot the black line of Fig.~\ref{phase_portrait2} is plotted along with the blue lines corresponding to a given value of the integration constant $C_2$. Thick black curves correspond to trajectories in phase space in $I_1$ and $I_3$, the direction being indicated by an arrow. The double arrow in the last plot indicates that a part of the curve is followed twice: once in $I_1$ and once in $I_3$. In $I_2$, the solution follows the closed blue line clockwise, starting from the first intersection with the thick one (materialized by a cross) at $z=-L$ and ending at the second intersection (box) at $z=L$. In between it can make an arbitrary number $n \in \mathbb{N}$ of turns.}\label{fig:9-traj}
\end{figure}

We are interested in solutions for which $f \rightarrow f_0$ as $z \rightarrow \pm \infty$. 
So, in Fig.~\ref{phase_portrait2} the solution must start on the black dot $f=f_0$, $f'=0$ at $z = - \infty$. When $z$ is increased, the solution either remains at that point (for the globally homogeneous solution) or moves along the black line until $z=-L$. It then follows the flow of the blue dashed lines until $z=L$. Finally, for $z>L$ it follows a black line again up to the black dot, which it reaches asymptotically. As described in subsection~\ref{App:single-h}, for a given value of the integration constant $C_2$, there are three possible trajectories in phase space for $z \in (-\infty,-L)$. The same is true for $z \in (L,\infty)$.
No restriction should be put \textit{a priori} on the solution in the central region $I_2$ since it is finite and will contribute to $E$ by a finite amount provided $f$ is regular in $I_2$. However, an inspection of the phase portrait in Fig.~\ref{phase_portrait2} reveals that, because of the matching conditions at $z = \pm L$, the solution in $I_2$ must lie in the central domain of the phase portrait in Fig.~\ref{phase_portrait2}.\footnote{In fact there exists one solution (type 3 in Fig \ref{fig:9-traj} for $n=0$) which can extend to values of $C_2$ giving solutions in the external domains. Whether it does so depends on the precise values of the parameters.}
As a result, the solution is characterized by the number of cycles in $I_2$, $n \in \mathbb{N}$, and the integration constant $C_2$. In total, for a given value of the discrete parameter $n \in \mathbb{N}$, there are nine different types of solutions. They are represented in Fig.~\ref{fig:9-traj}. For each of them, the value of the parameter $C_2$ is fixed by $L$. Also, the minimum value of $L$ at which solutions exist goes to infinity as $n \rightarrow \infty$. Hence, for a fixed $L$, there exists only a finite number of solutions. 

When $L$ is smaller than $L_0$ of \eq{eq:Lcrit}, only two solutions exist: the homogeneous one and 
another one of type 3 in Fig.~\ref{fig:9-traj} with $n=0$. 
(The latter becomes a dark soliton in the limit $L \to 0$.)
As shown in \eq{eq:E-os}, the energy density change in $I_j$ (with respect to the homogeneous solution) is
\be 
\Delta\mathcal{E}_j = -\frac{1}{2}g_j \lp f^4-f_0^4 \rp - J^2 \lp \frac{1}{f^2} - \frac{1}{f_0^2} \rp .
\ee
For $L < L_0$, the nonuniform solution has a positive energy. 
Hence the homogeneous configuration is stable. 
When $L > L_0$, the inhomogeneous solution is replaced by that corresponding to plot 1 in Fig.~\ref{fig:9-traj}, which has a negative energy. 
Therefore the homogeneous solution becomes energetically unstable at $L=L_0$. 
This confirms the results of our linear analysis presented in Section~\ref{Slt}, where the first dynamical instability was found for $L > L_0$. 
As expected from \cite{Jackson,PhysRevA.72.032101}, the dynamical instability appears together with a static instability when a solution becomes thermodynamically more favorable than the uniform one. Notice that the transition at $L=L_0$ is a second order one
since the amplitude of the oscillations in $I_2$ goes to zero as $L \rightarrow L_0$. Notice also that when $|v|<c_1,c_2$, the uniform solution is always stable, while if $|v|>c_1$ it is always unstable. 

Figure~\ref{fig:ene-9-eq} shows $\Delta E$ as a function of $L/L_0$. 
The formulas we used are given in subsection~\ref{App:eqs_G_L} [see Eqs. (\ref{eq:GtoE}-\ref{eqs-num-end})].
This figure first establishes that the type 1 solution with $n=0$ is indeed the lowest energy state. We also see that at large $L$, $\Delta E(L)$ becomes linear for all solutions with a negative slope $\frac{1}{2} g_2 (f_0^2 - f_{2,b}^2) + J \lp f_0^{-2} -f_{2,b}^{-2} \rp$, where $f_{2,b}$ is the subsonic uniform solution in $I_2$ given in \eq{eq:fb}.
Note that for $n \neq 0$ and $L$ slightly smaller than its critical values there are two solutions 
of type 3. This is because the length $L$ associated with this series of 
solutions is not monotonic in the integration constant $C_2$. It decreases close to its minimum value but then increases with $C_2$.

\begin{figure}
\centering
\includegraphics[width=0.49 \linewidth]{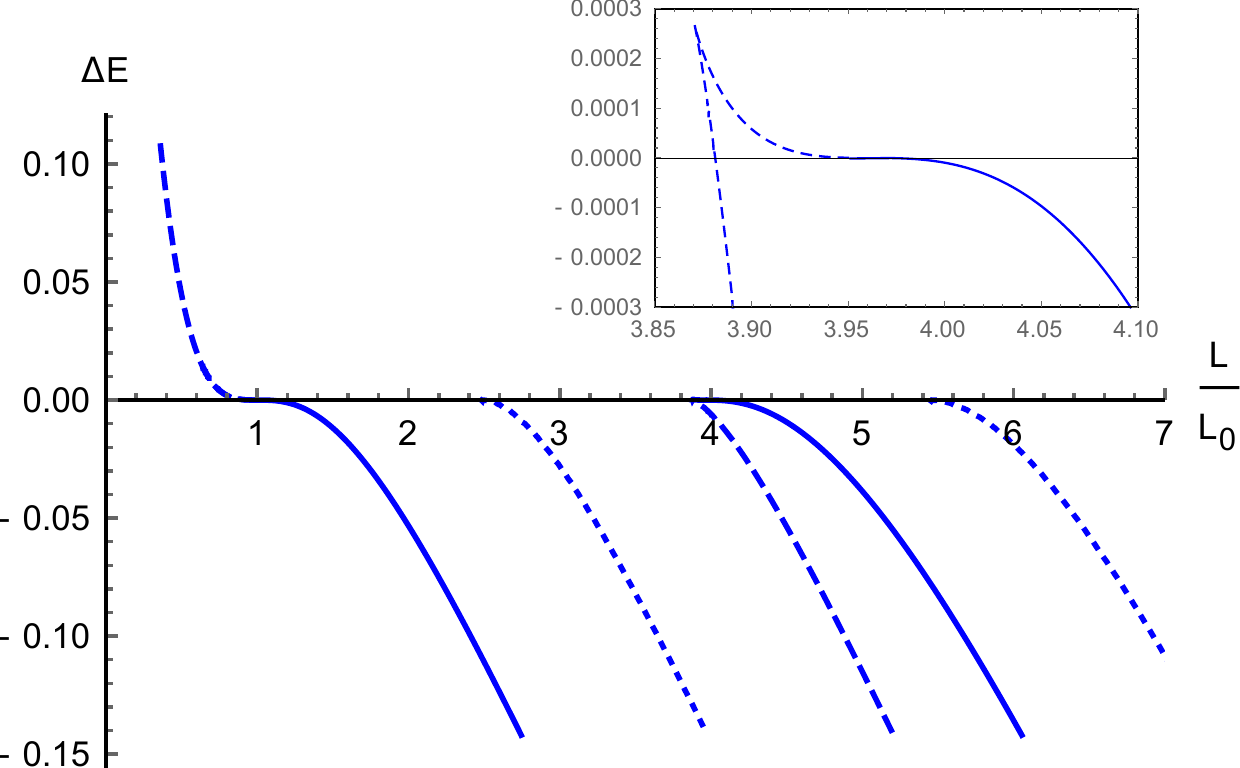} 
\includegraphics[width=0.49 \linewidth]{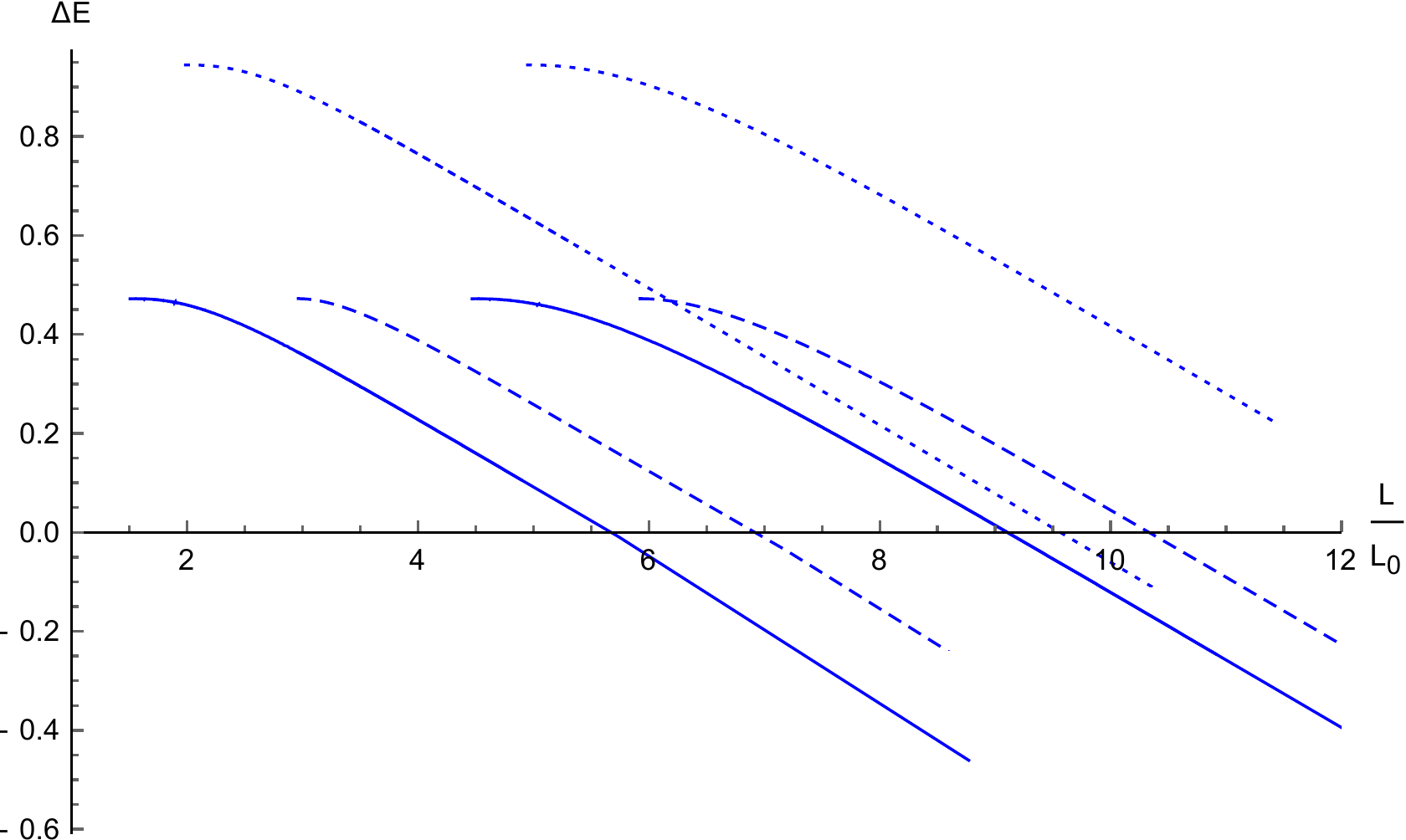} 
\caption{Left panel: Energy differences $\Delta E$ between the four different types of solutions with no soliton and the homogeneous one 
as functions of $L/L_0$. The number of cycles $n$ is equal to $0$ and $1$.
Solid lines: 
Type 1 in Fig.~\ref{fig:9-traj}; dotted lines: 
Types 2 and 4 (degenerate for $c_1 = c_3$),
and dashed lines: 
type 3. We set $c_3^2=c_1^2=8$, $c_2^2=1$, $v^2=8/3$ and $f_0=1$. 
The inset shows a zoomed-in picture of the beginning of the curve for types $1$ and $3$ when $n=1$.
As explained in the text, for $n= 0$, the type 3 solution exists from $L=0$ to $L = L_0$, with a larger energy than the homogeneous one, and type 1 from $L=L_0$ to $L \rightarrow \infty$ with a smaller energy than the homogeneous one. The situation is similar in the case $n=1$, except the first branch makes a U-turn, giving two parallel lines at large $L$.
Right panel: thermodynamic potential difference $\Delta E$ of the five different types of solutions with one or two solitons 
as functions of $L/L_0$.
Solid lines: 
Types 5 and 7 of Fig.~\ref{fig:9-traj} (which are degenerate when $c_3=c_1$);
dashed lines: 
Types 6 and 8 (also degenerate for $c_1 = c_3$), and dotted lines: 
type 9. These solutions have a larger energy than the homogeneous one when they appear. Their energy is also always larger than that of 
type 1. }\label{fig:ene-9-eq}
\end{figure} 

As can also be expected, solutions with one or two solitons, corresponding to types 5 to 9 in Fig.~\ref{fig:9-traj}, have a larger energy than the other solutions for a given value of $L$. 
It is therefore unlikely that they play an important role in the time evolution of the system. 
Solutions of type 1 exist for arbitrarily large values of $L$, while the solution of type 3 with $n=0$ does not exist beyond $L = L_0$. 
All other solutions can extend to $L=\infty$ or not depending on the parameters of the black hole laser. 
A straightforward calculation shows they actually extend to infinity if and only if the inequality of \eq{eq:inf} is satisfied. 
When it is not, as explained in subsection~\ref{App:single-h}, series of solutions terminate at a finite value of $L$ by merging with one another. A series of type 2 solutions will merge with one of type 6 and one of type 4 with one of type 8. The four series types 3, 5, 7 and 9 all merge.  
Instead, series of solutions of type 1 never terminate. This is important because the type 1 with $n=0$ gives the ground state of the system. We now study this case in more details.

\begin{figure}
\centering
\includegraphics[width = 0.49 \linewidth]{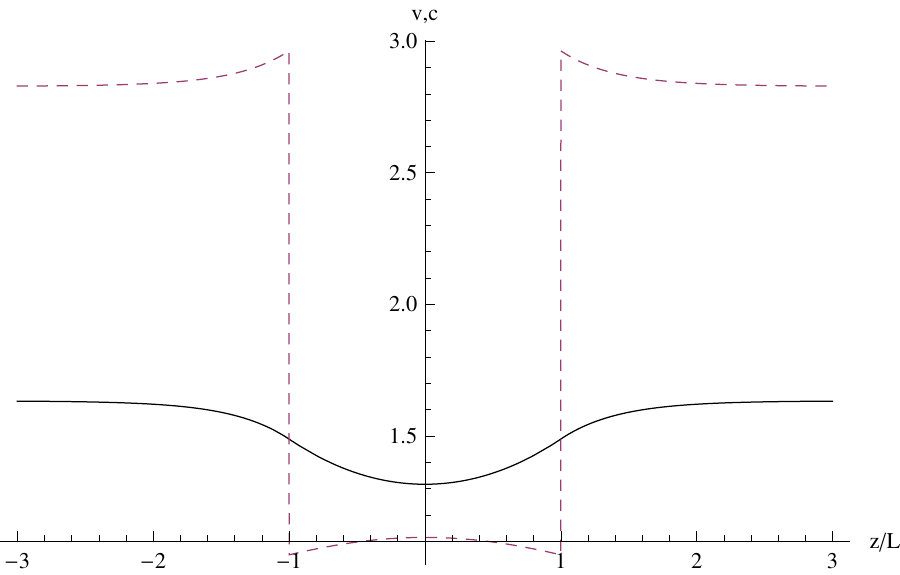}
\includegraphics[width = 0.49 \linewidth]{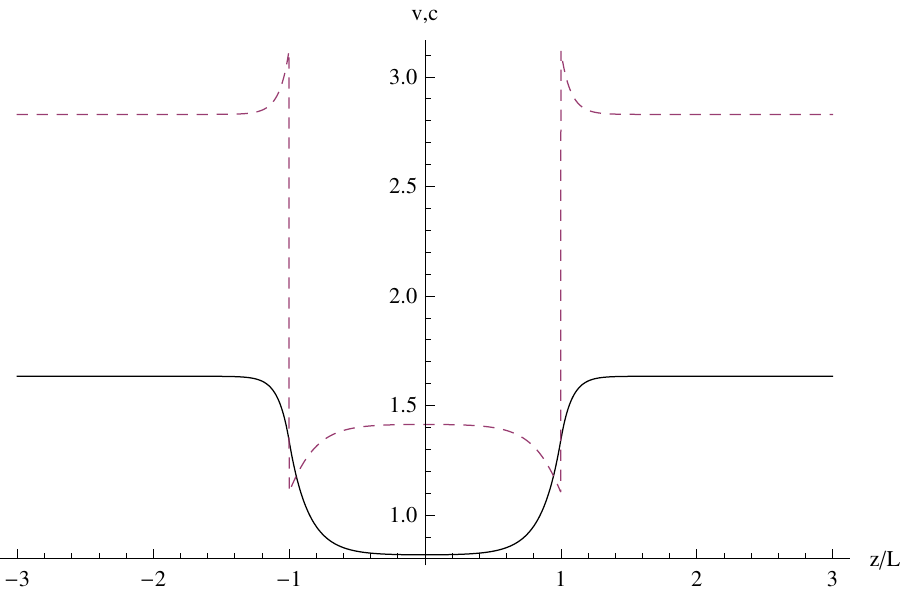}
\caption{Velocity (solid line) and sound speed (dashed line) as functions of $z/L$ for the solution with the lowest thermodynamic potential $E$ for a distance $L$ slightly above the threshold, $L =1.29\, L_0$ (left diagram), and well above the threshold $L = 7.0 L_0$ (right diagram). 
The parameters are $c_1=c_3=2 \sqrt{2}$, $c_2 = 1$, $v=\sqrt{8/3}$, and $f_0=1$. On the right panel, one clearly sees the saturation of the solution with a flat 
profile in the central region which corresponds to a subsonic flow. Notice that the density profile $f^2(z)$ can be deduced from that of $v$ since 
the current $J = f^2(z) v(z)$ is uniform. 
}\label{fig:ground-state}
\end{figure}

In this state, for $L \gg L_0$, the amplitude $f$ and velocity $v$ become nearly piecewise constant
with two transition regions at $z \approx \pm L$ with extensions of the order of the healing length; see the right panel of Fig.~\ref{fig:ground-state}. In addition, the condensate is subsonic outside the two transition regions. Since negative-energy fluctuations only exist when the supersonic flow has a sufficiently large extension, it is clear that this configuration is energetically stable, and represents the end point evolution of the black hole laser effect when the dynamics leads to stationarity and minimization of the thermodynamic potential. 
This hypothesis is investigated numerically in subsection~\ref{sub:BHLIaddpaper} and in~\cite{2015arXiv150900795D}. 
We now understand that the physical mechanism which stabilizes the laser effect is the accumulation of atoms in the central region. 
Indeed, the associated increase of the density reduces the velocity of the flow $v$, and increases the sound speed,
thereby removing the supersonic character of the flow. 
Obtaining the profile of the ground state is the main result of this chapter. 
It can be done by using the following procedure. 
The trajectory in phase space $(f,p \equiv f')$ is given by \eq{keyE}, 
with the constants in $I_1$ and $I_3$ equal to
\be 
C_i = (2 \, v^2 + c_i^2) f_0^2. 
\ee
The third constant, $C_2$, is fixed by the value of $L$ through \eq{eqs-num-beg}. 
The profile is then obtained by integration of \eq{keyE}
which is a first-order ordinary differential equation. A convenient 
initial condition is the value of $f$ at $z=- L$, given by $f_{\text{inter},+}$ of \eq{finter}.

So far we have worked with an idealized description where the parameters $g$ and $V$ entering \eq{GPE1} are piecewise constant with two discontinuities. However, in a realistic setup, $g$ and $V$ will change over some finite 
length scale, which we assume to be equal and call $\lambda_g$. 
To determine the validity range of results obtained with the steplike approximation, we look for the leading deviations of our results due to a small $\lambda_g \neq 0$. 
To this end, we replaced the piecewise constant $g$ and $V$ by various smooth profiles and solved \eq{eq:f} numerically using an imaginary-time evolution. To leading order in $\lambda_g/L_0$, where $L_0$ is given in \eq{eq:Lm}, the only effect is to change the critical values of $L$ where new unstable modes appear.   
Here $L$ is still defined as half the length of the supersonic region. In particular, for the  ground state of
the system, we checked that the relation between the maximum value of $f$ and $L-L_0$, written below in the symmetric
case $c_1^2-v^2 = v^2-c_2^2$ for simplicity,\footnote{\eq{dvtf} can be straightforwardly derived  from \eq{eqs-num-beg} in the case $\lambda_g=0$.}
\begin{align} \label{dvtf}
\frac{f_{\max }}{f_0} -1  = 2 \sqrt{2} \frac{\left(L-L_0\right) \left(v^2-c_2^2\right)^{3/2}}{2 v^2+c_2^2} + O(\left(L-L_0\right)^2),
\end{align}
is unchanged to lowest order in $\lambda_g$, although the value of $L_0$ changes. 
We should thus analyze how the latter is affected by $\lambda_g \neq 0$. 
In the general case, when $\lambda_g/L_0 \lesssim 1/10$, we found that the leading deviation of $L_0$ is linear in $\lambda_g$. 
For profiles which are symmetric between the subsonic and supersonic regions, we found that the differences are quadratic in $\lambda_g$.   
This robustness is in agreement with the spectral analysis of~\cite{Finazzi:2012iu} performed in the case of a single horizon. In that case it was found that the Bogoliubov coefficients encoding the scattering across a supersonic transition are well approximated by their steplike approximate values when $\lambda_g$, i.e., roughly speaking the inverse of surface gravity, is a tenth of the healing length; see Fig.~4 in~\cite{Finazzi:2012iu} for more details. With the observation that \eq{dvtf} remains unchanged at leading order, we have established that the robustness of the step-like approximation extends to the saturation process. 

To end this section we briefly comment on the changes brought about by different sound velocities in $I_1$ and $I_3$. The analysis is very similar to that in the case $c_1 = c_3$ with three phase portraits instead of two. 
The set of solutions is qualitatively similar. In particular, solutions are characterized by the same set of parameters. There is one additional solution for a limited range of $L$ with a larger energy than that of the uniform solution. The other differences are that the first nonuniform solution does not extend to $L=0$ anymore and that previously degenerate solutions now have different energies.

\subsection{Stationary solutions in the detuned case}
\label{sub:statsoldet}

It is worth verifying that a small detuning of the parameters, i.e., a deviation from \eq{eq:cond_tun}, does not significantly affect the main conclusions of the above analysis. To this end we follow the same method as in the ``tuned'' case with minor modifications, see subsection~\ref{NLsol}. 
To this end, it will be useful to define the two positive quantities $f_{b, \text{ext}}$ and $f_{p, \text{ext}}$ (respectively $f_{b, \text{int}}$ and $f_{p, \text{int}}$) solutions of the equation in $f$: $2 g_1 \, f^6 - 2 \mu_1 \, f^4 + J^2 = 0$ (respectively $2 g_2 \, f^6 - 2 \mu_2 \, f^4 + J^2 = 0$) and such that $0 < f_{p, \text{ext}} < f_{b, \text{ext}}$ (respectively $0 < f_{p, \text{int}} < f_{b, \text{int}}$). 
These correspond to the amplitudes of the two homogeneous solutions in the exterior (respectively interior) region. 
The one labeled with an index ``$b$'' gives a subsonic solution, while the other one gives a supersonic solution. 
The ``tuned'' case where a globally homogeneous, transcritical solution subsonic for $z \to \pm \infty$ exists thus corresponds to $f_{b, \text{ext}} = f_{p, \text{int}}$.   

\begin{figure}
\centering
\includegraphics[width=0.49\linewidth]{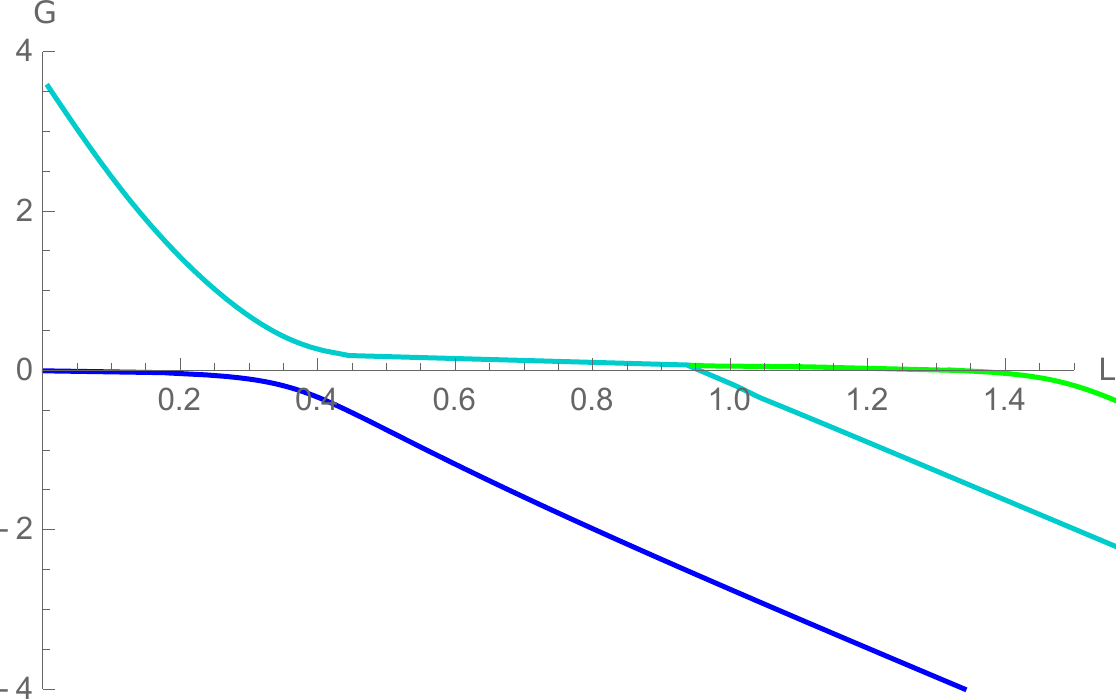}
\includegraphics[width=0.49\linewidth]{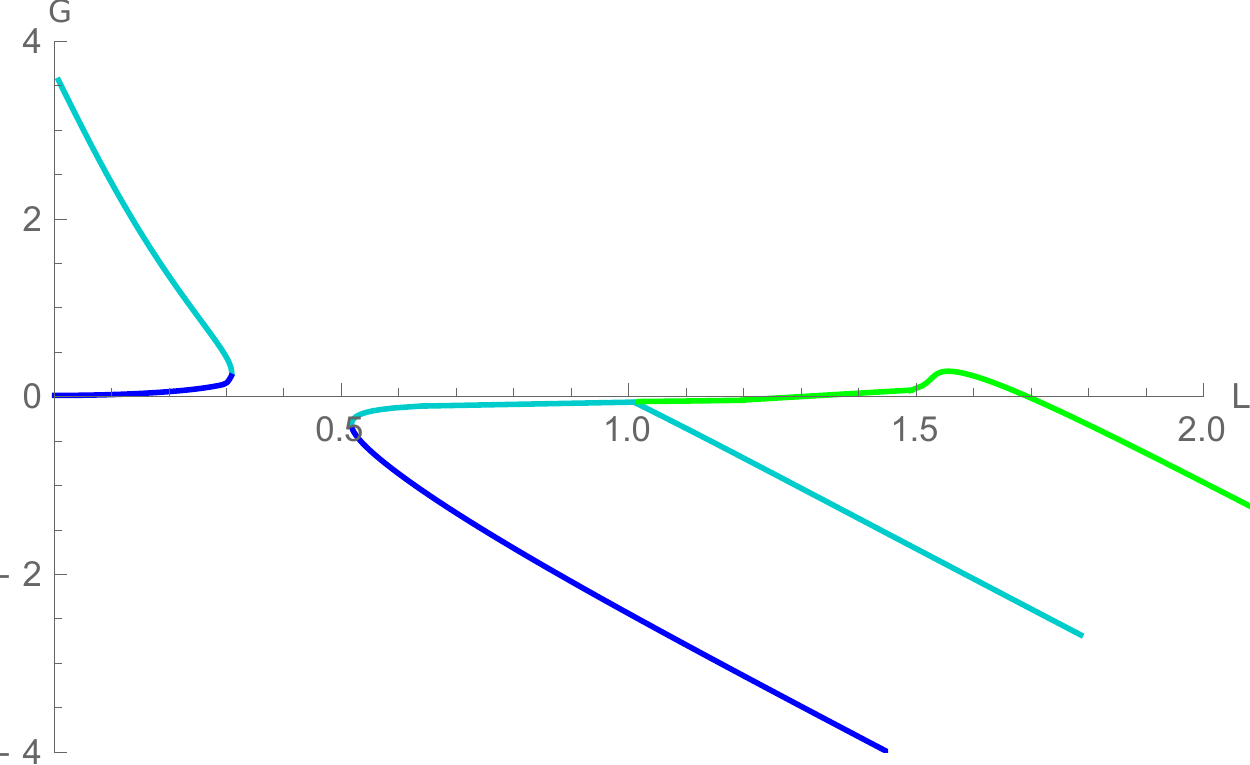}
\includegraphics[width=0.49\linewidth]{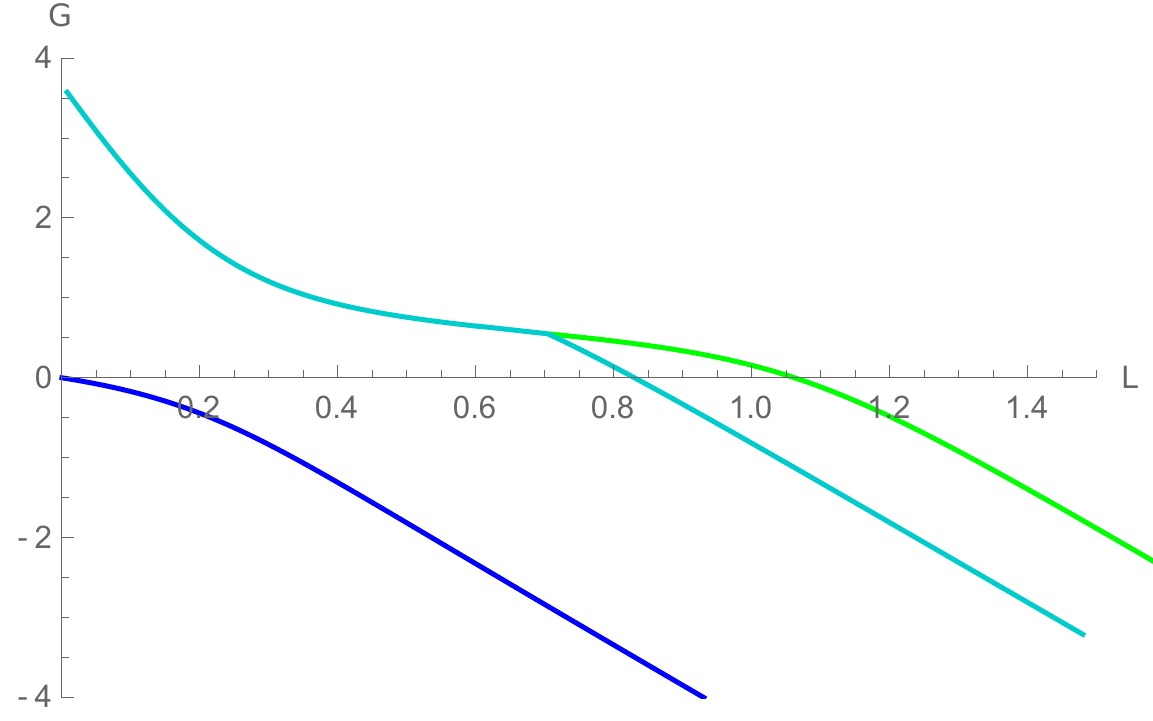}
\includegraphics[width=0.49\linewidth]{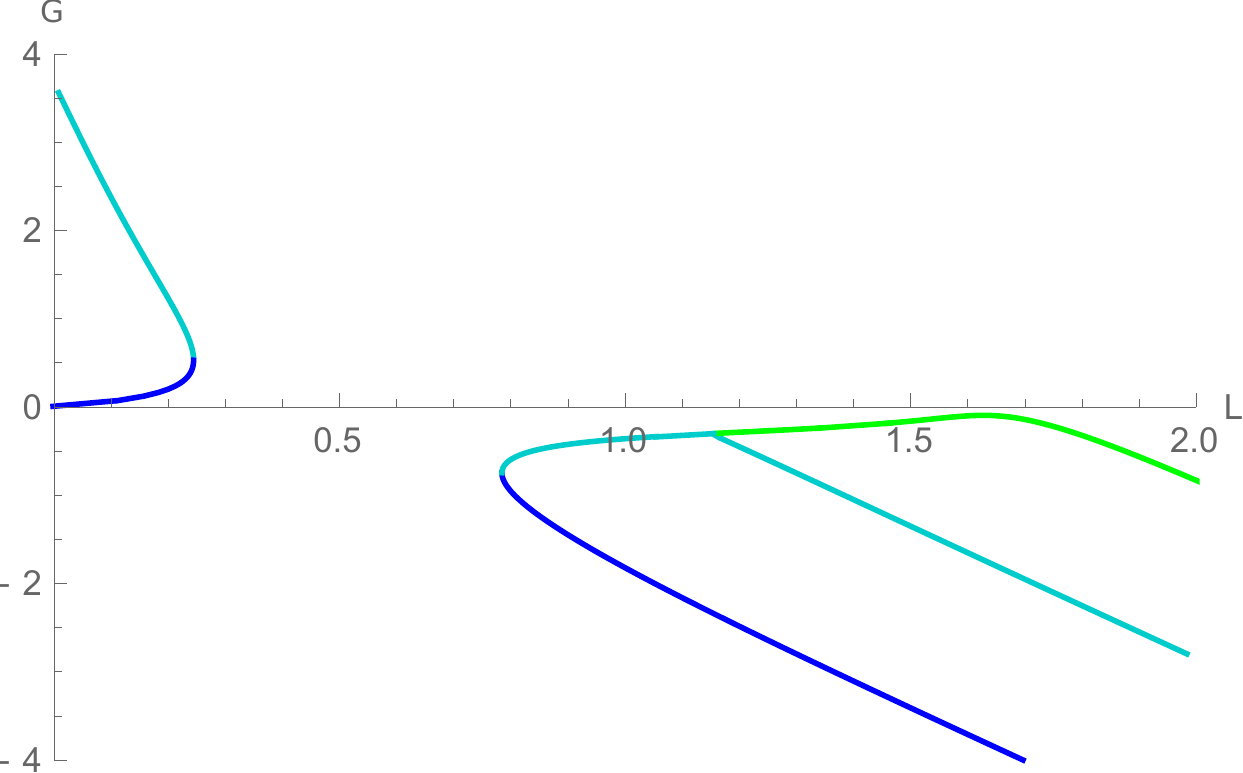}
\caption{Thermodynamic potentials $G$ of the first nonlinear stationary solutions of the GPE as functions of the half-distance $L$ between the two discontinuities of the potential, for four different ``detuned'' sets of parameters which do not satisfy \eq{eq:cond_tun}. The parameters $J = \sqrt{8/3}$, $f_{b,{\rm ext}}=1$, $f_{p,{\rm ext}}=0.7$, and $f_{b,{\rm int}} = 1.5$ take the same values for these four plots. The two plots on the left correspond to a positive detuning, with $f_{p,{\rm int}} = 0.99$ (top) and $0.9$ (bottom). The two plots on the right correspond to a negative detuning, with $f_{p,{\rm int}}=1.01$ (top) and $1.05$ (bottom).
Colors have the same meaning as in \fig{fig:Soltun}.
}\label{fig:NLsol}
\end{figure} 
The energy $G$ \eq{eq:G_off} of the first solutions is shown in~\fig{fig:NLsol}.\footnote{We use $G$ instead of $E$ as the results are more easily seen on this quantity. The properties of the solutions we here consider are invariant when replacing $G$ with $E$.} 
For definiteness, we assume the detuning is small, in the sense that
\be 
f_{p,{\rm int}}^2 \sqrt{\frac{2}{f_{p,{\rm int}}^2+f_{b,{\rm int}}^2}} < f_{b,{\rm ext}} < f_{b,{\rm int}}.
\ee
In other words, the subsonic homogeneous density in the external region is between the two extremal densities of the stationary soliton in the internal region. We also impose that
\be 
f_{p,{\rm ext}}^2 \sqrt{\frac{2}{f_{p,{\rm ext}}^2+f_{b,{\rm ext}}^2}} < f_{p,{\rm int}}.
\ee
These two conditions are always satisfied provided $f_{b, \rm ext}$ is sufficiently close to $f_{p, \rm int}$, i.e., for small detunings. 
Then, the main difference with respect to the tuned case is the following: When $f_{b,{\rm ext}} = f_{p,{\rm int}}$, the homogeneous solution $f = f_{b,{\rm ext}}$ exists for all values of $L$ and is connected to an infinite number of series of solutions. When $f_{b,{\rm ext}} \neq f_{p,{\rm int}}$, each series of solutions is now connected only to a finite number of other series when varying $L$, as can be seen in \Fig{fig:NLsol} for the first few solutions. We checked this remains true when including all solutions, as can be easily deduced from plots of the phase portrait of \eq{keyE}; see \Fig{fig:PP}. 
Continuity of the set of solutions in the limit of a tuned black hole laser is recovered when noticing that, for very small detunings, different series of solutions are alternatively very close to being homogeneous. This can be seen in the two upper panels of the figure: For most of the represented values of $L$, there exists a solution whose energy is close to zero. The spatial profiles of $f$ for such solutions are nearly homogeneous. 

Let us first consider the case of a positive detuning (left panels of the figure). For $L \approx 0$, there are two stationary solutions. The one with highest energy is analogous to the first type 3 solution in the tuned case, in that it contains fractions of solitons attached at $x=-L$ and $x=+L$. The one with lowest energy is close to being homogeneous. When increasing $L$ the type 3 solution moves towards the other one, as the fraction of soliton at $x = \pm L$ decreases. Increasing $L$, we see a transition similar to an avoided crossing in quantum mechanics, and the series of solutions close to the type 3 one becomes nearly homogeneous while the other solution shows fractions of shadow solitons at $x = \pm L$, therefore becoming analogous to the type 1 solution. 
Two new series of solutions with the same energy, related by a parity transformation, appear at a critical value of $L$ close to $1$ for the parameters of the figure. These series correspond to type 2 and type 4 solutions, in that both of them have part of a soliton at one end of the internal region and part of a shadow soliton at the other end. With increasing $L$ again, the nearly homogeneous solution goes to the second type 1 solution. 
We find that the same pattern repeats itself periodically: One branch of solutions appears corresponding to a type 3 one, which becomes more homogeneous, generates one series of type 2 and type 4 solutions, and turns continuously to a type 1 solution. The set of stationary solutions is thus very similar to that of the tuned case, except that no series of solution goes continuously from one type 1 to the next type 3 solution. Instead, an avoided crossing separates the two series. 

The case of a negative detuning is very similar, except that the avoided crossing is replaced by a merging: The two initial solutions merge at a critical value of $L$ and two new solutions appear at a second, larger critical value. Interestingly, even when considering higher-energy solutions not represented in the figure, no stationary solution exists between the first two critical values of $L$. In that case, numerical simulations using the same code as those presented in the next section always show the emission of infinite soliton trains. We thus recover a situation similar to the one found in Ref.~\cite{Hakim1997}, in that soliton trains arise from the absence of stationary solution. 

In brief, even though a small detuning introduces some modifications of the series of stationary solutions, such as some avoided crossing or merging, it does not significantly affect the physical properties of the set of stationary solutions.

\section{Next-to-quadratic effects and saturation}
\label{thermo}

There exists a close correspondence between the linear analysis of Section~\ref{Slt} and the nonlinear solutions of Section~\ref{stat_sol}. 
Indeed, when $c_1 = c_3$, each degenerate ABM appears at $L = L_m$ for an integer value of $m$, together with a series of stationary solutions of type 1 which possess a smaller thermodynamic potential than the homogeneous one.
Moreover, for $L < L_m$, the QNM which turns into this ABM when its frequency crosses the real axis corresponds to a solution of type 3, with a larger thermodynamic potential. In addition, each nondegenerate ABM appears at $L = L_m$ for a half-integer value of $m$ with a series of stationary solutions of types 2 and 4. 

However, this correspondence is not manifest when using the exact treatment  of Section~\ref{stat_sol}. 
Here, we introduce a simplified energy functional $E_s$ which displays  
more clearly the correspondence near $L \approx L_m$ 
for integer values of $m$. For half-integer values of $m$, the analysis is more complicated, 
as briefly explained at the end of the section. 
To construct the functional we use an expansion at lowest nonquadratic order of the relevent thermodynamic potential. 
This perturbative treatment, being rather general, might allow for extensions to other cases where zero-frequency waves with large amplitudes are also found, for instance, in hydrodynamics~\cite{Coutant:2012mf,Johnson}, in massive theories of gravity~\cite{2013arXiv1309.0818B}, and in the presence of extra dimensions~\cite{Gauntlett_helical_BHs,Gauntlett_QNM}.
This construction was inspired by the analysis of \cite{Pitaevskii1984,Baym}, which describes the occurrence of spatially modulated phases in superfluids with a roton-maxon spectrum when the flow velocity slightly exceeds the Landau velocity. 
In that case, the effective energy functional which governs the saturation of the amplitude is quartic, as in standard second-order phase transitions. In the present case instead, the stabilizing term 
is cubic, as in a $\lambda \phi^3$ theory. This odd term is due to the breaking of the $\mathbb{Z}_2$ symmetry $f \rightarrow 2 f_0 - f$ discussed in subsection~\ref{App:single-h}. In what follows, we concentrate on even solutions without a soliton, corresponding to types 1 and 3 in Fig.~\ref{fig:9-traj}. For definiteness, we set $c_3 = c_1$.
Then local extrema of a simplified energy functional allow us to recover the change of stability occurring at all $L=L_n$, $n \in \mathbb{N}$. 

\begin{figure}
\centering
\includegraphics[width = 0.49 \linewidth]{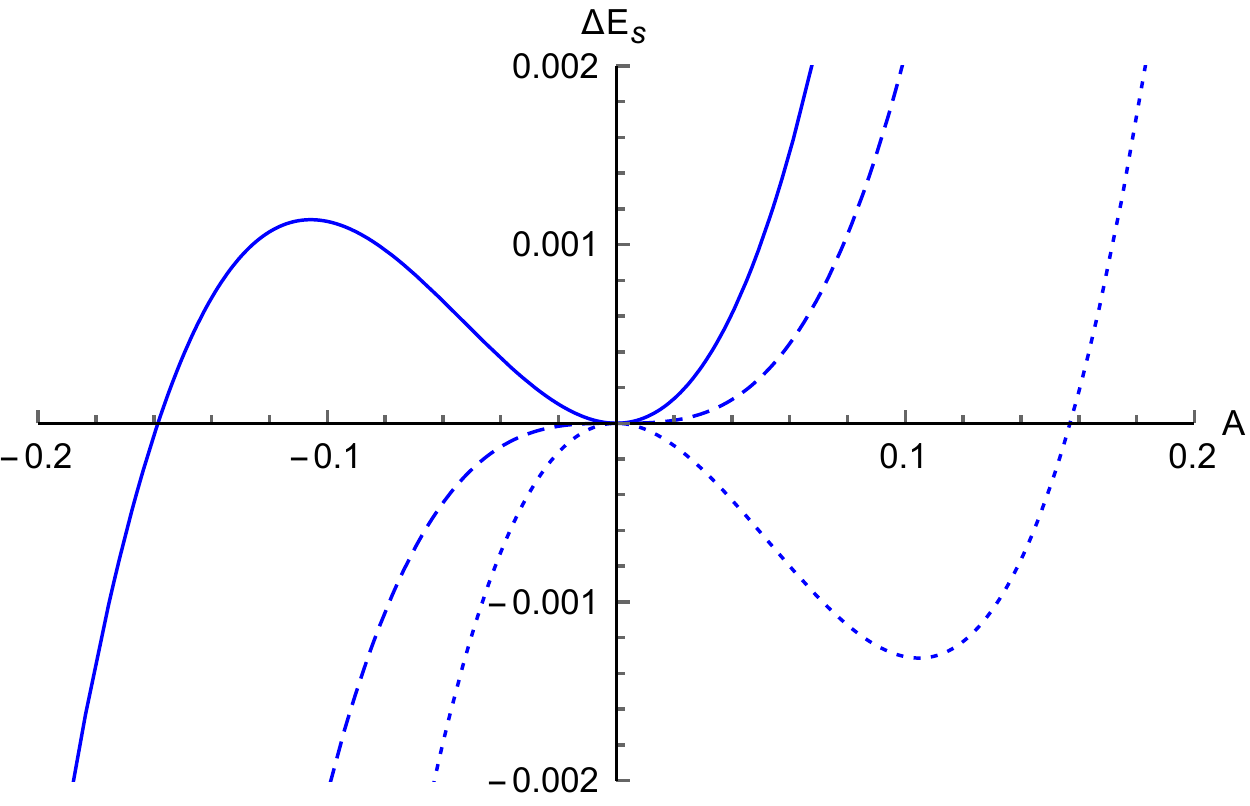}
\includegraphics[width = 0.49 \linewidth]{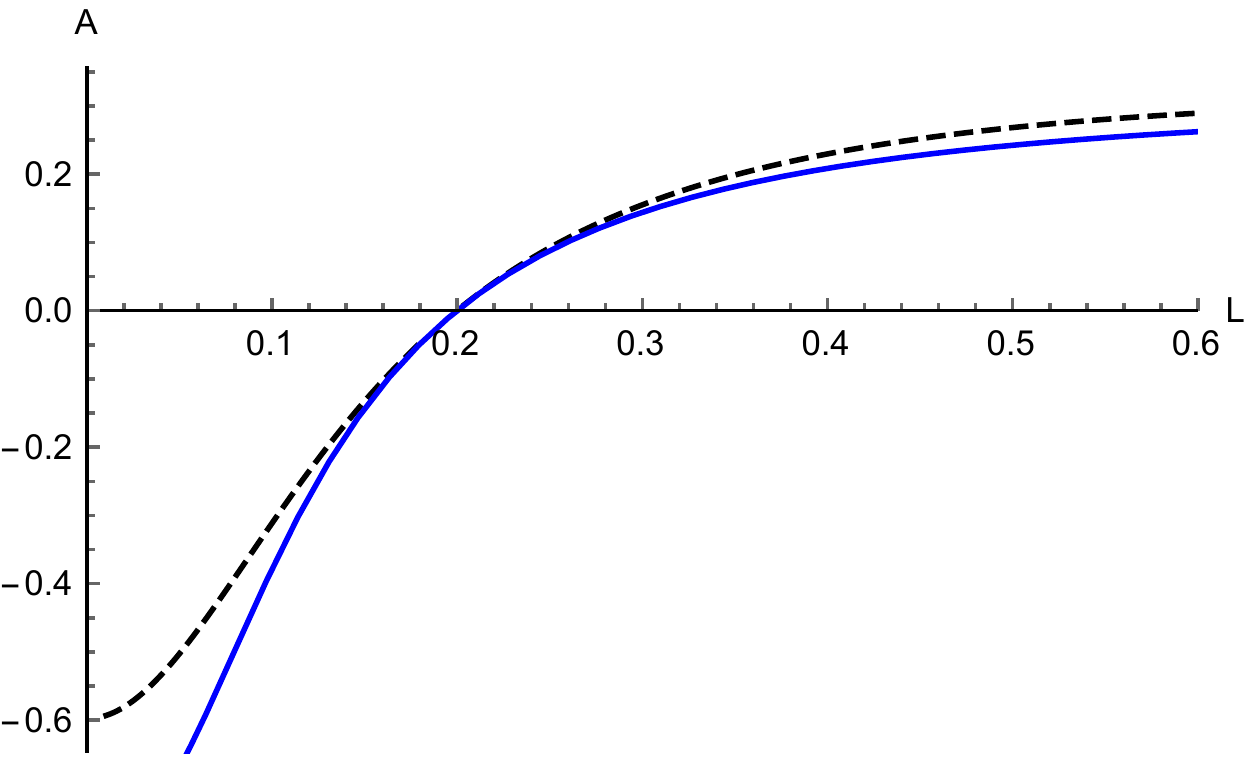}
\caption{Left panel: Simplified thermodynamic potential (\ref{eq:en_A}) 
as a function of the amplitude $A$ for $L=0.85 L_0$ (solid line), $L= L_0$ (dashed line), and $L=1.2 L_0$ (dotted line). The wave vector is $k_2=2 \sqrt{v^2 - c_2^2}$; see \eq{eq:lambda}. One clearly sees that the change of stability of the homogeneous solution occurs for $L= L_0$. 
Right panel: Amplitude $A$ of the inhomogeneous solution (type 3 for $A<0$ and type 1 for $A>0$) as a function of $L$. The solid line is the result from the simplified treatment of (\ref{eq:AB}), and the dashed line is from the full nonlinear solution. The two methods give the same values for $A$ and $\pd_L A$ at $L=L_0$. The parameters are: $c_1=2.0$, $c_2=0.5$, $v=1.0$, and $f_0=1.0$.} \label{fig:EBaym}
\end{figure}

As in subsection~\ref{App:single-h}, 
we write $f(z)=f_0 + \delta f (z)$, where $f_0$ is the globally homogeneous solution and $\norm{\delta f}_\infty \ll f_0$.  
We assume that $\delta f$ belongs to the Sobolev space $W^{1,2}(\mathbb{R})$. 
To third order in $\delta f$, the thermodynamic potential $E$ reads (up to a constant term)~\footnote{Notice that for stationary solutions with densities not going to $f_0^2$ at infinity $\delta f$ is not square integrable. The energy difference $\Delta E$ is thus infinite. On the other hand, one can show that all functions belonging to $W^{1,2}(\mathbb{R})$ go to zero at both infinities and are in $L^p(\mathbb{R})$ for any integer $p \geq 2$.
Each term in the expansion of $\Delta E$ in powers of $\delta f$ is therefore finite. 
In addition, the series thus obtained is absolutely convergent provided 
\begin{align}
\exists \eta \in \left] 0, 1 \right[ , \, \exists p_c \in \mathbb{N}, \, \forall p \in \mathbb{N}, \, p \geq p_c \Rightarrow \norm{\frac{\delta f}{f_0}}_p \leq \eta,
\end{align}
which is satisfied for the ansatz \eqref{ansatz} provided $\abs{A}, \abs{A_1} < f_0$. 
\label{ft:DVE}}
\be \label{eq:third_order}
\Delta E=\int_{-\infty}^{\infty}  \left(\frac{1}{2}\left(\frac{\partial \delta f}{\partial z}\right)^2+2\left(c(z)^2-v^2\right)\delta f^2\right) \dd z+2\int_{-\infty}^{\infty}   \left(c(z)^2+v^2\right) \frac{\delta f^3}{f_0} \, \dd z+\text{...}
\ee
The idea is now to choose an ansatz for $\delta f$ which depends on
some parameters, and extremize $E$ with respect to them.  
If the ansatz is well chosen, the solution will be close to the exact solution.
To optimize the choice near $L_n$, we work with an ansatz compatible with the linear even solutions of \eq{linearfth}:
\be \label{ansatz}
\delta f(z)=
 \left\lbrace  
\begin{array}{ll}
 A_1 \, \e^{-k_1|z|} , & \text{for} \, |z|>L , \\
 A \cos \left(k_2 z\right),  & \text{for} \, |z|<L .
\end{array}
\right.
\ee
Continuity and differentiability at $|z|=L$ give
\be 
k_1=k_2 \tan \left(k_2 L\right) , 
\ee
and 
\be 
A_1=A\cos \left(k_2 L\right)e^{k_1 L}.
\ee 
Performing the integrals explicitly, \eq{eq:third_order} becomes
\be \label{eq:en_A}
\Delta E_s = W_2(k_2,L)\,  A^2 + W_3(k_2,L) \, A^3+ O(A^4),
\ee
where 
\be \label{eq:W2}
W_2=2\left(\left(\frac{k_2^2}{4}+c_2^2-v^2\right) L +\left(c_2^2-c_1^2\right)\frac{\sin \left(2 k_2 L\right)}{2 k_2}+\left(c_1^2-v^2\right)\frac{\cos \left(k_2 L\right) }{k_2 \sin \left(k_2 L\right)}\right) 
\ee
and
\be \label{eq:W3}
W_3= \frac{4}{f_0} \lp\frac{c_2^2+v^2}{k_2}\sin \left(k_2 L\right)\left(1-\frac{\sin \left(k_2 L\right)^2}{3}\right)+ \frac{c_1^2+v^2}{3 k_2} \frac{\cos \left(k_2 L\right)^4}{\sin \left(k_2 L\right)}\rp . 
\ee
Because of the term of order 3, the simplified thermodynamic potential (\ref{eq:en_A}) seen as a function of $A$ at fixed $k_2$ is not bounded from below (see Fig.~\ref{fig:EBaym}). Adding higher-order terms would not solve this issue. 
Indeed, a straightforward calculation shows that in spite of the positive contribution from $\frac{1}{2} g f^4$, the quartic term is always negative. In addition, all higher even-order terms have negative coefficients because they are all obtained from the expansion of $-J^2/(2 f^2)$. 
It is currently unclear to us what information can be drawn from this behavior, which seems to be an artifact of the ensemble in which we are working. 
Indeed, comparing the thermodynamic potentials $G$ and $E$ of subsection~\ref{App:single-h}, one sees that the former is bounded from below while the latter is not. 
The arbitrarily large values that $-E$ can reach are thus due to our working at fixed $J$, i.e., to the boundary conditions imposed at $z \to \pm \infty$. 
While this choice is convenient to classify the nonlinear solutions, it may not accurately describe actual experimental setups. 
For this reason, the absence of lower bound on $E$ should not be interpreted as signaling an instability. 
On the contrary, results drawn from the behavior of solutions in the bulk, i.e., the characterization of nonlinear solutions and the presence of ABM and/or QNM with complex frequencies, are valid independently of the precise boundary conditions imposed on $f$, provided the condensate is long enough to be able to neglect the contributions of spatially growing modes in $\delta f$. 
(The relations between the discrete spectra in finite and infinite condensates are detailed in~\cite{Coutant:2016bgk}.) 
Moreover, the lowest-energy stationary solution does not change when using $G$ instead of $E$. 

The extremization proceeds in two steps. First we extremize \eq{eq:en_A} with respect to the amplitude.
Then the optimal value of $k_2$ is found by extremizing the result with respect to $k_2$.
We start by examining the situation for $L$ near $L_0$. 
At fixed $k_2$, Eq. (\ref{eq:en_A}) has two extrema (see Fig.~\ref{fig:EBaym}, left panel): 
a local minimum and a local maximum, which can be interpreted as a metastable and an unstable solutions respectively. 
One extremum corresponds to 
$A=0$, \textit{i.e.} to the homogeneous solution. It is metastable if $W_2>0$ and unstable if $W_2<0$. The other extremum describes 
an inhomogeneous solution. 
Its amplitude is 
\be \label{eq:AB}
A=-\frac{2 W_2}{3 W_3} , 
\ee
and its thermodynamic potential is 
\be \label{GEA}
\Delta E_s^{\rm inhom} = \frac{4 W_2^3}{27 W_3^2}.
\ee 
On the right panel of Fig.~\ref{fig:EBaym}, we compare the value of \eq{eq:AB} for $k_2 =2 \sqrt{v^2 - c_2^2}$
with the exact value of the amplitude, defined as $f(z=0)-f_0$. 
Near $L = L_0$ we have a very good agreement between the two which demonstrates 
that \eq{eq:en_A} correctly describes the relevant field configurations involved 
in the destabilization of the homogeneous solution.
This agreement is guaranteed by the 
facts that, to quadratic order, our ansatz \eq{ansatz} is exact  
and that the third-order term does not vanish. Indeed, it is easily shown that terms coming from a more accurate ansatz would be at least fourth order in the amplitude. 

\begin{figure}
\centering
\includegraphics[width = 0.49 \linewidth]{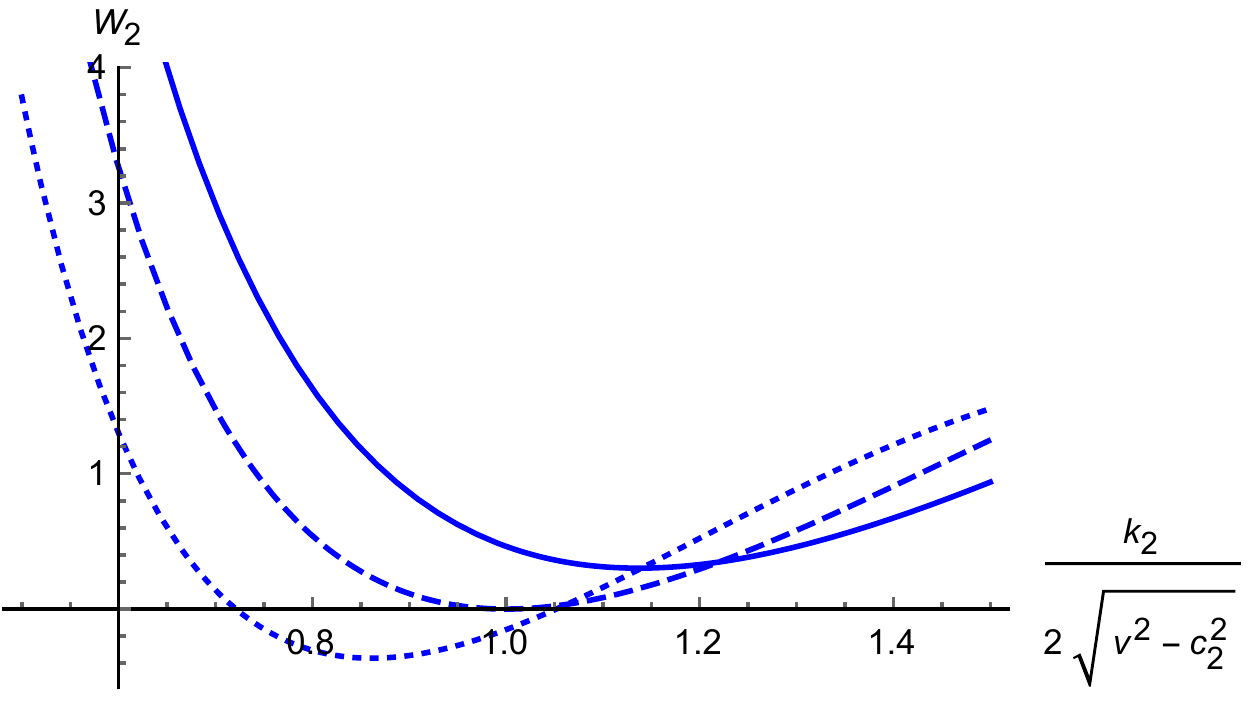}
\includegraphics[width = 0.49 \linewidth]{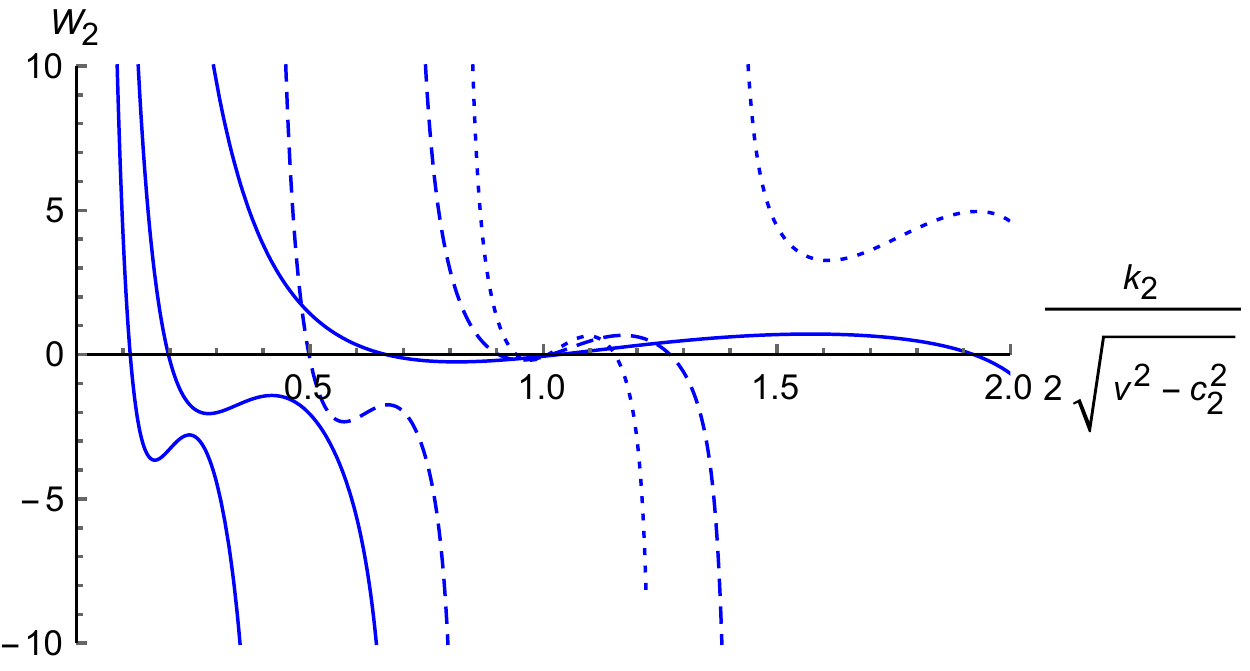}
\caption{Left panel: Coefficient $W_2$ of the quadratic term in the thermodynamic potential $E$ as a function of $k_2/(2 \sqrt{v^2 - c_2^2})$ for $L=0.85 L_0$ (solid line), $L= L_0$ (dashed line), and $L=1.2 L_0$ (dotted line). 
We see that the instability occurs for $L > L_0$, and that 
the wave vector of the unstable mode is exactly $k_2 = 2 \sqrt{v^2 - c_2^2}$ in the limit $L \to L_0$. Right panel: The value of $W_2$ as a function of $k_2/(2 \sqrt{v^2-c_2^2})$ along the branches $n=0$ (solid line), $n= 1$ (dashed line), and $n=2$ (dotted line) for $L=1.2 L_0$ , $L=1.2 L_0+\lambda_0/4$ and $1.2 L_0+\lambda_0/2$. We see that $W_2$ has an infinite series of local minima for $k_2 \to \infty$. 
They describe solutions of type 3 when the local minimum is positive, and type 1 when it is negative.
As $L$ increases, these minima migrate to lower values of $E$ and of $k$.
For $L_n < L < L_{n+1}$, $n+1$ minima have a lower energy than the homogeneous solution. 
The parameters of both panels are: $c_1=2.0$, $c_2=0.5$, $v=1.0$, $f_0=1.0$.} \label{fig:w2}
\end{figure}

A complementary point of view is provided by the dependence of $W_2$ in $k_2$. 
It is shown in Fig.~\ref{fig:w2} for three values of $L$, slightly below, equal to, and above $L_0$. 
For $L<L_0$, one sees that $W_2$ remains positive for all values of $k_2$, which confirms that the homogeneous solution is stable for all these perturbations.
We also see that the first mode which becomes unstable corresponds to $k_2 = 2 \sqrt{v^2-c_2^2}$, in agreement with \eq{eq:lambda}. 
The sign change of  $W_2$ at $L = L_0$ precisely corresponds to the transition from type 3 for $L< L_0$ to type 1 for $L> L_0$. 

It is rather easy to consider the other sectors with $n > 0$. As the right panel of Fig.~\ref{fig:w2} shows, $W_2$ has an infinite set of local minima in $k_2$. 
The minima increase with $n$ and decrease with $L$. 
For any positive integer $n$, the $n$th minimum becomes negative when increasing $L$ above $L_n$ of \eq{eq:Lm}.
Notice that the corresponding value of $k_2$ is always $2 \sqrt{v^2 - c_2^2}$ irrespective
of the value of $n$. This signals the birth of a new instability of the homogeneous solution as well as 
the beginning of a new series of metastable nonlinear solutions. 
This can be understood from the behavior of \eq{eq:W2} and \eq{eq:W3} under a change of $n$. 
Indeed, when  $k_2 = 2 \sqrt{v^2 - c_2^2}$, the first term in \eq{eq:W2} vanishes.
As a result, $W_2$ is unchanged under $L \rightarrow L +\lambda_0/2$, while $W_3$ and $\pd_{k_2} W_3$ 
flip signs.
This simply reflects that adding one wavelength to the solution in $I_2$
replaces a minimum at $z = 0$ by a maximum. A straightforward calculation shows that $\pd_{k_2} W_2$ is also invariant.
So, for all $n \in \mathbb{N}$, $k_2 = 2 \sqrt{v^2-c_2^2}$ remains the value of $k_2$ where 
a change of stability occurs for $L = L_{n}$, as was the case for $L=L_0$.

It is also possible to use the initial velocity $v$ as a control parameter instead of $L$. The analysis is then very similar. If $v = c_2$ there is no unstable mode, which translates as the absence of a negative local minimum in $W_2(k_2)$. This is because $L_0$ (as well as $\lambda_0$) is infinite, so that no ABM can be sustained by a finite supersonic region. 
When $v$ is increased from $c_2$ to $c_1$, $L_0$ decreases monotonically from $\infty$ to $0$. The first unstable mode appears when $L_0$ becomes equal to $L$. Then other unstable modes arise each time 
$L_0 + n \lambda_0/2 = L$ for some integer $n$. $\lambda_0$ is also monotonically decreasing in $v$ but remains finite in the limit $v \rightarrow c_1$, with a limiting value given by
\be 
\lambda_{0, \rm min} = \frac{\pi}{\sqrt{c_1^2 - c_2^2}}.
\ee 
The number of stable or metastable inhomogeneous solutions at fixed $L$ thus goes 
from $0$ for $v=c_2$ to 
\be 
\left\lfloor \frac{2 L}{\pi} \sqrt{c_1^2 - c_2^2} \right\rfloor +1 
\ee
for $v=c_1$.

So far we have discussed the transition occurring for integer values of $m$.
The stability changes associated with $L \approx L_m$ with a half-integer $m$ are more subtle 
for the following reasons.
When $c_1 \neq c_3$, one series of solutions (of type 2 if $c_1 > c_3$ or type 4 if $c_1 < c_3$) extends up to $L = L_m' < L_m$ with a larger thermodynamic potential than the homogeneous solution. The other one (type 4 if $c_1 > c_3$ or type 2 if $c_1 < c_3$) then exists only for $L > L_m$ with a smaller thermodynamic potential. When $c_3 \rightarrow c_1$, $L_m' \rightarrow L_m$ and the two series of solutions become degenerate. This change of behavior has important consequences for the analysis presented above. If $c_1 \neq c_3$, it is still the third-order term which governs the saturation and the expansion of the energy functional accurately describes the change of stability. However, if $c_1 = c_3$ the third-order term vanishes. One must then include contributions which are of order 4 in the amplitude and choose a more accurate ansatz than that provided by linearized solutions. This makes the analysis technically more involved and hides the intrinsic simplicity of the procedure. Similarly, the study of types 5 to 9 requires expanding the thermodynamic potential functional around a solution with one or two solitons. 

\section{Time evolution of black hole lasers} 
\label{Timeevolution}

In this section, we study the time evolution of the density in black hole laser configurations. We first analyze individual histories associated with specific initial conditions. In order to relate these histories to the ensemble averaged density observed in~\cite{BHLaser-Jeff}, we study the evolution of the mean value over several solutions with different initial conditions. 

For definiteness, we present numerical simulations with tuned parameters. We explicitly checked that a small detuning does not change the main results. All numerical simulations presented below have been done on a torus of length $480 \pi/v$, where $v$ is the velocity of the flow in the homogeneous solution. 
This is much larger than all other relevant length scales, so that the torus can be considered as infinite provided the amplitude of waves making a full turn is small enough. 
The integration was done on a uniform grid with $8196$ space points and a time step of $5 \times 10^{-3}$. We checked that dividing the space step by 2 and the time step by 4 did not change the numerical solutions in a noticeable way. 
The initial conditions consist in a superposition of two waves of constant amplitude $\delta f/f = 10^{-4}$, with a  wave vector $k = 2 \sqrt{v^2 - c_2^2}$  (the dispersive zero-frequency root in the initially homogeneous supersonic region). 
These give a significant initial amplitude of the most unstable laser mode, and a relatively small amplitude to the real-frequency modes. 
We adopted these initial conditions for practical convenience, and we checked that similar results are obtained when using different ones. 

\subsection{Nonlinear effects on individual configurations} 
\label{NLEOBHL}

\begin{figure}
\centering
\includegraphics[width=0.42\linewidth]{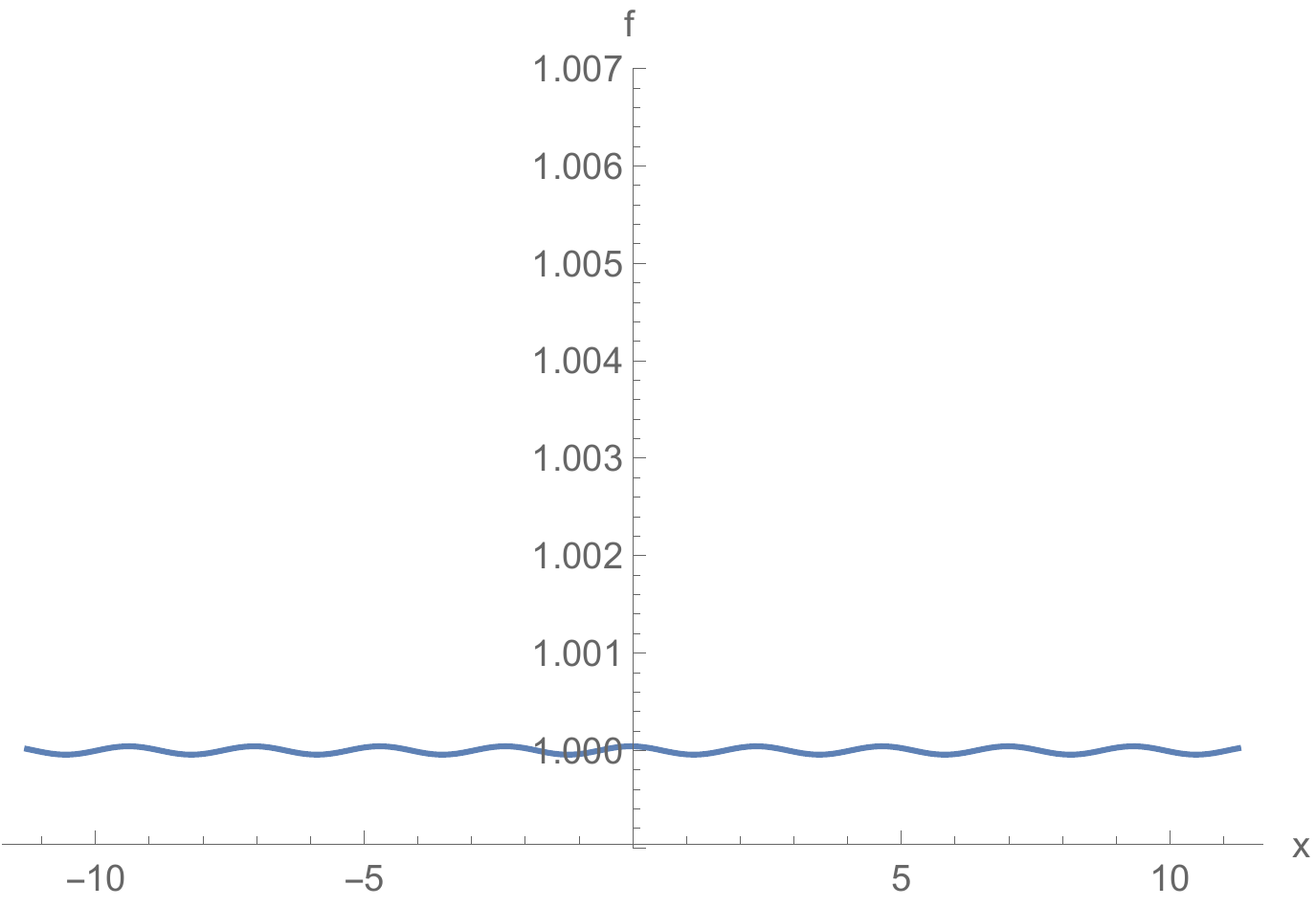}
\includegraphics[width=0.42\linewidth]{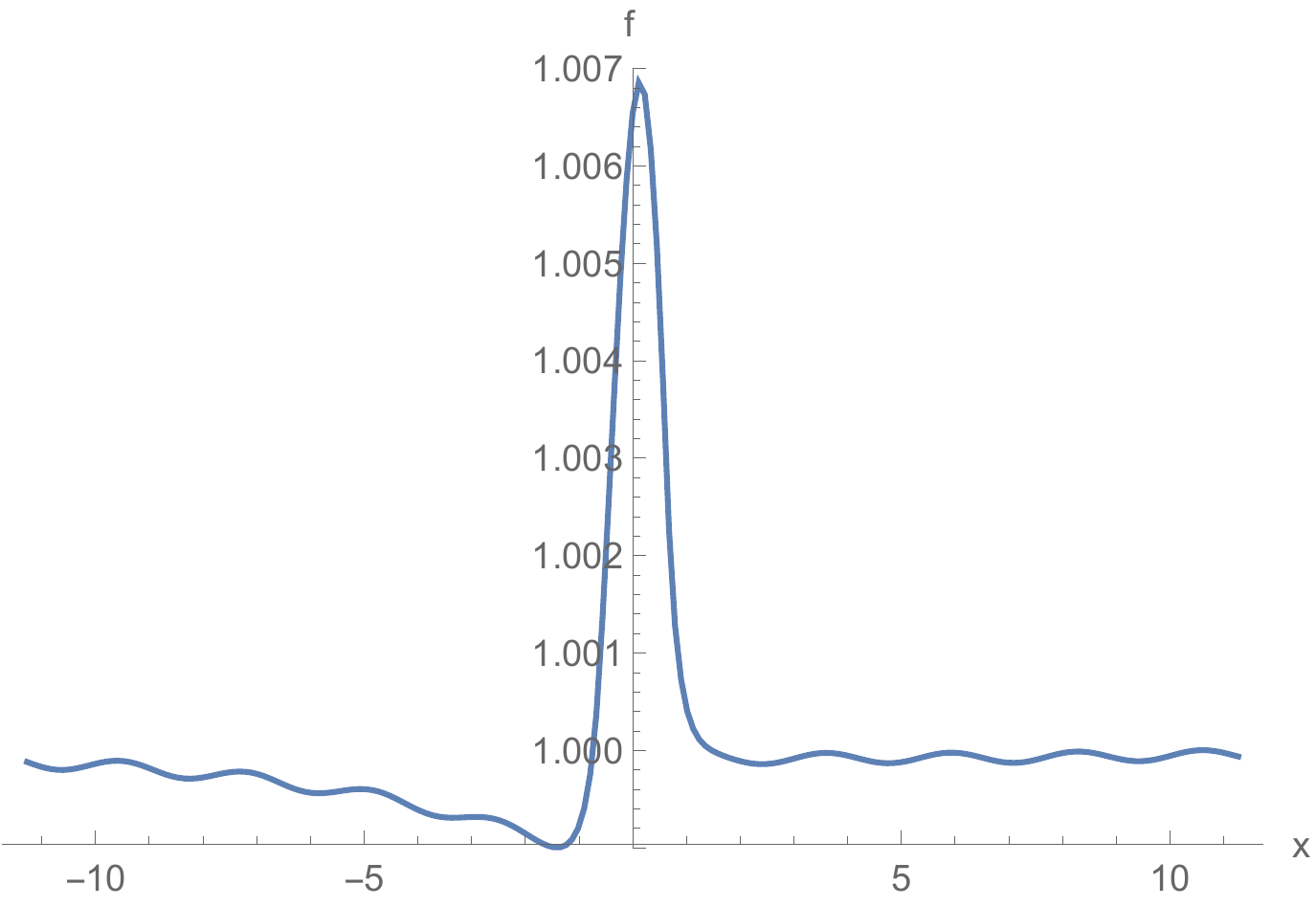}
\includegraphics[width=0.42\linewidth]{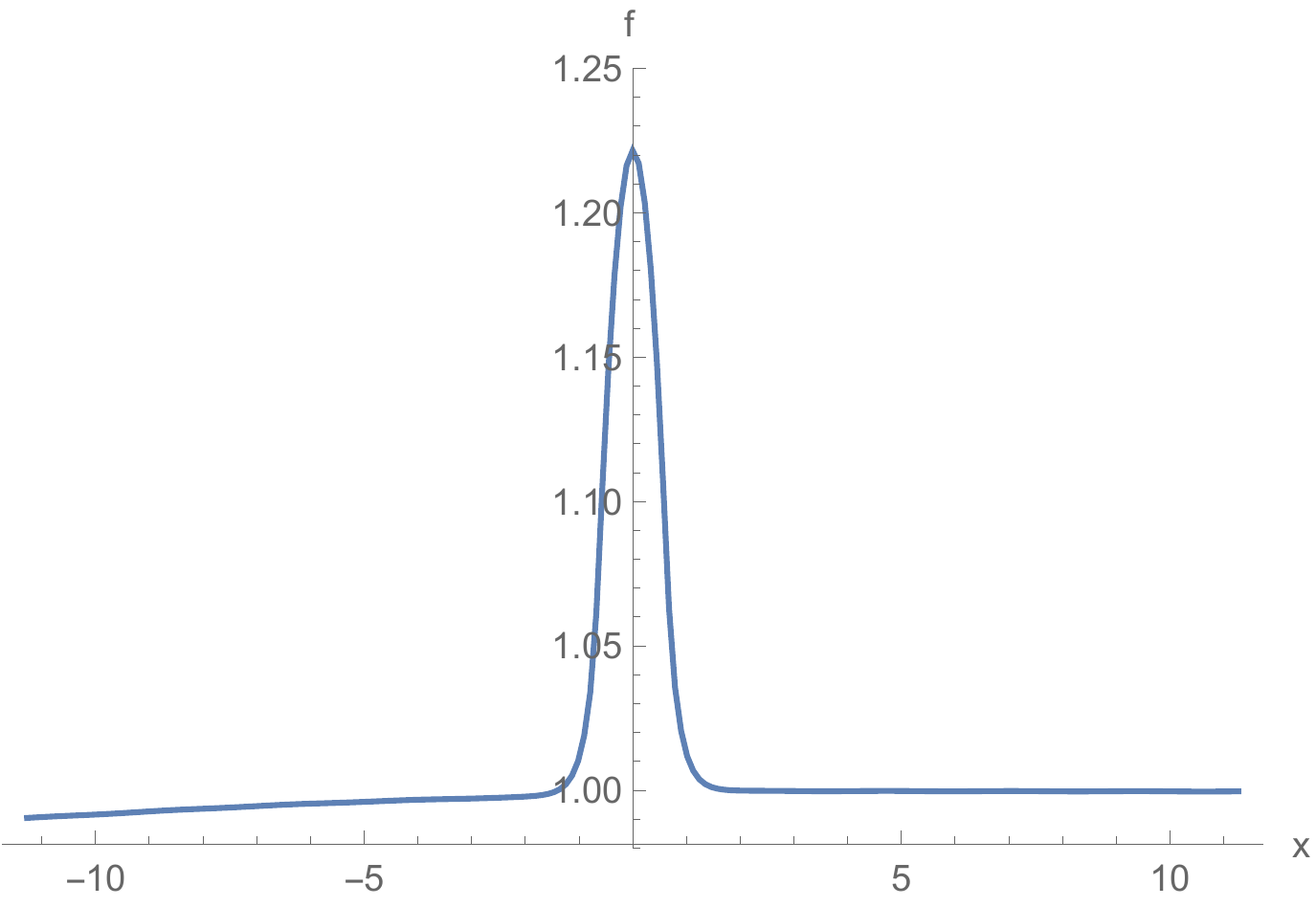}
\includegraphics[width=0.42\linewidth]{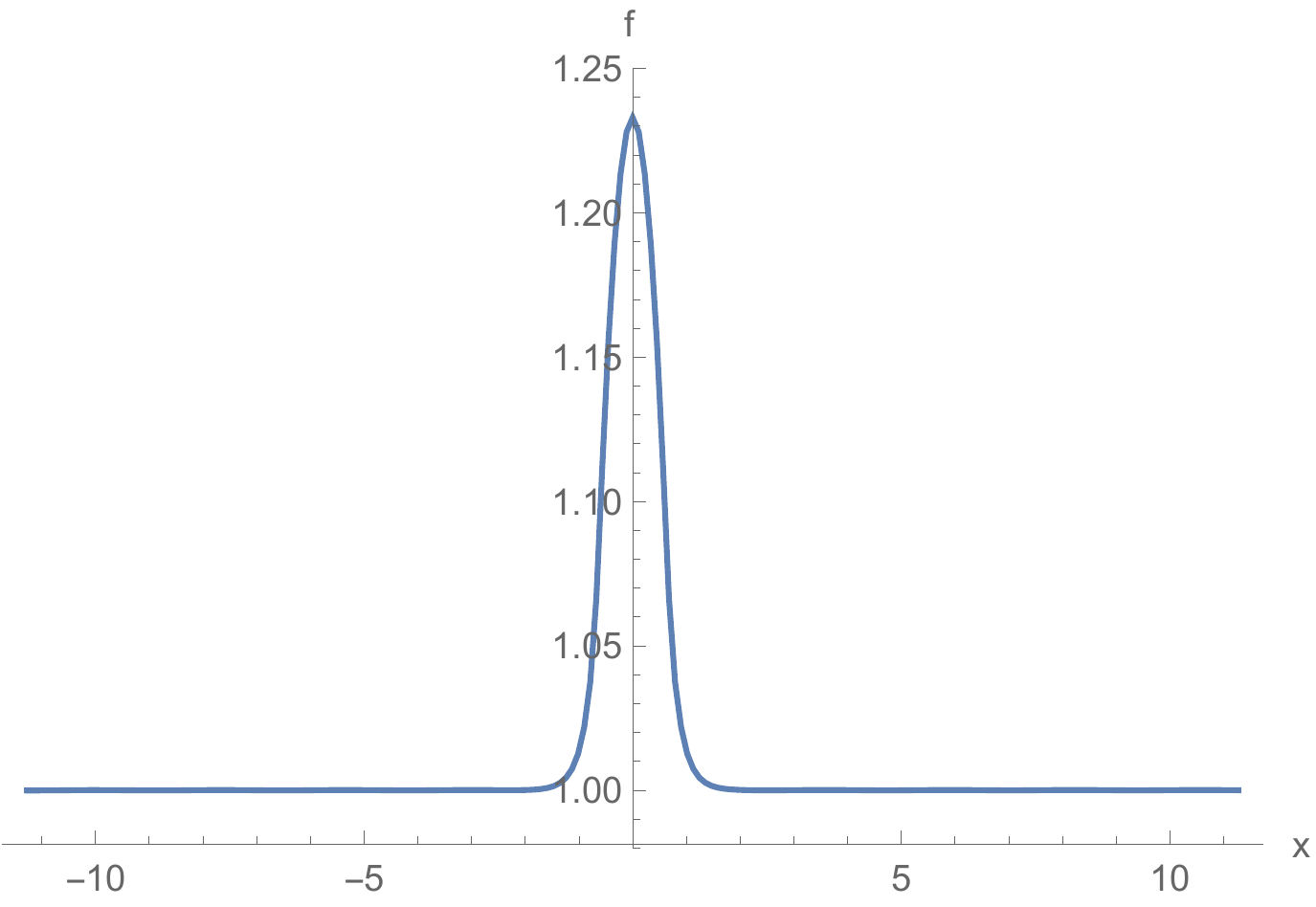}
\caption{Plot of $f$ as a function of $x$ for different times: $t=0$ (top, left), $25$ (top, right), $50$ (bottom, left), and $75$ (bottom, right). 
The parameters are $J = \sqrt{8/3}$, $f_{p,{\rm int}} = f_{b,{\rm ext}} =1$, $f_{p, {\rm ext}} = 0.7$, $f_{b,{\rm int}} = 1.5$, and $L=L_0 + \frac{\lambda_0}{8} \approx 0.68$. 
The initial conditions are such that $f(t=0) - 1 > 0$ for $-L<x<L$. 
Notice that the range of $\delta f = f - 1$ in the first two plots is $[-0.001,0.007]$, whereas it is $[-0.02, 0.25]$ for the last two. 
Nonlinear effects typically become important when the maximum of $\abs{f-1}$ becomes of the order of $0.1$. 
} \label{fig:simu5p}
\end{figure}

To start, we ran simulations with $L < L_0$. As expected since there is no ABM, the amplitude of the perturbation remained of order $10^{-4}$. 
As expected as well, we observed a richer behavior for $L >  L_0$. Figure~\ref{fig:simu5p} shows the case where $L_0 < L = L_0 + {\lambda_0}/{8} < L_1$, and when $f(x\approx0, t=0) - 1>0$. 
This difference acts as a positive detuning in the sense that it sends the solution towards the stable type 1 solution with $n=1$. At early times\footnote{We use ``early times'' for times large enough for the unstable mode with the largest growth rate to dominate, but small enough for nonlinear effects to be negligible. ``Late times'' will refer to times large enough for nonlinear effects to be important. In contrast, in Ref.~\cite{BHLaser-Jeff} ``late times'' refers to times where the most unstable mode dominates.} (though larger than $1/\Gamma_{\rm hom.}$, where $\Gamma_{\rm hom.}$ is the imaginary part of the complex frequency of the ABM on top of the homogeneous solution), the evolution is dominated by the laser mode which dictates both the shape of $\delta f(x,t) = f(x,t)-1$, and its exponential growth. 
We verified that the growth rate is equal to $\Gamma_{\rm hom.}\simeq 0.27$,  computed by solving \eq{eq:det}.  At later times, of order $t \sim 30$, the solution approaches the type 1 solution. This process is smooth, in the sense that no large-amplitude perturbation is emitted away from the horizons, and $\delta f(0,t)$ is a monotonically growing function of time. At late time the flow is stationary and the spatial profile is exactly given by that of the type 1 solution, independently of the initial conditions. 
\begin{figure}
\centering
\includegraphics[width=0.42\linewidth]{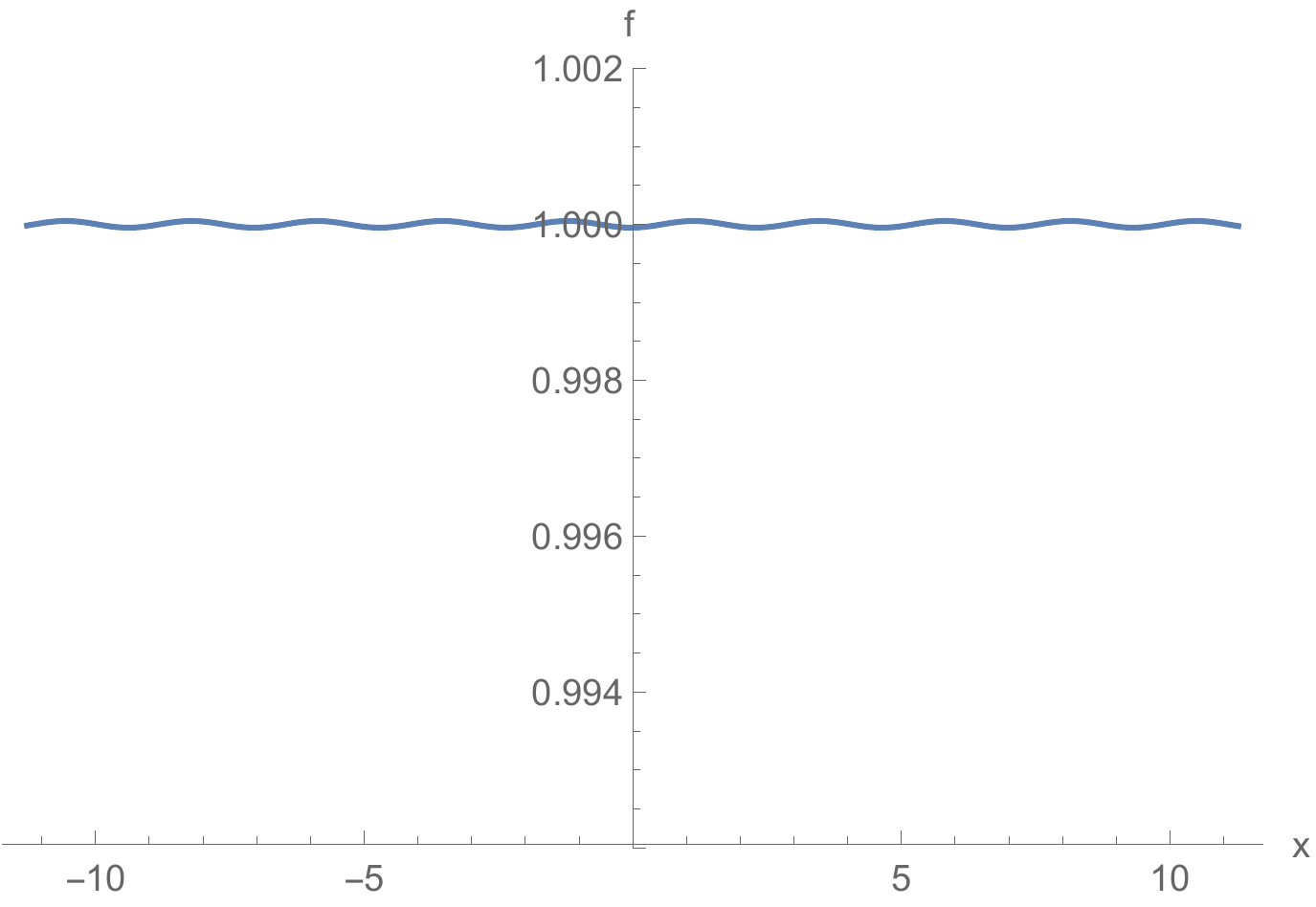}
\includegraphics[width=0.42\linewidth]{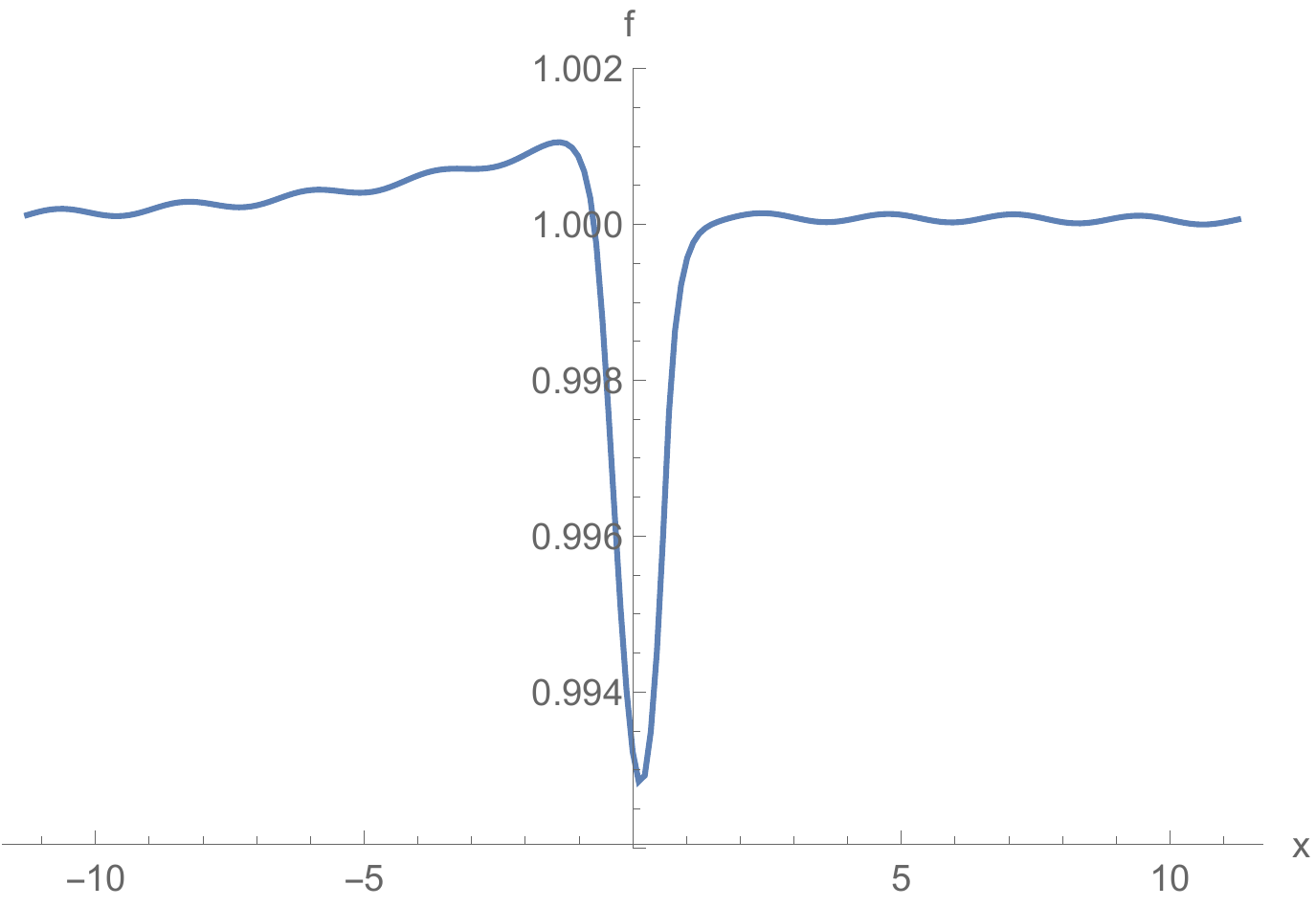}
\includegraphics[width=0.42\linewidth]{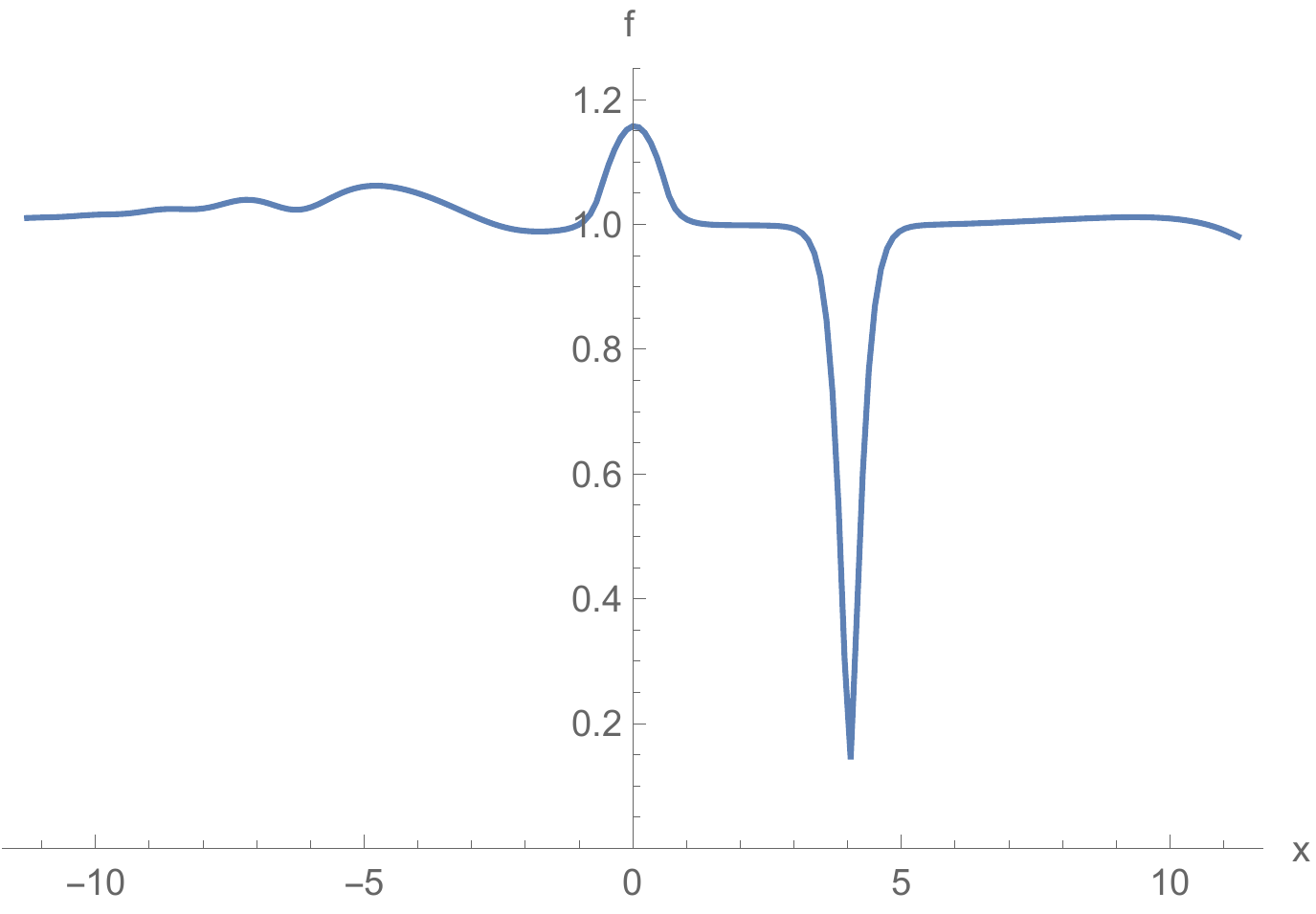}
\includegraphics[width=0.42\linewidth]{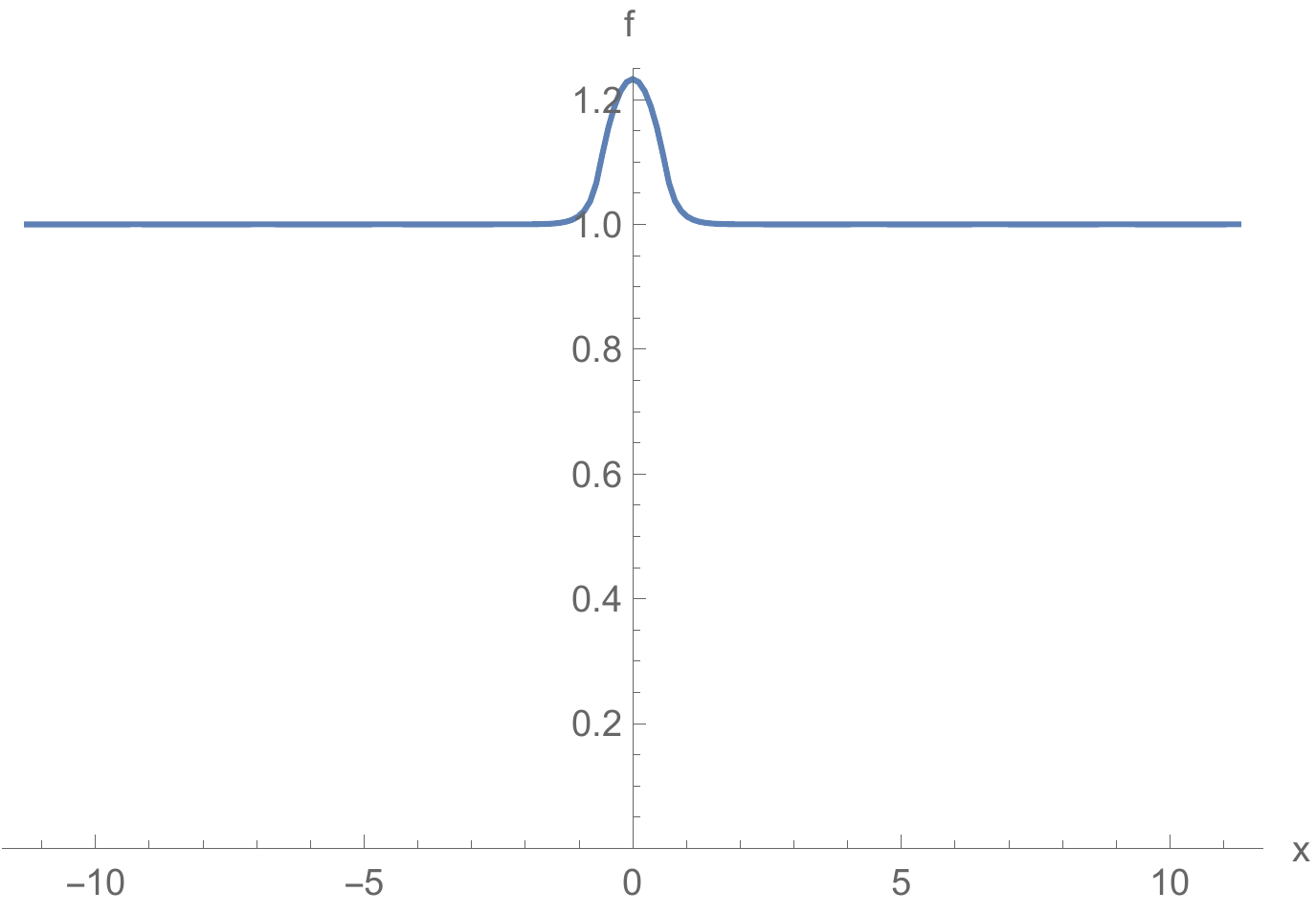}
\caption{Plot of $f$ as a function of $x$ for different times: $t=0$ (top, left), $25$ (top, right) $50$ (bottom, left), and $75$ (bottom, right). The parameters are $J = \sqrt{8/3}$, $f_{p,{\rm int}} = f_{b,{\rm ext}} =1$, $f_{p, {\rm ext}} = 0.7$, $f_{b,{\rm int}} = 1.5$, and $L=L_0 + \frac{\lambda_0}{8} \approx 0.68$. The initial conditions are such that $f(x,t=0) - 1$ has the opposite value as that of Fig.~\ref{fig:simu5p}. 
One verifies that the final profile, obtained after having emitted the soliton, is identical to that of the former plot. 
} \label{fig:simu5m}
\end{figure}

Figure~\ref{fig:simu5m} shows the evolution for the same value of $L = L_0 + {\lambda_0}/{8}$ when the initial value of $\delta f(x, t=0) $ has the opposite sign. In this case one has the equivalent of a negative detuning. At early times, $|\delta f|$ grows exponentially with the same rate $\Gamma_{\rm hom.}$, exactly as predicted by the Bogoliubov-de Gennes equation. 
However, instead of making it saturate, nonlinear effects now turn the hollow of the type 3 solution with $n=0$ into a soliton which is emitted towards $x \to + \infty$. 
The residual value of $\delta f(x \approx 0)$ is now positive and saturates on the same solution as above: the type 1 solution with $n=1$. Therefore, at very late times, the two solutions obtained by flipping the sign of the initial value of $\delta f$ both asymptote to the type 1 solution, and this despite their different behaviors at intermediate times. Moreover, we verified that the convergence is exponential with a decay rate given by the imaginary part of the frequency of the quasinormal mode (QNM) on this solution. We performed simulations with other different initial conditions, and always found the same end state, which thus acts as a local attractor, in a sense which will be made more precise in Chapter~\ref{ch:nohair}.  
It should also be noticed that the $\mathbb{Z}_2$ symmetry, which is present at early times when nonlinear effects are negligible, is thus completely broken at late times. This has important consequences on observables which are odd in $\delta f$, as shall be shown in the next subsection. 

We found similar results when choosing $L_1 < L < L_2$. In this case, we also found that the end state corresponds to the ground state, the type 1 solution with $n = 1$. At early times $\delta f$ grows exponentially, with a sign which depends on the initial conditions. The difference with respect to the previous case is that the frequency of the laser mode now has a non-vanishing real part, so that $\delta f$ periodically changes sign in the linear regime. As a result, the sign of $\delta f$ for $t=0$ in the internal region is no longer directly related to the emission of a soliton. Importantly, we here observe the first manifestation of a general tendency. When increasing the distance $2L$ between the discontinuities of $V$ and $g$, the set of lasing modes gets larger. 
Consequently, the behavior of nonlinear time-dependent solutions becomes more intricate, and less straightforwardly related to the initial conditions. It is possible that this complexity will lead to a chaotic, i.e., unpredictable, behavior when there are several lasing modes. 
It would be interesting to validate or invalidate this conjecture. 

\begin{figure}
\centering
\includegraphics[width=0.6 \linewidth]{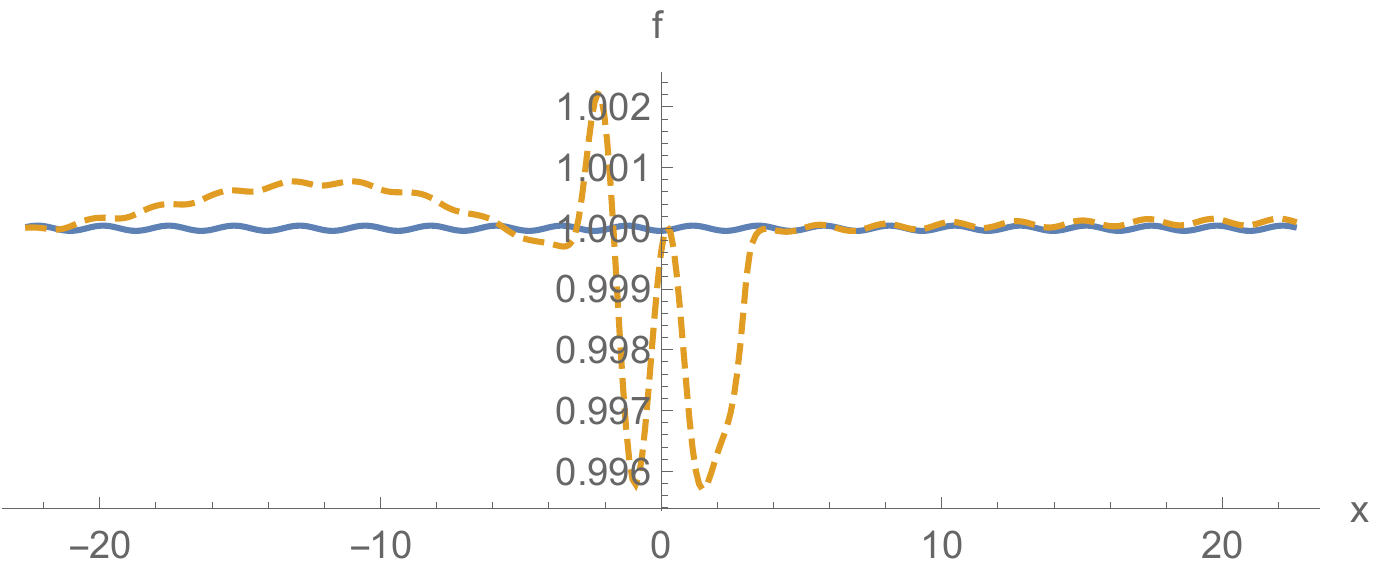}\\
\includegraphics[width=0.6 \linewidth]{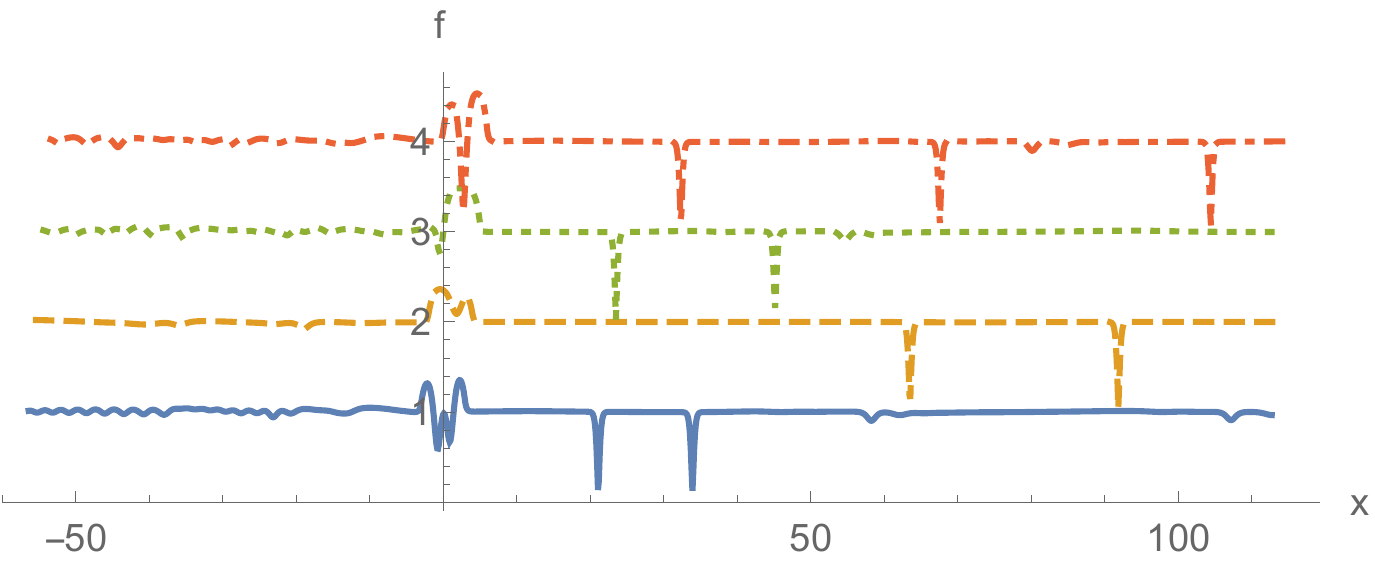}
\caption{Plot of $f$ as a function of $x$ for different times: $t=0$ (top, solid), $50$ (top, dashed), $100$ (bottom, solid), $200$ (bottom, dashed), $300$ (bottom, dotted), and $400$ (bottom, dot-dashed). To ease the reading, the last three curves on the bottom plot have been displaced upwards by $1$ (dashed), $2$ (dotted), and $3$ (dot-dashed). The parameters are $J = \sqrt{8/3}$, $f_{p,{\rm int}} = f_{b,{\rm ext}} =1$, $f_{p, {\rm ext}} = 0.7$, $f_{b,{\rm int}} = 1.5$, and $L=L_0 + \frac{9}{8} \lambda_0 \approx 3.0$. At early times, for $t = 50$, in the internal region $|x| < 3$ one observes the laser mode with the highest growth rate. At intermediate times, one observes some solitons which are not equally spaced due to a transient behavior. At late times, they seem to be equally spaced.
} \label{fig:simu8}
\end{figure}
To verify that complexity increases with the number of laser modes, we ran simulations with $L_{2n} < L < L_{2(n+1)}$ for $n=1$, $n=2$, and $n=6$. In the three cases, at early times, we found that the laser mode whose frequency has the largest imaginary part $\Gamma_m$ dominates. As a result, $f(x,t)$ goes very close to the $m$th type 1 solution for times of a few $1/\Gamma_m$. 
However, because this stationary solution is not stable, the time-dependent solution then emits one or several solitons which escape to $x \to \infty$ and then approaches the $l$th type 1 solution, with $l<m$. In our simulations, we observed that the solution quickly evolves, and saturates close to the type 1 solution with $n = 2$. (At present we do not clearly understand this observation. We suspect it is due to the fact that mode mixing across the horizons is large, as the transition is sharp since we are using discontinuous parameters. We thus conjecture that the solution will evolve more slowly when using smooth profiles for $V$ and $g$.) After having approached  the type 1 solution with $n = 2$, we observe in \fig{fig:simu8} the  emission of solitons in an apparently periodic way. 
At present we have not been able to identify any criterion able to distinguish the solutions that shall emit soliton trains, from those which shall not.\footnote{I.~Carusotto, S.~Finazzi, and J.-R.~de Nova~\cite{PVcomm} observed emission of solution trains in their numerical simulations of black hole lasers. Later, J.-R.~de Nova informed us that he had numerically found that (for a fixed initial uniform density $f$) there exists a $L$-dependent threshold value of $f_{b,{\rm int}} - f_{p,{\rm ext}}$ above which infinite soliton trains are emitted. This is in agreement, and completes, our own findings. The threshold was studied in~\cite{2015arXiv150900795D}.}  Let us here note that similar soliton trains have been observed in Ref.~\cite{Hakim1997}.

In any case, these intricate behaviors result from an interesting interplay between the linear instabilities governing early time dynamics and nonlinear effects at later times. While linear instabilities trigger the cascading between the various type 1 solutions with decreasing values of $n$, going from one solution to the next one is always accompanied by the emission of solitons. Depending of the solution, either a finite number of solitons is emitted, or an infinite number of solitons, as for white hole flows~\cite{Michel:2015pra}, so that a stationary solution is apparently never reached. A categorization of the set of possible behaviors and their respective domains in parameter space is probably possible but beyond the scope of this work. 

\subsection{Time evolution of the mean density, breaking the \texorpdfstring{$\mathbb{Z}_2$}{Z2} symmetry}

In his experiment~\cite{BHLaser-Jeff}, J.~Steinhauer observed that the averaged value of the density (taken over 80 realizations) develops a clear spatial pattern with a rapidly growing amplitude; see Fig.~2 in~\cite{BHLaser-Jeff}. The nodes of the profiles seem compatible with those of the most unstable lasing mode, which dominates the growth of the two-point correlation function; see Fig.~4 in~\cite{BHLaser-Jeff}. However, a more precise comparison is needed to determine whether or not they are precisely equal.  
In the following we show that a behavior similar to that of his Fig.~2 can be obtained from the breakdown of the $\mathbb{Z}_2$ symmetry by non linear effects. 
Indeed, as the Bogoliubov-de Gennes equation contains only linear and antilinear terms, its set of solutions is invariant under multiplication by $-1$. 
Moreover, this operation does not change the physical properties of the perturbation, such as its energy or momentum, as it amounts to a change of phase by $\pi$. 
As a relative perturbation of the condensate wave function $\phi$ gives a relative density perturbation $\delta f/f = \Re \phi$, we thus have a $\mathbb{Z}_2$ symmetry $\delta f \to -\delta f$ leaving the energy of the solution unchanged to linear order. 
Therefore, when working with a  thermal state (or any other state which does not break this symmetry), the average of $\delta f$, or of any observable which is odd in $\delta f$, is and remains identically equal to zero. 
This applies both to the ensemble average value of the undulation amplitude emitted by white hole flows and to density fluctuations associated with the black hole laser instability. 
This $\mathbb{Z}_2$ symmetry will be broken by nonlinear terms in the GPE, the first of which are quadratic in $\delta f$. 
So, the ensemble average value of $\delta f$ will generally develop an expectation value of order $\delta f^2$. 
\begin{figure}
\centering
\includegraphics[width=0.4\linewidth]{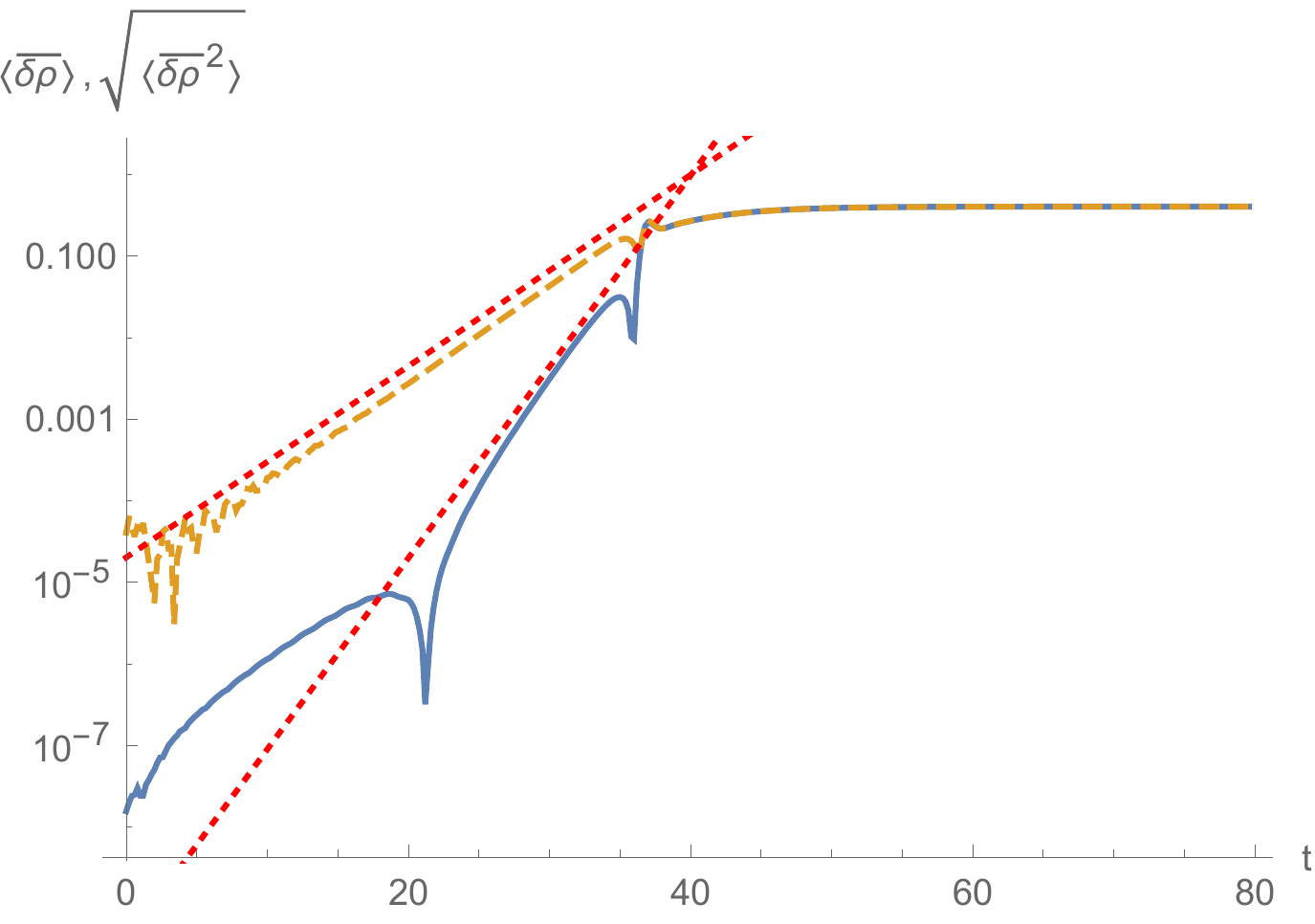}
\includegraphics[width=0.4\linewidth]{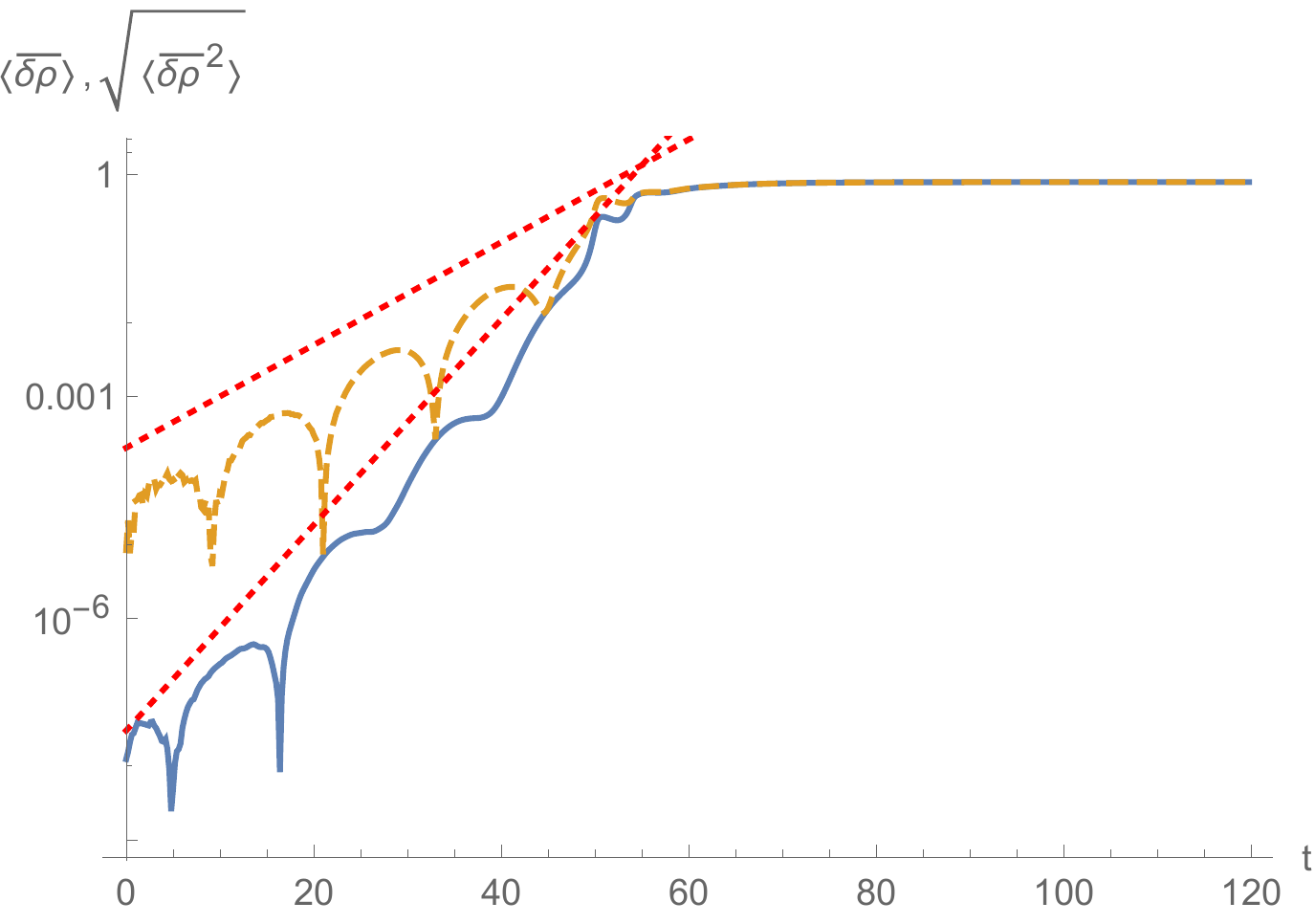}
\caption{Time evolution of the ensemble average of $\delta \rho$ (with respect to the homogeneous solution) for the two simulations of Figs~\ref{fig:simu5p} and~\ref{fig:simu5m} (left), and for two simulations with the same parameters except that $L = L_0 + \frac{3}{8} \lambda_0$ (right) so as to have a nondegenerate dynamical instability. 
The solid blue lines represent the mean $\left\langle \overline{\delta \rho} \right\rangle$, whereas the dashed orange lines represent the rms ${\left\langle \lp\overline{\delta \rho}\rp^2 \right\rangle}^{1/2}$. A bar means space average over the internal domain $-L<x<L$,  and $\langle \rangle$ means average over the two simulations with opposite perturbations at $t=0$. 
The dotted red lines show exponentials with growth rates $\Gamma$ and $2 \Gamma$, where $\Gamma$ is the imaginary part of the frequency of the laser mode. One clearly sees that the mean grows with a rate which is twice that of the rms value. One also sees that the two quantities coincides at late times.
} \label{fig:deltarho1}
\end{figure}
\begin{figure} 
\centering
\includegraphics[width=0.4\linewidth]{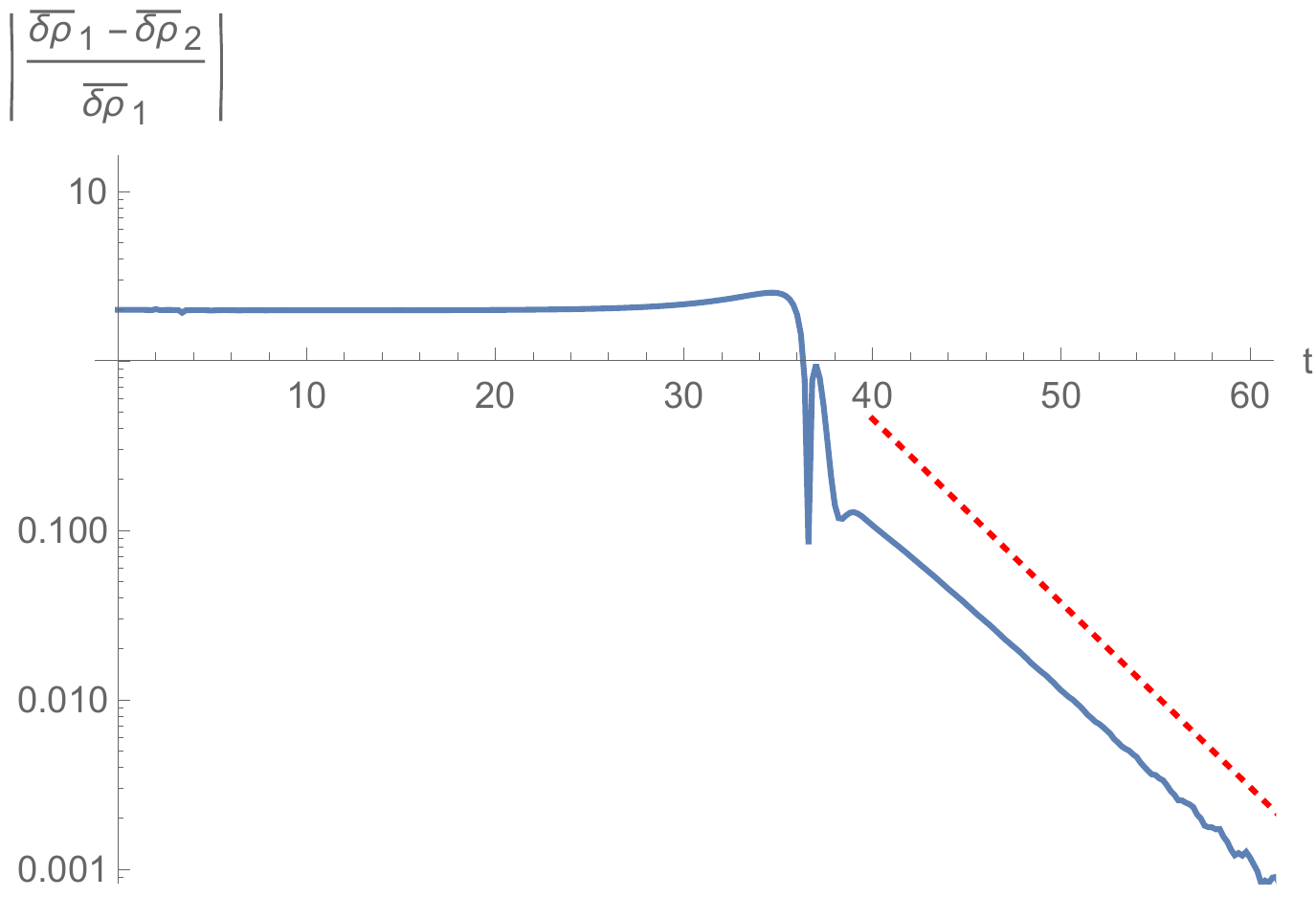}
\includegraphics[width=0.4\linewidth]{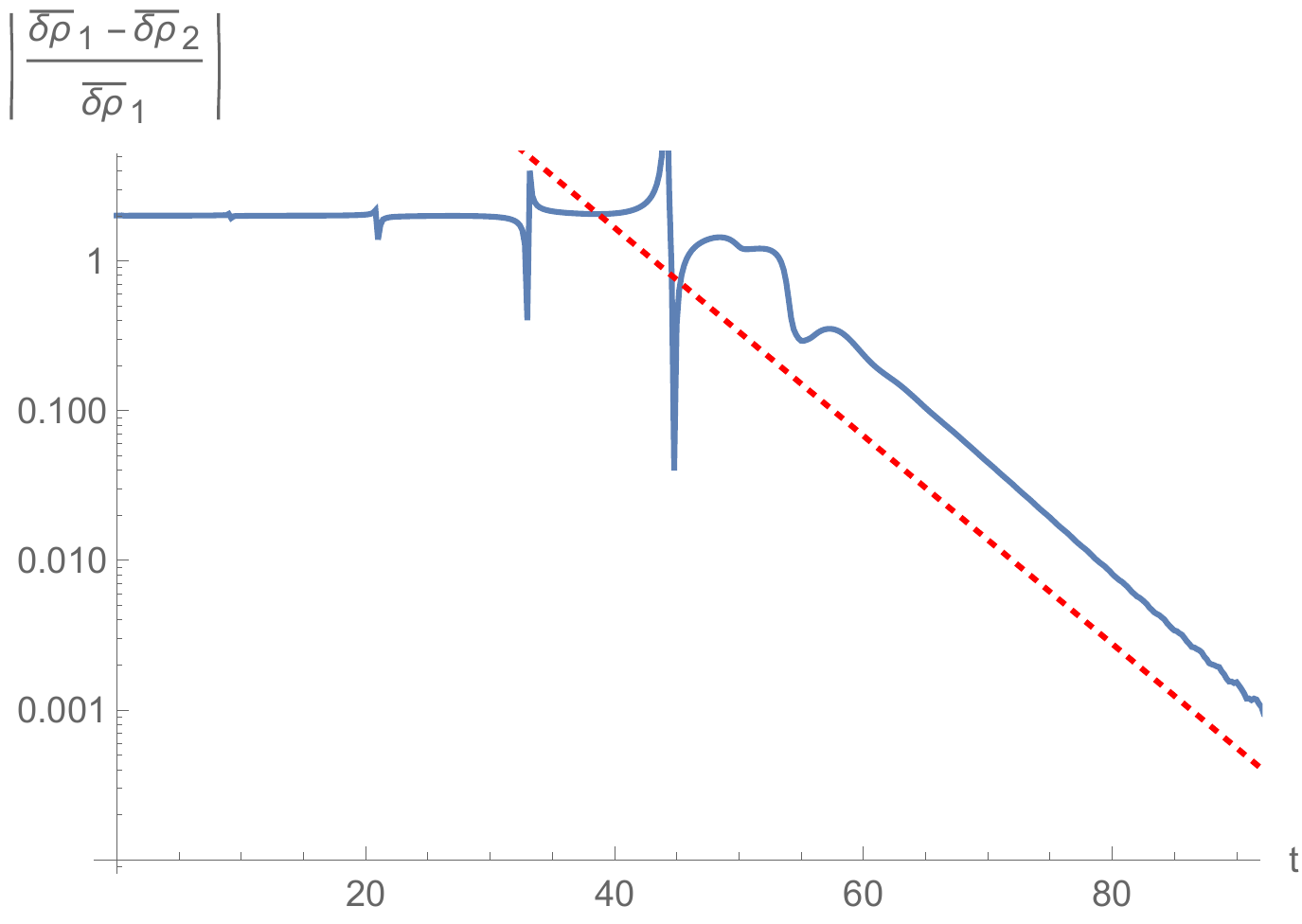}
\caption{Relative differences of the space averaged density perturbation, as functions of time. The left (respectively, right) panel corresponds to the left (respectively, right) panel of \fig{fig:deltarho1}. 
The dotted red lines show exponentials with growth rates equal to the imaginary part of the frequency of the QNM closest to the real axis over ground state. 
At early times, the flat plateau of height equal to 2 reveals that both perturbations $\delta \rho_i$, $i = 1,2$ grow at the same rate, and stay equal and opposite to each other. It ends when nonlinearities become significant. At late time, the difference of $\delta \rho_i$ exponentially decreases, as both configurations approach the same ground state. Hence this difference is governed by the time dependence of the QNMs on that ground state.
 } \label{fig:deltarho3}
\end{figure}

To illustrate the roles of the $\mathbb{Z}_2$ symmetry and its breaking, we show in \fig{fig:deltarho1} the space and ensemble average of $\delta \rho \equiv \rho - 1 = f^2 - 1$ over the internal region. 
We consider a very simple ensemble consisting of two realizations with opposite initial perturbations, i.e., related by the $\mathbb{Z}_2$ symmetry. We denote by $\left\langle \cdot \right\rangle$ the ensemble average over the two realizations and by $\overline{\cdot}$ the space average over the region $-L<x<L$. 
In \fig{fig:deltarho1}, the solid lines represent $\left\langle \overline{\delta \rho} \right\rangle$. For comparison, we also represent the root mean square (rms) value $\sqrt{\left\langle \overline{\delta \rho}^2 \right\rangle}$ (dashed) whose behavior is closely related to that of the density-density correlation unction, see~\cite{Michel:2015pra,Michel:2016tog}.  
These plots are obtained in flows possessing one degenerate instability (left) and one nondegenerate instability (right). 

At early times, we notice that $\left\langle \overline{\delta \rho} \right\rangle$ grows like $\left\langle \overline{\delta \rho}^2 \right\rangle$. This is due to the suppression of $\left\langle \delta \rho \right\rangle$ by the $\mathbb{Z}_2$ symmetry to linear order: $\left\langle \overline{\delta \rho} \right\rangle$ is of order 2 in the amplitude of the perturbation, while the rms, which is not suppressed by the symmetry, remains linear in this amplitude. At late times instead, for both the left and the right panels, the two curves giving the mean and the rms values are indistinguishable. This is a direct consequence of the fact that the profiles of $f(x)$, and thus those of $\rho(x)$, of the two simulations become identical, so that $\left\langle \overline{\delta \rho}^2 \right\rangle  = \left\langle \overline{\delta \rho} \right\rangle^2= \overline{\delta \rho}^2$, where in the last expression $\delta \rho$ is the common value of the density perturbation for the two solutions. In the nondegenerate case, because of the real part of the frequency of the lasing mode, the two density perturbations periodically vanish at linear order, hence the hollows in the plot of the rms.   

A complementary view of this symmetry breaking is shown in \fig{fig:deltarho3}. In this figure we show the relative difference between the space averaged values of the density perturbations $\overline{\delta \rho_i}$ as a function of time. At early times, its value remains close to $2$, as the density perturbations remain opposite to each other while increasing exponentially. Its value deviates from $2$ when nonlinear effects become important, and exponentially decreases at late times. We verified that the late time decay rate is equal to the imaginary part of the frequency of the QNM defined on top of the type 1 solution.

In brief, in this subsection, we saw that the ensemble averaged value of the density fluctuation $\delta \rho$ grows exponentially, even when its initial value is zero. The departure from zero is due to the breaking of the $\mathbb{Z}_2$ symmetry by nonlinear effects. In this case, its growth rate  is twice larger than that of the rms value of $\delta \rho$, which is also that of $\delta \rho_i$ of each particular realization. The doubling of the growth rate of the mean value can thus serve as an unambiguous test to determine whether the non-vanishing value of $\delta\rho$ is due to nonlinear effects or to classical initial conditions. 
(J.~Steinhauer's observations can also be explained by the excitation of waves with a non-vanishing ensemble average close to the white hole horizon. 
This mechanism actually seems to play a dominant role in the simulations of~\cite{Tettamanti:2016ntx, Wang:2016jaj}.)

\section{Discussion}

To analyze their stability at the linear and nonlinear levels, we used a simple model of black hole lasers in one-dimensional infinite Bose-Einstein condensates. 
The simplicity is due to the use of a piecewise constant potential which is tuned in such a way that there exists an exact solution with a uniform flow velocity, while the sound velocity has two discontinuities. 
Using the linearized mode equation, the set of complex-frequency modes that are responsible for the dynamical instability has been explicitly obtained. 
In particular we showed that each new unstable mode arises in two steps. For a finite interval of the distance $2L$ between the two discontinuities, we found that the unstable mode has a purely imaginary frequency. 
For larger values we recovered the situations found in~\cite{Coutant:2009cu,Finazzi:2010nc}, see Fig.~\ref{fig:QNM}. We claim (and verified numerically on a few examples) that this two-step process will also apply to smooth profiles, at least when the gradients of the potential $V$ and the coupling $g$, are sufficiently large in the units of the inverse healing length. 
This was expected: in this limit, on the first hand, it has been shown~\cite{Finazzi:2012iu,Macher:2009nz} that the Bogoliubov coefficients encoding the mode mixing at each sonic horizon are in close agreement with those derived from the matching conditions we used. Hence the solutions of the equation $\det M = 0$ should continuously depend on the gradients, with a smooth step-like limit. 
On the other hand, we found that the dimensionality of the unstable sector is 1 when the frequency is purely imaginary, and not 2 as is the case when the frequency is complex. 
This discrepancy can not be eliminated by an infinitesimal, smooth change in the coefficients of $M$. 
The frequencies of degenerate ABM thus remain purely imaginary. 

To find the end-point of the evolution of this dynamical instability, we characterized the stationary nonlinear solutions of the GPE with a finite thermodynamic potential. 
We showed that a set of nine nonlinear solutions corresponds to each unstable mode, and we explained the origin of this multiplicity, see Fig.~\ref{fig:9-traj}. 
We also showed that, in each set, one solution can be conceived as the end-point of the evolution (in the mean field approximation since we work with solutions of the GPE) of the corresponding instability, see Fig.~\ref{fig:ene-9-eq}. When considering the whole set of solutions at fixed $L$, we identified the lowest-energy state and studied its properties. 
In particular, we numerically verified that the maximum value of the density is, at leading order, unchanged when replacing our discontinuous profiles by continuous ones which are sufficiently steep. 
In the steplike regime, we analytically constructed the exact solutions by pasting building blocks consisting of exact solutions of the GPE associated with each homogeneous region, 
see subsection~\ref{App:single-h}. 
To explicitly relate the onset of instability described by the complex-frequency modes of Section~\ref{Slt} to the nonlinear solutions of Section~\ref{stat_sol}, we presented in Section~\ref{thermo} a treatment based on a Taylor expansion of the energy functional and a simplified ansatz which displays the second-order transition between the homogeneous solution and a spatially structured one. 
Finally, in Section~\ref{Timeevolution} we numerically solved the GPE to see whether a stationary solution is reached at late times. 
We found that this is actually not always the case: when $L$ is large, the solution generally has no limit $t \to \infty$. 
This behavior was first observed by J.~R. de~Nova, S.~Finazzi, and I.~Carusotto, whose results are reported in~\cite{2015PhRvL.115b5301B}. 
To identify the validity domain of our findings, it would be interesting to work beyond the mean-field approximation, and to consider in more detail smooth profiles in which the initial sound velocity is continuous. 

%

\section{Additional remarks}
\label{AR1}

\subsection{Structure of the equation on complex frequencies}
\label{App:M-matrix}

In this subsection we detail the procedure we used to find the ABM and QNM over the homogeneous solution. 
The explicit form of the matrix $M$ whose determinant encodes the matching conditions is shown and the results are compared with the Bohr-Sommerfeld approximation used in \cite{Finazzi:2010nc}.

\subsubsection{Complex-frequency modes}
We use the following procedure to find ABM and QNM. 
First we solve the linearized Gross-Pitaevskii equation (GPE) in each of the three regions $I_1$, $I_2$, and $I_3$ and impose boundary conditions at $z \rightarrow \pm \infty$ to retain the solutions which are ``outgoing'' in a generalized sense.
We then impose matching conditions at the two horizons $z = \pm L$ to find the globally-defined modes. 

\begin{figure}
\centering
\includegraphics[width = 0.49 \linewidth]{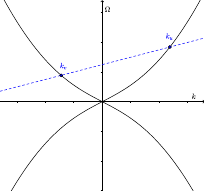}
\includegraphics[width = 0.49 \linewidth]{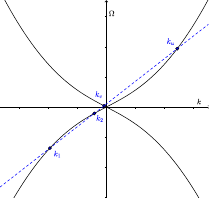}
\caption{Graphical resolution of the dispersion relation in a subsonic (left diagram) or supersonic (right diagram) flow. The solid curve represents $\Omega(k)$ of \eq{eq:disprel1} and the dashed blue line shows $\omega - v k$. Here $\om$ is real and positive. In the subsonic case there are two real roots: a left mover $k_v$ and a right mover $k_u$. In the supersonic case and if $\om$ is small enough there are two additional real roots with $\Omega < 0$: $k_1$ and $k_2$. $k_1$ is a right mover while $k_2$ is a left mover.}\label{fig:disprel1}
\end{figure}

\begin{figure}
\centering
\includegraphics[width = 0.6 \linewidth]{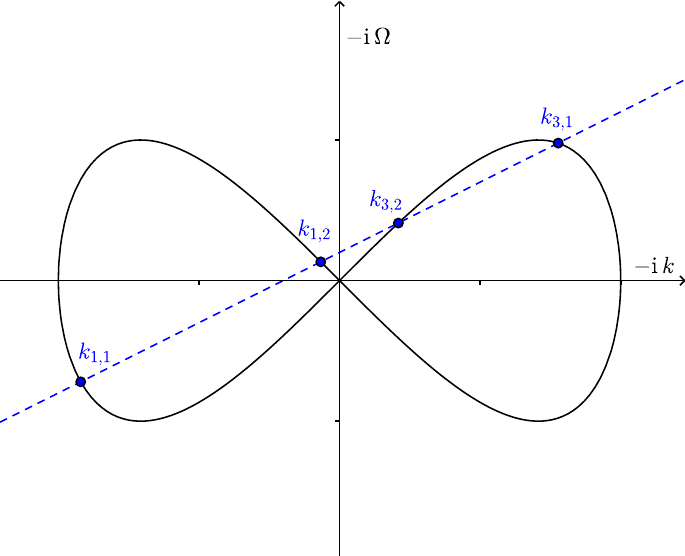}
\caption{Graphical resolution of the dispersion relation for $\om \in \ii \mathbb{R}$. Solid curve: $- \ii \, \Omega$ as a function of $- \ii \, k$ for $k \in \ii \mathbb{R}$. Dashed blue line: $- \ii \, (\omega - v k)$ as a function of $\ii \, k$ for a subsonic flow and $0<-\ii \, \om<\Gamma_0$, where $\Gamma_0$ is the positive value of $-\ii \, \om$ at which two roots merge.
In that case there are four purely imaginary roots to the dispersion relation. We set $c_3 = c_1$. $k_{j,1}$ and $k_{j,2}$ are the two modes we use in region $j$ to build ABM.}
\label{fig:disprel_im_sub}
\end{figure}

We write $\psi(t,z)= (f_0 + \delta f(t,z))\, \e^{\ii (\theta_0(z)+\delta \theta(t,z))}$, where $f_0 \, \e^{\ii \theta_0}$ is the solution of \eq{GPE1} 
with a uniform amplitude. To first order in $(\delta f, \delta \theta)$, \eq{GPE1} gives
\be \label{linearfth}
\left\lbrace 
\begin{array}{ll}
 \partial_t \, \delta f + v \, \partial_z \delta f + \frac{1}{2} \, f_0 \, \partial_z^2 \delta \theta= 0 &   \\
  - \frac{1}{2} \, \partial_z^2 \delta f +2 \, c_j^2 \, \delta f \,+ \,f_0 \, v \, \partial_z \delta \theta \, + \, f_0 \, \partial_t \delta \theta=0 &  
\end{array}
\right. .
\ee
The solutions given in \eq{eq:fandtheta} determine the dispersion relation \eq{eq:disprel1}. 
For a given $\om$, there are four solutions. In a subsonic flow, for $|c|>|v|$ and $\om$ real, 
$k_u$ describes the right mover, and $k_v$ the left mover, see the left panel of 
Fig.~\ref{fig:disprel1}. The two other roots are complex: $k_+$ gives the exponentially decreasing mode at $z \rightarrow + \infty$, whereas $k_-$ is the decreasing one at $z \rightarrow - \infty$.  In the central supersonic region, as can be seen from the right panel of
Fig.~\ref{fig:disprel1}, the four roots are real if $-\om_{\rm max}<\om<\om_{\rm max}$, where
\be \label{eq:omMax}
\om_{\rm max}=2 \sqrt{2} \sqrt{|v| +\sqrt{ v^2 +8 \, c_2^2}} \lp \frac{v^2-c_2^2}{3 \, |v| +\sqrt{ v^2 +8 \, c_2^2}} \rp^{3/2} .
\ee

When looking for ABM, one must keep only the wave vectors with a negative imaginary part in $I_1$,
and a positive imaginary part in $I_3$. When considering the ABM which grows in time, i.e. for $\Im \om = \Gamma > 0$, 
in $I_1$, the two wave vectors respectively correspond to the analytical continuations
of the left-moving mode $k_v$ and the evanescent mode $k_-$.  In $I_3$ instead, they correspond to the right-moving mode $k_u$ and the evanescent mode 
$k_+$; see Fig.~\ref{fig:disprel1} for $\om \in \mathbb{R}$. 
The modes selected in this way are outgoing in that the analytical continuation of the roots $k_\om$
which are real for real $\om$ possess an outgoing group velocity.
Notice that this definition also applies to the degenerate case characterized by
a purely imaginary $\om$. Indeed, as long as $\vert \Im \om \vert < \Gamma_0$, where $\Gamma_0$   
is given by 
\be\label{eq:Gamma0} 
\Gamma_0=\sqrt{8 \frac{ |v| +\sqrt{ v^2 +8 \, c_1^2}}{\left(3 \,  |v| +\sqrt{ v^2 +8 \, c_1^2}\right)^3}\left(c_1^2-v^2\right)^3} ,
\ee
the various roots do not cross each other; see Fig. \ref{fig:disprel_im_sub}. Hence, in that interval, 
the complex roots can be viewed as analytical extensions of their ancestors defined at $\om \approx 0$. 

We {\it define} the QNM by the same outgoing condition, but this time
in the complex lower half-plane $\Gamma<0$. We thus also retain $k_v$ and $k_-$ in $I_1$, and $k_u$ and $k_+$ in $I_3$.\footnote{N.B. These conditions differ from those used in Ref.~\cite{Garay}.} 
It is therefore not so surprising that all ABM appear as some QNM cross the real axis. 
Yet, there exists additional QNM which are not the analytical continuation of ABM. 
The link between these QNM and the retarded Green function was explored in~\cite{Coutant:2016bgk}.

It should be noticed that the matrix $M$ defined below 
possesses a smooth limit $\Re \om \rightarrow 0$. So, the procedure to find purely imaginary frequencies does not differ from the general case. Yet, in this case, the instability is described by a real degree of freedom, instead of a complex one as in the case $\Re \om \neq 0$. This reduction can be seen by considering 
the solutions of the Bogoliubov-de Gennes equation~\cite{Shlyapnikovcourse,pitaevskii2003bose}.
Whether $\Re k = 0$ or not, when $\Re \om = 0$, the complex-frequency modes
obey $\phi_k=\phi_{-k^*}$. 
The number of degrees of freedom is thus halved with respect to the case $\Re \om \neq 0$. However, the number of matching conditions is also halved, which explains why the equation $\det M = 0$ still gives a discrete set of modes with purely imaginary frequencies. 

\subsubsection{Structure of the matching matrix \texorpdfstring{$M$}{M}}

Continuity and differentiability of $\delta f$ and $\delta \theta$ at the two horizons give eight matching conditions which can be written as eight linear relations between the coefficients of the modes for a given $\om$. A nontrivial solution exists if and only if the determinant of the 8-by-8 matrix $M$ defined below vanishes. 

Lines of $M$ correspond to each of the eight matching conditions, while its columns correspond to the eight modes: the two modes in $I_1$ in the first two columns, the four modes in $I_2$ in the next four columns and the modes in $I_3$ in the last ones. 
The coefficients of the first line of $M$ are the values of $\e^{\ii k z}$ evaluated at $z=-L$ for the corresponding mode, multiplied by $k^2$. The same factor $k^2$ multiplies all the coefficients of a given column, so it does not change the equation $\det M = 0$. It is introduced to avoid important numerical errors when $k$ is close to zero. The last two coefficients of the first line are set to zero because the modes in $I_3$ do not contribute at $z=-L$. 
The second line of $M$ contains the derivative  of $\e^{\ii k z}$ evaluated at $z=-L$, multiplied by $k^2$. 
As for the first line, the last two coefficients are set to zero. 
The third and fourth lines are built in the same way, except that $\e^{\ii k z}$ is replaced by $\frac{\delta \Theta}{\delta F} \, \e^{\ii k z}$. So, the first four lines encode the matching conditions at $z=-L$.
The last four lines are constructed similarly, except $-L$ is replaced by $L$ and the first two coefficients are set to zero instead of the last two, since the relevant regions are then $I_2$ and $I_3$. 

Explicitly, the first two columns of $M$ have the form
\be 
\left(
\begin{array}{c}
 k_1^2\e^{-\ii k_1L} \\
 k_1^3\e^{-\ii k_1L} \\
 \Omega_1 \e^{-\ii k_1L} \\
 k_1 \Omega_1 \e^{-\ii k_1L} \\
 0 \\
 0 \\
 0 \\
 0
\end{array}
\right),
\ee 
where $k_1$ is the wave vector of either one of the two modes in $I_1$: $k \in \left\lbrace k_{1,1}, k_{1,2} \right\rbrace$, and $\Omega_1 = \om - v k_1$. 
The seventh and eighth columns have the same structure, with the first four lines replaced by the last four, evaluated with the appropriate roots. The four central columns have no zero, and contain twice the above structure of four entries.

\subsubsection{Comparison with the Bohr-Sommerfeld approach of \texorpdfstring{\cite{Finazzi:2010nc}}{Finazzi:2010nc}}

\begin{figure}
\centering
\includegraphics[scale=0.7]{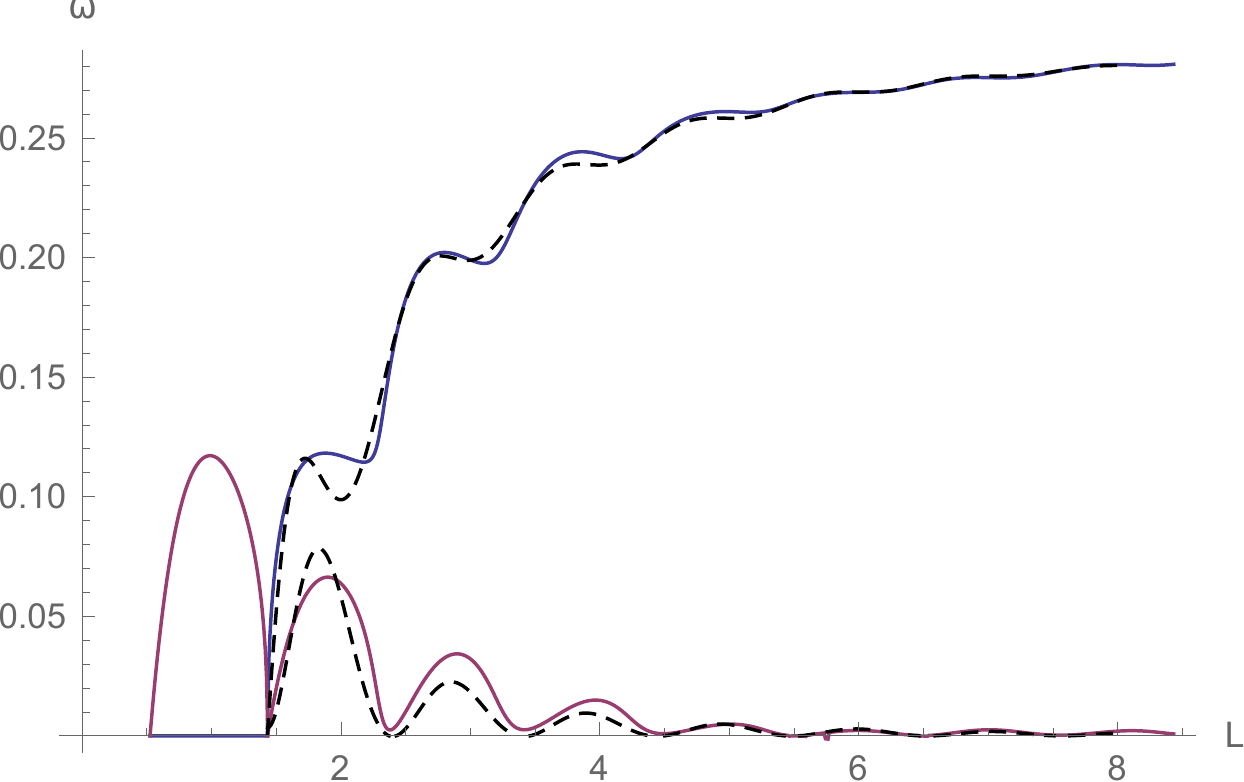}
\caption{Frequencies of the ABM with $n=0$ obtained using the Bohr-Sommerfeld approach,
and by solving $\det(M)=0$ as functions of $L$. The solid lines show the root of $\det(M)$ (blue: real part; purple: imaginary part) and the dashed lines are from the Bohr-Sommerfeld approximation. The parameters are: $c_1=c_3=1.5$, $c_2=0.5$, $v=1.0$ and $f_0=1$.}\label{fig:BS}
\end{figure}

In~\cite{Finazzi:2010nc} a semiclassical (Bohr-Sommerfeld) 
approach was used to study the set of ABM. As usual in this kind of treatment, the set of single-valued solutions is characterized by a positive integer $n_{\rm BS}$ equal to the integrated phase shift when making a round trip between the two horizons, divided by $2 \pi$. This approach was shown to be in good agreement with the numerical data when $L$ is large enough for a fixed $n$, as expected because corrections to the semiclassical approximation decrease in this limit. In fact, our exact treatment agrees both qualitatively and quantitatively with~\cite{Finazzi:2010nc} in this limit; see Fig.~\ref{fig:BS}. In particular, one verifies that the parameter $n_{\rm BS}$ plays exactly the role of $n$ defined in Section~\ref{Slt}. There is thus a one-to-one correspondence between the set of ABM found using the two methods. However, while it correctly predicts that new complex-frequency ABM appear at $L_{n+1/2}$, the Bohr-Sommerfeld approach can not describe the existence of the degenerate ABM with imaginary frequency which exists for each $n \in \mathbb{N}$ for $L_{n}<L<L_{n+1/2}$. This is not surprising since corrections to the semiclassical approximation are large in this case.

\subsection{Stationary solutions in the presence of one single horizon}
\label{App:single-h}

In this subsection we describe the stationary solutions of the GPE in the presence of a single discontinuity of $V$ and $g$, for a black- or white-hole configuration. We focus on solutions for which $f$ goes to a constant $f_0$ at infinity in the subsonic region, which serve as building blocks for the black hole laser solutions. 

We consider a one-dimensional, infinite Bose-Einstein condensate whose two-body coupling $g$ and external potential $V$ are piecewise constant with a single discontinuity at $z=0$. We assume repulsive interactions, i.e., $g>0$, and $J \neq 0$. 
We denote as $g_1$, $V_1$ the parameters in the region $z<0$ and as $g_2$, $V_2$ the parameters in the region $z>0$. For simplicity, they are tuned so that a globally uniform stationary solution exists. 
 
\subsubsection*{Homogeneous case}
\label{homo}

Many properties of the solutions can be derived in the homogeneous case without discontinuity. We thus momentarily assume that $g$ and $\mu$ are uniform and discuss the solutions of \eq{eq:f}. We begin with the homogeneous solutions. Setting $f''=0$ in (\ref{eq:f}) gives
\be \label{poly1}
  2 g f^6-2 \mu  f^4+ J^2=0 .
\ee
\eq{poly1} is a third-order polynomial in $f^2$, which can be solved exactly. There are obviously no real solutions in $f$ for $\mu \leq 0$. We therefore assume $\mu > 0$. A straightforward calculation shows that there are then real solutions if and only if $|J| \leq J_{\max}$, where 
\begin{align} \label{eq:Jmax} 
J_{\max} \equiv \sqrt{\frac{8}{27}\frac{\mu^3}{g^2}} .
\end{align}
There are then two homogeneous solutions with positive $f$: a subsonic one $f_b$ and a supersonic one $f_p$, with $f_b \geq f_p$. 
To see this, consider the behavior of the function $f \mapsto 2 g f^6 - 2 \mu f^4 + J^2$ for $f > 0$. 
This polynomial goes to $J^2 > 0$ for $f \to 0$ and to $+ \infty$ for $f \to +\infty$. 
Moreover, its derivative is positive if and only if $f^2 > f_m^2$, where $f_m \equiv \sqrt{2 \mu / (3 g)}$ is the point where the function reaches its minimum. 
Direct calculation gives
\begin{align*}
2 g f_m^6 - 2 \mu f_m^4 + J^2 = J^2 - J_{\rm max}^2.
\end{align*}
So,
\begin{itemize}
\item If $J^2 > J_{\rm max}^2$, \eq{poly1} has no real solutions;
\item If $J^2 < J_{\rm max}^2$, it has two strictly positive solutions $f_p$ and $f_b$, satisfying $f_p < f_m < f_b$.
\end{itemize}
To determine the sub- or supersonic character of these solutions, we compute $v^2 - c^2$:
\begin{align*}
v^2 - c^2 = \frac{J^2}{f^4} - g f^2 = 2 \mu - 3 g f^2 = 3 g \lp f_m^2 - f^2 \rp,
\end{align*}
which is positive for $f_b$ and negative for $f_p$. 

It is convenient to treat \eq{eq:f} as a system of coupled first-order equations for $f,p$ with $p \equiv f'$:
\be 
\left(
\begin{array}{c}
 f \\
 p
\end{array}
\right)'=\left(
\begin{array}{c}
 p \\
  -2 \mu  f + 2 g f^3+\frac{J^2}{f^3}
\end{array}
\right) .
\ee 
Each stationary solution draws a trajectory in phase space $(f,p)$. 
As can be seen in the left panel of Fig.~\ref{phase_portrait2}, they divide the phase space into three regions. The two external ones correspond to solutions which go to infinity at a finite distance from the origin, so they cannot be solutions in an infinite or semi-infinite interval. They also turn out to be irrelevant for the black hole laser case despite the presence of a finite supersonic region. For this reason we will not consider them.
We will instead focus on solutions in the middle domain and at the boundaries.
The former contains periodic solutions which oscillate around the supersonic homogeneous 
one. Their wavelength varies between a finite value in the limit of small amplitudes (solutions which remain close to the homogeneous supersonic one) and infinity close to the boundary. The minimum wavelength is given by \eq{eq:lambda}. The boundary of this region is the dark soliton. At the boundary between the two external domains one finds solutions which are asymptotically divergent on one side and go to a constant on the other side, called \textit{shadow solitons}. 
Periodic solutions are shown in \fig{im:NL_sol}. 

\subsubsection*{One horizon: The $2+1$ inhomogeneous solutions.}
\label{2+1sol}

When taking the discontinuity into account, we have two phase diagrams: one for $z<0$ and one for $z>0$. There are \textit{a priori} many qualitatively distinct global solutions. But only few of them are relevant for the problem at hands. An important technical simplification comes from the assumption that there exists a globally homogeneous solution. We want this solution to be subsonic for $z<0$ and supersonic for $z>0$. Depending on the sign of the velocity, this is either a 
black- or white-hole horizon. Since this sign does not affect the stationary solutions, we will not specify it. Our analysis will thus be directly applicable to the black hole laser case, the corresponding solutions being obtained by gluing one solution describe below with the translated mirror image of another one or itself. 

We are interested in solutions that go to the subsonic homogeneous solution as $z \rightarrow - \infty$.  
The qualitative properties of the solutions can be seen by superimposing the two phase portraits associated with the two choices of parameters (see \ref{phase_portrait2}, right panel). Relevant solutions start on the black dot at $z \rightarrow - \infty$ (this is equivalent to saying that they go to $f_0$ at $-\infty$). 
There are then four possibilities when increasing $z$:
\begin{itemize}
\item the homogeneous solution with $f=f_0$, $p = 0$;
\item the solution following the black line with increasing $f$ and $p$;
\item or the two solutions with (initially) decreasing $f$ and $p$.
\end{itemize} 

In any case the black line 
 is followed until $z=0$. Then the trajectory changes and follows a blue line in Fig.~\ref{phase_portrait2}. If the solution was homogeneous for $z<0$, it remains so for $z>0$. Other solutions are periodic for $z>0$: $f$ shows oscillations around $f_0$ with a finite amplitude $a \equiv (f_\textrm{max}-f_0)/f_0$. In the limit of small amplitudes, the wave vector is then given by the nonvanishing root of \eq{eq:disprel1} with $\om = 0$. The wave length goes to infinity when the maximum value of $f$ approaches the subsonic solution $f_{2,b}$ given by 
\be \label{eq:fb} 
f_{2,b} = \frac{1}{2}\frac{|v|}{c_2} \sqrt{1+\sqrt{1+8 \frac{c_2^2}{v^2}}} .
\ee

\begin{figure}
\centering
\includegraphics[width=0.49 \linewidth]{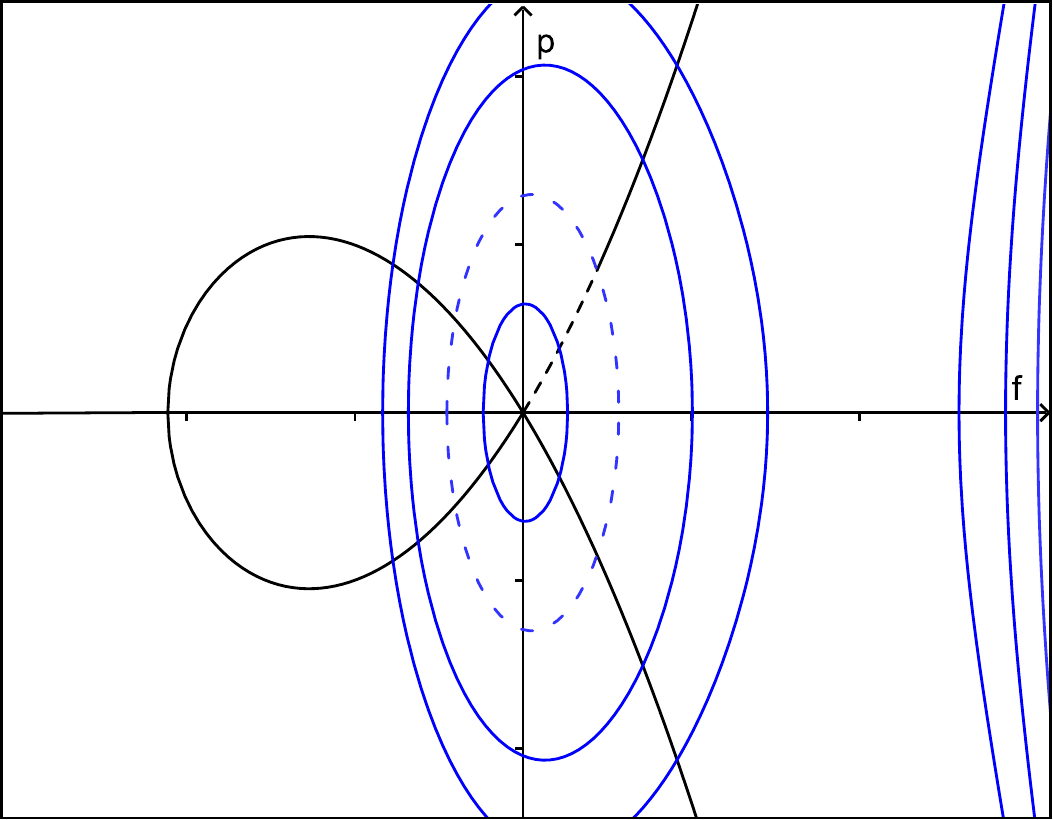}
\includegraphics[width=0.49 \linewidth]{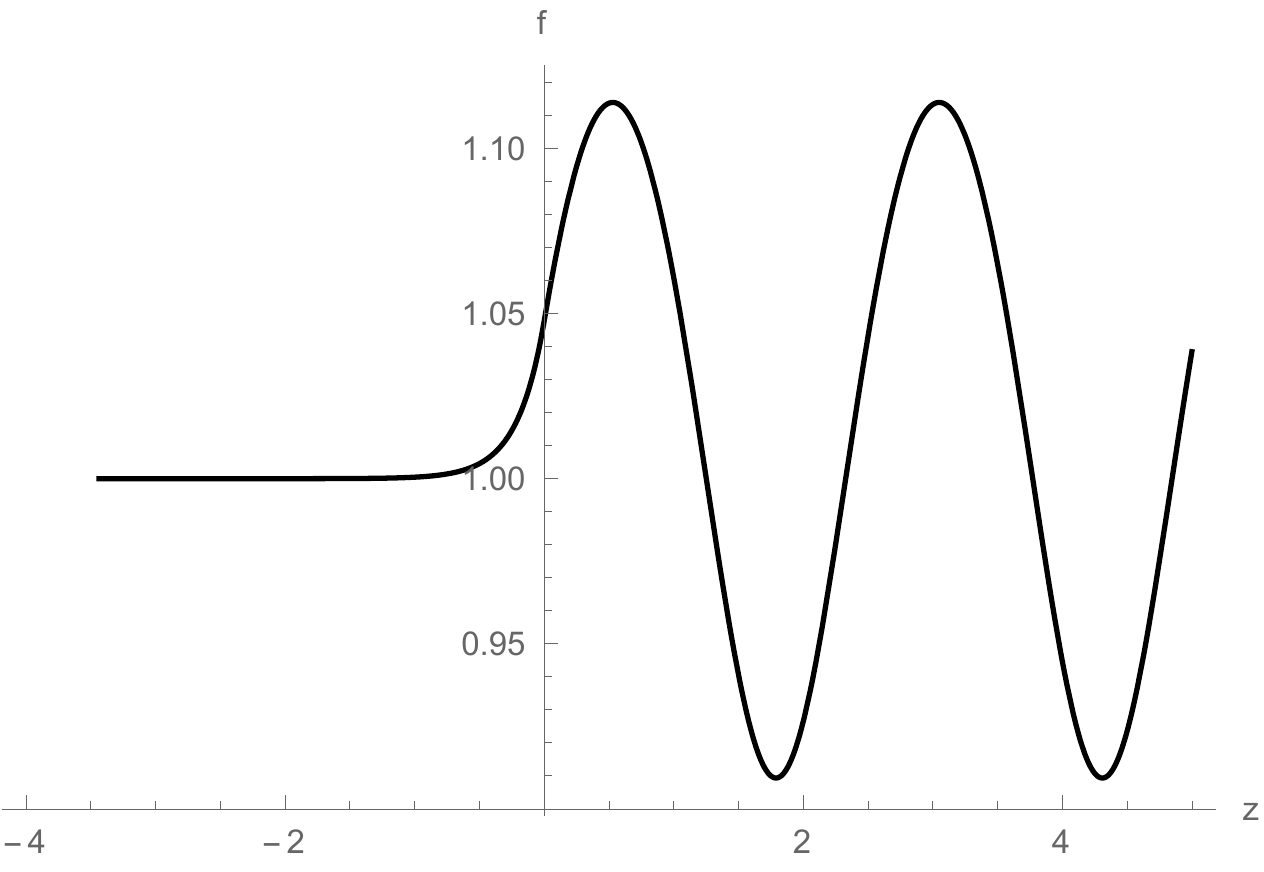}
\includegraphics[width=0.49 \linewidth]{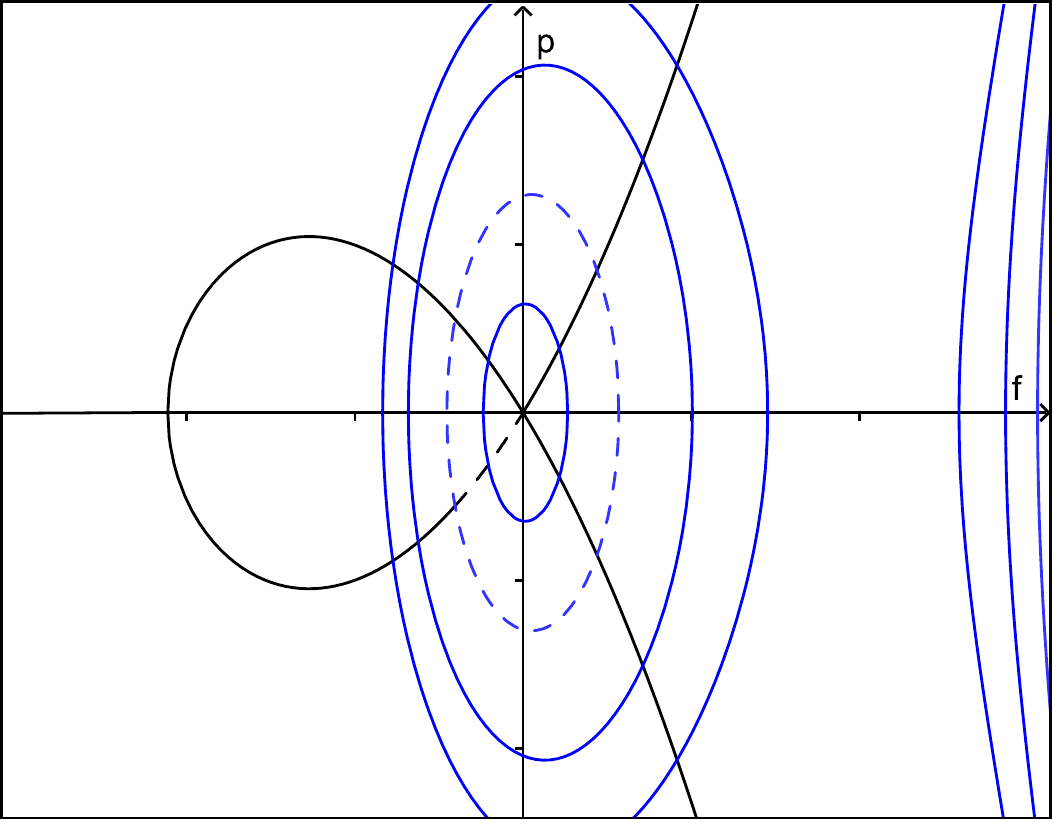}
\includegraphics[width=0.49 \linewidth]{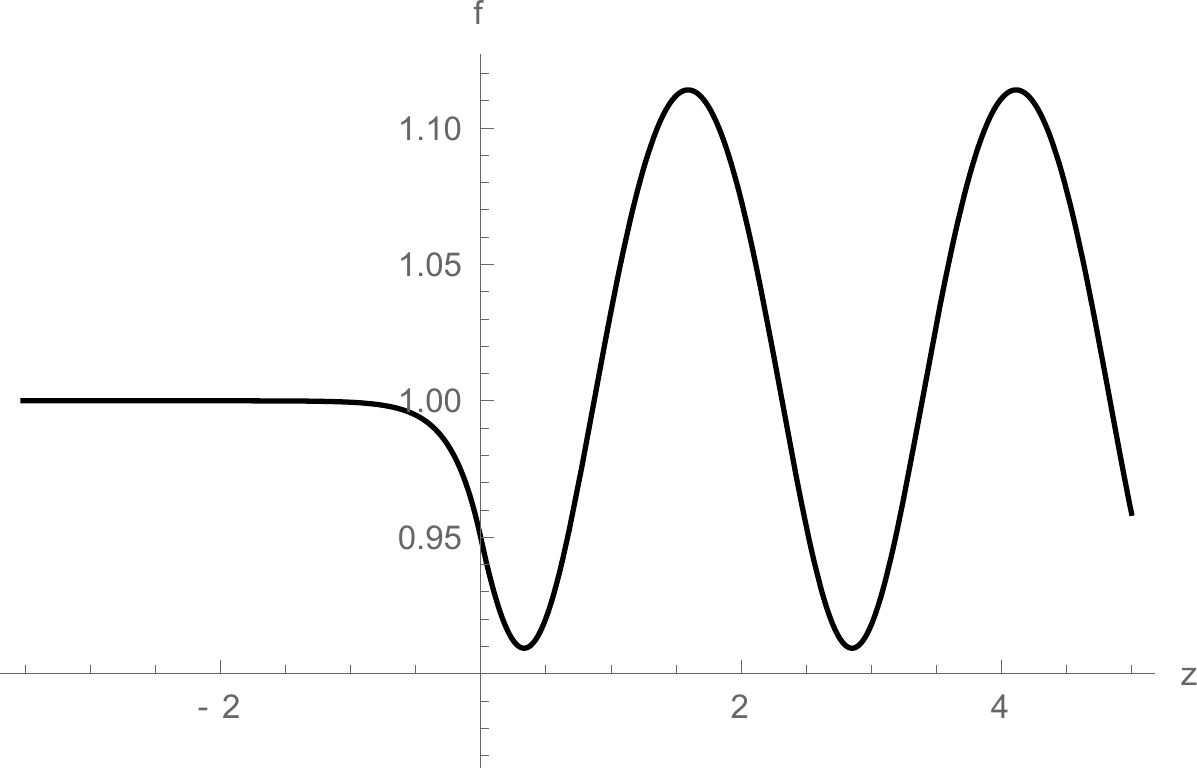}
\includegraphics[width=0.49 \linewidth]{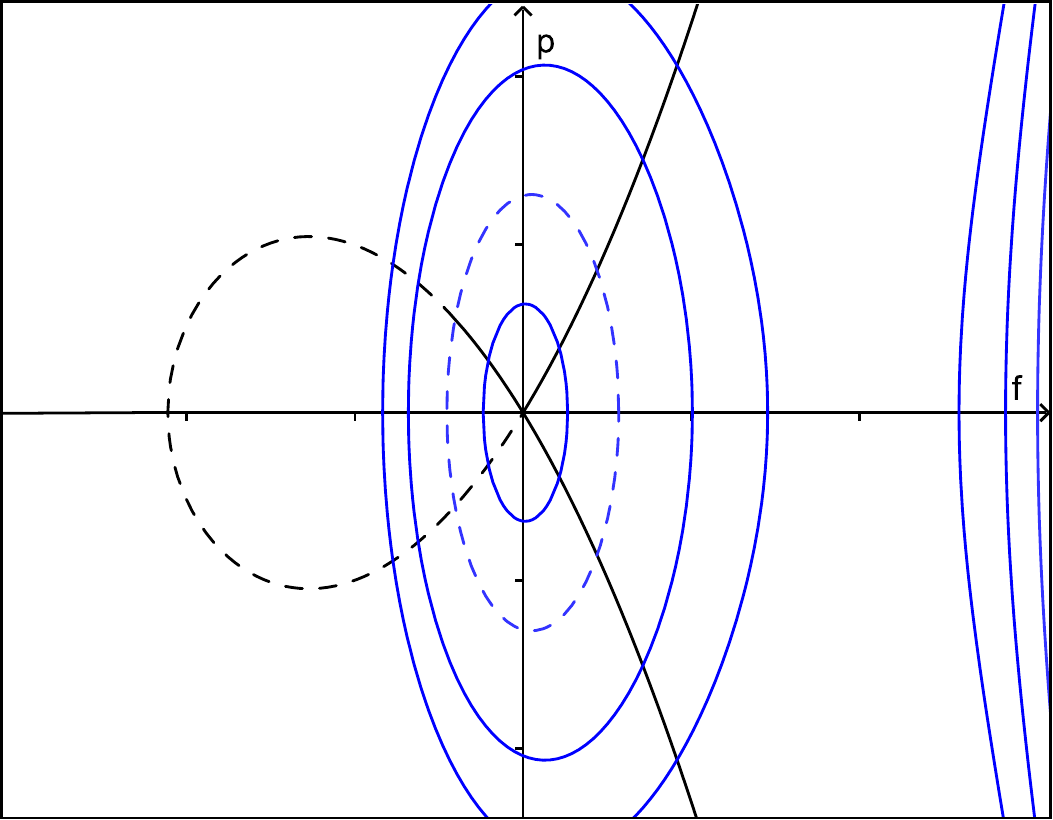}
\includegraphics[width=0.49 \linewidth]{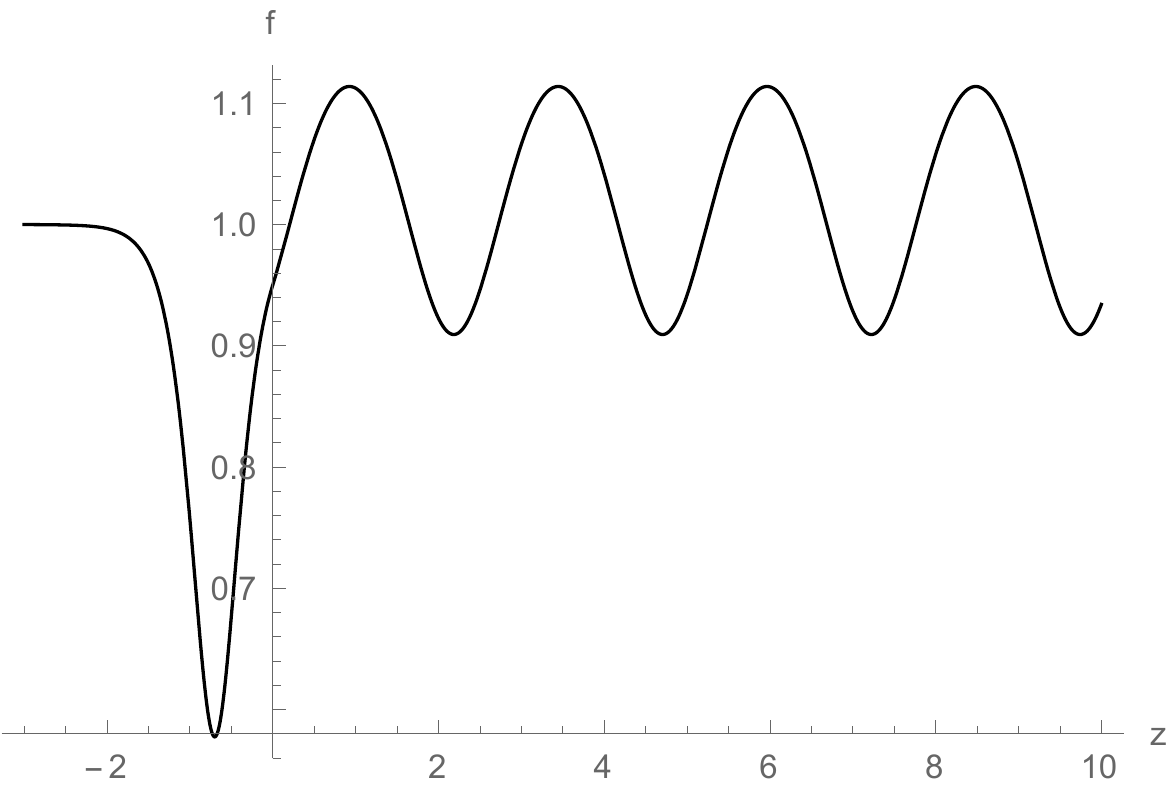}
\caption{Trajectories in phase space (left) and $f$ as a function of $z$ (right) for $a \approx 0.115$. Left: Simplified phase portrait where only three blue lines and the black line of Fig.~\ref{phase_portrait2} are represented. The dashed curve corresponds to a solution going to $f=f_0$ at $z=- \infty$. We show the three solutions giving the same amplitude in the region $z>0$. The parameters are $g_1=8$, $g_2=1$, $\mu_1=28/3$, $\mu_2=7/6$ and $J^2=8/3$. 
Right: $f/f_0$ as a function of $z$ for these three solutions. The soliton can be seen on the lower plot for $z<0$. 
The upper plots correspond to the shadow-soliton solution, the middle ones to the first soliton solution and the lower ones to the second soliton solution.}\label{fig:2+1}
\end{figure}

The first nonlinear solution is obtained by following the black line in the direction of increasing $f$ and $p$.
The last two are found by following it in the direction of decreasing $f$ and $p$ (the black loop in Fig.~\ref{phase_portrait2}). There are two of them because, if the black loop crosses a blue line, it does it twice by symmetry $p \rightarrow -p$. The two intersection points give two solutions. If the amplitude of the oscillations is small, one solution corresponds to a very small path on the black loop, hence a tiny fraction of the soliton, while the other one has a nearly complete soliton in the region $z<0$.
The three trajectories in phase space and their corresponding profiles $f(z)$ are represented in Fig.~\ref{fig:2+1}. The second soliton solution (lower plots) is physically less interesting since it has a larger, finite 
energy and cannot be continuously deformed into the homogeneous solution while keeping the oscillations for $z>0$ small. This is the meaning of "$2+1$" in the title of this subsection: For a given (small) amplitude there are three nonlinear solutions, but one of them has a much larger energy than the other two. To linear order, the latter are related by a $\mathbb{Z}_2$ symmetry $\delta f \rightarrow - \delta f $ and correspond to the undulation (zero-frequency wave) described in \cite{Coutant:2011in}.

For larger amplitudes there may be only one solution if the corresponding blue line does not cross the black loop. A straightforward calculation shows the three solutions persist up to the maximum amplitude (at which $f$ reaches $f_{2,b}$ asymptotically) if and only if
\be
\frac{2 f_0^4}{f_{2,b}^2+f_0^2} \geq  \frac{2 f_{1,p}^4}{f_{1,p}^2+f_0^2},
\ee 
where $f_{1,p}$ is the homogeneous supersonic solution for $z<0$. This can be rewritten as
\begin{align} \label{eq:inf}
\frac{v^2}{c_1 \, c_2} \lp 1+\sqrt{1+8 \, \frac{c_2^2}{v^2}} \rp \leq 4,
\end{align}
where $c_1$ and $c_2$ are, respectively, the sound velocities for $z<0$ and $z>0$.

The same analysis applies at each horizon of the black hole laser. The decomposition $2+1$ then becomes $(2+1) \times (2+1)=4+5$, \textit{i.e.} $4$ solutions can be 
arbitrarily close to the homogeneous one and can be studied at linear order, while five of them contain at least one soliton. The first four solutions are analogous to those described in~\cite{PhysRevB.53.6693}.

\subsubsection*{Thermodynamic considerations}
\label{thermo_app}

We use the grand-canonical ensemble: The temperature (set to zero) and chemical potential are fixed while the energy and number of particles depend on the solution. 
We will study two cases: fixed mean velocity or fixed current $J$. 
At fixed velocity\footnote{More precisely, the fixed quantity is the difference in the phases $\theta$ evaluated at two points $z_+ \gg L$ and $z_- \ll -L$.}, 
the (off-shell) energy functional is the grand potential \eq{eq:G_off}. 
It is defined up to a constant, which we choose so that $G=0$ for the global homogeneous solution. 
Using (\ref{eq:f}), one finds the on-shell function
\be \label{eq:G_on}
G(a,J,i) = \int \frac{\dd z}{2} \, \lp \frac{\dd}{\dd z} \lp f_{a,J,i}(z) f_{a,J,i}'(z) \rp - g(z) \lp f_{a,J,i}(z)^4 - f_0^4 \rp \rp, 
\ee
where $a$ is the amplitude of the solution for $z>0$, $J$ the current, 
and $i$ a discrete parameter 
telling which of the above three solutions $f_{a,J,i}$ is considered. 

The first term in (\ref{eq:G_on}) is a boundary term.
The only contribution comes from $z \rightarrow + \infty$ since 
we assume $f' \rightarrow 0$ at $z \rightarrow - \infty$.
For definiteness we suppose (for a moment) the supersonic region $z>0$ is finite, although arbitrarily large, with a length $l$ such that $f_{a,J,i}(l) = 0$. In that case the on-shell grand potential reduces to
\be \label{GE}
G = - \int_{-\infty}^{0} \frac{\dd z}{2} \,   g_1 \lp f(z)^4 - f_0^4 \rp - \int_{0}^{l} \frac{\dd z}{2} \,   g_2 \lp f(z)^4 - f_0^4 \rp .
\ee 
\eq{GE} can be divided into two contributions. That of the region $z<0$ (first term) is finite and comes from the deformation of the solution close to the horizon. The contribution of the region $z>0$ (second term) is proportional to $l$ and thus to the number of periods of $f_{a,J,i}$ in the supersonic region. The proportionality coefficient, \textit{i.e.} the difference in thermodynamic potential per period, is always negative and diverges at the maximum amplitude. 

\eq{eq:G_off} 
is the thermodynamic potential in the grand-canonical ensemble if the mean value of the condensate velocity is fixed. This can be seen by computing the total on-shell variation of $G$
\be 
\delta G = -N \delta \mu  +\left[ \delta f \, \partial_z f \right]_{-\infty }^{\infty }+J \left[ \delta \theta \right]_{-\infty }^{\infty },
\ee
where $\left[ X \right]_{-\infty }^{\infty } \equiv \text{lim}_{z \rightarrow \infty} \lp X(z) - X(-z) \rp$ and $N$ is the total number of atoms. 
To characterize exact solutions, we found it is more convenient to work at fixed current $J$. The relevant thermodynamic potential is then the Legendre transform of (\ref{eq:G_off}) 
\be \label{eq:E}
E \equiv G - \int \, J \pd_z \theta \, \dd z = \int  \left(\frac{1}{2}f'^2-\frac{J^2}{2f^2}-\mu  f^2+\frac{1}{2}g f^4\right) \dd z.
\ee 
Since $\int \, J \pd_z \theta \, \dd z$ is a boundary term, the equations of motion are unchanged.
Up to a constant term chosen so that the energy of the homogeneous solution vanishes, (\ref{eq:E}) can be written as
\be \label{eq:E-os} 
\Delta E = \int  \left(-\frac{1}{2}g \lp f^4-f_0^4 \rp - J^2 \lp \frac{1}{f^2} - \frac{1}{f_0^2} \rp \right) \dd z.
\ee
The contribution of the deformation of the solution in the region $z<0$ to \eq{eq:E} is always positive. Notice also that the change of $E$ per period in the region $z>0$ with respect to the homogeneous solution is {\it third order} in the oscillation amplitude. The reason is that the first- and second-order terms in the expansion of 
\be \label{eq:Ebis}
\int_{z > 0}  \left(\frac{1}{2}f'^2-\frac{J^2}{2f^2}-\mu  f^2+\frac{1}{2}g f^4\right) \dd z
\ee 
in $f/f_0 - 1$ vanish for $f/f_0 - 1 \propto \cos(k z)$ with $k = 2 \sqrt{v^2 - c_2^2}$. This property is directly related to the stationarity assumption, as we now show. Let us write the condensate wave function as $\psi = \psi_0 + \phi_\om$, where $\psi_0$ is the homogeneous solution 
and $\phi_\om$ is a perturbation with frequency $\om$. 
Then the first-order term in \eq{eq:Ebis} automatically vanishes as $\psi_0$ satisfies the GPE, and the second-order term is
\[
E_2 = \int \, \om \, \phi^*(t,z) \, \phi(t,z) \, \dd z,
\]
which also vanishes if $\om = 0$.

To understand the nonlinear results of this chapter, it is of interest to determine the behavior of the integrand of \eq{eq:E} for a homogeneous solution $f' = 0$. 
To do so, we compute its derivative with respect to $f^2$:
\begin{align*}
\frac{\dd}{\dd f^2} \lp - \frac{J^2}{2 f^2} - \mu f^2 + \frac{1}{2} g f^4 \rp = 
\frac{J^2}{2 f^4}  - \mu + g f^2. 
\end{align*}
This expression vanishes when using \eq{eq:f} with $f'' = 0$, which was expected as $E$ is the energy functional associated with this equation.\footnote{Doing the same calculation using $G$ instead of $E$, one would obtain an expression which does not vanish for homogeneous stationary solutions. The reason is that, as explained above, $G$ is the relevent energy functional at fixed $v$, not at fixed $J$. To perform the same analysis with $G$ thus requires rewriting $J$ as $v f^2$ before differentiating.}
So, the two homogeneous stationary solutions are local extrema of the energy density. 
To go further, we compute
\begin{align*}
\frac{\dd^2}{\dd (f^2)^2} \lp - \frac{J^2}{2 f^2} - \mu f^2 + \frac{1}{2} g f^4 \rp = 
- \frac{J^2}{f^6} + g = \frac{c^2 - v^2}{f^2}.  
\end{align*}
We conclude that the homogeneous subsonic solution is a local minimum of the energy density while the homogeneous supersonic solution is a local maximum. 
From this one can expect that the solution minimizing the energy in the black hole laser case with a large intermediate region $I_2$ is the one which remains closest to the local subsonic solution. 
This is indeed what is found in Section~\ref{stat_sol} and subsection~\ref{sub:BHLIaddpaper}: both in the tuned and detuned cases, the ground state is the solution which interpolates between the subsonic homogeneous solutions in the three regions.

\subsubsection*{Characterization of the solutions for $z>0$}
\label{char_sol}

\begin{figure}
\centering
\includegraphics[scale=1.0]{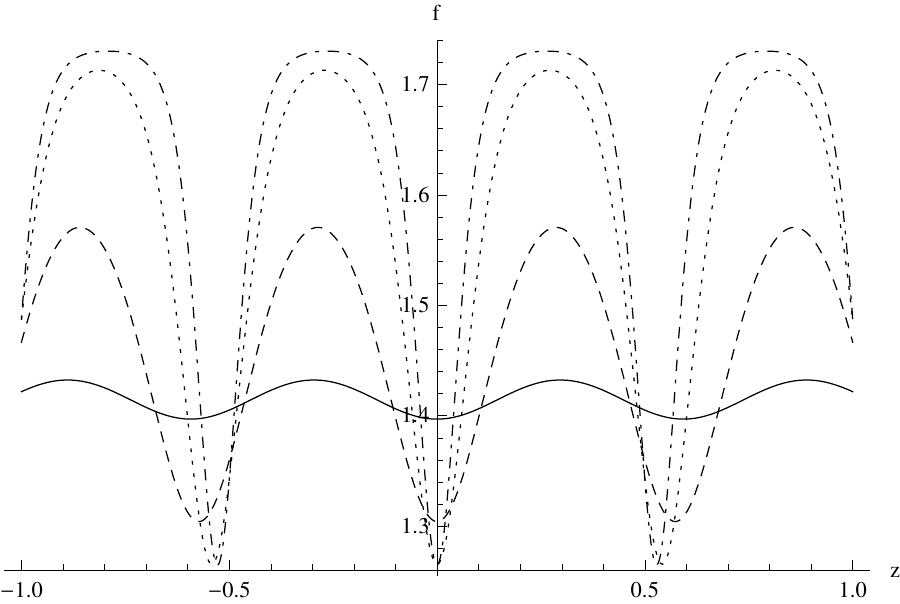}
\caption{Evolution of the shape of the solutions in the region $z>0$ when the wavelength is increased. The parameters are: $g_1=27$, $g_2=5$, $\mu_1 = 63$, $\mu_2 = 19$, $J=6 \sqrt{2}$. The coordinate $z$ is rescaled for each curve so that $z \in [-1,1]$ corresponds to a fixed number of periods. The values of $\lambda / \lambda_0 - 1$ (where $\lambda_0$ is the minimum wavelength for periodic solutions) are $0.0015$ (solid line), $0.11$ (dashed line),  $1.1$ (dotted line)  and $1.9$ (dot-dashed line).}\label{im:NL_sol}
\end{figure}

In the previous subsections we characterized a solution in the region $z>0$ by its amplitude $a=(f_{\rm max}-f_0)/f_0$. This definition is justified because, for large amplitudes, the solution remains mostly close to $f_{\rm max}$; see Fig.~\ref{im:NL_sol}. In order to get a better understanding of these solutions, here we relate the profile to the wavelength.  

In general solutions of \eq{eq:f} are Weierstrass elliptic functions, with a complex argument for the periodic ones
 \cite{belokolos1994algebro}. 
 The latter can also be expressed in terms of Jacobi elliptic functions \cite{Kamchatnov}.
 Close to the minimum wavelength 
of \eq{eq:lambda}, a straightforward calculation gives the following expansion for the amplitude:
\be \label{eq:avsla}
\frac{f_{\max }-f_0}{f_0} = 2\left(A+\frac{v^2+c_2^2}{v^2-c_2^2}A^2\right) + O \lp A^3 \rp,
\ee
where 
\be 
A\equiv \frac{1}{\sqrt{3}}\frac{v^2-c_2^2}{c_2 \sqrt{4v^2+c_2^2}}\sqrt{\frac{\lambda }{\lambda_0}-1}.
\ee
The minimum value of $f$ for a given solution, $f_{\rm min}$, is related to $f_{\rm max}$ by 
\be 
f_{\min }^2 &=& 
\frac{2\mu _2f_{\max }^2 -g_2f_{\max}^4-\sqrt{\left(g_2f_{\max}^4 -2\mu _2f_{\max}^2\right)^2-4g_2J^2f_{\max}^2}}{2 g_2f_{\max}^2}  
.
\ee 

We represent the profile of the solutions in Fig.~\ref{im:NL_sol} for increasing values of their amplitudes. 
For small amplitudes, $f$ is very close to a sinusoid, in accordance with the results of Section~\ref{Slt} and subsection~\ref{App:M-matrix}. To linear order, the $\mathbb{Z}_2$ symmetry $f \rightarrow 2 f_0 - f$ sends the shadow soliton solution to the first soliton solution and \textit{vice-versa}.
This invariance is broken at nonlinear orders as can be seen by the fact that solutions remains close to $f_{\rm max}$ over large intervals in order to minimize the energy $E$ by approaching the subsonic density of \eq{eq:fb}.

\subsection{Evaluation of \texorpdfstring{$E$, $G$ and $L$}{E, G, and L}} 
\label{App:eqs_G_L}

For the interested readers, we give the expressions of the on-shell thermodynamic function $G$ and relate the distance between the discontinuities $2 L$ to the integration constants characterizing the solutions. For definiteness we assume $c_1 = c_3$. The generalization to $c_1 \neq c_3$ is straightforward but makes the expressions longer. 
Because of the discontinuities, we must choose three integration constants: $C_1$ in $I_1$, $C_2$ in $I_2$ and $C_3$ in $I_3$. The requirement that the solution goes to $f=f_0$ as $z \rightarrow \pm \infty$ imposes $C_1=C_3 =\lp 2 v^2 + c_1^2 \rp f_0^2$. 
We define
\be \label{finter}
f_{\text{inter},\pm} \equiv f_0\sqrt{1\pm \sqrt{1-\frac{C_1-C_2}{f_0^2\left(c_1^2-c_2^2\right)}}},
\ee
which corresponds to the two possible values of $f$ at $z=\pm L$, and 
\be 
f_s \equiv \frac{v}{c_1} f_0 , 
\ee
which gives the value of $f$ at the bottom of the soliton in $I_1$ or $I_3$. The minimum and maximum values of $f$ for a periodic solution in $I_2$, denoted by $f_{\rm min}$ and $f_{\rm max}$, are the first and second positive roots of the polynomial
\be 
g_2 \, X^6 - 2 \mu_2 \, X^4 + C_2 \, X^2 - J^2 = 0 .
\ee
We also define the two functions $p_1$ and $p_2$ by
\be 
p_i(f) \equiv \frac{1}{f} \sqrt{g_i \, f^6-2 \mu_i \, f^4+C_i f^2-J^2},
\ee
where $i \in \left\lbrace 1,2 \right\rbrace$.

Using these definitions, the expressions of $G$ and $L$ for the nine types of solutions of Fig.~\ref{fig:9-traj} are given below, where $n$ denotes the number of wavelengths in $I_2$. The corresponding value of $E$ can be deduced from:
\be \label{eq:GtoE}
E = G - \int \frac{\dd f}{p_j(f)} J^2 \lp f^{-2} - f_0^{-2} \rp \dd f .
\ee

\underline{Type 1:}
\be \label{eqs-num-beg}
L=\int _{f_{\text{inter},+}}^{f_{\max }}\frac{\dd f }{p_2(f)}+ n \int _{f_{\min }}^{f_{\max }}\frac{\dd f }{p_2(f)},
\ee
\begin{align*}
G=-g_1\int _{f_0}^{f_{\text{inter,+}}}\frac{\dd f}{p_1(f)}\left(f^4-f_0^4\right)-g_2\int _{f_{\text{inter,+}}}^{f_{\max }}\frac{\dd f}{p_2(f)}\left(f^4-f_0^4\right)
-n g_2\int_{f_{\min }}^{f_{\max }}\frac{\dd f}{p_2(f)}\left(f^4-f_0^4\right).
\end{align*}

\underline{Type 3:}
\be 
L=-\int _{f_{\text{inter},-}}^{f_{\min }}\frac{\dd f}{p_2(f)}+ n \int _{f_{\min }}^{f_{\max }}\frac{\dd f }{p_2(f)},
\ee
\begin{align*}
G=g_1\int _{f_0}^{f_{\text{inter},-}}\frac{\dd f}{p_1(f)}\left(f^4-f_0^4\right)+g_2\int _{f_{\text{inter},-}}^{f_{\min }}\frac{\dd f}{p_2(f)}\left(f^4-f_0^4\right) 
-n g_2\int_{f_{\min }}^{f_{\max }}\frac{\dd f}{p_2(f)}\left(f^4-f_0^4\right)
\end{align*}.

\underline{Types 2 and 4:}
\be 
L=\frac{1}{2}\int _{f_{\text{inter},+}}^{f_{\max }}\frac{\dd f}{p_2(f)}+\frac{1}{2}\int _{f_{\min }}^{f_{\text{inter},-}}\frac{\dd f}{p_2(f)}+2 \lp n+\frac{1}{2}\right) \int _{f_{\min }}^{f_{\max }}\frac{\dd f}{p_2(f)},
\ee
\be 
G&=&-\frac{g_1}{2}\int _{f_{\text{inter},-}}^{f_{\text{inter},+}}\frac{\dd f }{p_1(f)}\left(f^4-f_0^4\right)- \frac{g_2}{2}\int _{f_{\text{inter},+}}^{f_{\max }}\frac{\dd f }{p_2(f)}\left(f^4-f_0{}^4\right)\nn &&-\frac{g_2}{2}\int _{f_{\min }}^{f_{\text{inter},-}}\frac{\dd f }{p_2(f)}\left(f^4-f_0^4\right)-\lp n+\frac{1}{2} \rp g_2\int _{f_{\min }}^{f_{\max }}\frac{\dd f }{p_2(f)}\left(f^4-f_0^4\right).
\ee

\underline{Types 5 and 7:}
\be 
L= (n+1) \int _{f_{\min }}^{f_{\max }}\frac{\dd f}{p_2(f)},
\ee
\be 
G=g_1\int _{f_0}^{f_s}\frac{\dd f}{p_1(f)}\left(f^4-f_0^4\right)- (n+1) g_2\int _{f_{\min }}^{f_{\max }}\frac{\dd f}{p_2(f)}\left(f^4-f_0^4\right).
\ee

\underline{Types 6 and 8:}
\be 
L=\frac{1}{2}\int _{f_{\text{inter},+}}^{f_{\max }}\frac{\dd f}{p_2(f)}+\frac{1}{2}\int _{f_{\text{inter},-}}^{f_{\max }}\frac{\dd f}{p_2(f)}+ n \int _{f_{\min }}^{f_{\max }}\frac{\dd f}{p_2(f)},
\ee
\be 
G& \hspace{-0.2 cm} = \hspace{-0.2 cm} &-\frac{g_1}{2}\int _{f_s}^{f_{\text{inter},+}}\frac{\dd f }{p_1(f)}\left(f^4-f_0{}^4\right)-\frac{g_1}{2}\int _{f_s}^{f_{\text{inter},-}}\frac{\dd f }{p_1(f)}\left(f^4-f_0^4\right)-\frac{g_2}{2}\int _{f_{\text{inter},+}}^{f_{\max }}\frac{\dd f }{p_2(f)}\left(f^4-f_0^4\right) \nn &&
-\frac{g_2}{2}\int _{f_{\text{inter},-}}^{f_{\max }}\frac{\dd f }{p_2(f)}\left(f^4-f_0^4\right)-ng_2\int _{f_{\min }}^{f_{\max }}\frac{\dd f }{p_2(f)}\left(f^4-f_0^4\right).
\ee

\begin{figure}
\centering
\includegraphics[width=0.32\linewidth]{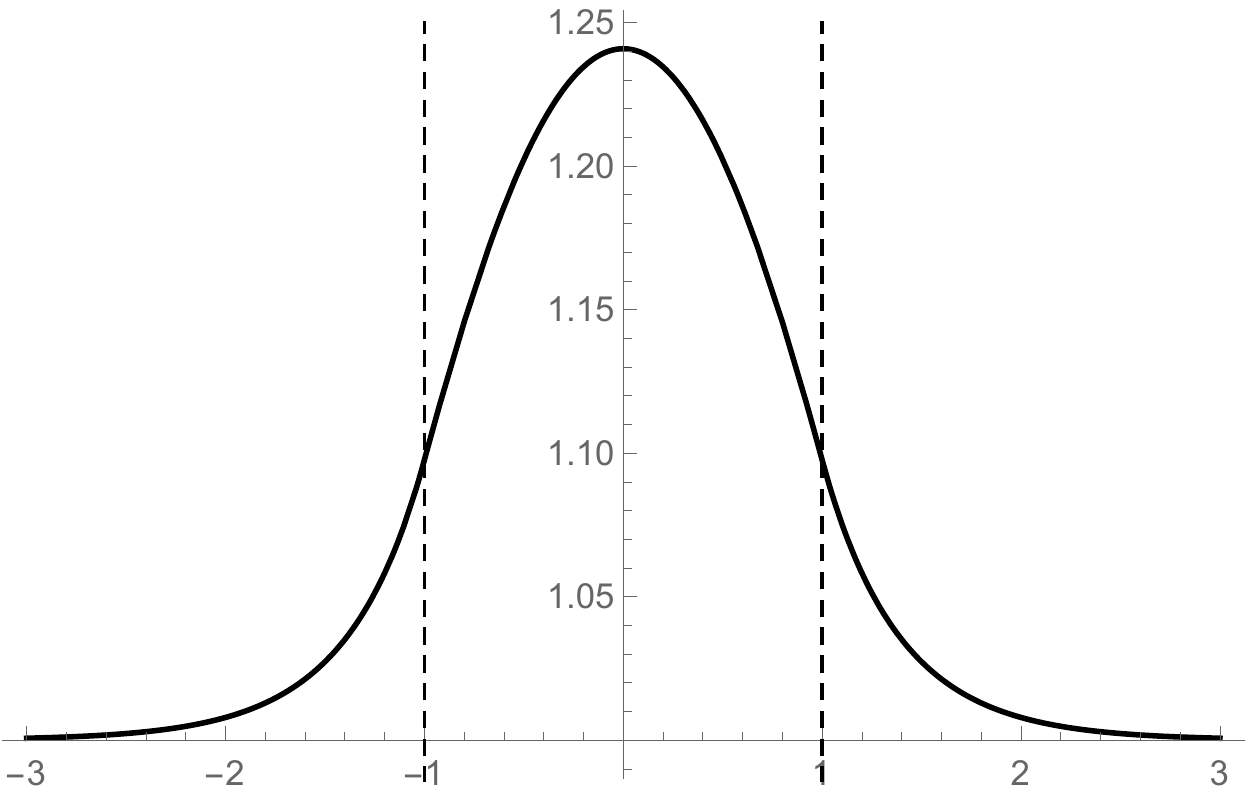}
\includegraphics[width=0.32\linewidth]{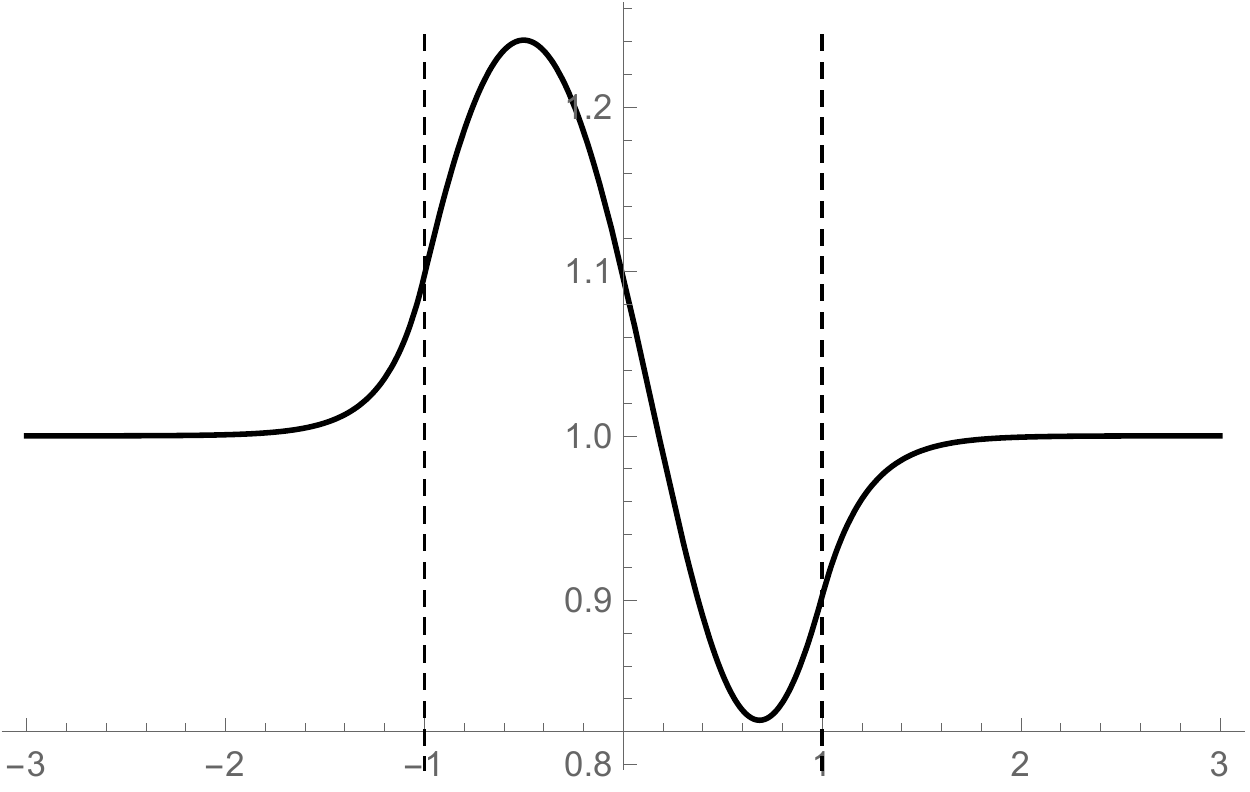}
\includegraphics[width=0.32\linewidth]{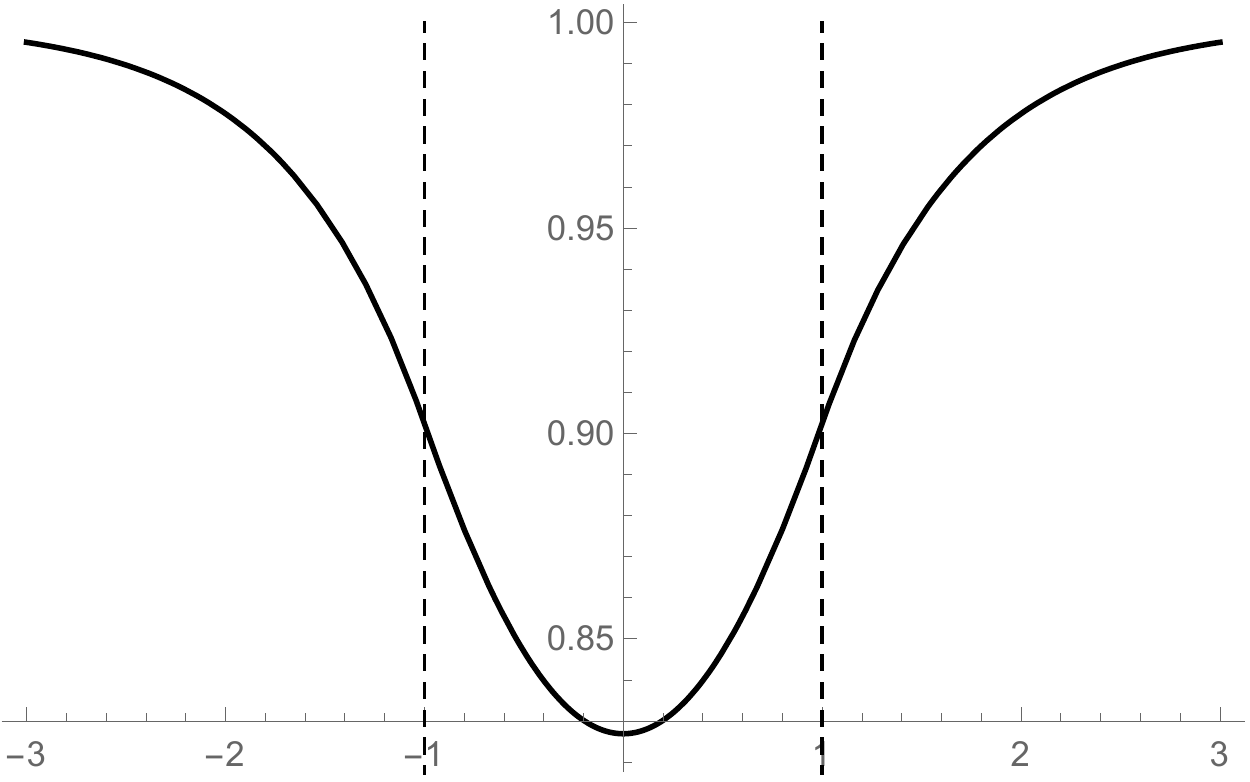}
\includegraphics[width=0.32\linewidth]{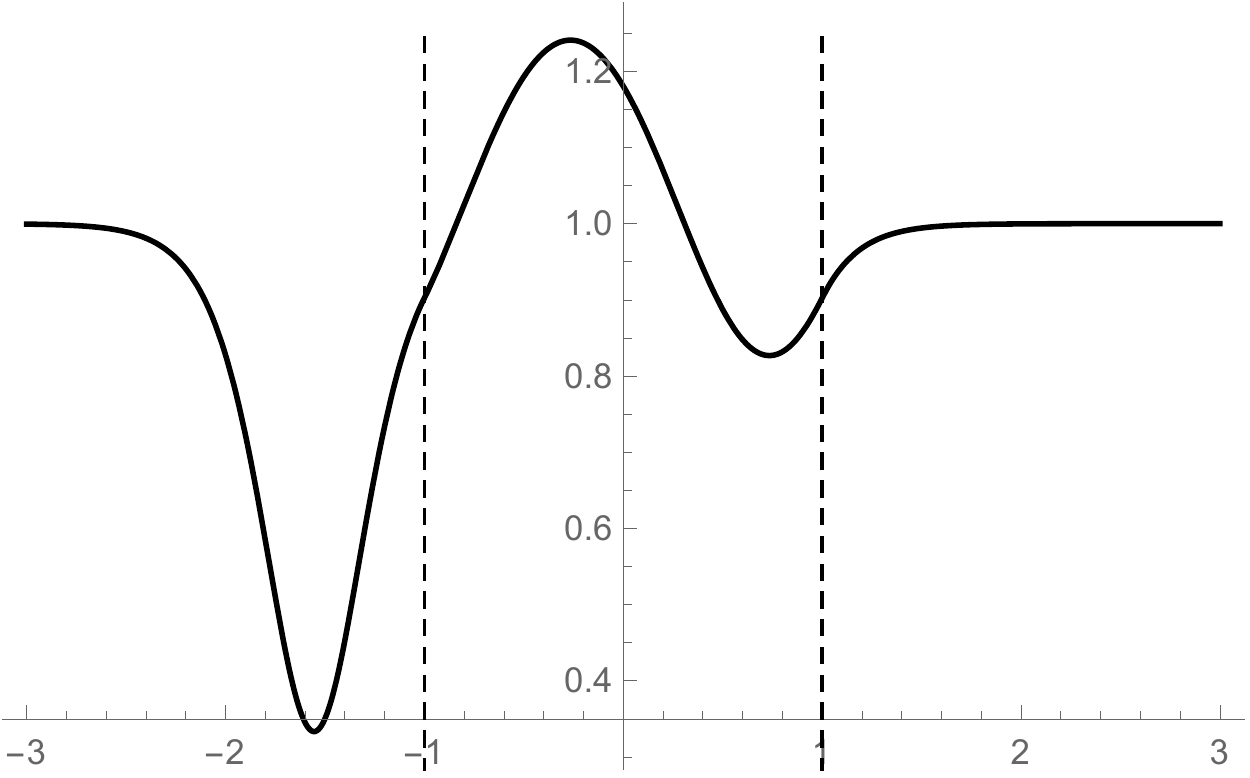}
\includegraphics[width=0.32\linewidth]{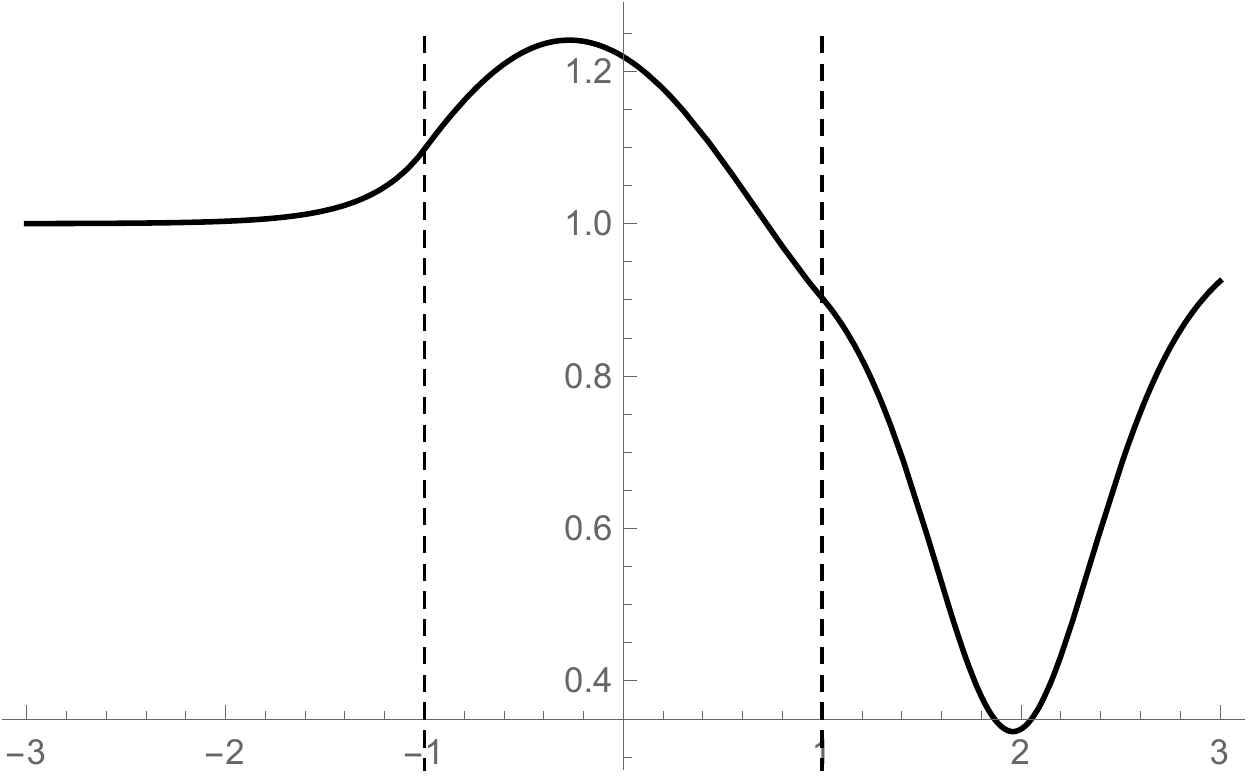}
\includegraphics[width=0.32\linewidth]{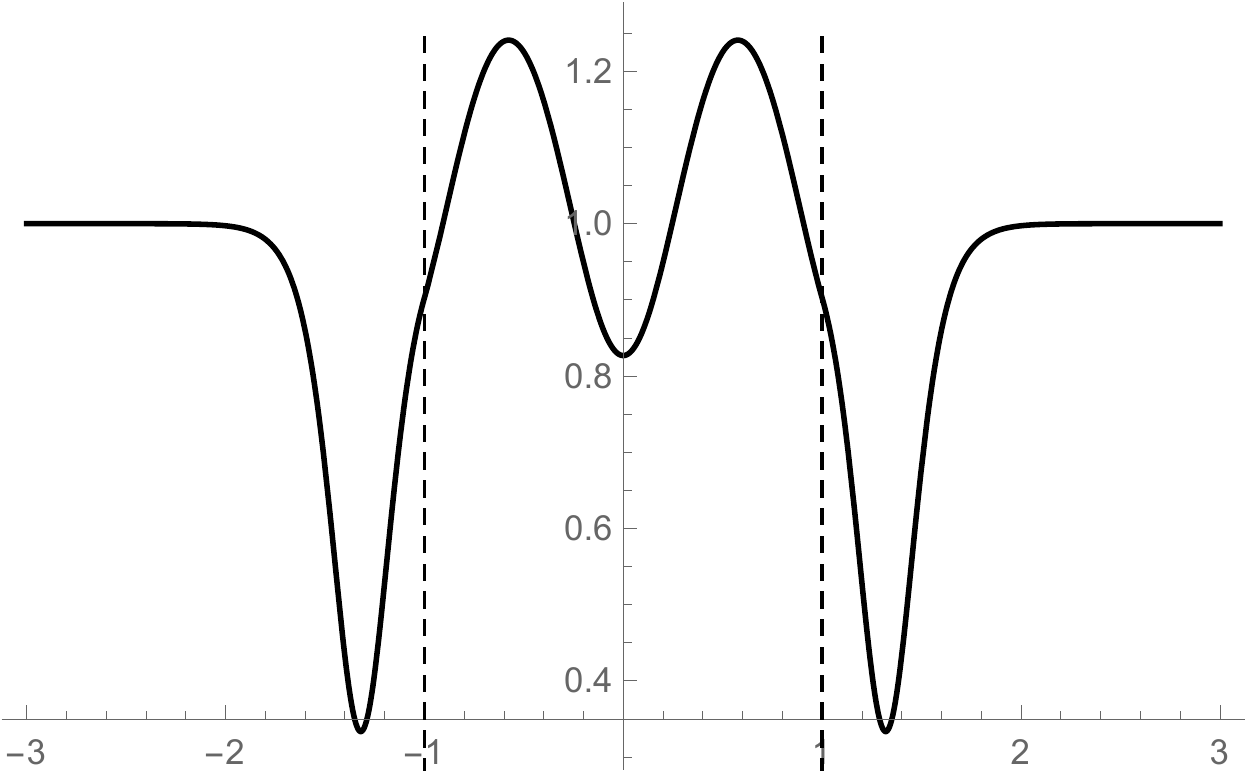}
\caption{Plots of $\rho = f^2$ as a function of $z/L$ for solutions of type 1 (top,left), type 2 (top, middle), type 3 (top,right), type 5 (bottom, left), type 6 (bottom, middle) and type 9 (bottom, right) with $n=0$. 
Dashed horizontal lines indicate the location of the discontinuities of $V$ and $g$.   
The parameters are $c_1 = 2 \sqrt{2}$, $c_2 = 1$, $v=\sqrt{8/3}$, $f_0=1$ and $C_2=6.4$. Solutions of types 4, 7 and 8 are obtained from those of types 2, 5 and 6 by the symmetry $z \rightarrow -z$.
}\label{fig:profiles}
\end{figure}

\underline{Type 9}
\be 
L=\int _{f_{\text{inter},-}}^{f_{\max }}\frac{\dd f }{p_2(f)}+ n \int _{f_{\min }}^{f_{\max }}\frac{\dd f }{p_2(f)},
\ee
\be \label{eqs-num-end}
G=g_1\int _{f_0}^{f_s}\frac{\dd f }{p_1(f)}\left(f^4-f_0^4\right)-g_1\int _{f_s}^{f_{\text{inter},-}}\frac{\dd f }{p_1(f)}\left(f^4-f_0^4\right)- g_2\int _{f_{\text{inter},-}}^{f_{\max }}\frac{\dd f }{p_2(f)}\left(f^4-f_0^4\right) \nn
- n g_2\int _{f_{\min }}^{f_{\max }}\frac{\dd f }{p_2(f)}\left(f^4-f_0^4\right).
\ee
The integration constant $C_2$ can take any value between $C_{2, \rm min}$ and $C_{2,\rm max}$. $C_{2,\rm min}$ is always equal to:
\be \label{eq:C2min} 
C_{2,\rm min} = \left(2 v^2+c_2^2\right) f_0^2 .
\ee  
In general, the maximum value of $C_2$ is:
\be \label{eq:C2max}
C_{2,\rm max}=\frac{v^2}{8 c_2^2}\left(-4c_2^2+v^2+v^2\left(1+8 \frac{c_2^2}{v^2}\right)^{3/2}\right) .
\ee
The only exception is type 3 with $n=0$, for which
\be 
C_{2,\rm max} (3,n=0)=\left(\frac{c_1^6+2 c_1^2c_2^2v^2-v^4c_2^2+c_1^2v^4}{c_1^4}\right)f_0^2 .
\ee
Figure~\ref{fig:profiles} show $f$ as a function of $z/L$ for the nine different types of solutions and $n=0$. 

\subsection{Finding the stationary solutions and ABM (detuned case)}

In this subsection we present the procedure we used to find the stationary solutions for detuned black-hole laser configurations and the discrete spectrum over inhomogeneous solutions.

\subsubsection{Stationary solutions in the detuned case}
\label{NLsol}

Following subsection~\ref{App:single-h}, we first find the stationary solutions in each of the three regions $I_1: x<-L$, $I_2: -L<x<L$, and $I_3: x>L$, and match them at $x = \pm L$ to find the global solutions on $\mathbb{R}$. We also assume that the flow is uniform and subsonic for $x \to \pm \infty$. In each region $I_i$, \eq{eq:f} gives
\begin{align}\label{eq:fCapp}
\frac{1}{2} \lp \pd_z f \rp^2 = -\mu_i f^2 + \frac{g_i}{2} f^4 - \frac{J^2}{2 f^2} + C_i,
\end{align}
where $C_i$ is an integration constant. 
As explained in subsection~\ref{App:single-h}, assuming $\abs{J} < J_{\rm max}$ there exist two homogeneous solutions $f_{b,i}$ and $f_{p,i}$ in each region, with $f_{p,i} < f_{b,i}$. 
Moreover, periodic solutions in $I_i$ exist for values of $C_i$ between a minimum one $C_{i, \rm min}$ and a maximum one $C_{i, \rm max}$. 
As the potential $V$ and two-body coupling $g$ take the same values in the two asymptotic regions, we can adopt more explicit notations and replace the index ``$i$'' by ``ext'' for $i \in \left\lbrace 1,3 \right\rbrace$ and by ``int'' for $i = 2$. 
That is, ``ext'' (respectively ``in'') denotes a quantity evaluated in the exterior region $I_1 \cup I_3$ (respectively in the interior region $I_2$). 

\begin{figure}
\centering
\includegraphics[width=0.49\linewidth]{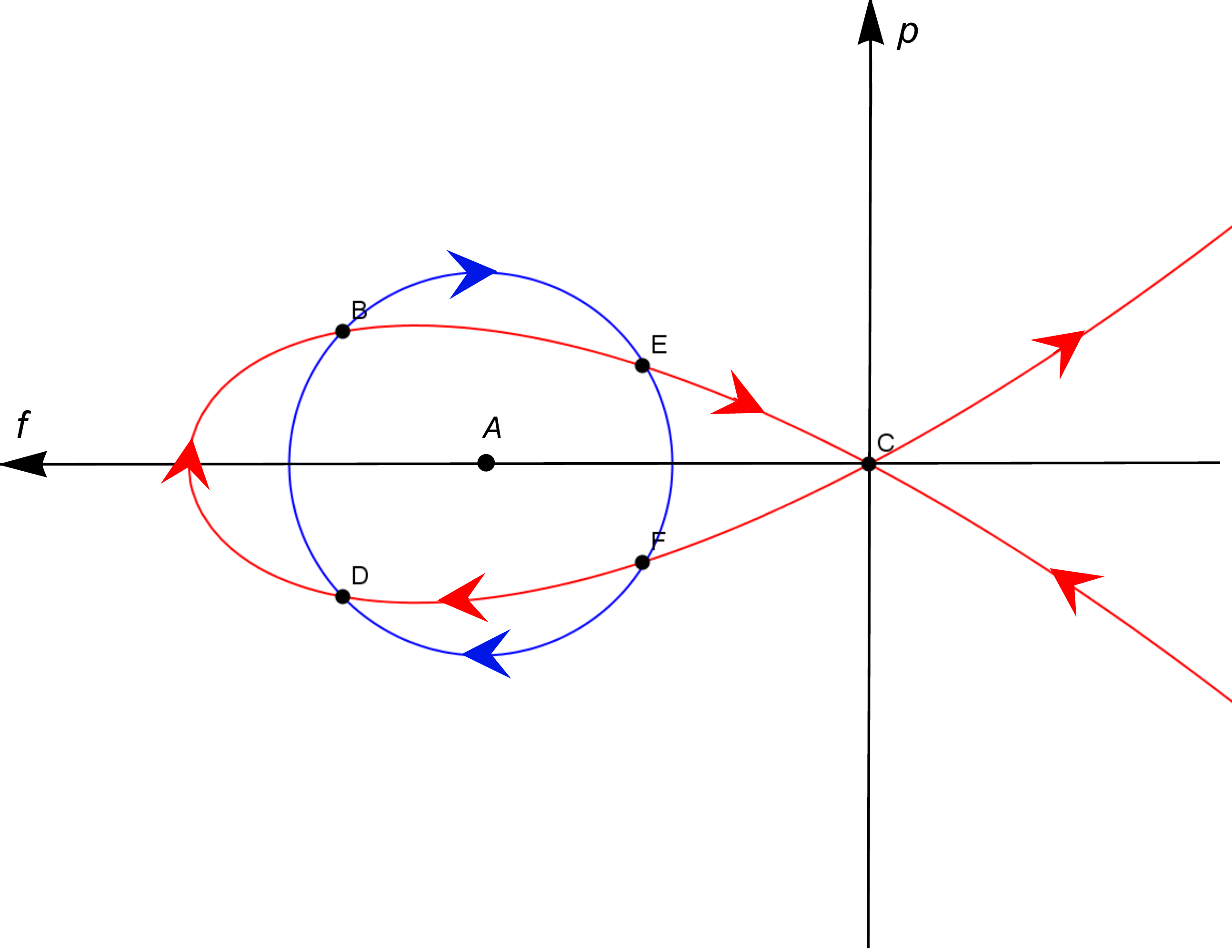}
\includegraphics[width=0.49\linewidth]{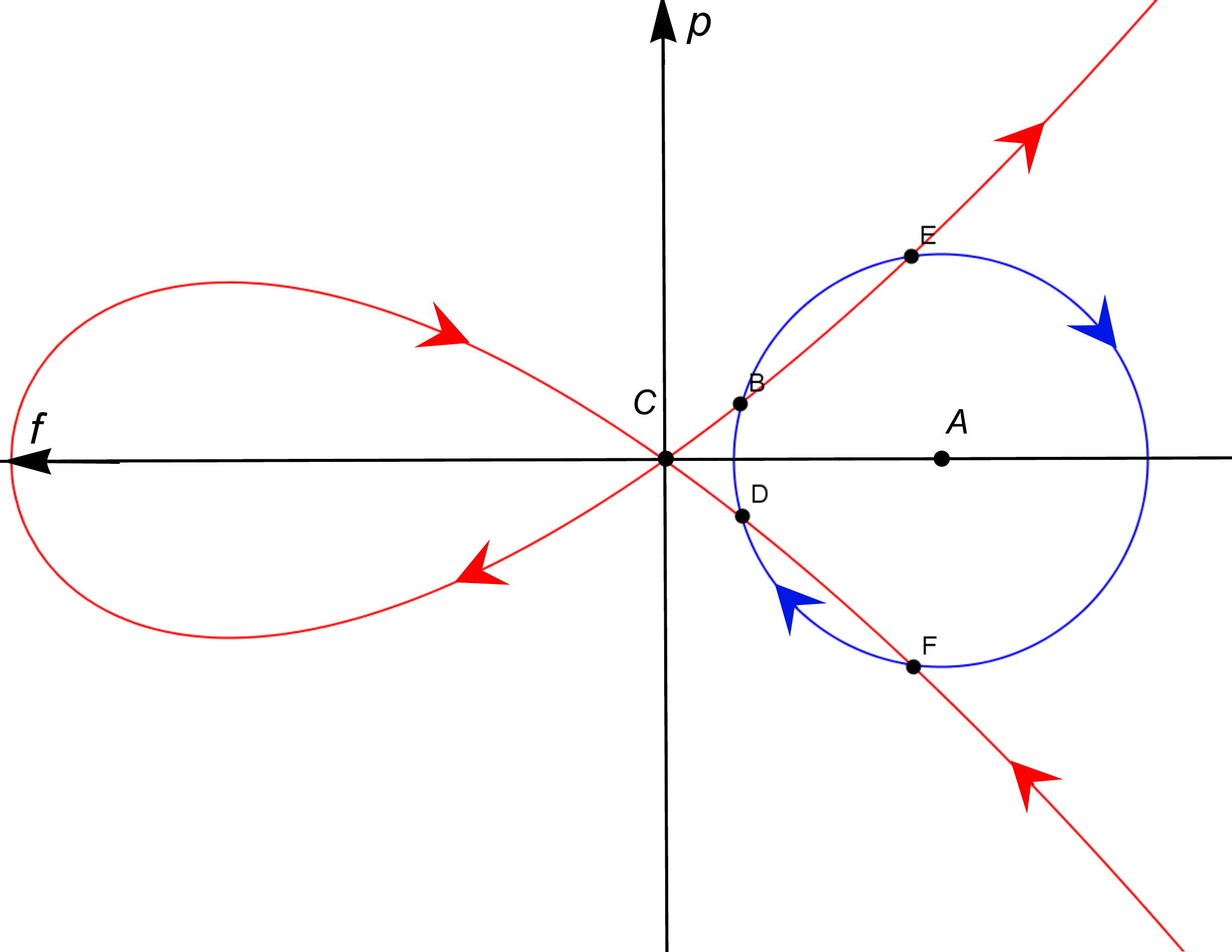}
\includegraphics[width=0.49\linewidth]{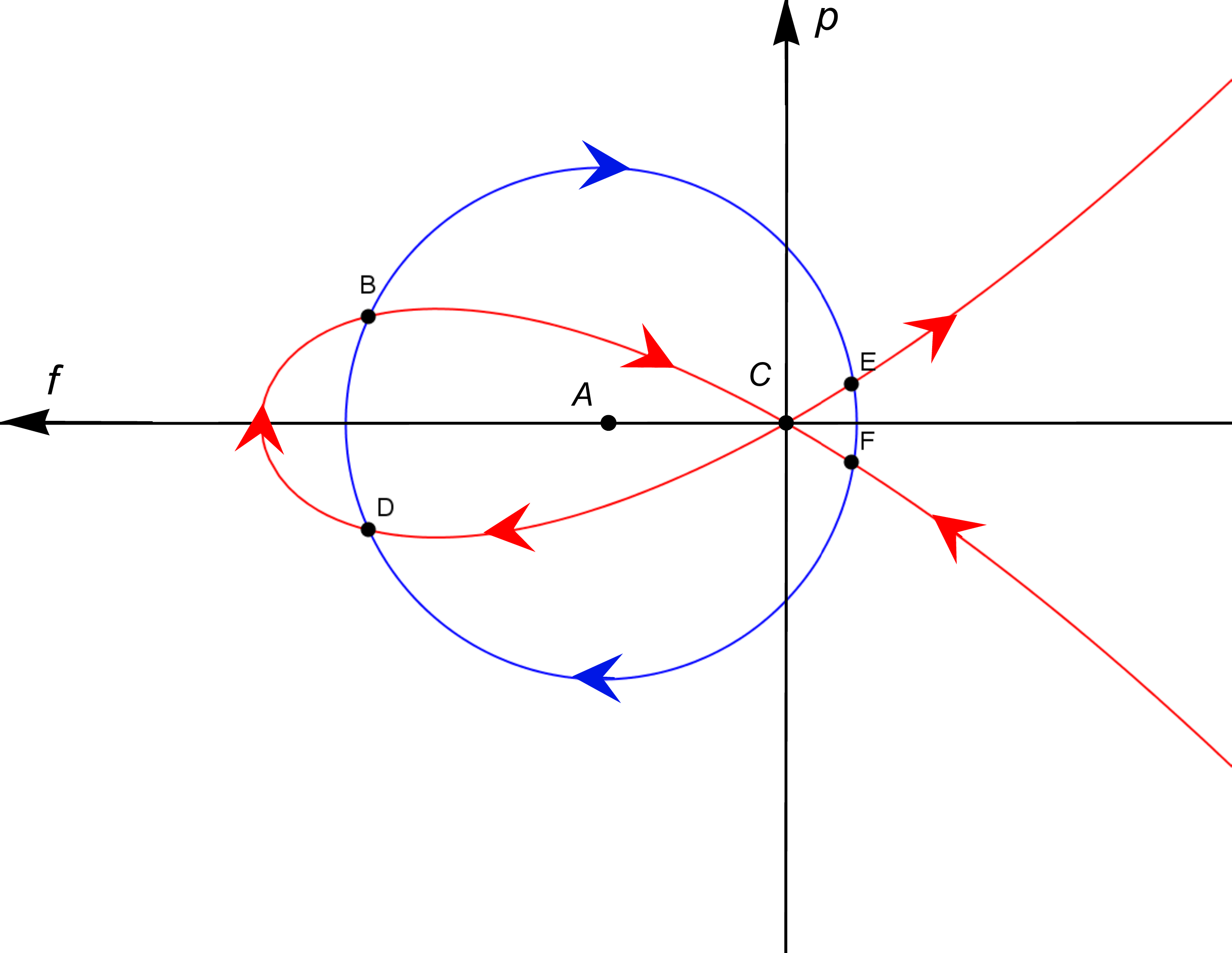}
\includegraphics[width=0.49\linewidth]{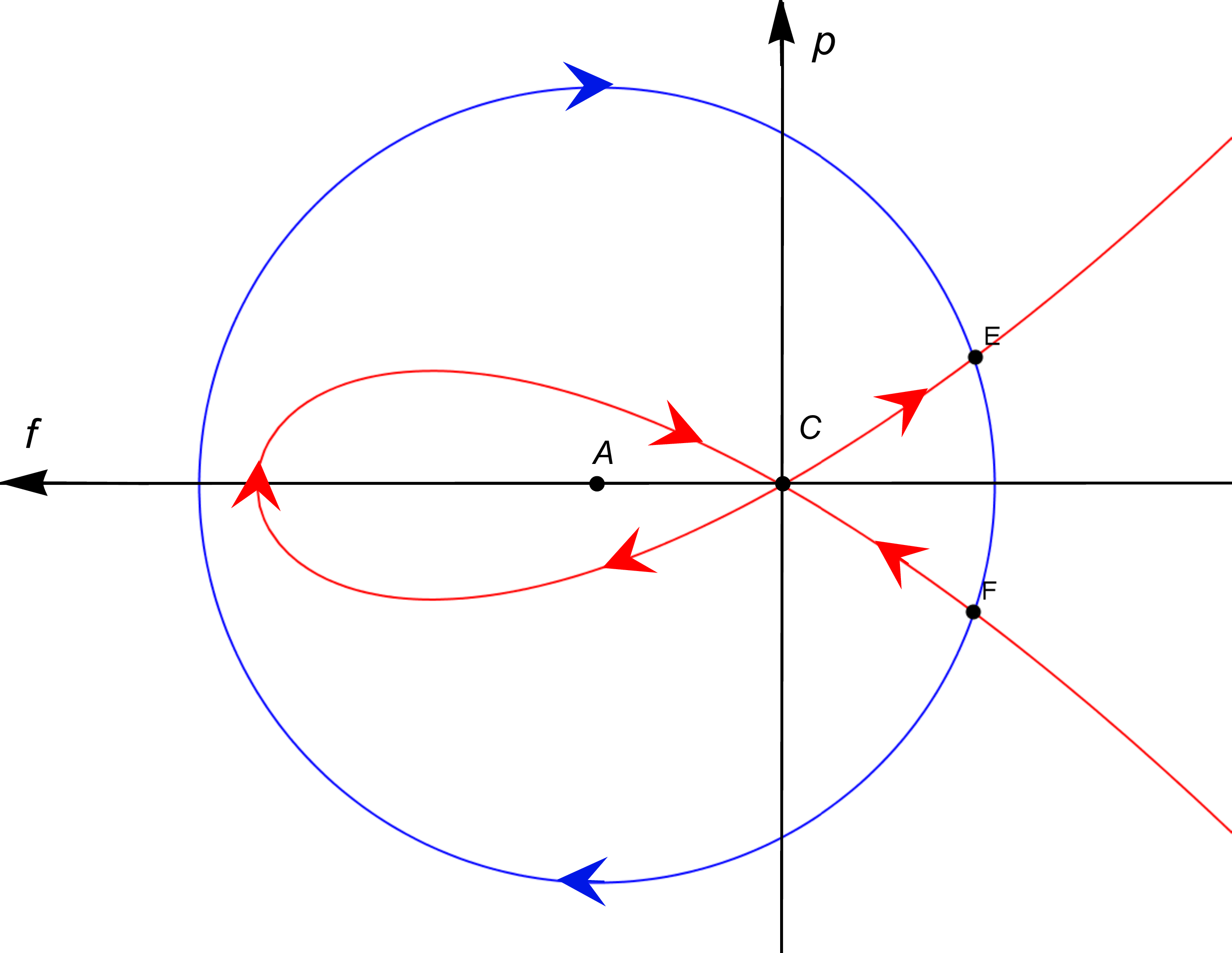}
\caption{Schematic drawings of the phase portraits $p = \pd_z f$ vs $f$ of~\eq{eq:fCapp}, restricted to $C = C_{\rm ext,max}$ in the external regions (red line) and one value of $C$ between $C_{\rm int, min}$ and $C_{\rm int, max}$ in the internal regions. Arrows give the direction in phase space for increasing values of $z$. The 4 panels show the trajectories in phase space for four different sets of parameters. The detuning manifests itself in the separation between the points $A$ (located at $f = f_{p,\rm{int}}, p=0$) and $C$ ($f = f_{b,\rm{ext}}, p=0$).}\label{fig:PP}
\end{figure}

Imposing that the solution be homogeneous and subsonic in the limits $z \to \pm \infty$ gives $C_1 = C_3 = C_{\rm max,ext}$. On the other hand, $C_2$ can be varied continuously. The phase portrait of~\eq{eq:fCapp} is shown schematically in~\fig{fig:PP} for different values of the parameters $g_{\rm int}$, $\mu_{\rm int}$, and $C_{\rm int}$. The red curve shows the trajectory in phase space $(f, p \equiv \pd_z f)$ of the solution in the external regions, while the blue one shows its trajectory in the internal region. Point $A$ corresponds to the homogeneous supersonic solution in the internal region and point $C$ to the homogeneous subsonic solution in the external regions. So, the tuned case corresponds to $A=C$. Global solutions are found by following the red line in the direction of the arrows from point $C$ to one of the intersections with the blue line ($B,D,E,F$), then to another one or the same intersection point following the blue line, and then back to $C$ following the red line. The first step corresponds to the region $I_1$, the second step to $I_2$, and the third one to $I_3$. For each solution, the length of the internal region is 
\be 
2 L = \int \frac{\dd f}{p},
\ee
the integral being evaluated over the path followed in the second step. The main difference between the tuned and detuned cases is that there is now no homogeneous solution. 
The set of solutions thus qualitatively changes for values of $C_{\rm int}$ at which the number of times the blue curve crosses the red one changes. To express these critical values, it is convenient to first define
\be 
f_{s,\rm{ext}} \equiv \sqrt{\frac{2 f_{p,\rm{ext}}^4}{f_{b,\rm{ext}}^2+f_{p,\rm{ext}}^2}}.
\ee 
$f_{s,\rm{ext}}$ is the value of $f$ at the bottom of the stationary soliton in the external regions. The first critical value of $C_{\rm int}$ is the one for which the blue line is tangent to the red one at point $E = B$. It is given by
\be 
C_{\rm{int},m} = C_{\rm{ext}} + \frac{\lp \mu_{\rm int} - \mu_{\rm ext} \rp^2}{2 \lp g_{\rm int} - g_{\rm ext} \rp}. 
\ee
$C_{\rm{int},m}$ is the minimum value of $C_{\rm int}$ for which the matching conditions, i.e., continuity of $f$ and $\pd_x f$ at $x = \pm L$, can be satisfied. (It is equal to $C_{\rm{int,min}}$ in the tuned case.) The second critical value of $C_{\rm int}$ is the one for which $E=F=C$ (for a positive detuning) or $B=D=C$ (for a negative detuning). It is given by
\be 
C_{{\rm int},0} = \frac{J^2 \lp f_{b,{\rm int}}^4 + f_{p,{\rm int}}^4 + f_{b,{\rm int}}^2 f_{p,{\rm int}}^2 \rp}{2 f_{b,{\rm int}}^4 f_{p,{\rm int}}^4} f_{b,{\rm ext}}^2 - \frac{J^2 \lp f_{b,{\rm int}}^2 + f_{p,{\rm int}}^2 \rp}{4 f_{b,{\rm int}}^4 f_{p,{\rm int}}^4} f_{b,{\rm ext}}^4 + \frac{J^2}{2 f_{b,{\rm ext}}^2}.
\ee
Another critical value is the one for which the blue and red lines are tangent at $B=D$. It is given by
\be 
C_{{\rm int},s} = \mu_{\rm ext} f_{s,\rm{ext}}^2 - \frac{g_{\rm int}}{2} f_{s,\rm{ext}}^4 + \frac{J^2}{2 f_{s,\rm{ext}}^2}.
\ee
Finally, the last critical value of $C_{\rm int}$ is $C_{\rm {int,max}}$. \Fig{fig:PP} shows the four cases $C_{\rm{int},m} < C_{\rm int} < C_{{\rm int},0}, C_{{\rm int},s}, C_{\rm{int,max}}$ (two upper panels), $C_{\rm{int},m} < C_{{\rm int},0} < C_{\rm int} < C_{{\rm int},s}, C_{\rm{int,max}}$ (bottom left panel), and $C_{{\rm int},m} < C_{{\rm int},0}, C_{{\rm int},s} < C_{\rm int} < C_{\rm{int,max}}$ (bottom right panel). 

\subsubsection{Complex-frequency modes}
\label{ABM}

We remind that the dispersion relation of perturbations in a region with uniform density $\rho$ and flow velocity $v$ is 
\begin{align*}
\lp \om - v k \rp^2 = g \rho k^2 + \frac{k^4}{4},
\end{align*} 
where $\om$ is the angular frequency and $k$ is the wave vector. 
When $\om \in \mathbb{C} - \mathbb{R}$, the four solutions in $k$ of~\eq{eq:disprel} are complex with non-vanishing imaginary parts.~\footnote{This can be seen by noticing that $\om = v k \pm \sqrt{g \rho k^2 + k^4 / 4}$ is real whenever $k \in \mathbb{R}$.}  
Two of them have a positive imaginary part while the other two have a negative imaginary part. 
This can be easily shown using the following arguments:
\begin{itemize}
\item For $\om \in \mathbb{R}$, we have either 4 real roots if $\abs{\om} \leq \om_{\rm max}$ of two real roots and two complex ones with opposite imaginary parts if $\abs{\om} > \om_{\rm max}$.
\item Half of the real roots have (strictly if $\abs{\om} \neq \om_{\rm max}$) positive group velocities while the others have negative group velocities. 
Since the group velocity is, by definition, equal to the inverse of $\partial_\om k_\om$, this implies that when adding a small imaginary part to $\om$ half of the real roots move in the upper complex half-plane while the other half move in the lower one. 
\item Since there is no real root when $\om \in \mathbb{C} - \mathbb{R}$, none of them crosses the real axis when increasing $\abs{\Im \om}$. 
\end{itemize}

Let us denote by $k_i$, $i \in \left\lbrace 1,2,3,4 \right\rbrace$ these four roots, with $\Im k_1 \leq \Im k_2 < \Im k_3 \leq \Im k_4$ and by $\phi_i$ the solution of the linearized GPE in a homogeneous background with wave vector $k_i$. In our black hole laser configuration, for $x \to -\infty$, $\phi_1$ and $\phi_2$ decay exponentially while $\phi_3$ and $\phi_4$ grow exponentially. On the other hand, for $x \to \infty$, $\phi_3$ and $\phi_4$ decay exponentially while $\phi_1$ and $\phi_2$ grow exponentially. Let $\Phi_{i,\pm}$ be the solution of the linearized GPE which is equal to $\phi_i$ for $x \to \pm \infty$. 
These define two bases of solutions at fixed angular frequency $\om$, which we now want to relate to each other.
To this end, we define the coefficients $A_{i,j}$, $(i,j) \in \left\lbrace 1,2,3,4 \right\rbrace^2$ through
\be 
\Phi_{i,-} = \sum_{j=1}^4 A_{i,j} \Phi_{j,+}.
\ee
An asymptotically bounded solution must be simultaneously a linear superposition of $\Phi_{1,-}$ and $\Phi_{2,-}$ (to be asymptotically bounded at $x \to -\infty$), and a linear superposition of $\Phi_{3,+}$ and $\Phi_{4,+}$ (to be asymptotically bounded at $x \to +\infty$). Such a solution exists if and only if 
\be \label{eq:det}
\left\lvert 
\begin{matrix}
A_{1,1} & A_{1,2} \\
A_{2,1} & A_{2,2}
\end{matrix}
\right\rvert
=0.
\ee
Our strategy is thus to look for zeros of this determinant in the complex plane. To this end, we found it convenient to compute its phase along closed lines. Indeed, this determinant is holomorphic in the complex plane, except for branch cuts which are easily identified and terminate at values of $\om$ for which the dispersion relation has a double root. Choosing a contour which does not cross a branch cut, the phase shift is equal to $2 \pi n$, where $n$ is the number of zeros of the determinant inside the contour, which can then be refined to locate them more precisely. In practice, we chose a rectangle with vertices at $\pm \om_{\rm max}$ and $\pm \om_{\rm max} + i \, \Gamma_0$, where $\om_{\rm max}$ and $\Gamma_0$ are given in \eq{eq:omMax} and \eq{eq:Gamma0} respectively. 
The frequencies of all the asymptotically bounded modes we found are well inside this contour, and we saw no evidence of complex frequencies outside it while looking at the evolution of the phase along the contour (as a nearby zero would give a rapid variation of the phase) or by extending it. When dynamical instabilities were found, the contour was then refined to locate them with an accuracy of $10 \%$ for both the real part and the imaginary part. 

Figure~\ref{fig:Soltun} shows the number of unstable modes of the 4 ``connected'' stationary solutions. We first remark that, apart from the homogeneous solution for $L < L_0$, the only connected dynamically stable solution is the first type 1 solution, i.e., the one with lowest energy. 
We checked numerically that the 5 ``non-connected'' solutions are all dynamically unstable. 
This partially proves the conjecture that was formulated in subsection~\ref{stat_sol}, namely that if the system evolves towards a stationary solution at late times, then the final state is the solution with lowest energy, i.e., the homogeneous solution for $L<L_0$ and the first type 1 solution for $L > L_0$. 
The underlying hypothesis, namely that the system generally becomes stationary at late times, is investigated in Section~\ref{Timeevolution}. The first type 3, type 4, and type 2 solutions all have a degenerate dynamical instability. We found that, in general, a nondegenerate dynamical instability appears when going from one type 1 (respectively, type 3, type 2, or type 4) solution to the next one. 
This could be expected from the results of Section~\ref{Slt}, where it was shown that the homogeneous solution gains one nondegenerate dynamical instability each time $L$ is increased by $\lambda_0 / 2$. Our present numerical calculations confirm that series of solutions which can be continuously deformed into the homogeneous one inherit these additional instabilities. The only exception we found is the second series of type 3 solutions, as for some values of $L$ two solutions of this series coexist. 
Then, as shown in the inset of~\fig{fig:Soltun}, the one with lowest energy has a degenerate and a nondegenerate dynamical instabilities, while the one with highest energy only has a nondegenerate instability. The same pattern repeats itself for the next series of type 3 solutions, with the addition of one nondegenerate dynamical instability when going from one series to the next one. 

As shown in subsection~\ref{sub:statsoldet}, a small detuning has little effect on the set of stationary solutions, except for $L \approx L_m$, $m \in \mathbb{N}$. 
Stationary solutions can thus be identified with the ones studied above. We found that the above results on linear stability continue to hold; see~\fig{fig:NLsol}. We also conjecture that they remain true for smooth variations of $g$ and $\mu$. Although such setups could in principle be examined using the method described above, we leave this to a later work.  

To end this subsection, let us compare the growth rates of the unstable modes. 
In subsection~\ref{res_ABM/QNM} it was shown that, unless $L$ is very close to one of its critical values, the most unstable ABM is the one which appears last. 
We found a similar result for the non-homogeneous solutions. Moreover, when considering an inhomogeneous solution which appears for $L = L_{m}$, $m \in \mathbb{N}$, we find that the set of complex frequencies on this solution is close to that on the homogeneous solution for $L$ slightly below $L_m$. In other words, the new series of solutions inherits the ABM that were present on the homogeneous solution for $L < L_m$. It is thus less unstable than the homogeneous solution, which has a new dynamical instability with a generally larger growth rate. This is different for series of solutions appearing at $L = L_{m + 1/2}$, as when crossing this critical value of $L$ no new unstable mode appears on the homogeneous solution. Instead, a degenerate instability is converted to a nondegenerate one, while on the type 2 solution it remains degenerate. In that case we found the growth rates have the same order of magnitude, with the growth rate of the mode on the inhomogeneous solution being in general larger than that on the homogeneous solution; see \fig{fig:Gammamax}. 

\subsection{Symplectic structure and instabilities} 
\label{sec:degvsnondeg}

In this subsection we show the key role played by the symplectic structure of the field theory when instabilities appear. This will allow us to establish the relationship between the (real) energies of field configurations and the (complex) frequencies of dynamically unstable modes (DIM) in a more general framework. In particular, we shall prove that, for systems described by a discrete number of degrees of freedom, DIM and energetic instabilities (EI) occur for the same value of the parameters describing the external potential, as shown in this chapter for the black hole laser. 
The symplectic structure will also allow us to explain why there exist two qualitatively different types of DIM, which we call degenerate and nondegenerate. Nondegenerate DIM, appearing in the ``second step'' of Section~\ref{Slt} (or the ``third step'' when $c_3 \neq c_1$), involve one \textit{complex} degree of freedom. Their frequencies in general have nonvanishing real and imaginary frequencies. 
A degenerate DIM instead, as the Gregory-Laflamme instability \cite{Gregory93,Gregory:2011kh} or the one appearing after the ``first step'', involves a \textit{real} degree of freedom and has a purely imaginary frequency. 
To show these links, it is useful to adopt a Hamiltonian description.

\subsubsection{Symplectic structure, EI, and DIM} 

The model we use is based on restriction of the space of solutions of the linear equation to a finite-dimensional sub-space of modes. A restricted Hamiltonian can then be defined to determine the time-evolution of their coefficients. To be more precise, we consider solutions of a scalar, linear field equation in the form 
\be \label{eq:fewmodes}
\phi(x,t) = \sum_{i = 1}^N q_i(t) \, \phi_i(x), \; N \in \mathbb{N}^*,
\ee 
where $\phi_i$ are known orthogonal and normalized (in the sense of the $L^2$ scalar product) functions. Our goal is to study the evolution of the coefficients $q_i$ when taking into account only the interactions between the $N$ modes appearing in \eq{eq:fewmodes}. 

In the following, we work with {\it real} time-dependent coefficients. (This analysis extends to the case of complex ones after decomposing them into real and imaginary parts, see subsection~\ref{sub:GeneHam}.) We then use the fact that the field equation under study has a canonical Hamiltonian structure. Plugging \eq{eq:fewmodes} into the expression of the Lagrangian $\mathcal{L}$, one can define the conjugate moments $p_i$ of the $q_i$ by
\begin{align*}
p_i \equiv \frac{\partial \mathcal{L}}{\partial \dot{q}_i},
\end{align*} 
satisfying the standard Poisson bracket relations $\left\lbrace q_i, q_j \right\rbrace = 0$, $\left\lbrace p_i, p_j \right\rbrace = 0$, and $\left\lbrace q_i, p_j \right\rbrace = \delta_{i j}$. 

Let us first consider the case with two real degrees of freedom $q_1, q_2$. We define the vector 
\be \label{eq:X} 
X \equiv 
\left(
\begin{array}{cc}
q_1 \\
q_2 \\
p_1 \\
p_2
\end{array}
\right).
\ee
As the field equation is linear, the Hamiltonian is quadratic in $(q_1,q_2,p_1,p_2)$:
\be 
H = \frac{1}{2} X^T M_H X,
\label{H1}
\ee
where the hamiltonian matrix $M_H$ is real and symmetric. 
($H$ is the energy of the mode, not to be confused with the time-translation operator which we denote by $H_S$.) 
Let us write the Hamilton equations 
\begin{align*}
\def\arraystretch{2.}
\left\lbrace
\begin{array}{l}
 \dfrac{\dd q_i}{\dd t}=\dfrac{\partial H}{\partial p_i}  \\
 \dfrac{\dd p_i}{\dd t}=-\dfrac{\partial H}{\partial q_i}  
\end{array}
\right. 
\end{align*}
in the matrix form 
\be 
\ii \frac{\dd}{\dd t}\left(
\begin{array}{c}
 q \\
 p
\end{array}
\right) = \ii \mathbb{J} \, M_H \left(
\begin{array}{c}
 q \\
 p
\end{array}
\right)= H_S
\lp
\begin{array}{c}
 q \\
 p
\end{array}
\rp .
\label{Schreq}
\ee
In this equation
\be 
\mathbb{J} \equiv \left( \begin{array}{cc}
 0 & \mathbf{1} \\
 -\mathbf{1} & 0
\end{array} \right), 
\ee
is the symplectic matrix and $\mathbf{1}$ is the $2$ by $2$ identity matrix. 
Indeed, the Poisson brackets between $q_i$ and $p_j$ are all encoded in 
\be \label{eq:scalXY}
\left\lbrace X,Y \right\rbrace \equiv X^T \, \mathbb{J} \, Y.
\ee
Because $\mathbb{J}^2 = - 1$, the matrix $H_S$ of \eq{Schreq},
\be \label{eq:HEtoHS}
H_S \equiv \ii \mathbb{J} M_H,
\ee 
gives back the Hamiltonian $H$ of \eq{H1} when using the symplectic scalar product of \eq{eq:scalXY}:
\be 
H = \frac{\ii}{2} \, \left\lbrace X,H_S \, X \right\rbrace .
\ee  
This identity is the restriction in the $(q_1,q_2)$ subspace of the relation obeyed by a complex scalar field:  
\be
H[\phi] = (\phi, \ii \pd_t \phi), 
\ee
between the field Hamiltonian $H[\phi]$ and the scalar product in a quadratic field theory~\cite{Coutant:2016bgk}. 

\begin{figure}[ht]
\centering
\includegraphics[width=0.49 \linewidth]{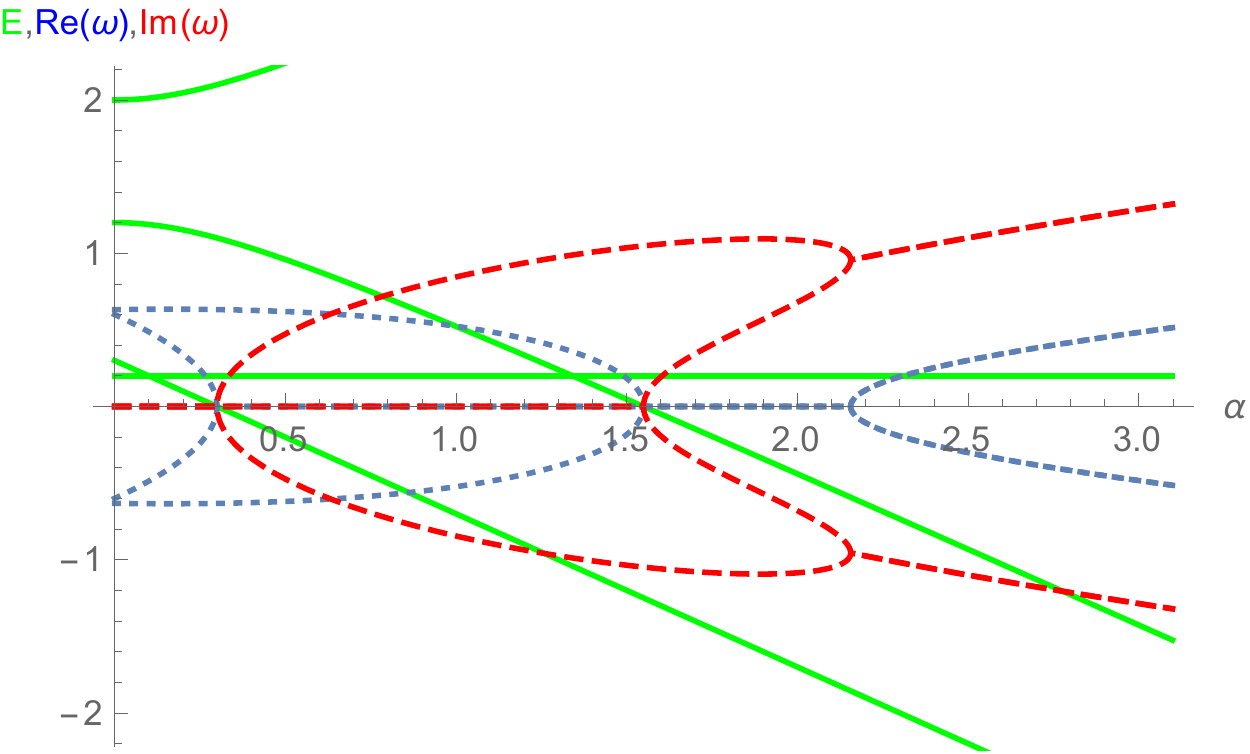} 
\includegraphics[width=0.49 \linewidth]{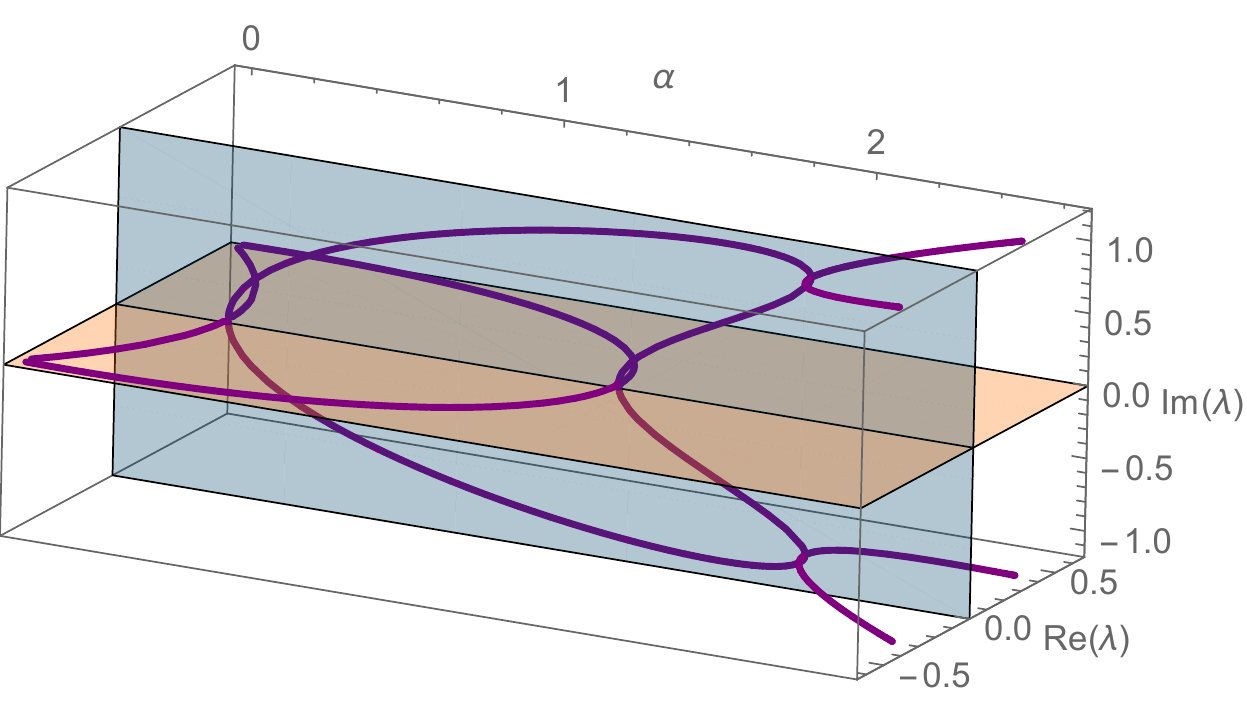} 
\caption{Left panel: The green, continuous lines represent the four eigenvalues of the Hamiltonian matrix $M_H$ (see \ref{eq:Hqq}), with $E_{q_1} = 2.0, E_{q_2} = 1.2$, $E_{p_1} = 0.2$, and $E_{p_2} = 0.3-\alpha$. These real numbers give the energy of the corresponding (real) field configurations. The blue, dotted (red, dashed) lines give the real (imaginary) parts of the four eigenvalues of $H_S$. (They are multiplied by 2 for an easier visualization.) 
These complex numbers give the eigen-frequencies of the complex modes $\propto \e^{- \ii \om t}$. We observe that two degenerate DIM appear with the two energetic instabilities at $\alpha \approx 0.3$ and $\alpha \approx 1.55$. They merge to give a nondegenerate DIM for $\alpha \approx 2.16$. 
Right panel: We show the 4 eigenvalues of $H_S$ in the complex plane as functions of $\alpha$. The orange plane corresponds to $\Im \la = 0$ and the blue one to $\Re \la = 0$. The 4 purple lines show the positions of the 4 eigenvalues when varying $\alpha$ from $0$ to $2.6$.  
}\label{Fig:omBHL}
\end{figure}

Since $M_H$ is a symmetric real matrix, its eigenvalues are necessarily real. 
A negative eigenvalue of $M_H$ indicates that there is an energetic instability. In that case the energy of the system is not bounded from below (in the quadratic approximation). Interestingly, the eigenvalues of $H_S$ may be complex. Since in full generality the spectrum is invariant under complex-conjugation, complex eigenvalues only arise in pairs. (In the present settings, it can be traced to the fact that $H_S$ is hermitian for the symplectic scalar product \eq{eq:scalXY}.) 
Each of them is associated with a DIM. 
In addition, since we work with real degrees of freedom $q_i$, the spectrum of $H_S$ is also invariant under $\la \to -\la^*$ from the invariance of the evolution equation under the (formal) transformation $(q_i, p_i) \to (q_i^*, p_i^*)$. (Another way to see this is to note that $H_S^* = -H_S$.) A distinction can then be made between two different types of DIM: 
\begin{itemize}
\item If only two eigenvalues of $H_S$ are complex, we speak of a degenerate DIM. In that case the two complex eigenvalues of $H_S$ are purely imaginary and opposite to each other. The associated subsystem contains one real degree of freedom: an 
upside-down harmonic oscillator. 
\item If the four eigenvalues of $H_S$ are complex, we speak of a nondegenerate DIM. These four eigenvalues are then $\la, \la^*, -\la$, and $-\la^*$ for some $\lambda \in \mathbb{C}$. The associated subsystem contains one complex degree of freedom: a rotating 
upside-down harmonic oscillator, where $\Re \la$ gives the angular velocity, and $\Im \la$ gives the growth rate.
\end{itemize}
We can now discuss the general relationships between the eigenvalues of $H_S$ and those of $M_H$. 
\begin{itemize}
\item First, since the energy $H$ of \eq{H1} is conserved in time, a DIM necessarily has a vanishing energy. A dynamical instability thus requires at least one energetic instability. More precisely, it requires that $M_H$ has two eigenvalues with opposite signs. When all the eigenvalues of $M_H$ have the same sign, there is thus no dynamical instability and all the eigenvalues of $H_S$ are real. 
\item One degenerate dynamical instability is either created or erased each time one eigenvalue of $M_H$ changes sign 
(this is shown more generally in the next subsection). So, when $M_H$ has three strictly positive eigenvalues and one strictly negative one (or conversely), $H_S$ has two complex-conjugate, purely imaginary eigenvalues, while its two other eigenvalues remain real. 
\item When a second energetic instability turns in, i.e., when $M_H$ has two positive and two negative eigenvalues, either the previous dynamical instability is erased or a new one arises. In the first case, the four eigenvalues of $H_S$ are again real. In the second case, they are divided into two sets of complex-conjugate, purely imaginary frequencies $\pm i \Gamma_1$ and $\pm i \Gamma_2$. 
\item Eventually $\Gamma_1$ and $\Gamma_2$ may merge, giving rise to a quartet $\la, \la^*, -\la, -\la^*$ of complex eigenfrequencies. 
\end{itemize}
To illustrate this, we consider a simple coupling between the two modes, described by a matrix $M_H$ of the form
\be \label{eq:Hqq}
M_H=\begin{pmatrix}
 E_{q_1} & \alpha  & 0 & 0 \\
 \alpha  & E_{q_2} & 0 & 0 \\
 0 & 0 & E_{p_1} & 0 \\
 0 & 0 & 0 & E_{p_2}
\end{pmatrix}.
\ee
In \fig{Fig:omBHL}, the evolution of eigenvalues of $M_H$ and those of $H_S$ are shown for increasing values of the off-diagonal term $\alpha$ in the case where the last eigenvalue follows $E_{p_2}=E_{p_2}^0 - \alpha $. 

For $\alpha = 0$ all eigen-values $E_i$ are taken positive. The system is thus stable. At a first critical value of $\alpha$, close to $0.3$ in the present case, an energetic instability appears as one eigenvalue of $M$ becomes negative. Simultaneously, two real eigenvalues of $H_S$ merge and become purely imaginary. 
When  $\alpha$ reaches $1.55$, a second  energetic instability arises as another eigenvalue of $M$ becomes negative. At that value, the other two real eigenvalues of $H_S$ merge and become purely imaginary. Finally, at a third critical value of $\alpha$ close to $2.16$, the two imaginary eigen-frequencies merge, giving a nondegenerate DIM. This third step is not associated as the appearance/disappearance of an energetic instability. 

This simple model thus captures all essential aspects of the ``three-step process'' observed in Section~\ref{Slt} when $c_3 \neq c_1$. 
(In the case $c_1 = c_3$ the two last steps occur for the same value of $L$ as $\lambda$ vanishes exactly when the second QNM frequency reaches the real axis.) 
We now understand that this sequence comes from two properties of the mode equation: its symplectic structure and the description using real degrees of freedom. For the Bogoliubov-de Gennes equation, the first property directly comes from the Lagrangian formulation, while the second one can be traced back to the symmetry of the spectrum under $\la \to -\la^*$ (see the last paragraph of subsection~\ref{sub:GeneHam}).  

\subsubsection{Generalizations}
\label{sub:GeneHam}

This analysis can be straightforwardly generalized to a larger number $N$ of degrees of freedom, and to an arbitrary symmetric matrix $M_H$. Then, 
\be 
H_S = 2 \ii \mathbb{J} M_H,
\ee
where 
\be 
\mathbb{J}=\left(
\begin{array}{cc}
0 & \textbf{1}_N \\
-\textbf{1}_N & 0
\end{array}
\right),
\ee
and $\textbf{1}_N $ is the $N$ by $N$ identity matrix. 
As in the previous case, an energetic instability corresponds to a negative eigenvalue of $M$, a degenerate DIM to a pair $(\la, \la^*)$ of purely imaginary eigenvalues of $H_s$, and a nondegenerate DIM to a quartet $(\la, \la^*, -\la, -\la^*)$ of eigenvalues of $H_s$. Moreover, the number of degenerate DIM must change each time a new energetic instability appears since the numbers of DIM and negative-energy modes have the same parity. Indeed, a degenerate DIM corresponds to an imaginary eigenvalue of $H_S$ with a positive imaginary part, i.e., to a positive eigenvalue of $\mathbb{J} M_H$. On the other hand, a negative-energy mode corresponds to a negative eigenvalue of $M_H$. Since the characteristic polynomial of $\mathbb{J} M_H$ is of even order $2N$, it as an even number of real roots. Among them, there are an even number of positive roots if $\det (\mathbb{J} M_H) > 0$ and an odd number of them if $\det (\mathbb{J} M_H) < 0$, as $\det \lp \mathbb{J} M_H -\lambda \rp$ goes to $+ \infty$ for $\la \to \pm \infty$. Similarly, there are an even number of energetic instabilities if $\det M_H > 0$ and an odd number of them if $\det M_H < 0$. Noticing that $\det (\mathbb{J} M_H) = \det M_H$ since $\det \mathbb{J} = 1$, we obtain the above result. In particular, if there is only one energetic instability, there is one degenerate DIM.~\footnote{An alternative derivation of this statement is to note that only dynamically unstable modes contribute negatively to the energy, see Eq.~(15) in \cite{Coutant:2009cu}.} To obtain a nondegenerate DIM requires (at least) a second energetic instability. 

The above naturally extends to complex degrees of freedom, as in the electric model of~\cite{Coutant:2016bgk} and the appendix of~\cite{Fullingbook}, after decomposing each of them into a pair of real ones. Doing this multiplies the sizes of $M_H$ and $H_S$ by two. So, two eigenvalues of $H_S$ correspond to a single eigenfrequency for the initial system. To identify these two frequencies, we note that $H_S$ expressed in terms of real degrees of freedom has a symmetry under complex conjugation sending  $\la$ to $-\la^*$,  which was not present in the initial model. So, two eigenfrequencies $\la$, $-\la^*$ in general correspond to the single eigenfrequency $\la$ for the initial model. The quartet $(\la, \la^*, -\la, -\la^*)$ associated with a nondegenerate DIM thus becomes a doublet $(\la, \la^*)$. A degenerate DIM, which involves only two real degrees of freedom, will not appear in general, except if the modes involved are described by real degrees of freedom, as in the case studied in Section~\ref{Slt} for modes with a purely imaginary frequency. 

\subsection{Quasi-quasinormal modes (Work in progress)}

\begin{figure}
\centering
\begin{tikzpicture}
\newcommand{\xa}{0.75}
\draw (0,0) -- (\xa,0);
\draw (2.*\xa,0) -- (3.*\xa,0);
\draw (4.*\xa,0) -- (5.*\xa,0);
\draw[dashed] (5.1*\xa,0) -- (5.9*\xa,0);
\draw (6.*\xa,0) -- (7.*\xa,0);
\draw (8.*\xa,0) -- (9.*\xa,0);
\draw (1.5*\xa,0) circle (0.5*\xa);
\draw (3.5*\xa,0) circle (0.5*\xa);
\draw (7.5*\xa,0) circle (0.5*\xa);
\draw (1.5*\xa,0) node{$S_1$};
\draw (3.5*\xa,0) node{$S_2$};
\draw (7.5*\xa,0) node{$S_n$};
\draw (0,0.25*\xa) -- (0,-0.25*\xa) node[below left]{$S_0 = B_L$};
\draw (9*\xa,0.25*\xa) -- (9*\xa,-0.25*\xa) node[below right]{$S_{n+1} = B_R$};
\draw[dashed,blue] (0.5*\xa,-0.5*\xa) -- (0.5*\xa,0.5*\xa) ;
\draw[dashed,blue] (2.5*\xa,-0.5*\xa) -- (2.5*\xa,0.5*\xa) ;
\draw[dashed,blue] (4.5*\xa,-0.5*\xa) -- (4.5*\xa,0.5*\xa) ;
\draw[dashed,blue] (6.5*\xa,-0.5*\xa) -- (6.5*\xa,0.5*\xa) ;
\draw[dashed,blue] (8.5*\xa,-0.5*\xa) -- (8.5*\xa,0.5*\xa) ;
\draw[blue,<->] (0.5*\xa,-0.9*\xa) -- (2.5*\xa,-0.9*\xa) node[midway,below]{$I_1$};
\draw[blue,<->] (2.5*\xa,-0.9*\xa) -- (4.5*\xa,-0.9*\xa) node[midway,below]{$I_2$};
\draw[blue,<->] (6.5*\xa,-0.9*\xa) -- (8.5*\xa,-0.9*\xa) node[midway,below]{$I_n$};
\end{tikzpicture}
\caption{Schematic representation of a scattering problem in $1+1$ dimensions with $n$ scattering regions ($n \in \mathbb{N}^*$), denoted by $S_i$ for $i \in [ \! [ 1, n ] \! ]$. Straight horizontal lines are regions where no scattering occurs. QQNM can be defined in each region $I_i$, $i \in [ \! [ 1, n ] \! ]$. The QNM are defined, if at all, only for the ``full'' scattering problem including the $n$ scattering regions and the boundaries $B_L$ and $B_R$ (which can be at a finite distance from the origin or at infinity).} \label{fig:saturation_n_scat_reg}
\end{figure}
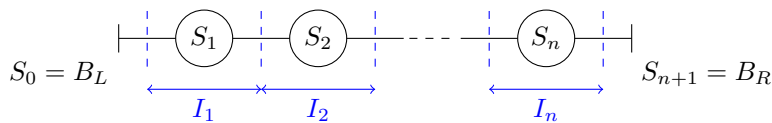

\begin{figure}
\centering
\begin{tikzpicture}
\newcommand{\xa}{0.75}
\draw[dashed] node[left]{$t=0: \;$} (0,0) -- (\xa,0);
\draw (2.*\xa,0) -- (3.*\xa,0);
\draw (4.*\xa,0) -- (5.*\xa,0);
\draw[dashed] (6.*\xa,0) -- (7.*\xa,0) node[right]{\phantom{$t=0: \;$}};
\draw (1.5*\xa,0) circle (0.5*\xa);
\draw (3.5*\xa,0) circle (0.5*\xa);
\draw (5.5*\xa,0) circle (0.5*\xa);
\draw (1.5*\xa,0) node{$S_{i-1}$};
\draw (3.5*\xa,0) node{$S_i$};
\draw (5.5*\xa,0) node{$S_{i+1}$};
\draw[samples=100,scale=1.,domain=2.5*\xa:4.5*\xa,smooth,variable=\x,blue] plot ({\x},{0.5*exp(-20.*(\x-3.5*\xa)^2)*sin(4000*\x)});
\end{tikzpicture} \\ \vspace*{0.5 cm}
\begin{tikzpicture}
\newcommand{\xa}{0.75}
\draw[dashed] node[left]{$t_{\textrm{dec},i} \ll t \ll t_{\textrm{out},i}: \;$} (0,0) -- (\xa,0);
\draw (2.*\xa,0) -- (3.*\xa,0);
\draw (4.*\xa,0) -- (5.*\xa,0);
\draw[dashed] (6.*\xa,0) -- (7.*\xa,0) node[right]{\phantom{$t_{\textrm{dec},i} \ll t \ll t_{\textrm{out},i}: \;$}};
\draw (1.5*\xa,0) circle (0.5*\xa);
\draw (3.5*\xa,0) circle (0.5*\xa);
\draw (5.5*\xa,0) circle (0.5*\xa);
\draw (1.5*\xa,0) node{$S_{i-1}$};
\draw (3.5*\xa,0) node{$S_i$};
\draw (5.5*\xa,0) node{$S_{i+1}$};
\draw[samples=100,scale=1.,domain=2.*\xa:5.*\xa,smooth,variable=\x,blue] plot ({\x},{0.05*exp(-2.5*(\x-3.5*\xa)^2)*sin(4000*\x)+0.5*exp(-80.*(\x-2.4*\xa)^2)*sin(4000*\x)+0.25*exp(-80.*(\x-4.6*\xa)^2)*sin(4000*\x)});
\end{tikzpicture} \\ \vspace*{0.5 cm}
\begin{tikzpicture}
\newcommand{\xa}{0.75}
\draw[dashed] node[left]{$t \gtrsim t_{\textrm{out},i}: \;$} (0,0) -- (\xa,0);
\draw (2.*\xa,0) -- (3.*\xa,0);
\draw (4.*\xa,0) -- (5.*\xa,0);
\draw[dashed] (6.*\xa,0) -- (7.*\xa,0) node[right]{\phantom{$t \gtrsim t_{\textrm{out},i}: \;$}};
\draw (1.5*\xa,0) circle (0.5*\xa);
\draw (3.5*\xa,0) circle (0.5*\xa);
\draw (5.5*\xa,0) circle (0.5*\xa);
\draw (1.5*\xa,0) node{$S_{i-1}$};
\draw (3.5*\xa,0) node{$S_i$};
\draw (5.5*\xa,0) node{$S_{i+1}$};
\draw[samples=200,scale=1.,domain=1.*\xa:6.*\xa,smooth,variable=\x,/pgf/fpu,/pgf/fpu/output format=fixed,blue] plot ({\x},{0.06*exp(-2.5*(\x-3.5*\xa)^2)*sin(4000*\x)+0.4*exp(-80.*(\x-1.5*\xa)^2)*sin(4000*\x)+0.2*exp(-80.*(\x-5.3*\xa)^2)*sin(4000*\x) + 0.2*exp(-10*(\x-2.2*\xa)^2)*sin(2000*\x) + 0.15*exp(-10*(\x-4.8*\xa)^2)*sin(8000*\x)});
\end{tikzpicture} \\ \vspace*{0.5 cm}
\begin{tikzpicture}
\newcommand{\xa}{0.75}
\draw[dashed] node[left]{$t \gg t_{\textrm{QNM}}: \;$} (0,0) -- (\xa,0);
\draw (2.*\xa,0) -- (3.*\xa,0);
\draw (4.*\xa,0) -- (5.*\xa,0);
\draw[dashed] (6.*\xa,0) -- (7.*\xa,0) node[right]{\phantom{$t \gg t_{\textrm{QNM}}: \;$}};
\draw (1.5*\xa,0) circle (0.5*\xa);
\draw (3.5*\xa,0) circle (0.5*\xa);
\draw (5.5*\xa,0) circle (0.5*\xa);
\draw (1.5*\xa,0) node{$S_{i-1}$};
\draw (3.5*\xa,0) node{$S_i$};
\draw (5.5*\xa,0) node{$S_{i+1}$};
\draw[samples=200,scale=1.,domain=0.*\xa:7.*\xa,smooth,variable=\x,/pgf/fpu,/pgf/fpu/output format=fixed,blue] plot ({\x},{0.05*exp(+0.125*(\x-3.5*\xa)^2)*sin(1500*\x))});
\end{tikzpicture} 
\caption{Schematic representation of the evolution of a perturbation initially localized in the region $I_i$ for some $i \in [ \! [ 1, n ] \! ]$. In $I_i$, the dynamics is dominated by QQNM in the second plot, and by QNM in the fourth one.
(This last step does not occur in a finite system without dissipation.)} \label{fig:saturation_int_times_QQNM}
\end{figure}
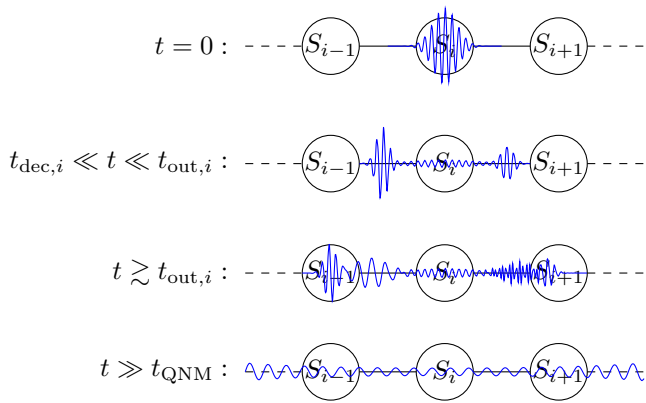

Let us take the opportunity to introduce the notion of ``quasi-quasinormal modes''. 
As seen above and discussed in more details in~\cite{Coutant:2016bgk} (see also Chapter~\ref{ch:concl}, Section~\ref{sec:DIQNM}), quasinormal modes (QNM) are solutions of a linear field equation which decay in time but are generally not spatially bounded. 
As such, they strongly depend on the boundary conditions and can be drastically modified when adding well-separated scattering regions. 
However, one can reasonably expect that the dynamics will \textit{not} be significantly affected by these changes over moderate time scales which are long enough to enter the regime of exponential decay but short enough that initially localized perturbations have not reached the additional scattering regions or the asymptotic ones. 
We now explain how a ``local'' definition of QNM-like solutions can be used to identify the main properties of the evolution over such ``moderate'' time scales. 
Our aim is not to give a precise or rigorous definition, but rather to convey the qualitative idea, which can be used in practical problems to determine the decay rates of perturbations. 
Although counter-examples to the scenarios detailed below probably exist (they can be constructed, for instance, by fine-tuning the initial conditions so that the amplitudes of QQNM vanish), we believe it is sufficiently general to be useful. 

We consider a general scattering problem in $1+1$ dimensions which can be divided into $n$ spatially separated and localized scattering regions (for some $n \in \mathbb{N}^*$) and two boundary conditions, see \fig{fig:saturation_n_scat_reg}. 
In the figure, the horizontal lines represent regions where the WKB modes propagate without being mixed with each others. 
The $S_i$, $i \in [\![1,n]\!]$, denote the $n$ scattering regions, and $B_L$ and $B_R$ denote the boundaries of the system (possibly at infinity; they then correspond to asymptotic conditions). 
We define an interval $I_i$ around each scattering region $S_i$. 
We assume there is no instability, and we denote by $t$ the time coordinate. 
Let us consider a perturbation localized at $t = 0$ in $I_i$ for some $i \in [ \! [ 1, n ] \! ]$, see \fig{fig:saturation_int_times_QQNM}. 
Intuitively, if the regions are sufficiently well separated from each others and from the boundaries, the perturbation will, in some time interval, decay following the QNM defined taking only $S_i$ into account. 
To be slightly more precise, let us define $t_{\textrm{dec},i}$, the typical time needed to enter the regime dominated by QNM for the problem defined with $S_i$ only, and $t_{\textrm{out},i}$, the typical time needed for a perturbation to move out of $I_i$. 
If $t_{\textrm{out},i} \gg t_{\textrm{dec},i}$, then the dynamics around $S_i$ will be dominated by QNM defined over $S_i$ only for $t_{\textrm{dec},i} \ll t \ll t_{\textrm{out},i}$, see the second plot from the top. 
Indeed, the perturbation having not yet reached the other scattering regions, the latter have no effect on the dynamics. 
We call these modes, which would be true QNM if there was only one scattering region and the boundary conditions were of the outgoing type, ``quasi-quasinormal modes'' (QQNM). 
In particular, in $I_i$ the perturbation will decay with the decay rate of QQNM associated with $S_i$.
For times of the order of or longer than $t_{\textrm{out},i}$, the evolution is more complicate since the perturbations scattered on $S_{i \pm 1}$ are, in general, partially reflected towards $S_i$, see the third plot in the figure. 
For still longer times, longer than the typical time $t_\textrm{QNM} > t_{\textrm{out},i}$ when the QNM of the full problem come into play (assuming they exist), the latter will determine the evolution. 
One thus generally expects a transition between (at least) two exponentially-decaying behaviors: one for ``intermediate'' times $t_{\textrm{dec},i} \ll t \ll t_{\textrm{out},i}$ with a decay rate given by the QQNM frequencies, and one for ``late'' times $t \gg t_\textrm{QNM}$ with a decay rate given by the QNM frequencies. 
If the boundaries are at a finite distance from the origin and there is no dissipation nor absorption, the late-time dynamics can be more complicate. But the QQNM are still expected to dominate at ``intermediate'' times.

\begin{figure}
\centering
\includegraphics[width = 0.49\linewidth]{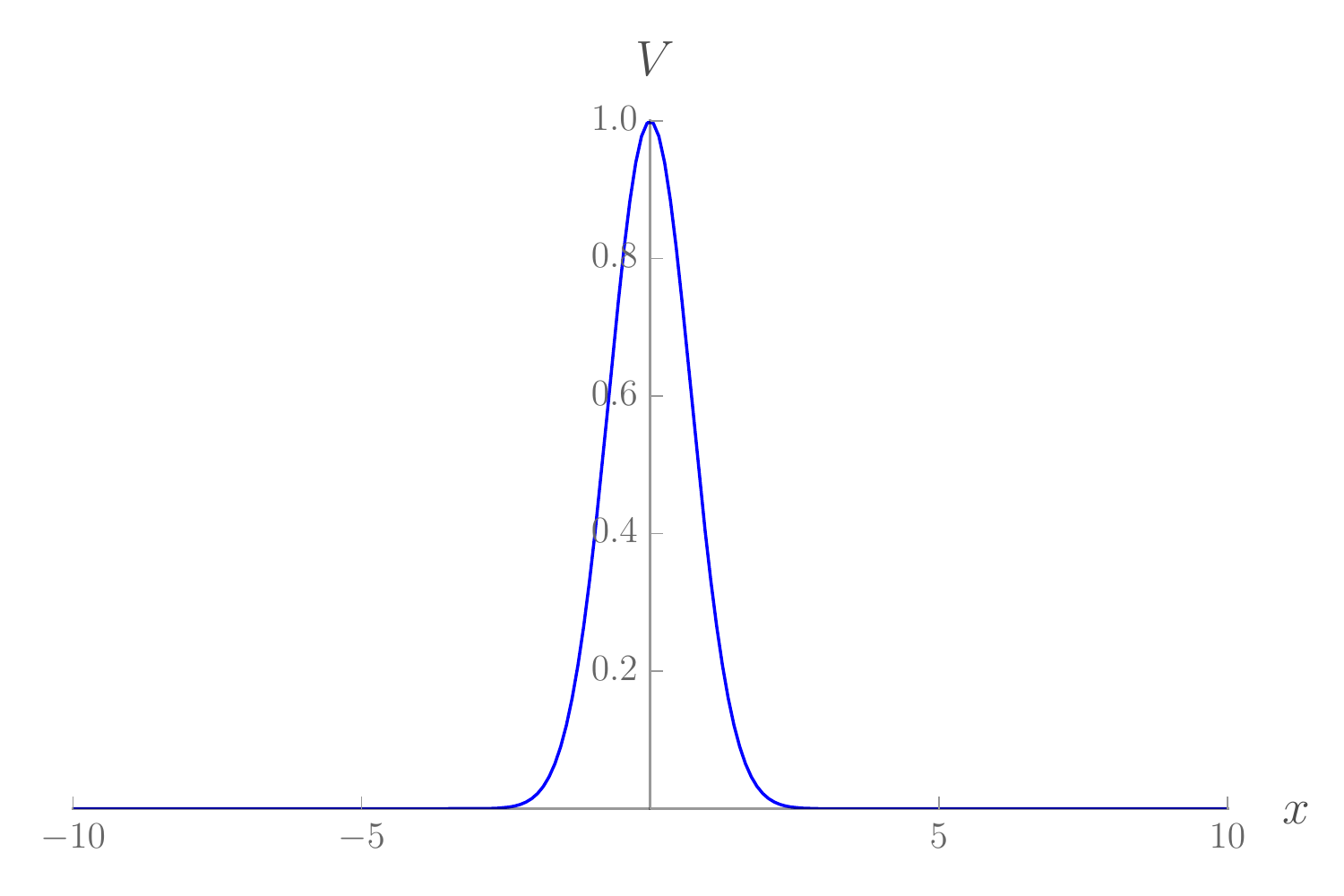}
\includegraphics[width = 0.49\linewidth]{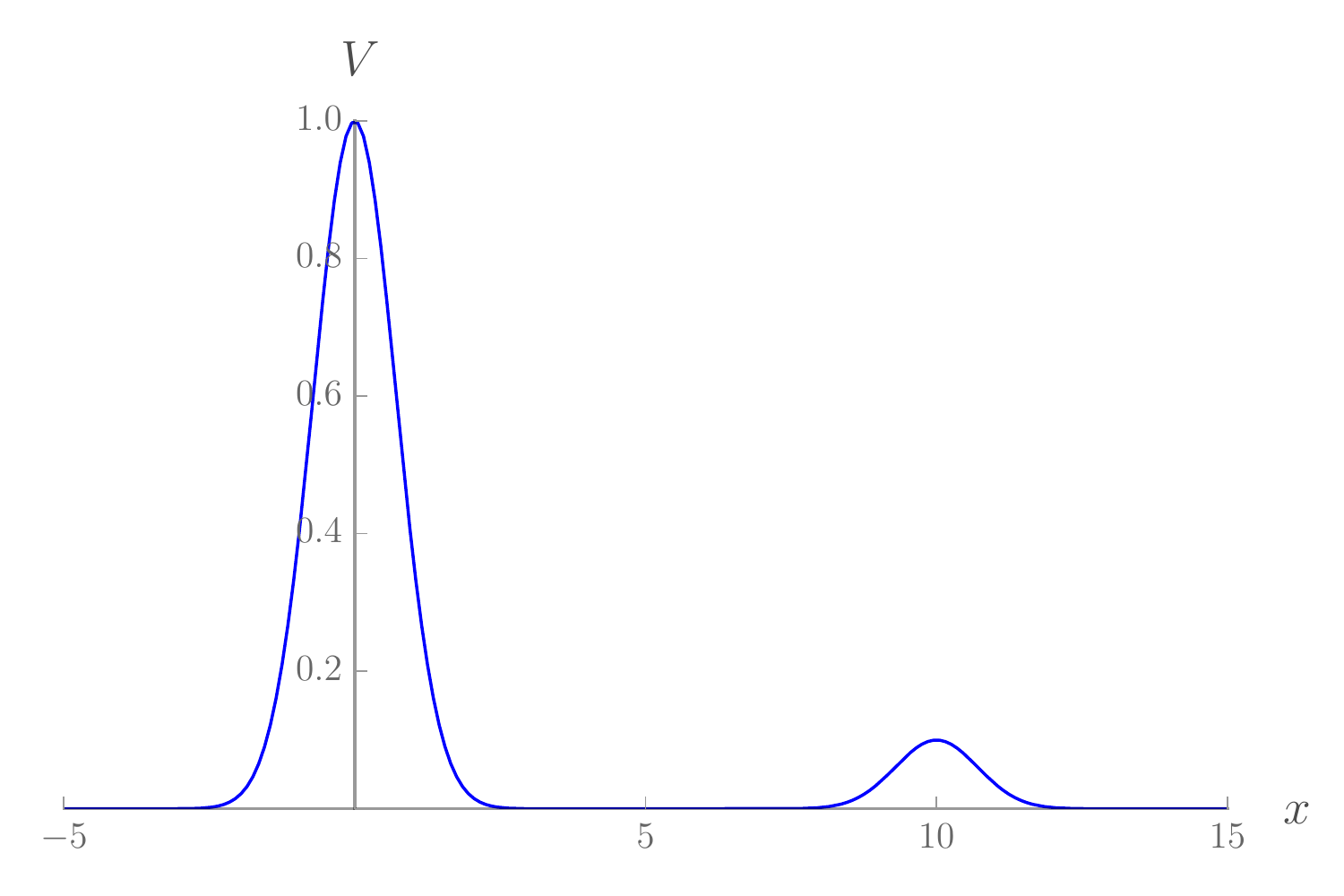}
\includegraphics[width = 0.49\linewidth]{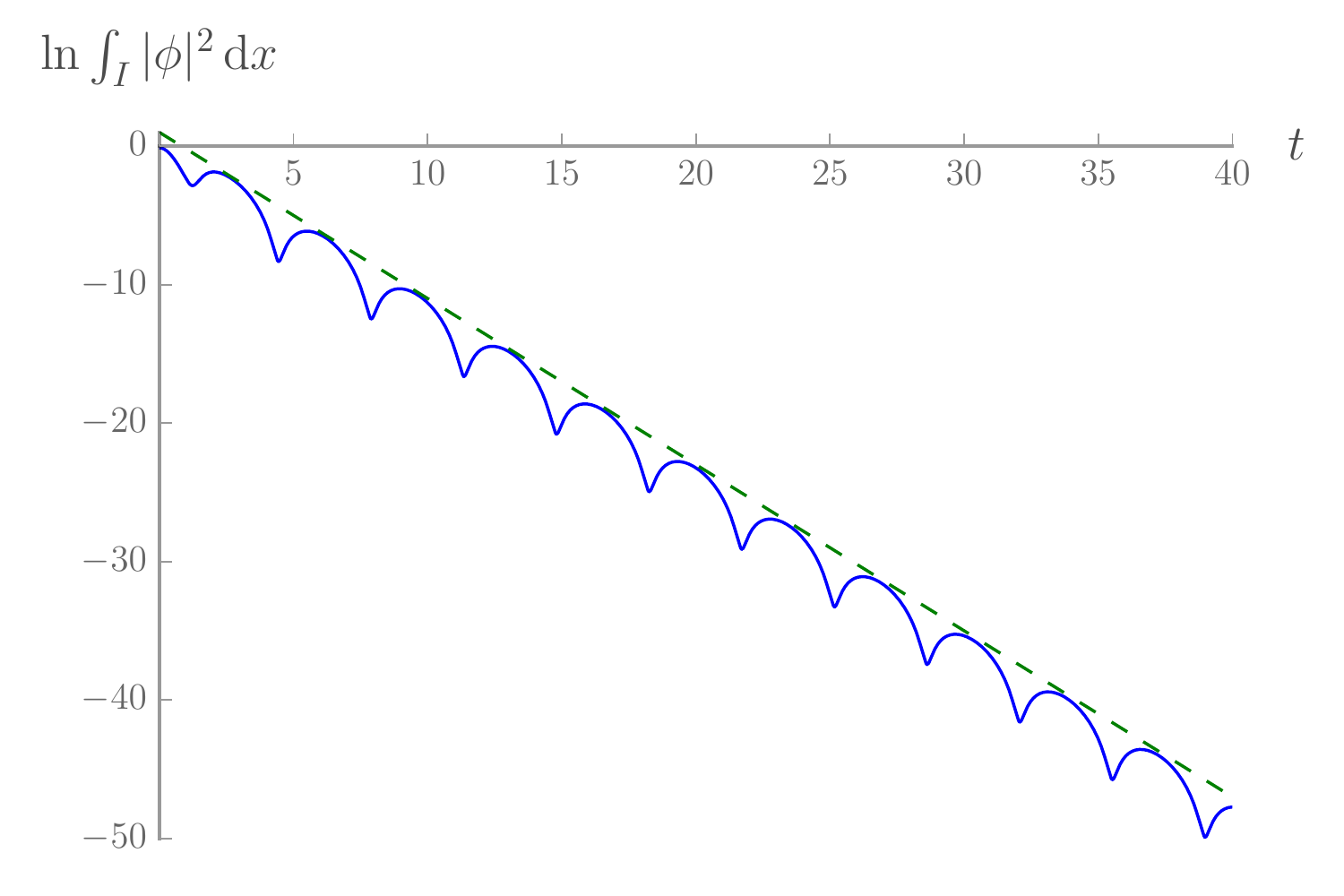}
\includegraphics[width = 0.49\linewidth]{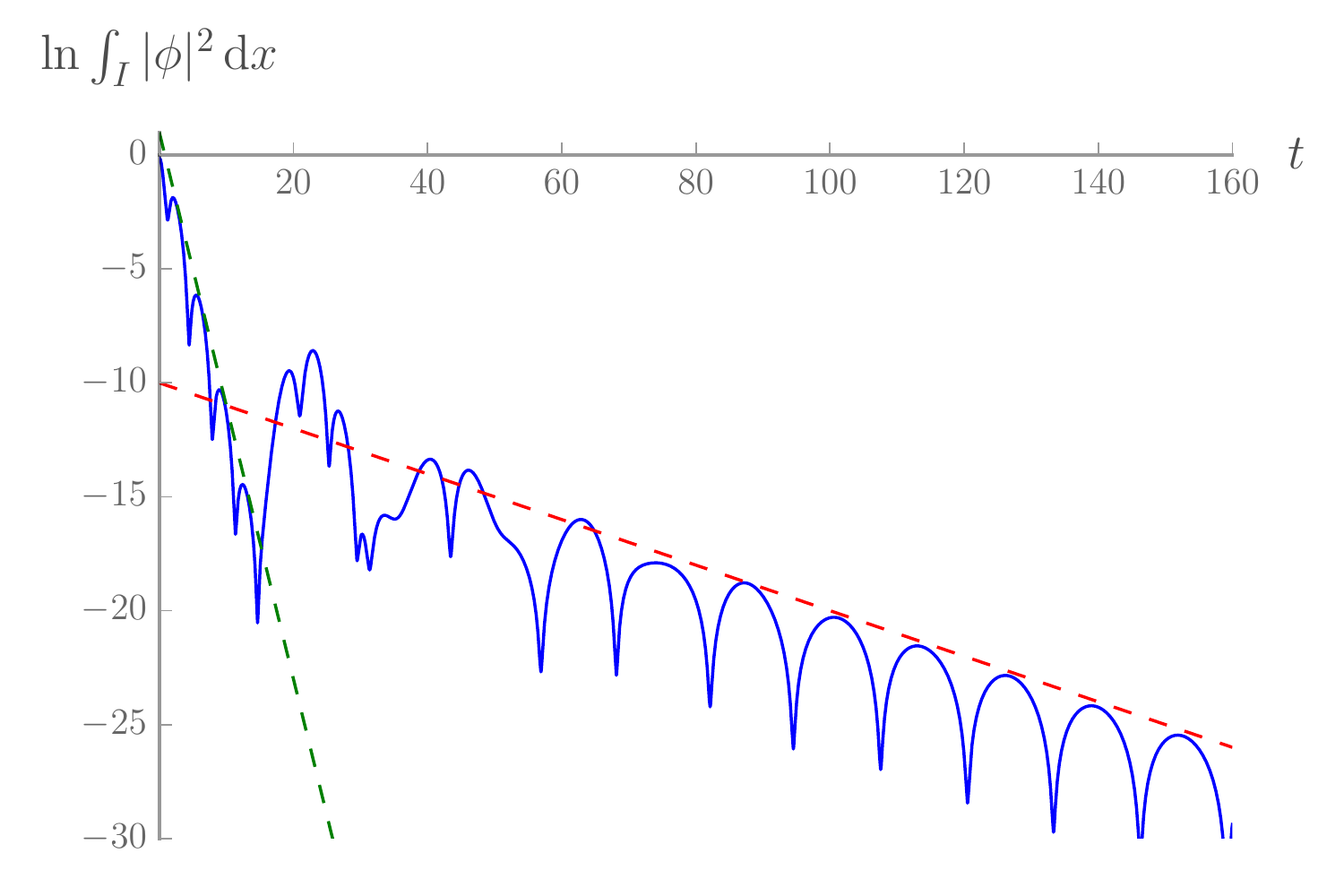}
\caption{Decay of an initially localized perturbation, solution of \eq{eq:saturation_KG}, in the presence of a single (left) or double (right) potential barrier, given by \eq{eq:satur_V_QQNM} with $A_1 = 1$, $\sigma = 1$, $x_2 = 10$, and $A_2 = 0$ (left panel) or $A_2 = 0.1$ (right panel). The initial data is given by \eq{eq:satur_data_QQNM} with $A_\phi = 1$ and $\la = 1 / \sqrt{2}$. The upper panels show the potential, and the lower ones show the evolution in time of $|\phi|^2$, integrated over the interval $I = [-0.5, +0.5]$. The dashed, straight lines are guides for the eye. The green ones have the same slope and value at the origin in the two panels.} \label{fig:QQNM_QNM}
\end{figure}

To illustrate this, let us consider the case of a massless relativistic scalar field $\phi \in C^2 \lp \mathbb{R}^2, \mathbb{C} \rp$ in the potential $V \in C^0 \lp \mathbb{R} \rp$, satisfying the Klein-Gordon equation
\begin{equation} \label{eq:saturation_KG}
\left[ \pd_t^2 - \pd_x^2 + V \right] \phi = 0.
\end{equation} 
For definiteness, we choose a potential of the form
\begin{equation} \label{eq:satur_V_QQNM}
V: x \mapsto A_1 \, \e^{- x^2 / \sigma^2} + A_2 \, \e^{- \lp x - x_2 \rp^2 / \sigma^2},
\end{equation}
with $\lp A_1, A_2, \sigma, x_2 \rp \in \mathbb{R}_+^4$, and the initial data
\begin{equation} \label{eq:satur_data_QQNM}
\forall \, x \in \mathbb{R}, \; \phi(0,x) = A_\phi \, \e^{- x^2 / \la^2} \; \wedge \; \pd_t \phi(0,x) = 0,
\end{equation}
with $A_\phi \in \mathbb{C}$ and $\la \in \mathbb{R}_+$. 
When $A_1$, $A_2$, $x_2$, and $A_\phi$ are all nonvanishing, assuming $x_2 \gg \sigma$ and $x_2 \gg \la$, the potential has two peaks centered at $x = 0$ and $x = x_2$, and $\phi$ is localized close to the first one for $t = 0$. 
We consider two cases, shown in \fig{fig:QQNM_QNM}. 
The first one corresponds to $A_2 = 0$ (left panels).  
The potential then has only one peak, i.e., there is only one scattering region. 
Correspondingly, the amplitude of the perturbation decreases exponentially in time, with a decay rate given by the imaginary part of the frequency of a QNM. 

The second case (right panels) corresponds to nonvanishing values of both $A_1$ and $A_2$. 
As revealed by the bottom-right panel, the decay of the perturbation follows two different exponential laws. 
At ``intermediate'' times ($t < 15$), it decays with the same rate as in the previous case, although the corresponding mode is not a QNM anymore but a QQNM. 
At late times ($t > 60$), it decays with a different rate, given by a QNM for the potential with two barriers. 

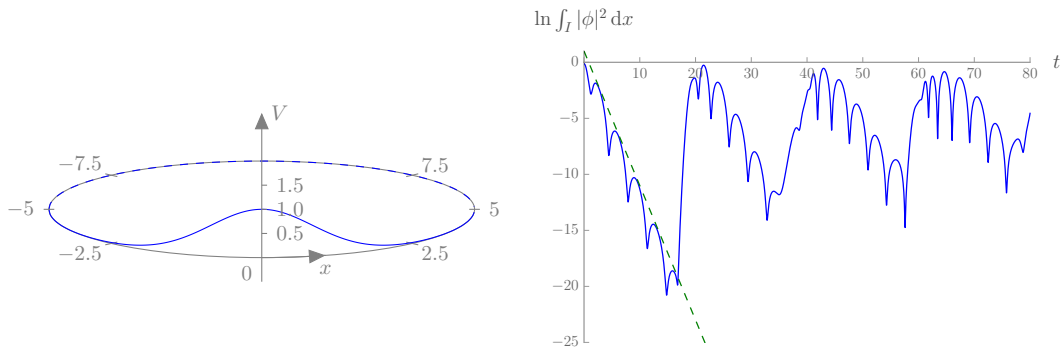
\begin{figure}
\centering
\raisebox{0.4\height}{
\begin{tikzpicture}
\newcommand{\xa}{1.4}
\newcommand{\ya}{0.32}
\draw[color=gray] (0,0) ellipse (2*\xa cm and 2*\ya cm);
\draw[color=gray] (0,-2*\ya) node[below left]{\scalebox{0.8}{$0$}};
\newcommand{\xp}{sqrt(2.*\xa*\xa)}
\draw[color=gray] ($({1.*1.04*\xp}, {-1.04*\ya*sqrt(4.*\xa*\xa-\xp*\xp)/\xa})$) node[below right, xshift=-0.8mm, yshift=1mm]{\scalebox{0.8}{$2.5$}} -- ($({1.*0.96*\xp}, {-0.96*\ya*sqrt(4.*\xa*\xa-\xp*\xp)/\xa})$);
\draw[color=gray] ($({-1.*1.04*\xp}, {-1.04*\ya*sqrt(4.*\xa*\xa-\xp*\xp)/\xa})$) node[below left, xshift=0.8mm, yshift=1mm]{\scalebox{0.8}{$-2.5$}} -- ($({-1.*0.96*\xp}, {-0.96*\ya*sqrt(4.*\xa*\xa-\xp*\xp)/\xa})$);
\draw[color=gray] ($({1.*1.04*\xp}, {1.04*\ya*sqrt(4.*\xa*\xa-\xp*\xp)/\xa})$) node[above right, xshift=-0.8mm, yshift=-1mm]{\scalebox{0.8}{$7.5$}} -- ($({1.*0.96*\xp}, {0.96*\ya*sqrt(4.*\xa*\xa-\xp*\xp)/\xa})$);
\draw[color=gray] ($({-1.*1.04*\xp}, {1.04*\ya*sqrt(4.*\xa*\xa-\xp*\xp)/\xa})$) node[above left, xshift=0.8mm, yshift=-1mm]{\scalebox{0.8}{$-7.5$}} -- ($({-1.*0.96*\xp}, {0.96*\ya*sqrt(4.*\xa*\xa-\xp*\xp)/\xa})$);
\draw[samples=400,scale=1.,domain=-2.*\xa:2.*\xa,smooth,variable=\x,blue,dashed] plot ({\x},{\ya*sqrt(4.*\xa*\xa-\x*\x)/\xa});
\draw[samples=400,scale=1.,domain=-2.*\xa:2.*\xa,smooth,variable=\x,blue] plot ({\x},{-\ya*sqrt(4.*\xa*\xa-\x*\x)/\xa+2.*\ya*exp(-\x*\x)});
\draw[samples=100,scale=1.,domain=0.*\xa:0.6*\xa,smooth,variable=\x,gray,-triangle 45] plot ({\x},{-\ya*sqrt(4.*\xa*\xa-\x*\x)/\xa}) node[below]{\scalebox{0.8}{$x$}};
\draw[color=gray,-triangle 45] (0,-3*\ya) -- (0,4*\ya) node[right]{\scalebox{0.8}{$V$}};
\draw[color=gray] (-2.05*\xa, 0) node[left]{\scalebox{0.8}{$-5$}} -- (-1.95*\xa, 0);
\draw[color=gray] (2.05*\xa, 0) node[right]{\scalebox{0.8}{$5$}} -- (1.95*\xa, 0);
\draw[color=gray] (0.05*\xa, -\ya) node[right]{\scalebox{0.8}{$0.5$}} -- (-0.0*\xa, -\ya);
\draw[color=gray] (0.05*\xa, -0.*\ya) node[right]{\scalebox{0.8}{$1.0$}} -- (-0.0*\xa, -0.*\ya);
\draw[color=gray] (0.05*\xa, 1.*\ya) node[right]{\scalebox{0.8}{$1.5$}} -- (-0.0*\xa, 1.*\ya);
\end{tikzpicture}}
\includegraphics[width = 0.49\linewidth]{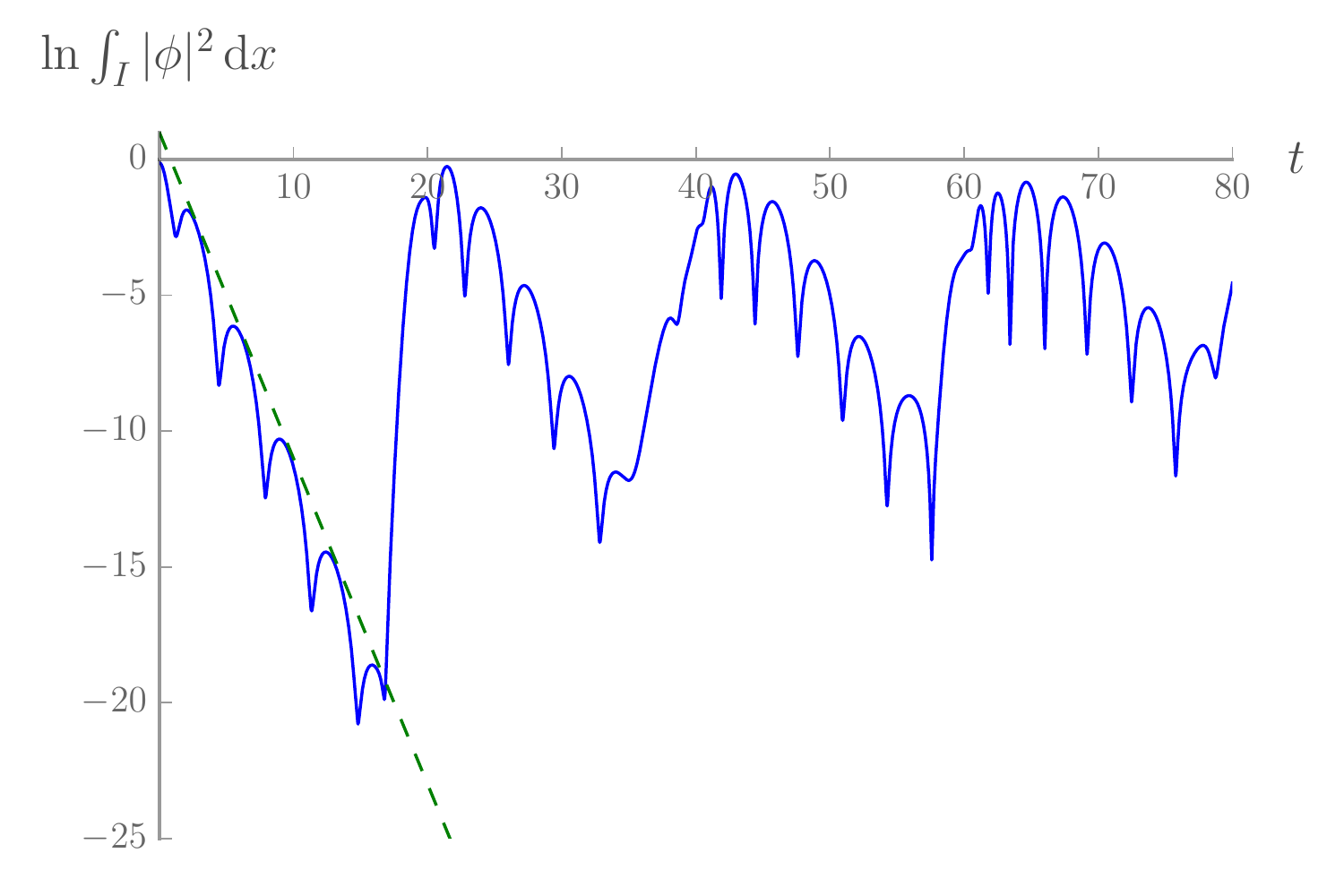}
\caption{Decay of an initially localized perturbation solution of \eq{eq:saturation_KG} in the presence of a single barrier on a torus (shown in the left panel). The parameters, potential, and initial data are the same as in the left panels of \fig{fig:QQNM_QNM} and periodic boundary conditions are imposed at $x = \pm 10$. The green, dashed line is the same as in that figure.} \label{fig:QQNM_T}
\end{figure}

Similarly, one can consider the evolution on a torus, see \fig{fig:QQNM_T}. 
In that case, there is no QNM. 
But there are still QQNM, defined by replacing the periodic boundary conditions far from the peak of the potential by outgoing ones. 
As seen in the figure, at early times, the dynamics is indistinguishable from that over a single barrier in infinite space, with a decay rate given by the imaginary part of the frequency of the QQNM. 

As these three examples show, QQNM are useful tools to determine the behavior of perturbations in systems with well-separated scattering regions and/or boundaries at a finite distance from the origin. 
We also verified that similar results apply to the model of~\cite{Coutant:2016bgk} on the torus. 
In the future, it would be interesting to see whether they can be defined in a more intrinsic manner\footnote{We are grateful to V. Cardoso for mentioning this point.} and what are their imprints on the retarded Green function, whose poles give the dynamical instabilities and QNM.

\newchapter{No-hair theorems for analogue black holes}
\label{ch:nohair}
\begin{tikzpicture}[overlay]
\newcommand*{\xA}{-0.2}
\newcommand*{\xB}{13.35}
\newcommand*{\yA}{5.5}
\newcommand*{\yB}{1.5}
\newcommand*{\epsil}{0.75}
\draw[overlay] (\xA-\epsil,\yA) -- (\xB-\epsil,\yA);
\draw[overlay] (\xA,\yA+\epsil) -- (\xA,\yB);
\draw[overlay] (\xB,\yB-\epsil) -- (\xB,\yA);
\draw[overlay] (\xB+\epsil,\yB) -- (\xA+\epsil,\yB);
\end{tikzpicture}
\begin{small}
One important question, both for experiments aimed at observing the analogue Hawking radiation and for understanding analogue black holes at the theoretical level, concerns the structure and stability of the space of black hole (or white hole) solutions in a given model. 
In~\cite{Michel:2015pra}, we numerically studied the evolution in time of their perturbations in Bose-Einstein condensates. 
Our main result was that black hole flows seem to generically expel away the initial perturbations at infinity, while white holes accumulate them close to the horizon. 
The former are thus very stable, in that the initially perturbed solution will tend to a black hole flow at late time provided the initial perturbation is not too large, and have a behavior reminiscent of the ``no-hair'' theorems of General Relativity. 

The main objective of the present chapter, based on~\cite{Nohair}, is to show this property using analytical techniques and more systematic numerical simulations. 
On the analytical side, we first work to linear order and show explicitly that initial perturbations decay in time. 
We then use Whitham's modulation theory to build explicit, approximate solutions on which the fate of (nonlinear) perturbations can be explicitly followed.  
An interesting by-product of this analysis is that the properties of the nonlinear waves emitted during the evolution can be characterized analytically. 
Numerical simulations are then used to verify that these properties remain when considering the full Gross-Pitaevskii equation. We also show that they extend to the KdV equation. 

From an experimental point of view, our study indicates that black hole flows can be realized without fine-tuning, as the perturbations which may be present during the formation of the analogue horizon will generally disappear at late times. 
White hole flows, on the other hand, will require more care. 
From a more abstract viewpoint, this work shows that the analogy between General Relativity and ``analogue'' models partially extends to the nonlinear domain. 

\end{small} 
\newpage

\renewcommand*{\theHsection}{\theHchapter.\the\value{section}}
\renewcommand\thesection{\arabic{section}}
\renewcommand\thesubsection{\arabic{section}.\arabic{subsection}}

\section{Introduction}

Since general relativity and hydrodynamics can both be formulated as nonlinear classical field theories, many analogies can be drawn between them. 
As was seen in Chapter~\ref{chap:intro}, one of these turns out to be a precise mathematical correspondence.  
Indeed, in the limit of long wavelengths, the linearized wave equation in a moving fluid is identical to the d'Alembert equation of a scalar field propagating in a four-dimensional curved space-time~\cite{Unruh:1980cg}. This remark led to many studies aimed at understanding the validity domain of this correspondence. It is clear that the strict equivalence is lost when including dispersive effects which affect short-wavelength modes~\cite{Jacobson:1991gr}. 
An interesting and nontrivial question is to identify the consequences of these dispersive effects: for instance, how they affect the asymptotic properties of the Hawking radiation emitted by an analogue black hole flow~\cite{Unruh:1994je}. After a thorough analysis, it was found that they do not significantly change the spectrum when the dispersive length is sufficiently short and the flow is smooth enough, see \cite{Coutant:2011in,Robertson:2012ku} for recent updates.
 
The present chapter aims at addressing similar questions when including nonlinear effects. It is clear that the equations of general relativity and hydrodynamics differ at this level. However, as we will show, both the linear and nonlinear stability properties of analogue black hole flows are closely similar to those of black holes in general relativity plus Maxwell theory~\cite{Misner1973}. 
Let us briefly recall the main properties of the latter. First, stationary black holes are fully characterized by a few macroscopic, conserved quantities: their mass, angular momentum, and electric charge. Second, when perturbed during a finite time, the evolution subsequently brings back the solution to one of those stationary black holes. 
These two properties are generally referred to as uniqueness and ``no-hair'' theorems. Their domains of validity is still the subject of investigations~\cite{Chrusciel:2012jk}. Moreover, several ``hairy'' black hole solutions evading these theorems have been found when including exotic matter fields, see the references in~\cite{Chrusciel:2012jk}. 

When studying the behavior of one-dimensional transonic flows, solutions of the Gross-Pitaevskii (GP) or Korteweg-de Vries (KdV) equation, we observe similar properties. We work with time-independent external potentials (shape of the obstacle for the KdV equation) which vary only in a finite domain of $x$. 
In this case, when considering the set of stationary flows which are asymptotically homogeneous (AH) as $x\to \pm \infty$ -- a condition analogous to asymptotic flatness for black holes -- a simple counting of integration constants reveals that there is at most a discrete set of solutions. Moreover, when the potential consists of a single step, we explicitly demonstrate that the solution is unique. 
In other words, we find a single series of solutions which can be parameterized by the conserved current characterizing the flow. When the potential is smooth, we numerically found a series of solutions which smoothly connect to this one in the step-like limit. For some values of the parameters, we also found a disconnected series of solutions, which can be considered as ``hairy'' black holes and contain a large fraction of a soliton attached to the sonic horizon. 

The analogy with general relativity is reinforced when considering the stability of the AH black hole solutions. At the linear level, numerical and analytical results establish that local perturbations decay in time.  
Hence the AH black hole solutions of these equations are linearly stable. 
It should be pointed out that the time-reversed flows, analogous to white holes, are not: in these flows, the scattering of linear perturbations generates a macroscopic undulation~\cite{Coutant:2012mf}. (When adding nonlinearities, several behaviors are found at late time. Our findings are in agreement with, and add to, the results of \cite{Mayoral:2010ck,Michel:2013wpa,Michel:2015pra,2015arXiv150900795D}.) 

Taking into account the nonlinearities of the GP or KdV equations, by a combination of analytical and numerical methods, we motivate that there is a large domain of initial black hole configurations which evolve towards an AH stationary solution. In particular, using G.~B.~Whitham's modulation theory applied to the GP equation~\cite{Kamchatnov}, we show that the initial perturbations are expelled away from the horizon by three nonlinear waves. At small amplitudes, these three waves become the three linear waves emitted by an analogue black hole flow. 
They can thus be considered as the nonlinear version of the stimulated Hawking effect~\cite{Rousseaux:2007is,Macher:2009nz,Weinfurtner:2010nu}. The predictions of Whitham's theory are confirmed and extended by a numerical analysis. 
Interestingly, it seems that the integrability of the equations plays an important role both in the analogue models and in general relativity. In the first case, it provides the invariants needed to apply Whitham's theory and eventually characterize the time-evolution from a perturbed configuration to an AH, stable, stationary solution. 
In the second case, an integrable nonlinear sigma model is used to constrain the set of asymptotically flat, stationary, axisymmetric solutions of the Einstein-Maxwell theory, leading to the uniqueness theorem~\cite{Chrusciel:2012jk}. 

This chapter is organized as follows. In Section~\ref{sec:2}, we first study the stationary solutions of the GP equation and show that the set of AH solutions is discrete. Then we study the linear stability of black hole flows and show that perturbations decay polynomially in time. In Section~\ref{sec:NLstab}, we consider the nonlinear stability of black hole flows. We first use Whitham's theory in subsection~\ref{sec:analytical}, and then present numerical results for the full GP equation in subsection~\ref{sec:numres}. White hole flows are considered in Section~\ref{Sec:WH}. We discuss our results in Section~\ref{sec:concl}. 
Section~\ref{App:Whitham} recalls the derivation of Whitham's equations and derives the properties of the nonlinear waves used in the main text. 
Section~\ref{sec:NHT:AR} sketches a few extensions of this study: 
subsections~\ref{App:KdV} and~\ref{App:linKdV} extend the analysis to KdV-like equations, while subsection~\ref{App:CQNLS} treats the case of a cubic-quintic GP equation. 
\section{Uniqueness and linear stability of black hole flows}
\label{sec:2}

\subsection{Asymptotically homogeneous transonic flows} 
\label{sub:Ahtf} 

We consider a one-dimensional flowing condensate whose wave-function $\psi$ satisfies the GP equation. In a unit system where the atomic mass and reduced Planck constant are equal to $1$, this equation reads 
\begin{align}\label{eq:GPE}
\ii \pd_t \psi = -\frac{1}{2} \pd_x^2 \psi +V(x) \psi + g(x) \left\lvert \psi \right\rvert^2 \psi.
\end{align}
Here $V$ is the external potential and $g$ is the effective one-dimensional coupling. 
In order to have stable homogeneous configurations for uniform $V$ and $g$, we shall consider only condensates with repulsive interactions between atoms, so that $g(x) > 0$. The stationary solutions can be written as
\begin{align}\label{eq:Madelung}
\psi(x,t) = \sqrt{\rho(x)} \, \e^{\ii \int^x v(x') \dd x'}\, \e^{- \ii \om t},
\end{align}
where $\omega$, $\rho(x)$, and $v(x)$ are real. $\rho(x)$ is the density of the condensate and $v(x)$ its local velocity. Plugging \eq{eq:Madelung} into \eq{eq:GPE} gives
\begin{align}\label{eq:eqf} 
\frac{\pd_x^2 \sqrt{\rho(x)}}{\sqrt{\rho(x)}} = 2 \lp V(x) - \om \rp + 2 g(x) \rho(x) + \frac{J^2}{\rho(x)^2}, 
\end{align}
where $J \equiv \rho(x) v(x)$ is the conserved current.

\subsubsection{Homogeneous potentials} \label{subsub:homo}

\begin{figure}
\centering
\includegraphics[width=0.5 \linewidth]{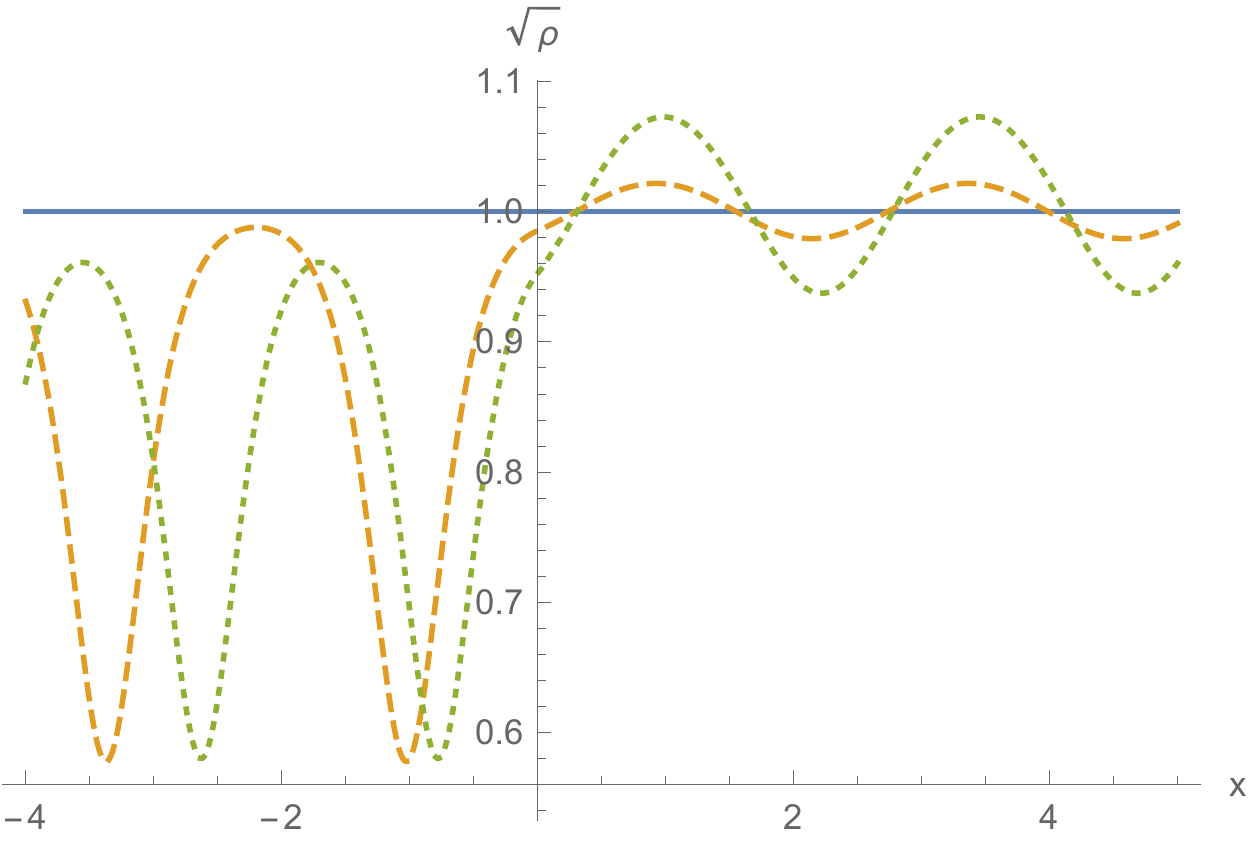}
\caption{We represent the profile of $\sqrt{\rho(x)}$ for three stationary transonic flows with $g_+ = 1$, $g_- = 8$, $\om - V_+ = 7/3$, $\om - V_- = 28/3$, and $J = \sqrt{8/3}$. 
For these values, \eq{eq:GPE} has a stationary, homogeneous, transonic solution with density $\rho_0 = 1$ and a moderately large difference between the sound velocities on the two sides of the horizon. 
The sonic horizon is located on the discontinuity, at $x=0$. The continuous line corresponds to the unique homogeneous solution, see~(\ref{eq:rho0}, \ref{omc}). 
The two other flows (represented by a green, dotted and an orange, dashed lines) contain a soliton train on the subsonic (left) side and an undulation on the supersonic (right) side. When the amplitude of the undulation goes to $0$, the distance between two consecutive solitons and the negative position of the first one go to infinity while their depth remains finite.
} \label{fig:gensol}
\end{figure}

To prepare the analysis of transonic flows in inhomogeneous potentials, it is useful to recall the main properties of the stationary solutions when $g$ and $V$ are independent of $x$. In this case, spatially bounded solutions of \eq{eq:eqf} exist if and only if
\begin{align}
\label{Jmax}
J^2 \leq J_{\rm max}^2, \, J_{\rm max} \equiv \sqrt{\frac{8 \lp \om - V \rp^3}{27 g^2}}.
\end{align}
All bounded solutions are periodic and can be written with elliptic functions~\cite{Kamchatnov}. Two of them are homogeneous, and given by the two positive roots $\rho_p < \rho_b$ of the right-hand side of \eq{eq:eqf}. 
(There is also a negative root, which will play no role in the following as the density is positive by definition.) 
They merge when $|J| = J_{\rm max}$. The smallest one, $\rho_p$, describes a supersonic flow since the condensate velocity $ |J |/ \rho_p$ is larger than the speed of long-wavelength perturbations $c_p = \sqrt{g \rho_p}$ (see Chapter~\ref{ch:saturation}). 
Instead, the flow described by $\rho_b$ is subsonic as $|J| / \rho_b < c_b = \sqrt{g \rho_b}$. 
The stationary perturbations over these two homogeneous solutions are radically different. Stationary flows with a density close to the supersonic value $\rho_p$ contain a zero-frequency modulation of the density (hereafter called ``undulation''). Instead, flows with mean densities close to $\rho_b$ contain a train of solitons, see Fig.~\ref{fig:gensol}. 

Let us briefly explain how to get the main properties of these solutions. More details can be found in Chapter~\ref{ch:saturation}. Multiplying \eq{eq:eqf} by $\sqrt{\rho} \, \pd_x \sqrt{\rho}$ and integrating the resulting equation gives 
\begin{align}\label{eq:GPErhpgen}
\frac{1}{4} \lp \pd_x \rho \rp^2 = g (\rho(x) - \rho_{1}) (\rho(x) -\rho_{2}) (\rho(x)-\rho_{3}),
\end{align}
where $\rho_{i}$, $i \in \left\lbrace 1,2,3 \right\rbrace$ are three constants. They are related to the current $J$ and frequency $\om$ through
\begin{align}
\rho_{1} + \rho_{2} + \rho_{3} = 2 \frac{\om - V}{g}, \; \rho_{1} \rho_{2} \rho_{3} = \frac{J^2}{g}.
\end{align}
When there is no double root, the range of $\rho(x)$ is a connected component of the domain where the polynomial on the right-hand side of \eq{eq:GPErhpgen} is positive. The density $\rho$ can therefore be asymptotically bounded only if that domain has a bounded component. This requires that the three constants $\rho_{i}$ are all real.~\footnote{After multiplication by $\rho(x)^2$, the right-hand side of \eqref{eq:eqf} is a third-order polynomial in $\rho(x)$ with real coefficients. It thus has either $1$ or $3$ real roots. In the first case, if $\rho_1$ denotes the real root, the accessible domain of $\rho$ is $\left[ \rho_1, +\infty \right[$, which has no bounded component.} 
We order them as $\rho_{1} \leq \rho_{2} \leq \rho_{3}$. Then, the two domains in which the right-hand side of \eq{eq:GPErhpgen} is positive are $\left[\rho_1, \rho_2 \right]$ and $\left[\rho_3, + \infty \right[$. The second one corresponds to a divergent solution. The only bounded one is the first interval, and $\rho(x)$ oscillates between $\rho = \rho_1$ and $\rho = \rho_2$. 
The bounded solution is also characterized by a fourth parameter giving the phase of the oscillations. 

In the limit $\rho_2 \to \rho_1$, $\rho$ becomes the homogeneous supersonic solution with density $\rho_p$ discussed after \eq{Jmax}. When $\rho_2$ is close to $\rho_1$, $\rho$ describes a small-amplitude undulation on top of this homogeneous solution, see Fig.~\ref{fig:gensol}. In the opposite limit $\rho_2 \to \rho_3$, one gets a soliton with asymptotically homogeneous subsonic flow of density $\rho_b$ (or a purely homogeneous solution when the center of the soliton is sent to infinity), and when $\rho_2$ is close to $\rho_3$ $\rho$ contains a train of widely-separated solitons. 

Interestingly, when $\rho_2 = \rho_3$ there exists another solution, called ``shadow soliton'', which diverges at a finite value of $x$ and also goes to the subsonic flow $\rho = \rho_3$ asymptotically. This last solution must be discarded when working with homogeneous functions $V$ and $g$ because of its divergence. However, it must be included in the forthcoming analysis 
of the steplike case as the divergence can be erased by the change of $g$ and $V$, as explained in Chapter~\ref{ch:saturation}.  

\subsubsection{Step-like potentials, unicity theorem}

To obtain stationary transonic flows, $g$ and/or $V$ must depend on $x$. To have simple solutions, we assume that $g$ and $V$ have the form 
\begin{align}\label{eq:step}
& g(x) = \theta(-x) \, g_-  + \theta(x)\, g_+ , \;
\\ \nonumber
 & V(x) = \theta(-x)\,V_-  + \theta(x) \, V_+ ,
\end{align}
with $0 < g_+ < g_- $ and $V_+ > V_- $. 
There are two possibilities to build a black hole solution, which must be stationary, asymptotically homogeneous\footnote{In this chapter we refer to configurations with homogeneous densities as ``homogeneous'' ones, although their phases in general depends on $x$.}, subsonic for $x \to -\infty$, and supersonic for $x \to + \infty$ if $J > 0$:
\begin{itemize}
\item One can match two solutions with strictly homogeneous densities on each side of the origin $x = 0$, giving a globally homogeneous solution (with a discontinuity of $c$ at the origin due to that of $g$);
\item One can match a half-soliton in the region $x < 0$ with a homogeneous supersonic solution in the region $x > 0$.  
\end{itemize}
For any real value of $J$, there exists a globally homogeneous solution of \eq{eq:eqf} given by
\begin{align}\label{eq:rho0}
\rho(x) = \rho_0, \, \rho_0 \equiv \frac{V_+ - V_-}{g_- - g_+}.
\end{align}
Its frequency $\om(J)$ varies with $J$ so that \eq{Jmax} is always satisfied 
on each side. 
Explicitly, one finds
\begin{align}
\om(J)= V_- + g_- \rho_0 + \frac{J^2}{2 \rho_0^2} = V_+ + g_+ \rho_0 + \frac{J^2}{2 \rho_0^2}. 
\label{omc}
\end{align}
This solution is transonic iff 
\begin{align}\label{eq:BHcond} 
c_p \equiv \sqrt{g_+ \rho_0} < \frac{|J|}{\rho_0} < \sqrt{g_- \rho_0} \equiv c_b. 
\end{align}
The flow is then subsonic for $x<0$ and supersonic for $x>0$. When $J > 0$, it thus corresponds to a black hole flow. Incoming counter-propagating waves experience a large reduction of their wave vectors, see next subsection and~\cite{Unruh:1980cg,Macher:2009nz,Barcelo:2005fc}. When $J$ (and thus $v$) is negative, the transonic flow is analogous to a white hole. In that case, long-wavelength incoming modes are converted into short-wavelength modes. 
This is where the above constraints on $g_-$, $g_+$, $V_-$, and $V_+$ are important. The condition that $g_+ - g_-$ and $V_+ - V_-$ have strictly opposite signs ensures that $\rho_0 > 0$ as required for a density. The condition $g_- > g_+$ ensures it corresponds to a black hole for $J > 0$. (With the opposite choice, one would find the same results up to a change of sign of $J$.) 

Interestingly, having fixed $g_+$, $g_-$, $V_+$, and $V_-$ of \eq{eq:step}, when $|J|$ obeys \eq{eq:BHcond}, the solution \eq{omc} is the {\it unique} AH transonic flow. Indeed, in this interval of $|J|$, all the other stationary bounded transonic solutions contain an undulation in the supersonic region and/or a train of solitons in the subsonic one, see Fig.~\ref{fig:gensol}. 
This can be understood as follows. 
When choosing the two (so far independent) stationary solutions for $x < 0$ and $x > 0$, we have 6 free parameters: $\om$, $J$, and 4 integration constants coming from the integration of the second-order differential equation \eqref{eq:eqf} on each side of the origin. 
To define a global solution, we impose two scalar constraints, namely continuity of $\rho$ and $\rho'$ at $x = 0$. 
The asymptotic conditions gives further constraints. 
On each side, to linear order, we have two independent solutions. 
For $x \to - \infty$, one of them decays exponentially while the other grows exponentially. The coefficient of the latter must thus be set to $0$, giving one constraint. 
In the opposite limit $x \to + \infty$, the two linear solutions are oscillating with a constant amplitude. We must thus set their two coefficients to $0$, giving two additional constraints. 
In total, we thus have only one remaining free parameter. If we now fix the value of the conserved current $J$, the set of solutions is of dimension $0$, i.e., discrete. 
As mentioned above, the possible black hole solutions are either homogeneous or contain half a soliton. 
The former is possible only if $\rho (x) = \rho_0$, which completely fixes the solution in $\psi$ up to a global phase. 
To see whether it is unique, it thus remains to study the (discrete) set of solutions with half a soliton for $x<0$, i.e. the ``waterfall'' configurations studied in~\cite{PhysRevA.85.013621}.

\begin{figure}
\centering
\includegraphics[width=0.49 \linewidth]{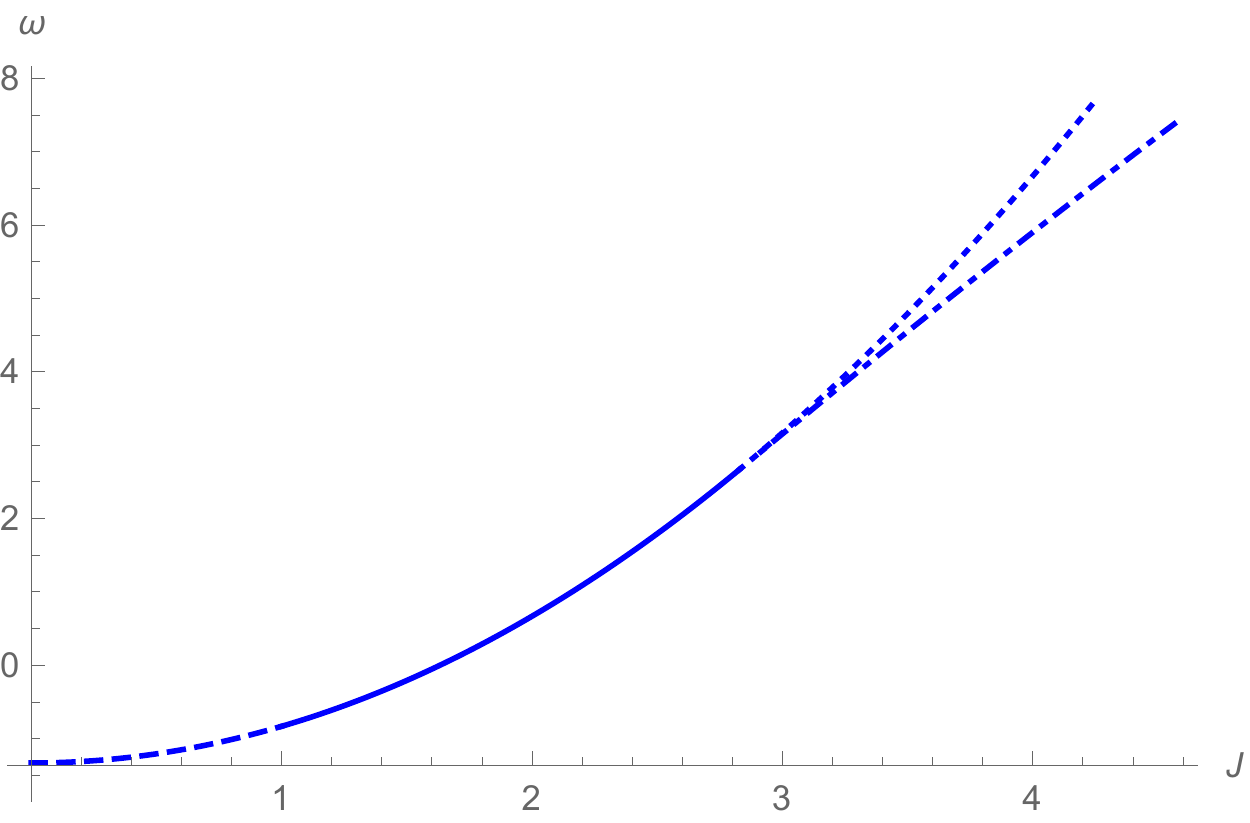}
\includegraphics[width=0.49 \linewidth]{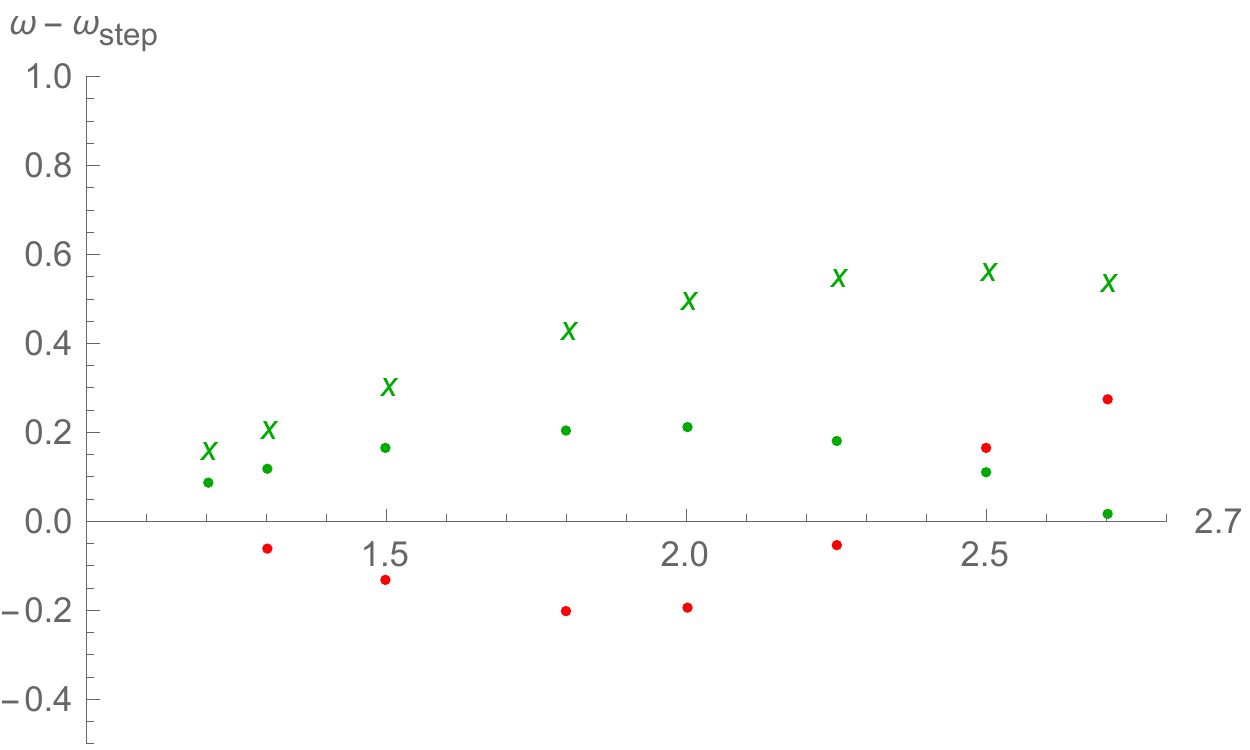} 
\caption{On the left panel, we show the angular frequency $\om$ of the homogeneous solution and the ``waterfall'' configuration as functions of the current $J$ for the steplike potentials of \eq{eq:step}. The continuous line, which extends from $J = 1$ to $J = 2 \sqrt{2}$, corresponds to the homogeneous transonic flows. The dashed (respectively dotted) line corresponds to homogeneous solutions which are globally subsonic (respectively supersonic). The dot-dashed line corresponds to transonic waterfall solutions. $g_-$, $g_+$, $V_-$, and $V_+$ are the same as in \fig{fig:gensol}. The right panel shows the deviations of $\om(J)$ from its value in the steplike case, for some numerical solutions found with functions $g$ and $V$ given by \eq{eq:vargV}. Red points correspond to $\sigma_V = 1.2, \,  \sigma_g = 0.5$, while green points and crosses correspond to $\sigma_g = 1.2,\,  \sigma_V = 0.5$. The eight green crosses show ``hairy'' solutions containing one soliton. Their density profiles are similar to that 
shown on the right panel of \fig{fig:flatsol}. 
} \label{fig:statsols}
\end{figure}

The value of $\om$ for a ``waterfall'' solution, $\om^{\rm wf}(J)$, is 
\begin{align}
\om^{\rm wf}(J) = V_- + g_- \rho_-^{\rm wf} + \frac{J^2}{2(\rho_-^{\rm wf})^2} = g_+ \rho_+^{\rm wf} + V_+ + \frac{J^2}{2(\rho_+^{\rm wf})^2}, 
\label{omcwf}\end{align}
where $\rho_\mp^{\rm wf}$ is the density for $x \in \mathbb{R}^\mp$. Remarkably, a straightforward calculations shows that these solutions exist only for values of $|J|$ larger than the upper bound of \eq{eq:BHcond}. When decreasing $\abs{J}$ along the branch of waterfall solutions, the difference between the densities on the two sides of the horizon decreases. It vanishes precisely when $J = \sqrt{g_- \rho_0^3}$, at which point the homogeneous and the waterfall solutions coincide. Decreasing $J$ further, only the homogeneous solution of \eq{eq:rho0} remains bounded. The series of \eq{omc} and \eq{omcwf} are represented in \fig{fig:statsols}.
In brief, in the case of steplike potentials, for all $|J|>  \sqrt{g_+ \rho_0^3}$, there is a unique AH transonic solution. Hence, in this case, stationary transonic flows obey an analogous version of the black hole uniqueness theorem~\cite{Israel}.

\subsubsection{Smooth potentials and hairy black holes}
 
When $V$ and $g$ smoothly vary with $x$, the above arguments are no longer sufficient to characterize the complete set of stationary bounded solutions of \eq{eq:GPE}. Yet, the set of AH stationary transonic flows remains discrete at fixed $J$ as the above counting of the degrees of freedom still applies. 
In particular, the frequency is restricted to a (possibly empty) discrete set of branches $\om^i(J)$. (Notice that the homogeneous solution of \eq{eq:rho0} still exists if the variations of $g$ and $V$ are tuned in such a way that $\pd_x g / \pd_x V$ is a constant.) 

\begin{figure}
\centering
\includegraphics[width=0.49 \linewidth]{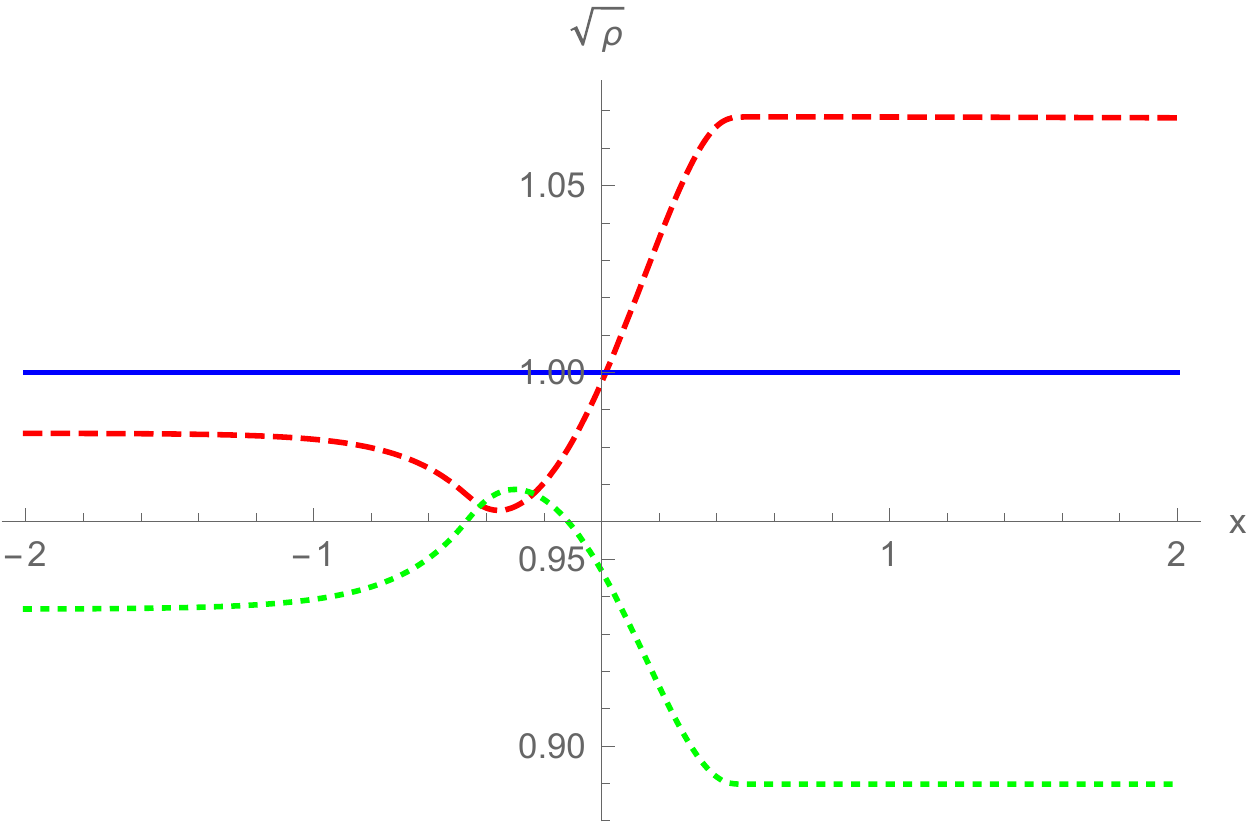}
\includegraphics[width=0.49 \linewidth]{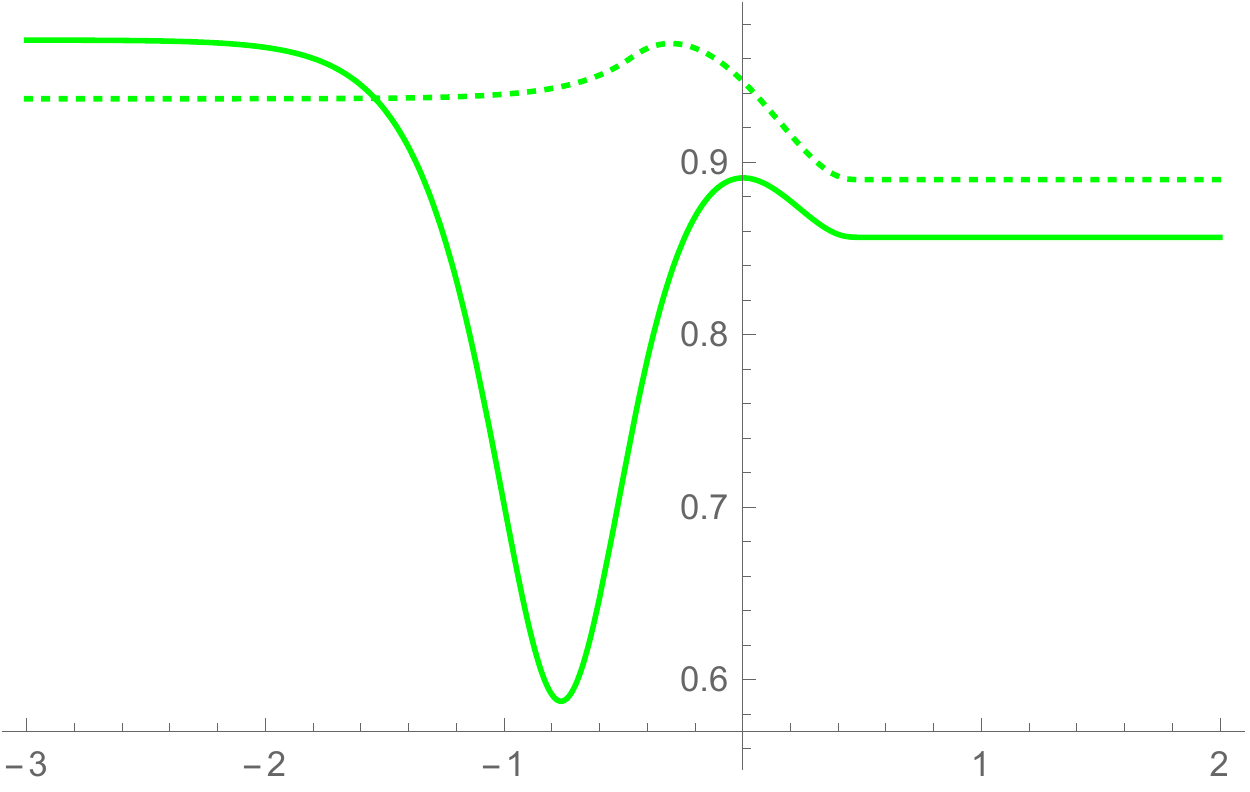} 
\caption{On the left panel we show the density profiles $\sqrt{\rho}$ of stationary AH transonic flows obtained using the continuous profiles of \eq{eq:vargV} with different slopes $1 / \sigma_g$ and $1 / \sigma_V$. The values of $g_\pm$, $V_\pm$, 
and $J$ are the same as those used in Fig.~\ref{fig:gensol}. Hence the modifications are only due to the finiteness of $\sigma_g$ and $\sigma_V$. The blue, continuous line corresponds to $\sigma_g = \sigma_V = 1$, the red, dashed one to $\sigma_V = 1.2, \,   \sigma_g = 1$, and the green, dotted one to $\sigma_V = 1, \,\sigma_g = 1.2$. On the right panel, we show the hairy solution (continuous) and the main solution (dotted) obtained with the last set of parameters. The hairy solution contains a nearly complete soliton attached to the horizon.} \label{fig:flatsol}
\end{figure}

To 
generalize the step-like potentials of \eq{eq:step}, we considered the continuous profiles 
\begin{align}\label{eq:vargV}
g(x) =& \, g_-  + (g_+ - g_-) \, \theta \lp x + \frac{\sigma_g}{2} \rp \left[ \theta \lp \frac{\sigma_g}{2} - x \rp \lp \frac{x}{\sigma_g}  + \frac{1}{2}\rp +  \theta \lp x - \frac{\sigma_g}{2} \rp \right] , 
\\ \nonumber
V(x) =& \, V_-  + (V_+ - V_-) \, \theta\lp x + \frac{\sigma_V}{2} \rp \left[ \theta\lp \frac{\sigma_V}{2} - x \rp \lp \frac{x}{\sigma_V} + \frac{1}{2}\rp+  \theta\lp x - \frac{\sigma_V}{2} \rp \right] , 
\end{align}
characterized by the domains $\sigma_g$ and $\sigma_V$ where $g$ and $V$ linearly interpolate 
between their asymptotic values. 
Stationary flows are obtained by solving \eq{eq:eqf} numerically using a finite difference method. As expected, we found AH solutions which are smoothly connected to those of the former subsection in the limit $\sigma_{g}, \sigma_{V} \to 0$. More interestingly, we also found a series of ``hairy'' AH solutions when $\sigma_V$ is smaller than $\sigma_g$. These solutions are disconnected from the previous ones (hereafter referred to as ``the main series'') because they contain an almost complete soliton which is attached to the sonic horizon, and which is mostly localized in the subsonic region, see \fig{fig:flatsol}. 
Our numerical results indicate that its center is rejected at $x \to -\infty$ in the steplike limit as well as in the limit where $\sigma_V = \sigma_g$. Our simulations also indicate that this solution is unstable under nonlinear perturbations, which can trigger the emission of the soliton at infinity. We hope to characterize this instability in a future work. 
On the right panel of \fig{fig:statsols},  we show the values of $\om(J)$ for the AH solutions obtained with the profile \eq{eq:vargV}. The red dots describe solutions of the main series with $\sigma_V = 1.2, \,  \sigma_g = 0.5$. The green dots (crosses) correspond to (hairy) solutions obtained with $\sigma_g = 1.2, \,  \sigma_V = 0.5$. It is clear that these solutions belong to two different series of $\om(J)$. In \fig{fig:flatsol} we represent the density profile of three solutions of the main series and one hairy solution with $\sigma_V / \sigma_g \in \left\lbrace 1, 1.2, 1/1.2 \right\rbrace$.

In conclusion, when dealing with smooth profiles for $g$ and $V$, one always finds the main series of AH solutions characterized by a smooth density profile across the horizon. In addition, a discrete number of hairy solutions can exist. When we found such solutions, we observed that they are disconnected from the solutions of the main series because they contain some solitonic configuration attached to the region where the potential varies. 

\subsection{Linear stability of black hole flows} 
\label{sec:lin}

We now study the stability of transonic flows to linear perturbations. It appears that those corresponding to black holes are much more stable than those describing white holes. This is nontrivial because the S-matrix describing the scattering of linear perturbations on a white hole horizon is the inverse of that describing the scattering on the time reversed black hole flow~\cite{Macher:2009nz}. This relation implies that white hole flows 
possess the same degree of stability as black holes: they are dynamically stable, i.e., the spectrum of linear perturbations contains no complex-frequency modes.  Yet, the late-time evolution of perturbations scattered on a sonic horizon 
strongly depends on whether the modes experience a red-shift (black hole), or a blue-shift (white hole).
White hole flows are studied in Section~\ref{Sec:WH} where it is shown that they exhibit a linear infra-red instability. 
In the present subsection we consider the linear stability properties of black hole flows and show that they are stable under linear perturbations. 

\begin{figure}[ht]
\centering
\includegraphics[width=0.49 \linewidth]{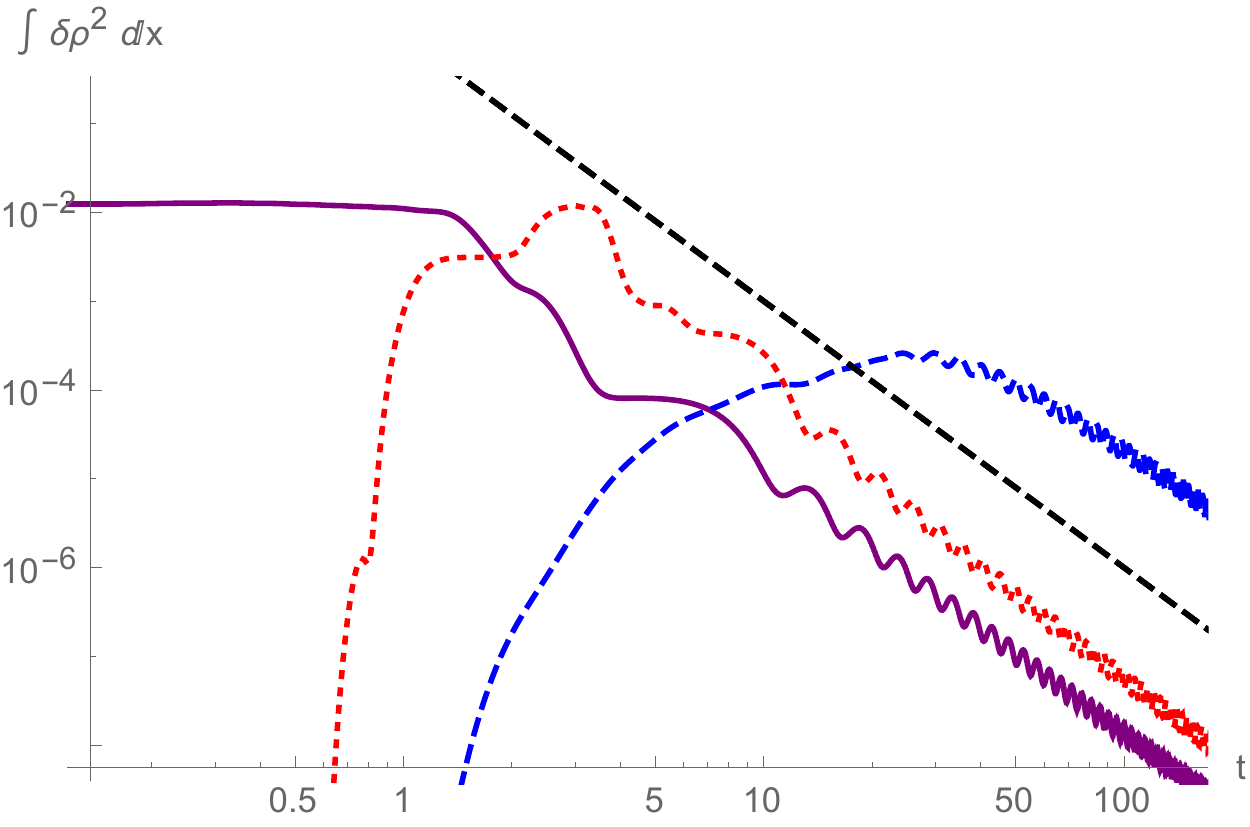} \includegraphics[width=0.49 \linewidth]{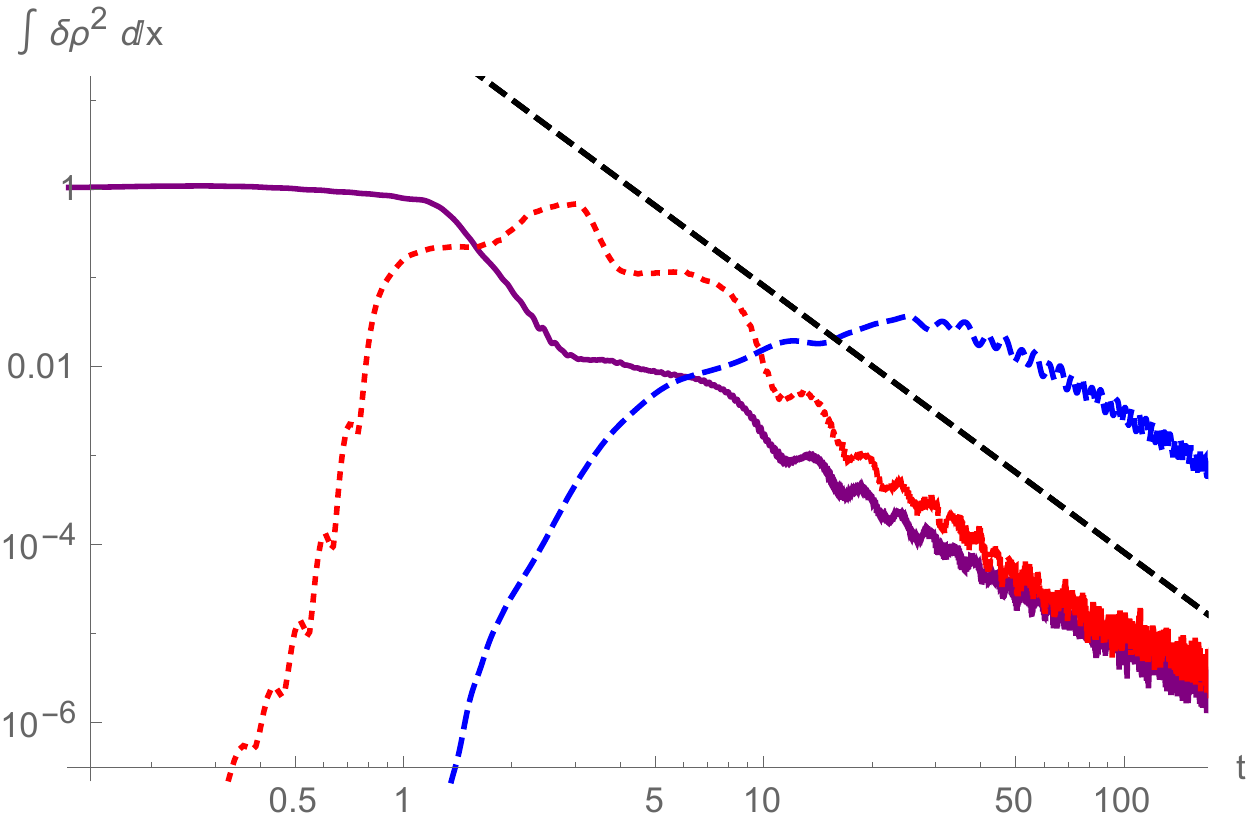} 
\caption{We show the evolution of the squared density perturbation, integrated between $x=-5$ and $x=5$, 
due to the scattering of three wave packets on a homogeneous black hole flow.
The perturbation is initially located in the supersonic region (red, dotted), in the subsonic one (blue, dashed), or centered on the horizon (continuous, purple). The initial value of their amplitude is $\delta \rho_{\rm max} = 0.1$ on the left plot, and $\delta \rho_{\rm max} = 0.9$ on the right plot. 
The oblique dashed line shows $t^{-3}$ 
(left) and $81 t^{-3}$ (right). When comparing the two plots, 
besides the factor of $9^2$ which relates the initial values of $\delta \rho^2$ one sees that the behaviors are very similar.
The late-time oscillations have an angular frequency of $2 \om_{\rm max}$, 
and are due to the vanishing of the group velocity at $\om=\om_{\rm max}$, see \fig{fig:DR}. The background flow parameters 
are $g_+ = 0.86$, $g_- = 8.27$, $\om - V_+ = 2.19$, $\om - V_- = 9.61$, 
and $J = \sqrt{8/3}$. Their values have been chosen so that $\rho_{2,-} = \rho_{3,+} = 1$, $\rho_{2,+} = 2.25$, and $\rho_{3,-} = 0.49$.} \label{fig:lin3/2}
\end{figure}

We first present the results of numerical simulations shedding light on the evolution of perturbations. We choose a spatial domain $I \subset \mathbb{R}$ containing the sonic horizon and look at the late-time evolution of the perturbations in $I$. Numerical simulations indicate that perturbations decay as $t^{-3/2}$ up to logarithmic factors. These observations hold for all transonic AH solutions, both in a steplike potential of \eq{eq:rho0} and in smooth $V(x)$ and $g(x)$. To represent the various typical cases,  we consider the scattering of a gaussian perturbation $\delta \rho$ which is initially localized in the subsonic region, in the supersonic region, or centered on the horizon. 
On the left panel of \fig{fig:lin3/2}, we study the evolution of a  perturbation of relative amplitude  $\delta \rho/\rho_0$ of order $0.1$ and of width equal to $1$ (in our system of units). 
On the right panel, to show that similar results hold for much larger perturbations, 
we work with $\delta \rho/\rho_0$ of order $1$ for the same spatial width. In both cases, we show the time dependence of the integral of $\delta \rho^2(t,x)$ over the segment $[-5,5]$, which contains the sonic horizon localized at $x=0$. At early times, the density fluctuation increases as the perturbation enters the integration region, or remains roughly constant when the perturbation is already in the integration region at $t=0$. At late times, in all cases, one sees that the average of $\delta \rho^2$ goes to zero as $t^{-3}$. 
The fact that the temporal behavior is very similar on the left panel (linear regime) and on the right panel (nonlinear regime since $\delta \rho/\rho_0$ reaches values of order $1$) reveals that the decay law  of $\delta \rho^2$ in $t^{-3}$ is robust. It implies that all perturbations are diluted away so as to give at late time an AH stationary solution.

This behavior is similar to that of linear perturbations propagating in a black hole metric~\cite{Misner1973}. (We briefly present at the end of this Section the nature of the correspondence between the present case and relativistic fields.) We now give the analytical elements needed to understand this behavior. The full calculation is done in subsection~\ref{App:linKdV} for the linearized KdV equation and a square perturbation.

We look for perturbed solutions of the form
\begin{align}
\psi(x,t) = \psi_0 (x) \, \e^{- \ii \omega t} \lp 1 + \phi(x,t) \rp, 
\label{relpert}
\end{align}
where $\psi_0 = \sqrt{\rho_0(x)} \, \e^{\ii \int^x v_0(x') \dd x'} $, see \eq{eq:Madelung}.
To first order in $\phi$, \eq{eq:GPE} gives
\begin{align}\label{eq:BdG} 
\ii \pd_t \phi(x,t)  = -\frac{1}{2} \pd_x^2 \phi(x,t) - \frac{\pd_x \psi_0(x)}{\psi_0(x)} \pd_x \phi(x,t) + g(x) \rho_0(x) \lp \phi(x,t) + \phi(x,t)^* \rp.
\end{align}
Asymptotically, on either side, the background quantities $\rho_0$ and $v_0$ are constant. Hence we can look for solutions of the form 
\begin{align}\label{eq:formlin}
\phi_k(x,t) = \mathcal{U}_k \, \e^{\ii (k x - \om t)} + \mathcal{V}_k^* \, \e^{-\ii (k x - \om t)}. 
\end{align}
Plugging this form into \eq{eq:BdG}, we obtain a system of two linear equations on $\mathcal{U}_k$ and $\mathcal{V}_k$, which has nontrivial solutions iif the dispersion relation
\begin{align}\label{eq:disprelBdG} 
(\om - v_0 k)^2 = g \rho_0 k^2 + \frac{k^4}{4}\, 
\end{align}
(shown in \fig{fig:DR}) is satisfied. For $\om \in \mathbb{R}$, there are two roots in $k$ in the subsonic region. In the supersonic region, when $\abs{\om}$ is smaller than a critical frequency $\om_{\rm max}$, see Eq.~(25) in \cite{Macher:2009nz}, there are two additional roots. For $0 < \om < \om_{\rm max}$, there are thus six asymptotic plane waves (two in the subsonic region and four in the supersonic one), which give the asymptotic behavior of the globally-defined bounded modes. There are three linearly independent bounded modes. The scattering process is characterized by the S-matrix relating the in and out bases of the vector space they span. (The in (out) basis is composed of the three modes which asymptotically contain exactly one wave with group velocity oriented towards (away from) the sonic horizon.) The behavior of its coefficients and the link with the Hawking effect can be found in~\cite{Macher:2009nz,Macher:2009tw}. 
\begin{figure}
\centering
\includegraphics[width=0.49 \linewidth]{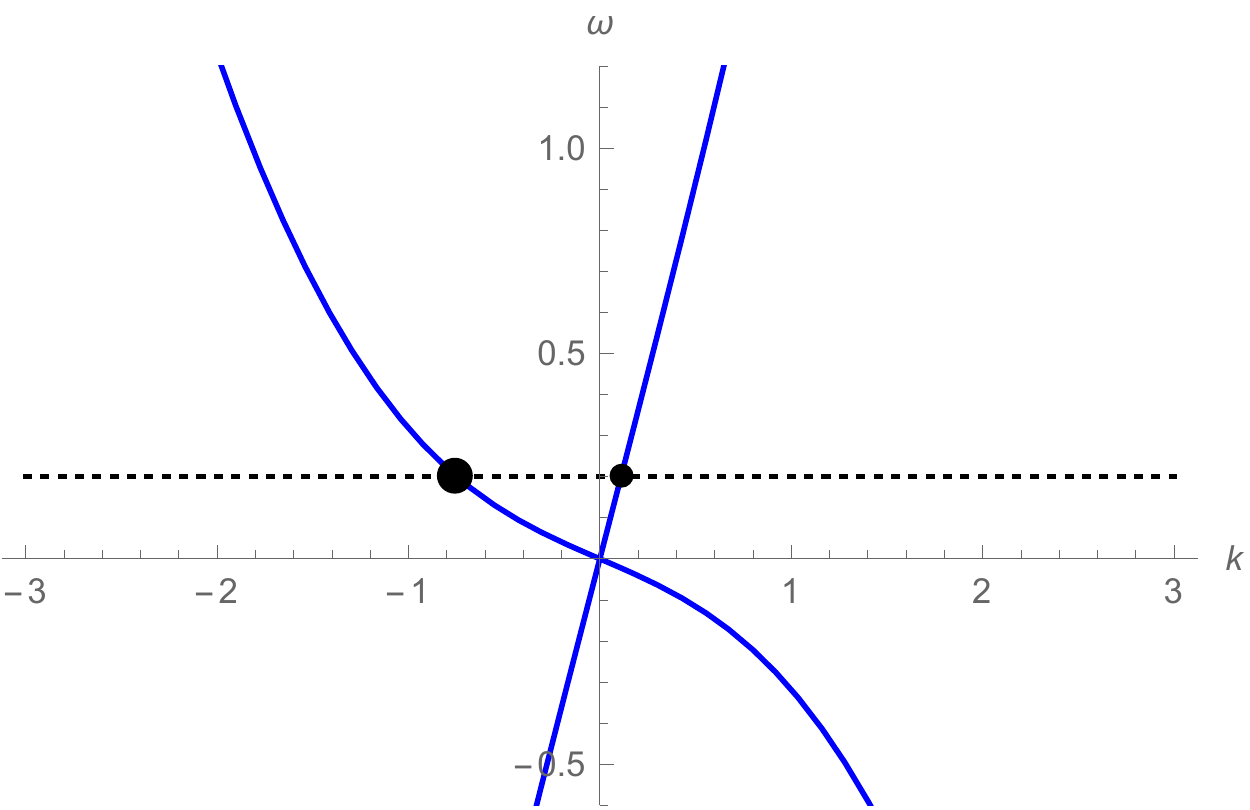} \, 
\includegraphics[width=0.49 \linewidth]{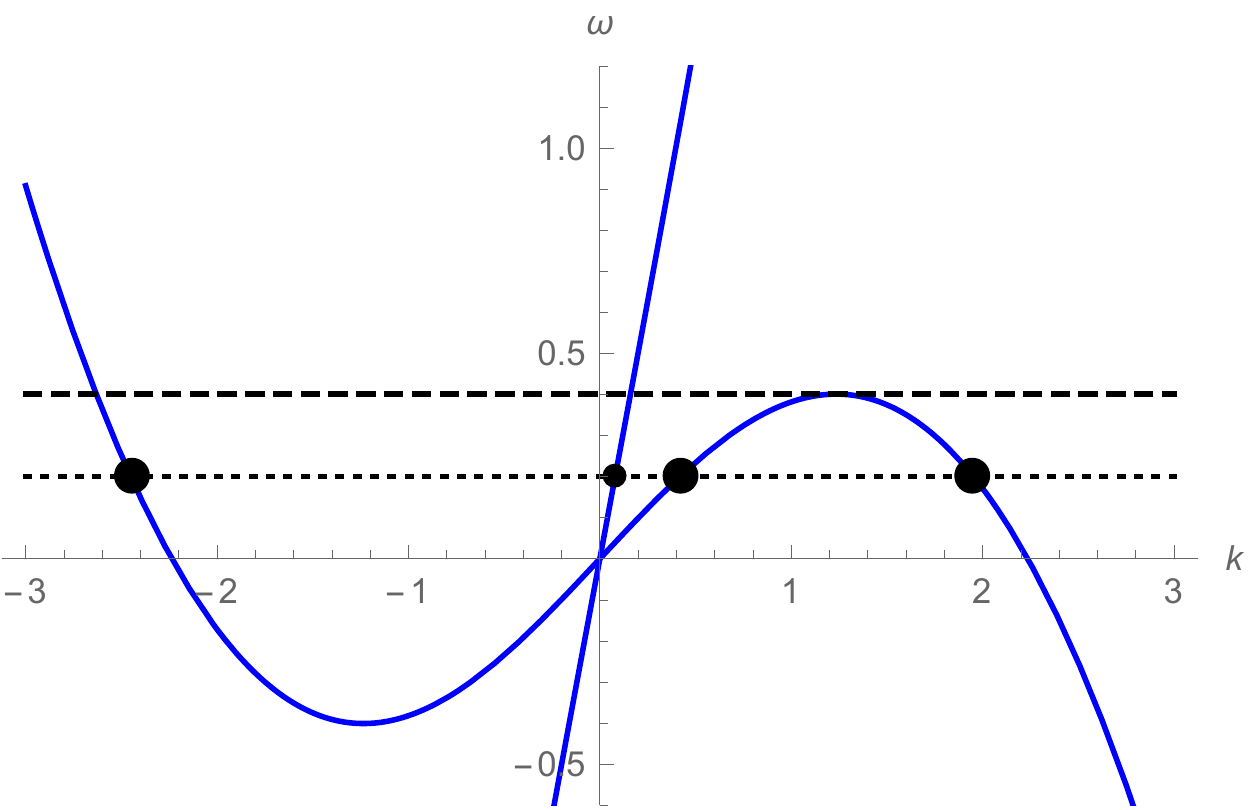}
\caption{Dispersion relation $\om$ versus $k$ of \eq{eq:disprelBdG} for $g \rho= 1$ and $v=0.8$ (left) and $1.5$ (right). The horizontal dashed line on the right plot shows the critical frequency $\om_{\rm max}$. The dotted line shows $\om = \om_{\rm max} / 2$. Large dots show the wave vectors of the counter-propagating modes (in the fluid frame), which give the dominant part of the scattering. The smaller dot on each panel shows that of the co-propagating mode, which only mildly affects the scattering~\cite{Macher:2009tw}. On the right panel, the two external roots with $k \sim -2.5$ and $k \sim 2$ describe dispersive (short-wavelength) waves. The 4 other roots describe long-wavelength phonic excitations, and propagate with a group velocity (in the lab frame) given by $v_{\rm gr} \approx v \pm c$ at small frequencies. 
} \label{fig:DR}
\end{figure}

We consider some initial perturbation $\phi_{\rm ini}(x) = \phi(x,t=0)$ and look for its late-time behavior. We assume that $\phi_{\rm ini}$ decays fast enough at infinity so that the coefficients of its expansion into incoming modes are finite. In each of the asymptotic regions, $\phi(x,t)$ has the form
\begin{align}
\phi(x,t) = \int \sum_i \lp \mathcal{U}_{k_i} \e^{\ii (k_i x - \om t)} + \mathcal{V}_{k_i}^* \e^{-\ii (k_i x - \om t)}  \rp \dd \om,
\end{align}
where the subscript $i$ labels the branches of the dispersion relation. The coefficients $\mathcal{U}_{k_i}$ and $\mathcal{V}_{k_i}$ are determined by the overlap between $\phi_{\rm ini}$ and incoming modes, and by the expansion of the latter into asymptotic plane waves. Possible divergences in the expansion into outgoing modes thus come only from those of the 9 Bogoliubov coefficients. These were computed analytically in the steplike regime in~\cite{Finazzi:2012iu} and numerically for a smooth flow in~\cite{Macher:2009nz}. In both cases, it was found that the only divergence occurs for $\om \to 0$, and only affects the coefficients relating long-wavelength to short-wavelength modes. However, as explained in subsection~\ref{App:linKdV}, the leading terms in these coefficients do not contribute to the amplitude of $\phi_0$  as they come in pairs with opposite signs which cancel each other. At late times, $\phi$ can be computed using a saddle-point approximation. The saddle points are located where
\begin{align}
\frac{\dd \varphi}{\dd \om} + \frac{\dd k_i}{\dd \om} x - t = 0.
\end{align}
In this expression, $\varphi$ represents the phase of the prefactor ($\mathcal{U}_{k_i}$ or $\mathcal{V}_{k_i}$). In the limit $t \to \infty$ at fixed $x$, saddle points correspond to $\left\lvert d k_i / d \om \right\rvert \to \infty$, 
i.e., where the group velocity vanishes. (If the initial conditions are smooth, divergences in $\frac{\dd \varphi}{\dd \om}$ can arise only from terms in $k_j x_{i,j}(\om)$ in the expression of $\varphi$, where $x_{i,j}(\om)$ is a smooth function coming from the integral giving the overlap, analogous to \eq{eq:A} for the KdV equation. Hence the divergences of  $\frac{\dd \varphi}{\dd \om}$ can only arise through divergences of $\frac{\dd k_i}{\dd \om}$.) Such divergence comes only at $\om = \pm \om_{\rm max}$, for the two roots which merge there. 
This confirms that the dominant waves at late times have a frequency near $\om_{\rm max}$, in accordance with 
the numerical results of \fig{fig:lin3/2}.
A straightforward calculation shows that close to this point $dk_i/d\om$ diverges as $\lp \om_{\rm max} - \abs{\om} \rp^{-1/2}$. Performing a Gaussian integration, we obtain that the dominant waves decrease as $t^{-3/2}$, hence the decay of $\delta \rho^2$ ads $t^{-3}$.   
In principle, this linear analysis is valid provided ${\rm sup}_{x \in\mathbb{R}}\abs{\phi(x)} \ll 1$ at all times. However, numerical simulations indicate that the late-time behavior remains the same even for ``large'' initial perturbations, with ${\rm sup}_{x \in\mathbb{R}}\abs{\phi(x)}$ close to $1$ (see Fig.~\ref{fig:DR}, right panel). This is due to the dilution of the perturbation: as explained in the next section, initial nonlinear perturbations decay in time over any finite interval of $x$. Hence the perturbation close to the horizon effectively enters the linear regime at late times. 
In brief, numerical simulations and the above discussion establish that the AH black hole flows are 
linearly stable and that, in the vicinity of the sonic horizon, all linear perturbations decay as $t^{-3/2}$. This is very reminiscent to the no-hair theorem of general relativity~\cite{Misner1973}.

To conclude this subsection, we briefly remind the link between the decay we obtained and that exhibited by free scalar fields propagating in a black hole metric. 
The correspondence between phonons in a Bose-Einstein condensate and excitations of a scalar field in a black-hole spacetime is derived in more details in Chapter~\ref{chap:intro}. 
Here we only stress the important points for understanding the aforementioned observations.
The comparison should be done at two different levels, namely when including or not the dispersive terms of \eq{eq:disprelBdG}.
When ignoring dispersion effects and the quantum pressure, the simplest way to proceed consists in separating $\phi$ into its real and imaginary parts~\cite{Barcelo:2005fc}. Straightforward algebra shows that the imaginary part of $\phi$ obeys
\begin{align}
\partial_\mu \lp F^{\mu \nu} \partial_\nu \Im \phi \rp = 0,
\label{KGe}
\end{align}
where $\lp \mu,\nu \rp \in \lb 0, 1 \rb^2$, and where Einstein's summation conventions are used. 
(See Chapter~\ref{chap:intro} for more details.) 
The matrix $F^{\mu \nu}$ is given by
\begin{align}
F^{\mu \nu}(x) = 
\begin{pmatrix}
1 & v_0(x) \\ 
v_0(x) & v_0^2(x) - c_0^2(x)  
\end{pmatrix},
\end{align}
where we introduced the local value of the sound speed $c_0(x) = \sqrt{g(x) \rho_0(x)}$. Ignoring some subtleties related to the conformal symmetry in 1+1 dimensions, \eq{KGe} is the d'Alembert equation in a stationary black hole with line element given by
\begin{align}
d s^2 = c_0^2(x) d t^2 - (d x - v_0(x) \, d t)^2. 
\end{align}
When the flow is transonic, it describes a black hole metric with an event horizon located where $|c_0(x)| = |v_0(x)|$. The analogy goes deeper as the ``analog'' surface gravity, given by $\kappa = \partial_x (c_0 - v_0)$ (evaluated on the horizon), governs the red-shifting of the solutions of \eq{KGe}
which propagate against the flow. That is, the wave number $k$ of a localized wave packet leaving the near-horizon region is red-shifted in time, following $k = k_0 \e^{- \kappa t}$~\cite{Macher:2009nz} (where $k_0$ is a constant), as found in General Relativity.
At the classical level, this guarantees that the decay of the solutions of \eq{KGe} will be similar to that 
found in general relativity~\cite{Misner1973}. Similarly, at the quantum level, 
the scattering of vacuum fluctuations on the sonic horizon should produce a thermal flux 
with a temperature given by the standard expression $k_B T_H =  \hbar \kappa/2 \pi$~\cite{Unruh:1980cg}. 

\begin{figure} \centering
\includegraphics[width=0.5 \linewidth]{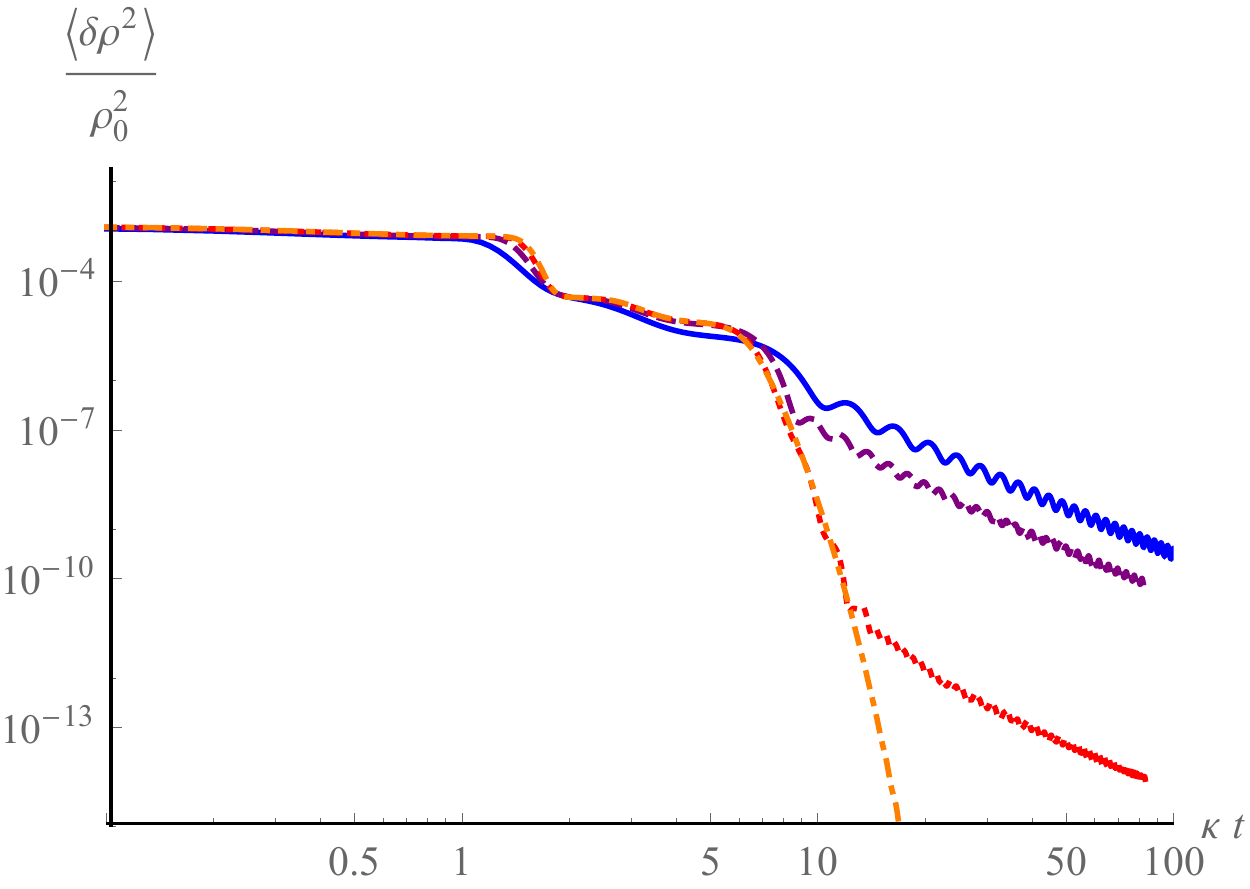}
\caption{We show the evolution of the squared relative density perturbation for a perturbation initially centered on $x = 0$ with the same parameters as in the left plot of \fig{fig:lin3/2}. We consider four different background flows
such that the quantities entering \eq{KGe} are held fixed. 
The four values of $\kappa/\omega_{\rm max}$ are $1.3$ 
(blue, continuous), $0.65$ (purple, dashed), $0.33$ (red, dotted), and $0.15$ (orange, dot-dashed). The functions $g$ and $\mu$ are given by \eq{eq:tanh}, where the asymptotic values are the same as in \fig{fig:lin3/2} and $\sigma = 0.56 c_0(x=0) / \kappa$.
}
\label{fig:varommax}
\end{figure}

Yet, to reproduce the late-time decay law in $t^{-3/2}$, the dispersive effects of \eq{eq:disprelBdG} must be taken into account, as they govern the behavior near the critical frequency $\omega_{\rm max}$.\footnote{When including them, the separation into real and imaginary parts of $\phi$ is no longer useful as they now couple to each other, see Eq.~(A6) in~\cite{Macher:2009nz}. This coupling complicates the relationship between the present settings and those characterizing dispersive fields in the context of alternative theories of gravity which break the local Lorentz invariance at short distances, see e.g.~\cite{Jacobson:2010mx}.
The interested reader might consult the appendixes A and B of \cite{Macher:2009nz} for further discussions.} 
To clarify the role of this frequency, it is interesting to consider the decay represented in Fig.~\ref{fig:lin3/2} when sending $\omega_{\rm max} / \kappa \to \infty$ while keeping fixed all quantities appearing in \eq{KGe}. In \fig{fig:varommax} we represent the decay of the averaged squared density fluctuations for four different values of $\kappa/\omega_{\rm max}$, namely $1.3$ 
(blue, continuous), $0.65$ (purple, dashed), $0.33$ (red, dotted), and $0.15$ (orange, dot-dashed). To compare the four cases, the time is given in units of $\kappa^{-1}$. For the same reason, the vertical axis represents the 
squared relative density perturbation $(\delta \rho/\rho_0)^2$, averaged over a domain which scales as $c_0/\kappa$ in our unit system, see \eq{eq:GPE}. One can see see that decreasing $\kappa/\omega_{\rm max}$ leaves the early evolution unchanged. This means that the decay 
of linear density perturbations at early times is governed by the relativistic equation \eqref{KGe}. 
At late times we find the decay law in $t^{-3}$. We verified that the value of $(\delta \rho/\rho_0)^2$ when it starts to decay in $t^{-3}$ is  proportional to the squared of the Fourier component of the initial value of the relative
density perturbation with the critical wave vector $k_{\omega_{\rm max}}$. Hence, even when working at fixed $\kappa/\omega_{\rm max}$, the evolution of very smooth perturbations (i.e., containing only Fourier component with $\abs{k} \ll k_{\omega_{\rm max}}$) is governed by \eq{KGe}. This is guaranteed by the fact that the propagation of the perturbation in the black hole flow redshifts the wave vectors $k$. In brief, \fig{fig:varommax} establishes that the dispersive effects of \eq{eq:disprelBdG} reduce the decay rate of perturbations near a black hole horizon when compared with the two-dimensional relativistic result, but does not eliminate it. 

\section{Nonlinear stability of black hole flows}
\label{sec:NLstab} 

\subsection{Analytical results}
\label{sec:analytical} 

We now turn to the nonlinear evolution of perturbations on black hole flows. In this subsection, we show analytical results obtained using Whitham's modulation theory~\cite{Whitham1}. Numerical results, which do not rely on the assumptions of this method, are shown afterwards. The reader interested in the derivation of Whitham's equations for the problem at hands may consult Section~\ref{App:Whitham} and the textbook~\cite{Kamchatnov}. 

Whitham's modulation theory gives a general framework for finding approximate quasiperiodic solutions of (quasi) integrable nonlinear partial differential equations. The solutions we consider oscillate on a short scale while their amplitudes, mean values, and wavelengths vary on much larger ones. The key idea is to average some of the conservation laws of the equation over the fast scale to obtain coupled equations for the slow evolution of a set of effective parameters, called \textit{Riemann invariants}, which describe the solution locally. It is particularly useful when the equation is integrable by the inverse scattering method: as explained in Section~\ref{App:Whitham}, the AKNS scheme then provides a general way to find the set of Riemann invariants. To the best of our knowledge, \eq{eq:GPE} with $x$-dependent functions $V$ and $g$ is not integrable. However, since Whitham's equations are local, when working with $V$ and $g$ of \eq{eq:step}, one can use the integrability for uniform $V,g$ to determine the solutions on each side of the discontinuity at $x=0$. The globally-defined solutions are then obtained by using the matching conditions, i.e., continuity of $\psi$ and $\pd_x \psi$. 

We look for solutions characterized by asymptotic densities $\rho_\pm$ and velocities $v_\pm$ as $x \to \pm \infty$ different from those of the homogeneous solution \eq{eq:rho0}. Our aim is to show that, for a wide range of values of $(\rho_\pm, v_\pm)$, the solution locally converges to that particular solution. Doing so we shall first extend the stability analysis of the previous section by including nonlinear effects, as well as perturbations extending to infinity. Second, we shall obtain the main features of the emission process which progressively replaces the initial configuration by the homogeneous and stationary black hole flow of \eq{eq:rho0}. We shall see that, if the initial perturbation is not too large, three macroscopic, nonlinear, scale-invariant waves are emitted. These can be seen as the result of a nonlinear stimulated Hawking radiation. As shown below, these features are well reproduced by numerical simulations and, along with the above linear analysis, they provide a precise description of the late-time behavior of the solution when approaching the homogeneous black hole flow. They strongly suggest that the set of solutions given by \eq{eq:rho0} acts as a local attractor. 

In this work we consider approximate solutions described by at most 4 Riemann invariants. Since the 
steplike $x$-dependence of $V$ and $g$ introduces no scale, one can further restrict our attention to scale-invariant solutions, for which the Riemann invariants depend only on $z \equiv x/t$. Our goal is to find the domain of parameter space in which the time-dependent solution interpolates between given asymptotic values of $\rho$ and $v$ as $x \to \pm \infty$ and a homogeneous black hole solution in some spatial interval $I_t$ which contains the origin $x=0$ and grows linearly in time. That is, we look for the solutions of the Whitham equations~(\ref{eq:Whz}) describing functions $\rho$ and $v$ such that
\begin{itemize}
\item $\displaystyle{\forall t \geq 0, \, \rho(x,t) \mathop{\to}_{x \to + \infty} \rho_+ \wedge v(x,t) \mathop{\to}_{x \to + \infty} v_+ }$,
\item $\displaystyle{\forall t \geq 0, \, \rho(x,t) \mathop{\to}_{x \to - \infty} \rho_- \wedge v(x,t) \mathop{\to}_{x \to - \infty} v_- }$,
\item $\displaystyle{\exists I \subset \mathbb{R}, \, 0 \in \mathring{I}}, \, \forall t > 0, \, \forall x \in \mathbb{R}, \, \frac{x}{t} \in I \Rightarrow \rho(x,t) = \rho_0$.~\footnote{We remind that $\mathring{I}$ denotes the interior of the set $I$. That is, $0 \in \mathring{I}$ is equivalent to saying that there exists $\epsilon > 0$ such that $\left[ -\epsilon, +\epsilon \right] \subset I$.}
\end{itemize} 
Such global solutions can be built using two types of nonlinear waves found when $g,V$ are uniform: dispersive shock waves (DSW) and simple waves (SW), see subsection~\ref{app:DSWNLS}. Two examples are shown in \fig{fig:solsNLS}. Along a SW, $\rho(t,x)$ and $v(t,x)$ are monotonic functions of the sole variable $z$. They are related through 
\begin{align}
z = v \pm \sqrt{g \rho},
\label{vgr}
\end{align}
where the sign $\pm$ is positive for a right-moving wave (in the fluid frame) and negative for a left-moving one. On the other hand, a DSW interpolates between small-amplitude oscillations (which vanish at the edge of the wave) and a soliton. 

When the variations of the density and velocity along a SW or a DSW are small, the wave has a small amplitude and propagates with a velocity (in the laboratory frame) close to the group velocity $v_{\rm gr} = v \pm \sqrt{g \rho}$ of long-wavelength linear perturbations, in agreement with \eq{vgr}. For a SW, this result remains true whatever the amplitude because it can be described locally by a superposition of non-dispersive waves. For a DSW instead, dispersion plays an important role when the amplitude becomes large, hence the more complicated expressions for the velocity, see Eqs.~(\ref{eq:NLSzb},\ref{eq:NLSzp}). As a result, in a black hole flow, provided their amplitudes are not too large, these waves propagate {\it away} from the horizon. They may thus be seen as a nonlinear version of outgoing wave-packets produced by the scattering of the initial perturbation on the horizon. 
These outgoing waves are governed by the hydrodynamic roots of the dispersion relation, see Fig.~\ref{fig:DR}. 
\begin{figure}[ht]
\centering
\includegraphics[width=0.49 \linewidth]{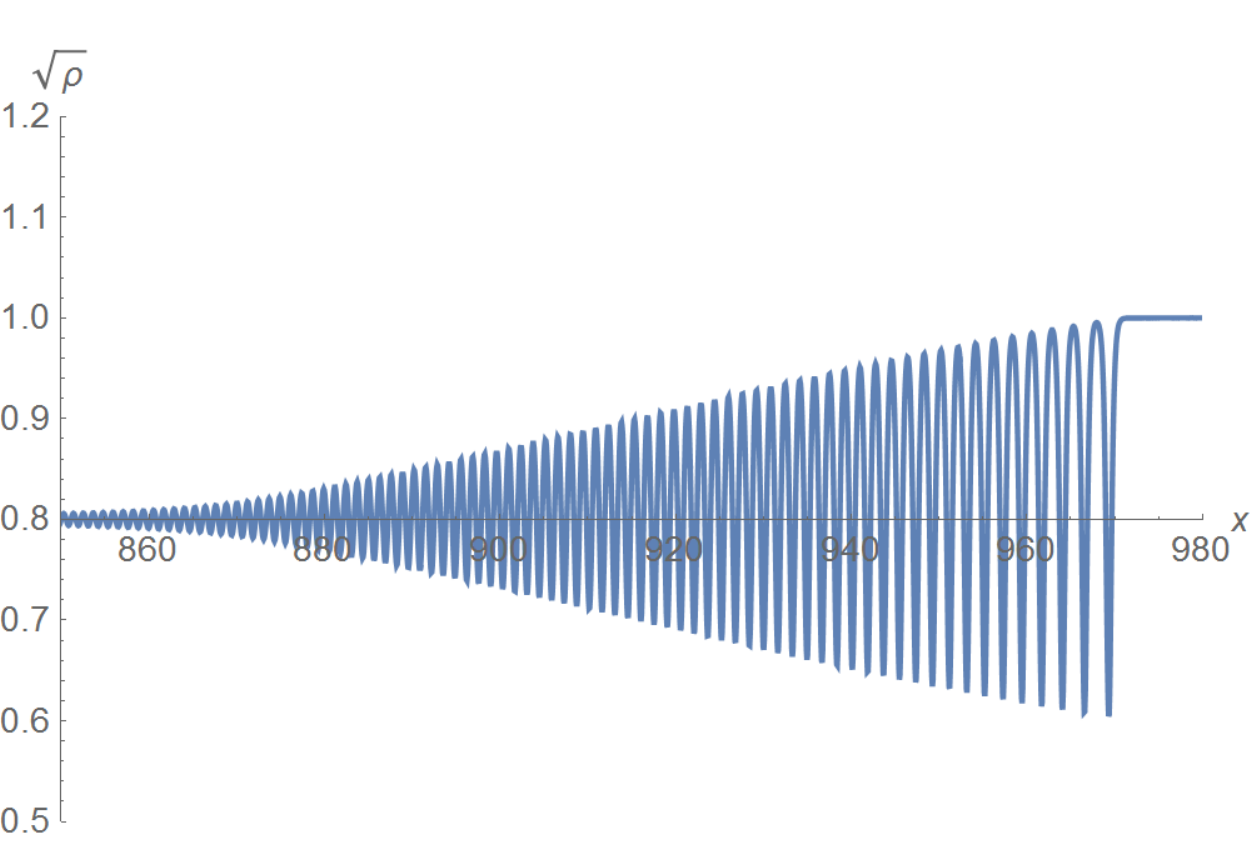}
\includegraphics[width=0.49 \linewidth]{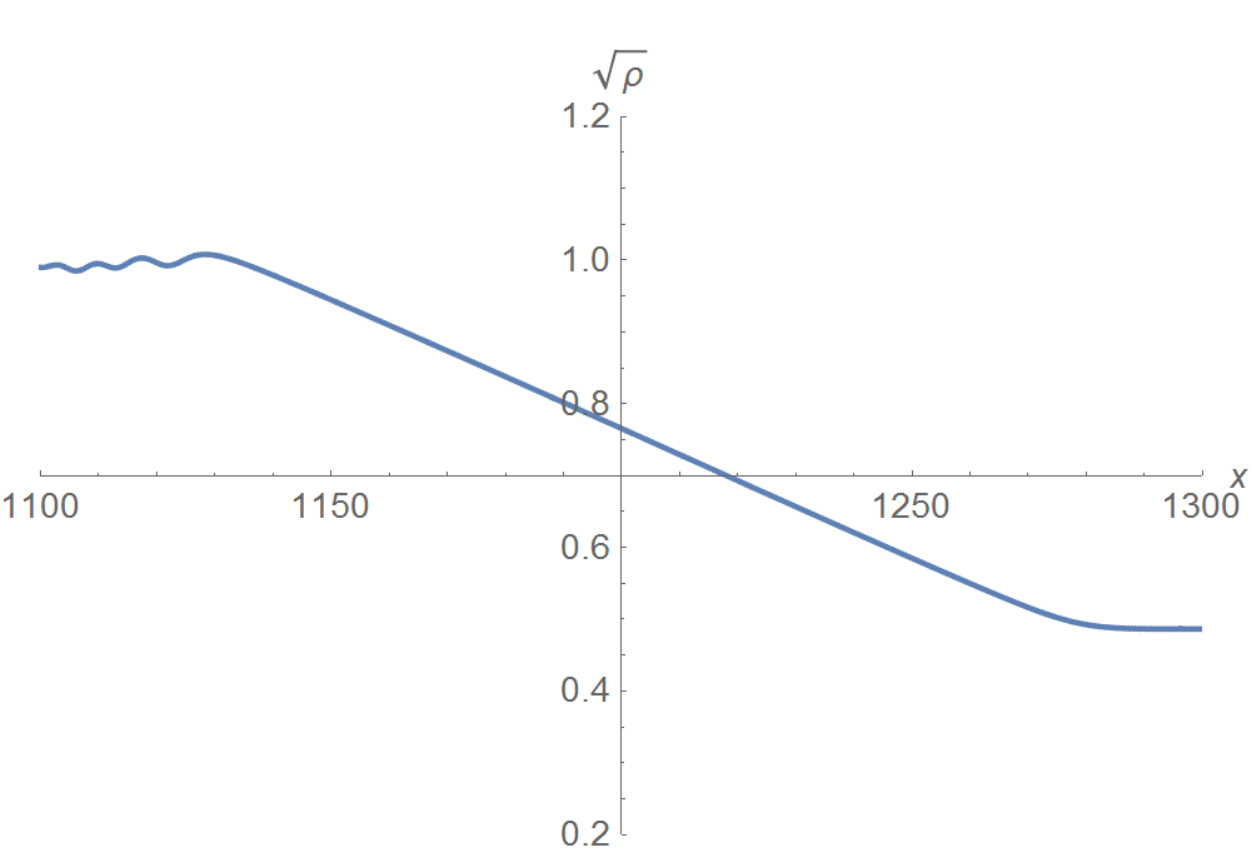}
\caption{We show the density profiles $\sqrt{\rho}$ of a DSW (left) and a SW (right). Both are computed numerically in a domain where $g$ and $V$ are constant. The SW is strictly scale-invariant, as along this solution $\rho$ and $v$ depend only on $x/t$ up to corrections not captured by the Whitham equations. For the DSW, the envelope and wavelength of the oscillations depend only on $x/t$, while the rapid oscillations move with velocity $s_1/2$, where $s_1$ is defined in \eq{eq:Pla}. 
The small oscillations on the right panel are due to residual finite-time effects.  
}\label{fig:solsNLS}
\end{figure} 

Global solutions can be obtained by matching exact solutions of the Whitham equation on each side of the point $x=0$, imposing continuity of $\psi$ and $\pd_x \psi$. Let us first study the case of symmetric asymptotic conditions, i.e., $\rho_+ = \rho_- \equiv \rho_i$ and $v_+ = v_- \equiv v_i$. We consider solutions with three waves (DSW or SW) in total, two of them propagating to the right and one moving to the left. This is motivated by the behavior of linear modes emitted by a black hole: at the linear level, two waves are emitted in the supersonic region and one in the subsonic region~\cite{Macher:2009nz}. Since the DSW and SW we are looking for are nonlinear versions of outgoing wave-packets, it is natural to assume they follow the same behavior. The validity of this hypothesis will come a posteriori from the existence of solutions when $\rho_i$ is sufficiently close to the density $\rho_0$ of \eq{eq:rho0}. 
\begin{figure}[ht]
\includegraphics[width=0.49 \linewidth]{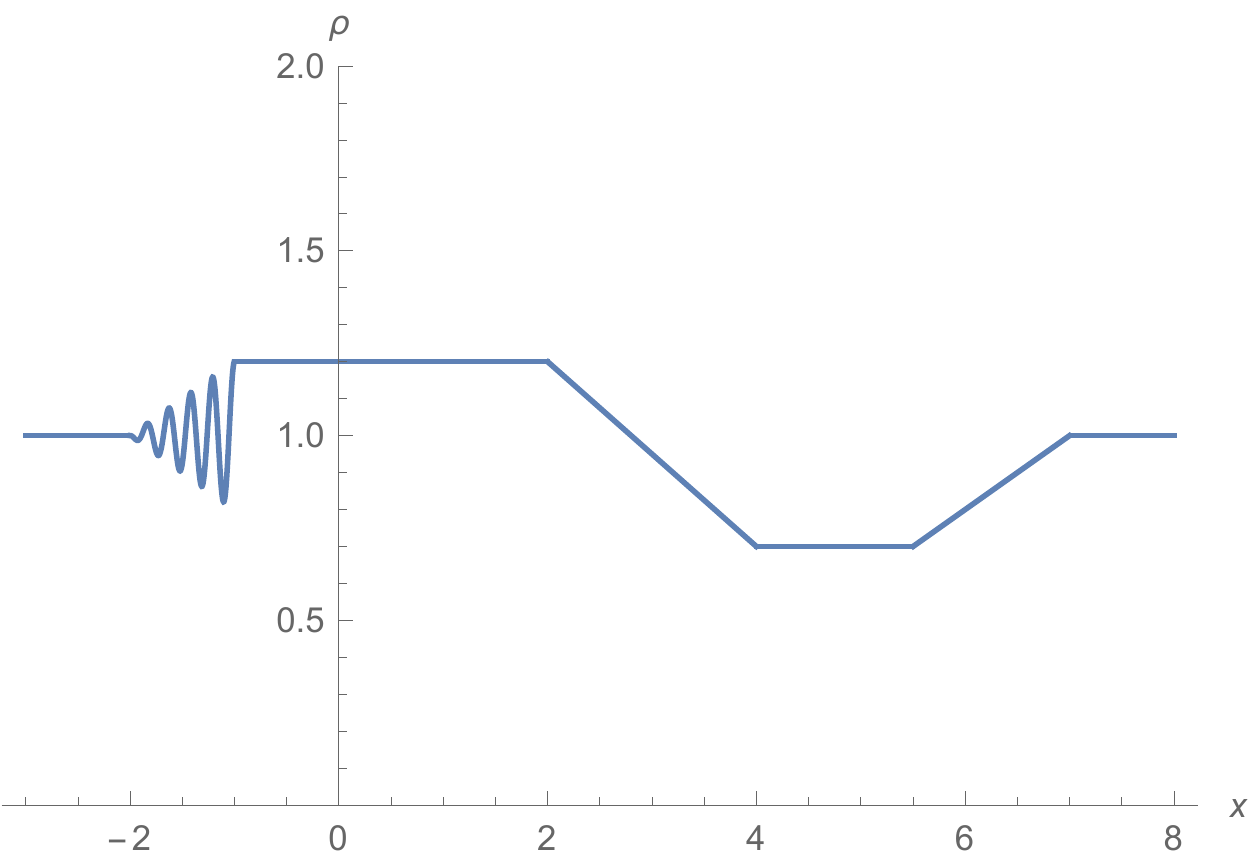}
\includegraphics[width=0.49 \linewidth]{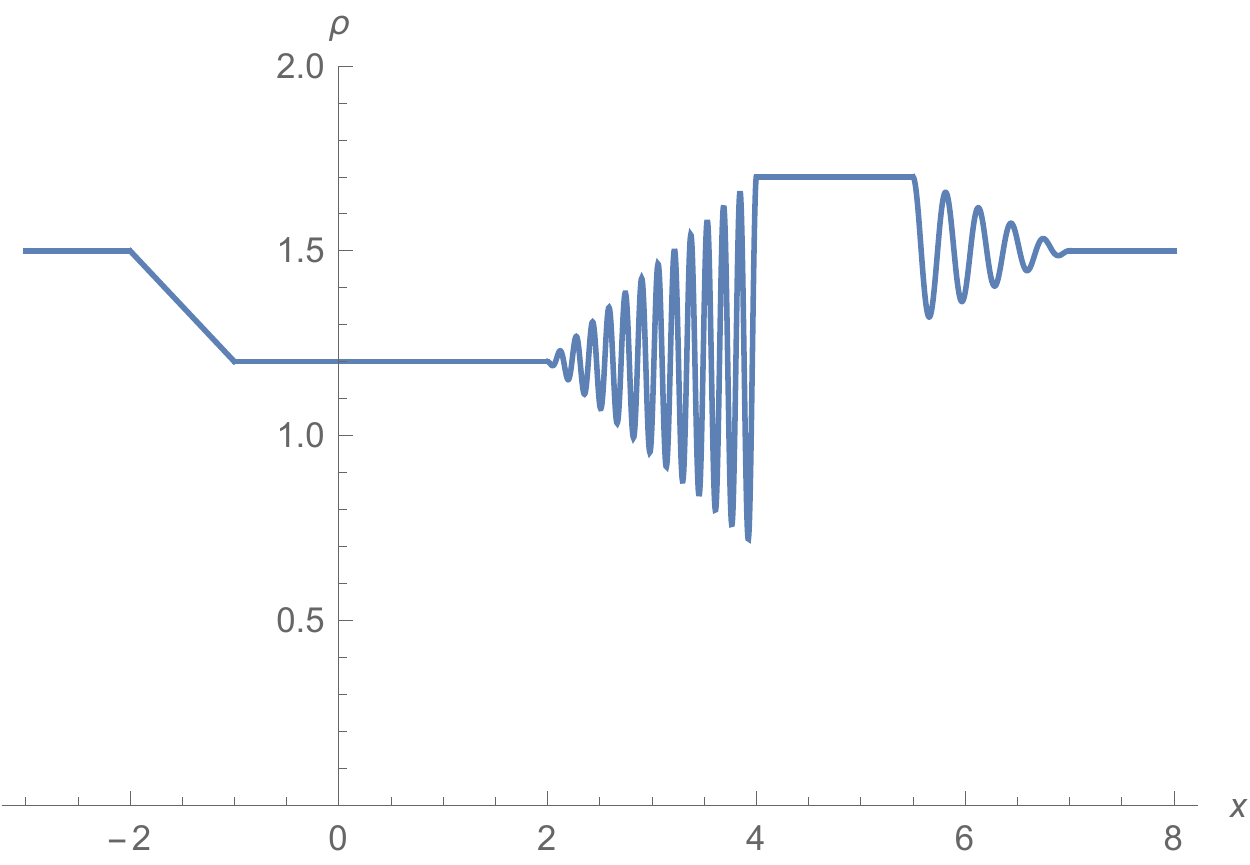}
\caption{Schematic drawing of two solutions of the Whitham equations corresponding to the emission of SW and DSW leaving behind them (in a region containing the horizon at $x=0$) the homogeneous black hole solution of \eq{eq:rho0}. On the left panel, we show the case with one DSW and two SW, which arises when $\rho_+ = \rho_- < \rho_0$. On the right panel, we show the other case with one SW and two DSW, arising when $\rho_+ = \rho_- > \rho_0$.}\label{fig:schema}
\end{figure}

Using the properties of SW and DSW outlined in subsection~\ref{app:DSWNLS}, we find two types of solutions depending on the sign of $\rho_i - \rho_0$. If $\rho_i - \rho_0 < 0$, the solution has one DSW for $z<0$ and two SW for $z>0$, separated by two homogeneous regions. A schematic plot is shown on the left panel of \fig{fig:schema}. On the left of the DSW and on the right of the two SW, the solution is (by construction) homogeneous with density $\rho_i$ and velocity $v_i$. Between the DSW and the leftmost SW, the density is equal to $\rho_0$ of \eq{eq:rho0}. This can be understood from the fact that $\rho_0$ is the only value of $\rho$ allowing to match two homogeneous solutions at $x=0$. The conservation law \eq{eq:cons2} fixes the final value $v_f$ of the velocity around $x=0$ in terms of the asymptotic conditions on the left side: 
\begin{align}\label{eq:vf} 
v_f = v_- + 2 \sqrt{g_-} \lp \sqrt{\rho_-} - \sqrt{\rho_0} \rp.  
\end{align}
In general, the late-time value of the current $J$ close to the horizon differs from the initial one, as a linearly-growing mass is carried by the nonlinear waves.

\begin{figure}
\centering
\includegraphics[scale=0.75]{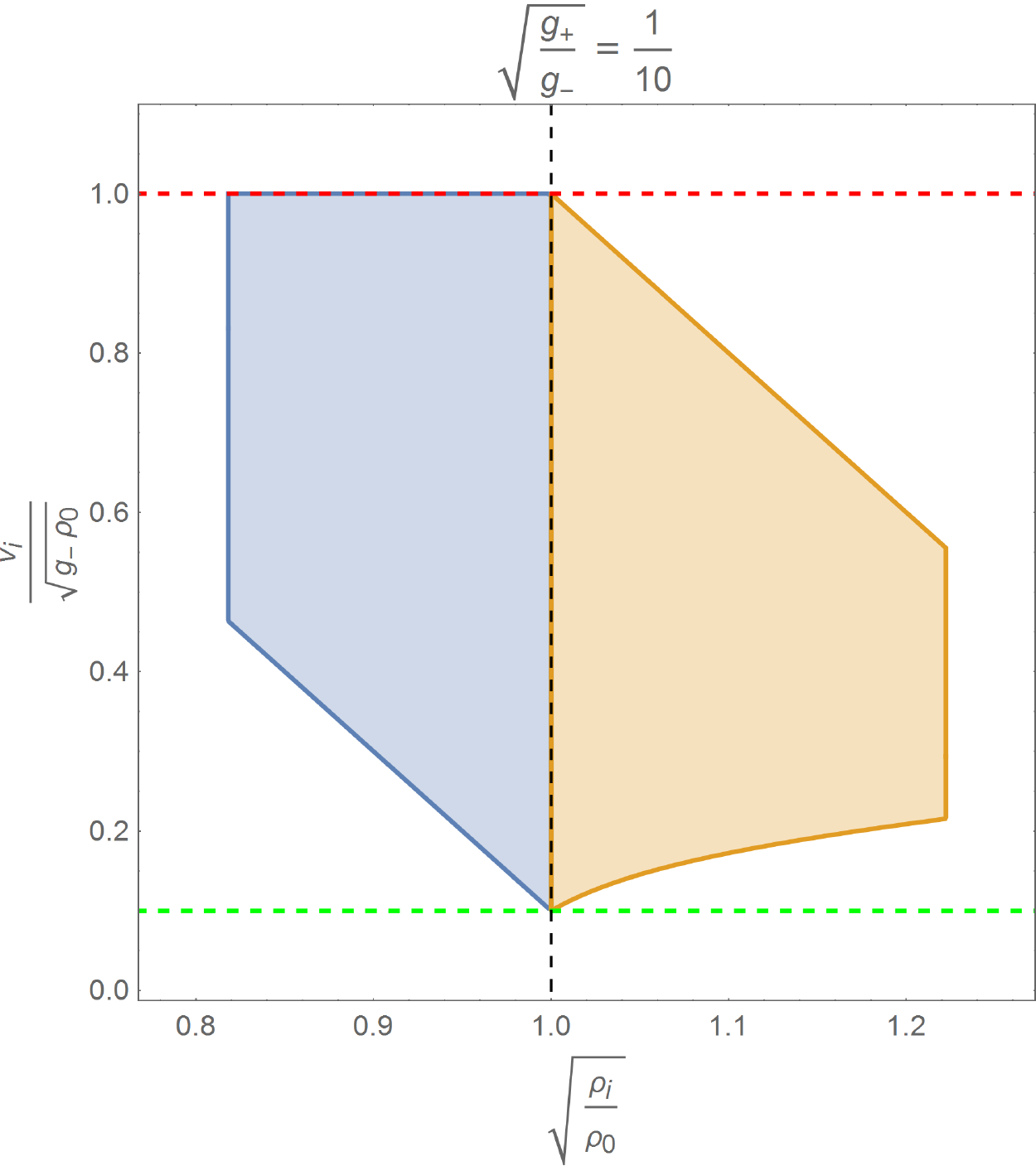}
\caption{We represent the domains of existence of the two solutions schematically shown in \fig{fig:schema}, for $g_- = 100 g_+$, in the plane $\lp \sqrt{\rho_i}, v_i \rp$ of asymptotic values of the perturbed solution. The large value of $g_- / g_+$ is chosen to show all the possible boundaries, the two vertical ones disappearing for $g_+ / g_- > 0.056$ (blue) and $g_+ / g_- > 0.02$ (orange). We work in adimensional units: $\rho_i$ is divided by the density of the homogeneous solution $\rho_0$ and $v_i$ by the sound velocity in the left subsonic region $c_b = \sqrt{g_-\, \rho_0}$. The blue region gives the domain of the solutions with one DSW and two SW, while the orange one gives that of the solutions with one SW and two DSW. The vertical dashed line shows the locus of $\rho_i = \rho_0$. The green horizontal dashed line gives $v_i = \sqrt{g_+\, \rho_0}$, while the red one gives $v_i = \sqrt{g_-\, \rho_0}$. They bound the domain where a homogeneous configuration $\rho = \rho_0$ has a black hole horizon at $x=0$. 
	}\label{fig:dom}
\end{figure}
We now consider the validity domain of this solution. We find that it exists if and only if the two following conditions are satisfied:
\be\label{eq:cond1DSW2SW} 
\frac{\sqrt{g_-}-\sqrt{g_+}}{\sqrt{g_-}+\sqrt{g_+}} &< &\sqrt{\frac{\rho_i}{\rho_0}} < 1 , \nonumber 
\\
\label{eq:cond1DSW2SWbis} 
\sqrt{\frac{g_+}{g_-}} + 2 \lp 1- \sqrt{\frac{\rho_i}{\rho_0}} \rp &<& \frac{v_i}{\sqrt{g_- \rho_0}} < 1.
\ee
This domain is shown in blue in \fig{fig:dom} for $g_- = 100 g_+$. (For steplike potentials, $V_+$ and $V_-$ only intervene in fixing the value of $\rho_0$.) The important point is that for $\rho_i \to \rho_0$ these inequalities reduce to $\sqrt{g_+ \rho_0} < v_i < \sqrt{g_- \rho_0}$. This condition is equivalent to saying that the left asymptotic region is subcritical and the right one is supercritical, see \eq{eq:BHcond}. We also notice that the blue domain pinches off when $v_i$ saturates its lowest bound. For lower values of $v_i$, the flow is globally subsonic. In brief, if the asymptotic conditions are such that the flow is transonic and of the black hole type, then there exists a unique solution with one DSW and two SW provided $\rho_0 - \rho_i > 0$ is not too large.

For $\rho_i >  \rho_0 $, one finds essentially the same results with one SW emitted to the left and two DSW emitted to the right. A schematic plot of the density profile is shown on the right panel of \fig{fig:schema}. The density and the velocity around $z=0$ are still given by $\rho = \rho_0$ and $v=v_f$ of \eq{eq:vf}. The conditions of existence of this solution are
\begin{align}\label{eq:cond1SW2DSW} 
1 < \sqrt{\frac{\rho_i}{\rho_0}} &< \frac{\sqrt{g_-} + \sqrt{g_+}}{\sqrt{g_-} - \sqrt{g_+}} ,
\nonumber
\\
2 \sqrt{\frac{g_+ \rho_i}{g_- \rho_0}} - \frac{g_+/g_-}{\sqrt{\frac{\rho_i}{\rho_0}}+ \sqrt{\frac{g_+ \rho_i}{g_- \rho_0}}-1} &< \frac{v_i}{\sqrt{g_- \rho_0}} < 3 - 2 \sqrt{\frac{\rho_i}{\rho_0}}.
\end{align}
The argument below \eq{eq:cond1DSW2SW} can also be applied here, 
giving that for $\sqrt{g_+ \rho_0} < v_i < \sqrt{g_- \rho_0}$ this solution exists provided $\rho_i - \rho_0 > 0$ is not too large. The corresponding domain is shown in orange in \fig{fig:dom}.

The same analysis can be carried out with different asymptotic conditions $\rho_+ \neq \rho_-$, $v_+ \neq v_-$. We must then consider 6 additional types of solutions. 4 of them are obtained from the above ones by replacing one of the two DSW by a SW or conversely. One solution has three SW. The last one has three DSW. The domain of existence of each of these solutions is given in subsection~\ref{App:domex}. There, it is also shown that one of these 8 solutions always exists in a neighbourhood of any set of asymptotic conditions compatible with a homogeneous black hole flow, i.e., $\rho_+ = \rho_- = \rho_0$ and $v_+ = v_- \in ]\sqrt{g_+ \rho_0},\sqrt{g_- \rho_0}[$.

To summarize, when working with the steplike $V$ and $g$ of \eq{eq:step}, we have shown that sufficiently small initial perturbations are expelled at infinity, leaving at late times the homogeneous flow $\rho(x) = \rho_0$. The set of all such solutions therefore acts as a \emph{local} attractor, in the sense that the solution $\rho(t,x)$ and the corresponding velocity profile $v(t,x)$ uniformly converge to $\rho_0$ and $v_f$ of \eq{eq:vf} over any bounded interval. To make this claim more precise, we propose the following \\ \\
{\bf Conjecture.} {\it
	There exist two sets of strictly positive real numbers $\{R_{n}\in\mathbb{R}_+^*:n=0,\dots, p\}$ and $\{V_{n}\in\mathbb{R}_+^*:n=0,\dots, p\}$ such that for every initial data $\rho_i,v_i \in C^p(\mathbb{R})$ satisfying the three conditions:
\begin{enumerate}
	\item $\rho_i$ and $v_i$ are homogeneous outside of some bounded interval, with the asymptotic values $v_\pm=\lim_{x\to\pm\infty}v_i(x)$ such that $\sqrt{g_+ \rho_0} > v_\pm > \sqrt{g_- \rho_0}$;
	\item $\|\rho_i-\rho_0\|_{\mathbb{R},\infty} < R_0$ and $\|v_i - v_f\|_{\mathbb{R},\infty}< V_0$;
	\item $\|\pd_x^n \rho_i\|_{\mathbb{R},\infty} < R_{n}$ and $\|\pd_x^n v_i\|_{\mathbb{R},\infty}< V_{n}$	for every $n\in\{1,2,\dots, p\}$;
\end{enumerate}
we have, for every bounded interval $I\subset\mathbb{R}$, $$\lim_{t\to+\infty}\|\rho(t)-\rho_0\|_{I,\infty}=0 \quad \text{and} \quad\lim_{t\to+\infty}\|v(t)-v_f\|_{I,\infty}=0.$$ } \\
We have used here the standard notation $\|f\|_{I,\infty} \equiv \sup_I|f|$ for any function $f:\mathbb{R}\to\mathbb{C}$ and any interval $I \subset \mathbb{R}$. While it is sufficient for having the analogue of the asymptotic flatness condition on the initial data, condition 1. may well be too restrictive. We expect that an exponential convergence, or even maybe polynomial ones, should be enough. Similarly, the sufficient level of regularity of the initial data -- i.e. the actual value of $p\in\mathbb{N}\cup\{\infty\}$ -- is not entirely clear to us at this stage, with $p=2$ and $p=\infty$ as natural candidates. Conditions 2. and 3. are, despite their dependence on the choice of $p$, the important point, providing a sense in which the initial perturbation is small enough. We hope to be able to sort these questions out in a future work. Let us nonetheless emphasize that all the numerical simulations presented in the next section support the above conjecture. They also indicate that the conjecture should hold when replacing the steplike functions of \eq{eq:step} by smooth ones, such as those of \eq{eq:tanh}. 
Finally we conjecture that, under the above three conditions, the late-time properties of the three nonlinear waves moving away from the horizon should only depend on the asymptotic initial conditions $v_\pm$ and $\rho_\pm$. In other words, the Fourier components of the smooth profiles $v_i(x)$ and $\rho_i(x)$ are diluted away at very late time, as was rigorously shown in the linear treatment in the former section, and as was also found in our 
simulations, see below.  

To conclude this subsection, we notice that the present analysis does not apply to white hole flows. Indeed, a crucial point in our calculations is that the three nonlinear waves move {\it away} from the discontinuity at $x=0$. For black hole flows, this is realized for the domain of initial conditions represented in Fig.~\ref{fig:dom}. In white hole flows instead, the nonlinear waves move towards $x=0$ in the supersonic region, see Section~\ref{Sec:WH}. When reaching this point, the Whitham theory breaks down. We then expect that white hole flows will show a more complex behavior than black hole ones, in accordance with~\cite{Michel:2015pra}. New results concerning white hole flows are presented in Section~\ref{Sec:WH}. 

\subsection{Numerical results}
\label{sec:numres} 

We numerically solved the GP equation for several reasons. First, by solving the GP equation directly, the results of the previous section can be checked without relying on the approximations of Whitham's modulation theory. Second, the differences in the emission process induced by smooth functions $V$ and $g$ can be studied. Finally, we wanted to study what happens outside the domain of existence of solutions with three waves. In all simulations, we used the same code as in~\cite{Michel:2015pra}. 

\begin{figure}[ht]
\centering
\includegraphics[width=0.49 \linewidth]{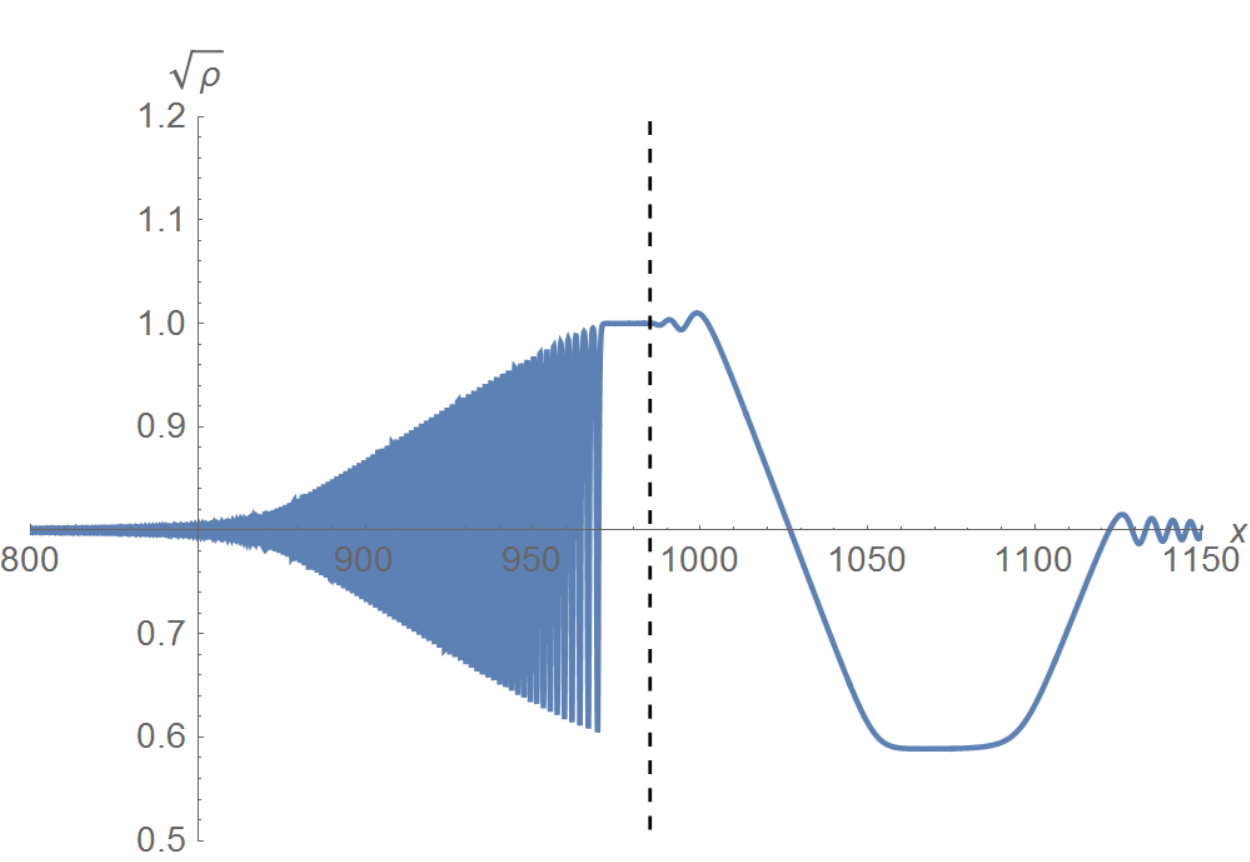} \
\includegraphics[width=0.49 \linewidth]{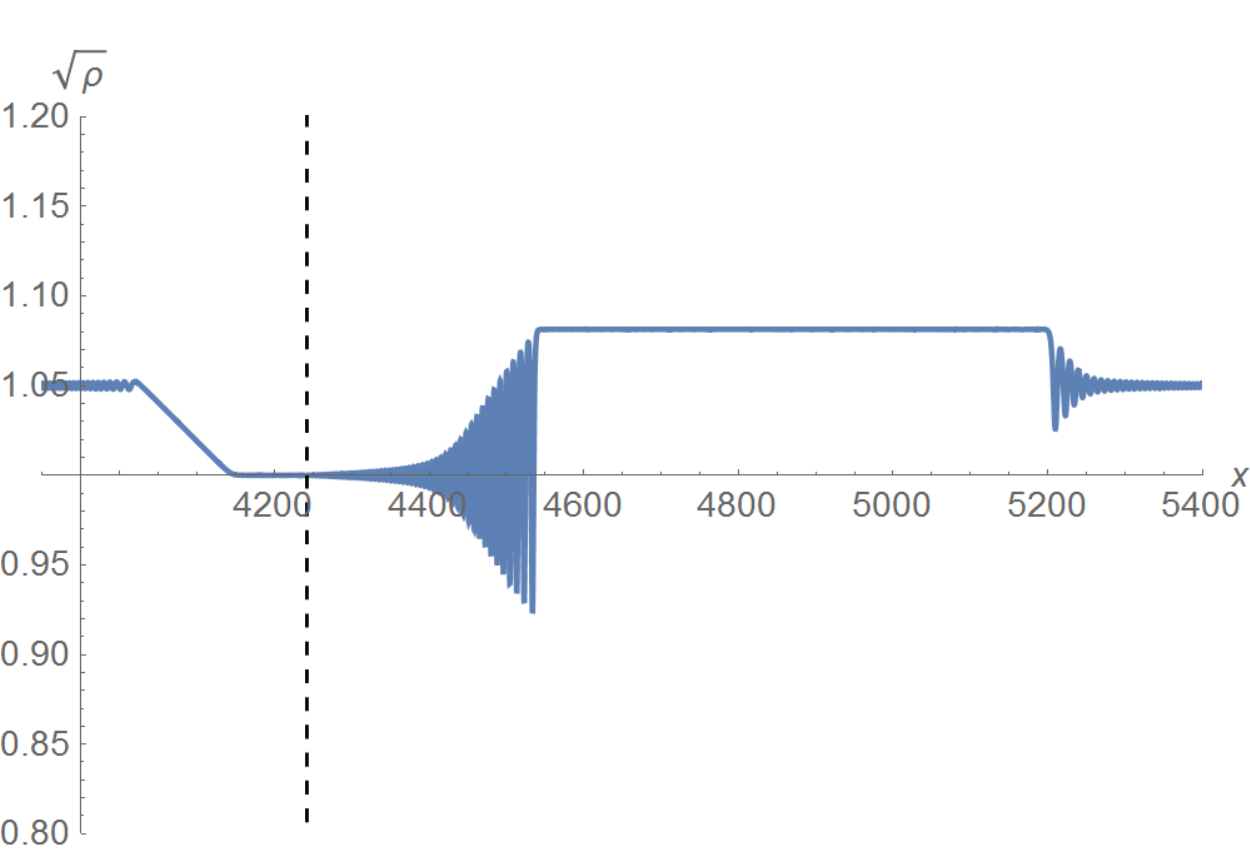}
\caption{Numerical solutions of the GP equation with one DSW and two simple waves (left) and one simple wave and two DSW (right). The vertical dashed line indicates the position of the horizon. In both cases, the initial configuration is homogeneous and $J=\sqrt{8/3}$. For the left plot, the initial density is $\rho_i=0.81$, $t=136$, and $g_-$, $g_+$, $\om - V_-$, and $\om - V_+$ are the same as in \fig{fig:lin3/2}. For the right plot, the initial density is $\rho_i =1.1$, the solution is shown at $t=400$, and the parameters are $g_- = 3.7$, $g_+ = 0.72$, $\om - V_- = 5$, and $\om - V_+ = 2.1$. They are chosen so that $\sqrt{\rho_{2,-}} = \sqrt{\rho_{3,+}} = 1$, $\sqrt{\rho_{2,+}} = 1.6$, and $\sqrt{\rho_{3,-}} = 0.9$.} \label{fig:solsNLSb} 
\end{figure}

In \fig{fig:solsNLSb}, we show two solutions obtained numerically when starting at $t=0$ from a configuration with homogeneous velocity $v_i$ and density $\rho_i \neq \rho_0 = 1$. On the left panel, $\rho_i$ and $v_i$ satisfy \eq{eq:cond1DSW2SW}, while on the right panel they satisfy \eq{eq:cond1SW2DSW}.
At late times, we observe that these solutions are superpositions of the three waves of Section~\ref{sec:analytical} plus perturbations whose amplitude decays in time as $t^{-{3/2}}$. We checked that the properties of the SW and DSW, in particular their amplitudes, the positions of their edges, and their domains of existence, agree with those given by Whitham's equations in the limit $t \to \infty$. 
We performed additional simulations by replacing the homogeneous initial conditions $v_i, \, \rho_i$ by smooth ones varying in a bounded domain, and we observed that this agreement was preserved. 
Moreover, the fact that perturbations still decay as $t^{-3/2}$ shows that the three macroscopic waves do not change the late-time behavior of perturbations obtained in Section~\ref{sec:lin} close to the horizon. 
It would be interesting to identify sufficient conditions on local variations of $v_i$ and $\rho_i$ ensuring that the late time properties of the three nonlinear waves only depend on the asymptotic values of $v_i$ and $\rho_i$. 

To complete our analysis, we replaced steplike $g$ and $V$ by smooth functions. We first chose a dependence in $x$ such that a solution with homogeneous density $\rho_0$ still exists, i.e., such that $V(x) + g(x) \rho_0$ is a constant. We worked with functions of the form
\begin{align}\label{eq:tanh}
& g(x) = \frac{g_+ + g_-}{2} + \frac{g_+ - g_-}{2} \tanh (x /\sigma), \nonumber 
\\
& V(x) = \frac{V_+ + V_-}{2} + \frac{V_+ - V_-}{2} \tanh (x /\sigma),
\end{align}
where $\sigma > 0$. When the asymptotic conditions are inside the domain described by Eqs.~(\ref{eq:cond1DSW2SW},\ref{eq:cond1SW2DSW}), we observed that the late-time properties of the solution are the same as in the steplike case: the homogeneous black hole solution is reached for $t \to \infty$ through the emission of three nonlinear waves similar to those of subsection~\ref{sec:analytical}.   
However, the typical formation time of the SW and DSW now depends on $\sigma$, and is linear in $\sigma$ for $\sigma \approx 0$. We verified that these results qualitatively extend to the case where $\sigma$ takes different values for $g$ and $V$. In that case, there exists no stationary solution with a homogeneous density. However, as discussed in subsection~\ref{sub:Ahtf}, there is a one-parameter family of solutions 
with angular frequencies $\omega(J)$  and asymptotically constant values of $\rho$. Our numerical results indicate that, if the initial conditions are sufficiently close to this series, one of its solutions is reached at $t \to \infty$ through the emission of three waves. In brief, our simulations confirm that initial configurations within the domain specified by Eqs.~(\ref{eq:cond1DSW2SW},\ref{eq:cond1SW2DSW}) all evolve, at late times, towards an AH solution.

We finally performed numerical simulations starting from initial conditions with asymptotic behaviors {\it outside} the domains described by Eqs.~(\ref{eq:cond1DSW2SW},\ref{eq:cond1SW2DSW}). These conditions can be violated in two different ways. In the first case, corresponding to crossing one of the two vertical boundaries in \fig{fig:dom}, the two waves in the supersonic region $x>0$ overlap each other, producing a complicate interference pattern. Yet, our simulations indicate that the solution still converges to a homogeneous black hole flow at late times, as the overlapping waves still escape to infinity. In the second case, one of the waves in the region $x>0$ (respectively $x<0$) has its left (respectively, right) edge moving to the left (respectively, right). [The wave with one edge moving towards the horizon lies in the region $x>0$ when crossing the lower boundary, or in the region $x<0$ when crossing the upper one. The sign of $\rho_i - \rho_0$ then gives its type (DSW or SW).] When the wave is a SW, the late-time solution contains part of a soliton or shadow soliton~\cite{Michel:2013wpa}, and asymptotes to an asymptotically homogeneous density different from $\rho_0$ on the corresponding side. 

When the corresponding wave is a DSW, the situation is more complicated. In the simplest case, the solution still becomes stationary at late times over any finite interval of $x$. It then contains a stationary density modulation attached to $x=0$. We also found cases where the solution apparently never reaches a stationary profile around $x=0$, instead emitting a density modulation with a non-vanishing phase velocity, which at the nonlinear level corresponds to a propagating soliton train. 
Our numerical investigation was not extensive enough to determine with confidence the conditions in which one or the other behavior occurs, although it seems that a stationary solution is reached when the DSW with an edge moving towards the horizon is on the left $x<0$, while the emission of soliton trains occurs when it is on the right $x>0$. 

\section{White hole flows}
\label{Sec:WH} 

In this section we briefly study the time evolution of perturbations on white hole flows. As discussed at the end of Section~\ref{sec:analytical}, there is an important difference between the evolutions of black and white hole flows: while the former expel perturbations at infinity, the latter have a tendency to accumulate them close to the horizon, 
as can be understood from the fact that white holes behave as the time-reversed of black holes. One can thus expect that the set of white hole flows will act as a ``repellor'' rather than an attractor. We here show that it is indeed the case. When starting from the homogeneous solution, depending on whether the perturbation gives rise to a (sufficiently large) decrease or an increase of the near horizon density, the flow is destabilized by nonlinear effects and either develops a single macroscopic undulation or sends a train of solitons accompanied by a macroscopic undulation. When working to linear order, one finds that
small perturbations generally leave at late times a stationary undulation with a large amplitude, 
thereby signaling an infra-red instability of the background flow. 
\subsection{Linear perturbations}

\begin{figure}
\centering
\includegraphics[width=0.49 \linewidth]{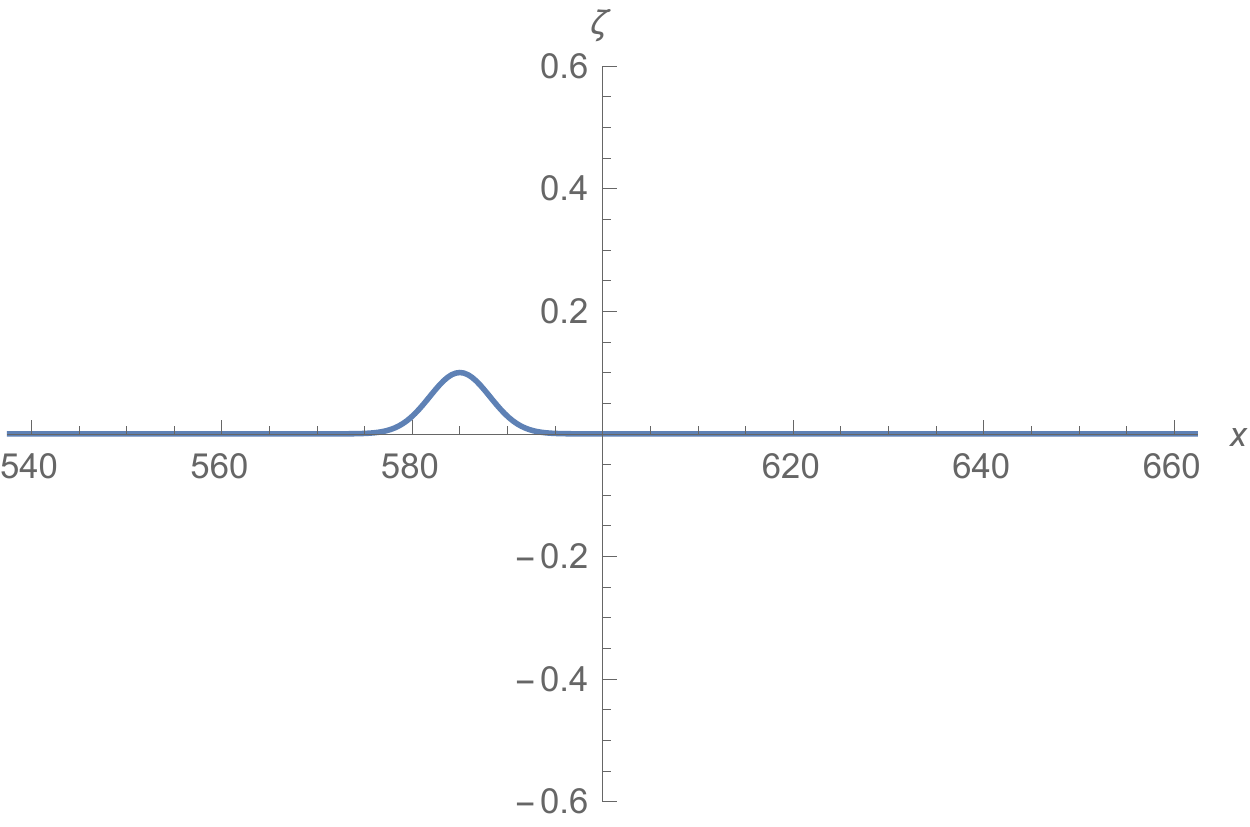} \,
\includegraphics[width=0.49 \linewidth]{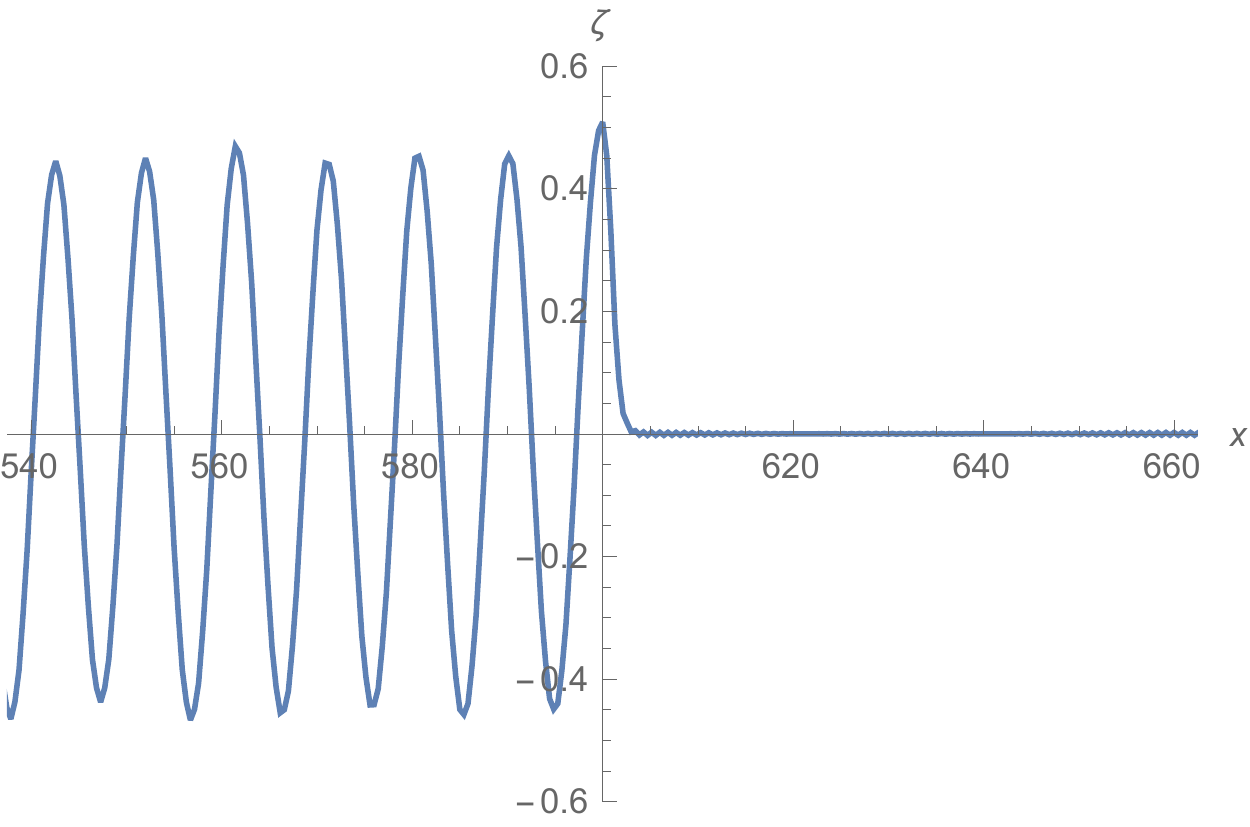}
\caption{Scattering of a localized perturbation on a white hole configuration for the linearized KdV equation. We work with $v = J / h$, where $J = -1$ (i.e., a flow to the left) and $h(x) = 2+\tanh(600 - x)$. The white hole horizon is located at $x = 600$. The initial perturbation is a gaussian with amplitude $0.1$ and width $10$. The left panel shows the initial perturbation of the water height and the right one shows the late-time undulation, which is here localized in the subsonic region because the dispersion relation is subluminal. Time-dependent effects are still visible through the small variations of amplitudes between two adjacent oscillations. }\label{fig:linKdVWH}
\end{figure}

Let us first consider white hole solutions of the KdV equation (see Section~\ref{App:KdV}),
as they are technically simpler to characterize. We work with functions $v$ and $h$ given by \eq{eq:ansatzvh}, with $v_-,v_+ < 0$, $v_- + \sqrt{g h_-} < 0$, and $v_+ + \sqrt{g h_+} > 0$. The trivial solution $\zeta = 0$ then corresponds to a white hole flow. We consider some initial perturbation $\zeta(x,t=0)$. The corresponding initial data on $\psi$ is 
\begin{align}
\psi(x,t=0) = \int_0^x \zeta(y,0) \dd y .
\end{align}
The evolution of $\psi$ in time can be determined by expanding it into out modes. Due to the symmetry between in modes of black hole flows and out modes of white hole flows~\cite{Macher:2009tw}, the calculation is similar to the one sketched in subsection~\ref{App:linKdV}. 
In particular, the structure of divergences is the same, except that those which multiplied incoming waves now multiply outgoing ones, and conversely. This introduces an important difference, as the divergence in $1/\om$ of the coefficients of dispersive waves for an extended perturbation, which did not contribute for a black hole as it multiplied incoming waves, now multiplies outgoing waves~\cite{Mayoral:2010ck}. 
As explained in \cite{Coutant:2012mf}, it adds a saddle-point contribution for $\om \approx 0$, which generates a stationary undulation with a large amplitude.\footnote{In~\cite{Coutant:2012mf} only perturbations localized in $\psi$ were considered. 
In that case, corresponding to perturbations in $\zeta$ with a vanishing integral, the amplitude of the undulation vanishes as $t \to \infty$. When the perturbation satisfies $\int_\mathbb{R} \zeta \dd x \neq 0$ instead, we verified that its amplitude goes to a finite, non-vanishing constant for $t \to \infty$.} 
Numerical simulations using the linearized KdV equation confirm  that this extends to smooth white hole configurations, see \fig{fig:linKdVWH}. Since the scattering coefficients on a white hole flow described by the GP equation have the same divergences for $\om \to 0$ as those obtained using the KdV equation, see~\cite{Macher:2009tw}, the same argument tells us that a stationary undulation shall also be produced in a condensate. 
In all cases, for extended perturbations the undulation amplitude predicted by the linear equation goes to infinity. 
This macroscopic character of the undulation amplitude indicates that nonlinear effects will play a crucial role, which implies that linear equations are unable to predict the late-time evolution.  

\subsection{Nonlinear evolution} 

The nonlinear evolution of perturbations on a white hole flow was studied numerically in~\cite{Mayoral:2010ck,Michel:2015pra}. In the present subsection we report some new numerical results. To make contact with the nonlinear analysis of black hole flows of the previous section, we focus on the case of the GP equation. We obtained qualitatively similar results for the KdV equation, with the signs of the perturbations to the boundary conditions reversed. A rule of thumb, which works for the GP and KdV equations as well as for a superluminal KdV equation obtained by changing the sign of the dispersive term, is that analogue white hole flows are most unstable to perturbations with the sign of the difference between the stationary soliton and the corresponding homogeneous solution. For the GP equation, the soliton is a local underdensity, so the strongest instabilities come from density perturbations with a negative sign. For the KdV equation instead, the soliton is a local surelevation of the free surface. Correspondingly, a positive perturbation on $\zeta$ leads to generally wilder behaviors than a negative one. 

\begin{figure}[ht]
\centering
\includegraphics[width=0.49 \linewidth]{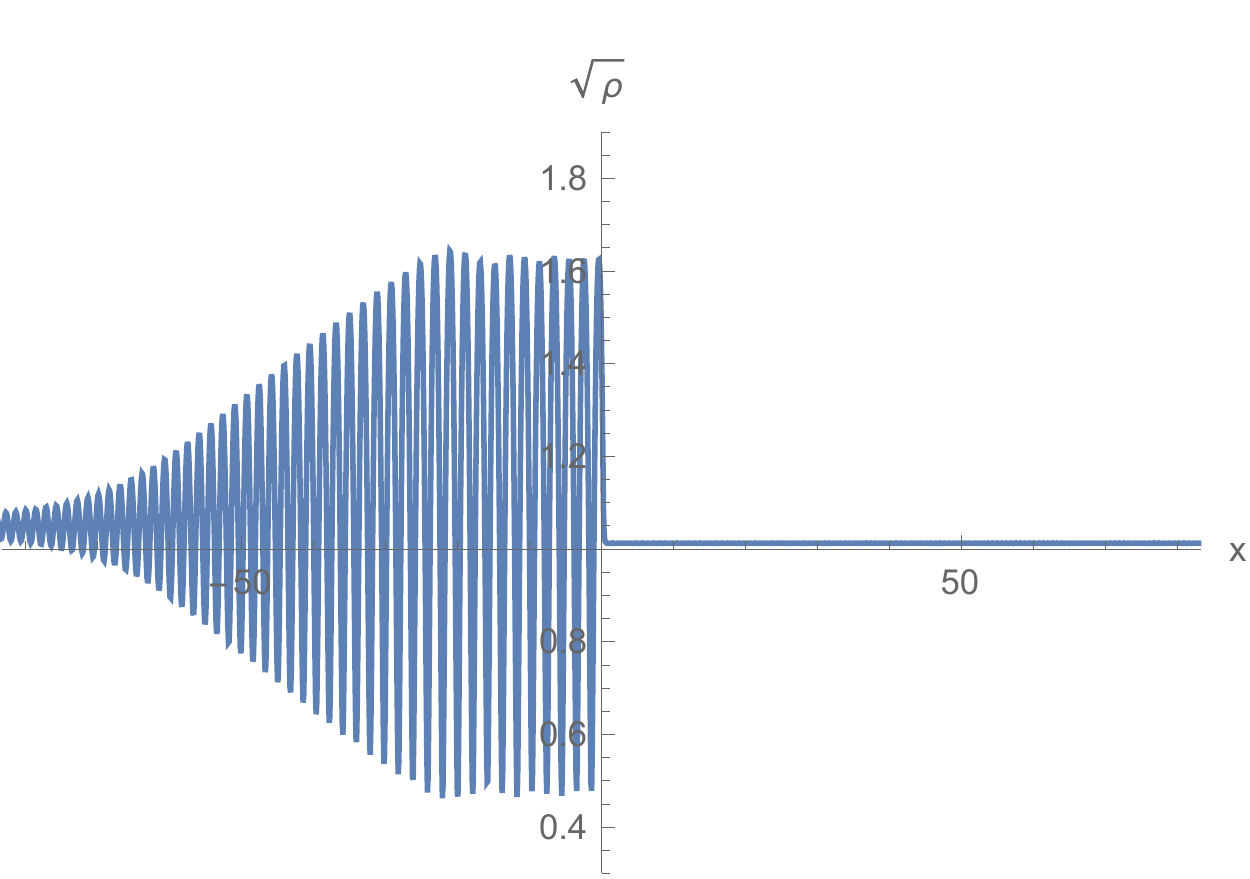} 
\includegraphics[width=0.49 \linewidth]{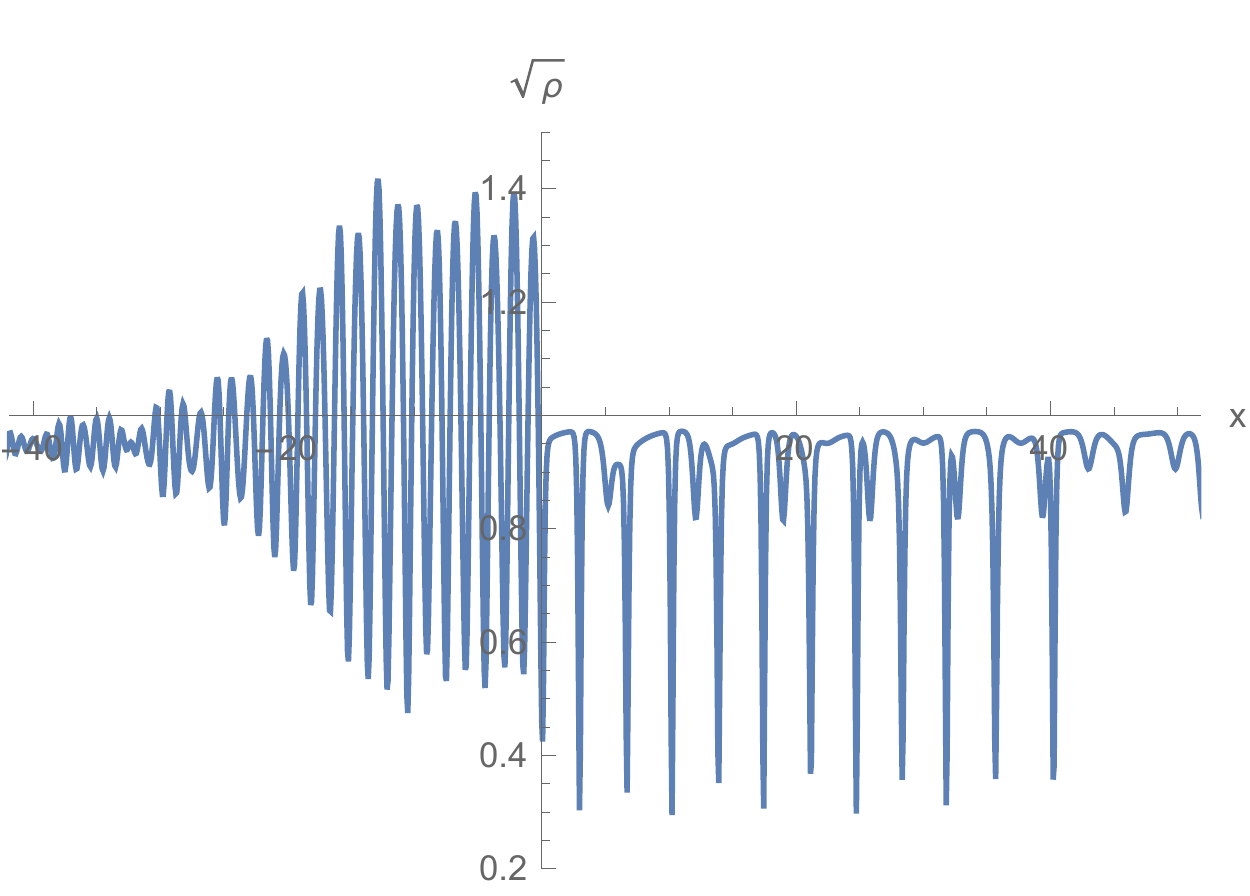}
\caption{We show the undulation (left) and soliton trains (right) produced when the initial homogeneous density differs from that of the homogeneous white hole flow with $\rho_0 = 1$. The vertical axis shows the horizon. 
For $x<0$ (supersonic side), the parameters are $g_-= 0.625$ and $V_-=-2.625$, while for $x > 0$ (subsonic side) they are $g_+= 40$ and $V_+=-42$.
The initial configuration has a homogeneous density equal to $1.1$ (left) and $0.9$ (right), and a current $J=2$. The solution is shown at $t=30$ (left) and $t=10$ (right).}\label{fig:solsNLSb2}
\end{figure}

To start the analysis, we work with steplike functions $g$ and $V$, given by \eq{eq:step} with $g_+ > g_-$ and $V_+ < V_-$. We first consider an initially homogeneous configuration with $\rho = \rho_i$ close to $\rho_0$ and $v \in ]\sqrt{g_- \rho_0}, \sqrt{g_+ \rho_0}[$. Two typical solutions at intermediate times are shown in \fig{fig:solsNLSb2}.
When $\rho_i > \rho_0$, a finite-amplitude undulation develops in the supersonic region. Its amplitude is linear in $\sqrt{\rho_i - \rho_0}$ and is constant at late time. When $\rho_i < \rho_0$ instead, the horizon emits superposed soliton trains in the subsonic region (three of them can be seen in the figure) and a perturbed undulation in the supersonic one. A large-amplitude perturbation is produced periodically close to the horizon, with a frequency linear in $\sqrt{\rho_0 - \rho_i}$, which then separates into several solitons with different velocities in the subsonic region, plus a perturbation propagating on top of the undulation in the supersonic one. It should be emphasized that for homogeneous initial configurations there is no threshold on the value of $\rho_i - \rho_0$. When it is positive (negative), one obtains a stationary undulation 
(non-stationary soliton train). For both signs of $\rho_i - \rho_0$, the nonlinear solution is characterized by the non-analytic parameter $|\rho_i - \rho_0|^{1/2}$. In other words, irrespectively of the perturbation amplitude, the Bogoliubov-de Gennes equation \eqref{eq:BdG} cannot be used to determine the late-time configuration. 

We now study the evolution of localized perturbations. To this end, we choose initial conditions of the form
\begin{align}
\psi(x,t=0) = \sqrt{\rho_0 + A \, \e^{- \lp x-x_0 \rp^2 / \lambda^2}} \, \e^{\ii v_i x}, 
\end{align}
where $\lp A, x_0, \lambda \rp \in \mathbb{R}^3$ and $\abs{A} < \rho_0$. 
For $A > 0$, we observed only the emission of an undulation in the supersonic region. For $A < 0$, we saw the emission of a finite number of solitons in the subsonic region, as well as a perturbed undulation in the supersonic one. In both cases, the amplitude of the undulation is roughly linear in $\abs{A}$ at early times provided $\abs{A} \ll \rho_0$. Interestingly, we observe that it slowly decreases in time due to nonlinear effects, apparently going to zero for $t \to \infty$. As in the case of homogeneous initial configurations, irrespectively of value of $|A|$, the linear equation \eqref{eq:BdG} cannot determine the late-time solution. This is due to the accumulation of the low frequency configurations on the sonic horizon~\cite{Coutant:2012mf}. 

Finally, we performed numerical simulations with functions $g$ and $V$ of the form \eq{eq:tanh} with $V_+ + g_+ \rho_0 = V_- + g_- \rho_0$, so that a solution with a homogeneous density $\rho_0$ exists. The main difference with the above results is that, when working with localized perturbations, solitons are emitted only if $A$ is below a negative threshold value $A_s < 0$. For $A$ not too close to $A_s$, we observed that the time needed to produce the first soliton scales as $( |A| \lambda /\sigma)^{-1/2}$. This can be understood as the condition for obtaining a sufficiently large underdensity so as to allow for the emission of a soliton. The simulations we performed were not precise enough to accurately determine the scaling of $A_s$ in $\sigma$ and $\lambda$, although we found that $\abs{A_s}$ decreases with $\lambda$ and $1/\sigma$, going to 0 for $\lambda \to \infty$ or $1/\sigma \to \infty$. It would be interesting to further investigate these questions along with the validity of the linear equation \eqref{eq:BdG} for $|A| < |A_s|$.

\section{Discussion}
\label{sec:concl} 

In this chapter we studied one-dimensional transonic solutions of the GP (and KdV in Section~\ref{sec:NHT:AR}) equations. We showed that they exhibit behaviors which are analogous to those of black hole solutions of general relativity. At the level of stationary solutions, we showed that the set of solutions which are asymptotically homogeneous (AH) on both sides is discrete at fixed value of a conserved quantity (the current $J$ for the GP equation). When considering steplike potentials, we demonstrated that the series of solutions parameterized by $J$ is unique. For smooth potentials, we numerically found a series of AH flows which is smoothly connected to the above series, as the solutions coincide in the high gradient limit. However, under some conditions, we also found a second series of AH solutions which is disconnected from the first one. These hairy solutions possess a large fraction of a soliton attached to the sonic horizon. Our preliminary investigations indicate that they are less stable than those of the first series because the soliton can be sent away from the horizon by an incoming perturbation. In the remainder of the chapter, we focused on the stability properties of the first series of AH solutions. 

At the level of linear perturbations, we found, both analytically and numerically, that the near-horizon amplitudes of all localized perturbations decay in time as a power law. This establishes that AH black hole flows are linearly stable. It should be noticed that the $S$-matrix which governs this linear scattering is the same as that encoding the Hawking effect in the present setting. One clearly sees here the close link between the (stimulated) Hawking process, i.e., the wave amplification upon scattering on the horizon, and the expulsion of all incident perturbations away from the horizon. 
Moreover, in the limit where the dispersive momentum scale is sent to infinity this expulsion follows the relativistic prediction. 

We used two different approaches to study the stability when including nonlinearities of the GP equation. We first worked with the approximate scheme of Whitham's modulation theory to characterize in analytical terms the late-time evolution of the solutions for steplike potentials. We showed that some field configurations are expelled from the horizon region to infinity by three nonlinear waves, known in the literature as dispersive shock waves and simple waves. Importantly, in the vicinity of the sonic horizon, the solution tends to one of the AH configurations that we formerly characterized. It should be pointed out that Whitham's theory also provides a characterization of the domain of (homogeneous) initial conditions which evolve at late times to an AH transonic flow. Finally, when taking the limit of small amplitude, it can be verified that these results give back those of the linear analysis.

We then performed numerical simulations. We first showed that the late time behavior of a much wider class of initial configurations agrees with that predicted by Whitham's theory, namely, an AH transonic flow is obtained by the emission 
of three nonlinear waves plus perturbations that decay in time. We verified that the properties of the three nonlinear waves are in agreement with those obtained using Whitham's theory. We also showed that the time-dependent perturbations, which are not accounted for in our nonlinear analytical approach, decay in time with the same power law as that found in the linear analysis. All these results indicate that the set of AH transonic flows is a local attractor for neighboring flows. In a future work, using the integrability of the GP equation, we hope to be able to demonstrate in a fully analytical way this property, which is important for experiments. Indeed, it motivates that these solutions can be produced without fine-tuning the initial conditions nor the potential $V$. This should help observing both the spontaneous and stimulated analogue Hawking emission. 

We also studied numerically the behavior of solutions when the initial conditions are outside the validity domain of the solutions obtained with Whitham's theory. In this regime of large deviations from the attractor, several behaviors have been observed. In some cases, we found that the emitted nonlinear waves can leave behind them an undulation which propagates backwards towards the black hole horizon. In other cases, we observed the emission of soliton trains. This variety of behaviors is similar to that observed in Section~\ref{Sec:WH} when studying the evolution of white hole flows. For these flows, the late-time properties are rather complicated even when the perturbations have a small amplitude. Yet, the observed behaviors can be separated into two types. In this respect, our analysis indicates that the set of AH white hole solutions is a kind of ``separator'', rather than attractor, as the type of the solution is determined by the sign of the density fluctuation in the near-horizon region. When there is a sufficiently large increase of the density, the solution displays a macroscopic undulation in the supersonic domain, whereas it gives rise to an emission of soliton trains when there is a sufficiently large density decrease. These two types of behaviors have been already found in the context of the dynamical instability (called the black hole laser) obtained when a stationary flow crosses twice the sound speed~\cite{2015arXiv150900795D,Michel:2015pra} (see also Chapter~\ref{ch:saturation}). Our study indicates that it is the white hole horizon which is responsible for the wide variety of temporal evolutions that was observed. 

The present work leaves open several questions which deserve further study. First, it would be interesting to prove rigorously our conjecture that black-hole flows are nonlinearly stable using inverse scattering techniques. In the same vein, investigating deformed GP or KdV equations including non-integrable terms could shed light on the relations between the mathematical properties of the equation and the ``no-hair'' results. As a first example, we study numerically the case of the cubic-quintic GP equation in subsection~\ref{App:CQNLS}. Another possible extension concerns higher-dimensional systems. In general relativity, no-hair and uniqueness results crucially depend on the dimensionality of space~\cite{Chrusciel:2012jk}. It would certainly be enlightening to see how the dimensionality affects black hole stability in systems described by the GP or hydrodynamic equations. 

\section{Whitham equations}
\label{App:Whitham}

In this Section we give the main steps in obtaining the Whitham modulation equations for \eq{eq:GPE} and their solutions used in the main text. The interested reader will find in~\cite{Kamchatnov} and references therein a full derivation. When possible we use the same notations and conventions as in this reference. 

Whitham's modulation theory~\cite{Whitham1} was developed to study oscillating solutions of partial differential equations with slowly varying parameters. It rests on the two following ideas. First, if there is a clear separation between the fast scale of oscillations and the slow scale of variation of the parameters, an averaging procedure can decouple them. Second, averaging conservation laws instead of the wave equation generally gives the most accurate and best-controlled results. This theory is then particularly useful when one has enough conservation laws to characterize the space of solutions one is interested in. As such, it is no surprise that deep links exist with integrability. A generic way to obtain the Whitham equations for an integrable system is to use the AKNS scheme, developped in~\cite{Ablowitz:1974ry} and applied to the Whitham theory in~\cite{Kamchatnov1994387}. Here we briefly review this procedure, following the presentation of~\cite{Kamchatnov}. 

\subsection{The AKNS scheme}

The idea of the AKNS (Ablowitz, Kaup, Newell, and Segur~\cite{Ablowitz:1974ry}) scheme is to reformulate a partial differential equation one wishes to solve as the compatibility condition of a linear system of the form
\begin{align}\label{eq:sys} 
\left\lbrace
\begin{array}{cc}
	\pd_x \phi_\lambda(x,t) = U(x,t;\lambda) \phi_\lambda(x,t), \\
	\pd_t \phi_\lambda(x,t) = V(x,t;\lambda) \phi_\lambda(x,t),
\end{array}
\right.
\end{align}
where $\phi_\lambda(x,t)$ is a two-component complex vector and $U(x,t;\lambda)$, $V(x,t;\lambda)$ are two matrices of the form
\begin{align}\label{eq:formAKNS}
U(x,t;\lambda) = 
\begin{pmatrix}
	F(x,t;\lambda) & G(x,t;\lambda) \\ 
	H(x,t;\lambda) & -F(x,t;\lambda)
\end{pmatrix}, \; 
V(x,t;\lambda) = 
\begin{pmatrix}
	A(x,t;\lambda) & B(x,t;\lambda) \\ 
	C(x,t;\lambda) & -A(x,t;\lambda)
\end{pmatrix},
\end{align}
where $A$, $B$, $C$, $F$, $G$, and $H$ are differential operators, analytic in $\la$. Here $\lambda$ is a complex number, called ``spectral parameter'', independent on $t$ and $x$. To simplify the notations, from now on the dependence in $\lambda$ will not be written explicitly when no confusion is possible. The compatibility condition of \eq{eq:sys} is $\pd_t U - \pd_x V + [U,V] = 0$, i.e.,
\begin{align}\label{eq:comp1}
\left\lbrace
\begin{array}{cc}
	\pd_t F - \pd_x A + C G - B H = 0 \\
	\pd_t G - \pd_x B + 2 \lp B F - A G \rp = 0 \\
	\pd_t H - \pd_x C + 2 \lp A H -C F \rp = 0
\end{array}
\right. .
\end{align}
Importantly, this system must be equivalent to our original partial differential equation for all values of $\lambda$. Then $\lambda$ will generate an infinite number of conserved quantities which can be used to solve the problem, either exactly using the inverse scattering method when it applies, or approximately using the Whitham equations. 

To see this, it is convenient to define two linearly independent solutions $\phi$ and $\varphi$ of the linear problem and the three scalar quantities
\begin{align}\label{eq:deffgh}
f \equiv -\frac{\ii}{2} \lp \phi_1 \varphi_2 + \phi_2 \varphi_1 \rp, \; g \equiv \phi_1 \varphi_1, \; \text{and} \; h \equiv -\phi_2 \varphi_2.
\end{align}
Partial derivatives of $f$, $g$, and $h$ can be computed straightforwardly. We find
\begin{equation}\label{eq:comp2}
\lb
\begin{aligned}
	\pd_x f & = \ii G h- \ii H g, \\
	\pd_t f & = \ii B h - \ii C g, \\
	\pd_x g & = 2 F g + 2 \ii G f, \\
	\pd_t g & = 2 A g + 2 \ii B f, \\
	\pd_x h & = -2 F h-2 \ii H f, \\
	\pd_t h & = -2 A h-2 \ii C f.
\end{aligned}
\right.
\end{equation}
Conversely, the compatibility conditions of the system \eq{eq:comp2} give back \eq{eq:comp1} provided $g f h \neq 0$. \eq{eq:comp2} can thus be seen as a rewriting of the original problem. This formulation is particularly useful for deriving conserved or slowly-varying quantities, as we now explain. 

We first notice that $f^2 - g h$ is directly related to the generalized wronskian of $(\phi,\varphi)$:
\begin{align}
f^2 - g h = -\frac{1}{4} W^2, \; W \equiv \phi_1 \varphi_2 - \phi_2 \varphi_1.
\end{align}
Since $U$ and $V$ are traceless, $\pd_t W = \pd_x W = 0$. So, $f^2 - g h$ depends only on the spectral parameter $\lambda$. As we shall see, for the solutions we are interested in this quantity is a polynomial in $\lambda$, which we denote by $P(\lambda)$. The crux of the construction is that $g$ satisfies the following conservation law: 
\begin{align}\label{eq:consrv}
\pd_t \lp \frac{G}{g} \rp - \pd_x \lp \frac{B}{g} \rp = 0.
\end{align}
\eq{eq:consrv} can be used to generate an infinite number of conservation laws after expanding $g$ in powers of $\lambda$. For our purposes, it is enough to retain only a finite number of terms, giving as many Whitham equations. 

Let us assume that we can find solutions of \eq{eq:comp2} where $g$ reads 
\begin{align}
g(x,t;\lambda) = g_0(x,t;\lambda) \lp \lambda - \mu(x,t) \rp,
\end{align}
where $g_0$ is a smooth function which does not vanish at $\lambda = \mu(x,t)$. Evaluating the partial derivatives of $g$ gives
\begin{align}
\pd_t \mu = -2 \ii \frac{B}{g_0} \sqrt{P(\mu)}, \; \pd_x \mu = -2 \ii \frac{G}{g_0} \sqrt{P(\mu)}.
\end{align}

So far we worked with exact solutions of the problem. Let us now consider a solution with two well-separated length scales: a ``fast'' scale on which it oscillates periodically and a ``slow'' one on which the parameters describing the local oscillations, such as their amplitude, mean value and wave vector, vary. We can then apply the above procedure in two different ways:
\begin{itemize}
	\item One can define global functions $f_g, g_g, h_g$ which describe the exact solutions but have in general no analytical expression.
	\item One can also define local functions $f_l, g_l, h_l$ by neglecting all variations on the ``slow'' scale.
\end{itemize}
Locally, these global and local solutions have the same form by definition, except for the normalization of the vectors $\psi$ and $\phi$ used to define them. For the global functions, this normalization must be independent of $x$ and $t$, say $P_g(\lambda) = 1$. However, in general $f_l, g_l, h_l$ will take a simple form when using a normalization such that $P_l$ depends on $x$ and $t$. Then,
\begin{align}
(f_l(x,t), g_l(x,t), h_l(x,t)) = \sqrt{P_l(x,t;\lambda)} (f_g(x,t), g_g(x,t), h_g(x,t)).
\end{align}
To avoid unnecessarily cumbersome notations, in the following we shall not write the index $l$ explicitly, as we shall work only with the ``local'' functions. The exact conservation law \eq{eq:consrv} applied to $g_g$ gives
\begin{align}
\pd_t \lp \sqrt{P(x,t,\lambda)} \frac{G(x,t)}{g(x,t)} \rp - \pd_x \lp \sqrt{P(x,t,\lambda)} \frac{B(x,t)}{g(x,t)} \rp = 0.
\end{align}
We now assume that $G/g_0$ is a constant, which is the case for the solutions we will consider. Then, at fixed $t$, $\dd \mu \propto \sqrt{P(\mu)} \dd x$. We can then average the conservation law by integrating over a few wavelengths and replacing the integral over $x$ by one over $\mu$, taken over a contour encircling its locus in the complex plane:
\begin{align} \label{eq:intcons} 
\pd_t \lp \oint \sqrt{P(x,t;\lambda)} \frac{G(x,t;\lambda)}{\sqrt{P(x,t;\mu)} g_0(x,t;\lambda) (\lambda - \mu)} \dd \mu \rp \hspace*{1 cm} 
\nn - \pd_x \lp \oint \sqrt{P(x,t;\lambda)} \frac{B(x,t;\lambda)}{\sqrt{P(x,t;\mu)} g_0(x,t;\lambda) (\lambda - \mu)} \dd \mu \rp \approx 0. \hspace*{-1 cm} 
\end{align}
The last step is to extract the singularities in $\lambda$ from \eq{eq:intcons}. These come from the simple roots $\lambda_i$ of $P(x,t;\lambda)$, which give after differentiation terms in $1/\sqrt{\lambda - \lambda_i}$. Cancelling them gives the general form of the Whitham modulation equations:
\begin{align}
\lp \oint \frac{G(x,t;\lambda_i)}{\sqrt{P(x,t;\mu)}g_0(x,t;\lambda_i) (\lambda_i - \mu)} \dd \mu \rp \pd_t \lambda_i \hspace*{1 cm} 
\nn - \lp \oint \frac{B(x,t;\lambda_i)}{\sqrt{P(x,t;\mu)}g_0(x,t;\lambda_i) (\lambda_i - \mu)} \dd \mu \rp \pd_x \lambda_i \approx 0. \hspace*{-1 cm} 
\end{align}
Notice that when neglecting the slow evolution, each of the two integrals depends only on the local parameters of the solution. If the ansatz chosen for $f$, $g$, and $h$ and leading to the polynomial form of $W$ is sufficiently general, these parameters can all be expressed in terms of the roots $\la_i$ of $P$. Then, for each root $\lambda_i$, the Whitham equations may be written in the more transparent form
\begin{align}
\lp \pd_t + v_i(\left\lbrace \lambda_j \right\rbrace) \pd_x \rp \lambda_i = 0,
\end{align}
where
\begin{align}\label{eq:vi}
v_i \equiv -\frac{\displaystyle{\oint} \frac{B(x,t;\lambda_i)}{\sqrt{P(x,t;\mu)}g_0(x,t;\lambda_i) (\lambda_i - \mu)} \dd \mu}{\displaystyle{\oint} \frac{G(x,t;\lambda_i)}{\sqrt{P(x,t;\mu)}g_0(x,t;\lambda_i) (\lambda_i - \mu)} \dd \mu}.
\end{align}
The roots $\la_i$ are called the Riemann invariants. 

\subsection{Applications and characteristic velocities}

Let us apply the above formalism to the GP equation. A direct calculation using \eq{eq:comp1} shows that the compatibility condition obtained when choosing
\begin{align}\label{eq:UVNLS}
U = \begin{pmatrix}
	- \ii \lambda & \ii a \psi \\
	- \ii a \psi^* & \ii \lambda
\end{pmatrix}, \; 
V = \begin{pmatrix}
	- \ii \lambda^2 - \ii a^2 \left\lvert \psi^2 \right\rvert / 2 & \ii \lambda a \psi - a \pd_x \psi / 2 \\
	- \ii a \lambda \psi^* & \ii \lambda^2 + \ii a^2 \left\lvert \psi^2 \right\rvert / 2
\end{pmatrix}
\end{align}
(which are of the form \eq{eq:formAKNS}) is
\begin{align}
\ii \pd_t \psi + \frac{1}{2} \pd_x^2 \psi -a^2 \left\lvert \psi^2 \right\rvert \psi = 0.
\end{align}
This is exactly \eq{eq:GPE} with a uniform two-body coupling\footnote{To avoid conflicts of notations, in this section we denote the two-body coupling by $a^2$, and use the symbol ``$g$'' only for the function defined in \eq{eq:deffgh}.} 
$a^2$  and a vanishing potential. Notice that a uniform potential $V$ can be absorbed in the redefinition $\psi \to \e^{- \ii V t} \psi$. So, the choice \eq{eq:UVNLS} allows us to find solutions in the presence of uniform $V$ and $a^2$. We look for solutions where $f$ is real-valued and $h = g^*$. The system \eq{eq:comp2} then becomes
\begin{align}\label{eq:compNLS1}
\left\lbrace 
\begin{array}{ll}
	\pd_x f \eg -a \lp \psi g^*+\psi^* g \rp \\
	\pd_t f \eg -a \lp \lp \lambda \psi + \frac{\ii}{2} \pd_x \psi \rp g^*+ \lp \lambda \psi^* -\frac{\ii}{2} \pd_x \psi^* \rp g \rp \\
	\pd_x g \eg -2 \ii \lambda g - 2 a \psi f \\
	\pd_t g \eg -2 \ii \lp \lambda^2 + \frac{a^2}{2} \abs{\psi^2} \rp - 2 a \lp \la \psi + \frac{\ii}{2} \pd_x \psi \rp f
\end{array}
\right.  .
\end{align}
We now further restrict to solutions of the form
\begin{align}\label{eq:ansatz}
	f(x,t;\la) \eg \la^2-f_1(x,t) \lambda + f_2, \nn
	g(x,t;\la) \eg \ii \, a \, \psi(x,t) \lp \la - \mu(x,t) \rp.
\end{align}
The justification of this ansatz will come {\it a posteriori} by obtaining all the periodic, stationary solutions. Expanding \eq{eq:compNLS1} in powers of $\la$ and eliminating the trivial equalities gives a system of 8 differential equations on $f_1$, $f_2$, and $\mu$. It is useful to parametrize the solution using the coefficients of $P(\lambda)$. In our case, the later is a fourth-order polynomial of the form
\be\label{eq:Pla}
P(\la) = \la^4 - s_1 \la^3 + s_2 \la^2 - s_3 \la + s_4.
\ee
$f_1$ and $f_2$ are related to $s_1$ and $s_2$ through
\begin{align}
f_1 = \frac{s_1}{2}, \; f_2 = \frac{s_2}{2} - \frac{s_1^2}{6}+\frac{a^2}{2} \abs{\psi^2}.
\end{align}
Variations of $\mu$ with $x$ and $t$ are given by
\begin{align}\label{eq:mu}
\left\lbrace
\begin{array}{rl}
	\pd_x \mu(x,t) \eg -2 \ii \sqrt{P(\mu(x,t))} \\
	\pd_t \mu(x,t) \eg \frac{s_1}{2} \pd_x \mu(x,t)
\end{array}
\right. .
\end{align}
The corresponding equations on $\psi$ are more conveniently written in terms of
\begin{align}
\tilde{\psi}(x,t) \equiv \e^{\ii \lp \frac{s_1^2}{4} - s_2 \rp t} \psi(x,t). 
\end{align}
One obtains
\begin{align}
\left\lbrace
\begin{array}{rl}
	\ii \pd_x \tilde{\psi}(x,t) \eg 2 \lp \frac{s_1}{2} - \mu(x,t) \rp \tilde{\psi}(x,t) \\
	\ii \pd_t \tilde{\psi}(x,t) \eg \frac{s_1}{2} \ii \pd_x \tilde{\psi}(x,t)
\end{array}
\right. .
\end{align}
Notice that $\mu$ and $\tilde{\psi}$ depend on $x$ and $t$ only through $\xi \equiv x + \frac{s_1}{2} t$. The density perturbations thus move with the velocity $-s_1 / 2$ without changing their shape. Using the two other conserved quantities $s_3$ and $s_4$ to determine the relationship between $\mu$ and $\psi$,  one obtains a closed equation on $\abs{\psi}$ of the form 
\begin{align}\label{eq:psiW}
\frac{\dd \abs{\psi^2}}{\dd \xi} = \pm \frac{2}{a^2} \sqrt{-\mathcal{R} \lp \abs{\psi^2} \rp},
\end{align}
where $\mathcal{R}$ is a third-order polynomial with roots at $(\lambda_4 + \lambda_3 - \lambda_1 - \lambda_2 )^2/(4 a^2)$, $(\lambda_4 - \lambda_3 - \lambda_1 + \lambda_2 )^2/(4 a^2)$, and $(\lambda_4 - \lambda_3 + \lambda_1 - \lambda_2 )^2/(4 a^2)$, and $\la_i$, $i =1..4$ are the four roots of $P(\la)$, i.e., the Riemann invariants. 
We assume they are all real and order them so that $\la_1 \leq \la_2 \leq \la_3 \leq \la_4$. Importantly, one can check that all the periodic solutions of the GP equation can be recovered from \eq{eq:psiW}, which justifies the ansatz \eq{eq:ansatz}.

Let us now turn to slowly-modulated solutions, for which the parameters $\la_i$ slowly vary in space and time. Using Eqs.~(\ref{eq:vi},\ref{eq:UVNLS},\ref{eq:Pla}), we find that the characteristic velocities are
\begin{align}
v_i = \frac{1}{2} \lp \frac{L}{\pd_{\lambda_i}L} - s_1 \rp,
\end{align}
where $L$ is the wavelength of the periodic solution, given by
\begin{align}
L = \frac{a^2}{\sqrt{(\la_4 - \la_2) (\la_3 - \la_1)}} K \lp \frac{(\la_4 - \la_3) (\la_2 - \la_1)}{(\la_4 - \la_2) (\la_3 - \la_1)} \rp,
\end{align}
where $K$ is the complete elliptic integral of the first kind~\cite{Abramowitz}.

\subsection{Dispersive shock waves and simple waves} 
\label{app:DSWNLS}

We now look for scale-invariant solutions depending only on $z \equiv x/t$. The Whitham equations then take the simple form 
\begin{align}\label{eq:Whz}
\lp v_i - z \rp \frac{\dd \la_i}{\dd z} = 0.
\end{align}
That is, each Riemann invariant is constant except in a domain where the associated velocity is equal to $z$. We further restrict to solutions for which $\abs{\psi}$ is asymptotically homogeneous at $z \to \pm \infty$. This means that $\mathcal{R}$ must have a double root in each asymptotic region, i.e., two Riemann invariants must be equal there. We thus have two possibilities to build a non-trivial solution:
\begin{itemize}
	\item If two Riemann invariants are equal and strictly homogeneous while a third one is varying in a bounded interval of $z$, the solution is a simple wave (SW);
	\item If one Riemann invariant $\lambda_{i_0}$ ($i_0 \in \lb 2, 3 \rb$) varies between $\lambda_{i_0-1}$ at $z=z_-$ and $\lambda_{i_0+1}$ at $z=z_+$ for two real numbers $z_-$ and $z_+$, the solution is a dispersive shock wave (DSW).
\end{itemize}
In the first case, the solution shows no oscillations as two Riemann invariants remain equal all along. In the second case, it shows oscillations which start in the linear regime on one side and become widely spaced solitons when approaching the other side. A SW and a DSW are shown in \fig{fig:solsNLS}. 

Let us first focus on SW. A careful analysis shows that there are only two possibilities, corresponding to SW propagating to the left or to the right in the fluid frame. They are characterized by
\begin{align}\label{eq:cons1} 
\left\lbrace
\begin{aligned}
	\pd_z \lp \pm a \sqrt{\rho} -\frac{v}{2} \rp & = 0 \\
	v \pm a \sqrt{\rho} &=  z
\end{aligned}
\right. ,
\end{align}
where $\rho \equiv \left\lvert \psi \right\rvert^2$ is the density and $v \equiv - \ii (\pd_x \psi) / \psi$ the velocity of the condensate. 

There are also two different DSW: one along which $\la_2$ varies between $\la_1$ and $\la_3$ and one along which $\la_3$ varies between $\la_2$ and $\la_4$. Each of them interpolates between a subsonic homogeneous solution at $z=z_b$ and a supersonic one at $z=z_p$. In the first case, these extremal values of $z$ are given by
\begin{align}\label{eq:NLSzb} 
z_b = v_p + a \sqrt{\rho_b}
\end{align}
and 
\begin{align}\label{eq:NLSzp}
z_p = \frac{8 a \sqrt{\rho_b} \lp v_p - v_b \rp + v_p^2 - \lp v_b + 2 a \sqrt{\rho_b} \rp^2}{2 \lp v_p-v_b-2 a \sqrt{\rho_b} \rp},
\end{align}
where an index $b$ (respectively, $p$) indicates a quantity evaluated at $z=z_b$ (respectively $z=z_p$). The constraints on this solution, coming from the conservation of $\lambda_4$ and the assumption $\lambda_1 < \lambda_3$, are 
\begin{align}\label{eq:cons2}
a \sqrt{\rho_p} - \frac{v_p}{2} = a \sqrt{\rho_b} - \frac{v_b}{2}, \; \rho_b > \rho_p.
\end{align}
Importantly for our purposes, a direct calculation gives $z_p - z_b>0$. The second DWS is obtained by flipping the signs in front of $a$. The constraints on this solution are
\begin{align}
a \sqrt{\rho_p} + \frac{v_p}{2} = a \sqrt{\rho_b} + \frac{v_b}{2}, \; \rho_b > \rho_p.
\end{align}
A direct calculation gives $z_p - z_b<0$. 

\subsection{Three-waves solutions of the Whitham equations in the single step configuration}
\label{App:domex}

We now give the domains of existence (in the space of asymptotic conditions) of the 8 solutions of the Whitham equations mentioned in subsection~\ref{sec:analytical}. They are given by the above conditions and those on the signs of the velocities of the nonlinear waves. The solution with 3 SW exists for
\begin{align}
\left\lbrace
\begin{array}{l}
	\rho_- \leq \rho_0 \\
	\sqrt{g_- \rho_0} - \sqrt{g_- \rho_-} - \sqrt{g_+ \rho_0} - \sqrt{g_+ \rho_+} \leq \frac{v_- - v_+}{2} \leq \sqrt{g_- \rho_0} - \sqrt{g_- \rho_-} - \left\lvert \sqrt{g_+ \rho_0} - \sqrt{g_+ \rho_+} \right\rvert \\
	\sqrt{g_+ \rho_0} + 2 \sqrt{g_- \rho_0} -2 \sqrt{g_- \rho_-} \leq v_- \leq 3 \sqrt{g_- \rho_0} - 2 \sqrt{g_- \rho_-}
\end{array}
\right. .
\end{align}
The solution with one DSW on the left exist if and only if
\begin{align}
\left\lbrace
\begin{array}{l}
	\rho_- \leq \rho_0 \\
	\sqrt{g_- \rho_0} - \sqrt{g_- \rho_-} - \sqrt{g_+ \rho_0} - \sqrt{g_+ \rho_+} \leq \frac{v_- - v_+}{2} \leq \sqrt{g_- \rho_0} - \sqrt{g_- \rho_-} - \left\lvert \sqrt{g_+ \rho_0} - \sqrt{g_+ \rho_+} \right\rvert \\
	2 \sqrt{g_- \rho_1} + \sqrt{g_+ \rho_1} -2 \sqrt{g_- \rho_-} \leq u_- \leq \sqrt{g_- \rho_1}
\end{array}
\right. .
\end{align}
The solution with one DSW between two SW exists if and only if
\begin{align}
\left\lbrace
\begin{array}{l}
	\rho_- \geq \rho_0 \\
	\sqrt{g_- \rho_0} - \sqrt{g_- \rho_-} + \sqrt{g_+ \rho_0} - \sqrt{g_+ \rho_+} \leq \frac{v_- - v_+}{2} \leq \sqrt{g_- \rho_0} - \sqrt{g_- \rho_-} - \sqrt{g_+ \rho_0} + \sqrt{g_+ \rho_+} \\
	v_- + 2 \sqrt{g_- \rho_-} - 3 \sqrt{g_- \rho_1} \leq 0
\end{array}
\right. .
\end{align}
The solution with one DSW on the right of two SW exists if and only if
\begin{align}
\left\lbrace
\begin{array}{l}
	\rho_- \geq \rho_0 \\
	\sqrt{g_- \rho_0} - \sqrt{g_- \rho_-} - \sqrt{g_+ \rho_0} + \sqrt{g_+ \rho_+} \leq \frac{v_- - v_+}{2} \leq \sqrt{g_- \rho_0} - \sqrt{g_- \rho_-} + \sqrt{g_+ \rho_0} - \sqrt{g_+ \rho_+} \\
	\sqrt{g_+ \rho_0} +2 \sqrt{g_- \rho_1} -2 \sqrt{g_- \rho_-} \leq u_- \leq 3 \sqrt{g_- \rho_1} -2 \sqrt{g_- \rho_-}
\end{array}
\right. .
\end{align}
The solution with one SW on the left of two DSW exists if and only if
\begin{align}
\left\lbrace
\begin{array}{l}
	\rho_- \geq \rho_0 \\
	\sqrt{g_- \rho_-} - \sqrt{g_- \rho_0} - \left\lvert \sqrt{g_+ \rho_+} - \sqrt{g_+ \rho_0} \right\rvert \leq \frac{v_- - v_+}{2} \leq \sqrt{g_- \rho_0} - \sqrt{g_- \rho_-} + \sqrt{g_+ \rho_+} + \sqrt{g_+ \rho_0} \\
	v_+^2 - v_+ v_- + 2 v_- \sqrt{g_+ \rho_+} - 2 v_+ \lp \sqrt{g_- \rho_-} - \sqrt{g_- \rho_1} + 2\sqrt{g_+ \rho_+} \rp + \\ 
	\hspace*{2 cm} + 4 \lp \sqrt{g_- \rho_-} - \sqrt{g_- \rho_0} + \sqrt{g_+ \rho_+} \rp \sqrt{g_+ \rho_+} -2 g_+ \rho_0 \leq 0\\
	v_- + 2 \sqrt{g_- \rho_-} - 3 \sqrt{g_- \rho_0} \leq 0 \\
\end{array}
\right. .
\end{align}
The solution with one SW between two DSW exists if and only if
\begin{align}
\left\lbrace
\begin{array}{l}
	\rho_- \leq \rho_0 \\
	\sqrt{g_+ \rho_+} - \sqrt{g_- \rho_-} - \sqrt{g_+ \rho_0} + \sqrt{g_- \rho_0} \leq \frac{v_- - v_+}{2} \leq \sqrt{g_+ \rho_0} + \sqrt{g_- \rho_0} - \sqrt{g_+ \rho_+} - \sqrt{g_- \rho_-} \\
	v_+ \geq -4 \sqrt{g_+ \rho_+} \\
	v_- + 2 \lp \sqrt{g_- \rho_-} - \sqrt{g_- \rho_0} \rp - \sqrt{g_+ \rho_0} \geq 0 \\
	v_- \geq \sqrt{g_- \rho_0}
\end{array}
\right. .
\end{align}
The solution with one SW on the right of two DSW exists if and only if
\begin{align}
\left\lbrace
\begin{array}{l}
	\rho_- \leq \rho_0 \\
	v_- \leq \sqrt{g_- \rho_0} \\
	\sqrt{g_- \rho_0} - \sqrt{g_- \rho_-} - \sqrt{g_+ \rho_+} + \sqrt{g_+ \rho_0}  \leq \frac{v_- - v_+}{2} \leq \sqrt{g_- \rho_0} - \sqrt{g_- \rho_-} + \sqrt{g_+ \rho_+} - \sqrt{g_+ \rho_0} \\
	v_+^2 - v_+ v_- + 2 v_- \sqrt{g_+ \rho_+} - 2 v_+ \lp \sqrt{g_- \rho_-} - \sqrt{g_- \rho_1} + 2\sqrt{g_+ \rho_+} \rp + 
	\\ \hspace*{2 cm} + 4 \lp \sqrt{g_- \rho_-} - \sqrt{g_- \rho_0} + \sqrt{g_+ \rho_+} \rp \sqrt{g_+ \rho_+} -2 g_+ \rho_0 \leq 0\\
\end{array}
\right. .
\end{align}
Finally, the solution with 3 DSW exists if and only if
\begin{align}
\left\lbrace
\begin{array}{l}
	\rho_- \leq \rho_0 \\
	v_- \leq \sqrt{g_- \rho_0} \\
	\sqrt{g_- \rho_0} - \sqrt{g_- \rho_-} + \left\lvert \sqrt{g_+ \rho_+} - \sqrt{g_+ \rho_0} \right\rvert \leq \frac{v_- - v_+}{2} \leq \sqrt{g_- \rho_0} - \sqrt{g_- \rho_-} + \sqrt{g_+ \rho_+} + \sqrt{g_+ \rho_0} \\
	v_+^2 - v_+ v_- + 2 v_- \sqrt{g_+ \rho_+} - 2 v_+ \lp \sqrt{g_- \rho_-} - \sqrt{g_- \rho_1} + 2\sqrt{g_+ \rho_+} \rp + 
	\\ \hspace*{2 cm} + 4 \lp \sqrt{g_- \rho_-} - \sqrt{g_- \rho_0} + \sqrt{g_+ \rho_+} \rp \sqrt{g_+ \rho_+} -2 g_+ \rho_0 \leq 0\\
\end{array}
\right. .
\end{align}

\begin{figure}[ht]
	\centering
	\includegraphics[width=0.49 \linewidth]{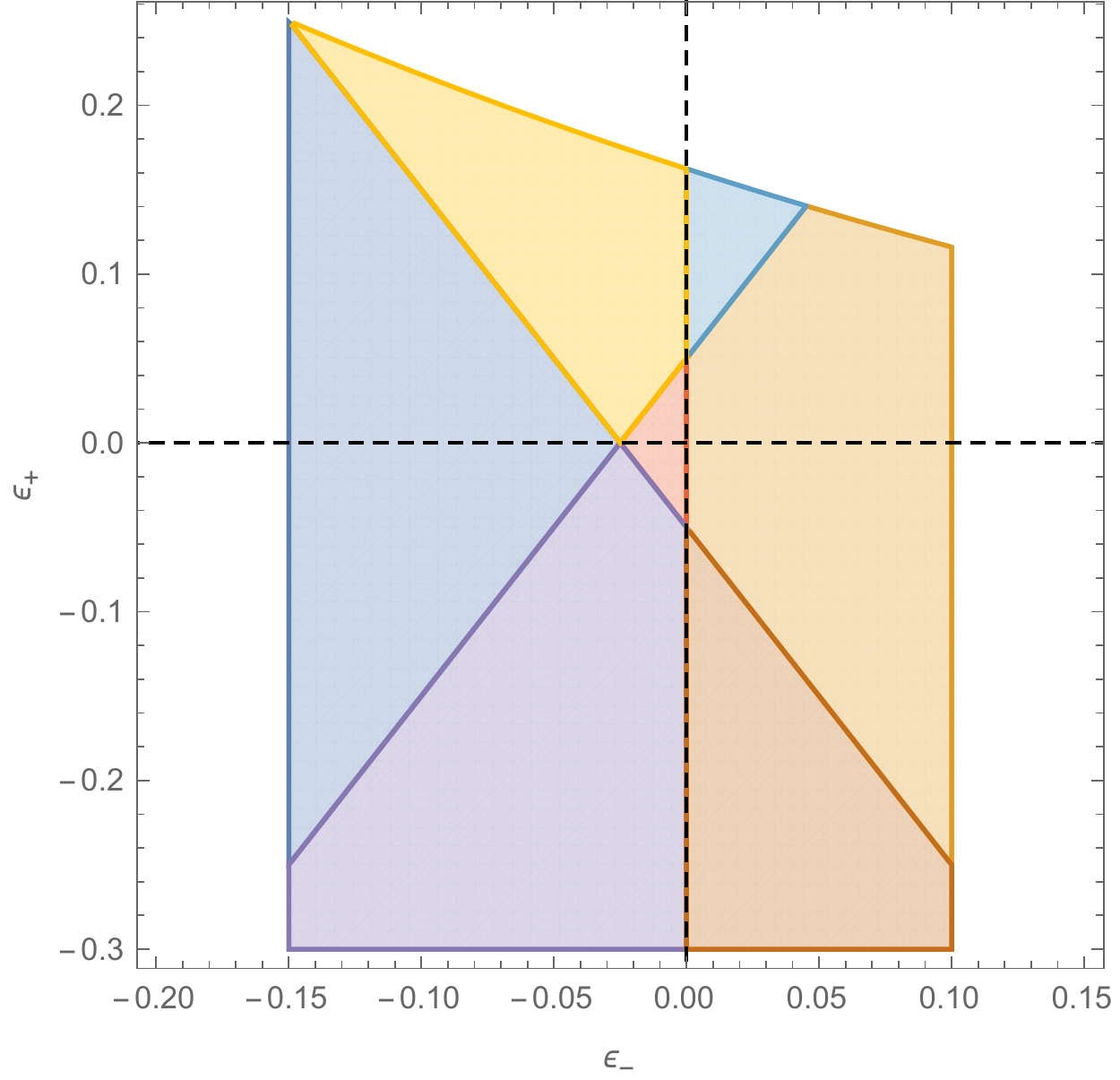} \,
	\includegraphics[width=0.49 \linewidth]{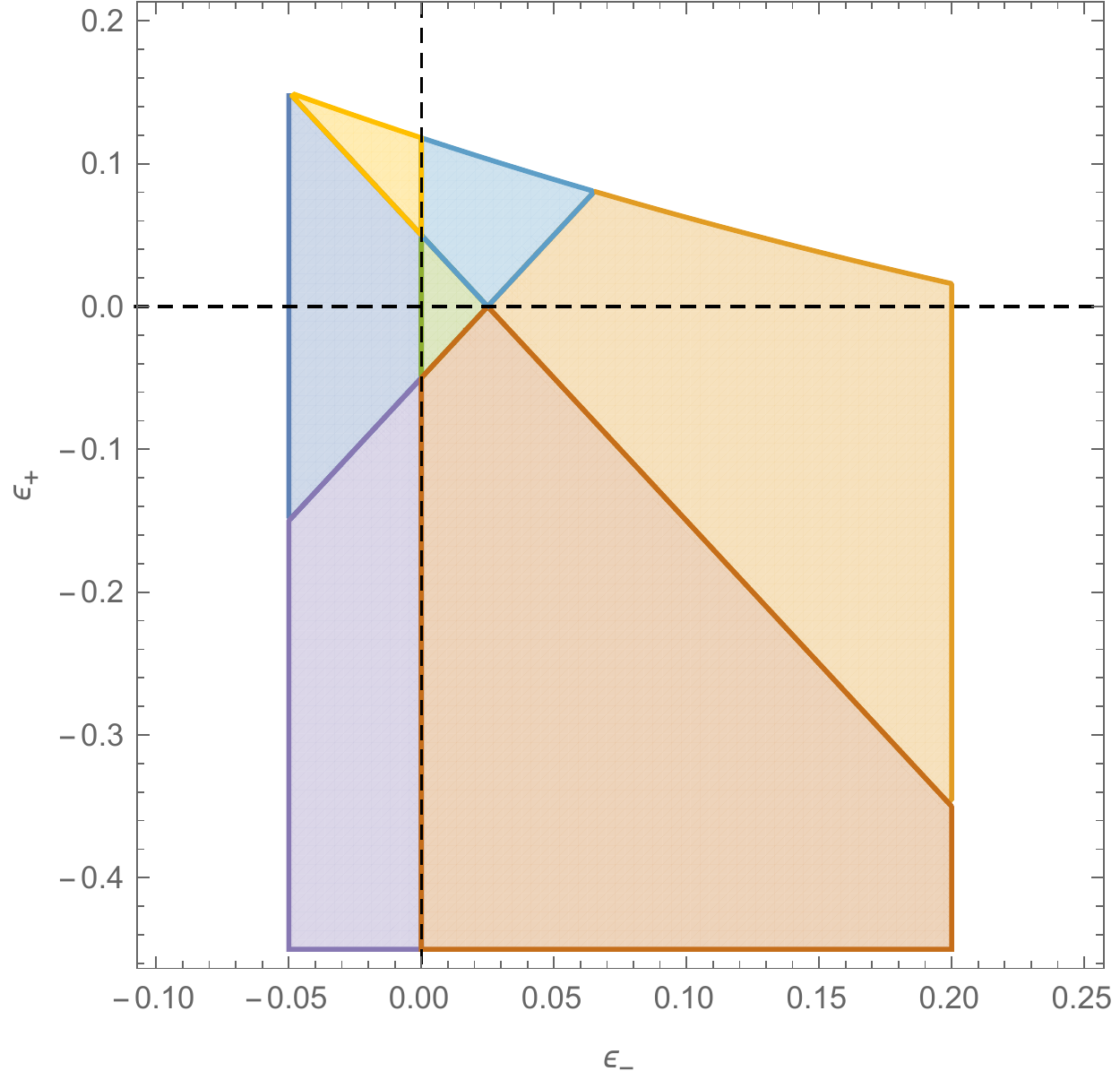}
	\caption{Domains of existence of the 8 solutions with 3 waves in the $(\ep_-, \ep_+)$ plane, where $\ep_i \equiv \sqrt{\rho_i / \rho_0} - 1$, for $g_- = 4 g_+$. The asymptotic velocities are given by $v_- = 0.8 \sqrt{g_- \rho_0}, \; v_+ = 3 \sqrt{g_+ \rho_0}$ (left) and $v_- = 0.6 \sqrt{g_- \rho_0}, \; v_+ = 2.6 \sqrt{g_+ \rho_0}$ (right). The horizontal dashed line shows the locus $\ep_+ = 0$, while the vertical one shows $\ep_- = 0$. The domains of the solutions described in the text are shown, in the same order, in green, blue, cyan, brown, orange, purple, yellow, and red.}\label{fig:8sols}
\end{figure}

The domains of existence of these solutions are shown in \Fig{fig:8sols} for fixed asymptotic velocities. 
Interestingly, one can show that one of them always exists provided the asymptotic conditions are sufficiently close to those obtained from a homogeneous black hole flow. 
To see this, let us assume that $\sqrt{g_+ \rho_0} < v_- < \sqrt{g_- \rho_0}$ 
(this condition is always satisfied for a flow close to a homogeneous black hole one). Then, 
\begin{itemize}
	\item if $-4 \sqrt{g_+ \rho_0} < v_- - v_+ < 0$ and $\rho_0 - \rho_- \geq 0$ is sufficiently small, the solution with three SW exists in an open interval of $\rho_+$ containing $\rho_0$;
	\item if $-2 \sqrt{g_+ \rho_0} < v_- - v_+ < 0$ and $\rho_- - \rho_0 \geq 0$ is sufficiently small, the solution with one DSW on the left of two SW exists in an open interval of $\rho_+$ containing $\rho_0$;
	\item if $\max \lp -4 \sqrt{g_+ \rho_0}, \frac{1}{2} \lp 4 \sqrt{g_+ \rho_0} - v_- - \sqrt{g_- \rho_0} \sqrt{8+ \frac{v_-^2}{g_+ \rho_0}} \rp \rp < v_+ - v_- < 0$ and $\rho_- - \rho_0 \geq 0$ is sufficiently small, then the solution with a SW on the left of two DSW exists in an open interval of $\rho_+$ containing $\rho_0$;
	\item if $\max \lp -4 \sqrt{g_+ \rho_0}, \frac{1}{2} \lp 4 \sqrt{g_+ \rho_0} - v_- - \sqrt{g_- \rho_0} \sqrt{8+ \frac{v_-^2}{g_+ \rho_0}} \rp \rp < v_+ - v_- < 0$ and $\rho_0 - \rho_- \geq 0$ is sufficiently small, then the solution with three DSW exists in an open interval of $\rho_+$ containing $\rho_0$.
\end{itemize}
(Notice that the left-hand sides in the two last double inequalities on $v_+ - v_-$ are strictly negative for $v_- > \sqrt{g_+ \rho_0}$.) 
To be complete, one must also consider the case $v_- = v_+$, $\rho_+ \neq \rho_-$. There are then six possibilities. Assuming $\rho_+$ and $\rho_-$ are both close to $\rho_0$,
\begin{itemize}
	\item If $\rho_- > \rho_0$,
	\begin{itemize}
		\item If $\sqrt{\frac{\rho_+}{\rho_0}} - 1 > \sqrt{\frac{g_-}{g_+}} \lp \sqrt{\frac{\rho_-}{\rho_0}} - 1 \rp$, we obtain a solution with one DSW between two SW; 
		\item If $ \left\lvert \sqrt{\frac{\rho_+}{\rho_0}} - 1 \right\rvert < \sqrt{\frac{g_-}{g_+}} \lp \sqrt{\frac{\rho_-}{\rho_0}} - 1 \rp$, we have one SW on the left of two DSW;
		\item If $\sqrt{\frac{\rho_+}{\rho_0}} - 1 <- \sqrt{\frac{g_-}{g_+}} \lp \sqrt{\frac{\rho_-}{\rho_0}} - 1 \rp$, we have one DSW on the right of two SW;
	\end{itemize}
	\item If $\rho_- < \rho_0$,
	\begin{itemize}
		\item If $\sqrt{\frac{\rho_+}{\rho_0}} - 1 > - \sqrt{\frac{g_-}{g_+}} \lp \sqrt{\frac{\rho_-}{\rho_0}} - 1 \rp$, we have one SW on the right of two DSW;
		\item If $ \left\lvert \sqrt{\frac{\rho_+}{\rho_0}} - 1 \right\rvert < -\sqrt{\frac{g_-}{g_+}} \lp \sqrt{\frac{\rho_-}{\rho_0}} - 1 \rp$, we have one DSW on the left of two SW;
		\item If $\sqrt{\frac{\rho_+}{\rho_0}} - 1 < \sqrt{\frac{g_-}{g_+}} \lp \sqrt{\frac{\rho_-}{\rho_0}} - 1 \rp$, we have one SW between two DSW.
	\end{itemize}
\end{itemize}
So, if the asymptotic conditions are sufficiently close to those from a black hole flow with a homogeneous density, there always exists a solution of the Whitham equations with three waves moving away from $x=0$ and leaving such a flow behind them.

\section{Additional remarks}
\label{sec:NHT:AR}
\subsection{Nonlinear evolution for KdV equations}
\label{App:KdV}

To complete the analysis of subsections~\ref{sub:Ahtf},~\ref{sec:analytical}, and~\ref{App:domex}, we here extend it to the case of the KdV equation with variable coefficients. Our first goal is to show explicitly that the main results are not restricted to the specific case of the GP equation. We also aim at unveiling the qualitative differences which arise when dealing with a subluminal dispersion relation. To disentangle these two aspects, we also consider a superluminal version of the KdV equation. 

\subsubsection{KdV equation}

\subsubsubsection{KdV equation with variable coefficients}

To draw a parallel with results obtained from the GP equation at both the linear and nonlinear levels, it is convenient to use a form of the KdV equation coming from an action principle giving a canonical Hamiltonian structure. Let us consider a real scalar field $\psi$ with Lagrangian density
\begin{align}\label{eq:LKdV}
\mathcal{L} = \lp \pd_x \psi \rp \lp \pd_t \psi \rp + \lp \sqrt{g h(x)} + v(x) \rp \lp \pd_x \psi \rp^2 + \frac{1}{2} \sqrt{\frac{g}{h(x)}} \lp \pd_x \psi \rp^3 - \frac{h(x)^2}{6} \sqrt{g h(x)} \lp \pd_x^2 \psi \rp^2,
\end{align}
where $g$ is the gravitational acceleration, $h(x)$ the background water depth, and $v(x)$ the background flow velocity. The momentum conjugate to $\psi$ is $\zeta \equiv \pd_x \psi$. The Euler-Lagrange equation from \eq{eq:LKdV} is
\begin{align}\label{eq:KdV}
\pd_t \zeta + \pd_x \lp \lp \sqrt{g h(x)} + v(x) \rp \zeta + \frac{3}{4} \sqrt{\frac{g}{h(x)}} \zeta^2 + \pd_x \lp \frac{h(x)^2}{6} \sqrt{g h(x)} \pd_x \zeta \rp \rp = 0.
\end{align}
We consider a flow to the left: $v < 0$. 
The dispersion relation for a linear perturbation $\delta \zeta \propto \e^{- \ii \om t + \ii k x}$ in a homogeneous region is
\begin{align}
\om - \lp \sqrt{g h} + v + \frac{3}{2} \sqrt{\frac{g}{h}} \zeta \rp k + \frac{h^2}{6} \sqrt{g h}k^3 = 0.
\end{align}
There is an analogue horizon where $\sqrt{g h} + v + \frac{3}{2} \sqrt{\frac{g}{h}} \zeta = 0$. It corresponds to a black hole horizon if this quantity passes from negative to positive when increasing $x$.  

\subsubsubsection{Stationary black hole solutions in the steep regime}

When considering stationary solutions, integrating \eq{eq:KdV} over $x$ gives
\begin{align}
\lp \sqrt{g h} + v \rp \zeta + \frac{3}{4} \sqrt{\frac{g}{h}} \zeta^2 + \pd_x \lp \frac{h^2}{6} \sqrt{g h} \pd_x \zeta \rp = C,
\end{align}
where $C$ is a real integration constant. One can show that if $h$ and $v$ are homogeneous, then at most two asymptotically homogeneous solutions (up to translations) exist for each value of $C$. These are the homogeneous, subcritical solution and the soliton, which is asymptotically supercritical. 

Let us assume that $v$ and $h$ are piecewise constant functions with only one discontinuity at $x=0$: 
\begin{align}\label{eq:ansatzvh}
v(x) = \left\lbrace 
\begin{array}{rl}
	v_+ & x>0 \\
	v_- & x<0
\end{array}
\right.
, \;
h(x) = \left\lbrace 
\begin{array}{rl}
	h_+ & x>0 \\
	h_- & x<0
\end{array}
\right. .
\end{align}
We look for transcritical AH solutions for $v<0$. That is, we impose that $\zeta$ becomes homogeneous for $x \to \pm \infty$, and that the solution is supercritical for $x \to -\infty$ and subcritical for $x \to + \infty$. Using the matching conditions at $x=0$, namely continuity of $\zeta$ and $h^{5/2} \pd_x \zeta$, we find that the solution must then either be strictly homogeneous (in that case, the soliton on the left is sent to $x \to -\infty$) or contain a half soliton (in which case its center is at $x=0$).  

A straightforward calculation shows that there exists two homogeneous solutions given by
\begin{align}
\zeta=0
\end{align}
and
\begin{align}
\zeta = \frac{4}{3} \frac{\sqrt{g h_-}-\sqrt{g h_+}+v_- - v_+}{\sqrt{\frac{g}{h_+}}- \sqrt{\frac{g}{h_-}}}.
\end{align}
We choose parameters such that the trivial solution $\zeta = 0$ is a black hole solution: $\sqrt{g h_-} + v_- < 0$ and $\sqrt{g h_+} + v_+ > 0$. Then, the second homogeneous solution is not transcritical. A long but straightforward calculation shows that a black hole solution with a half-soliton exists if and only if $h_+ > 16 h_-$. So, for $h_+ < 16 h_-$ the trivial solution is the only AH transcritical stationary solution. The uniqueness result demonstrated in Section~\ref{sub:Ahtf} for the GP equation thus also applies to the KdV equation.

\subsubsection{Whitham modulation equations} 
\label{WKdV}

To apply the general formalism presented in Section~\ref{App:Whitham}, it is convenient to rewrite the KdV equation \eq{eq:KdV} in a canonical form. We assume the background flow is homogeneous and define the non-dimensional variables $Y$, $T$, and $u$ through
\begin{align}
Y \equiv \frac{1}{h} \lp x- \lp \sqrt{g h} + v \rp t \rp, \; 
T \equiv \sqrt{\frac{g}{h}} \frac{t}{6}, \; \text{and} \;
u (Y,T) \equiv \frac{3}{2 h} \zeta (Y,T).
\end{align}
\eq{eq:KdV} then becomes
\begin{align}
\pd_T u + 6 u \pd_Y u + \pd_Y^3 u = 0.
\end{align}
A direct calculation shows that it is the compatibility condition of the system \eq{eq:sys} with the choice
\begin{align}\label{eq:UVKdV}
U = \begin{pmatrix}
	-\ii \la & -1 \\
	u & \ii \la
\end{pmatrix}, \; 
V = \begin{pmatrix}
	-4 \ii \la^3 + 2 \ii u \la + \pd_Y u & -4 \la^2 + 2 u \\
	4 u \la^2 - 2 \ii \la \pd_Y u-2 u^2 - \pd_Y^2 u & 4 \ii \la^3 - 2 \ii \la u - \pd_Y u
\end{pmatrix} . 
\end{align}
The procedure is similar to the one we used for the GP equation, except that we can not restrict to solutions where $h = g^*$. Writing down the system \eq{eq:comp2} explicitly gives, after a few lines of algebra, 
\begin{align}
\left\lbrace 
\begin{array}{ll}
	f + \la g - \ii  \pd_Y g / 2  = 0  \\
	h + \pd_Y^2 g / 2 + \ii \la \pd_Y g + u g = 0 \\
	\pd_Y f^2 - g h = 0 \\
	\pd_T f^2 - g h = 0 \\
	\pd_T g - 2 \lp g \pd_Y u - \lp 2 \la^2 + u \rp \pd_Y g \rp = 0
\end{array}
\right. .
\end{align}
We now look for solutions where $g$ has the form
\begin{align}
g(Y,T; \la) = \la^2 - \mu(Y,T).
\end{align}
Notice that $P(\la) = f^2 - g h$ is now a third-order polynomial in $\la^2$. A straightforward calculation gives 
\begin{align}
\left\lbrace 
\begin{array}{ll}
	u+s-2 \mu = 0 \\
	\lp \pd_T - 2 s \pd_Y \rp \mu = 0 \\
	\lp \pd_Y \mu \rp^2 + P(\la^2 = \mu) = 0
\end{array}
\right.,
\end{align}
where $s = \mu_1 + \mu_2 + \mu_3$ and $\mu_1 \leq \mu_2 \leq \mu_3$ are the three roots of $P(\la)$ in $\la^2$. One can check that this system describes all the periodic solutions which are stationary in a Galilean frame.

Turning to modulated solutions, the parameters $\mu_i$ become slowly-varying functions of $Y$ and $T$ obeying the Whitham equations:
\begin{align}\label{eq:WhithamKdV}
\lp \pd_T + v_i \pd_Y \rp \mu_i = 0,
\end{align}
where
\begin{align}
v_i \equiv 2 \lp \frac{L}{\pd_{\mu_i} L} - s \rp 
\end{align}
and $L$ is the local wavelength, given by
\begin{align}
L = \frac{1}{\sqrt{\mu_3 - \mu_1}} K \lp \frac{\mu_3 - \mu_2}{\mu_3 - \mu_1} \rp.
\end{align}
We now look for scale-invariant solutions depending only on $Z \equiv Y/T$. Then \eq{eq:WhithamKdV} becomes
\begin{align}
\lp v_i - Z \rp \frac{\dd \mu_i}{\dd Z} = 0. 
\end{align}
There are two non-trivial solutions of this system giving asymptotically homogeneous values of $u$. The (unique) simple wave, where one Riemann invariant varies over a finite interval while the two other ones are equal and constant, has $u = Z/6$. In the original variables of \eq{eq:KdV}, it becomes 
\begin{align}\label{eq:SWKdV}
\zeta(x,t) = \frac{2}{3} \sqrt{\frac{h}{g}} \lp \frac{x}{t} - \lp \sqrt{g h} + v \rp \rp
\end{align}
over the domain of variation of $\zeta$. 

There is also one dispersive shock wave, along which $\mu_1$ and $\mu_3$ are constant while $\mu_2$ varies from $\mu_1$ to $\mu_3$. Its two edges are located at \begin{align}
\frac{x}{t} = -\frac{\sqrt{g h}}{3} \lp 2 \mu_1 + \mu_3 \rp + \sqrt{g h} +v \equiv z_1
\end{align}
and
\begin{align}
\frac{x}{t} = -\sqrt{g h} \lp 2 \mu_3 - \mu_1 \rp + \sqrt{g h} +v \equiv z_3.
\end{align}
Since $\mu_3 > \mu_1$, we have $z_1 > z_3$. $\mu_1$ and $\mu_3$ are related to the values of $\zeta$ outside the shock wave through $\zeta(x/t < z_3) = -2 h \mu_1 / 3$ and $\zeta(x/t > z_1) = -2 h \mu_3 / 3$. Existence of this solution thus requires $\zeta(x/t < z_3) > \zeta(x/t > z_1)$. 
These two solutions, shown in \fig{fig:solsKdV},  will now be used to build global ones. 

\subsubsection{Solving the Whitham equations}

\begin{figure}
	\centering
	\includegraphics[width=0.49 \linewidth]{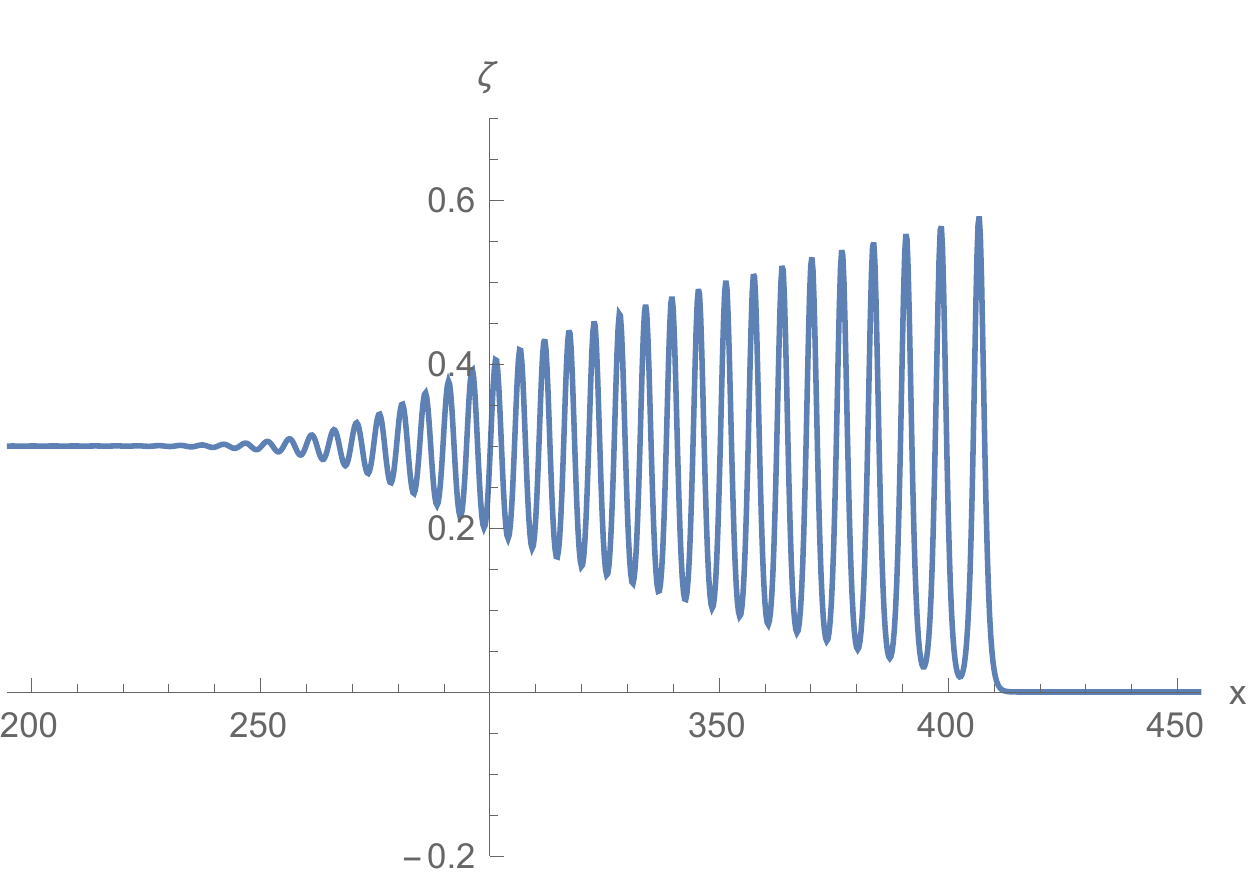}
	\includegraphics[width=0.49 \linewidth]{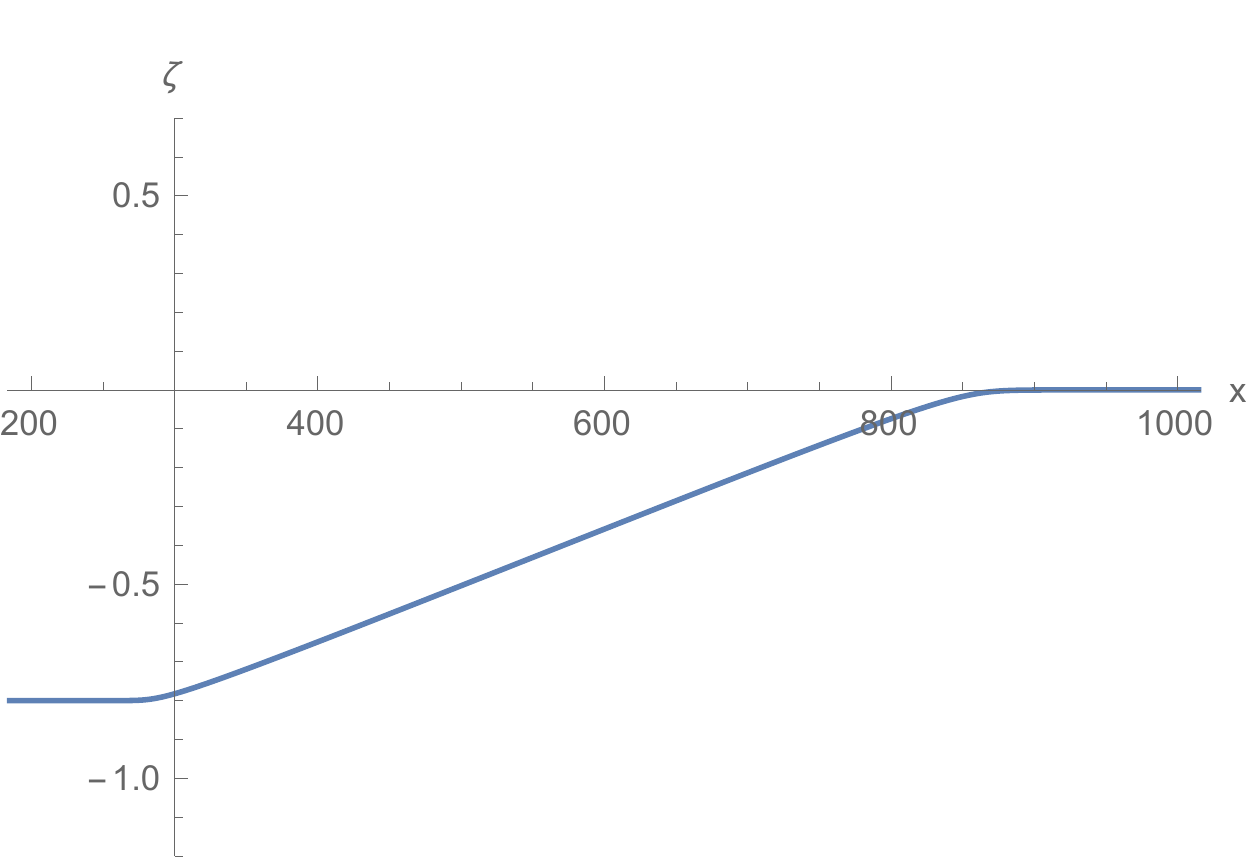}
	\caption{We show a DSW (left) and a SW (right) for the KdV equation. The SW is strictly scale-invariant, as along this solution $\zeta$ depends only on $x/t$ up to corrections not captured by the Whitham equations. For the DSW, the envelope and wavelength of the oscillations depend only on $x/t$.}\label{fig:solsKdV}
\end{figure}

We now turn to the resolution of the Whitham modulation equations for \eq{eq:KdV}. Specifically, we look for self-similar solutions which interpolate between the homogeneous solution $\zeta = 0$ around $x=0$ and arbitrary asymptotic values $\zeta_\pm$ of $\zeta$ at $x \to \pm \infty$. Since the solution is independent on $\zeta_\pm$ around $x = 0$, we can consider independently the two regions $x<0$ and $x>0$.\footnote{Hence we avoid the additional complication of the GP equation, where the velocity of the solution around $x=0$ is determined by the asymptotic conditions.} On each side, we look for a DSW or simple wave with the two following properties:
\begin{itemize}
	\item It has correct values of $\zeta$ at its two edges, i.e., $\zeta = 0$ at its edge closest to the horizon $x=0$, and $\zeta = \zeta_\pm$ at its other edge;
	\item It moves away from $x=0$, i.e., both edges have a strictly positive velocity in the region $x > 0$ and a strictly negative velocity in the region $x<0$.  
\end{itemize}
Examples of SW and DSW are shown in \Fig{fig:solsKdV}. 
Using the results of above subsection, we find the following necessary and sufficient conditions to obtain 
\begin{itemize}
	\item an SW for $x<0$: 
	\begin{align}
	\zeta_- < 0 \wedge \sqrt{g h_-} + v_- < 0;
	\end{align}
	\item an SW for $x>0$: 
	\begin{align}
	\zeta_+ > 0 \wedge \sqrt{g h_+} + v_+ > 0;
	\end{align}
	\item a DSW for $x<0$:
	\begin{align}
	\zeta_- > 0 \wedge \sqrt{g h_-} + v_- + \sqrt{\frac{g}{h_-}} \zeta_- < 0;
	\end{align}
	\item a DSW for $x>0$:
	\begin{align}
	\zeta_+ < 0 \wedge \sqrt{g h_+} + v_+ + 3 \sqrt{\frac{g}{h_+}} \zeta_+ > 0.
	\end{align}
\end{itemize}
Hence, if the parameters $h_+$, $h_-$, $v_+$, and $v_-$ are such that the trivial homogeneous solution $\zeta = 0$ has a black hole horizon, then there exists a solution which is identically equal to zero in a finite, time-dependent neighborhood $I_t$ of $x=0$ provided
\begin{align}\label{eq:fincondKdV}
\zeta_- < - \sqrt{\frac{h_-}{g}} \lp \sqrt{g h_-} + v_- \rp \wedge \zeta_+ > - \frac{1}{3} \sqrt{\frac{h_+}{g}} \lp \sqrt{g h_+} + v_+ \rp.
\end{align}
Moreover, in this case, $\lim_{t \to \infty} I_t = \mathbb{R}$. The solution thus uniformly converges to the trivial one $\zeta = 0$ in any bounded interval. 
In conclusion, as in the case of the GP equation, when  \eq{eq:fincondKdV} is satisfied, the set of AH solutions acts as a local attractor.  

When \eq{eq:fincondKdV} is not satisfied, one dispersive shock wave computed using the Whitham equations has one edge moving towards the horizon. The wave is then scattered by the discontinuity of $h$ and $v$ at $x=0$. This process is not described by the modulation theory.\footnote{However, the resulting pattern should be describable using two-phase solutions of the Whitham equations~\cite{doublecnoidal,multiphase}.} Numerical simulations indicate that a stationary undulation is 
generically obtained at late times.

\subsubsection{Superluminal KdV equation}

A superluminal version of the KdV equation can be obtained by changing the sign of the last term in \eq{eq:LKdV}:
\begin{align}\label{eq:sKdV}
\pd_t \zeta + \pd_x \lp \lp \sqrt{g h(x)} + v(x) \rp \zeta + \frac{3}{4} \sqrt{\frac{g}{h(x)}} \zeta^2 - \pd_x \lp \frac{h(x)^2}{3} \sqrt{g h(x)} \pd_x \zeta \rp \rp = 0.
\end{align}
Like the GP equation, \eq{eq:sKdV} has a superluminal dispersion relation: 
\begin{align}
\om - \lp \sqrt{g h} + v + \frac{3}{2} \sqrt{\frac{g}{h}} \zeta \rp k - \frac{h^2}{6} \sqrt{g h}k^3 = 0.
\end{align}
However, like the KdV equation it describes only right-moving modes in the fluid frame where $v=0$. Solutions for the superluminal KdV equation \eq{eq:sKdV} are in one-to-one correspondence with those of \eq{eq:KdV} through the transformation 
\begin{align}\label{eq:KdVtosKdV}
\begin{array}{rl}
	x & \hspace*{-0.1 cm} \to -x, \\
	\zeta & \hspace*{-0.1 cm} \to -\zeta, \\
	\sqrt{g h} + v & \hspace*{-0.1 cm} \to - \lp \sqrt{g h} + v \rp,
\end{array}
\end{align}
which preserves the black- or white hole nature of the flow. The uniqueness result mentionned above for the KdV equation thus extends to the present case. The conditions of existence of the simple waves are unchanged. The conditions for the DSW become
\begin{itemize}
	\item DSW for $x<0$:
	\begin{align}
	\zeta_- > 0 \wedge \sqrt{g h_-} + v_- + 3 \sqrt{\frac{g}{h_-}} \zeta_- < 0;
	\end{align}
	\item DSW for $x>0$:
	\begin{align}
	\zeta_+ < 0 \wedge \sqrt{g h_+} + v_+ + \sqrt{\frac{g}{h_+}} \zeta_+ > 0.
	\end{align}
\end{itemize}
Hence, if the parameters $h_+$, $h_-$, $v_+$, and $v_-$ are such that the trivial homogeneous solution $\zeta = 0$ has a black hole horizon, then there exists a solution which is identically equal to zero in a finite, time-dependent neighborhood $I_t$ of $x=0$ such that $\lim_{t \to \infty} I_t = \mathbb{R}$ provided
\begin{align}\label{eq:fincondsKdV}
\zeta_- < - \frac{1}{3} \sqrt{\frac{h_-}{g}} \lp \sqrt{g h_-} + v_- \rp \wedge \zeta_+ > - \sqrt{\frac{h_+}{g}} \lp \sqrt{g h_+} + v_+ \rp .
\end{align}
As in the previous case, this condition is always satisfied provided the asymptotic conditions are sufficiently close to those compatible with a homogeneous black hole flow (where $\zeta_+ = \zeta_- = 0$). 
\subsection{Linear no-hair theorem for KdV}
\label{App:linKdV}

In this subsection we consider the linear stability of a black hole flow of the KdV equation (\ref{eq:KdV}). For definiteness and simplicity, we consider the trivial solution $\zeta = 0$. We follow a procedure similar to that of the Appendix C in~\cite{Michel:2015pra}, although we here use the exact scattering coefficients to match the solutions at $x=0$. The physical picture underlying the calculation is the following: an initially localized perturbation will generally split into two parts. 
One of them moves away from the horizon and is diluted due to dispersion. The other one is scattered on the horizon and stimulates the Hawking effect, that is, pairs of of modes carrying opposite energies are emitted on both sides of the horizon. 
At late times, provided the amplification from the scattering is not strong enough to compensate for the dilution, only the modes with very small group velocities remain, with an amplitude which decreases polynomially in time, as the range of frequencies of modes which remains in the near-horizon region decreases. 

To make this idea more precise, we first briefly discuss the structure of the modes of the linearized KdV equation. We then determine the mode content of a square perturbation and compute its late-time evolution using a saddle-point approximation. To carry out the explicit calculation, we consider a steplike black hole flow. 

\subsubsection{Modes in a steplike black hole flow}

We consider the setup of \eq{eq:ansatzvh}, with $\sqrt{g h_-} + v_- < 0$ and $\sqrt{g h_+} + v_+ > 0$. Linearising the KdV equation and integrating over $x$ gives (up to a real constant which can be absorbed by a redefinition of $\psi$ through $\psi \to \e^{\ii \mu t}$, $\mu \in \mathbb{R}$)
\begin{align} \label{eq:lKdV}
\pd_t \psi + \lp \sqrt{g h} + v \rp \pd_x \psi + \pd_x \lp \frac{h^2}{6} \sqrt{g h} \pd_x^2 \psi \rp = 0.
\end{align}
The inner product $\lp \cdot, \cdot \rp$ of two solutions of this equation is defined by
\begin{align}
\lp \psi_1, \psi_2 \rp \equiv \frac{\ii}{2} \int_\mathbb{R} \lp \psi_1^* \pd_x \psi_2 - \psi_2 \pd_x \psi_1^* \rp \dd x.
\end{align}
A straightforward calculation shows that $\pd_t \lp \psi_1, \psi_2 \rp = 0$. 
In the following we shall loosely call $(\psi_1, \psi_1)$ the ``norm'' of the mode $\psi_1$. 

\begin{figure}
	\centering
	\includegraphics[width = 0.5 \linewidth]{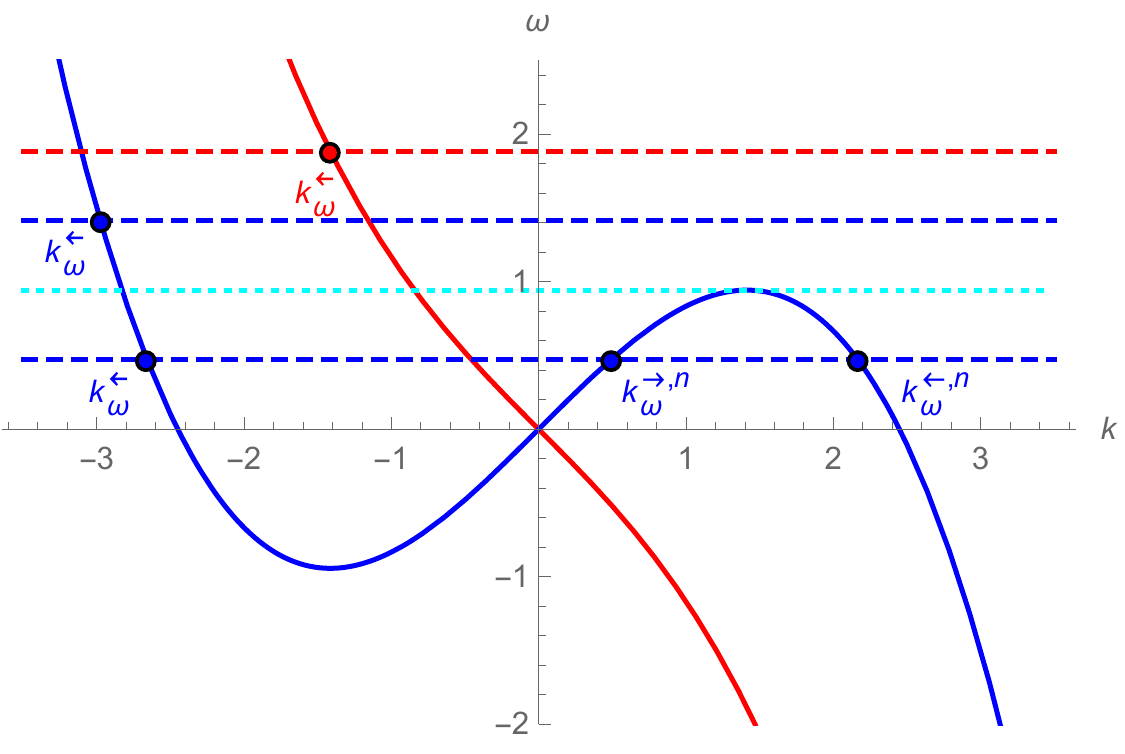}
	\caption{Dispersion relation for the linearized KdV equation~(\ref{eq:lKdV}) in subcritical (blue) and supercritical (red) regions. Points give the real roots of the dispersion relation at fixed frequencies. The dotted, cyan line shows the value $\om_c$ of $\om$ at which two roots in the subcritical region merge and become complex. Arrows indicate the direction of the group velocity, and a superscript ``$n$'' labels the modes with negative norms.} \label{fig:DRKdV}
\end{figure}

The dispersion relation relating the angular frequency $\om$ of a stationary mode to its wave vector $k$ in a homogeneous region is shown in \Fig{fig:DRKdV}. At fixed $\om \in \mathbb{R}$, we have only one real wave vector $k_\om^{\leftarrow}$ on the supercritical side. The corresponding mode has a negative group velocity and a positive norm for $\om > 0$. We also have two complex roots, giving exponentially increasing and decreasing modes for $x \to -\infty$. We shall call the wave vector of the decreasing mode $k_\om^-$. In the subcritical region, the two complex wave vectors become real when $\om \in [-\om_c, \om_c]$, where
\begin{align}
\om_c = \frac{2 \sqrt{2}}{3} \sqrt{\frac{g}{h_+}} \lp  1 + \frac{v_+}{\sqrt{g h_+}} \rp^{3/2}.
\end{align}
In the following it will be convenient to work with globally-defined incoming modes. For $\om \in [- \om_c, \om_c]$, there are two of them, which we call $\psi_{\om}^{\rm in}$ and $\psi_{\om}^{{\rm in},n}$.  If $\om > 0$, $\psi_{\om}^{\rm in}$ has a positive norm while $\psi_{\om}^{{\rm in},n}$ has a negative norm. They are given by
\begin{align}
\psi_{\om}^{{\rm in}, (n)} =
\left\lbrace 
\begin{array}{ll}
	b_{\om,\leftarrow}^{(n)} \e^{\ii k_{\om,L}^\leftarrow x} + b_{\om,-}^{(n)} \e^{\ii k_{\om,L}^- x} & x<0 \\
	e^{\ii k_{\om,R}^{\leftarrow,(n)} x} + b_{\om, \rightarrow}^{(n)} \e^{\ii k_{\om,R}^{\rightarrow,n}x} & x>0
\end{array}
\right.,
\end{align}
where an index ``$L$'' or ``$R$'' indicates the region (left or right) where the wave vector is evaluated. The coefficients $b_{\leftarrow}^{(n)}$, $b_-^{(n)}$, and $b_{\rightarrow}^{(n)}$ are given by
\begin{align}\label{eq:epxplicitcoeff}
b_{\om,\rightarrow}^{(n)} &= \frac{h_-^{5/2} \lp k_{\om,L}^- k_{\om,R}^{\leftarrow,(n)} + k_{\om,L}^{\leftarrow} k_{\om,R}^{\leftarrow,n} -k_{\om,L}^\leftarrow k_{\om,L}^- \rp - h_+^{5/2} \lp k_{\om,R}^{\leftarrow,(n)} \rp^2}{h_-^{5/2} \lp k_{\om,L}^- k_{\om,L}^{\leftarrow} - k_{\om,L}^\leftarrow k_{\om,R}^{\rightarrow,n} -k_{\om,R}^{\leftarrow,n} k_{\om,L}^- \rp + h_+^{5/2} \lp k_{\om,R}^{\rightarrow,n} \rp^2}, \nn
b_{\om,\leftarrow}^{(n)} &= \frac{ \lp k_{\om,R}^{\leftarrow,(n)} - k_{\om,R}^{\rightarrow,n} \rp \lp h_+^{5/2} \lp k_{\om,R}^{\rightarrow,(n)} k_{\om,R}^{\leftarrow,(n)} + k_{\om,L}^- k_{\om,R}^{\rightarrow,n} -k_{\om,L}^- k_{\om,R}^{\leftarrow,(n)} \rp + h_-^{5/2} \lp k_{\om,L}^{-} \rp^2 \rp}{ \lp k_{\om,L}^- - k_{\om,L}^\leftarrow \rp \lp h_-^{5/2} \lp k_{\om,L}^- k_{\om,L}^{\leftarrow} - k_{\om,L}^\leftarrow k_{\om,R}^{\rightarrow,n} -k_{\om,R}^{\leftarrow,n} k_{\om,L}^- \rp + h_+^{5/2} \lp k_{\om,R}^{\rightarrow,n} \rp^2 \rp}, \nn
b_{\om,-}^{(n)} &= \frac{ \lp k_{\om,R}^{\leftarrow,(n)} - k_{\om,R}^{\rightarrow,n} \rp \lp h_+^{5/2} \lp k_{\om,L}^\leftarrow k_{\om,R}^{\rightarrow,(n)} + k_{\om,L}^{\leftarrow} k_{\om,R}^{\leftarrow,n} -k_{\om,R}^{\leftarrow,n} k_{\om,R}^{\rightarrow,(n)} \rp + h_{\om,R}^{5/2} \lp k_{\om,L}^{\leftarrow} \rp^2 \rp}{ \lp k_{\om,L}^- - k_{\om,L}^\leftarrow \rp \lp h_-^{5/2} \lp k_{\om,L}^- k_{\om,L}^{\leftarrow} - k_{\om,L}^\leftarrow k_{\om,R}^{\rightarrow,n} -k_{\om,R}^{\leftarrow,n} k_{\om,L}^- \rp + h_+^{5/2} \lp k_{\om,R}^{\rightarrow,n} \rp^2 \rp}.
\end{align}
When $|\om| > \om_c$, only the mode $\psi_{\om}^{\rm in}$ remains, with $k_{\om,R}^{\rightarrow,n}$ replaced by the wave vector with a positive imaginary part. It can be shown that the only divergences of these coefficients occur for $\om \to 0$, where $b_{\om,\leftarrow}^{(n)}$ and $b_{\om,\rightarrow}^{(n)}$ diverge as $1/\om$ while $b_{\om,-}^{(n)}$ remains finite. Importantly, the diverging coefficients multiply exponentials with wave vectors which vanish linearly in $\om$ for $\om \to 0$. So, 
this divergence is regularized when expressing the results in terms of $\zeta = \pd_x \psi$. 

\subsubsection{Late-time evolution of a steplike perturbation}

Let us consider some initial condition $\psi(x,t=0) = \psi_0(x)$. One may expand it on the basis of incoming modes as
\begin{align}\label{eq:intmodes} 
\psi_0(x) = \int_{-\om_c}^{\om_c} \lp A_\om \psi_\om^{\rm in} (x,0) + A_\om^n \psi_\om^{{\rm in},n} (x,0) \rp \dd \om + \int_{\mathbb{R} \backslash [-\om_c, \om_c]} A_\om \psi_\om^{\rm in}(x,0) \, \dd \om,
\end{align}
where
\begin{align}\label{eq:A}
A_\om^{(n)} = - \frac{\lp \psi_\om^{\rm in,(n)}, \psi_0 \rp}{2 \pi k_{\om,R}^{\leftarrow,(n)} \left\lvert \pd_\om k_{\om,R}^{\leftarrow,(n)} \right\rvert^{-1}}.
\end{align}
For definiteness, we work with a perturbation of the form
\begin{align}
\psi_0 (x) = C \theta(x-x_-) \theta(x_+-x),
\end{align}
where $x_- < x_+$ are two real numbers. Notice that using a sum of such terms we can approximate any localized, smooth initial perturbation. Then,
\begin{align}\label{eq:explicitsc}
\lp \psi_\om^{\rm in,(n)}, \psi_0 \rp = \ii \, C \lp \psi_\om^{\rm in,(n)}(x_-,0) - \psi_\om^{\rm in,(n)}(x_+,0) \rp.
\end{align}
From then on, the analysis is long but straightforward. Since it does not present any particular difficulty, we shall only sketch it and give its main results. Using the linearity of \eq{eq:lKdV} we can assume without loss of generality that $x_-$ and $x_+$ have the same sign. Using Eqs.~(\ref{eq:epxplicitcoeff}, \ref{eq:A}, \ref{eq:explicitsc}), \eq{eq:intmodes} may be written as a sum of integrals of plane waves multiplied by prefactors which depend polynomially on $\om$. Writing everything explicitly, one checks that all divergences of the prefactors cancel each other when working with $\zeta$. At late times, the only contributions thus come from the saddle-points. These occur only when $\pd_\om k_\om = (x+...) / t$, where the three dots indicate the possible addition of $\pm x_+$ and/or $\pm x_-$. In the limit $t \to \infty$ at fixed $x$, only the waves with vanishing group velocities thus contribute, i.e., those of frequency $\om = \pm \om_c$. Using that the corresponding group velocity goes to zero like $\sqrt{\om_c - |\om|}$, we obtain after the Gaussian integration
\begin{align}\label{eq:zetat3/2}
\zeta(x,t) \mathop{\approx}_{t \to \infty} O (t^{-3/2}),
\end{align}
where the symbol $\approx$ is used to emphasize that logarithmic corrections not captured by the Gaussian integration may be present.

The same analysis can be done for a steplike perturbation where either $x_-$ or $x_+$ is sent to infinity. This is important as a localized perturbation on $\psi$ corresponds to a perturbation on $\zeta$ with a vanishing mean, and it is not a priori clear that the limits $t \to \infty$ and $x_\pm \to \pm \infty$ commute. To be complete, we must thus also consider one localized perturbation on $\zeta$ such that $\int \zeta \dd x \neq 0$. Choosing for instance $\psi_0(x) = C \theta(x)$, we obtain
\begin{align}
\lp \psi_\om^{{\rm in},(n)}, \psi_0 \rp = \ii \, C (1+b_{\om,\rightarrow}^{(n)}).
\end{align}
The calculation then follows the same lines as the previous case. 
An additional divergence is present for $\om \to 0$, but does not contribute to the late-time solution as it multiplies an incoming wave with a non-vanishing group velocity. We thus find again \eq{eq:zetat3/2}.

\subsubsection{Case of the BdG equation~\eq{eq:BdG}} 

As mentioned in the main text, the above calculation can be done for \eq{eq:BdG}.
The only point which does not directly extend to that case concerns the cancellation of divergences for $\om \to 0$. For the KdV equation, their cancellation when considering physical waves is due to two properties. First, all diverging terms multiply waves with a linearly vanishing wave vector. Second, the particular relationship between the solution $\phi$ of the linear equation and the physical observable introduces an additional factor $\om$. The first point carries on to the case of the BdG equation. The second one is replaced by the fact that the solution of \eq{eq:formlin} linearly vanishes when $\om \to 0$ for the corresponding waves.

\subsection{black hole solutions of the cubic-quintic GP equation}
\label{App:CQNLS} 

Both the GP equation \eqref{eq:GPE} and the KdV equation possess the specific property of being integrable by inverse scattering in a homogeneous background. Since this property played a crucial role in the derivation of DSW and SW solutions, it is of interest to consider the case of non-integrable equations. As an example, we consider the cubic-quintic GP (CQGP) equation 
\begin{align}\label{eq:CQGPE}
\ii \pd_t \psi(x,t) = -\frac{1}{2} \pd_x^2 \psi(x,t) +V(x) \psi(x,t) + g(x) \left\lvert \psi(x,t) \right\rvert^2 \psi(x,t) + \la \abs{\psi(x,t)}^4 \psi(x,t),
\end{align}
where $\la$ is a real parameter. To our knowledge, \eq{eq:CQGPE} is not integrable for $\la \neq 0$, even with uniform $V$ and $g$. It may thus be used to determine the effect of a small non-integrable deformation of \eq{eq:GPE}. 

To obtain the linearized equation, we consider a solution given by \eq{relpert}. Plugging it into \eq{eq:CQGPE} and extracting the first order in $\phi$ gives
\begin{align}
\ii \lp \pd_t + v \pd_x \rp \phi = -\frac{1}{2} \pd_x^2 \phi - \frac{\pd_x \rho}{2\rho} \pd_x \phi + \lp g + 2 \la \rho \rp \rho \lp \phi + \phi^* \rp.
\end{align}
In a region where $\rho$ and $v$ are constant, we can again look for solutions of the form of \eq{eq:formlin}. The resulting system has solutions if and only if the dispersion relation:
\begin{align}
\lp \om - v k \rp^2 = \lp g \rho + 2 \la \rho^2 \rp k^2 + \frac{k^4}{4}
\end{align}
is satisfied. The sound velocity, i.e., the velocity of long wavelength perturbations in the fluid frame, is now given by $c_S(\rho) = \sqrt{g \rho + 2 \la \rho^2}$. 

Let us briefly discuss the asymptotically homogeneous, stationary solutions of \eq{eq:CQGPE}. Rewriting $\psi$ as $\psi(x,t) = \sqrt{\rho(x,t)} \e^{\ii \theta(x,t)}$, where $\rho$ and $\theta$ are real, \eq{eq:CQGPE} becomes
\begin{align}
\left\lbrace 
\begin{array}{l}
	\pd_t \rho + \pd_x \lp \rho \pd_x \theta \rp = 0 \\
	\pd_t \theta - \frac{1}{2} \lp \frac{\pd_x^2 \sqrt{\rho}}{\sqrt{\rho}} - \lp \pd_x \theta \rp^2 \rp + V(x) + g(x) \rho + \la  \rho^2 = 0
\end{array}
\right. .
\end{align}
We consider a stationary solution where $\rho$ is independent on $t$ and $\pd_t \theta = \om$ is a constant. Then, $J \equiv \rho \pd_x \theta$ is also a constant and 
\begin{align}
\pd_x^2 \sqrt{\rho} = \frac{J^2}{\rho^{3/2}} + 2 (V - \om) \sqrt{\rho} +2 g \rho^{3/2} + 2 \la \rho^{5/2}. 
\end{align}
In a region where $V$ and $g$ are homogeneous, solutions with homogeneous densities exist iff 
$J^2 < J_c^2$, where $J_c$ is given by
\begin{align}
J_c & = \sqrt{-\lp 2 (V-\om) \rho_c^2 + 2 g \rho_c^3 + 2 \la \rho_c^4 \rp}, 
\nonumber \\
\rho_c & \equiv \frac{\sqrt{9 g^2 + 32 (\om - V) \la} -3 g}{8 \la}.
\end{align}
Then, as was the case for the standard GP equation, there exists two homogeneous solutions: a supersonic one with density $\rho_p$ and a subsonic one with density $\rho_b$. The latter can be seen as the limit of a stationary soliton solution when the center of the soliton is sent to infinity. 

We now focus on the steplike regime, with functions $g$ and $V$ given by \eq{eq:step}. If $g_+ < g_-$ and $V_+ > V_-$, there exist AH transonic solutions with density $\rho = \rho_0$, a conserved current satisfying $\sqrt{g_+ \rho_0 + 2 \la \rho_0^2} < (J/\rho_0) < \sqrt{g_- \rho_0 + 2 \la \rho_0^2}$, and a frequency 
\begin{align}
\om = \frac{J^2}{2 \rho_0^2} + \frac{V_+ + V_-}{2} + \frac{g_+ + g_-}{2} \rho_0 + \la \rho_0^2.
\end{align}
The other possible black hole solution with an asymptotically homogeneous density on both sides contains half a soliton in the subsonic region. Like in the case of the standard GP equation, one can show that it does not exist for $J/\rho_0$ inside the above interval.  

\begin{figure}[ht]
	\centering
	\includegraphics[width=0.49 \linewidth]{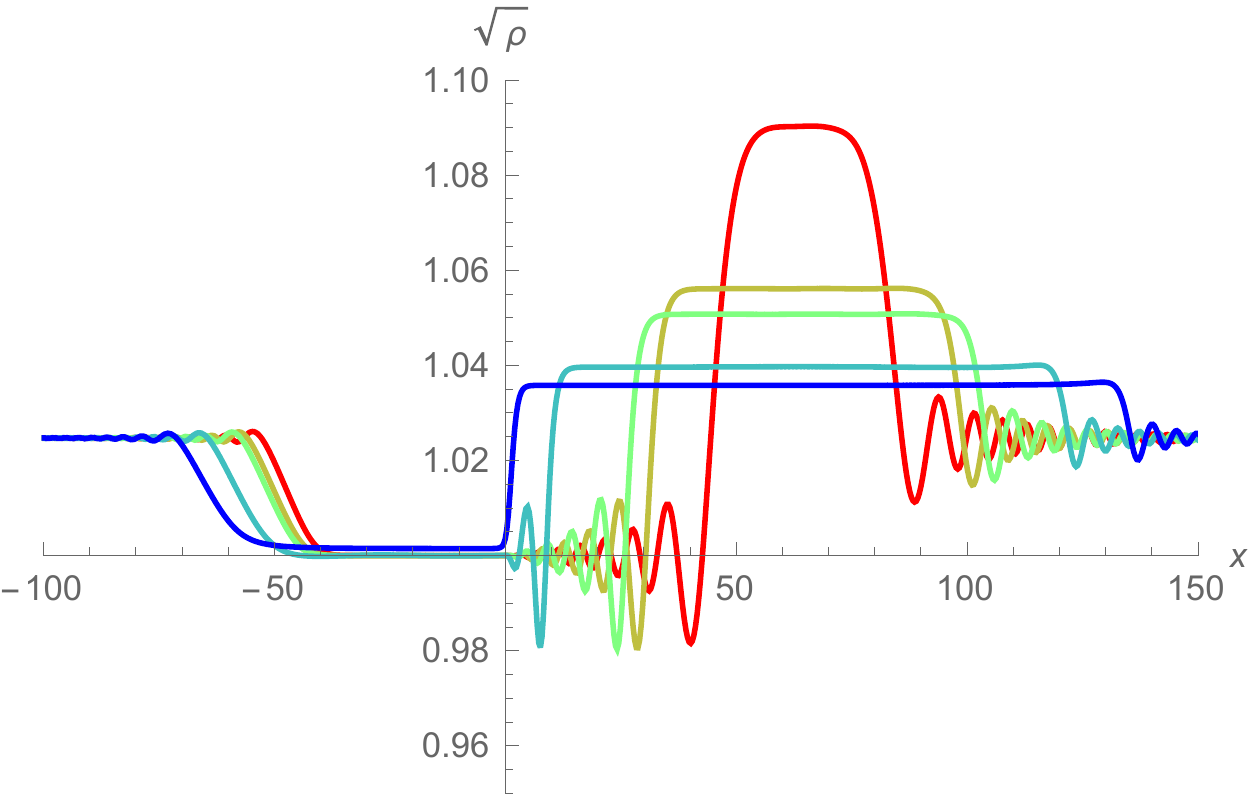} 
	\includegraphics[width=0.49 \linewidth]{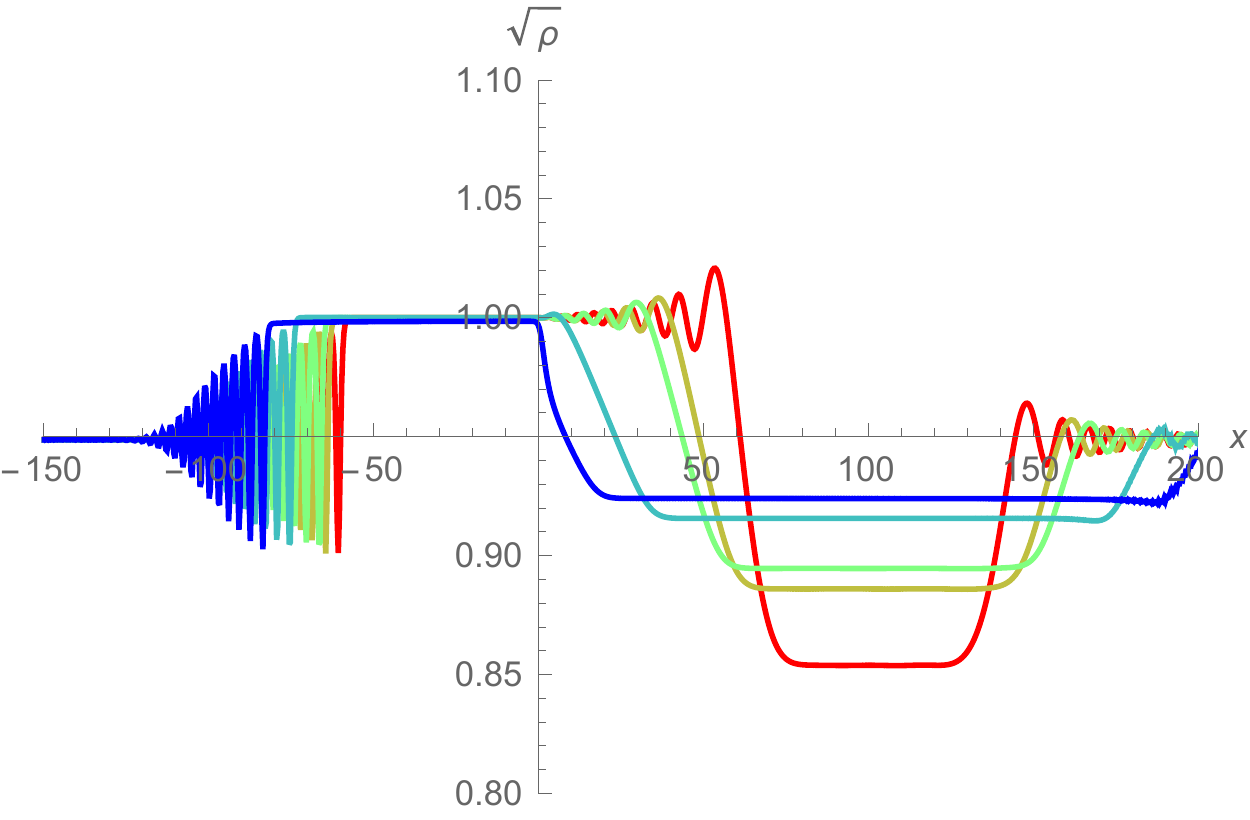}
	\caption{Solutions obtained with the CQGP \eq{eq:CQGPE} in the steplike regime. The parameters $g_+$, $g_-$, $V_+$, $V_-$, and $J$ are the same as in the left panel of \fig{fig:solsNLSb}. The curves correspond to different values of $\la$: from red to blue, $-0.3$, $-0.1$, $0$, $0.5$, and $1$. The profiles are shown for  $t=40$. The initial configuration at $t=0$ is homogeneous with a density $\rho_i = 1.05$ (left) and $\rho_i = 0.9$ (right). One clearly sees that increasing $\lambda$ smoothly deforms the solutions without affecting their qualitative properties. In particular, the set of homogeneous flows still seems to act as a local attractor.} \label{fig:CQGPE}
\end{figure}
In \fig{fig:CQGPE} we show results of numerical simulations for homogeneous initial conditions. For small values of $\la$, the solution still has three waves of the same type (SW or DSW) and with the same direction of propagation as for $\la = 0$. The main effects of $\la$ are to change the velocities of the waves and the value of the density between the two rightmost ones. A positive value of $\la$ seems to increase the velocity of the rightmost wave and reduce that of the two other ones, while bringing the density between the two rightmost waves closer to $\rho_0$. A small, negative value of $\la$ has the opposite effect. When increasing $\abs{\la}$, we observe a qualitative change in the solution. For $\la$ smaller than a critical value, close to $-0.4$ for the parameters of the figure, the solution diverges after a finite time. When $\la$ is larger than another critical value, close to $1$ in the figure, the velocity of the middle wave becomes negative and the solution develops a hairy black hole flow.

\newchapter{Probing the thermal character of analogue Hawking radiation for shallow-water waves}
\label{ch:probing}
\begin{tikzpicture}[overlay]
\newcommand*{\xA}{-0.2}
\newcommand*{\xB}{14.55}
\newcommand*{\yA}{6.7}
\newcommand*{\yB}{1.5}
\newcommand*{\epsil}{0.75}
\draw[overlay] (\xA-\epsil,\yA) -- (\xB-\epsil,\yA);
\draw[overlay] (\xA,\yA+\epsil) -- (\xA,\yB);
\draw[overlay] (\xB,\yB-\epsil) -- (\xB,\yA);
\draw[overlay] (\xB+\epsil,\yB) -- (\xA+\epsil,\yB);
\end{tikzpicture}
\begin{small}
This chapter deals more specifically with the third axis of the thesis, namely the design of experiments aimed at observing the analogue Hawking radiation. 
As was mentioned in the general introduction, one of the conceptually simplest analogue models is a transcritical water flow over an obstacle. 
Its main advantage over the original model involving sound waves is that, since the velocity of surface water waves is typically much smaller than that of sound waves, realizing a horizon requires flows with smaller velocities. 
Moreover, as surface waves can be seen with the naked eye they provide a visual and intuitive model where the Hawking radiation can be studied. 
However, below this apparent simplicity are several nontrivial problems which can severely complicate the interpretation of experimental observations. 

First, while a theoretical model may deal with arbitrarily long-wavelength modes, in practice the latter are very difficult to measure accurately. 
This is an important limitation for black hole flows: because of the strong redshift of incoming waves close to the horizon, the modes produced by the Hawking effect have a much longer wavelength than the incoming ones, making them very difficult to detect at low frequency. 
For this reason, experiments are usually done with white hole flows instead, in which the Hawking effect generates waves with much smaller wavelengths.  
However, flows which are asymptotically subcritical in the downstream region, and in particular white-hole flows, usually give rise to an undulation of the free surface. 
The latter has the same wavelength as the dispersive zero-frequency modes, and may thus induce a resonant scattering on top of the Hawking effect, from which it may be difficult to separate. We shall briefly comment on this at the end of this chapter. 

Another issue, which motivated the work~\cite{Michel:2014zsa} on which the present chapter is based, is that while the outgoing waves have a short wavelength and can be accurately measured, the incoming ones are much longer. 
It is thus difficult to measure the scattering coefficients independently, which would require a good control of both the incoming and outgoing waves. 
Instead, in the seminal experiment~\cite{Weinfurtner:2010nu}, the group of S.~Weinfurtner determined the ratio of the two coefficients involving dispersive outgoing waves. 
They found that the behavior of this quantity was consistent with the thermal spectrum predicted by S.~Hawking.  
This raises the question of whether, or in which sense, the sole ratio of these scattering coefficients can give information about the thermal character of the spectrum. 
With this in mind, the first objective of this chapter is to see under which conditions thermality in the sense of~\cite{Weinfurtner:2010nu} is equivalent to usual statistical definition of temperature. 
We shall see that, in the present context, a necessary and sufficient condition for the two definitions to be equivalent is that the flow be significantly transcritical. 
Another, deeply related, aim is to determine the evolution of the spectrum when going from a transcritical to a globally subcritical flow. 
This is also important for experiments because it is in practice difficult to realize a stable, transcritical flow with a small undulation. 
For this reason, so far all analogue gravity experiments using water waves seem to work with subcritical flows, without any analogue Killing horizon. 
A precise understanding of the spectral properties in subcritical flows is thus required to analyze the experimental data and use them as input to progressively move towards actual transcritical flows with a strong connection to the standard Hawking process.

As we shall see, our numerical results show a good agreement with the observations of~\cite{Weinfurtner:2010nu}. 
We also find that their interpretation in terms of thermal spectrum depends on the precise definition of the effective temperature. 
Using the definition of~\cite{Weinfurtner:2010nu}, the emitted spectrum is clearly thermal in a wide range of frequencies. 
When using the ``statistical'' definition, however, the spectrum does not seem to be thermal for parameters expected to model this experiment with a relatively good fidelity.  
\end{small}
\newpage

\renewcommand*{\theHsection}{\theHchapter.\the\value{section}}
\renewcommand\thesection{\arabic{section}}

\section{Introduction}

As shown by B.~Unruh in~\cite{Unruh:1980cg}, it is possible to use non-relativistic fluids to test S.~Hawking's prediction that black holes spontaneously emit a steady thermal flux~\cite{Hawking:1974sw}. 
This is because the wave equation governing the propagation of long-wavelength density perturbations in an inhomogeneous flow has the form of the d'Alembert equation in a curved space-time. 
As a result, in a transonic stationary flow where the velocity $v$ crosses the speed of low-frequency waves $c$, the wave equation is identical to that of a scalar field in a black hole metric. 
The coefficients governing the scattering of density perturbations should thus show the mode amplification at the root of the Hawking effect. 
However, this strict correspondence breaks down because the scattering also involves short-wavelength modes~\cite{Jacobson:1991gr}, the propagation of which is dispersive and thus not governed by the d'Alembert equation. 
To identify the consequences of such dispersive effects, B.~Urunh~\cite{Unruh:1994je} numerically solved a dispersive wave equation which governs the propagation of sound waves in an analogue black hole flow. 
He found that, when there is a neat separation between the short dispersive length scale and the surface gravity scale which fixes the Hawking temperature, dispersive effects do not significantly affect the spectral properties of the scattering coefficients. 
This second work therefore indicates that one may experimentally test the Hawking prediction in dispersive media in which these two scales are well separated.

This analogy is not restricted to density perturbations. 
For instance, surface waves propagating over a water flow in a flume can also be seen as an analogue model, as was shown in~\cite{Schutzhold:2002rf}. 
Following this work, several experiments have been conducted to observe the conversion of shallow-water waves (i.e., long wavelengths) into deep-water waves (i.e., short wavelengths) which occurs near a blocking point~\cite{Rousseaux:2007is,Weinfurtner:2010nu}. 
This can be seen as the time reversed of the standard Hawking effect, and the effective space-time metric near the blocking point is that of a white hole. 
To have a close analogy with the Hawking effect in astrophysics, the background flow that engenders this metric should be transcritical.  
Rephrased in the hydrodynamic language, the Froude number $F = v/c$ should be smaller than $1$ in some region and larger in another one. However in the experiments~\cite{Rousseaux:2007is,Weinfurtner:2010nu}, the flows were, apparently, subcritical. 
Yet, some mode conversion was clearly observed. 
In addition, when measuring the relative amplitudes of the scattered waves for different frequencies, S.~Weinfurtner \textit{et al.} observed a “thermal law”, in agreement with Hawking's prediction. 
These observations seem to indicate that dispersion plays an important role. 

Following~\cite{Unruh:1994je} the consequences of short-distance dispersion have received a lot of attention~\cite{Brout:1995wp,Corley:1996ar,Corley:1997pr,Balbinot:2006ua, Macher:2009tw,Robertson:2012ku, Coutant:2011in}, and by now there is a fair understanding of the spectral deviations due to dispersion when the flow is {\it significantly transcritical}, i.e., when the maximum value of the Froude number is above, and not too close to, unity. 
Comparatively, much less attention has been devoted to the cases where $F$ barely crosses 1 or does not cross it at all. 
In~\cite{finazziRP-proceedings}, it was shown that the Planckianity of the spectrum is progressively lost when the maximum value of $F$ approaches $1$ from above. 
When $F$ no longer crosses 1, it was also found that there is a critical frequency $\omega_{\rm min}$ below which a new scattering channel opens up, and above which the spectrum closely resembles that found when $F$ barely crossed 1. 
These results have been derived with a superluminal dispersion relation, as that found in atomic Bose gases, but also apply to subluminal dispersion because of the symmetry between sub- and superluminal dispersion, detailed in the subsection~III~E of~\cite{Coutant:2011in} (verified by direct calculation in~\cite{Robertson:2011xp,Robertson:2012ku}).

The main objective of the present chapter is to complete these analyses by focusing on the class of flows used in the recent experiments~\cite{Rousseaux:2007is,Weinfurtner:2010nu} so as to obtain a better understanding of what has been observed. To this end, we first consider monotonic flows in which $F$ either barely crosses $1$ or remains subcritical. We then study the scattering in nonmonotonic flows which either possess a pair of black and white horizons, or where the maximum value of $F < 1$ is reached at the top of an obstacle. The last case is the closest to that realized in~\cite{Weinfurtner:2010nu}, and our numerical results concerning the scattering coefficients closely reproduce what has been observed. However, our analysis also confirms the aforementioned results of~\cite{finazziRP-proceedings,Robertson:2012ku} that the Planckianity is lost for these subcritical flows, whereas a thermal law was observed in~\cite{Weinfurtner:2010nu}. 
This apparent contradiction triggered our interest and the forthcoming analysis. As we shall see, its resolution involves hydrodynamic modes which dominate below the critical frequency $\omega_{\rm min}$.

The effects of dispersion shall be computed in two different manners, along the lines of~\cite{Finazzi:2012iu,Robertson:2012ku,Euve:2015vml,Euve:2014aga}. 
First, we numerically obtain the spectral properties of the scattering in flows where the spatial gradient of the water height $h(x)$ is small: $\partial_x h \ll 1$. 
Second, by algebraic techniques, we compute the Bogoliubov coefficients in the steep regime where the water depth is piecewise constant. Even though this regime is \textit{a priori} far from the experimental setups, we shall see that it governs some of the spectral properties in the small-frequency limit of smooth profiles. 
The background flows will also be described at two different levels. In most of this chapter, for simplicity and clarity, we work with water height profiles $h(x)$ chosen from the outset. Amongst these, we shall briefly consider profiles that are modulated by an undulation~\cite{Coutant:2012mf}, i.e., a zero-frequency mode with a large amplitude, since these were systematically observed in~\cite{Rousseaux:2007is,Weinfurtner:2010nu}. 
We shall see that the main properties of the scattering coefficients are not significantly affected by this additional feature of the background flow provided the undulation is sufficiently short. 
In subsection~\ref{sub:NL}, we study profiles which result from integrating the nonlinear hydrodynamical equations. We shall see that the resulting spectra closely resemble those obtained by the first approach, thereby justifying it \textit{a posteriori}. 

A word of caution is perhaps necessary to conclude this Introduction. Our treatment is based on two main approximations: that of an ideal and irrotational fluid, and that based on a low-order expansion of the dispersion relation. To estimate the errors induced by these approximations is nontrivial, as it would require a precise description of the background flow including the effects of viscosity and vorticity, and using as well the full dispersion relation, perhaps including surface tension, see~\cite{Germain5, Germain7}. Yet, we believe our description captures the essential aspects of the scattering in the flows of~\cite{Rousseaux:2007is,Weinfurtner:2010nu}. We thus expect that its main predictions will be qualitatively correct, in particular the strong suppression of the low-frequency spectrum.  

This chapter is organized as follows. In Section~\ref{sec:GPAS}, we present the wave equation and background flows we use. We also compute the critical frequencies which separate the various regimes. In Section~\ref{sec:SP}, we solve the wave equation numerically for transcritical and subcritical flows, and determine which observables are, or are not, sensitive to the fact that $F$ crosses one.  We discuss our results in Section~\ref{sec:conclPR}. In subsection~\ref{sub:NL} we solve the nonlinear hydrodynamic equations to relate the shape of the free surface to that of the obstacle, before solving the wave equation in the resulting flow. Subsection~\ref{App:gradino} is devoted to the steep horizon limit. 
In subsection~\ref{sub:waveeqder}, we give a derivation of the wave equation. 
Finally, in Section~\ref{sec:scat_und} we discuss a few properties of the scattering on the undulation.

\section{General properties and settings} 
\label{sec:GPAS}

In this section we review the key concepts and methods used in the calculation of the scattering coefficients of shallow-water waves. Since these concepts are now well established, we shall be rather brief. The wave equation and its main properties are derived in ~\cite{Schutzhold:2002rf,Unruh:2012ve,Coutant:2012mf} and in Section~\ref{sub:waveeqder}. The general behavior of the scattering coefficients of dispersive waves in transcritical flows are explained in detail in Refs.~\cite{Macher:2009tw,Robertson:2012ku,Coutant:2011in}.

\subsection{Wave equation and dispersion relation} 
\label{sub:Weadr}

We consider irrotational laminar flows of an inviscid, ideal, incompressible fluid in an elongated flume. All dependences in the horizontal direction perpendicular to the flow are neglected. 
Perturbations of the velocity potential satisfy 
\begin{equation}\label{eq:waveeq}
\left[ \left(\partial _t+\partial _xv\right)\left(\partial _t+v\partial _x\right) -\ii g \partial _x \tanh \left(-\ii h \partial _x\right)\right] \phi = 0,
\end{equation}
where $v(x,t)$ is the horizontal component of the flow velocity, $h(x,t)$ is the background fluid depth, and $g$ is the gravitational acceleration. 
$\phi$ is related to the linear variation  of the water depth $\delta h$ through
\begin{equation}\label{eq:dh}
\delta h(t,x) = -\frac{1}{g}\left(\partial _t+ v \partial _x\right)\phi.
\end{equation}
For the sake of simplicity, in \eq{eq:waveeq} we neglected the contributions of the vertical velocity at the free surface. 
Taking it into account would add terms associated with the centrifugal acceleration of a fluid particle at the surface~\cite{Unruh:2012ve,Coutant:2012mf}. 
For the flow of \cite{Weinfurtner:2010nu}, we estimate that this additional term does not exceed $\sim 0.03 \, g$. 
For the flows considered in subsection~\ref{sub:NL}, it is smaller than $\sim 0.01 \, g$. We therefore expect that neglecting this term will not significantly affect the scattering coefficients. 

We also assume that the background flow is stationary, so that we can work with (complex) stationary waves $\e^{- \ii \omega t}\phi_\omega(x)$ with fixed frequency $\om$ in the laboratory frame. We then expand the dispersive term of~\eq{eq:waveeq} to fourth order in $h \partial_x$, assuming higher-order terms play no significant role in the determination of the scattering coefficients. When this is the case, these can be correctly obtained by solving the truncated equation
\begin{equation}\label{eq:om}
\left[ 
\left(-\ii \omega +\partial _xv\right)\left(-\ii \omega +v\partial_x\right) 
- g \partial_x 
h \partial_x 
-\frac{g}{3}\partial_x 
\left(h \partial_x \right)^3 
\right]
\phi_\omega 
= 0 .
\end{equation}
Notice that the ordering of $h(x)$ and $\partial_x$ has been preserved. This is important when considering the steep regime limit where $\partial_x h \gg 1$. (This interesting limit, where the scattering coefficients can be computed analytically, is discussed in subsection~\ref{App:gradino}.) 
This ordering also ensures that the expression of the conserved inner product is unchanged, see subsection~\ref{sub:sm}.

As we shall see, key properties of the scattering coefficients rely on the existence of turning points. Their location, and other properties of the geometrical optic approximation, are governed by the dispersion relation associated with \eq{eq:om}:
\begin{equation}\label{eq:disprel}
\Omega_\om^2 = c^2 k_\om^2 \lp 1- \frac{1}{3} h^2 k_\om^2 \rp,
\end{equation}
where $\Omega_\om \equiv \omega -v k_\om$ is the comoving angular frequency and $c^2 = g h(x)$ is the local group velocity of waves with low wave vector $k_\om(x)$ in the reference frame of the fluid. 
As in~\cite{Macher:2009tw,Robertson:2012ku,Coutant:2011in}, we will consider only positive values of $\om$, since the potential $\phi$ of \eq{eq:waveeq} is invariant under complex conjugation.

\subsection{Subcritical and transcritical flow profiles}

\begin{figure}
\centering 
\includegraphics[width = 0.49 \linewidth]{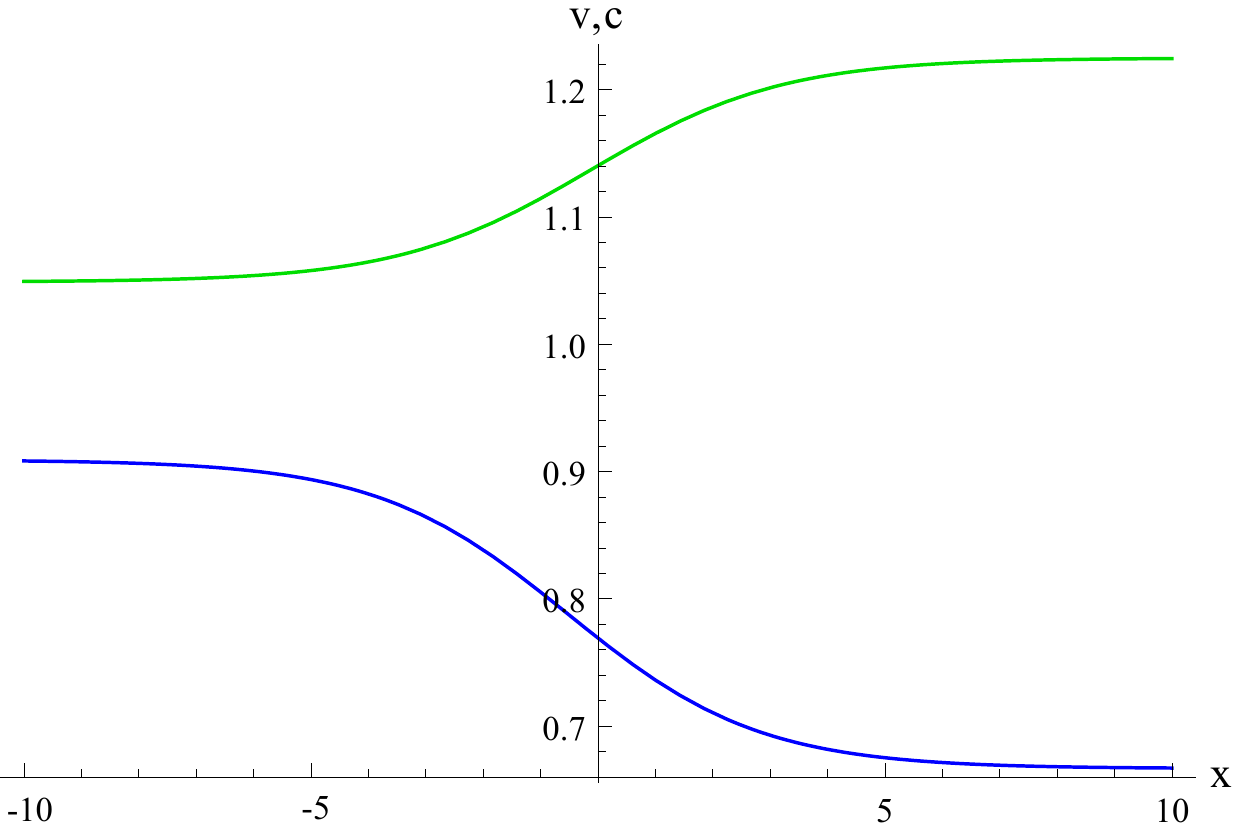} \,
\includegraphics[width = 0.49 \linewidth]{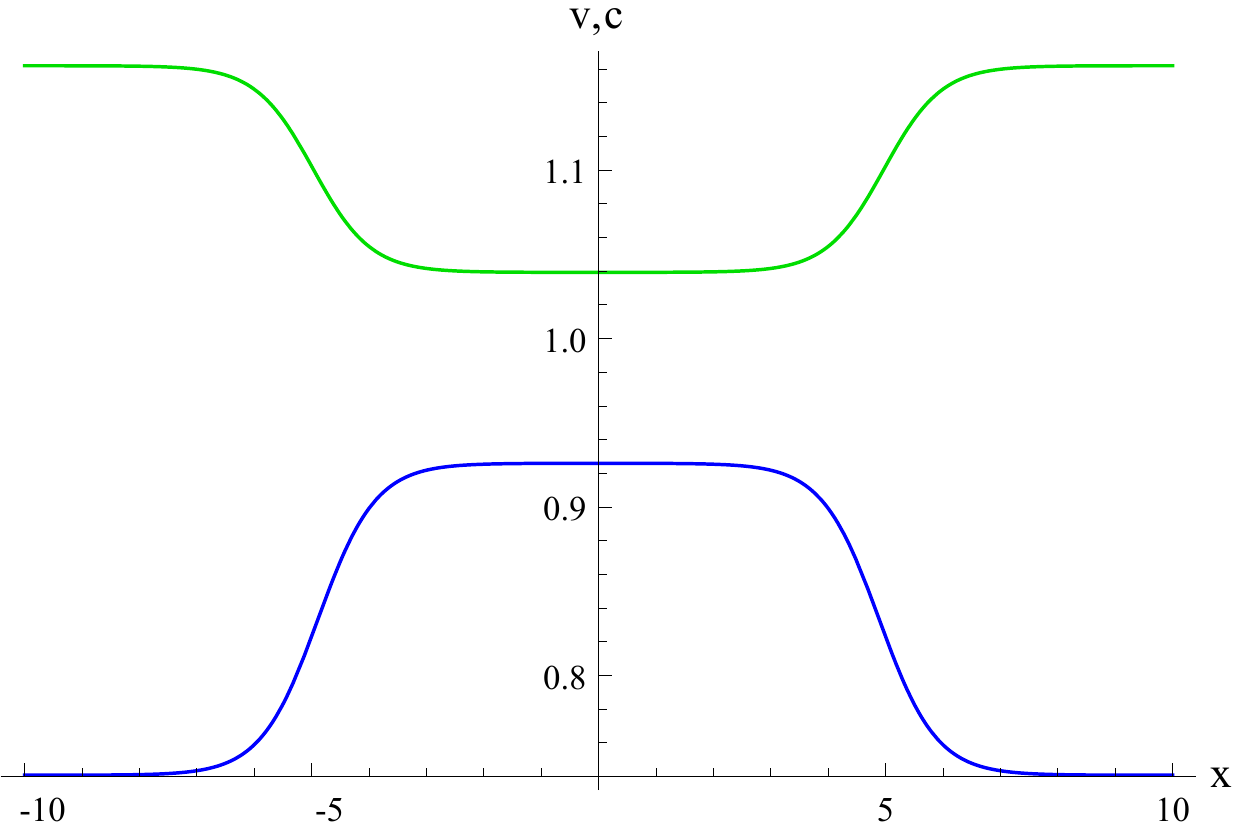}
\includegraphics[width = 0.49 \linewidth]{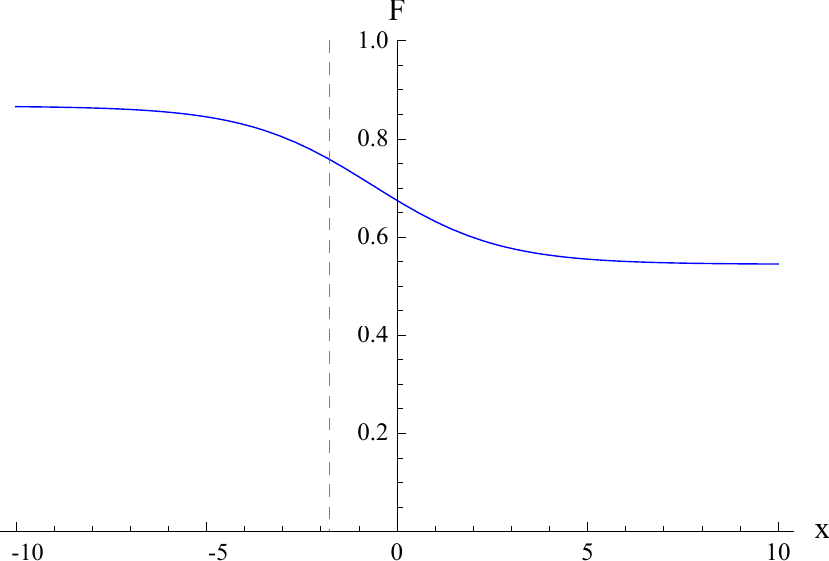} \,
\includegraphics[width = 0.49 \linewidth]{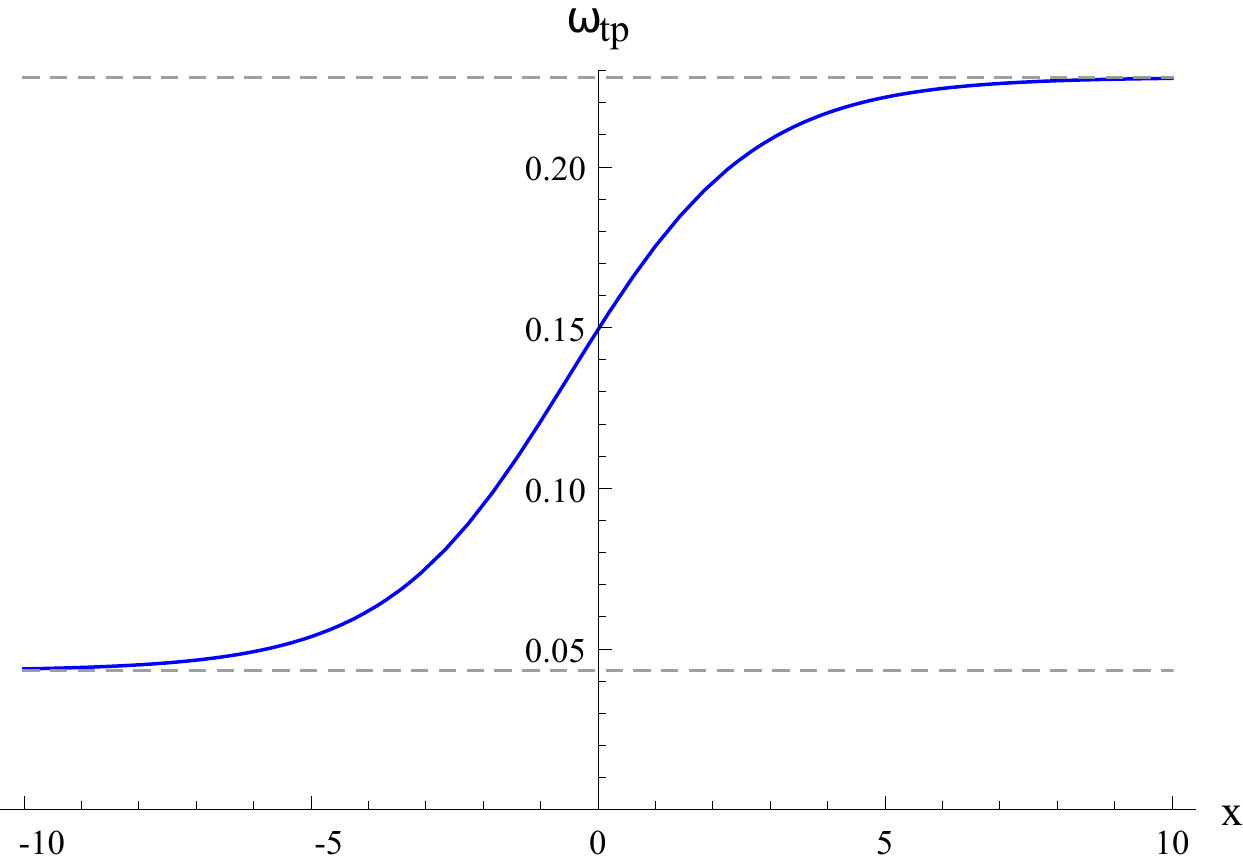}
\caption{
Top: Flow velocity $v$ (blue) and speed of long-wavelength waves $c$ (green) as functions of $x$ for a monotonic flow (left), and a nonmonotonic one (right), respectively given by \eq{eq:wdepth} and \eq{eq:wdepth_2}. 
The parameters are  $\sigma = \sigma_1 = \sigma_2 = 0.06$, $L=5$, $h_0 = 1.3$, and $D=0.2$, in units where $g=J=1$. 
Both flows are subcritical since $v(x) < c(x)$ for all $x \in \mathbb{R}$. The transcritical cases are similar, except that $v$ and $c$ cross each other, once or twice. 
Bottom: As functions of $x$, we show the Froude number $F = v/c$ (left) and the angular frequency $\om_\textrm{tp}(x)$ (right) for which the turning point is located at $x$, see \eq{eq:omtp}, for the flow of the top-left panel.  
The vertical dashed line in the left plot shows the location of turning point when $\om = 10^{-1}$, corresponding to the characteristics shown in \fig{fig:modes}. The maximal and minimal values of $\om_\textrm{tp}$, indicated by dashed horizontal lines on the right plot, give the two critical frequencies of \eq{eq:om-max-min}.}\label{fig:typical_flow_1}
\end{figure}

In this chapter, the sign of the flow velocity $v$ is taken positive, so that counterpropagating shallow water waves are coming from the right side. In addition, $v$ reaches its minimum value at $x \to +\infty$, see the upper panels of \Fig{fig:typical_flow_1}. Hence, when $F = v/c$ crosses 1, counterpropagating waves are all blocked, in analogy to what happens near a white hole horizon. The locus where $F = 1$ is sometimes referred to as a “phase velocity horizon”, as in \cite{Weinfurtner:2010nu}. 

For simplicity, in the body of the text we use background profiles for the water depth $h(x)$ with a simple analytical description. In subsection~\ref{sub:NL} we verify that our results remain valid for more complicate profiles which obey hydrodynamical equations over known obstacles. 
To unravel the various aspects of the scattering, we shall consider two classes of flows. The first one contains flows with monotonic function $x \mapsto v(x)$ which are asymptotically uniform on both sides. They shall be parametrized by water depths of the form
\begin{equation}\label{eq:wdepth}
h(x)= h_0 + D \, \tanh \lp \frac{\sigma x}{D} \rp, \, D >0, \, \sigma > 0,
\end{equation}
and $v = J / h$, where $J$ is the 2D current.
The maximum slope of $h$ is located at $x=0$, and given by ${\rm Max} \, \partial_x h= \sigma$, irrespectively of the parameter $D$ which fixes the asymptotic height change $\Delta h = 2 D$ between the asymptotic values for $x \rightarrow \pm \infty$: $h_{\rm  as.}^{\pm} \equiv h(\pm \infty) = h_0 \pm D$. 
Most of our results will be expressed in the system of units where $g = J = 1$. Then the Froude number is simply given by
\begin{equation}
F = h^{-3/2}. 
\end{equation}
In these units, a “phase velocity horizon” corresponds to a point where $v = c = h = F = 1$. Notice also that the surface gravity $\kappa_G = \left\lvert \partial_x( c -v)\right\rvert_{v=c}$~\cite{Unruh:1980cg, Unruh:1994je} is given by 
\begin{equation} \label{eq:SG} 
\kappa_G = \left\lvert \pd_x F \right\rvert_{F=1}.
\end{equation}
We assume $h_0 + D > 1$. 
The monotonic flows of \eq{eq:wdepth} then split into two subclasses. For $h_{\rm  as.}^- = h_0 - D < 1$, $F$ crosses $1$ and the flow is transcritical, whereas it remains globally subcritical when $h_{\rm  as.}^- > 1$. To study the transition between these two cases, we shall work with highly ``asymmetric'' profiles, where the minimum Froude number $F_{\rm min} = F (x \rightarrow + \infty)$ is always significantly smaller than $1$, whereas its maximum value $F_{\rm max} = F (x \rightarrow -\infty)$ is close to 1, see~\fig{fig:typical_flow_1} bottom-left panel.

The second class contains nonmonotonic flows where the maximum value of $F$ is reached at $x =0$, and where $F$ is asymptotically constant on both sides. These shall be parametrized by
\begin{equation}\label{eq:wdepth_2}
h(x) =h_0 + 
D  \tanh \lp \frac{\sigma_1}{D} (x + L) \rp 
\tanh \lp \frac{\sigma_2}{D} (x - L) \rp, 
\end{equation}
where $2 L$ characterizes the spatial extension of the domain where the height $h$ is close to its minimum value. When $F$ remains smaller than 1, these subcritical flows are qualitatively similar to those experimentally realized in Nice~\cite{Rousseaux:2007is} and Vancouver~\cite{Weinfurtner:2010nu}.

\subsection{Turning points and characteristics}
\label{Tpc}

For definiteness, in this subsection we assume the flow is monotonic. 
The discussion also applies to nonmonotonic flows described by \eq{eq:wdepth_2} with minor differences. 
For instance, quantities evaluated at $x = -\infty$ must then be evaluated where $h$ reaches its minimum value. 

Let us explain why the presence of a turning point directly affects the scattering of shallow-water waves. On the one hand, when there is a turning point in a monotonic flow as that of \eq{eq:wdepth}, one of the solutions of \eq{eq:om} becomes exponentially divergent in the ``forbidden region'' behind the turning point. 
On the other hand, scattering coefficients only relate the solutions of \eq{eq:om} which are asymptotically bounded, i.e., whose modulus remains finite at $x \rightarrow \pm \infty$~\cite{Macher:2009tw}. As a result, for the flows of \eq{eq:wdepth}, the number of linearly independent asymptotically bound modes is three when there is one turning point and four when it is absent.~\footnote{There exists a critical frequency $\om_{\rm max}$, defined below, above which the dispersion relation has only two real roots in both asymptotic regions. The number of independent modes is then equal to $2$. In this chapter we restrict our attention to positive frequencies smaller than $\om_{\rm max}$.}

\begin{figure}
\centering
\includegraphics[width = 0.5 \linewidth]{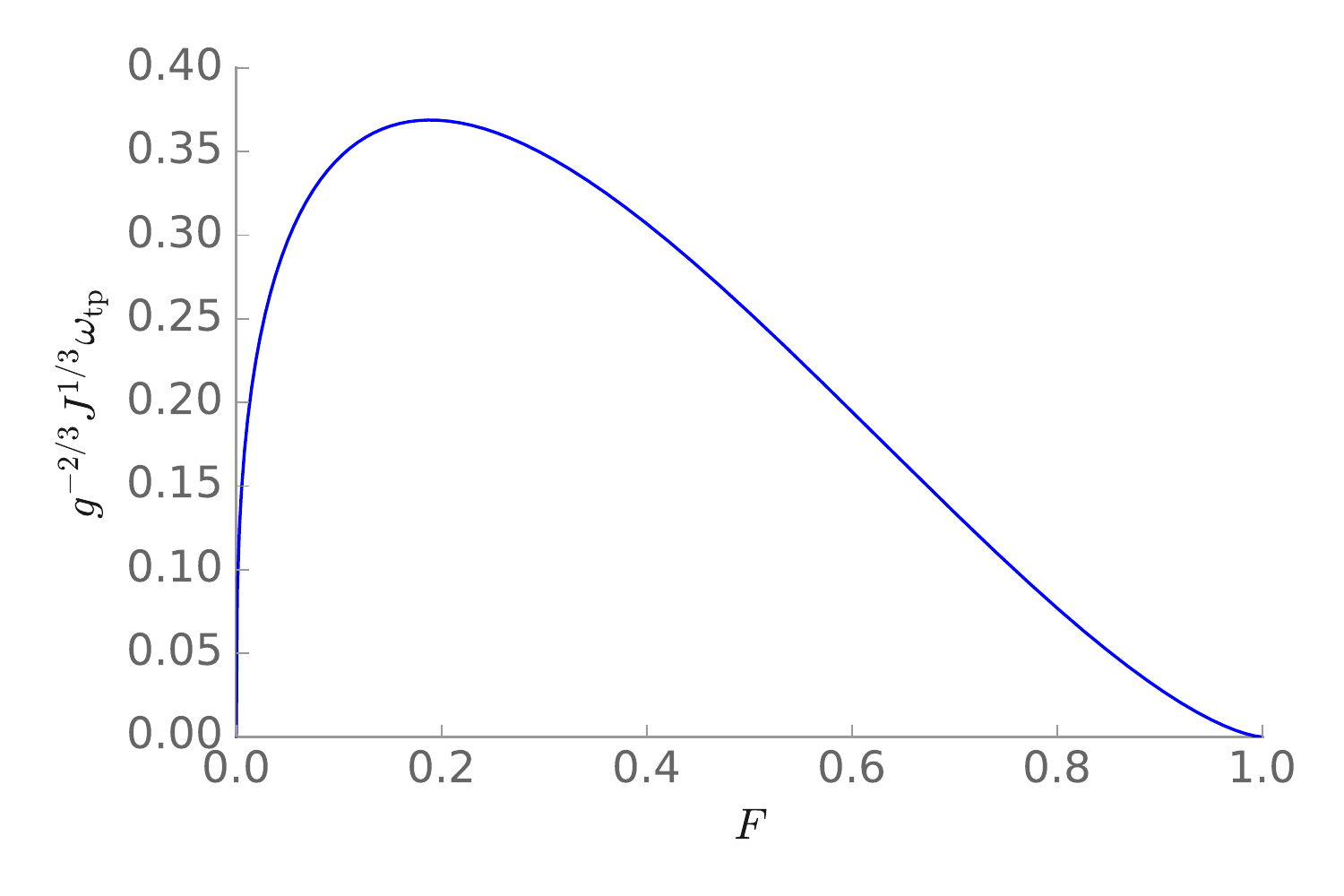}
\caption{Plot of $\om_{\rm tp}$, adimensionalized by $g^{2/3} J^{-1/3}$, as a function of $F$.}\label{fig:omtp}
\end{figure}

A turning point corresponds to a double root of the dispersion relation, where the group velocities of the corresponding modes vanish.  
Using the quartic law of \eq{eq:disprel}, double roots exist for the angular frequency
\begin{equation}\label{eq:omtp}
\om_{\rm tp} = \frac{c}{h} \frac{\sqrt{3}}{16} \lp \sqrt{F^2 + 8} - 3\abs{F} \rp^{3/2} \lp \abs{F} + \sqrt{F^2 + 8} \rp^{1/2}, 
\end{equation}
where $\om_{\rm tp}$, $c^2 = g h$ and $F = J/(g h^3)^{1/2}$ are functions of $x$ through the profile $h(x)$. 
In this expression, one can see that $c/h$ plays the role of the dispersive frequency $\Lambda$ of~\cite{Macher:2009tw}. 
It goes to zero in the two limits $F \to 0$ and $F \to 1$ at fixed $J^2 / g$, see \fig{fig:omtp}.~\footnote{Notice that while the behavior for  $F \approx 1$ is unchanged when adding the other dispersive terms, the limit $F \to 0$ is strongly affected by the quartic approximation.} 
In this figure, we also see that $\om_{\rm tp}$ is a monotonically decreasing function of $F$ when the latter is larger than $0.2$. In this chapter we will always assume this is the case. 
Notice also that $\om_{\rm tp}$ no longer exists as a real root when $F > 1$. In this chapter, only real positive frequencies will be considered. 

Given an anglar frequency $\omega$, \eq{eq:omtp} implicitly gives the location of the turning point $x_{\rm tp}$ through
\begin{equation}\label{eq:xtp}
\om_{\rm tp}(x_{\rm tp}) = \om.
\end{equation}
For monotonic flows, the minimum and maximum values of $F$ are $F_{\rm min/max} = F (x \rightarrow \pm \infty)$. Assuming the flow is subcritical, the maximum and minimum values of $\om_{\rm tp}$ respectively are
\begin{equation}\label{eq:om-max-min}
\om_{\rm max} = \om_{\rm tp}(x = + \infty), 
\quad
\om_{\rm min} = \om_{\rm tp}(x = - \infty),
\end{equation}
as clearly seen in the lower right panel of \Fig{fig:typical_flow_1}. (One can treat the subcritical and transcritical cases together by setting $\om_{\rm min} = 0$ for transcritical flows.) When $\om_{\rm min} < \om < \om_{\rm max}$, the trajectory associated with the left-moving root $k_\om^\leftarrow$ is blocked at the locus given by \eq{eq:xtp}, see \Fig{fig:modes}, lower left panel. (See~\cite{Brout:1995wp,Balbinot:2006ua,Coutant:2011in} for more details about the calculation of these trajectories.) For frequencies higher than $\om_{\rm max}$, there are only two real roots of the dispersion relation in each asymptotic region, see \Fig{fig:modes}, and thus only two modes.~\footnote{The number of modes can be determined as follows. To be slightly more generic, we consider a dispersion relation which is an arbitrary polynomial with real coefficients, of degree $d$.  Let us call $n_L$ (respectively $n_R$) the number of real roots in the left (resp. right) asymptotic region. On the left, we have $n_L$ real roots. We also have $d - n_L$ complex roots, half of them having a strictly positive imaginary part while the others have strictly negative ones. The space of modes which are asymptotically bounded on the left is thus of dimension $n_L + (d - n_L)/2$, i.e., $(d + n_L)/2$. Imposing that a mode contains no exponentially growing mode in the right asymptotic region gives $(d - n_R)/2$ constraints, in general independent from those found in the left region. Taking them into account, the space of solutions which are globally bounded is of dimension $(n_L + n_R)/2$. (Notice that, since the coefficients of the dispersion relation are all real, $n_L$ and $n_R$ both have the same parity as $d$, so that $(n_L + n_R)/2$ is always an integer.)} 
This high frequency regime will no longer be considered as it presents no direct relationship with the Hawking effect. 
The low-frequency regime $0 < \om < \om_{\rm min}$ is much more interesting. This regime only exists when the flow is subcritical, as $\om_{\rm tp}$ is real only for $F < 1$. In this regime, the four real roots $k_\om$ define four trajectories which are followed by the corresponding waves packets (in the WKB approximation), see \Fig{fig:modes}, bottom right panel.  
\begin{figure}
\centering
\includegraphics[width = 0.5 \linewidth]{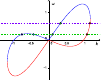} \\
\includegraphics[width = 0.49 \linewidth]{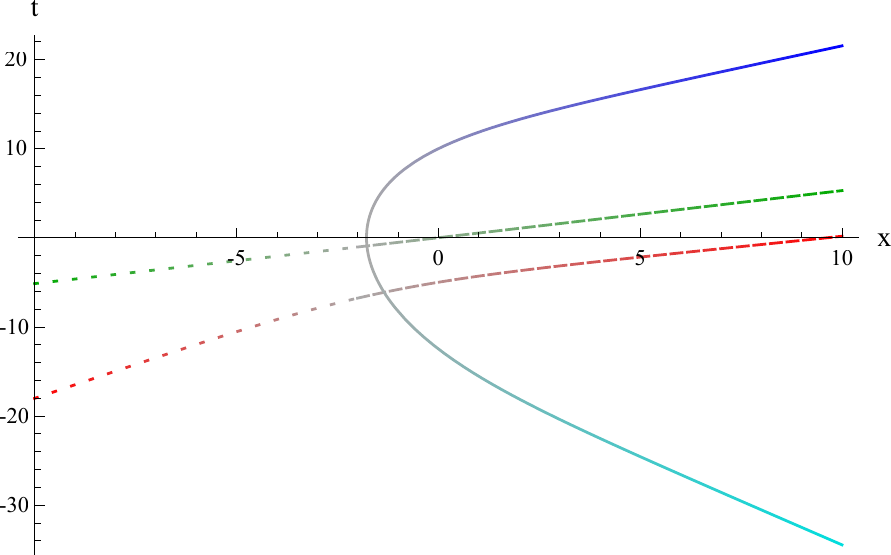}
\includegraphics[width = 0.49 \linewidth]{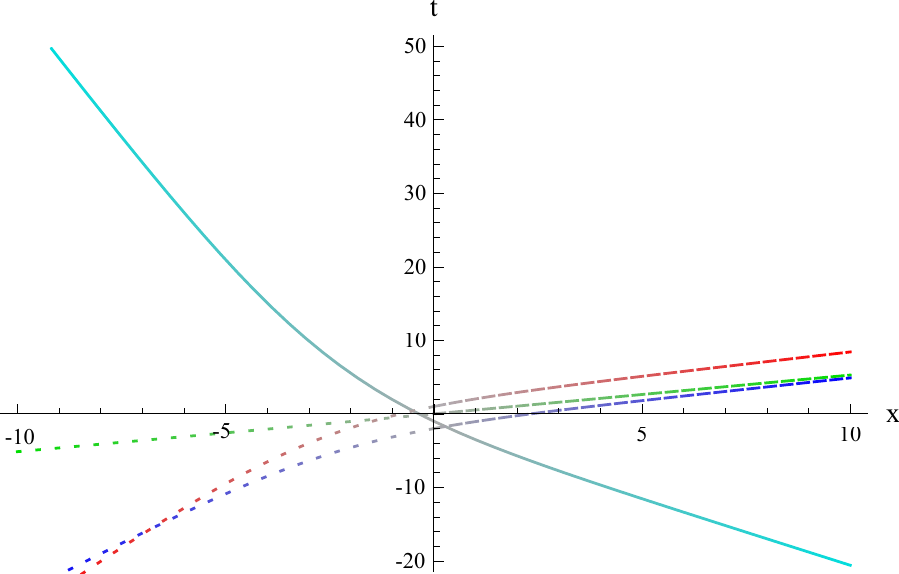}
\caption{
Top: Graphical resolution of \eq{eq:disprel} in the left asymptotic region, for a globally subcritical flow with $F_{\rm max} \approx 0.54$. The continuous lines show the frequency $\om$ as a function of the wave vector $k$ for positive (blue) and negative (red) value of the comoving frequency $\Omega$. The parameters are $g=J=1$ and $h_{\rm as}^- = 1.5$. Dashed lines show two values of $\om$ below (green) and above (purple) $\om_{\rm min}$. Dots show the real roots $k_\om$ of the dispersion relation at fixed $\om$. 
Bottom: Characteristics for $\om_{\rm min} < \om = 10^{-1} < \om_{\rm max}$ (left) and $\om = 10^{-2}< \om_{\rm min}$ (right). The water depth is given by \eq{eq:wdepth} with $h_0=1.3$, $D=0.2$, and $\sigma=0.06$. The solid line shows the trajectory of the low momentum incoming root propagating initially against the flow from the right side. On the left panel, there is a turning point, whereas on the right one there is none. The dashed lines indicate that these asymptotic waves are produced by the incoming mode of \eq{eq:Btrans} (left) and \eq{eq:Bsub} (right), whereas the dotted lines indicate that the waves are absent. 
In the asymptotic region, the color of each line corresponds to that of the corresponding root in the upper panel. 
} 
\label{fig:modes}
\end{figure}

Figure~\ref{fig:modes} upper panel shows a graphical resolution of \eq{eq:disprel} in the left asymptotic region of a subcritical flow for two frequencies $\om$ above and below $\om_{\rm min}$. The latter is given by the lowest horizontal tangent to the curve of $\om(k)$. (For a transcritical flow, the plain lines would be more tilted so that this horizontal tangent disappears.) The blue line corresponds to positive values of the comoving frequency $\Omega$ and the red line to negative values. For small frequencies, the largest root $k_\om$ is the only one which lives on the negative branch. As we shall see below, the norm of the corresponding mode $\phi_\om$ has the opposite sign as that of the three other waves. Above $\om_{\rm min}$, only two real roots exist. A similar plot can be drawn for the right asymptotic region. In that case, the horizontal tangent defines $\om_{\rm max}$. 

In the lower panels of \Fig{fig:modes}, we have represented the characteristics in the monotonic subcritical flow of \Fig{fig:typical_flow_1} for $\om_{\rm min} < \om < \om_{\rm max}$ (left) and $0< \om < \om_{\rm min}$ (right). On the left, there are three characteristics and hence three linearly independent modes. Two of them are co-propagating in the laboratory frame and the third one has a turning point. 
On the right there are four characteristics, and hence four linearly independent modes, because there is no turning point. Three of them are co-propagating and the fourth one is counterpropagating. The use of dotted and dashed lines schematically represents the scattering process of a counterpropagating wave packet (represented by a continuous line) with a nearly well-defined frequency $\om$ sent from $x = \infty$: the right-moving modes are initially absent (dotted) but are populated by the scattering (dashed). When the flow is monotonic and $\om < \om_{\rm min}$ the reflection of the wave packet (left panel) is total, whereas it is only partial when it is nonmonotonic as some wave can tunnel through the effective potential. Notice that both the left and the right panels of~\fig{fig:modes} are relevant for interpreting the observations of~\cite{Rousseaux:2007is,Weinfurtner:2010nu}.

\subsection{Scattered modes} 
\label{sub:sm}

The above geometrical properties are reflected in those of the various modes $\phi_\om$, solutions of \eq{eq:om}. 
In this chapter, we focus on the scattering coefficients relating asymptotic modes, which can be used to test the Hawking prediction. 
(We refer to~\cite{2014arXiv1402.2514C} for a recent analysis of the local properties of the modes $\phi_\om$.) 

The analytical properties of the scattering coefficients stem from the conserved inner product~\cite{Unruh:1994je} associated with \eq{eq:waveeq}. It is given by
\begin{equation}\label{eq:Kprod}
\lp \phi_1, \phi_2 \rp \equiv \ii \int \lp \phi_1^* \, (\pd_t + v \pd_x) \phi_2 - \phi_2 \, (\pd_t + v \pd_x) \phi_1^* \rp \, \dd x ,
\end{equation}
where $\phi_1$ and $\phi_2$ are two solutions. 
Then, $\pd_t \lp \phi_1, \phi_2 \rp = 0$. 
Note that the ``norm''~\footnote{Although this quantity does not satisfy the criteria of the usual definition of a norm, we shall use in the following this abuse of language as it is common in publications on analogue gravity.} 
$\lp \phi_1 , \phi_1 \rp$ is not positive definite. This important property allows for mode amplification (sometimes also called over-reflection~\cite{OnOverReflection} or super-radiance~\cite{GeneralSRR}) as positive-norm modes can be amplified alongside with the appearance of negative-norm ones while preserving the total norm. 
This over-reflection is at the root of the Hawking effect.

Since the flows we consider are asymptotically constant on both sides, the asymptotic solutions of \eq{eq:om} are plane waves (when the root $k_\om$ is real). When the asymptotic flow is subcritical, for low frequencies, the 4 wave vectors $k_\om$ are real, and the corresponding waves are, for decreasing $k_\om$, 
\begin{itemize}
\item $\phi_\om^{\rightarrow,d}$ is dispersive\footnote{The corresponding weave vector exists only because of the subluminal character of the dispersion relation of \eq{eq:disprel}, as can be seen in \fig{fig:modes}, top panel. We shall call a mode ``dispersive'' when this is the case, and ``hydrodynamic'' when its corresponding root still exists in the limit where the dispersion length scale is sent to 0. 
Similarly, we shall call ``hydrodynamic sector'' the set of scattering coefficients between two hydrodynamic modes, and ``dispersive sector'' the set of coefficients involving at least a dispersive mode.} 
and right moving in the laboratory frame; 
\item $\phi_\om^\leftarrow$ is hydrodynamic and left moving;
\item $\phi_\om^\rightarrow$ is hydrodynamic and right moving; 
\item $\lp \phi_{-\om}^{\rightarrow,d} \rp^*$ is dispersive and right moving\footnote{When using the quartic dispersion relation of \eq{eq:disprel} in place of the full one, $\Omega^2 = gk \tanh (hk)$, this root becomes left moving for $\om > \sqrt{3} v / h$. 
To avoid considering this spurious effect, we restrict our analysis to frequencies smaller than $\sqrt{3} v / h$.}. 
\end{itemize}
Unlike the first three waves, the last one has a negative norm. This can be seen in \fig{fig:modes} where the corresponding root lies on the negative $\Omega$ branch. It can be easily verified that the sign of $\Omega$ and that of the norm are always identical.
{As a result, when working with a positive frequency $\om = \ii \partial_t$, positive-norm modes describe waves carrying positive energy in the laboratory frame, whereas negative-norm modes describe {\it negative-energy waves}, see Ref.~\cite{Coutant:2011in}. The latter exist only in the presence of a counterflow, and their presence signals that the system under study is energetically unstable.} Because the inner product changes sign under complex conjugation, the mode $\lp \phi_{-\om}^{\rightarrow,d} \rp^*$ is conventionally written as the complex conjugate of a positive-norm one with negative frequency, in virtue of the invariance of \eq{eq:om} under complex conjugation and $\om \to - \om$. 

In addition, each of the above four modes possesses a well-defined group velocity given by $v_{\rm gr} = (\partial_\om k_\om)^{-1}$. As a result, on each side, each of them can be identified as an incoming $in$ mode (or as a reflected $out$ mode) when $v_{\rm gr}$ is pointing towards (away from) the central region $x \approx 0$. When $F < 1$ on both asymptotic sides, for $0 < \om < \om_{\rm min}$, this identification applies both at $x \rightarrow -\infty$, and at $x \rightarrow +\infty$. However, because the flow is inhomogeneous, the modes mix with each other. The four globally-defined incoming (in) modes -- defined by the requirement that the asymptotic weights of the 3 other incoming modes vanish -- determine 4 different superpositions of the four reflected asymptotic out modes. For instance, the incoming left-moving hydrodynamical mode is defined by the requirement that the 3 right-moving waves at $x \rightarrow -\infty$ vanish, see \Fig{fig:modes}. This mode describes the shallow water waves that have been studied in~\cite{Rousseaux:2007is,Weinfurtner:2010nu}. For this reason, we shall only consider this mode in what follows. For a more complete description of the scattering matrix, we refer to~\cite{Macher:2009tw,Robertson:2016ocv}.

Using the inner product \eqref{eq:Kprod} to normalize all modes, the scattering of this mode is fully described by the existence of a solution of the form
\begin{align}
\label{eq:Bsub}
\phi_\omega^{\leftarrow, \textrm{in}} (x) = \left\lbrace
\begin{array}{ll}
\phi_\om^{\leftarrow}(x) + \alpha_\om \phi_\om^{\rightarrow,d}(x) + \beta_\om \lp \phi_{-\om}^{\rightarrow}(x) \rp^* + A_\om \phi_\om^{\rightarrow}(x) & x \to +\infty \\
\tilde{A}_\om \phi_\om^{\leftarrow} (x) & x \to - \infty
\end{array}
\right. ,
\end{align}
where the coefficients obey 
\begin{equation}\label{eq:4coef}
\left\lvert \alpha_\omega \right\rvert^2 - \left\lvert \beta_\omega \right\rvert^2 + \left\lvert A_\omega \right\rvert^2 + \left\lvert \tilde{A}_\omega \right\rvert^2 = 1. 
\end{equation}
For higher frequencies, i.e., for $\om_{\rm min} < \om < \om_{\rm max}$, or if $F>1$ in the left asymptotic region, the transmitted mode $\phi_\omega^{\leftarrow}$ no longer exists. Instead we have an exponentially decaying mode $\phi_\om^{\rm dec}$. (There is also an exponentially increasing one, which as such can not be used to build asymptotically bounded solutions.) 
As a result, there are only three independent asymptotically bounded modes~\cite{Macher:2009tw}.
In this case, \eq{eq:Bsub} becomes
\begin{align}
\label{eq:Btrans}
\phi_\omega^{\leftarrow, \textrm{in}} (x) = \left\lbrace
\begin{array}{ll}
\phi_\om^{\leftarrow}(x) + \alpha_\om \phi_\om^{\rightarrow,d}(x) + \beta_\om \lp \phi_{-\om}^{\rightarrow}(x) \rp^* + A_\om \phi_\om^{\rightarrow}(x) & x \to +\infty \\
\breve{A}_\om \phi_\om^{\rm dec} (x) & x \to - \infty
\end{array}
\right. ,
\end{align}
and conservation of the norm implies
\begin{equation}\label{eq:3coef}
\left\lvert \alpha_\omega \right\rvert^2 - \left\lvert \beta_\omega \right\rvert^2 + \left\lvert A_\omega \right\rvert^2 = 1.
\end{equation}

In what follows, we shall compute these coefficients numerically with particular attention given to $\left\lvert \beta_\omega \right\rvert^2$ as this quantity allows to test the Hawking prediction. Indeed, in quantum settings, the mean occupation number of particles spontaneously emitted (i.e., emitted when the initial state is vacuum), is given by $n_{\om}^{\textrm{out}} = \left\lvert \beta_\omega \right\rvert^2$. 
In the relativistic settings used by Hawking, ignoring the gray body factor~\cite{Page76,Sandro_2014}, 
one finds a Planckian spectrum: $|\beta_\om|^2 =  (e^{\om/T_H} - 1)^{-1}$, governed by the Hawking temperature (or rather frequency~\footnote{The temperature $T$ associated to a frequency 
$\om$ being $T = \hbar\om/k_B $, in units $\hbar = k_B = 1$, one has $T = \om$.}) $T_H = \kappa/2\pi$, where $\kappa$ is the surface gravity of \eq{eq:SG}.

\section{Numerical analysis and spectral properties}
\label{sec:SP}

Following~\cite{Macher:2009tw,Finazzi:2012iu}, we wrote a Mathematica~\cite{Mathematica7} code which solves the wave equation \eq{eq:om} and identifies the full set of Bogoliubov coefficients, namely 16 when $0 < \om< \om_{\rm min}$, and 9 when $\om_{\rm min} < \om < \om_{\rm max}$.\footnote{We are grateful to J. Macher and S. Finazzi for providing C++ and Mathematica codes which were an appreciated source of inspiration. We also thank X. Busch for explaining us how the code written by S. Finazzi works.}   
As in these earlier works, the code computes (from right to left) a set of 4 linearly independent solutions of \eq{eq:om}, which are plane waves at the right boundary of the integration domain. For each of these solutions, it then uses the asymptotic values of $\phi_\om$ and its first three derivatives to extract the decomposition of $\phi_\om$ into plane waves at the left of the integration domain. Finally, a direct identification of the incoming and outgoing modes gives the Bogoliubov coefficients of \eq{eq:Bsub} and \eq{eq:Btrans}. When considering nonmonotonic flows, since the asymptotic values of $F$ are equal to each other and smaller than 1, the four modes are plane waves on both asymptotic sides and their identification is straightforward for all frequencies $0< \om < \om_{\rm max}$. For monotonic flows, when there is a turning point, i.e., for $\om_{\rm min} < \om< \om_{\rm max}$, one must work with superpositions of solutions which do not contain the growing mode on the left side to compute the three coefficients of \eq{eq:Btrans}. 

In all cases we have estimated the numerical errors by computing the “unitarity” relations of \eq{eq:4coef} and \eq{eq:3coef}. 
For the numerical calculations reported here, the corresponding errors are smaller than $10^{-5}$. When the coefficient $\beta$, defining the temperature, is smaller than $10^{-3}$, we imposed a better accuracy, so that the estimated relative error on $\beta$ is always smaller than $10^{-2}$. In practice, for all but a few numerical points \eq{eq:4coef} and \eq{eq:3coef} were satisfied to a much better precision, with deviations smaller than $10^{-5} \left\lvert \beta^2 \right\rvert$. 
As a result, the main sources of imprecision of our results seems to come from the approximations discussed in subsection~\ref{sub:Weadr} rather than the numerical errors. 

\subsection{Transcritical flows}

\subsubsection{Monotonic flows}

\begin{figure} 
\centering \color{white!30!black}\renewcommand\color[2][]{}
\def\svgwidth{0.49 \linewidth}
{\scriptsize 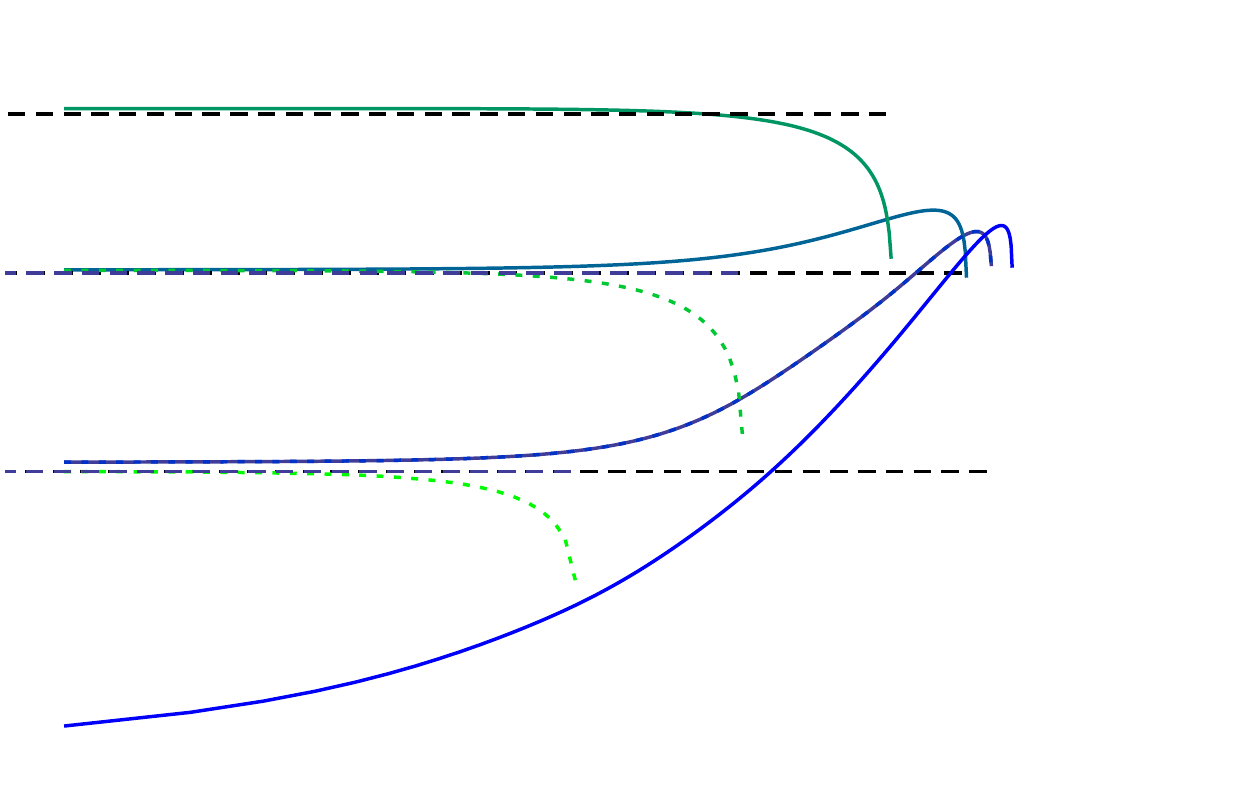}
\def\svgwidth{0.49 \linewidth}
{\scriptsize 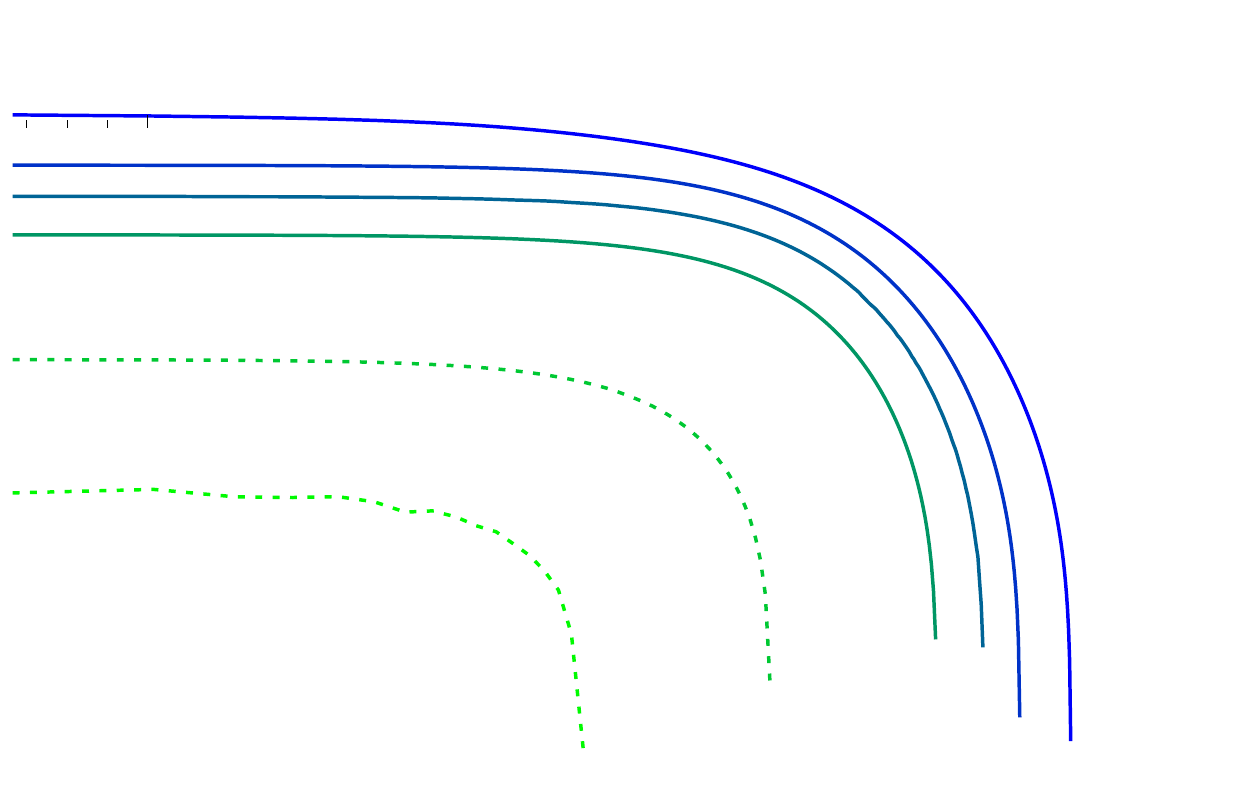}
\caption{Left: Effective temperature $T_\om$ of \eq{eq:effT} as a function of $\ln \om$ for a transcritical flow of the form \eq{eq:wdepth} with fixed values of $\sigma = 0.06$ and $D=0.2$, and for 6 different values of $h_0$. The values of $F_{\rm max}$ are increasing from top to blue, and fixed by $h_0=1.2, 1.15, 1.1, 1, 0.9$, and $0.85$. The three horizontal dashed lines give the Hawking temperatures of \eq{eq:TH} for the corresponding flow, and $T_{H,0}$ gives the reference value defined for $h_0=1$. In units where $g=J=1$, $T_{H,0} \approx 0.014$. For the last flow, $T_H$ vanishes, as $F_{\rm max}= 1$. The two dotted curves correspond to flows where the maximum Froude number is larger than in the symmetric case $h_0=1$. These two curves have been included to show that the spectra still follow the thermal prediction when increasing $F_{\rm max}$. Right: Logarithm of the transmission coefficient $|A_\om|^2$ of \eq{eq:Btrans} for the same flows.} \label{fig:1}
\end{figure}

When the flow is monotonic and transcritical, \eq{eq:Btrans} applies to all frequencies $\om \in \left] 0, \om_{\rm max} \right[$, since counterpropagating shallow-water waves are blocked irrespectively of their frequency $\om$. In this case, one expects to recover the standard results for the emitted flux to a good accuracy. To ease the comparison with S.~Hawking's Planckian prediction, we represent on the left panel of \fig{fig:1} the effective temperature $T_\om$ defined by
\begin{equation}\label{eq:effT}
\left\lvert \beta_\om \right\rvert^2 = \frac{1}{\e^{\om/T_\om}-1}.
\end{equation}
In accordance with the results of \cite{Macher:2009tw,Finazzi:2012iu,Robertson:2012ku}, when the maximum value of $F$ is significantly larger than 1, we first observe that $T_\om$ is constant in a large range of adimensional frequencies $\om/T_\om$, i.e., the spectrum is Planckian to a good accuracy, see \fig{fig:1}. 
Secondly, we observe that the height of the flat plateau closely follows S.~Hawking's prediction~\cite{Unruh:1980cg}
\begin{equation}\label{eq:TH}
T_\om \approx T_H, \, T_H \equiv \frac{1}{2 \pi} \left\lvert \partial_x \lp v-c \rp_{v=c} \right\rvert.
\end{equation}
The values of $T_H$ are represented by dashed lines in the figure. The agreement confirms that, for low frequencies with respect to $\om_{\rm max}$, the effective temperature $T_\om$ only depends on the local properties of the flow where $F$ reaches unity. This is the Hawking regime of~\cite{Finazzi:2012iu}. A closer study reveals that the relative deviations between $\mathop{\rm lim}_{\om \rightarrow 0} T_\om$ and $T_H$ scale as the square of the maximum slope of $h(x)$. This observation is completed by \fig{fig:Scott} of subsection~\ref{App:gradino}, where the validity range of the Hawking regime is established when increasing the slope $\sigma$. 

For all curves in \fig{fig:1}, the range of $\om$ is $\left] 0, \om_{\rm max} \right[$. As can be seen from \eq{eq:omtp} and \eq{eq:om-max-min}, this range shrinks when $h_0$ decreases, i.e., when the minimum Froude number increases. 
Let us focus on the limit where the maximal value of $F$ diminishes and approaches 1 from above. 
We clearly observe that the range of adimensional frequencies $\om/T_\om$ of the flat plateau shrinks as $F_{\rm max} \to 1$. This means that the Planckianity of the spectrum is progressively lost in this limit. We also observe that, for high frequencies (close to $\om_{\rm max}$), the effective temperature remains approximatively independent of the value of $F_{\rm max}$. (It is worth mentioning the similarity between the present curves and those obtained with a superluminal dispersion relation~\cite{finazziRP-proceedings}. 
The origin of this correspondence is explained in~\cite{Coutant:2011in}.) In conclusion, the Hawking spectrum is found only if $F_{\rm max}- 1$ is not too small. It would be interesting to determine with more precision the role of $F_{\rm max} - 1 > 0$ in limiting the validity domain of the Hawking prediction. 
Since the primary aim of this chapter is to study the case $F_{\rm max} < 1$, we shall not detail here the precise characterization of this domain, which was studied in \cite{Macher:2009tw,Finazzi:2012iu,finazziRP-proceedings, Robertson:2016ocv}.

On the right panel of \fig{fig:1}, we plot the squared norm of the coefficient $A_\om$ of \eq{eq:Btrans} which governs the elastic scattering between the incoming mode and the spectator mode $\phi_\om^{\rightarrow, \textrm{out}}$. We observe that $|A_\om|^2 \lesssim \e^{-5}$. 
This means that the (transmission) gray body factor $\Gamma_\om$~\cite{Page76} is close to 1 since $\Gamma_\om^2 \sim 1 - |A_\om|^2 \sim 1$. This is unlike what is found in the case of Schwarzschild black holes, where $\Gamma_\om^2 \propto \om^2$. 
In brief, for transcritical flows, it is legitimate to neglect $|A_\om|^2 $ as it hardly affects the unitarity relation \eq{eq:3coef}. As a result, in this regime, the Planckianity of the spectrum can be studied either from $|\beta_\om|^2$, as shown in \eq{eq:effT}, or from the ratio
\begin{equation}\label{eq:R}
R_\om \equiv \left| \frac{\beta_\om}{\alpha_\om} \right|^2 , 
\end{equation}
which satisfies $R_\om \approx \e^{- \om/T_\om}$, since $|\alpha_\om|^2\approx 1 +|\beta_\om|^2$.

Instead, when the condition $|A_\om|^2\ll 1$ is no longer satisfied, the relation between $R_\om$ and $T_\omega$ is no longer clear because $|\alpha_\om|^2 - |\beta_\om|^2 = 1 - |A_\om|^2 \neq 1$. On the contrary, irrespectively of the value of $|A_\om|^2$, $|\beta_\om|^2$ of \eq{eq:Bsub} (or \eq{eq:Btrans}) always gives the mean number $n^{BH}_\om$ of asymptotic particles spontaneously emitted by the corresponding black hole flow (as explained in~\cite{Macher:2009tw}, the emission spectrum of black holes and white holes slightly differ when the S-matrix mixes more than two 
modes.~\footnote{For this reason, it would be interesting to explicitly study the scattering of shallow water waves in black hole flows, as pointed out by S. Robertson.})  

\begin{figure}
\centering \color{white!30!black}\renewcommand\color[2][]{}
\def\svgwidth{0.49 \linewidth}
{\scriptsize 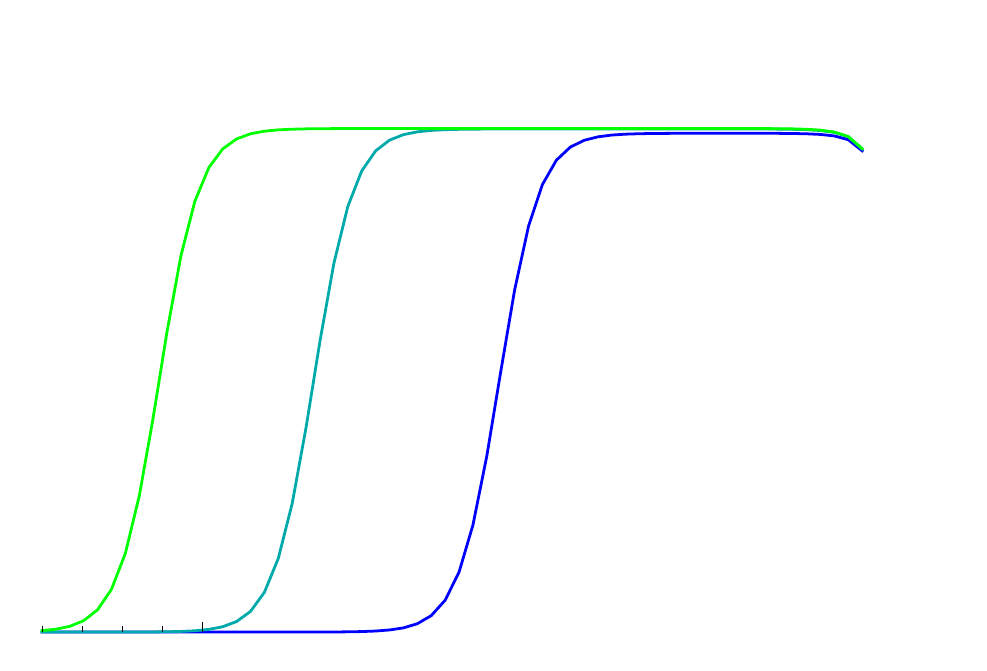}
\def\svgwidth{0.49 \linewidth}
{\scriptsize 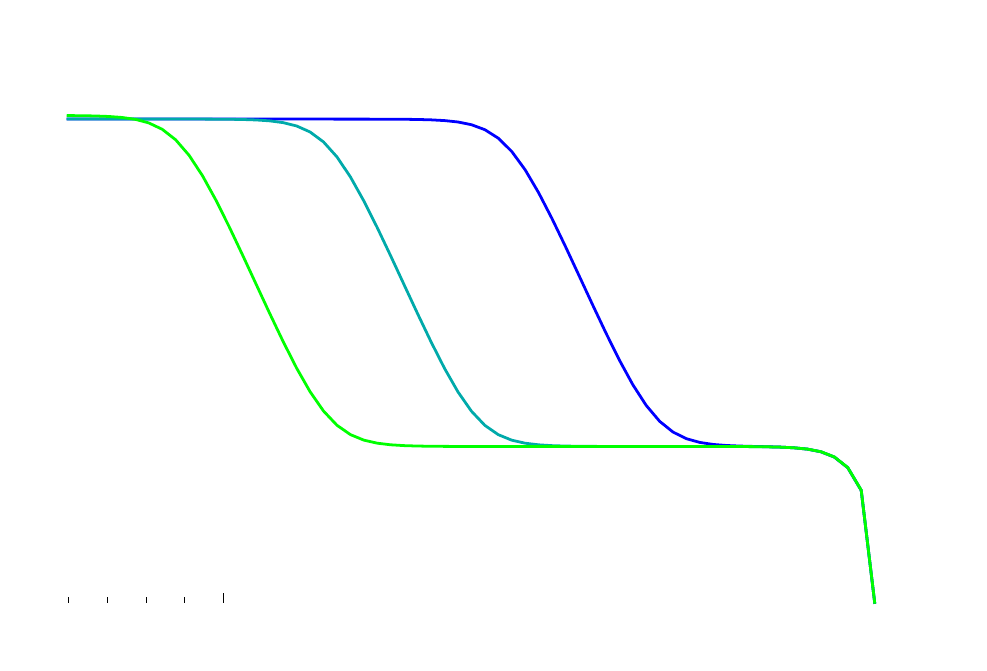}
\caption{Effective temperature (left) and logarithm of the squared hydrodynamic coefficients (right) as functions of the angular frequency $\om$ for a localized obstacle giving a transcritical flow with $F_{\rm max} \sim 1.2$. (This value has been chosen so that $\om_c$ of \eq{eq:omc} is not too small.) The background water depth is given by \eq{eq:wdepth_2}, with $\sigma_1 = \sigma_2 = 0.03$, $h_0=1.0$, $D=0.1$, and for extensions $L=5$ (blue), $7$ (cyan), and $10$ (green) in units where  $g=J=1$. The corresponding values of $\om_c$ are $7.0 \, 10^{-4} T_H,  1.2 \, 10^{-5} T_H$, and $2.6 \, 10^{-8} T_H$ respectively, where $T_H \approx 1.4 \, 10^{-2}$. On both panels, one clearly sees that the finite extension $L$ of the transcritical flow only affects the Planckian behavior of the spectrum for ultra low-frequencies $\om \lesssim \om_c$. In this regime, the effective temperature vanishes (left panel),  and the hydrodynamical sector completely dominates the scattering (right panel).} \label{fig:3}

\end{figure}

\subsubsection{nonmonotonic transcritical flows, adding a black hole horizon}
\label{App:aBHL}

We now consider profiles of the second class, see~\eq{eq:wdepth_2}, when $F_{\rm max}$ is significantly larger than 1 and $F(x \to \pm \infty) = F_{\rm min} < 1$. In these profiles, there is a second analogue horizon, which is that of a black hole since $v/c$ increases along the direction of $v$. We shall be rather brief and only focus on the new aspects with respect to the former case due to the presence of this second horizon.

The first important difference stems from the fact that the flows are now asymptotically subcritical on both sides. As a result, the four-mode mixing of \eq{eq:Bsub} here applies for all frequencies $\om \in \left] 0, \om_{\rm max} \right[$. However, because the flows are transcritical, in the WKB approximation, incoming modes from the right side will still be reflected for all these frequencies. As a result, we do not expect that the presence of the extra transmitted mode $\phi_\omega^\leftarrow$ will significantly modify the scattering coefficients for not-too-low frequencies. To verify this, in \Fig{fig:3} we show the effective temperature $T_\om$ and the sum of squared norms of the hydrodynamic coefficients for a background water depth of the form \eq{eq:wdepth_2}, with $F_{\rm max} \approx 1.2$. On the left panel we recover an approximately Planckian spectrum in a wide range of angular frequencies, which is still bounded by $\om_{\rm max}$ on the right, as was found in the former section. The novel feature concerns the low-frequency regime. The Hawking regime is now bounded from below by 
\begin{equation}\label{eq:omc}
\om_c \approx\frac{c(0)}{h(0)} (F_{\rm max} - 1) 
(\Lambda_{\text{dec}} h(0))
\e^{-2 \kappa_{\text{dec}}L},
\end{equation}
where the critical inverse length is 
\begin{equation}
\Lambda_{\text{dec}} \equiv \frac{1}{h (0)} \sqrt{3 \left(F_{\rm max}^2-1\right)}.
\end{equation}
The critical frequency $\om_c$ gives the value of $\om$ below which the existence of the second horizon significantly affects the scattering. The exponential factor in \eq{eq:omc} may be understood as follows. The left-moving mode is exponentially decaying in the interhorizon region. Therefore the amplitude of the waves scattered on the black hole horizon back towards the white hole horizon are exponentially suppressed, unless the frequency is exponentially small, as then the matching conditions of modes on the two sides of the horizon give an additional exponentially large factor. (Notice also that the above formula for $\om_c$ is valid provided $F_{\rm max}$ is significantly larger than unity, so that the exponential factor of \eq{eq:omc} is much smaller than unity, and $\om_c$ exponentially small. In the other limit, when $F_{\rm max}$ decreases and approaches 1, the range of frequencies in which the radiation is thermal shrinks as $\om_c$ is increased.) For $0 < \om \lesssim \om_c$, the coefficients $\alpha_\om$ and $\beta_\om$ both become proportional to $\om^{1/2}$. In addition, the reflection coefficient $A_\om$ vanishes like $\om$. So, in the limit $\om \rightarrow 0$ we have a {\it total transmission} in the hydrodynamic sector, i.e., $|\tilde{A}_\om| \to 1$.

From this brief study we learned that, in the low-frequency limit, the scattering displays a new regime which, first, only involves the hydrodynamic modes, and, second, where the effective temperature vanishes. These two observations will be also found below when studying subcritical flows. 
However, the suppression of $|\beta_\om|^2$ will be much more significant, as it will apply to a much larger range of frequencies. Indeed, when $F_{\rm max}$ is significantly smaller than $1$, $\om_{\rm min}$ of \eq{eq:om-max-min}, which will play the role of $\om_c$, is  of the same order as $\om_{\rm \max}$, and is thus in general much larger than the typical values of the gradient of $v-c$ which fixes the effective temperature. 

\subsection{Subcritical flows} 
\label{sec:sub}

We now address the properties of the scattering coefficients when $F_{\rm max}$ is lower than 1. As in the former subsection, we first consider monotonic flows described by \eq{eq:wdepth}.

\subsubsection{Monotonic subcritical flows}

To start with, we emphasize that the main modification introduced by considering subcritical flows concerns $\om_{\rm min}$ of \eq{eq:om-max-min}. As already mentioned in subsection~\ref{Tpc}, for $\om \in \left] 0, \om_{\rm min} \right[$ the scattering of incoming shallow water waves now involves four waves, as described in \eq{eq:Bsub}.
Instead, for $\om \in \left] \om_{\rm min}, \om_{\rm max} \right[$ one recovers the former situation involving only 3 outgoing modes, see \eq{eq:Btrans}.  As a result we expect that the scattering coefficients behave very differently below and above $\om_{\rm min}$. 

\begin{figure}
\centering \color{white!30!black}\renewcommand\color[2][]{}
\def\svgwidth{0.49 \linewidth}
{\scriptsize 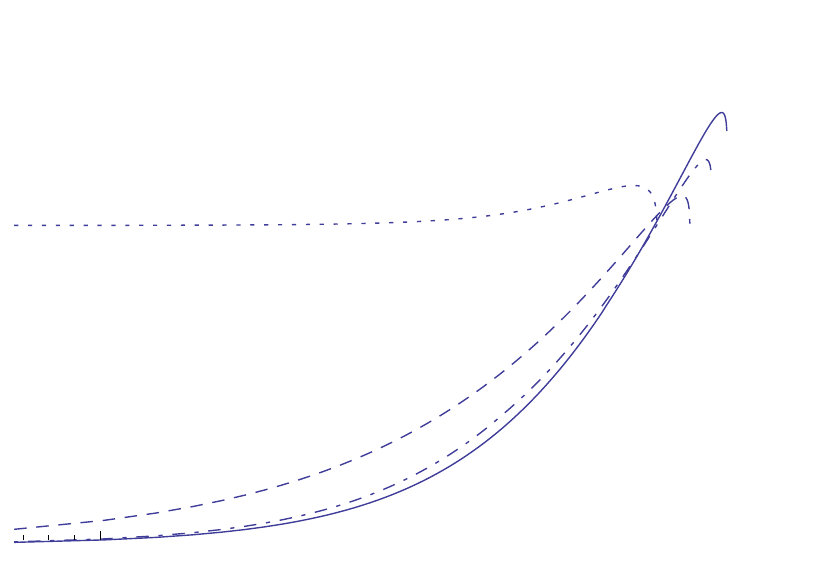}
\def\svgwidth{0.49 \linewidth}
{\scriptsize 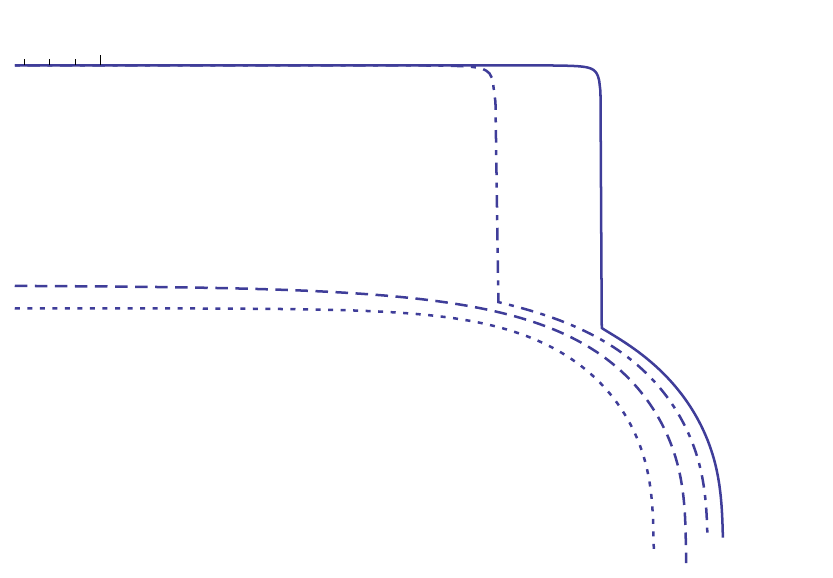}
\def\svgwidth{0.49 \linewidth}
{\scriptsize 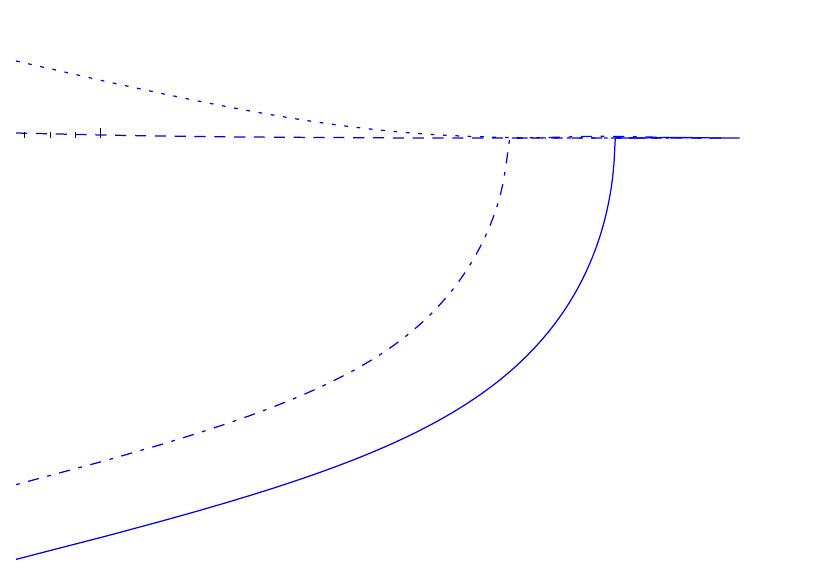}
\def\svgwidth{0.49 \linewidth}
{\scriptsize 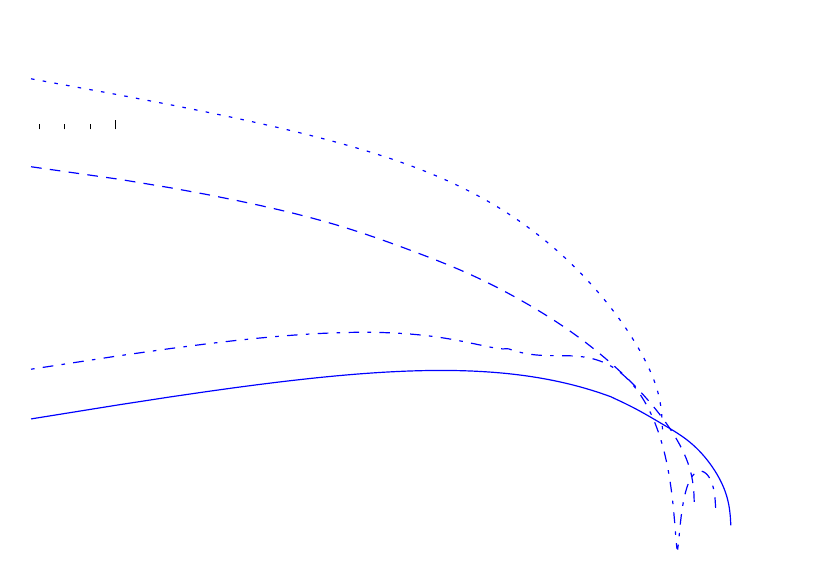}
\caption{Top, left: Effective temperature \eq{eq:effT} for monotonic flows of \eq{eq:wdepth} with $\sigma=0.06$, $D=0.2$, and four different values for $F_{\rm max}$, namely, $0.75$ (solid), $0.87$ (dot-dashed), $1.0$ (dashed), and $1.17$ (dotted). The temperature $T_{H, 0}\simeq 0.014$ is here used to ease the comparison with \fig{fig:1}. In the limit $\om \rightarrow 0$, the radical change between sub- and transcritical flows is easily seen, as $T_\om$ goes to zero in the former cases, whereas it remains finite for the latter. Top, right: Logarithm of the sum of the squared transmission and reflection coefficients.  One clearly notices that these coefficients are very small above $\om_{\rm min}$, i.e., when there is a turning point, but dominate below $\om_{\rm min}$. The two values of $\ln \lp \om_{\rm min}/T_{H, 0} \rp$ are $2.0$ and $1.1$. 
Bottom: Logarithm of the squared norms of the coefficients $\alpha_\om$ (left) and $\beta_\om$ (right). The sharp transition of $|\alpha_\om|^2$ occurring at $\om = \om_{\rm min}$ is clearly visible for the two subcritical flows. 
For these flows, one also notices that $| \beta_\om |^2$ remains much smaller than $1$. Hence the scattering is essentially elastic, without significant mode amplification.} 
\label{fig:2}
\end{figure}

This can be seen in the upper right panel of \fig{fig:2} where we represented the sum $|\tilde A_\om|^2 + |A_\om|^2$. Since this quantity is equal to $|\alpha_\om|^2- |\beta_\om|^2 - 1$ in virtue of \eq{eq:4coef}, it determines the relative importance of the hydrodynamic and dispersive sectors. 
(As no counterpropagating wave exists in the left asymptotic region when $\om > \om_{\rm min}$, we extended the definition of $\tilde{A}$ by setting $\tilde{A}(\om > \om_{\rm min})=0$. $\tilde{A}$ is then continuous across $\om = \om_{\rm min}$.) 
For each of the two subcritical profiles, we notice a sharp transition which precisely occurs at the corresponding value of $\om_{\rm min}$. 
Above this frequency, the reflexion coefficient $|A_\om|^2$ behaves essentially like for the critical and the transcritical flows, as can also be verified by comparison with the right panel of \fig{fig:1}. This was expected from the fact that, above $\om_{\rm min}$, the characteristics of the modes possess the same structure (given in the left panel of \fig{fig:modes}) whether $F_{\rm max}$ is above or below 1. 
On the contrary, below $\om_{\rm min}$,  $|\tilde A_\om|^2 + |A_\om|^2$ is close to 1. Hence, the low-frequency regime is dominated by the hydrodynamical sector, which moreover is purely elastic, i.e., involves no mode amplification. This is the first important result of this chapter. 

This conclusion is corroborated by the upper left panel where we observe that the effective temperature of \eq{eq:effT} vanishes as $\om \to 0$. In addition, we observe that it hardly changes when passing from $F_{\rm max}=1$ down to $0.87$ and $0.75$. In all cases, it displays no flat plateau, which would indicate a Planckian behaviour. We therefore conclude that in these monotonic subcritical flows, the Planckianity that was present for transcritical flows (see \fig{fig:1}, left panel) is completely lost. 
In fact, the vanishing of the effective temperature reflects something more fundamental: whereas $|\beta_\om|^2$ 
was growing as  $T_H/\om$ in transcritical flows, for subcritical ones, it remains much smaller than $1$ for all frequencies, as can be seen in the lower right panel of \fig{fig:2}. This key observation can be understood as follows. For $\om > \om_{\rm min}$, 
$\beta_\omega$ is exponentially suppressed (which is typical of nonadiabatic mode mixing~\cite{Massar:1997en}) because $\om_{\rm min}$, proportional to the dispersive frequency $c/h$ (see \eq{eq:omtp}), is typically much larger that the surface gravity scale. 
For $\om < \om_{\rm min}$, there is another mechanism at play: the incoming mode is essentially transmitted as there is no turning point. As a result, deviations from the WKB approximation, which predicts $\beta_\om = 0$, remain very small. 
 
Interestingly, the disappearance of the turning point has even a stronger consequence, namely  both $|\alpha_\om|^2$ and $|\beta_\om|^2$ vanish like 
\begin{equation}
\label{eq:bato}
|\beta_\om|^2 \sim |\alpha_\om|^2 \sim \om /\om_b , \; \om_b > 0,
\end{equation}
as can be seen from the two lower panels of \fig{fig:2}. This is our second important result.
The critical frequency $\om_b$ is found to be roughly proportional to
\begin{equation}
\label{omb} 
\om_b \propto \exp\left( \lp \sigma \frac{h_0}{D} F_{\rm max}^{1/3} \rp^{-2} \right), 
\end{equation}
for large and moderate values of $\sigma_1 \approx \sigma_2 = \sigma$, and if $D \ll h_0$. As a consequence, for $\om < \om_b$, the effective temperature behaves as
\begin{equation}\label{eq:Tomsub}
T_\om \approx -\frac{\om}{\ln \lp \frac{\om}{\om_b} \rp}  \lp 1+ O \lp \frac{\om}{\om_b} \rp \rp .
\end{equation}

\begin{figure} 
\centering \color{white!20!black}\renewcommand\color[2][]{}
\def\svgwidth{0.49 \linewidth}
{\scriptsize 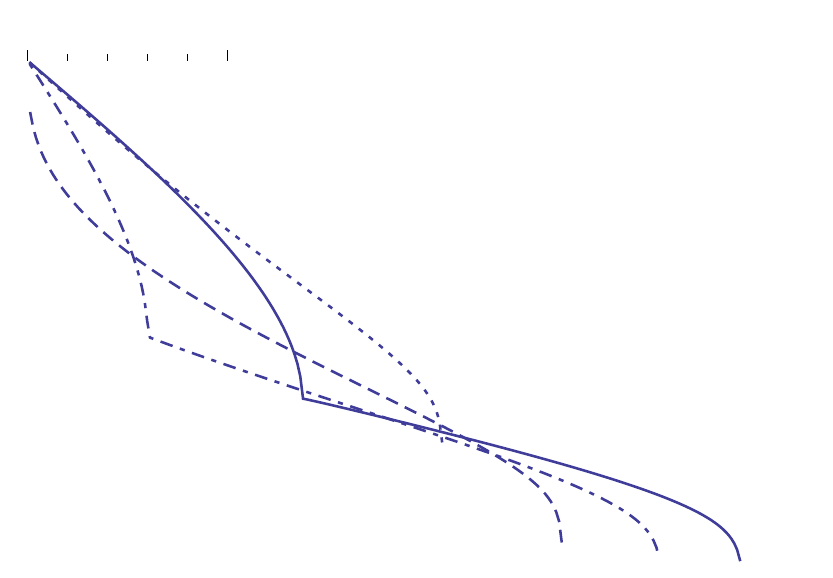} 
\def\svgwidth{0.49 \linewidth}
{\scriptsize 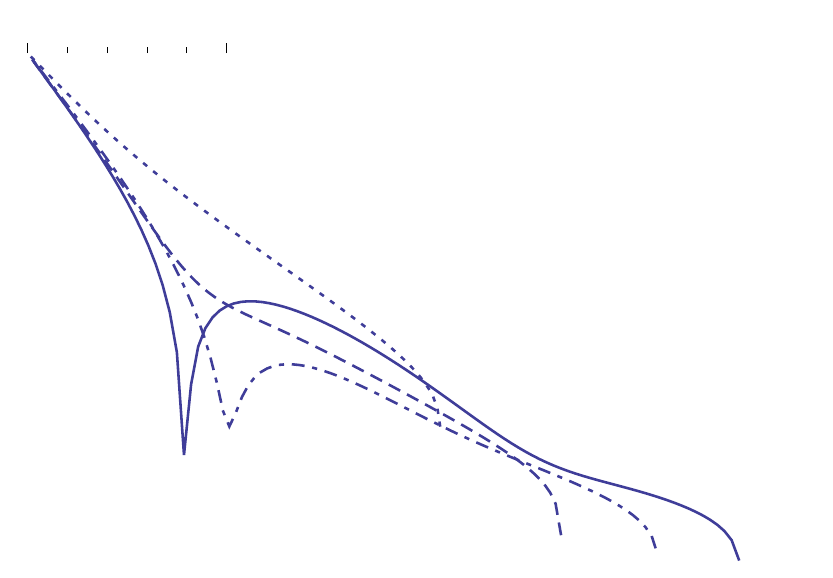}
\caption{Left: Logarithm of the parameter $R$ of \eq{eq:R} for the monotonic flows of \fig{fig:2}, using the conventions of that figure to designate the four cases. Notice that it is roughly linear in both the subcritical and transcritical cases. Notice also that the mean slopes are quite similar above and below the values of $\om = \om_{\rm min}$ where the two curves corresponding to subcritical flows show a kink. 
Right: Logarithm of $R$ for the nonmonotonic flows of \eq{eq:wdepth_2} with the same parameters as those of the monotonic ones, and with $L=4$. We see that the slope of $\ln R$ remains mostly unchanged. The kinks associated with $\om = \om_{\rm min}$ have now disappeared. Notice also that the sharp hollows here are related to a different phenomenon, namely resonance effects in the cavity formed by the two would-be horizons. 
The overall slope of $\ln R$ appears to be very robust, making it difficult to extract information from this sole quantity.}
\label{fig:lnR_subsub}
\end{figure}

We thus see that $|\alpha_\om|^2$, $|\beta_\om|^2$, and $ \left\lvert A_\om \right\rvert^2 + | \tilde{A_\om} |^2$ are all highly sensitive to the disappearance of the turning point. Surprisingly, the ratio of \eq{eq:R}, which was used in \cite{Weinfurtner:2010nu}, is not significantly affected by this disappearance, as can be seen in \fig{fig:lnR_subsub}, left panel. 
In fact the behavior of $R_\om$ is rather similar for the four flows considered in \fig{fig:2}. In particular, the limiting value of $R_\om$ when $\om \rightarrow 0$ is 1 in all cases. This can be explained as follows. 
When $\om < 0$, the roles of $\alpha_\om$ and $\beta_\om$ are exchanged with respect to the case $\om > 0$ because of the symmetry of the wave equation \eq{eq:om} under $\om \rightarrow - \om$, $\pd_x \rightarrow -\pd_x$ (known as ``crossing symmetry''~\cite{PhysRev.130.436}).~\footnote{This is a general property, see for instance the scattering of light waves on a mirror following a nonuniform trajectory, explained in Section 2.5.1 of \cite{Brout:1995rd}.} So, when $\om = 0$ the absolute values of $\alpha_\om$ and $\beta_\om$ must be equal. When there is a horizon, 
they both diverge since $|\beta_\om|^2 \sim |\alpha_\om|^2 \sim T/\om$. When there is none, we numerically observed that for $\om \to 0$,
they both vanish as $|\beta_\om|^2 \sim |\alpha_\om|^2 \sim \om /\om_b $. Hence in both cases $\ln R_\om$ is indeed linear for small values of $\om$ (if one assumes $|\beta_\om|^2 /|\alpha_\om|^2$ possess a regular Taylor expansion). It should be noticed that \eq{eq:bato} is compatible with the unitarity relations of \eq{eq:4coef}, or \eq{eq:3coef}, precisely because the hydrodynamic sector dominates the scattering ($|A_\om|^2 + |\tilde A_\om|^2 \to 1$) in the small-frequency limit. In brief, there is no contradiction between a Boltzmann-like behavior $\ln R_\om \propto - \om$ and \eq{eq:bato}. This offers a solution to the apparent contradiction (mentioned in the Introduction) between the observations of~\cite{Weinfurtner:2010nu}, where $\ln R_\om \propto - \om$ was observed at small $\om$, and the results of~\cite{finazziRP-proceedings} which established that the asymptotic spectrum is nonPlanckian, 
as they follow \eq{eq:bato}.~\footnote{{\it A priori}, \eq{eq:bato} could have been explained by some gray body factor $\Gamma_\om$. 
Indeed, for Schwarzschild black holes in four dimensions, for $\om \to 0$, one gets $\Gamma_\om \propto \om^2$~\cite{Page76}, which also gives that the asymptotic coefficient scales as $|\beta_\om|^2\sim \om$, without affecting the thermality of the Hawking radiation. In the present case, we do not think this explanation applies because $ \phi_{-\om}^{\rightarrow,d,\textrm{out}}$ of \eq{eq:Bsub} cannot be elastically reflected as it is the only negative-energy mode. We are grateful to W.~Unruh for interesting discussions about this issue.}


\subsubsection{Nonmonotonic subcritical flows, generalities}

\begin{figure}
\centering \color{white!25!black}\renewcommand\color[2][]{}
\def\svgwidth{0.49 \linewidth}
{\scriptsize 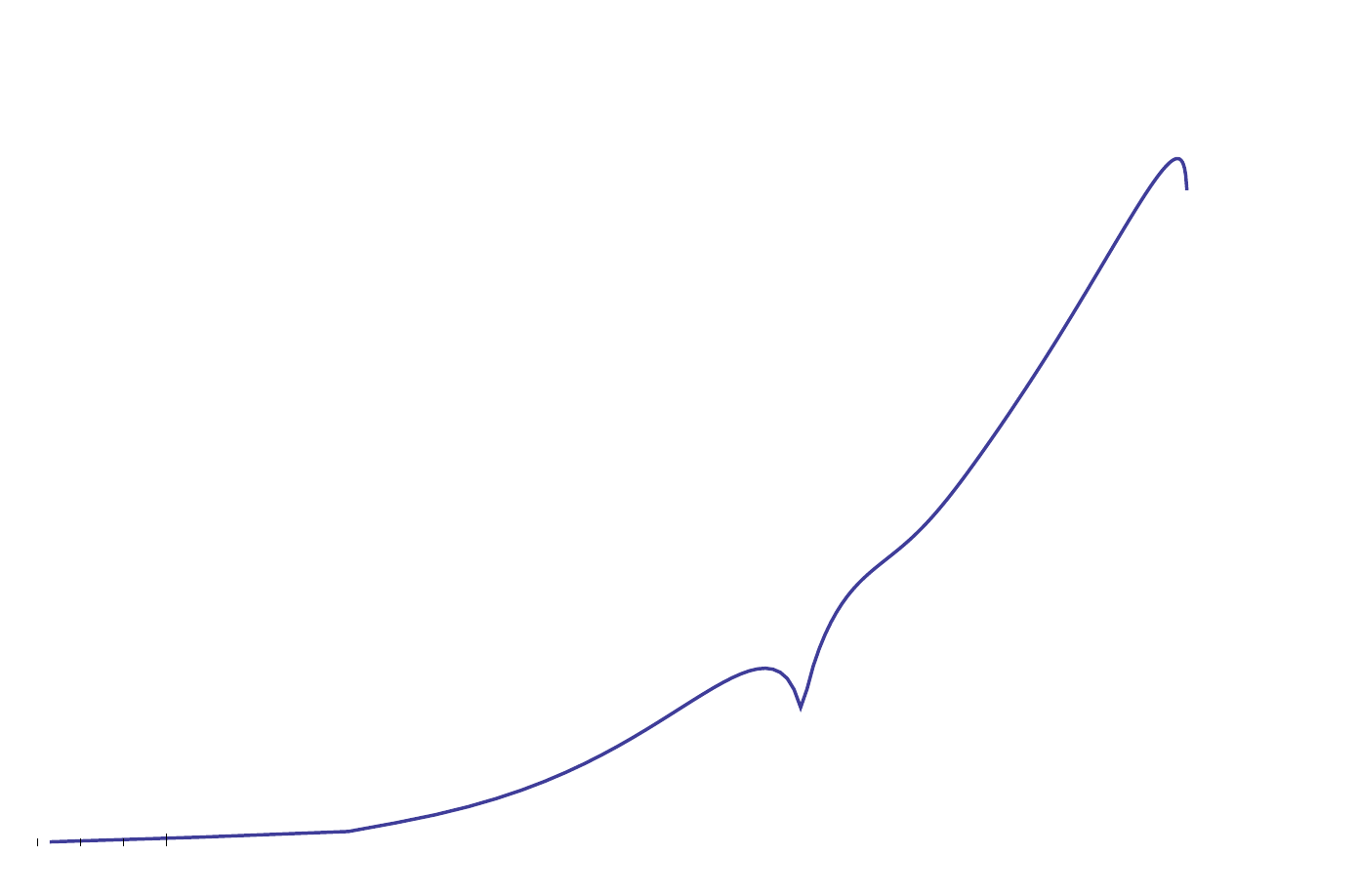}
\def\svgwidth{0.49 \linewidth}
{\scriptsize 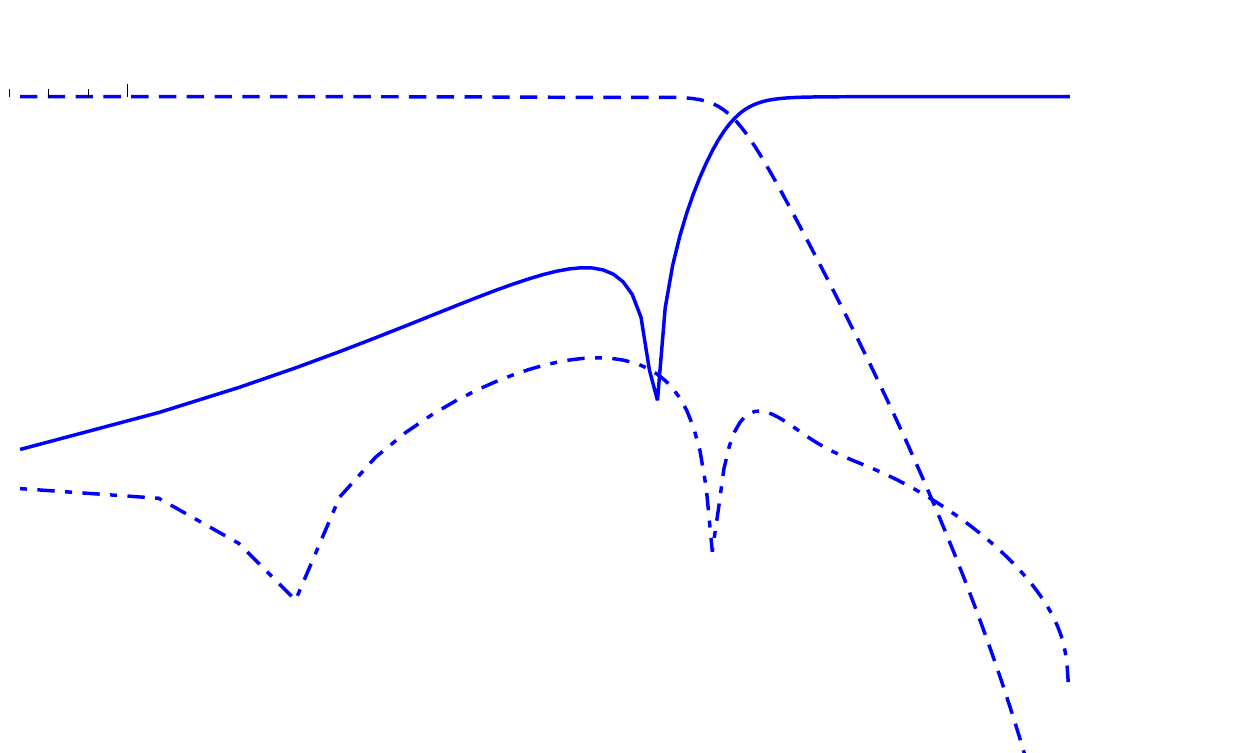}
\caption{Left: Effective temperature for a nonmonotonic subcritical flow \eq{eq:wdepth_2} (blue, continuous) with $F_{\rm max}= 0.87$, $\sigma_1 = \sigma_2 =0.06$ and $L = 10$, and for the corresponding monotonic one of \eq{eq:wdepth} (green, dashed) which coincides with the second subcritical flow of \fig{fig:2}. 
Apart from the hollows due to resonances, the effective temperature behaves in the same manner. Right: Logarithms of the squared scattering coefficients $|A_\om|^2 + |\tilde{A}_\om |^2$ (dashed), $|\alpha_\om|^2$ (solid), and $|\beta_\om|^2$ (dot-dashed) as functions of $\ln \om$, for the same nonmonotonic flow. 
At $\ln \lp \om_{\rm min} /T_{H,0} \rp \approx 1.1$, both $|\alpha_\om|^2$ and $|A_\om|^2 + |\tilde{A}_\om|^2$ display a transition.  While it was sharp in \fig{fig:2}, the transition is now smoothed out. Apart from this, the coefficients behave similarly to those of the corresponding monotonic flow.} \label{fig:3bis}
\end{figure}

\begin{figure}
\centering 
\includegraphics[width=0.49\linewidth]{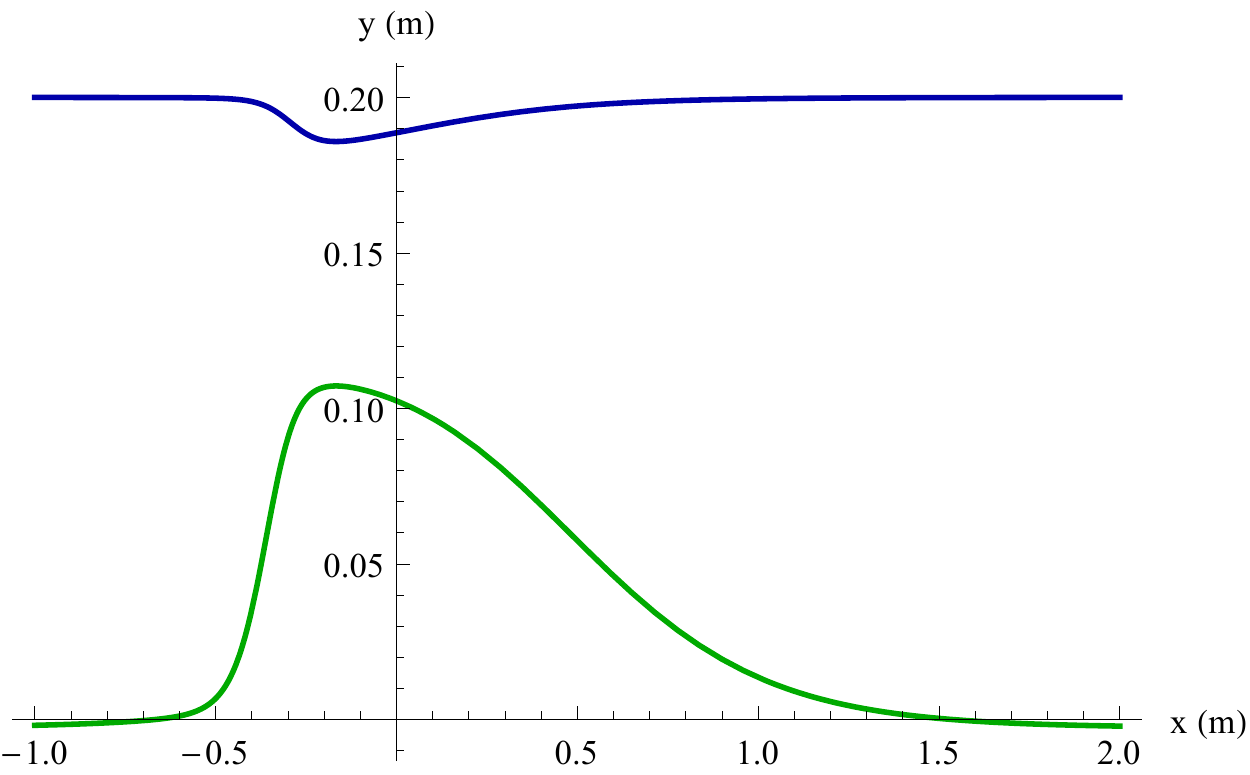}
\includegraphics[width=0.49\linewidth]{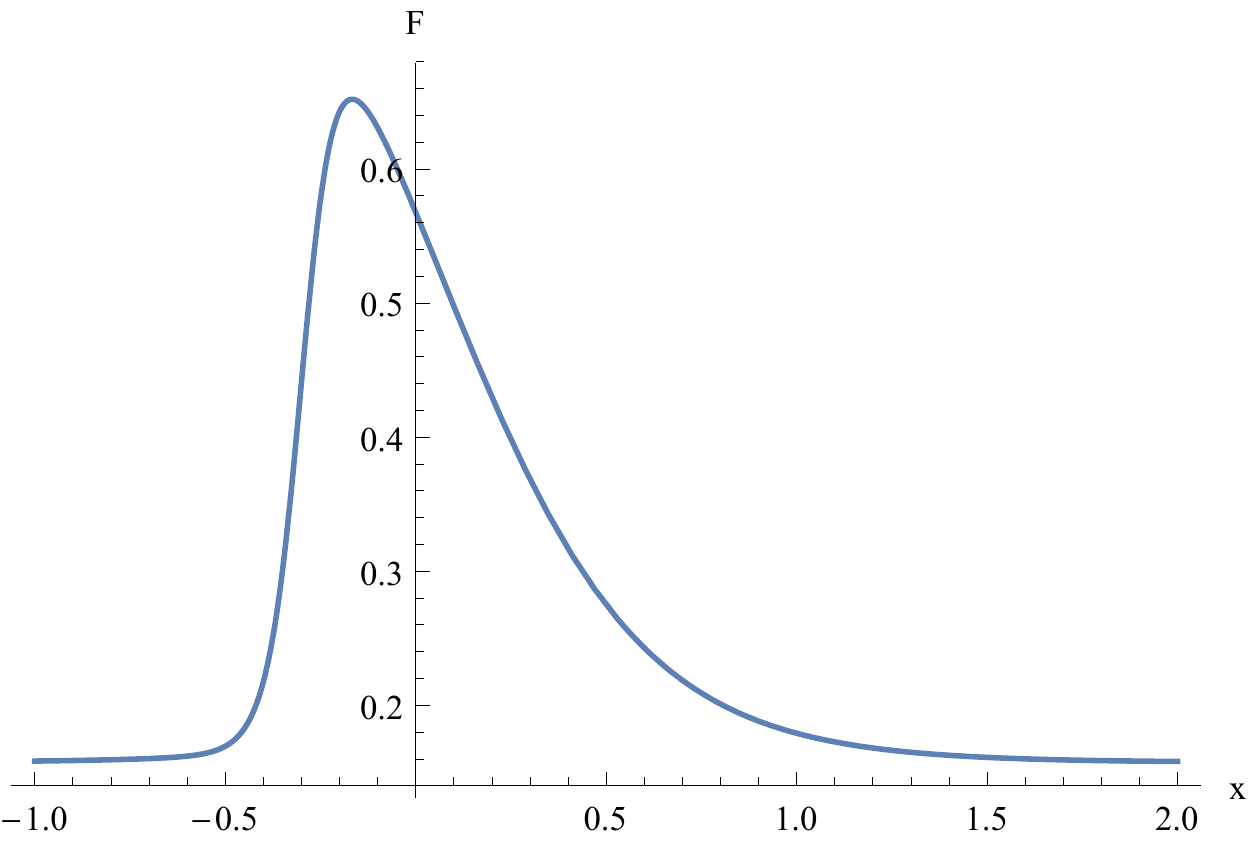}
\includegraphics[width=0.49\linewidth]{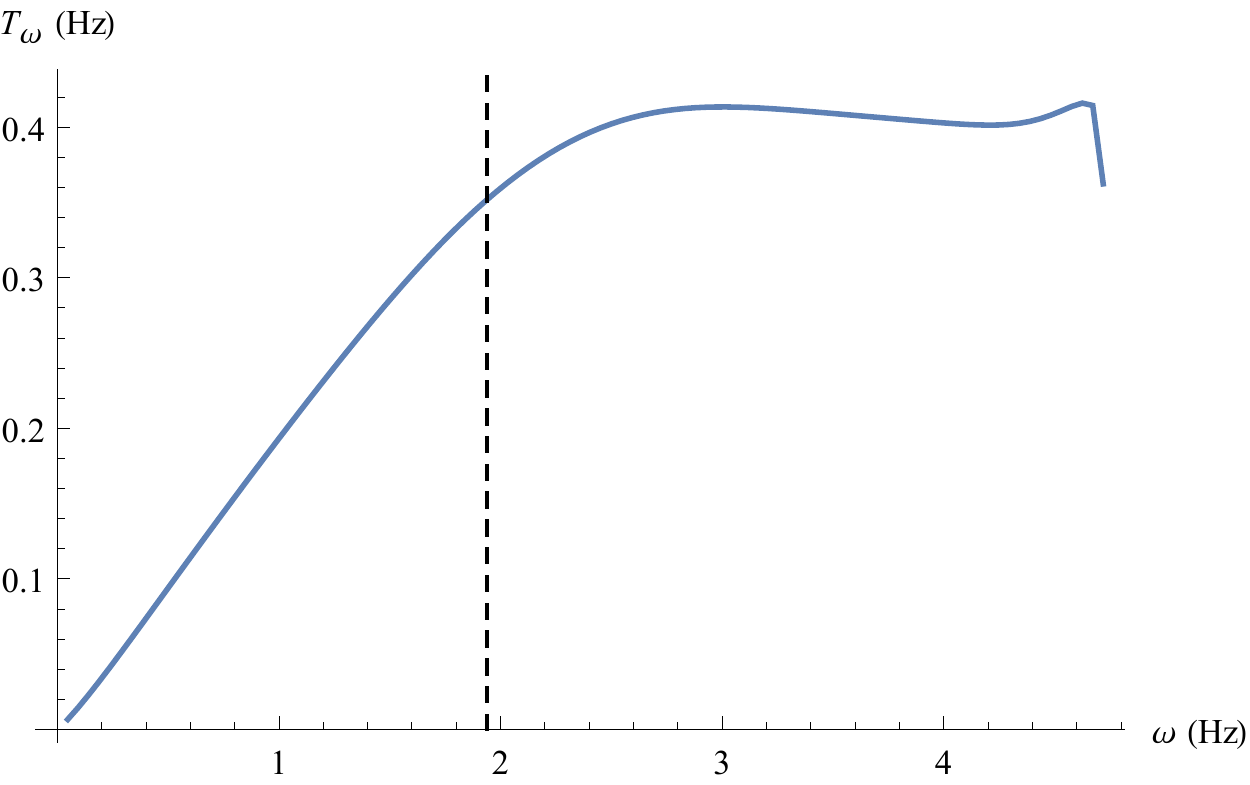}
\includegraphics[width=0.49\linewidth]{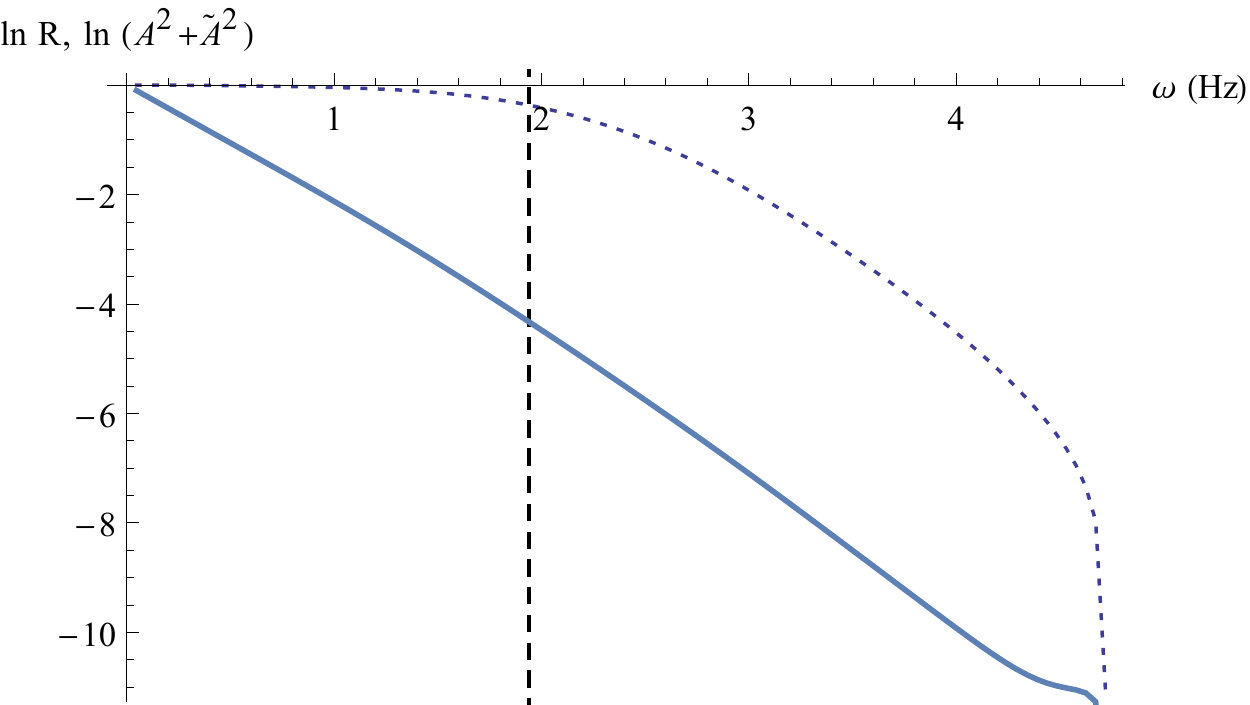}
\caption{Top, left: Free surface (blue) and obstacle (green) for a flow of the form \eq{eq:wdepth_2} resembling the one used in \cite{Weinfurtner:2010nu}. We took $g=9.8 m \cdot s^{-2}$ and $J=0.045 m^2 \cdot s^{-1}$. 
Top, right: Froude number as a function of $x$. 
Bottom, left: Effective temperature of \eq{eq:effT} as a function of $\om$. Bottom, right: $\ln R_\om $ (plain) and 
logarithm of $| A_\om|^2 + |\tilde{A}_\om|^2 $ (dotted) as functions of $\om$. Vertical dashed lines indicate the value of $\om_{\rm min}$. 
For $\om < \om_{\rm min}$, we see that the effective temperature linearly vanishes, and that the hydrodynamical coefficients dominate the scattering. We also see that $\ln R_\om $ is linear to a very good approximation, as was observed in~\cite{Weinfurtner:2010nu}. 
}\label{fig:smoothedVancouver}
\end{figure}

An extra ingredient must be added to make comparison with the observations of \cite{Weinfurtner:2010nu}. One should indeed consider nonmonotonic flows since the flow that was used had essentially the same velocities in the upstream and downstream regions. 
Unlike what we found when studying transcritical flows, for subcritical ones, we find that the replacement of monotonic flows by the corresponding nonmonotonic one does not significantly affect the results, see \fig{fig:3bis}. Indeed, for $\om \rightarrow 0$, the behavior of the Bogoliubov coefficients is only mildly affected, as one still finds \eq{eq:bato}, and $| \tilde{A}_\om |^2 + \left\lvert A_\om \right\rvert^2 \to 1$.~\footnote{There is a small difference between monotonic and nonmonotonic flows: in the latter, the reflection coefficient $A_\om$ goes to $0$ in the limit $\om \to 0$, whereas the limit of $|{A}_\om /\tilde A_\om |$ is generally finite in the former.} 
The absence of major difference with respect to the monotonic case reflects the fact that the value of $\om_{\rm min}$ matters more than the shape of the profile $h(x)$ in the upstream region far from the would-be white hole horizon. This can also be understood as follows. For transcritical flows, a qualitative change of behavior occurs as the left asymptotic region is supercritical for a flow of the form \eq{eq:wdepth}, but subcritical for a flow of the form \eq{eq:wdepth_2}. For subcritical flows instead, no such qualitative change occurs when going from \eq{eq:wdepth} to \eq{eq:wdepth_2}. In particular, a closer study reveals that \eq{eq:Tomsub} and \eq{eq:bato} are still valid for nonmonotonic subcritical flows.

Two relatively minor differences between monotonic and nonmonotonic flows are nevertheless worth mentioning. First, in~\Fig{fig:3bis} we observe hollows in $T_\om$, $|\alpha_\om|$ and $|\beta_\om|$, which correspond to resonances. Their presence is to be expected, as the high velocity central region acts as a resonant cavity, see~\cite{Zapata:2011ze}. In fact, their frequency strongly depends on $L$, which defines the length of the effective cavity. In particular, they disappear when $2 L \lesssim {\rm min} \lp D/\sigma_1, D / \sigma_2 \rp$, as can be verified in \Fig{fig:smoothedVancouver}. The second difference can be seen on the right panel of \fig{fig:lnR_subsub}. It concerns the disappearance of the sharp kinks observed for monotonic flows (see left panel), and associated with the presence of $\om_{\rm min}$. 
This disappearance can be understood from the fact that the transmission coefficients progressively vanish above $\om_{\rm min}$ when the flow is nonmonotonic, whereas they strictly vanish for monotonic flows.
 
\subsubsection{Comparison with the Vancouver experiment} 
 
Having clarified these points, we now consider a profile similar to the one used in \cite{Weinfurtner:2010nu}, save for the fact that we do not include the zero-frequency mode which modulated their background flow. At the end of this section, we shall briefly consider its impact on the scattering coefficients, and show that the modifications are not significant. 
A more detailed study of its effects is sketched in Section~\ref{sec:scat_und}.  

Ignoring the undulation, the water depth has the form \eq{eq:wdepth_2} where the parameters are chosen to fit the profile of~\cite{Weinfurtner:2010nu} using a least-squares method. In the international system of units, the optimum parameters are
\begin{equation}
\label{opt}
h_0 \approx 0.13 \, m , \quad 
D \approx 0.07 \, m , \quad 
\sigma_1 \approx 0.13,\quad
\sigma_2 \approx 0.76, \quad
2 L \approx 0.79 \, m.
\end{equation}
We did not use the exact description of the profile because its slope is discontinuous, making the numerical integration difficult. We believe this replacement has no significant consequences on our main results. 

Our description of the profile is represented in the upper left panel of \fig{fig:smoothedVancouver}. On the upper right one, we show the associated profile of the Froude number $F(x)$. 
In the lower plots, we represent the effective temperature, the squared norm of the hydrodynamic coefficients, and $\ln R$ of \eq{eq:R} as functions of $\om$. Vertical dashed lines indicate the value of $\om_{\rm min}$. 
By making series of simulations we observed that the values of the scattering coefficients can significantly depend on the precise shape of the profile, so we do not expect a good quantitative agreement. However, we also observed that three important features are not sensitive to the details of the profile shape. 

\begin{figure}
\centering
\def\svgwidth{0.49 \linewidth}
{\scriptsize 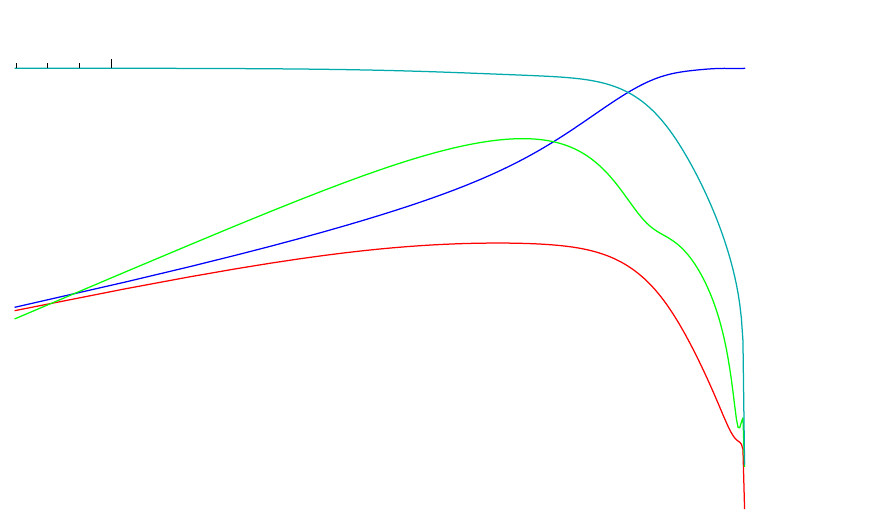}
\def\svgwidth{0.49 \linewidth}
{\scriptsize 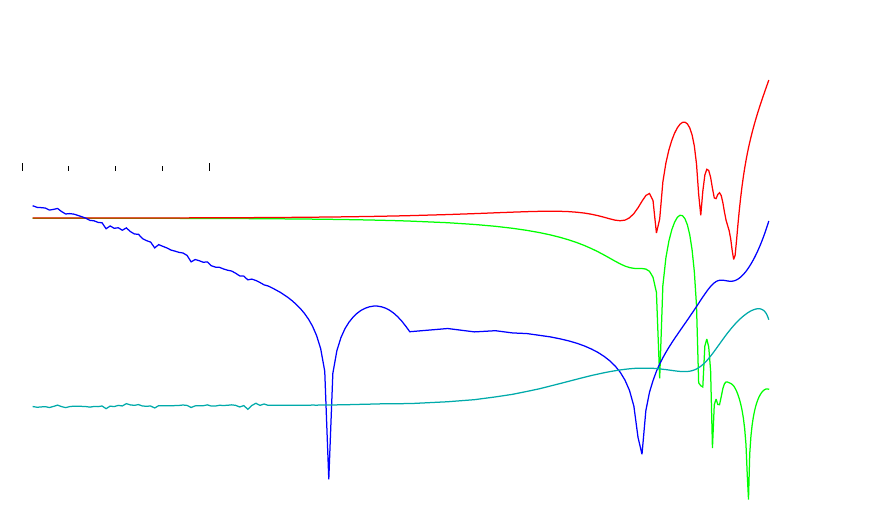}
\caption{Left: Logarithms of  $|\alpha_\om|^2$ (blue), $|\tilde{A}_\om|^2$ (cyan), $|A_\om|^2$ (green),
and $|\beta_\om|^2$ (red), as functions of $\om$ for the flow of \fig{fig:smoothedVancouver}. 
We see that $|\alpha_\om|^2$ becomes smaller than $|\tilde{A}_\om|^2$ 
for frequencies smaller than $\om_{\rm min} \approx 2 \, {\rm Hz}$, when there is no turning point, 
exactly as was seen in the lower right panel of \fig{fig:3bis}. 
Right: Logarithm of the relative differences of the norm of the 4 Bogoliubov coefficients  introduced by including an undulation of the form of \eq{eq:undul}, with the parameters of \eq{eq:paraun}. The rapid oscillations seen for $\ln \lp \om / \om_{\rm max} \rp < -6$ are due to numerical errors. 
We observe that the relative modification of the transmission coefficient $\tilde{A}_\om$ is very small, whereas those of the other three coefficients remain of the order of $2 \delta h_u/(h_0-D)\sim 0.06$, the relative change of the water height due to the undulation. 
}\label{fig:coefVanc}
\end{figure}

First, the hydrodynamic channels always dominate the scattering for $\om < \om_{\rm min}$, as can be seen in the bottom right panel. 
From the left panel of \Fig{fig:coefVanc}, we find that it is the transmission coefficient which dominates: 
\begin{equation}
\left\lvert \tilde{A}_\om \right\rvert^2 \mathop{=}_{\om \rightarrow 0} 1 + O \lp \frac{\om}{\om_{\rm min}} \rp .
\label{eq:blo}\end{equation}
Using the experimental data available, we estimate that $\om_{\rm min} \approx 2.7 \, Hz$ for the setup of~\cite{Weinfurtner:2010nu}\footnote{This value is computed with the full dispersion relation $\Omega_\om^2 = g k \tanh \lp h k \rp$, whereas in \fig{fig:smoothedVancouver} $\om_{\rm min}$ follows from the quartic law of \eq{eq:disprel}.}, corresponding to a linear frequency $f_{\rm min} \approx 0.42 \, Hz$. 
The second feature concerns the vanishing of $T_\om$ as $\om \to 0$, as can be seen in the lower left panel of \fig{fig:3bis}. In fact, we found that $\beta_\om$ and $\alpha_\om$ still obey \eq{eq:bato}. The third feature is a consequence of \eq{eq:bato}, and concerns the linearity of $\ln R_\om$ for low-frequency. In fact, $\ln R_\om$ is remarkably linear throughout the domain $\om \in \left] 0, \om_{\rm max} \right[$.

We now discuss a potentially important aspect that we so far neglected. It concerns the zero-frequency mode with a large amplitude that was observed in the downstream region. To investigate its effects on the scattering, we added various undulations to our profiles, along the lines of subsection~IV.B. of~\cite{Finazzi:2010yq}. 
To be able to distinguish the asymptotic modes on the left side, the amplitude of the undulation is exponentially suppressed at large values of $x$. To make the numerical integration simple, and to incorporate the information on the undulation we have, we worked with a profile of the form
\begin{equation}\label{eq:undul}
\delta h_u = \delta h_{u,0} \, \cos \lp k_u x \rp \lp 1 - \tanh \lp \kappa_l ( x-x_l) \rp \tanh \lp \kappa_r ( x-x_r) \rp \rp,
\end{equation}
where $k_u$ is the asymptotic wave number of the zero-frequency mode:
\begin{equation}
k_u = \frac{\sqrt{ 3 (1-F_{\rm min}^2) }}{h_0+D},
\end{equation}
and where 
\begin{equation}\label{eq:paraun}
\delta h_{u,0} = 0.002 \, m, \, \kappa_l = 1.0 \, m^{-1}, \, \kappa_r = 0.1 \, m^{-1}, \, x_l = 1.0 \, m, \, x_r = 50.0 \, m.
\end{equation}
To illustrate the various effects introduced by undulations, we show in the left panel of \fig{fig:coefVanc} the norms of the four coefficients of \eq{eq:Bsub} for the flow of \eq{opt} without undulation and, in the right panel, the relative variations of these coefficients when including the undulation parametrized by \eq{eq:undul} and \eq{eq:paraun}.
As can be seen in the figure, the relatives corrections are smaller than $\e^{-2}$ for $\om < \e^{-2} \om_{\rm min}$. So, the undulation does not change the qualitative behavior of the 4 scattering coefficients. In particular the key behaviors of \eq{eq:blo} and \eq{eq:bato} are recovered. 
Having done series of simulations with different amplitudes for the undulation, we found that the relative deviations are linear in the amplitude. These relative differences, evaluated for $\alpha$ and $\beta$ in the limit $\om \rightarrow 0$, are of the order of $2 \delta h_{u,0} / (h_0 - D)$. 
A more systematic study of the effects of zero-modes is beyond the scope of this chapter. 
Numerical results obtained in~\cite{Busch:2014hla} suggest it can drastically reduce the Hawking flux when an analogue horizon is present. 
Analytical results for an undulation without obstacle are given in Section~\ref{sec:scat_und}. 

With the above estimation of $\om_{\rm min}$, we notice that $5$ of the $9$ data points shown in Fig.~5 of \cite{Weinfurtner:2010nu}, left panel, correspond to frequencies below $\om_{\rm min}$, for which the squared norm of the transmission coefficient, $| \tilde{A}_\om |^2$, should be close to 1 since there is no turning point. Hence, we conjecture that the linearity of $\ln R_\om$ observed in the Vancouver experiment is probably not due to the fact that the incoming waves were blocked.~\footnote{Since the transmission coefficient $\tilde A_\om$ cannot be neglected for $\om < \om_{\rm min}$, we do not think it is legitimate to use, even as an approximation, $|\alpha_\om|^2 - |\beta_\om|^2 = 1$, as done between Eqs. (15) and (16) in~\cite{Unruh2014}. We are grateful to the referee of~\cite{Michel:2014zsa} to suggest us to discuss this recent work.}
Together with the absence of blocking, it would be interesting to see whether the low-frequency behavior of \eq{eq:bato} can be validated (or invalidated) by the experimental data of \cite{Weinfurtner:2010nu}. (The behavior of the norms of $\beta_\om$ and $\alpha_\om$ in the left panel of Fig.~5 of this reference indicates that \eq{eq:bato} could apply.) 
It was verified in two slightly different setups in~\cite{Euve:2014aga,Euve:2015vml}. 
It should be stressed that \eq{eq:bato} and \eq{eq:blo} will not be easily accessible when measuring the changes of the free surface associated with the 4 outgoing waves resulting by sending shallow water waves, see \eq{eq:Bsub}. Indeed, if we denote as $\delta h_\om^{\rm hydro}$ the variation associated with the transmitted wave, and $\delta h_\om^{\rm disp}$ that associated with the dispersive reflected wave of with negative $k_\om$, their ratio scales as
\begin{align}
\frac{\delta h_\om^{\rm hydro}}{\delta h_\om^{\rm disp}} \mathop{=}_{\om \to 0} O(1),
\end{align}
in spite of the fact that the ratio of the corresponding coefficients diverges as $|\tilde A_\om /\alpha_\om | \sim \om^{- 1/2}$, as implied by \eq{eq:bato} and \eq{eq:blo}.
The origin of the additional factor of $\om^{1/2}$ comes from, first, the action of the derivative operators in \eq{eq:dh}
which brings a factor of $\om - vk \propto \om$, and, second from the normalization factors, see for instance~\eq{eq:B8}.  
There is another property which probably further complicates the measurement of the transmitted wave, namely that its wavelength diverges like $1/\om$: for $\om = \om_{\rm min} \sim 2 \, Hz$, it is close to $5 {\rm m}$, and can become larger than the length of the flume used in~\cite{Weinfurtner:2010nu} if $\om$ is decreased. 

To conclude this section, we would like to discuss the status of the relationship between the effective temperature and the surface gravity, as this is a key feature of the Hawking effect. 
Since the limit $\om \to 0$ of the effective temperature $T_\om$ of \eq{eq:effT} vanishes, there is no unique way to associate a temperature to the system. A first possibility is to use the value of $T_\om$ at the plateau seen in the bottom left panel of \Fig{fig:smoothedVancouver}. This gives an effective temperature of approximately $0.4 \, Hz$. A second possibility is to use the inverse slope of $\ln R$ as a function of $\om$, giving a temperature of $0.5 Hz$, which is rather close to the previous one. 
(For comparison, the temperature obtained from the inverse slope of $\ln R$ in Fig. 5 of~\cite{Weinfurtner:2010nu} is $T_H \approx 0.70 Hz$. The relative good agreement between these numbers confirms that our numerical simulations correctly captures the key properties of the observations made in Vancouver.) 
The third possibility is to use the gradient of $h$ to define a pseudo-Hawking temperature as 
\begin{equation}\label{eq:Tps}
T_{\rm pseudo-H} \equiv \frac{1}{2 \pi} {\rm max} \left\lvert \pd_x (v-c) \right\rvert. 
\end{equation}
 If the maximum is taken over the descending slope, where the scattering is supposed to occur, and where one would find the white hole horizon if $F$ crossed 1, we find $T_{\rm pseudo-H} \approx 0.15  \, Hz$, smaller  than the previous ones by a factor 3. If the maximum is taken instead on the steeper ascending slope, we find $T_{\rm pseudo-H} \approx 0.77  \, Hz$.  
We believe this discussion gives a fair idea of the difficulties to relate the gradient of $v - c$ to an effective temperature. It seems to us that it is pointless to try to identify a precise relationship in subcritical flows. On the contrary, when the flow is sufficiently transcritical, the standard relationship of \eq{eq:TH} works very well, as can be seen in \Fig{fig:iVhF:r}.

\section{Discussion}
\label{sec:conclPR}

In this chapter, we first recalled the basic elements governing the scattering of shallow-water waves. We showed that, when the Froude number is significantly larger than 1, in which case the analogue of a relativistic (Killing) horizon is clearly present, the scattering coefficients quantitatively follow Hawking's thermal prediction, and this despite the fact that dispersion is included in the wave equation and strongly affects the characteristics of the waves. 

Turning to subcritical flows, we explained the important role played by the critical frequency $\om_{\rm min}$ in governing the behavior of the scattering coefficients. For frequencies above $\om_{\rm min}$, incoming counter-propagating modes are blocked, and one essentially recovers the behaviour found for slightly transcritical flows. In particular, the Planckianity of the spectrum is already lost. For frequencies below $\om_{\rm min}$, we observed a decrease of the effective temperature, which vanishes in the limit $\om \to 0$. This reflects the fact that both $|\beta_\om|^2$ and $|\alpha_\om|^2$, which were the dominant coefficients for  significantly transcritical flows, now both linearly decrease to 0 in the low-frequency limit. At the same time, we saw that the sum of the hydrodynamic (elastic) coefficients, $|A_\om|^2 + |\tilde A_\om|^2$, tends to $1$, which means that they dominate the scattering in this low-frequency regime. We then showed the consistency between these facts and the linearity of $\ln R_\om = \ln |\beta_\om/\alpha_\om|^2$ for small $\om$, as if the spectrum were still Planckian.

Besides comparing the scattering in sub- and transcritical flows, we also identified the consequences of considering nonmonotonic flows which are subcritical on both sides of an obstacle. For transcritical flows, this amounts to adding an analogue black hole horizon. Whereas the high-frequency regime is hardly affected, there is a new critical frequency $\om_c$ which governs the ``tunneling'' across the region where $F > 1$. When the latter is long enough, $\om_c$ is exponentially small. Below $\om_c$ a new regime is found where $|\beta_\om|^2$ and $|\alpha_\om|^2$ again linearly decrease to 0 as $\om \to 0$. 

We combined these aspects by considering nonmonotonic subcritical flows. We found that the nonmonotonic character of the flow does not significantly modify the scattering coefficients. Hence the spectral properties are similar to those found for monotonic flows. In particular for $\om < \om_{\rm min}$, the saturation of $|A_\om|^2 + |\tilde A_\om|^2$ to $1$, the vanishing of $|\beta_\om|^2$ and $|\alpha_\om|^2$ linearly in $\om$, and the linearity of $\ln R_\om$  appear to be very robust features of the scattering. Moreover, these three features have also been found when including an undulation with a macroscopic amplitude and finite extension, and when considering a subcritical nonmonotonic flow, solution of the nonlinear hydrodynamical equations, see subsection~\ref{sub:NL}. We therefore conclude that these properties should apply to the experiment of~\cite{Weinfurtner:2010nu}. In fact, when comparing the observed behavior of 
$R_\om = |\beta_\om/\alpha_\om|^2$ to that predicted by our analysis, we found a good qualitative agreement in that both the linearity of its logarithm and the value of the slope are well approximated. 
These properties have been observed in~\cite{Euve:2014aga} and~\cite{Euve:2015vml}. 

We should also remind the reader that our predictions have been obtained using a simplified version of the wave equation derived in~\cite{Unruh:2012ve,Coutant:2012mf}. Further comparison with detailed experimental data might allow one to determine the validity range of this simplified equation. 

Finally, in subsection~\ref{sub:NL}, we considered a transcritical nonmonotonic flow over an obstacle which is a solution of the nonlinear hydrodynamical equations. Our aim was to show that in this more ``realistic'' case the scattering coefficients closely follow, in quantitative terms, Hawking's prediction, i.e. $|\beta_\om|^2 \sim |\alpha_\om|^2\sim T_H/\om$ for low-frequencies. This indicates that, by a careful choice of the obstacle, one could engender a  transcritical background flow hardly contaminated by an undulation, which could then be used to experimentally test the thermal prediction.~\footnote{Notice that a trans-critical flow was clearly realized in the settings of~\cite{PhysRevE.83.056312} involving a circular jump. However, it remains unclear to us how to generate stationary waves in a controlled way so as to probe the mode mixing at the sonic horizon.}
We hope that this analysis may persuade an experimental team to pick up the gauntlet. 

\section{Additional remarks}
\subsection{Link with the nonlinear hydrodynamic equations}
\label{sub:NL}

So far our analysis was restricted to the linear wave equation \eq{eq:waveeq} in a background flow specified from the outset by the profile of the water depth $h(x)$. Since \eq{eq:waveeq} comes from the linearization of (nonlinear) hydrodynamical equations~\cite{LLhydro,Batchelor,Unruh:2012ve} (see also subsection~\ref{sub:waveeqder}), it is worth verifying that our results still apply to background flows which solve these equations. To this end, we use the hodograph transform method described in~\cite{Unruh:2012ve}. Given a flow with prescribed free surface, asymptotic water depth, and velocity, this method allows to find an explicit parametrization of the obstacle shape. We shall consider two typical examples, one transcritical and one subcritical, so as to be able to compare the resulting scattering coefficients with those obtained in Section~\ref{sec:SP}. 
We stress that these two examples may not be suitable for an experimental realization. (They were chosen to show that the results of the main text apply when using solutions of the hydrodynamic equations with a simple shape of the bottom.) In particular, their descending slopes may well be too large to maintain a laminar flow. However, the general method that we present here can be applied to find smoother obstacles, with smaller slopes. 
It was used to design the obstacle of~\cite{Euve:2015vml}. 

We remind the reader that an ideal, incompressible, inviscid, irrotational, 2D flow may be described using the velocity potential $\varphi$,~\footnote{Here $\varphi$ is the ``full'' velocity potential, whose gradient gives the velocity of the background flow, while we denote by $\phi$ the linear perturbation on this potential describing waves, see \eq{eq:waveeq}.} 
defined as
\begin{equation}
\nabla \varphi = \vec{v},
\end{equation}
and the stream function $\psi$, satisfying
\begin{equation}
\nabla \psi = \vec{e}_z \wedge \vec{v}.
\end{equation}
Here, $\vec{e}_z$ is the unit vector in the horizontal direction orthogonal to the mean flow velocity. In order to find a localized obstacle shape centered close to the origin for a flow with $F(\infty) < 1$, the free surface must have a hollow. A simple choice, which we shall use to illustrate the procedure, is
\begin{equation}\label{eq:Fs}
y_s(\varphi) = \frac{h_0}{(1+A \, \e^{-\sigma^2(\varphi-\varphi_0)^2}) (1+A \, \e^{-\sigma^2 (\varphi+\varphi_0)^2})},
\end{equation}
where $y_s$ denotes the vertical coordinate of the free surface, $A, \sigma, \varphi_0$, and $h_0$ are positive real numbers..~\footnote{Notice that this function is holomorphic in the domain of $\mathbb{C}$ defined by $\abs{\Im \lp \varphi \rp} < \sqrt{\pi/2} / \sigma$. With the notations defined just below, the procedure is thus consistent provided $\abs{v_0 \, h_0} < \sqrt{\pi / 2} \, \abs{\sigma}$.} 
The parametric representations of the free surface and the obstacle in real space are then obtained once the asymptotic velocity is chosen, assuming the height of the obstacle goes to zero at infinity, as we now briefly explain. More details can be found in \cite{Unruh:2012ve}. 

The two potentials $\varphi, \psi$ can be used as coordinates. Then the former Cartesian coordinates $x$ and $y$ are seen as functions of $\varphi$ and $\psi$. It is convenient to unify them in a single complex-valued function $z \equiv x + \ii \, y$. Performing the change of coordinates from $(x,y)$ to $(\varphi, \psi)$, one finds 
\begin{equation}
\pd_\varphi x = \pd_\psi y, \nn
\pd_\psi x = - \pd_\varphi y.
\end{equation}
These are exactly the Cauchy-Riemann conditions, showing that $z$ is a holomorphic function of $\Phi$. The stream function $\psi$ being constant along the free surface (since the latter is a streamline), an ansatz of the form \eq{eq:Fs} entirely determines the imaginary part of $z$ at $\psi = \psi_s$, where $\psi_s$ is the value of $\psi$ at the free surface. We choose the convention that $\psi = 0$ at the bottom. Then $\psi_s$ is equal to the 2D conserved current $J$ \cite{Unruh:2012ve,Coutant:2012mf}. The real part of $z$ at $\psi = \psi_s$ is found using the Bernouilli boundary condition, which reads 
\begin{equation}
\pd_\varphi \lp g y\left(\varphi ,\psi _s\right)+\frac{1}{2 \left(\left(\frac{\partial x}{\partial \varphi }\right)^2+\left(\frac{\partial y}{\partial \varphi }\right)^2\right)}\rp = 0.
\end{equation}
This gives a first-order ordinary differential equation on the function $\varphi \mapsto x(\varphi, \psi_s)$:
\begin{equation}\label{eq:probing_eqx}
\partial _{\varphi }x\left(\varphi ,\psi _s\right)=\sqrt{\frac{1}{v_0^2+2g \left(h_0-y\left(\varphi ,\psi _s\right)\right)}-\left(\partial _{\varphi }y\left(\varphi ,\psi _s\right)\right)^2},
\end{equation}
where $h_0$ is the asymptotic water depth and $v_0$ the asymptotic velocity, so that $\Re z(\varphi, \psi_s)$ is uniquely determined up to a constant. Changing this constant has the effect of translating the free surface and the obstacle in $x$ by the same amount. The obstacle can then be parametrized by making use of the holomorphic properties of $z$:
\begin{equation}
z(\varphi,0) = z(\varphi- \ii \psi_s,\psi_s), \nn
x_{bottom}(\varphi) = \Re z(\varphi,0), \nn
y_{bottom}(\varphi) = \Im z(\varphi,0).
\end{equation}

To summarize, the procedure used to find an obstacle shape is the following:
\begin{enumerate}
\item Choose $\psi_s \in \mathbb{R}$, defining the domain $D \equiv \lb z \in \mathbb{C}, \; \abs{\Im(z)} \leq \abs{\psi_s} \rb$. 
\item Choose a holomorphic function $y_o: D \to \mathbb{C}$ such that $\forall \Phi \in \mathbb{R}, \; y_o(\Phi + \ii \, \psi_s) \in \mathbb{R}$, going to a finite limit $h_0$ as $\Phi \to \pm \infty$ (for instance, by holomorphic continuation of a smooth function $y_s: \mathbb{R} \to \mathbb{R}$ similar to \eqref{eq:Fs}). 
\item Define the holomorphic function $x_o: D \to \mathbb{C}$ satisfying \eq{eq:probing_eqx} on the line $\Im \Phi = \psi_s$.
\item Defining the holomorphic function $z_o \equiv x_o + \ii \, y_o$, the Cartesian coordinates associated with $\Phi \in D$ are $x = \Re z_o(\Phi)$ and $y = \Im z_o (\Phi)$. 
In particular, the line $\psi_s = 0$ defining the obstacle shape is parametrized by
\begin{equation}
\lb 
\begin{aligned}
x_{bottom}: \varphi \mapsto \Re z_o (\varphi) = \Re x_o (\varphi) - \Im y_o (\varphi) \\
y_{bottom}: \varphi \mapsto \Im z_o (\varphi) = \Re y_o (\varphi) + \Im x_o (\varphi)
\end{aligned}
\right. .
\end{equation}
\end{enumerate}

\begin{figure}
\centering
\includegraphics[width = 0.49\linewidth]{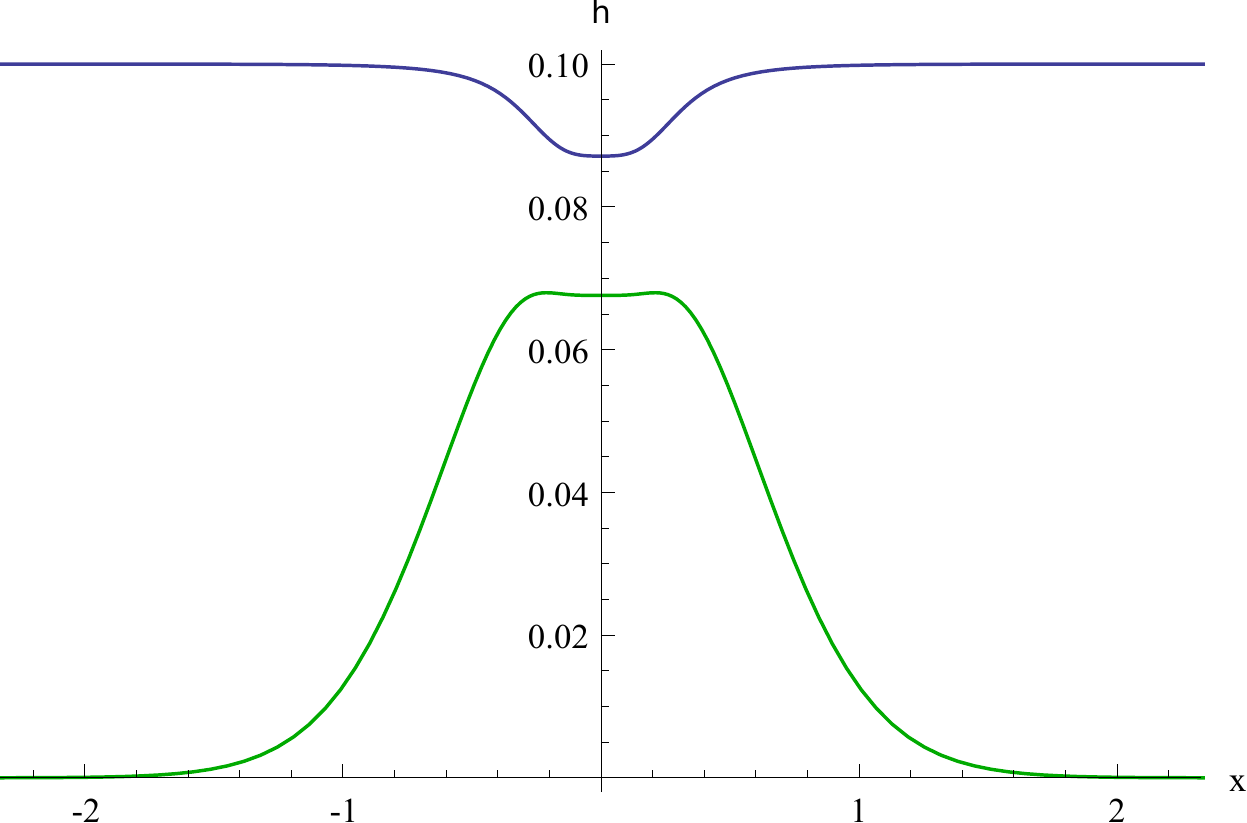}
\includegraphics[width = 0.49\linewidth]{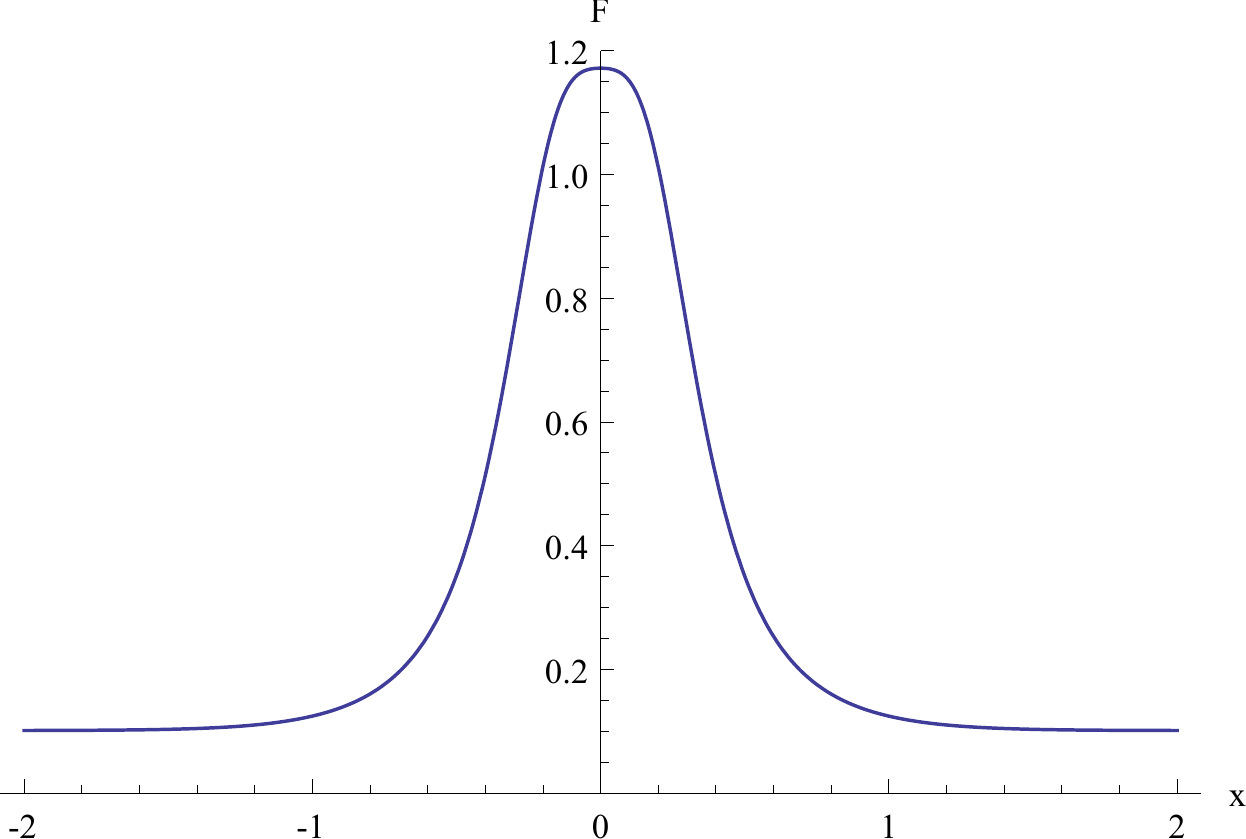}
\caption{Left: Heights of the free surface (blue) and of the obstacle (green) as functions of $x$ for the supercritical flow obtained by solving the hydrodynamical equations with the free surface specified by \eq{eq:Fs}. The units of both axes is the meter. Right: Froude number for the same flow. The maximum value of $F$ is $1.17$ and the length of the supercritical region is $0.41$ meters.}
\label{fig:iVhF}
\end{figure}

\begin{figure}
\centering
\includegraphics[width = 0.49\linewidth]{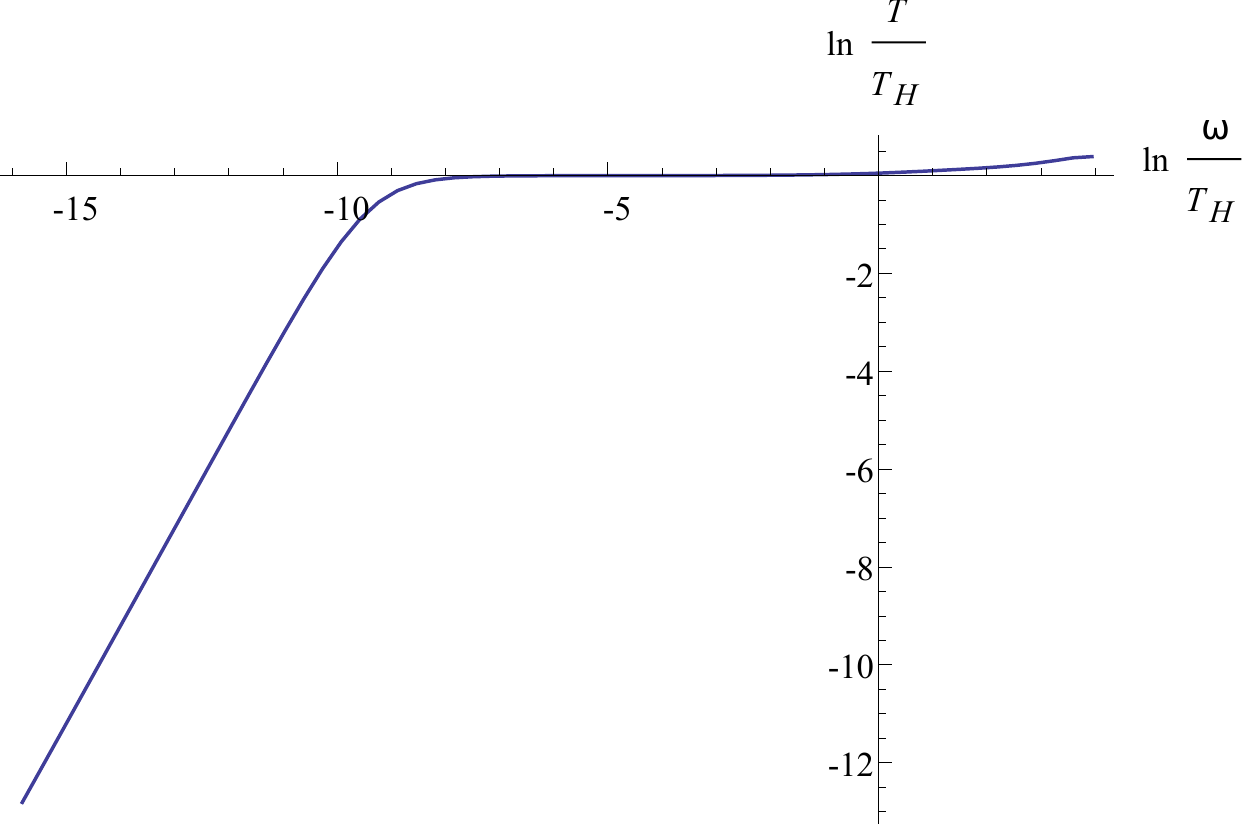}
\includegraphics[width = 0.49\linewidth]{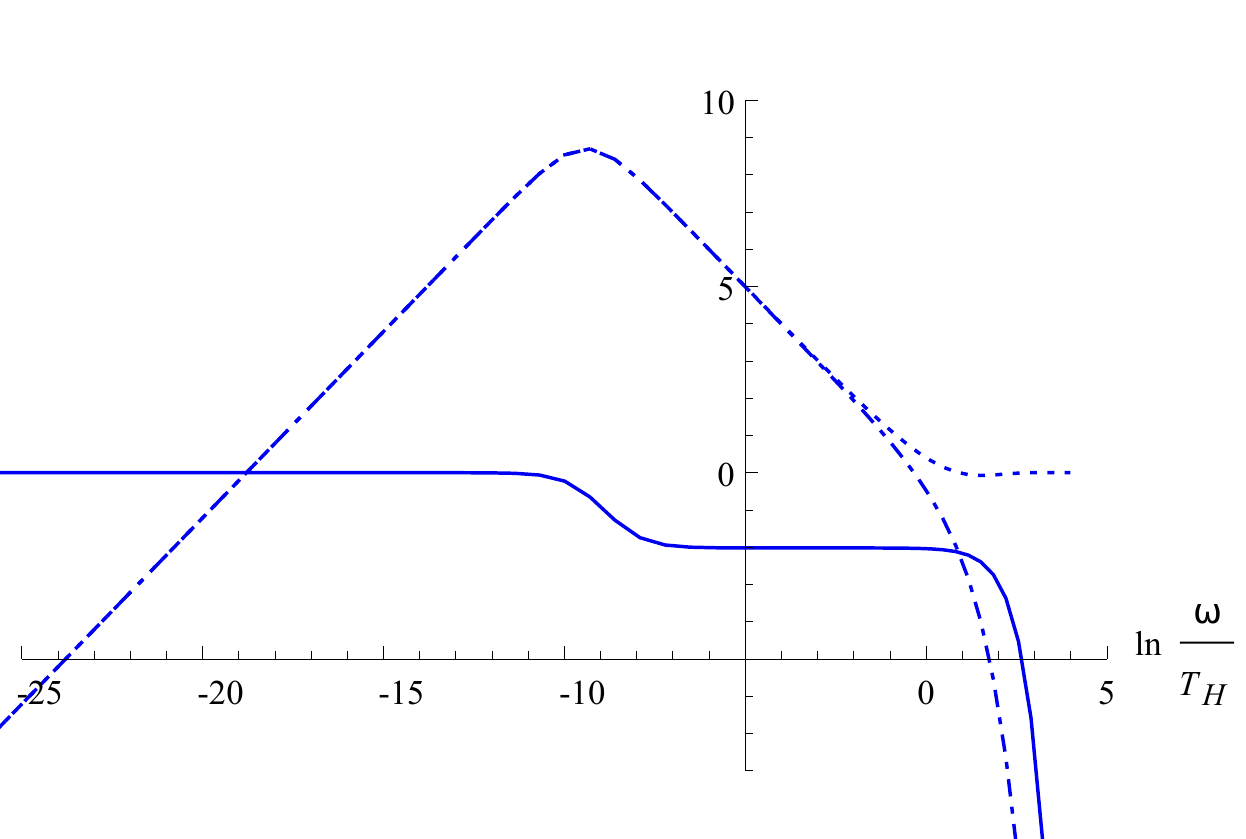}
\caption{On the left panel, we represent the logarithm of the effective temperature as a function of $\ln (\om/T_H)$ for the flow of \fig{fig:iVhF}. The Hawking frequency $T_H$ is approximately $0.164\, {\rm Hz}$. The good agreement with Hawking's prediction 
is clearly visible by the long extension of the plateau of relative height equal to 1. 
This plateau is bordered by the lower critical frequency $\om_c \approx \e^{-9} T_H$ of \eq{eq:omc} and the higher one $\om_{\rm max}$. On the right panel, we represent the logarithm of the squared norms of the Bogoliubov coefficients $|\alpha_\om|^2$ (dashed),  $|\beta_\om|^2$ (dot-dashed), and $|A_\om|^2 + |\tilde{A}_\om|^2$ (continuous) for the same flow.
It is clear that the hydrodynamic coefficients can be completely neglected for all frequencies larger than  $\om_c \approx \e^{-9} T_H$, thereby confirming the Hawkingness of this regime. 
}\label{fig:iVhF:r}
\end{figure}

The first example we considered describes a nonmonotonic transcritical flow. To be explicit, we now express quantities in the international system of units. The flow is characterized by $A=0.12$, $\sigma=10 \, {\rm s \cdot m^{-2}}$, $\psi_s=0.072 \, {\rm m^2 \cdot s^{-1}}$, and an asymptotic velocity $v_0=0.1 \, {\rm m \cdot s^{-1}}$. The resulting water depth and Froude number are shown in \fig{fig:iVhF}. 
The two small bumps at the top of the obstacle are fine-tuned to prevent the appearance of the undulation. One verifies that the flow is transcritical since $F_{\rm max} \simeq 1.17$. The main properties of the scattering coefficients are shown in \fig{fig:iVhF:r}. The comparison with \Fig{fig:3} shows a good correspondence between the two cases. In particular, for the left plot, we recover the extended flat plateau indicating a Planckian spectrum, with a value of the effective temperature $T_\om$ close to the Hawking frequency of \eq{eq:TH}, 
here given by $T_H = 0.143 {\rm Hz}$. We also observe the signature of the high-frequency cutoff $\om_{\rm max}$ of \eq{eq:om-max-min}, and that of the low-frequency one, $\om_c$ of \eq{eq:omc}. The approximate values of these critical frequencies are respectively $7.4 {\rm Hz}$, and $5. \, 10^{-6} {\rm Hz}$.  From the right panel, we also verify that below $\om_c$ the scattering is dominated by the hydrodynamic coefficients $A_\om$ and $\tilde A_\om$ of \eq{eq:Bsub}. 

\begin{figure}
\centering
\includegraphics[width = 0.49\linewidth]{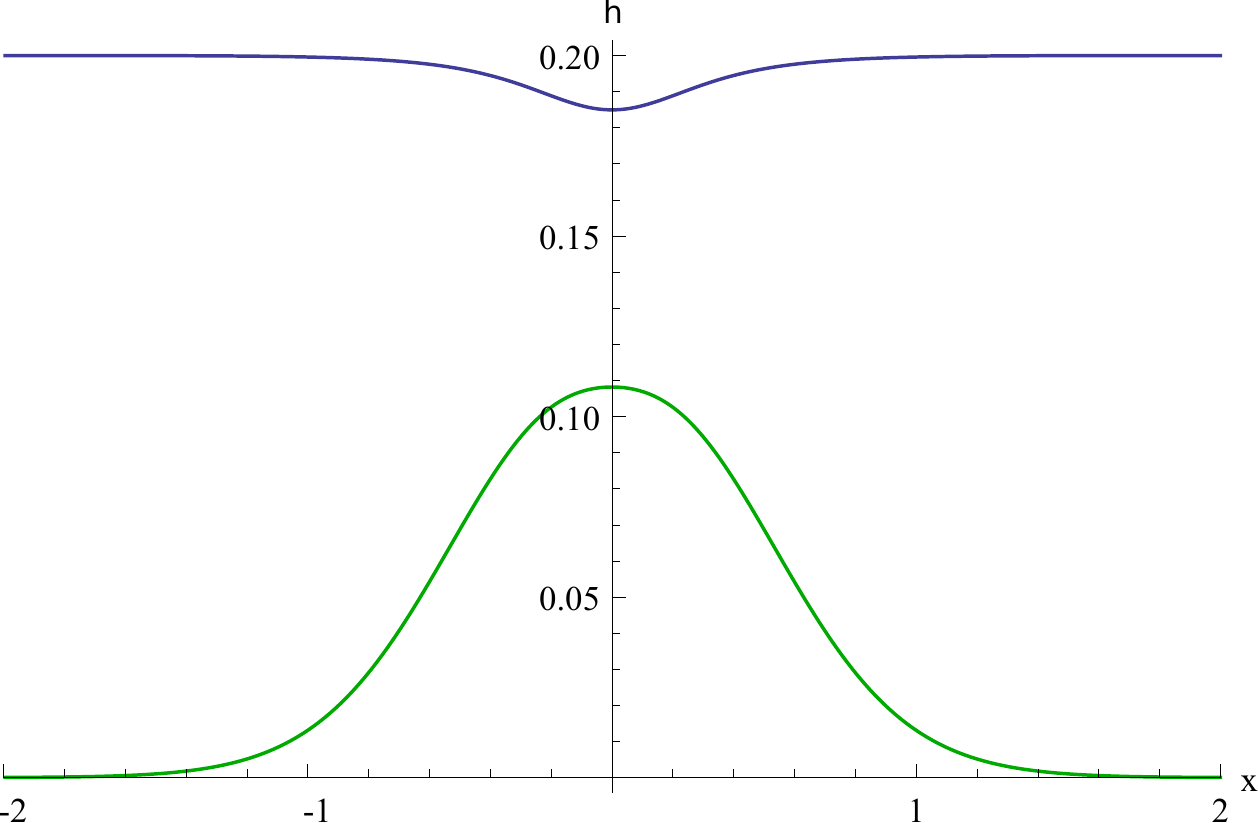}
\includegraphics[width = 0.49\linewidth]{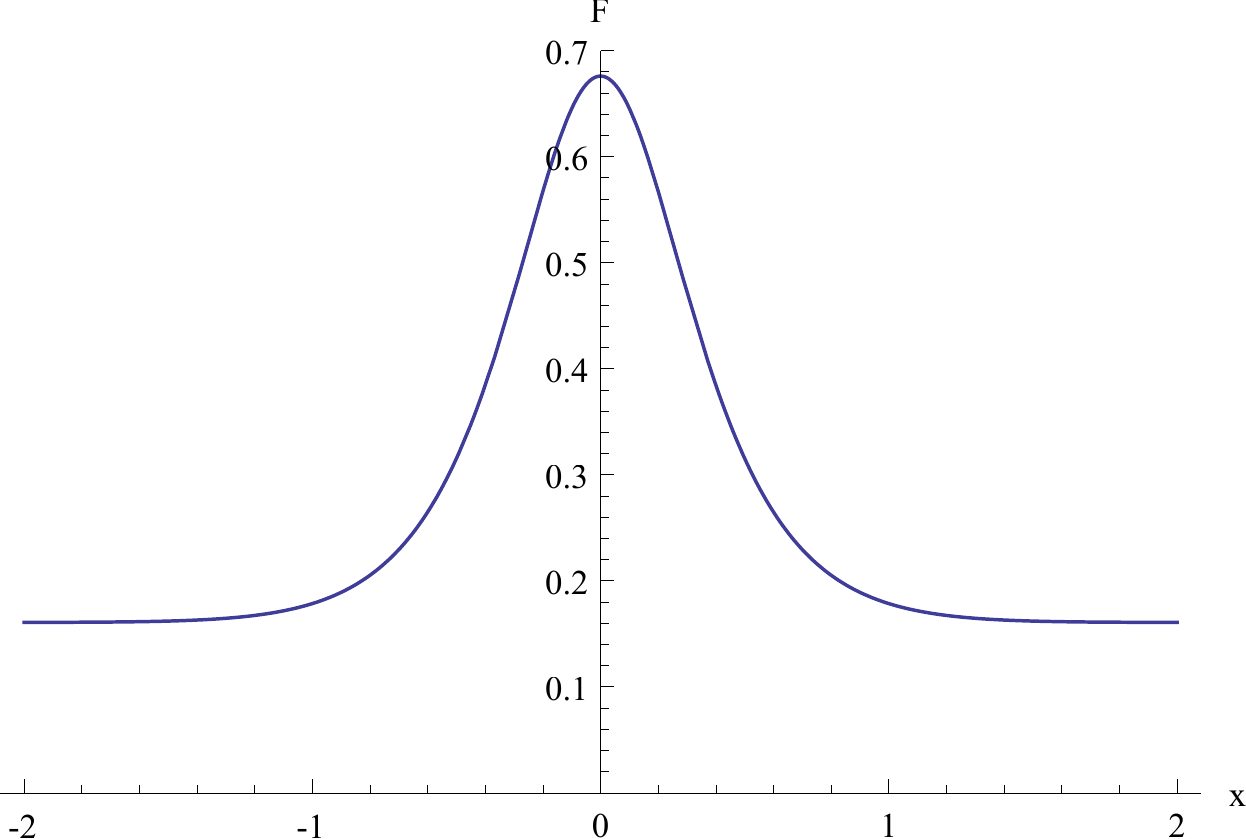}
\caption{Left: Free surface and obstacle for the second flow we obtained by solving the hydrodynamical equations with a known free surface \eq{eq:Fs}. Right: Froude number for the same flow.}\label{fig:iVhFbis}
\end{figure}

Our second example describes a subcritical flow. The parameters are $A=0.04$, $H_0 = 0.2 {\rm m}$, $v_0 = 0.0225 {\rm m \cdot s^{-1}}$, $\sigma = 5 {\rm s \cdot m^{-2}}$, and $\varphi_0 = 0.01 {\rm m^2 \cdot s^{-1}}$. They have been chosen to give a profile relatively close to the one used in~\cite{Weinfurtner:2010nu} for the downstream part $x > 0$ where the scattering and wave blocking occur. 
The water depth and the Froude number are shown in \fig{fig:iVhFbis}.
The maximum Froude number for this profile is close to $0.68$, the critical frequency $\om_{\rm min}$ to  $1.9 \, {\rm Hz}$, 
and the effective temperature of \eq{eq:Tps} is $0.21\,  {\rm Hz}$. The profile and the Froude number are similar to those of \Fig{fig:iVhF} as far as the downstream side of the flow is concerned. At this point, we consider that trying to reproduce more precisely the profile of~\cite{Weinfurtner:2010nu} is unjustified, as we have neither a good enough control of the various approximations we used, nor enough experimental data. 

\begin{figure}
\centering
\includegraphics[width = 0.49\linewidth]{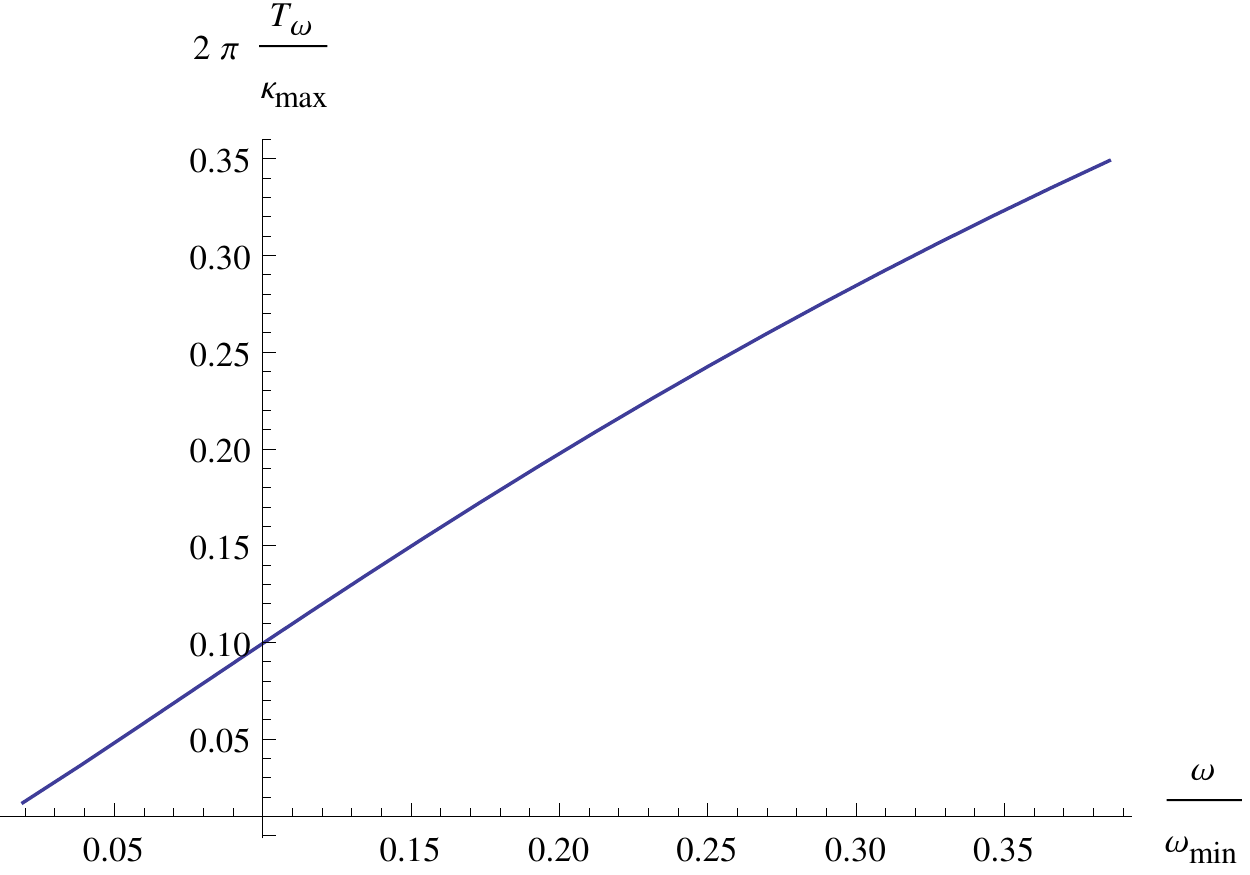}
\includegraphics[width = 0.49\linewidth]{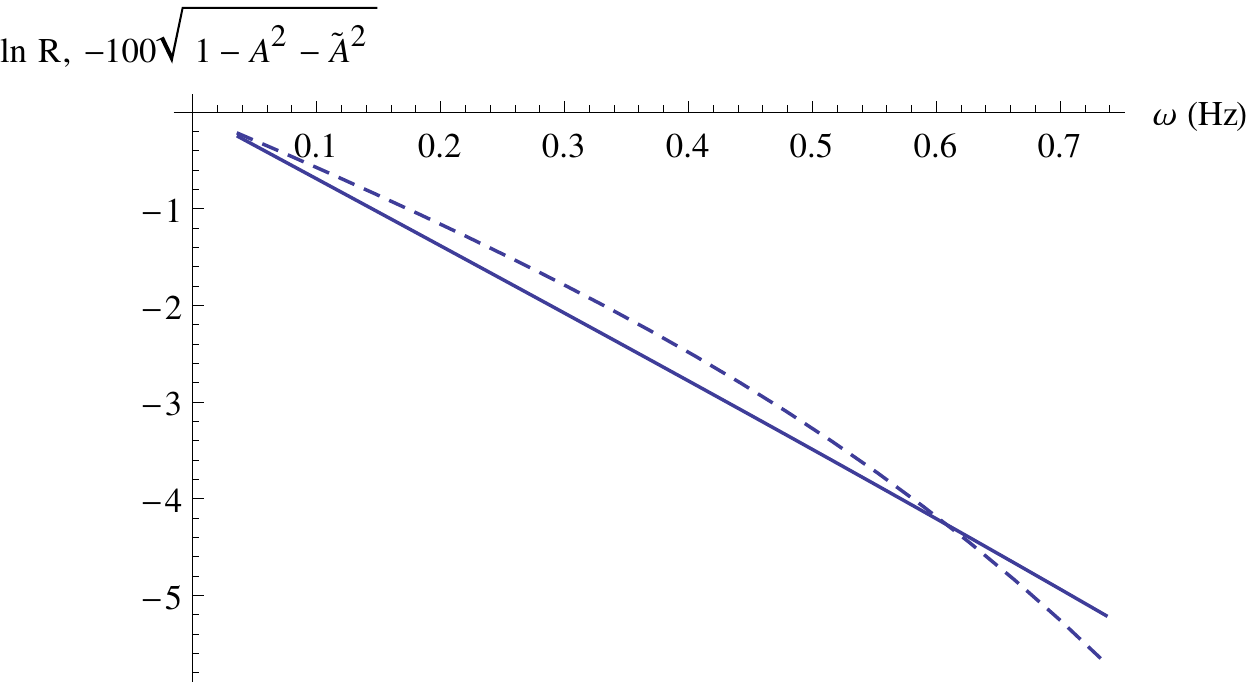}
\caption{Left: Effective temperature, adimentionalized by making use of the parameter of \eq{eq:Tps}, as a function of the adimensional frequency $\om/\om_{\rm min}$ for the flow of \fig{fig:iVhFbis}. Right: $\ln R_\om$ 
(plain) and $\sqrt{|A_\om|^2 - |\tilde{A}_\om|^2 - 1}$ (dashed) for the same flow. The square root and the factor $100$  have been used so as to clearly see the linear behaviors of both quantities for small values of $\om$.} 
\label{fig:iVhF:rbis}
\end{figure}

The properties of the scattering coefficients of our second flow are shown in \fig{fig:iVhF:rbis}. 
We see a good correspondence with those of Figs.~\ref{fig:2} and~\ref{fig:smoothedVancouver}. 
Namely, first, the effective temperature goes to 0 as $\om \to 0$, which confirms that Planckianity is lost; and second, the hydrodynamic elastic coefficients  $A_\om$ and $\tilde A_\om$ dominate the scattering for low frequencies. A few comments are in order. First, the range of frequencies we represented is smaller than the one in \Fig{fig:smoothedVancouver}. The reason is that obtaining a good numerical accuracy is more difficult in the present case because we no longer have a closed analytical formula for $h(x)$. Although our code can provide accurate results at higher values of $\om$, this becomes time consuming. We thus only present here results for small values of $\om$. For the same reason, the deviations from \eq{eq:4coef} (due to numerical errors) are larger than in the other cases, going from $10^{-3} \left\lvert \beta \right\rvert^2$ to $10^{-1} \left\lvert \beta \right\rvert^2$. 

From the left panel, we verify that the slope of ${2 \pi T_\om}/{\kappa_{\rm max}}$ versus $\om / \om_{\rm \min}$ is close to one. In addition, we also computed the effective temperature $T_\om$ for a few larger values of $\om$ and checked that the
qualitative agreement with \Fig{fig:smoothedVancouver} remains. 
Hence, we expect to get a rough plateau for $T_\om$ with a height close to the pseudo-Hawking temperature of \eq{eq:Tps}.
We checked that it is indeed the case: this plateau is at $T_\om \approx 0.17 \, {\rm Hz}$ while \eq{eq:Tps} gives $\approx 0.21 \, {\rm Hz}$. 

\subsection{Analytical calculation in the steplike limit}
\label{App:gradino}

In this subsection, following \cite{Finazzi:2012iu, Robertson:2012ku}, we consider the limit where the background water depth is piecewise constant, with one single discontinuity at $x=0$. This limit is rather unrealistic as in a real fluid the effects of viscosity, vorticity and compressibility are expected to become important when the slope of the obstacle is large~\cite{LLhydro,Batchelor}. Its interest lies in its mathematical simplicity, allowing a straightforward calculation of the spectrum. In spite of this, interestingly, one recovers the following important results of the main text
\begin{itemize}
\item in a transcritical flow, the effective temperature goes to a finite, nonvanishing constant when $\om \rightarrow 0$;
\item in a subcritical flow, it goes to zero like $\om / \ln \om$;
\item still for $F_{\rm max} < 1$, the coefficients $A$ and $\tilde{A}$ dominate the scattering for $\om < \om_{\rm min}$, but become small before $\alpha$ and $\beta$ for $\om > \om_{\rm min}$. 
\end{itemize} 

In each region $x < 0$ or $x > 0$, the solutions are proportional to $\e^{\ii k_\om x}$, where $k_\om$ is a root of the dispersion relation \eq{eq:disprel}. Equation~\ref{eq:disprel} in general has 4 solutions in $k$ at fixed $\om$. We denote them as $k_1, k_2, k_3 , k_4$. If they are all real, we order them as $k_1 \leq k_2 \leq k_3 \leq k_4$, see \fig{fig:modes}. 
If two of them are real and two are complex, we call $k_1 \leq k_2$ the two real roots, $k_3$ the root with a positive imaginary part, and $k_4$ the root with a negative imaginary part. 
We restrict our attention to these two cases, i.e., to $\om < \om_{\rm max}$, see \eq{eq:om-max-min}.

The modes computed in the two regions are matched at $x = 0$. To derive the matching conditions, it is convenient to use the variable $\xi$ defined by
\begin{equation}
\xi \equiv \int_0^x \frac{\dd y}{h(y)}.
\end{equation}
The wave equation \eq{eq:om} then takes the simpler form
\begin{equation}
\left(-\ii \om+\frac{1}{h}\partial _{\xi }\frac{J}{h}\right)\left(-\ii \om+\frac{J}{h^2}\partial _{\xi }\right)\phi - \frac{g}{h} \partial _{\xi }^2\phi -\frac{g}{3 h}\partial _{\xi }^4\phi = 0,
\end{equation}
where $J \equiv v h$.
The strongest singularities are now delta functions from $\partial_\xi$ acting on $h$. So, $\phi$ and its first and second derivatives with respect to $\xi$ are continuous across $\xi = 0$. The discontinuity in $\partial_\xi^3 \phi$ is given by
\begin{equation}
\left[\partial _{\xi }^3\phi \right]_{0^-}^{0^+}=3 \ii \frac{\omega }{g} [v]_{0^-}^{0^+}\phi (0)+3 \left[\frac{v^2}{c^2}\right]_{0^-}^{0^+}\partial _{\xi }\phi (0).
\end{equation}
We consider modes of the form
\begin{equation}
\phi (t,x)=\e^{-\ii \omega  t}
\left\lbrace
\begin{array}{ll}
 L_1\e^{\ii k_{1,L}x}+L_2\e^{\ii k_{2,L}x}+L_3\e^{\ii k_{3,L}x}+L_4\e^{\ii k_{4,L}x} & x<0 \\
 R_1\e^{\ii k_{1,R}x}+R_2\e^{\ii k_{2,R}x}+R_3\e^{\ii k_{3,R}x}+R_4\e^{\ii k_{4,R}x} & x>0
\end{array}
\right. ,
\end{equation}
where a subscript $L$ (respectively $R$) indicates a quantity evaluated for $x < 0$ (respectively $x > 0$).
The matching conditions at $x=0$ give a system of 4 linear equations on the coefficients $L_1$, $L_2$, $L_3$, $L_4$, $R_1$, $R_2$, $R_3$, and $R_4$. So, in general there are 4 linearly independent solutions. 
 
We now compute these coefficients, then the Bogoliubov coefficients, for white hole and subcritical flows. We restrict our attention to the left-moving incoming mode, with
$\om < \sqrt{3}\frac{v}{h}$ in each region, so that the sign of the group velocity computed with \eq{eq:disprel} agrees with that computed from the full dispersion relation $(\om - v k)^2 = g k \tanh(hk)$. 
We first assume there is a turning point, i.e.,  $\om_{\rm min} < \om < \om_{\rm max}$. The left-moving in mode satisfies 
\begin{equation}
\phi _{\text{in},v}
\text{: }
L_1=L_2=L_4=0.
\end{equation}
We find
\begin{equation}\label{eq:bigformula}
\left\lbrace
\begin{array}{ll}
 R_1=\frac{\left(h_Rk_{2,R}-h_Rk_{3,R}\right)\left(h_Rk_{4,R}-h_Rk_{2,R}\right)+\left(h_Lk_{3,L}-h_Rk_{4,R}\right)\left(h_Lk_{3,L}-h_Rk_{3,R}\right) L_3}{\left(h_Rk_{3,R}-h_Rk_{1,R}\right)\left(h_Rk_{4,R}-h_Rk_{1,R}\right)} R_2 &   \\
 R_3=\frac{\left(h_Rk_{2,R}-h_Rk_{1,R}\right)\left(h_Rk_{4,R}-h_Rk_{2,R}\right)+\left(h_Lk_{3,L}-h_Rk_{4,R}\right)\left(h_Lk_{3,L}-h_Rk_{1,R}\right) L_3}{\left(h_Rk_{1,R}-h_Rk_{3,R}\right)\left(h_Rk_{4,R}-h_Rk_{3,R}\right)} R_2 &   \\
 R_4=\frac{\left(h_Rk_{2,R}-h_Rk_{1,R}\right)\left(h_Rk_{3,R}-h_Rk_{2,R}\right)+\left(h_Lk_{3,L}-h_Rk_{1,R}\right)\left(h_Lk_{3,L}-h_Rk_{3,R}\right) L_3}{\left(h_Rk_{4,R}-h_Rk_{1,R}\right)\left(h_Rk_{4,R}-h_Rk_{3,R}\right)} R_2 &   \\
 L_3=\frac{\left(h_Rk_{2,R}-h_Rk_{1,R}\right)\left(h_Rk_{2,R}-h_Rk_{3,R}\right)\left(h_Rk_{2,R}-h_Rk_{4,R}\right)}{\left(h_Lk_{3,L}-h_Rk_{1,R}\right)\left(h_Lk_{3,L}-h_Rk_{3,R}\right)\left(h_Lk_{3,L}-h_Rk_{4,R}\right)-3 \left[\frac{v^2}{c^2}\right]_{0^-}^{0^+}h_Lk_{3,L}+3 \frac{\omega  }{g}[v]_{0^-}^{0^+}} R_2 &  
\end{array}
\right. .
\end{equation}
When the flow is transcritical, $\om_{\rm min} = 0$ and the limit $\om \rightarrow 0$ can be taken. In this limit, all coefficients remain finite. We denote by $\varphi_{\textrm{in},v}$ the normalized in mode, and $\varphi_{\textrm{out},u,1}$, $\varphi_{\textrm{out},u,3}$, and $\varphi_{\textrm{out},u,4}$ the three out modes. The numeral denotes the outgoing wave for the corresponding mode. 
In agreement with \eq{eq:Btrans}, we define the Bogoliubov coefficients as 
\begin{equation}
\varphi_{\textrm{in},v} = \alpha_\om \varphi_{\textrm{out},u,1} + \beta_\om \varphi_{\textrm{out},u,4} + A_\om \varphi_{\textrm{out},u,3}.
\end{equation}
Then,
\begin{align}\label{eq:B8}
\alpha_\om &= \sqrt{\left|\frac{\left(\omega -k_{R,1}\right)v_{g,R}\left(k_{R,1}\right)}{\left(\omega -k_{R,2}\right)v_{g,R}\left(k_{R,2}\right)}\right|} R_1, \nn
\beta_\om &= \sqrt{\left|\frac{\left(\omega -k_{R,4}\right)v_{g,R}\left(k_{R,4}\right)}{\left(\omega -k_{R,2}\right)v_{g,R}\left(k_{R,2}\right)}\right|} R_4 , \nn
A_\om &= \sqrt{\left|\frac{\left(\omega -k_{R,3}\right)v_{g,R}\left(k_{R,3}\right)}{\left(\omega -k_{R,2}\right)v_{g,R}\left(k_{R,2}\right)}\right|} R_3 . \nn
\end{align}
where $v_g$ denotes the corresponding group velocity. Note in particular that $\alpha$ and $\beta$ diverge like $\om^{-1/2}$ in the limit $\om \rightarrow 0$. 
In the transcritical case where $\om_{\rm min} = 0$
the $\om \rightarrow 0$ limit of the effective temperature of \eq{eq:effT} is
\begin{equation}\label{eq:Tgradino}
T^{\rm step}_{\om = 0}= \frac{\sqrt{3}\left(c_R^2-v_R^2\right)^{3/2}\left(v_L^2-c_L^2\right) h_R}{\left(\frac{v_L^2c_R^2}{c_L^2}-v_Rc_R+v_L\left(c_R-v_R\right)\right)^2\left(c_R+v_R\right)^2 h_L}\left|\sqrt{\frac{c_L^2-v_L^2}{h_L}}+\sqrt{\frac{c_R^2-v_R^2}{ h_R}}\right|^2.
\end{equation}

We now turn to subcritical flows. In this case, there is no turning point for $\om < \om_{\rm min}$. As a result, the left-moving in mode is defined by
\begin{equation}
\phi _{\text{in},v}
\text{: }
L_1=L_3=L_4=0.
\end{equation}
We find that \eq{eq:bigformula} is modified only through the replacements of $L_3$ by $L_2$ and $k_{3,L}$ by $k_{2,L}$. We define the Bogoliubov coefficients as 
\begin{equation}
\varphi_{\textrm{in},v} = A_\om \varphi_{\textrm{out},u,3} + \tilde{A}_\om \varphi_{\textrm{out},v} + \alpha_\om \varphi_{\textrm{out},u,1} + \beta_\om \varphi_{\textrm{out},u,4}.
\end{equation}
Taking the normalization into account, in the limit $\om \rightarrow 0$, the effective temperature behaves as
\begin{equation}
T^{\rm step}_\om = -\frac{\om}{\ln \lp \frac{\om}{\om_b} \rp} \lp 1+ O \lp \frac{\om}{\om_b} \rp \rp,
\end{equation}
where 
\begin{equation}\label{eq:omb}
\om_b = \left(\left(\frac{h_L}{v_L-c_L}-\frac{h_R}{v_R+c_R}\right)\frac{ c_R}{c_R+c_L}+\frac{h_Rc_R}{c_R^2-v_R^2}\right)^{-2}\frac{\sqrt{3} h_R}{\left(c_R^2-v_R^2\right)^{1/2}}.
\end{equation}
The limiting values of $A$ and $\tilde{A}$ for $\om \rightarrow 0$ are 
\begin{equation}
A \mathop{\rightarrow}_{\om \rightarrow 0} \frac{c_R - c_L}{c_R + c_L}, \quad \tilde{A} \mathop{\rightarrow}_{\om \rightarrow 0} \frac{2 \sqrt{c_R c_L}}{c_R + c_L}.
\end{equation}
As was found for smooth flows in Section~\ref{sec:SP}, the hydrodynamic sector dominates the scattering as $\left\lvert A_\om \right\rvert^2 + \left\lvert A_\om \right\rvert^2  \to 1$ for $\om \to 0$, while $\alpha_\om$ and $\beta_\om$ both go to zero like $\om^{1/2}$. The only important difference is that for a steplike discontinuity $\tilde{A}_\om$ does not vanish at $\om = \om_{\rm min}$, even though it still vanishes above $\om_{\rm min}$.

\begin{figure}
\centering
\includegraphics[width = 0.49 \linewidth]{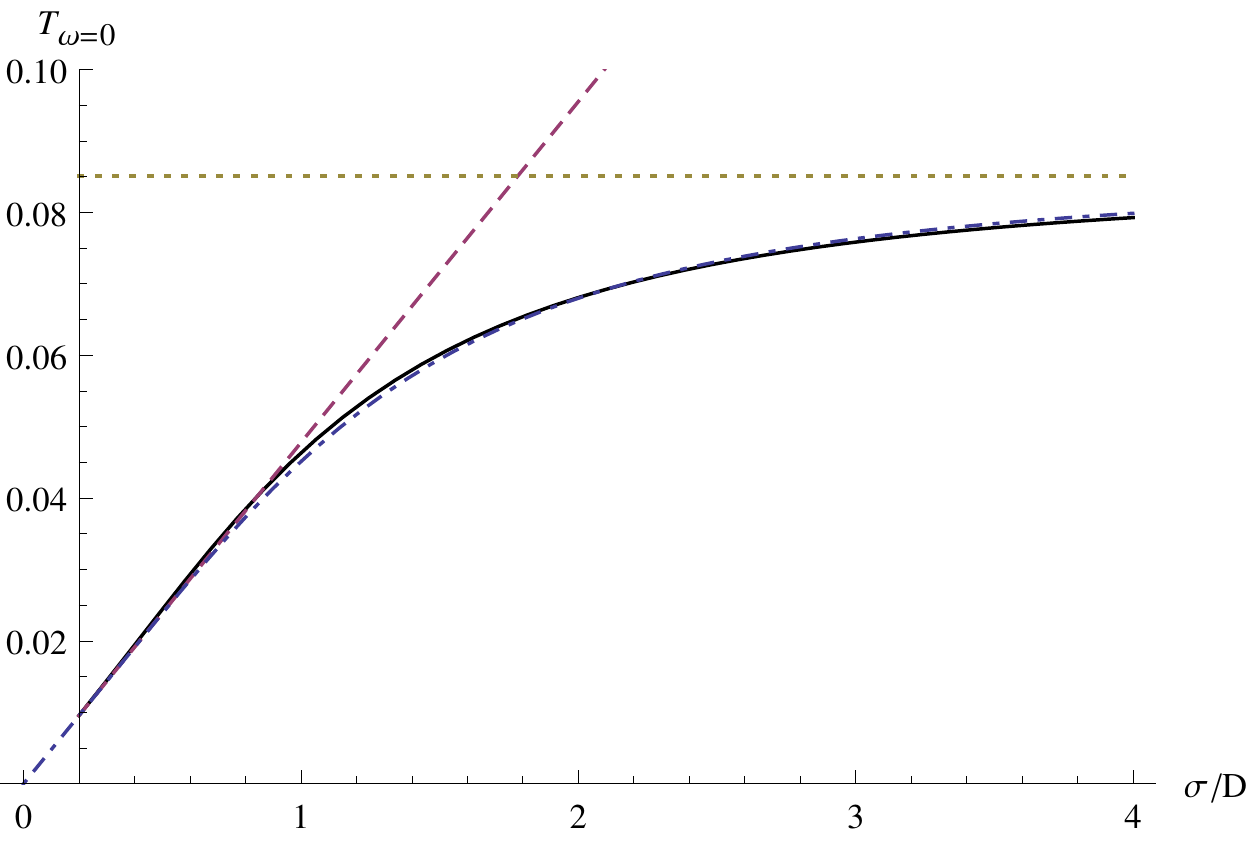}
\caption{Limit $\om \rightarrow 0$ of the effective temperature (solid), Hawking temperature (dashed), and steplike result of \eq{eq:Tgradino} (dotted) as functions of $\sigma$, for $g=J=1$, $h_0=1$, and $D=0.2$. The dot-dashed line shows the prediction \eq{eq:Scott_interpolation} of~\cite{Robertson:2011xp}. Note the very good agreement between the exact numerical calculation and \eq{eq:Scott_interpolation}.} \label{fig:Scott}
\end{figure}

To complete the analysis, we studied the transition between the Hawking regime where the surface gravity is small enough in units of the dispersive scale and the steplike regime studied above, for $\om \rightarrow 0$. 
Although we work in a slightly different case since the ordering of $h(x)$ and $\partial_x$ in the dispersive term 
of the wave equation is different from that of \eq{eq:om}, we found a very good agreement with the formula given in \cite{Robertson:2011xp}:
\begin{equation}\label{eq:Scott_interpolation}
T_{\om = 0} \approx T_H \, \tanh \lp \frac{T^{\rm step}_{\om = 0} }{T_H} \rp,
\end{equation}
where $T_{\om= 0}$ is the zero-frequency limit of the temperature of \eq{eq:effT} numerically evaluated in the
smooth flow, $T_H$ is the corresponding Hawking temperature, and $T^{\rm step}_{\om = 0}$ the zero-frequency limit of the temperature for the corresponding steplike profile, given by \eq{eq:Tgradino}. The agreement is illustrated in \fig{fig:Scott}.

\subsection{Derivation of the wave equation}
\label{sub:waveeqder}

In this subsection we give a step-by-step derivation of the wave equation used in this chapter. 
While it was already obtained in~\cite{Unruh:2012ve,Coutant:2012mf}, following their proof required some familiarity with hydrodynamical concepts. 
We here do the calculation in a more pedestrian way, which may be useful for readers not familiar with this field. 
To proceed, it will be convenient to use notations which are slightly different than those used in this chapter. 
In particular, in this subsection $\phi$ will denote the full velocity potential (background plus perturbation), while its perturbation is denoted by $\delta \phi$. 
The velocity field s denote by $\vec{v}$, the vertical position of the free surface by $y_s$, and the profile of the bottom by $y_b$.

We consider the flow of an \higt{inviscid, ideal fluid in an infinitely long flume, in irrotational, laminar motion in $(2+1)$} \higt{dimensions}. We also \higt{neglect the surface tension} and assume the atmospheric pressure above the flume is uniform. 
One then obtains the 5 following constraints:
\begin{itemize}
\item Since the flow is irrotational, one can define a \deft{velocity potential $\phi$} such that $\vec{\nabla} \phi = \vec{v}$.
\item The continuity equation (conservation of mass) gives ${\rm div} \, \vec{v} = 0$, i.e.,  
\begin{align} \label{eqD:Cont}
\forall t \in \mathbb{R}, \, \forall x \in \mathbb{R}, \, \forall y \in \left] y_b(x), y_s(x,t) \right[, \,  \Delta \phi(x,y,t) = 0.
\end{align}
\item The bottom is assumed to be unpenetrable. Then the velocity $\vec{v}$ at $y = y_b(x)$ must be parallel to its profile, i.e., 
\begin{align} \label{eqD:bott}
\forall t \in \mathbb{R}, \, \forall x \in \mathbb{R}, \, v_y(x, y_b(x),t) = y_b'(x) \, v_x(x, y_b(x),t) .
\end{align}
\item Since the flow is assumed to be laminar, a fluid particle which is at the free surface at a time $t$ remains at the free surface during the evolution. 
This means that its vertical displacement $v_y \, dt$ during an infinitesimal time $dt$ has only two contributions: one from the elevation of the free surface itself, $\pd_t y_s \, dt$, and one from the displacement of the fluid particle along the free surface, $v_x \, dt \, \pd_x y_s$. So, 
\begin{align} \label{eqD:surf}
\forall t \in \mathbb{R}, \, \forall x \in \mathbb{R}, \, \pd_t y_s(x,t) + \pd_x y_s(x,t) \, v_x(x, y_s(x),t) = v_y(x, y_s(x),t).
\end{align}
\item Assuming the only external force acting on the fluid is gravity, the second Newton's law applied to a fluid particle gives
\begin{align}
\rho \lp \pd_t + \vec{v} \cdot \vec{\nabla} \rp \vec{v} = - \vec{\nabla} P - g \vec{e}_y,
\end{align}
where $P$ is the pressure, $\rho$ the fluid density, and $\vec{e}_y$ the vertical, upward-pointing unit vector.  
Using that $\vec{v}$ is irrotational and $\rho$ is uniform, this may be rewritten as
\begin{align} \label{eqD:int01}
\vec{\nabla} \lp \pd_t \phi + \frac{\lp \vec{\nabla} \phi \rp^2}{2} + g y + \frac{P}{\rho} \rp = 0.
\end{align}
Finally, using that $P$ is uniform along the free surface and that one can add an arbitrary function of $t$ to $\phi$ without changing $\vec{v}$, integrating this equation gives~\footnote{To obtain this, we first use that direct integration of \eqref{eqD:int01} gives \eqref{eqD:Bern} with a right-hand side of the form $C(t) + P_0 / \rho$, where $C$ is some function of time only and $P_0$ is the atmospheric pressure. Since $\phi$ is defined only up to a function of time, one can replace it by $\bar{\phi}(x,y,t) \equiv \phi(x,y,t) + \int^t (C(u) + P_0/\rho) \dd t$. Doing this gives \eqref{eqD:Bern} without right-hand side, and with $\phi$ replaced by $\bar{\phi}$. Since the latter also satisfies the equation  defining $\phi$, i.e., $\vec{\nabla} \bar{\phi} = \vec{v}$, in the following we shall write $\phi$ instead of $\bar{\phi}$.} 
\begin{align} \label{eqD:Bern} 
\forall t \in \mathbb{R}, \, \forall x \in \mathbb{R}, \, 
\pd_t \phi(x, y_s(x,t),t) + \frac{\lp \vec{\nabla} \phi(x, y_s(x,t),t) \rp^2}{2} + g y_s(x,t) = 0.
\end{align}
\end{itemize}

Let us assume we know a stationary solution with velocity potential $\defe{\phi^{(0)}}$ and free surface $\defe{y_s^{(0)}}$. More generally, a superscript ``${(0)}$'' denotes a quantity evaluated on this background solution. We look for a solution of the form
\begin{align}
\defe{
\left\lbrace
\begin{array}{l}
\phi(x,y,t) = \phi^{(0)}(x,y) +\delta \phi(x,y,t)\\
y_s(x,t) = y_s^{(0)}(x) + \delta y_s(x,t)
\end{array}
\right.
}
\end{align}
and work to \higt{first order in $(\delta \phi, \delta y_s)$}. 

Linearizing \eqref{eqD:Bern} over the stationary background solution gives
\begin{align} \label{eqD:int1}
\forall t \in \mathbb{R}, \, \forall x \in \mathbb{R}, \, 
	\lp \pd_t \delta \phi + \vec{v}^{(0)} \cdot \vec{\nabla} \delta \phi \rp \lp x, y_s^{(0)}(x) \rp
	+ \lp \frac{1}{2} \pd_y \lp \vec{v}^{(0) 2} \rp \lp x, y_s^{(0)}(x) \rp + g \rp \delta y_s(x,t) = 0 .
\end{align}
Evaluating \eqref{eqD:surf} on the background solution gives
\begin{align}
\lp \pd_x y_s^{(0)}(x) \rp v_x^{(0)} (x, y_s^{(0)}(x), t) = v_y^{(0)} (x, y_s^{(0)}(x), t).
\end{align}
So, when everything is evaluated at the unperturbed free surface,
\begin{align}
\vec{v}^{(0)} \cdot \vec{\nabla} \delta \phi & = 
	v_x^{(0)} \pd_x \delta \phi + v_y^{(0)} \pd_y \delta \phi \\
	& = v_x^{(0)} \lp \pd_x \delta \phi + \lp \pd_x y^{(0)} \rp \pd_y \delta \phi \rp \\
	& = v_x^{(0)} \frac{\dd}{\dd x} \delta \phi (x, y_s^{(0)} (x,t),t),
\end{align}
where $\frac{\dd}{\dd x}$ denotes the total derivative with respect to $x$.
\eqref{eqD:int1} thus becomes
\begin{align}
\forall t \in \mathbb{R}, \, \forall x \in \mathbb{R}, \, 
	\lp \pd_t \delta \phi \lp x, y_s^{(0)}(x,t) \rp
		+ v_x^{(0)}\lp x,y_s^{(0)}(x) \rp \frac{\dd}{\dd x} \lp \delta \phi \lp x, y_s^{(0)}(x,t) \rp \rp \rp  
	\nn + \lp \frac{1}{2} \lp \pd_y \vec{v}^{(0)2} \rp \lp x, y_s^{(0)}(x,t) \rp + g \rp \delta y_s(x,t) = 0 .
\end{align}
This may be rewritten as
\begin{align} \label{eqD:deltay}
\delta y_s  = \left.
	-\frac{\pd_t \delta \phi
			+ v_x^{(0)} \frac{d \delta \phi}{dx}}
	{g + \frac{1}{2} \pd_y \vec{v}^{(0)2}}
\right\rvert_{y = y_s^{(0)}}
.s
\end{align}

Let us now use \eqref{eqD:surf}. To linear order, it reads
\begin{align} \label{eqD:int2}
\forall t \in \mathbb{R}, \, \forall x \in \mathbb{R}, \,
	\pd_t \delta y_s(x,t) 
	+ \lp \pd_x \delta \phi \rp \lp x, y_s^{(0)}(x,t),t \rp \, \pd_x y_s^{(0)}(x,t)
	\nn +  \thisisblue{\lp \delta y_s(x) \pd_x \pd_y \phi^{(0)} \rp \lp x, y_s^{(0)}(x,t),t \rp \, \pd_x y_s^{(0)}(x,t)}
	\nn + \thisisblue{\lp \pd_x \phi^{(0)} \rp \lp x, y_s^{(0)}(x,t),t \rp \, \pd_x \delta y_s(x,t)} 
	= \lp \pd_y \delta \phi \rp \lp x, y_s^{(0)}(x,t),t \rp  
	\nn + \thisisblue{ \delta y_s(x,t) \lp \pd_y^2 \phi^{(0)} \rp \lp x, y_s^{(0)}(x,t),t \rp} . 
\end{align}
Using the Laplace equation ~\eqref{eqD:Cont}, the sum of the three terms inside blue boxes becomes (after putting the third one on the left-hand side)
\begin{align}
\lp \pd_y v_x^{(0)} \rp \lp \pd_x y_s^{(0)} \rp \delta y_s 
	+ v_x^{(0)} \pd_x \delta y_s 
	+ \lp \pd_x v_x^{(0)} \rp \delta y_s
= \frac{\dd}{\dd x} \lp v_x^{(0)} \delta y_s \rp,
\end{align}
where everything is evaluated at the free surface $y = y_s^{(0)}(x,t)$. 
\eqref{eqD:int2} thus becomes
\begin{align}
\pd_t \delta y_s + \frac{\dd}{\dd x} \lp v_x^{(0)} \delta y_s \rp + \lp \lp \pd_x y_s^{(0)} \rp \pd_x - \pd_y \rp \delta \phi = 0.
\end{align}
Using \eqref{eqD:deltay}, we obtain
\begin{align} \label{eqD:int3}
\left[
- \lp \pd_t + \frac{\dd}{\dd x} \times v_x^{(0)} \rp \frac{1}{g+\frac{1}{2} \pd_y \vec{v}^{(0)2}} \lp \pd_t + v_x^{(0)} \times \frac{\dd}{\dd x} \rp \delta \phi
+ \lp \lp \pd_x y_s^{(0)} \rp \pd_x - \pd_y \rp \delta \phi
\right]_{y = y_s^{(0)}}
=0,
\end{align}
where \deft{the operations are evaluated from right to left} and where \deft{``$\times$'' denotes the multiplication in the sense of operators acting} \deft{on functions}.
That is, if $f$ and $h$ are functions of some variables including $\lambda$, 
\begin{align}
\lp \pd_\lambda \times f \rp h = \pd_\lambda f h = \pd_\lambda \lp f h \rp.
\end{align}

To go further, it will be useful to work in hodograph coordinates $\lp \phi^{(0)}, \psi^{(0)} \rp$, where $\psi^{(0)}$ is the stream function of the background solution, defined by
\begin{align}
\left\lbrace
\begin{array}{l}
\pd_y \psi^{(0)} = v_x^{(0)} \\
\pd_x \psi^{(0)} = - v_y^{(0)}
\end{array}
\right. .
\end{align}
The Laplace equation ensures that it is a good definition.
Notice that $\psi^{(0)}$ is constant along a streamline, i.e., $\psi^{(0)}(x + \delta x, y + \delta y, t) = \psi^{(0)}(x, y, t) + O \lp \delta x^2 \rp$ if $\delta y / \delta x = v_y^{(0)} / v_x^{(0)}$. 
$\psi^{(0)}$ is thus a constant along the unperturbed free surface and along the bottom.
Without loss of generality (as $\psi^{(0)}$ is only defined up to an additive constant), we can choose $\defe{\psi^{(0)}(x,y_b(x)) = 0}$. 
We call $\defe{\psi_s^{(0)} \equiv \psi^{(0)}(x,y_s^{(0)}(x))}$. 
The variation of $\phi^{(0)}$ along the free surface satisfies
\begin{align}
\delta \phi^{(0)} = v_x^{(0)} \delta x + v_y^{(0)} \delta y = v_x^{(0)} \delta x + \frac{v_y^{(0)2}}{v_x^{(0)}} \delta x = \frac{\vec{v}^{(0)2}}{v_x^{(0)}} \delta x.
\end{align}
So, 
\begin{align}
\frac{\dd}{\dd x} = \frac{\vec{v}^{(0)2}}{v_x^{(0)}} \pd_{\phi^{(0)}}.
\end{align}
We also have the relations 
\begin{align} \label{eq:probing:rel_der}
\left\lbrace
\begin{array}{l}
\pd_x = v_x^{(0)} \pd_{\phi^{(0)}} - v_y^{(0)} \pd_{\psi^{(0)}} \\
\pd_y = v_y^{(0)} \pd_{\phi^{(0)}} + v_x^{(0)} \pd_{\psi^{(0)}}
\end{array}
\right. .
\end{align}
So,
\begin{align}
\lp \pd_x y_s^{(0)} \rp \pd_x - \pd_y =
\frac{v_y^{(0)}}{v_x^{(0)}} \pd_x - \pd_y =
- \lp \frac{v_y^{(0)2}}{v_x^{(0)}} + v_x^{(0)} \rp \pd_{\psi^{(0)}} = 
- \frac{\vec{v}^{(0)2}}{v_x^{(0)}} \pd_{\psi^{(0)}}. 
\end{align}
The wave equation \eqref{eqD:int3} thus becomes
\begin{align}
\left[
\lp \pd_t + \frac{\vec{v}^{(0)2}}{v_x^{(0)}} \pd_{\phi^{(0)}} \times v_x^{(0)} \rp \frac{1}{g+\frac{1}{2} \pd_y \vec{v}^{(0)2}} \lp \pd_t + \vec{v}^{(0)2} \pd_{\phi^{(0)}} \rp \delta \phi
+ \frac{\vec{v}^{(0)2}}{v_x^{(0)}} \pd_{\psi^{(0)}} \delta \phi
\right]_{\psi = \psi_s^{(0)}}
=0.
\end{align}
Multiplication by $v_x^{(0)}/\vec{v}^{(0)2}$ gives
\begin{align} \label{eqD:int4}
\smallres{
\left[
\lp \pd_t + \pd_{\phi^{(0)}} \times \vec{v}^{(0)2} \rp \frac{v_x^{(0)}/\vec{v}^{(0)2}}{g+\frac{1}{2} \pd_y \vec{v}^{(0)2}} \lp \pd_t + \vec{v}^{(0)2} \times \pd_{\phi^{(0)}} \rp \delta \phi
+ \pd_{\psi^{(0)}} \delta \phi
\right]_{\psi = \psi_s^{(0)}}
=0}.
\end{align}
To obtain a $(1+1)$-dimensional partial differential equation, we now need to relate $\pd_{\psi^{(0)}}$ to $\pd_{\phi^{(0)}}$. 
To this end, we use that 
\begin{align}
\pd_x^2 + \pd_y^2 = \vec{v}^{(0)2} \lp \pd_{\phi^{(0)}}^2 + \pd_{\psi^{(0)}}^2 \rp,
\end{align}
which can be derived from \eqref{eq:probing:rel_der} using the irrotationality and incompressibility conditions. 
Assuming \higt{the velocity does not vanish}, $\delta \phi$ thus still satisfies the Laplace equation in hodograph coordinates. 
Looking for solutions which are bounded for $\phi^{(0)} \in \mathbb{R}$, we can thus write
\begin{align}
\delta \phi \lp \phi^{(0)}, \psi^{(0)}, t \rp = 
	\int_{\mathbb{R}} \lp \mathcal{A}_k(t) \cosh \lp k \psi^{(0)} \rp + \mathcal{B}_k(t) \sinh \lp k \psi^{(0)} \rp \rp \e^{\ii k \phi^{(0)}} \dd k.
\end{align}
The boundary condition \eqref{eqD:bott} gives $\pd_{\psi^{(0)}} \delta \phi \lp \phi^{(0)}, 0 \rp = 0$, so
\begin{align}
\forall \phi^{(0)} \in \mathbb{R}, \, \int_{\mathbb{R}} k \, \mathcal{B}_k \, \e^{\ii k \phi^{(0)}} \dd k = 0. 
\end{align}
Taking the Fourier transform, and assuming \higt{$\mathcal{B}_k$ is regular at $k = 0$} gives~\footnote{Notice that a $\delta$ singularity would not contribute to $\delta \phi$ as $\sinh \lp k \psi^{(0)} \rp$ identically vanishes for $k = 0$.}
\begin{align}
\forall k \in \mathbb{R}, \, \mathcal{B}_k = 0. 
\end{align}
So,
\begin{align}
\delta \phi \lp \phi^{(0)}, \psi^{(0)}, t \rp = 
	\int_{\mathbb{R}} \mathcal{A}_k(t) \cosh \lp k \psi^{(0)} \rp \e^{\ii k \phi^{(0)}} \dd k.
\end{align}
From this,
\begin{align}
\pd_{\psi^{(0)}} \delta \phi = 
	\int_{\mathbb{R}} k \mathcal{A}_k(t) \sinh \lp k \psi^{(0)} \rp \e^{\ii k \phi^{(0)}} \dd k 
	= \ii \pd_{\phi^{(0)}} \tanh \lp \ii \psi^{(0)} \pd_{\phi^{(0)}} \rp \delta \phi. 
\end{align}
Using this, \eqref{eqD:int4} becomes
\begin{align}
\rese{
\left[
\lp \pd_t + \pd_{\phi^{(0)}} \times \vec{v}^{(0)2} \rp \frac{v_x^{(0)}/\vec{v}^{(0)2}}{g+\frac{1}{2} \pd_y \vec{v}^{(0)2}} \lp \pd_t + \vec{v}^{(0)2} \times \pd_{\phi^{(0)}} \rp \delta \phi
+ \ii \pd_{\phi^{(0)}} \tanh \lp \ii \psi_s^{(0)} \pd_{\phi^{(0)}} \rp \delta \phi
\right]_{\psi = \psi_s^{(0)}}
=0},
\end{align}
which can be re-expressed in terms of the $x,y$ coordinates as
\begin{align}
\lp \pd_t + \frac{\dd}{\dd x} \times v_x^{(0)} \rp \lp g+\frac{1}{2} \pd_y \vec{v}^{(0)2} \rp^{-1} \lp \pd_t + v_x^{(0)} \times \frac{\dd}{\dd x} \rp \delta \phi_s
+ \ii \frac{\dd}{\dd x} \tanh \lp \ii \psi_s^{(0)} \frac{v_x^{(0)}}{\vec{v}^{(0)2}} \frac{\dd}{\dd x} \rp \delta \phi_s = 0,
\end{align}
where all the quantities are evaluated along the unperturbed free surface $y = y_s^{(0)}$ and 
\begin{align}
\delta \phi_s: \binom{\mathbb{R}^2 \to \mathbb{R}}{(x,t) \mapsto \delta \phi \lp x, y_s^{(0)}(x), t \rp}.
\end{align}
Under the approximations made at the beginning of subsection~\ref{sub:Weadr}, $\pd_y \vec{v}^{(0) 2}$ is negligible before $g$, $\psi_s^{(0)} v_x^{(0)} / \vec{v}^{(0) 2}$ is approximately equal to the water depth $h$, and $v_x^{(0)} \approx v^{(0)}$. 
We thus obtain \eqref{eq:waveeq} (where the superscript ``$(0)$'' was dropped to simplify the notations).

\section{Linearly growing modes on a modulated subcritical water flow (Work in progress)}
\label{sec:scat_und}

In this section we present a few preliminary results on the propagation of water waves over an undulation. 
As shown in subsection~\ref{sec:sub}, the latter does not seem to strongly affect the scattering coefficients when its extension is sufficiently short, say of a few wavelengths, at least in part of the relevant domain of parameter space. 
However, we showed numerically in~\cite{Busch:2014hla} for white hole flows in Bose-Einstein condensates that the scattering on the undulation can qualitatively change the results when it is longer and when the flow is transcritical. 
In particular, our study suggests that, in the limit where it is infinitely long, the undulation completely suppresses the low-frequency Hawking effect. 
The choice of Bose-Einstein condensates was motivated by the simplicity of the nonlinear theory: while the nonlinear, stationary solutions of the Gross-Pitaevskii equation can be easily obtained (either analytically in simple models like the step-like one of Chapter~\ref{ch:saturation} or numerically for smooth potentials), computing nontrivial nonlinear solutions of the hydrodynamic equations is significantly more involved. 
However, we verified numerically that the aforementioned result still hold for water waves.  
The reason seems to be that the undulation has (to leading order) the same wavelength as the zero-frequency dispersive modes. 
This induces a resonance between hydrodynamic and dispersive modes, which is apparently strong enough to suppress the Hawking effect. 
More recently, in~\cite{Euve:2015vml} we found that a long undulation in a subcritical flow could strongly enhance the mode conversion for $\om < \om_{\rm min}$, which can explain why the observed values of $\abs{\beta_\om}$ are larger than those predicted by numerical simulations done with a flat downstream region.  

In this section, we show a few analytical results which may serve as a first step to understand these observations. 
We shall see that there is an interesting interplay between the structure of non-linear solutions and the propagation of linear waves over an undulation, leading to a {\it linearly growing oscillating (LGO) mode} which becomes dominant in the limit of long undulations. 
We first give a general argument for the existence of LGO modes and discuss their relations with the nonlinear stationary solutions. 
Then we focus on the KdV equation, for which the structure of the latter solutions is well known and the LGO mode can be easily determined. 
Finally, we give the structure of the calculation for the ``full'' hydrodynamic equations under the assumptions of subsection~\ref{sub:waveeqder}. 
Our aim here is only to show the existence of at least one LGO. 
The detailed calculations, as well as a discussion of its effects on the scattering, will be given in a cfuture article, which we hope to complete soon. 

\subsection{General argument}
\label{subS:gen}

Let us first explain why LGO modes are expected in a large class of systems, including water waves and Bose-Einstein condensates. 
For simplicity, let us assume the system under consideration can be described by a scalar field $\phi$ in ($1+1$) dimensions, solution of a known partial differential equation. 
We also assume the latter has a series of stationary solutions of the form
\begin{align} \label{eqS:ansatz}
\phi_A(x,t) = f_A \lp k(A) x \rp,
\end{align} 
where $A$ van take values in some open interval of $\mathbb{R}$ and where, for each value of $A$, $f_A$ is a smooth, nonuniform, periodic function with period $2 \pi$ and $k(A) \in \mathbb{R}$. 

To proceed, we first notice that $\pd_A \phi_A$ is locally a solution of the linear problem over the solution $\phi_A$. 
Indeed, we have
\begin{align}
\forall x, \, \forall t, \, \phi_{A + \delta A}(x,t) \mathop{=}_{\delta A \to 0} \phi_A(x,t) + \delta A \, \pd_A \phi_A(x,t) + O \lp \delta A^2 \rp, 
\end{align}
which shows that $\pd_A \phi_A$ satisfies the definition of a linear solution. 
Using \eq{eqS:ansatz}, one finds
\begin{align}
\pd_A \phi_A(x,t) = \lp \pd_A f_A \rp \lp k(A) x \rp + k'(A) x \, f_A' \lp k(A) x \rp. 
\end{align}
The first term in this expression is periodic with period $2 \pi / k(A)$ (or constant if $k(A) = 0$), and thus bounded. 
But the second one oscillates with an amplitude linearly growing in $x$ (provided $k'(A) \neq 0$). 
It is thus a LGO mode. 
From this simple argument, one expects LGO modes to exists whenever the wavelength of oscillating, stationary, nonlinear solutions depends on their amplitude (or any other parameter of the solution, such as the mean water depth). 
This is the case, for instance, for the Gross-Pitaevskii equation (see Chapter~\ref{ch:saturation}) as well as the KdV equation (see subsection~\ref{subS:KdV} below). 
We shall see in subsection~\ref{subS:Ww} that this still holds for a more realistic model of water waves. 

In a sense, the presence of these LGO modes signals a breakdown of the linear approximation far from $x = 0$. 
This can be traced back to the fact that the difference between two nearby nonlinear solutions with different wave vectors can not remain in the linear domain on the whole real axis. 
Indeed, if their wave vectors differ by $\delta k \neq 0$, and if their phases are chosen to coincide for $x \approx 0$ (so that their relative difference is small in this region), the latter will differ by a term of order $\pi$ for values of $x$ of the order of $\pi / \delta k$. 
The difference between these solutions is thus not small anymore and requires nonlinear terms to be accurately described.

\subsection{Case of the KdV equation}
\label{subS:KdV}

In this subsection we consider a simple model based on the KdV equation to see how the above argument works in practice. 
Let us denote by $\eta$ the elevation of the free surface with respect to some reference background solution. 
We assume it satisfies the KdV equation
\begin{align} \label{eqS:KdV}
\pd_t \eta  + v \pd_x \eta + \pd_x^3 \eta + 6 \eta \pd_x \eta = 0,
\end{align}
with $v > 0$. 
We will first determine the LGO mode using the above argument on stationary solutions. 
We shall then recover it using only the linearized KdV equation, to show explicitly that it is a solution of this equation. 

Looking for stationary solutions $\pd_t \eta = 0$, \eq{eqS:KdV} becomes
\begin{align}
\pd_x \lp  v \eta + \pd_x^2 \eta + 3 \eta^2 \rp = 0.
\end{align}
Integrating over $x$ gives
\begin{align}
\pd_x^2 \eta + C_1 + v \eta + 3 \eta^2 = 0,
\end{align}
where $C_1$ is an integration constant. 
At fixed $C_1 < v^2/12$, there exist two real homogeneous solutions, given by
\begin{align}
\eta_{\pm} = \frac{-v \pm \sqrt{v^2 - 12 C_1}}{6}. 
\end{align}
As only $\eta_+$ goes to zero as $C_1 \to 0$, we look for ``small'' solutions of the form
\begin{align}
\eta(x) = \eta_+(x) + \delta \eta(x).
\end{align}
The equation on $\delta \eta(x)$ reads
\begin{align}
\pd_x^2 \delta \eta + \lp v + 6 \eta_+ \rp \, \delta \eta + 3 \delta \eta^2 = 0.
\end{align}
For the present purposes, it will be enough to work at linear order in $\delta \eta$. 
We thus obtain the following two-parameters series of solutions:
\begin{align}\label{eqS:int1}
\eta_{C_1,A}(x) = \frac{\sqrt{v^2 - 12 C_1} - v}{6} + A \, \cos \lp \lp v^2 - 12 C_1 \rp^{1/4} x \rp + O \lp A^2 \rp.
\end{align}
We want to find a LGO mode over the simple undulation solution $\eta_{0,A}$. 
To this end, we differentiate \eqref{eqS:int1} with respect to $C_1$ or, as will be more convenient here, some function of this parameter:~\footnote{One can also obtain an important LGO mode through differentiation with respect to $A$, but this requires going to third order in the undulation amplitude. This will be explained in details in the forthcoming article.}
\begin{align}
\partial_{\lp v^2 - 12 C_1 \rp^{1/4}} \eta_{C_1, A} = \frac{\lp v^2 - 12 C_1 \rp^{1/4}}{3} - A \, x \, \sin \lp \lp v^2 - 12 C_1 \rp^{1/4} x \rp + O \lp A^2 \rp.
\end{align}
Taking the limit $C_1 \to 0$ and multiplying by $3/\sqrt{v}$, one obtains the LGO mode
\begin{align} \label{eqS:LGOKdV}
\delta \eta_{\rm LGO}(x) = 1 - \frac{3 A}{\sqrt{v}} x \sin \lp \sqrt{v} x \rp + O \lp A^2 \rp. 
\end{align}

Let us check explicitly that it is indeed a solution of the linear KdV equation over $\eta_{0,A}$. 
This equation reads
\begin{align} \label{eqS:int2}
\lp v \pd_x + \pd_x^3 \rp \delta \eta (x) + 6 \pd_x \lp \eta_{0,A}(x) \, \delta \eta(x) \rp = 0.  
\end{align}
We look for a solution of the form
\begin{align}
\delta \eta (x) = 1 + \delta \delta \eta(x), \, \abs{\delta \delta \eta} = O(A).
\end{align}
\eqref{eqS:int2} then becomes 
\begin{align}
\lp v \pd_x + \pd_x^3 \rp \delta \delta \eta (x) = - 6 \eta_{0,A}'(x) + O \lp A^2 \rp
\end{align}
\begin{align} \label{eqS:int3}
\lp v \pd_x + \pd_x^3 \rp \delta \delta \eta (x) = 6 A \sqrt{v} \sin \lp \sqrt{v} x \rp + O \lp A^2 \rp.
\end{align}
We look for a solution of the form
\begin{align}
\delta \delta \eta (x) = B x \sin \lp \sqrt{v} x \rp.
\end{align}
\eqref{eqS:int3} then becomes
\begin{align}
B \lp v \sin \lp \sqrt{v} x \rp - 3 v \sin \lp \sqrt{v} x \rp \rp = 6 A \sqrt{v} \sin \lp \sqrt{v} x \rp + O \lp A^2 \rp.
\end{align}
\begin{align}
- 2 v B \sin \lp \sqrt{v} x \rp = 6 A \sqrt{v} \sin \lp \sqrt{v} x \rp + O \lp A^2 \rp.
\end{align}
\begin{align}
B = - 3 \frac{A}{\sqrt{v}} + O \lp A^2 \rp.
\end{align}
We thus recover \eqref{eqS:LGOKdV}. 

\begin{figure}
\centering
\includegraphics[width=0.49\linewidth]{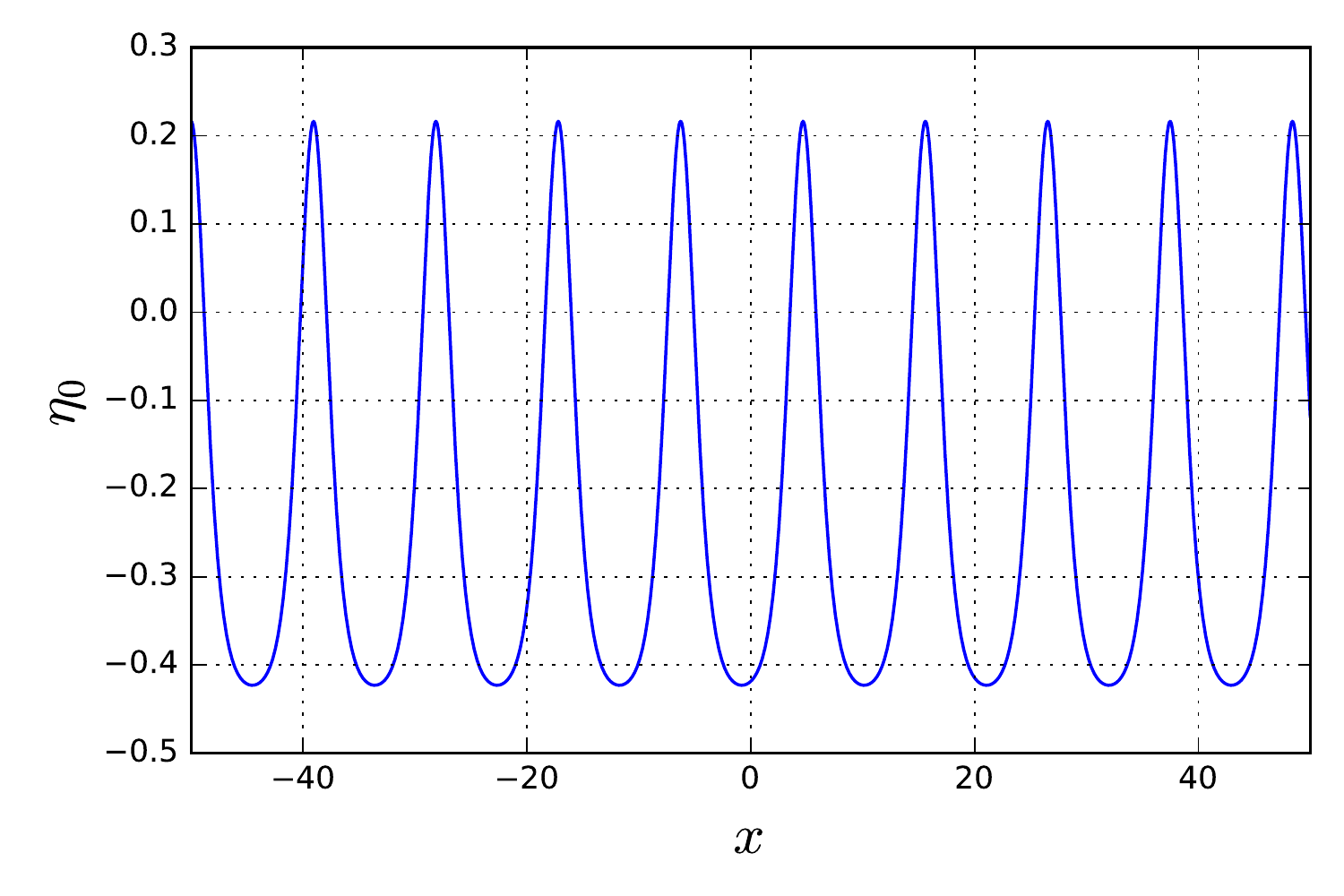}
\includegraphics[width=0.49\linewidth]{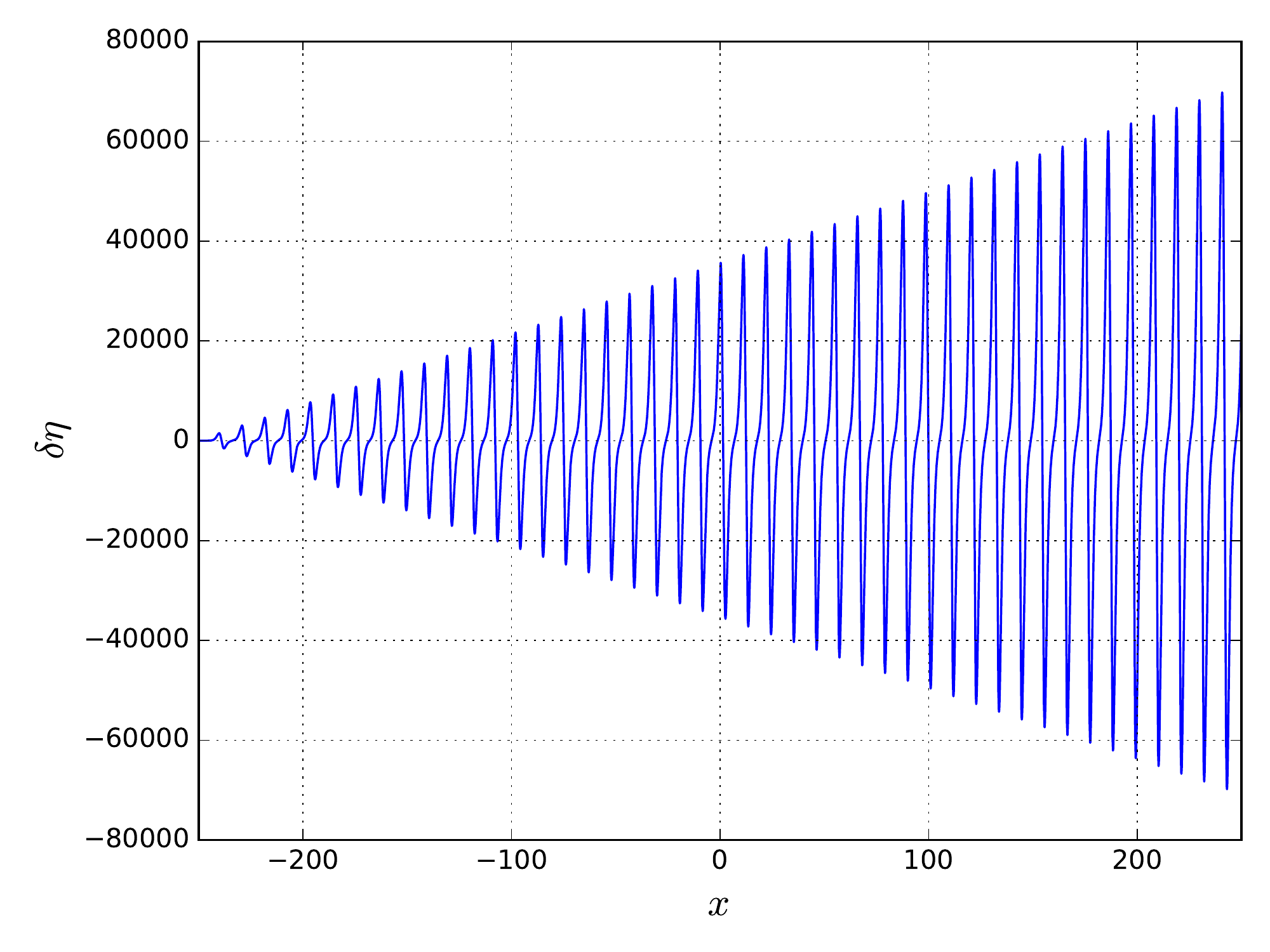}
\caption{Left: A nonlinear, stationary, periodic solution with a large amplitude for $v = 1.3$.
Right: Linearly growing, oscillating mode over this periodic solution.  
}
\label{fig:nlkuun}
\end{figure}

The existence of this mode has deep implications for the scattering of incoming waves on the undulation. 
In particular, one can show that at small frequencies an incident hydrodynamic wave over an undulation of size $L \gg 1 / \sqrt{v}$ generates dispersive waves with an amplitude proportional to $A \, L$. 
Performing the same calculation to next order in $A$ (which actually requires the shape of the undulation to third order) shows that sending a dispersive waves will also generate a larger one on the other side, with an amplitude proportional to $A^2 L$. This will be explained in detail in the future publication.
The LGO mode over a large-amplitude undulation is shown in \fig{fig:nlkuun}. 

\subsection{Water waves}
\label{subS:Ww}

We now consider the model of Section~\ref{sub:waveeqder}. 
To avoid technical difficulties, we shall give only the fully linear calculation, which is here simpler than using the argument of subsection~\ref{subS:gen}. 
Let us define $\delta \eta \equiv \dd \delta \phi / \dd \phi_0$. It satisfies the equation:
\begin{align}
\frac{\dd}{\dd \phi} \lp \frac{v_x v^2}{g + \frac{1}{2} \pd_y v^2} \delta \eta \rp
	+ \ii \tanh \lp \ii J \frac{\dd}{\dd \phi} \rp \delta \eta = 0,
\end{align}
where $J$ is the conserved 2D current.
We consider an undulation of the form 
\begin{align}
h(x) = h_0(x) + A \, \sin \lp k_0 x \rp + O \lp \frac{A^2}{h_0} \rp,
\end{align}
where $k_0$ is the strictly positive root of the dispersion relation for $\om = 0$,
and look for solutions of the form
\begin{align}
\delta \eta(\phi) = 1 + \delta \delta \eta(\phi), \, \delta \delta \eta(\phi) = O \lp \frac{A}{h_0} \rp.
\end{align}
The above equation becomes
\begin{align}
\frac{\dd}{\dd \phi} \lp \frac{v_x v^2}{g + \frac{1}{2} \pd_y v^2} \delta \delta \eta \rp
	+ \ii \tanh \lp \ii J \frac{\dd}{\dd \phi} \rp \delta \delta \eta 
	= - \frac{\dd}{\dd \phi} \frac{v_x v^2}{g + \frac{1}{2} \pd_y v^2}.
\end{align}
After a straightforward calculation, this may be rewritten as
\begin{align}
\frac{v_0^3}{g}  \frac{\dd}{\dd \phi} \delta \delta \eta
	+ \ii \tanh \lp \ii J \frac{\dd}{\dd \phi} \rp \delta \delta \eta 
	= \lp 3 k_0 - \frac{v_0^4 k_0^3}{g^2} \rp A \cos \lp \frac{k_0}{v_0} \phi \rp
	+ O \lp \frac{A^2}{h_0^2} \rp.
\end{align}
We look for a solution of the form
\begin{align}
\delta \delta \eta(\phi) = B \, \phi \cos \lp \frac{k_0}{v_0} \phi \rp.
\end{align}
Then,
\begin{align}
\lp \frac{v_0^3}{g} \frac{\dd}{\dd \phi} + \ii \tanh \lp \ii J \frac{\dd}{\dd \phi} \rp \rp \delta \delta \eta( \phi ) = 
	- B \lp \frac{v_0^3}{g} \frac{k_0}{v_0} - \tanh \lp J \frac{k_0}{v_0} \rp \rp \phi \sin \lp \frac{k_0}{v_0} \phi \rp
	\nn + B \lp \frac{v_0^3}{g} - J \lp 1 - \tanh^2 \lp J \frac{k_0}{v_0} \rp \rp \rp \cos \lp \frac{k_0}{v_0} \phi \rp.
\end{align}
To get this expression, we used that for $f_n: X \mapsto X^n$, $n \in \mathbb{N}^*$,
\begin{align}
f_n \lp \frac{\dd}{\dd \phi} \rp \lp \phi \e^{i k \phi} \rp = f_n \lp \ii k \rp \phi \, \e^{\ii k \phi} + n (\ii k)^{n-1} \e^{\ii k \phi}
= f_n \lp \ii k \rp \phi \, \e^{\ii k \phi} + f_n' \lp \ii k \rp \e^{\ii k \phi},
\end{align}
so that, for any polynomial function $f$, 
\begin{align}
f \lp \frac{\dd}{\dd \phi} \rp \lp \phi \, \e^{\ii k \phi} \rp 
= f \lp \ii k \rp \phi \, \e^{\ii k \phi} + f' \lp \ii k \rp \e^{\ii k \phi}.
\end{align}
Assuming $f$ is odd, this gives
\begin{align}
f \lp \frac{\dd}{\dd \phi} \rp \lp \phi \cos \lp k \phi \rp \rp 
= \ii f \lp \ii k \rp \phi \sin \lp k \phi \rp + f' \lp \ii k \rp \cos \lp k \phi \rp.
\end{align}
Using 
\begin{align}
\tanh \lp J \frac{k_0}{v_0} \rp = \frac{v_0^2 k_0}{g}
\end{align}
gives
\begin{align}
\lp \frac{v_0^3}{g} \frac{\dd}{\dd \phi} + \ii \tanh \lp \ii J \frac{\dd}{\dd \phi} \rp \rp \delta \delta \eta( \phi ) = 
	B \lp \frac{v_0^3}{g} - J + \frac{J v_0^4 k_0^2}{g^2} \rp \cos \lp \frac{k_0}{v_0} \phi \rp
	+ O \lp A^2 \rp.
\end{align}
We thus obtain to leading order
\begin{align}
B = \frac{3 k_0 - \frac{v_0^4 k_0^3}{g^2}}{\frac{v_0^3}{g} - J + \frac{J v_0^4 k_0^2}{g^2}} A
\end{align}
Using this, one obtains
\begin{align}
\rese{
\delta \eta_{\rm LGO}(\phi) = 1 + \frac{3 - \frac{v_0^4 k_0^2}{g^2}}{\frac{v_0^3}{g} - v_0 h_0 + \frac{v_0^5 h_0 k_0^2}{g^2}} k_0 A \phi \cos \lp \frac{k_0}{v_0} \phi \rp + O \lp \frac{A^2}{h_0^2} \rp
} \, . 
\end{align}
As was mentioned in the previous subsection for the KdV equation, this LGO mode has important implications for the scattering of incident waves on a long undulation. 
These will be explained in the future publication. 

\newchapter{Black hole radiation in the presence of a universal horizon}
\label{ch:universal}
\begin{tikzpicture}[overlay]
\newcommand*{\xA}{-0.2}
\newcommand*{\xB}{13.45}
\newcommand*{\yA}{5.5}
\newcommand*{\yB}{1.5}
\newcommand*{\epsil}{0.75}
\draw[overlay] (\xA-\epsil,\yA) -- (\xB-\epsil,\yA);
\draw[overlay] (\xA,\yA+\epsil) -- (\xA,\yB);
\draw[overlay] (\xB,\yB-\epsil) -- (\xB,\yA);
\draw[overlay] (\xB+\epsil,\yB) -- (\xA+\epsil,\yB);
\end{tikzpicture}
\begin{small}
The previous three chapters dealt with ``analogue'' models, where dispersive effects come from the breakdown of an effective local Lorentz invariance at high energy. 
In the present one, based on~\cite{Michel:2015rsa}, we consider modified theories of gravity which include dispersion explicitly. 
We focus on two such theories, namely Ho\v{r}ava gravity and Einstein-\AE ther theory. 
As we shall see, most of the basic concepts, such as the dispersion relation, conserved inner product, and relations between in and out modes at the root of the Hawking effect, are closely related to those of ``analogue'' models. 
In fact, the only qualitative difference comes from the presence of a new type of horizon in the inside (``supercritical'' in the analogue language) region, at which the local dispersive frequency scale goes to zero. 
This horizon, called ``universal'', acts as separatrix for outgoing modes. 
Indeed, because of the vanishing dispersive scale, all the modes either become relativistic -- and thus unable to move against the supercritical flow -- or get infinitely blueshifted when approaching the universal horizon. 

Another way to reach the same conclusion is to look at the structure of the wave equation. 
To retain the standard Hamiltonian structure, it is common to impose the existence of a ``preferred'' coordinate system where the equations of motion remain second-order in time. 
The ``preferred time'' thus defined can be shown to be always increasing when crossing the universal horizon from the outside to the inside, and decreasing when crossing it in the opposite direction. 
As the theory must be causal in this preferred time, an event localized inside the universal horizon can not influence the outside region. 

From these observations, universal horizons seem to play a role very similar to that of null horizons in General Relativity (without dispersion). 
It is thus tempting to expect that they should also determine the properties of Hawking radiation in such theories, i.e., that the late-time radiation should be essentially due to the universal horizon, not the Killing one. 
However, two arguments put this conjecture into question:
\begin{itemize}
\item First, numerous works in analogue gravity have shown that a null horizon does produce an approximately thermal radiation at low energy, even in the presence of high-frequency dispersion.
\item Second, while the wave vector of outgoing modes experience an infinite blueshift close to the universal horizon, as do relativistic modes close to a null horizon, their wave vectors have different quantitative behaviors.  
Indeed, in the relativistic case, the divergence is logarithmic in the affine coordinate $r$, while it follows an inverse power law for dispersive fields close to the universal horizon.
\end{itemize}

The aim of the present chapter is to clarify this issue by an explicit calculation of the late-time radiation emitted by a static black hole with a universal horizon in the presence of superluminal dispersion. 
We consider the case where the group velocity is unbounded, i.e., $\abs{\pd_\om k_\om}$ goes to infinity when $\om \to \infty$. 
To define the vacuum state of the field unambiguously, we work with a collapsing shell geometry, the radius of which is sent to infinity at early times $t \to - \infty$. 
Inside the shell, the metric is flat and the state of the field is vacuum. 
As we shall see, in this model, low-frequency radiation from the universal horizon is exponentially suppressed by a positive power of the frequency $\omega$ inside the shell. 
This suppression is directly related to the behavior of the wave vector close to the horizon. 
Because of the blue-shift at the universal horizon, this implies that the emission is not stationary, and that it vanishes in the late-time limit.  
We also compute numerically the emission spectrum from the null horizon, and we find that it is approximately thermal at low energy. 
This establishes that, under the approximations and hypotheses of our model, it is the null horizon, not the universal one, which determines the properties of the Hawking radiation. 

These hypotheses have been questioned in~\cite{Ding:2015fyx,Lin:2016myf,Cropp:2016gkn}, where it is pointed out that more regular models might give a different result. 
To our knowledge, the question of whether this is the case or not is currently unsolved. 
We hope to return to this interesting problem in the future, which could shed light both on the Hawking radiation and on the notion of regularity of the quantum state in the presence of a universal horizon.

\end{small}
\newpage

\renewcommand*{\theHsection}{\theHchapter.\the\value{section}}
\renewcommand\thesection{\arabic{section}}

\section{Introduction}
 
The laws of black hole thermodynamics are firmly established in Lorentz-invariant theories, and they play a crucial role in our understanding of black hole physics~\cite{wald1994quantum}. 
An important point is that the entropy and temperature are both governed by the properties of the event horizon, which leads to the second law. 
In Lorentz-violating theories, the status of these laws is less clear because essential ingredients of their derivations are no longer present~\cite{Jacobson:2001kz, Dubovsky:2006vk, Jacobson:2008yc, Betschart:2008yi, Blas:2011ni, Busch:2012ne}. 
For instance, the thermality of the Hawking flux is inevitably lost in the presence of high-frequency dispersion, although it is approximatively recovered for large black holes for which the surface gravity $\kappa$ is much smaller than the high-energy scale $\Lambda$ of the dispersion~\cite{Macher:2009tw}. 

These difficulties can be traced to the fact that the event horizon no longer separates the outgoing field configurations into two disconnected classes as in Lorentz-invariant theories: when the dispersion is superluminal, this horizon can be crossed by outgoing radiation. However, it was recently discovered that in some theories of modified gravity such as Ho\v{r}ava gravity~\cite{Horava:2009uw, Sotiriou:2010wn, Janiszewski:2014iaa} and Einstein-\ae ther~\cite{Jacobson:2000xp, Eling:2004dk, Eling:2006ec, Barausse:2011pu}, spherically-symmetric black hole solutions possess a second, inner horizon.~\footnote{However, it was recently found in~\cite{Barausse:2015frm} that this second horizon is absent in rotating black holes in Ho\v{r}ava gravity.}
This horizon, called {\it universal}, can not be crossed by outgoing modes, even for superluminal dispersion relations which allow for arbitrarily large group velocities. (The difficulty mentioned in~\cite{Jacobson:2001kz} is thus evaded.~\footnote{In that work it was shown that, for generic Lorentz-violating theories with superluminal dispersion relations, particles produced by unknown quantum gravity effects close to the singularity of a black hole could propagate outside the horizon and possibly fill the whole universe, making the theory nonpredictive. 
In the presence of a universal horizon, however, such particles must remain in the inside region.}) 
Following this discovery, it has been argued that the universal horizon should play a key role in the thermodynamics of these black holes, and indeed it appears to obey a first law~\cite{Berglund:2012bu, Horavasummary}. 
However, an important remaining question concerned the temperature of the Hawking radiation these black holes emit, and its relation with the first law. 
Could it be essentially governed by the (higher) surface gravity of the universal horizon? Or is it fixed by the null Killing horizon?  

Two recent works suggested that the universal horizon should emit a steady radiation with a temperature governed by the surface gravity of the universal horizon. 
Because of the complicated nature of the field propagation near that horizon, this conclusion was indirectly obtained, in~\cite{Berglund:2012fk}, by making use of a tunneling method and, in~\cite{Cropp:2013sea}, by analyzing the characteristics of the radiation field. 
In the present chapter, we reexamine this question by performing a direct calculation and reach the opposite conclusion that no radiation is emitted from the universal horizon at late time.

We proceed as follows. As in the original derivation of Hawking~\cite{Hawking:1974sw}, we identify the boundary conditions on the outgoing modes in the near vicinity of the universal horizon by considering a simple collapsing shell geometry and by assuming that the state of the field is vacuum inside it. 
We then compute the mode mixing across the shell between inside modes $\phi_\omega^{\rm in}$ propagating outwards and outside stationary modes $\psi_\lambda$ with a fixed Killing frequency. The late-time behavior is obtained by sending the inside frequency $\omega$ to $\infty$. 
In this limit, we show that the scattering coefficients involving modes with opposite norms vanish. This result can be understood from the fact that the modes $\psi_\lambda$ are accurately described by their WKB approximation in the immediate vicinity of the universal horizon. In other words, the pasting across the shell is adiabatic in the limit $\omega \to \infty$. Hence, for large outgoing radial momenta, the state of the field outside the shell is the usual vacuum, as explained in~\cite{Brout:1995wp}.

It then remains to propagate these high-momentum dispersive modes from the universal horizon to spatial infinity. This propagation has already been studied in detail (see~\cite{Coutant:2011in} for a recent update). 
These works have established that large black holes emit a stationary flux which is nearly thermal, with a temperature approximately given by its relativistic value. 
In a sense, the present chapter extends the robustness of the Hawking process, i.e. its insensitivity to high-frequency dispersion -- first established in~\cite{Unruh:1994je} -- to black holes with a universal horizon.

This suggests that the statistical properties of the emitted radiation and the second law of black hole thermodynamics are robust in the low-energy limit, and involve the properties of the null Killing horizon rather than the universal one.~\footnote{At finite energy, however, one can expect deviations from the second law, see \cite{Dubovsky:2006vk,Eling:2007qd,Jacobson:2008yc}.}
An important point to note is that the field configurations propagating on either side of a universal horizon come from two disconnected Cauchy surfaces and are highly blueshifted. 
Hence, it is not clear whether Hadamard condition of regularity~\cite{Busch:2012ne} could be satisfied on the universal horizon. 
This raises the question of the fate of the universal horizon; see~\cite{Blas:2011ni}. This difficult question will not be discussed in the present chapter.

This chapter is organized as follows. 
In Section~\ref{sec:standH}, we briefly review the derivation of~\cite{Hawking:1974sw,Brout:1995rd} and reformulate it to ease the generalization to the case of Lorentz-violating fields. 
In Section~\ref{sec:Horava} we determine the spontaneous emission from the universal horizon in a simple collapsing shell geometry. 
Its results are discussed in Section~\ref{uni:disc}. 
Section~\ref{sec:app_uni} gives additional details on the calculation of Section~\ref{sec:Horava} (see subsection~\ref{app:details}) and some additional remarks: 
In subsection~\ref{app:acc} we compare our model with previously-studied dispersive ones without a universal horizon, and show the role of the acceleration of the preferred frame; subsection~\ref{app:Hawking} shows the results of numerical simulations confirming the approximately thermal character of the emission at infinity governed by the surface gravity of the null Killing horizon. Throughout this chapter, we work in Planck units: $c = \hbar = G = 1$. 

\section{Massless relativistic scalar field in a collapsing shell geometry}
\label{sec:standH}

In this section, we briefly review the computation of the Hawking radiation emitted at late time in a collapsing geometry~\cite{Hawking:1974sw}. 
Although these concepts are well known, we believe it is useful to present them as this will prepare the calculation of the late-time flux when dealing with a dispersive field in the presence of a universal horizon. As explained in the Introduction, we perform a direct calculation which consists of pasting the modes across the infalling shell. We closely follow the derivation of~\cite{Massar:1997en}. 
Here we only underline the main steps of the calculation. More details are given in Chapter~\ref{chap:intro} and in the two above references.

\begin{figure}[ht!]
\centering
\includegraphics[width=0.5\linewidth]{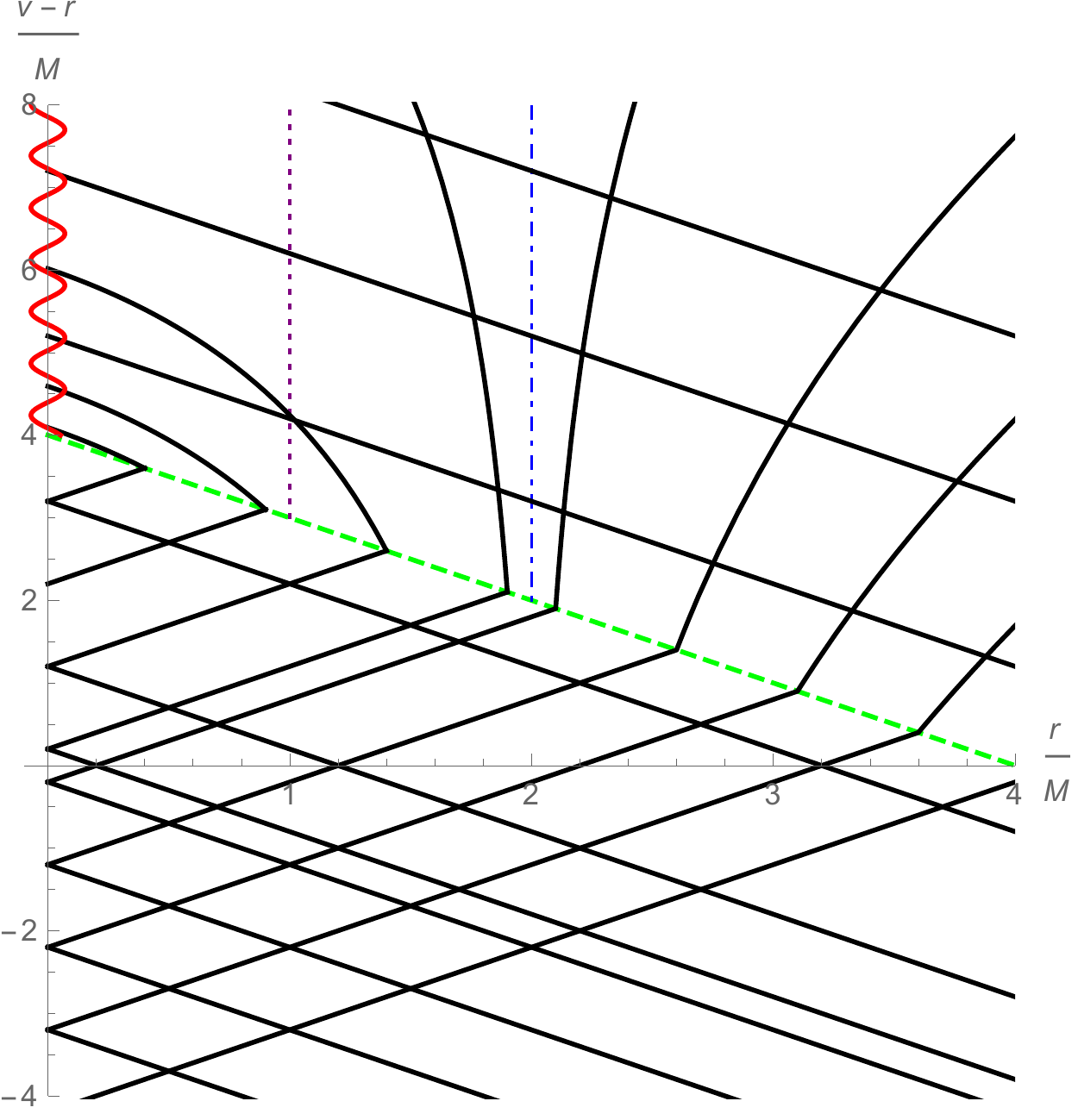} 
\includegraphics[width=0.35\linewidth]{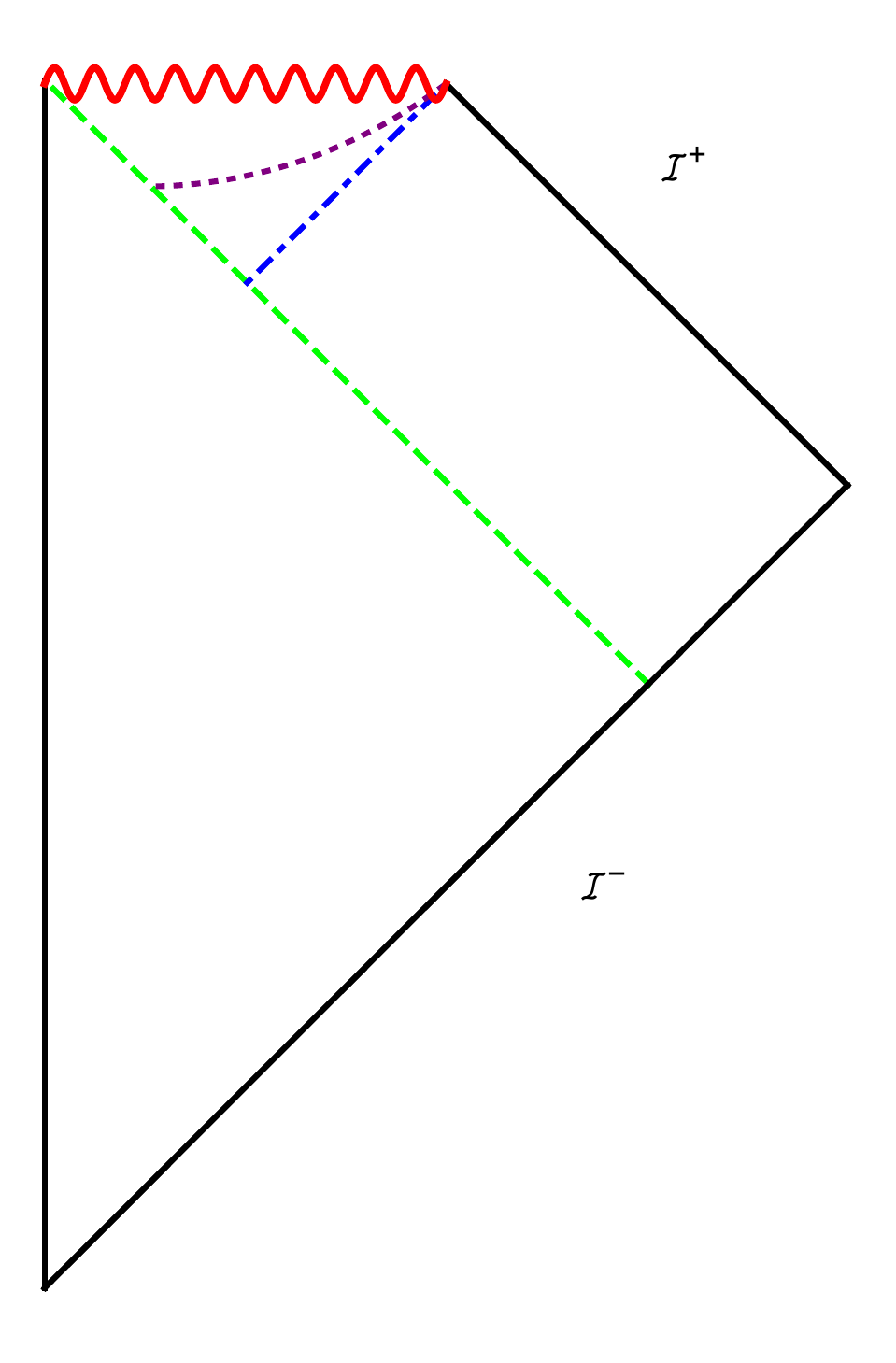}
\caption{Left panel: Null radial geodesics in the $(v-r,r)$ plane in units of $M$. The coordinate $v-r$ coincides with the Minkowski time $T$ inside the mass shell, and with the Schwarzschild time $t_S$ for $r \to \infty$. The solid black lines are null radial geodesics, some of which are reflected on $r=0$. The dashed green line represents the trajectory of the null shell $v = 4M$. The blue, dot-dashed one shows the null horizon at $r=2M$ outside the shell. The dotted purple line shows the locus $r=M$, $v>4M$ which will play a crucial role in Section~\ref{sec:Horava}. The wavy line shows the singularity, located at $r=0, \, v>4M$. Right panel: Penrose-Carter diagram of the collapsing shell geometry. The vertical line corresponds to $r=0, \, v<4M$. $\mathcal{I}^-$ corresponds to $u,U \to -\infty$, and $\mathcal{I}^+$ to $v \to \infty$. 
} \label{fig:RelST} 
\end{figure}

For simplicity, we consider an infalling, spherically symmetric, lightlike thin shell. It is useful to work with advanced Eddington-Finkelstein (EF) coordinates $v,r$, where $v$ is the advanced Killing time. At fixed $r$, one has $\dd v / \dd t_S = 1$, where $t_S$ is the usual Schwarzschild time. Hence, outside the shell, the stationary Killing field $K^\mu \partial_\mu$ is simply $\pd_v$. On both sides of the shell, chosen to be at $v=4M$, the line element reads
\begin{equation}\label{eq:met}
ds^2 = \lp 1-\frac{2M}{r} \Theta (v - 4 M) \rp dv^2 - 2 dv \, dr-r^2 \lp d \theta^2 + \sin(\theta)^2 d \varphi^2 \rp. 
\end{equation}
These coordinates cover the entire space-time, shown in the right panel of \fig{fig:RelST}. On the left panel, the infalling and outgoing null radial geodesics are represented in the $(v-r,r)$ plane. One clearly sees that the null Killing horizon (where $K^\mu K_\mu$ vanishes, hereafter simply referred to as the ``null horizon'') divides the outgoing geodesics into two separate classes. 

Let $\Phi$ be a massless real scalar field with the action
\begin{equation}\label{eq:relativisticaction}
S = \int \dd^4 x \sqrt{-g} \lp \pd_\mu \Phi \rp \lp \pd^\mu \Phi \rp. 
\end{equation}
We define $\psi \equiv r \Phi$ and consider radial solutions independent of $\lp \theta, \varphi \rp$. Inside the shell, for $v < 4M$, we introduce the null outgoing (affine) coordinate $U \equiv v-2r $. Outside the shell and for $r > 2M$,  we introduce the null coordinate
\begin{equation}
u \equiv v-2r_K^*, 
\end{equation}
where $r^*_K \equiv r + 2 M \ln \left\lvert {r}/{2M} - 1 \right\rvert$ is the usual tortoise coordinate, which diverges on the null horizon. To cover the region inside the null horizon, one needs another coordinate $u_L \equiv - (v - 2r_K^*)$.~\footnote{The overall sign guarantees that $dU/du_L$ is positive. As we shall see in Section~\ref{sec:Horava}, a similar sign is needed when studying a dispersive field on both sides of a universal horizon.} The field equation then reads
\begin{equation}
\left\lbrace 
\begin{array}{cc}
\pd_U \pd_v \psi = 0,  & v<4M , \\
\lp \pd_u \pd_v + \lp 1- \frac{2M}{r} \rp \frac{2M}{r^3} \rp \psi = 0 , & v>4M .
\end{array}
\right. 
\end{equation}
For simplicity, we neglect the potential engendering the grey body factor and work with the conformally invariant equations $\pd_U \pd_v \psi = \pd_u \pd_v \psi = 0$. 
Its solutions can be decomposed as
\begin{equation}
\psi(u,v) = \psi^{u}(u) + \psi^v(v), \, v > 4M, 
\end{equation}
and similarly for $v<4M$ with $u$ replaced by $U$. The infalling $v$ sector and the outgoing $u$ sector thus completely decouple. Moreover, the $v$ modes $\psi^v$ are regular across the horizon and play no role in the Hawking effect. We thus consider only the $u$ modes and, to lighten the notations, we no longer write the upper index $u$. 

To compute the global solutions, we need the matching conditions across the null shell. In the present case, $\psi$ must be continuous along $v=4M$. Hence $\psi_{\rm inside}(U) = \psi_{\rm outside}(u(U))$, where the relation between the null coordinates is
\begin{equation}  
u(U) = U - 4 M \ln \lp \frac{ - U }{4M} \rp 
\label{uU}
\end{equation}
for $r > 2M$ ($U < 0$). For $r < 2M$ ($U > 0$), one has $u_L(U) =  - u(|U|)$.

To obtain the Hawking flux one needs to relate the incoming modes $\phi_\omega^{\rm in}$, characterizing the vacuum inside the shell, to the outgoing modes $\phi_\lambda^{\rm out}$ characterizing the asymptotic outgoing quanta with Killing frequency $\lambda$. In the internal region, a complete orthonormal basis of positive-norm modes is provided by the plane waves
\begin{equation}
\phi_\omega^{\rm in} \equiv \frac{\e^{-\ii \omega U}}{2 \sqrt{\pi \omega}},
\label{relat-in-mode}
\end{equation}
where $\omega \in \mathbb{R}^+$ is the inside frequency (eigenvalue of $\ii \partial_U$). In the external region, the (positive-norm) stationary modes for $r > 2M$, are
\begin{equation} 
\phi_\lambda^{\rm out} 
\equiv \Theta(r-2M)\, \frac{\e^{-\ii \lambda u}}{2 \sqrt{\pi \lambda}} , \; \lambda \in \mathbb{R}^+.
\label{relat-out-mode}
\end{equation}
A similar equation defines $\phi_\lambda^{(L)}(u_L)$ in the trapped region $r< 2M$. One easily verifies that the conserved inner product for the $u$ modes can be written as 
\begin{equation}
\lp \psi_1, \psi_2 \rp = \ii \int^\infty_{- \infty} \dd u \lp \psi_1^* \pd_u \psi_2 - \psi_2 \pd_u \psi_1^* \rp. 
\end{equation}
The modes $\phi_\lambda^{\rm out} , \phi_{\lambda'}^{(L)}$ and their complex conjugates form a complete orthonormal basis.

The Bogoliubov coefficients encoding the Hawking flux are then given by the overlaps between the two sets of modes:
\begin{align} 
\alpha_{\lambda, \omega} &= \lp \phi_{\lambda}^{\rm out}, \phi_{\omega}^{\rm in}  \rp, \nn 
\beta_{\lambda, \omega} &= \lp (\phi_{\lambda}^{\rm out} )^{*}, \phi_{\omega}^{\rm in}  \rp.
\label{BogC}
\end{align}
They can be computed explicitly using the relation \eq{uU} between $u$ and $U$, see~\cite{Massar:1997en} for details. The late-time behavior is obtained by sending the inside frequency $\omega$ to $\infty$. In this limit, one recovers the standard thermal result
\begin{equation}
\left\lvert \frac{\beta_{\lambda, \omega}}{\alpha_{\lambda, \omega}} \right\rvert^2 \mathop{\sim}_{\omega \to \infty} \e^{-8 \pi M \lambda},
\label{betovera}
\end{equation}
with
\begin{align} 
\abs{\beta_{\la, \om}}^2 \mathop{\sim}_{\omega \to \infty} \, \frac{M}{2 \pi \om} \frac{1}{\e^{4 \pi r_S \la} - 1}. 
\end{align}

To prepare for the forthcoming analysis, it is instructive to compute the Bogoliubov coefficients by the saddle-point method~\cite{Parentani:1992me,Brout:1995rd}. For the $\alpha_{\lambda, \omega}$ coefficient, when $\omega \gg \lambda$, i.e., at late time, the location of the saddle is given by
\begin{equation}
\lambda = \omega \, 
\lp \e^{-\kappa (u - u_0)} + O(\e^{- 2 \kappa (u- u_0)})\rp 
,  
\label{CarterL}
\end{equation}
where $u_0$ is a constant which drops out of the late-time flux. (With the above conventions, $u_0$ vanishes.) From this equation we recover the time-dependent redshift relating $\omega$, the large frequency of the mode emitted from the collapsing star, to $\lambda$, the frequency received at infinity and measured using the proper time of an observer at rest. In particular, we recover the characteristic exponential law governed by the surface gravity $\kappa = \pd_u \ln U(u) =  \frac{1}{4M}$. Had we considered a collapsing shell following a (regular) infalling timelike curve, \eq{CarterL} would still have been obtained at late $u - u_0$ time. 

This is the kinematical root of the universality of Hawking radiation in relativistic theories. Indeed, when studying the coefficient $\beta_{\lambda, \omega}$, one finds that the saddle point is now located at $\lambda = - \omega \, \e^{-\kappa u_{\rm s.p.}}$. Taking into account the fact that the integration contour should be deformed in the lower $u$-complex plane, one finds that $u_{\rm s.p.}$ has an imaginary part $\Im \lp u_{\rm s.p.} \rp = - \pi / \kappa$, whereas its real part in unchanged. This gives a relative factor $\exp \lp  - \pi \lambda / \kappa \rp$ with respect to the coefficient $\alpha_{\lambda, \omega}$. 
Squaring their ratio, we recover \eq{betovera}. We also recover that the Hawking temperature $\kappa/2\pi$ is fixed by the late-time exponential decay rate entering \eq{CarterL}. We finally notice that the stationarity of the flux is nontrivial. It follows from the fact that the ratio of \eq{betovera} is independent of $\omega$, and from the fact that $|\beta_{\lambda, \omega}|^2 \propto 1/\omega$ for $\omega \to \infty$~\cite{Brout:1995rd}. 

\section{Emission from a universal horizon}
\label{sec:Horava}

\subsection{The model}

We aim to compute the late-time radiation of a dispersive field propagating in a collapsing geometry. In principle, the radiation and the background fields should both obey the field equations of some extended theory of gravity, such as Ho\v{r}ava-Lifschitz gravity~\cite{Horava:2009uw} or Einstein-\ae ther theory~\cite{Jacobson:2000xp, Eling:2004dk}. 
However, since our purpose is to study the radiation rather than the collapse, the latter shall be described by a simplified model. At the end of the calculations, we shall argue that our main result remains valid in more general ones. 

For reasons of simplicity, we assume that the collapsing object is a null thin shell and that the external geometry is still Schwarzschild. In this case, the metric is again given by \eq{eq:met} and the Penrose diagram of \fig{fig:RelST} still covers the whole space-time. To describe the (unit timelike) \ae ther field $u^\mu$ in the external region outside the shell, we adopt the solution of~\cite{Berglund:2012bu} (also used in~\cite{Cropp:2013sea}) with $c_{123}=0$, $r_0=2M$, and $r_u=0$.  The null horizon is still at $r=2M$, whereas the universal horizon, where $u^\mu K_\mu = 0$, is located at $r=M$. Inside the shell, we assume that the \ae ther field is at rest. To our knowledge, this configuration has not been shown to be a solution of the field equations. However, as explained in Section~\ref{sub:gen}, small deviations from this configuration should not modify our conclusions.

In EF coordinates, on both sides of the shell, the \ae ther field $u^\mu$ and its orthogonal spacelike unit field $s^\mu$ are given by
\begin{align}\label{eq:uands}
u^\mu \pd_\mu = & \pd_v - \frac{M}{r} \Theta (v - 4 M ) \pd_r, \nonumber\\
s^\mu \pd_\mu = & \pd_v + \lp 1-\frac{M}{r} \Theta (v - 4 M ) \rp \pd_r
\end{align}
We introduce the ``preferred'' coordinates $t,X$ by imposing that $u_\mu dx^\mu \propto dt$ and $s^\mu \pd_\mu = {\rm sgn} (r-M) \pd_X$. Their precise definition is given in Section~\ref{app:PC}.~\footnote{The ``preferred'' time coordinate $t$ defined by \eq{eq:pref_t} is discontinuous across the mass shell. However, this discontinuity is only due to the choice of parametrization of the constant-time hypersurfaces, which is here made independently inside and outside the shell to obtain simple expressions. Since we will not use the parameter $t$ in the calculation of the spectrum, working instead with the continuous coordinates $r$ and $v$, the discontinuity of $t$ is benign. \label{foot:universal_dis}} 
In these coordinates, the metric takes the Painlev\'e-Gullstrand form: 
\begin{equation}
ds^2 = c^2 dt^2 - \lp dX - V dt \rp^2, 
\end{equation}
where
\begin{align}\label{eq:vandc}
V &= - \frac{M}{r} \Theta (v - 4 M ), \nonumber \\
c &= 
\left\lvert
K^\mu u_\mu 
\right\rvert = \left\lvert 1 - \frac{M}{r} \Theta (v - 4 M ) \right\rvert . 
\end{align}
At fixed $t$, outside the shell, $V$ and $c$ only depend on $X$. We have
\begin{equation}
u_\mu dx^\mu = c \, dt. 
\end{equation}
The factor $c$ ensures that $dt$ is a total differential. Moreover, as explained in Section~\ref{app:acc}, $c$ is constant when the \ae ther field is geodesic. Here we work with an accelerated \ae ther, which is a necessary condition to have a universal horizon. Importantly, $c$ vanishes on the universal horizon.~\footnote{In an analogue gravity perspective~\cite{Unruh:1980cg, Barcelo:2005fc}, to reproduce such a situation one needs a medium in which the group velocity of low-frequency waves vanishes locally. From \eq{eq:DR}, we see that the effective dispersive scale $\Lambda / c$ must be divergent at the point where $c \to 0$. It would be interesting to find media which could approximatively reproduce this behavior. A promising proposal was made in~\cite{Cropp:2016teb}.} In fact, the novelties of the present situation with respect to the standard case studied in~\cite{Macher:2009tw} only arise from the vanishing of $c$, and the associated divergence of the dispersive scale $\Lambda / c$.

\begin{figure}
\centering
\includegraphics[width=0.5 \linewidth]{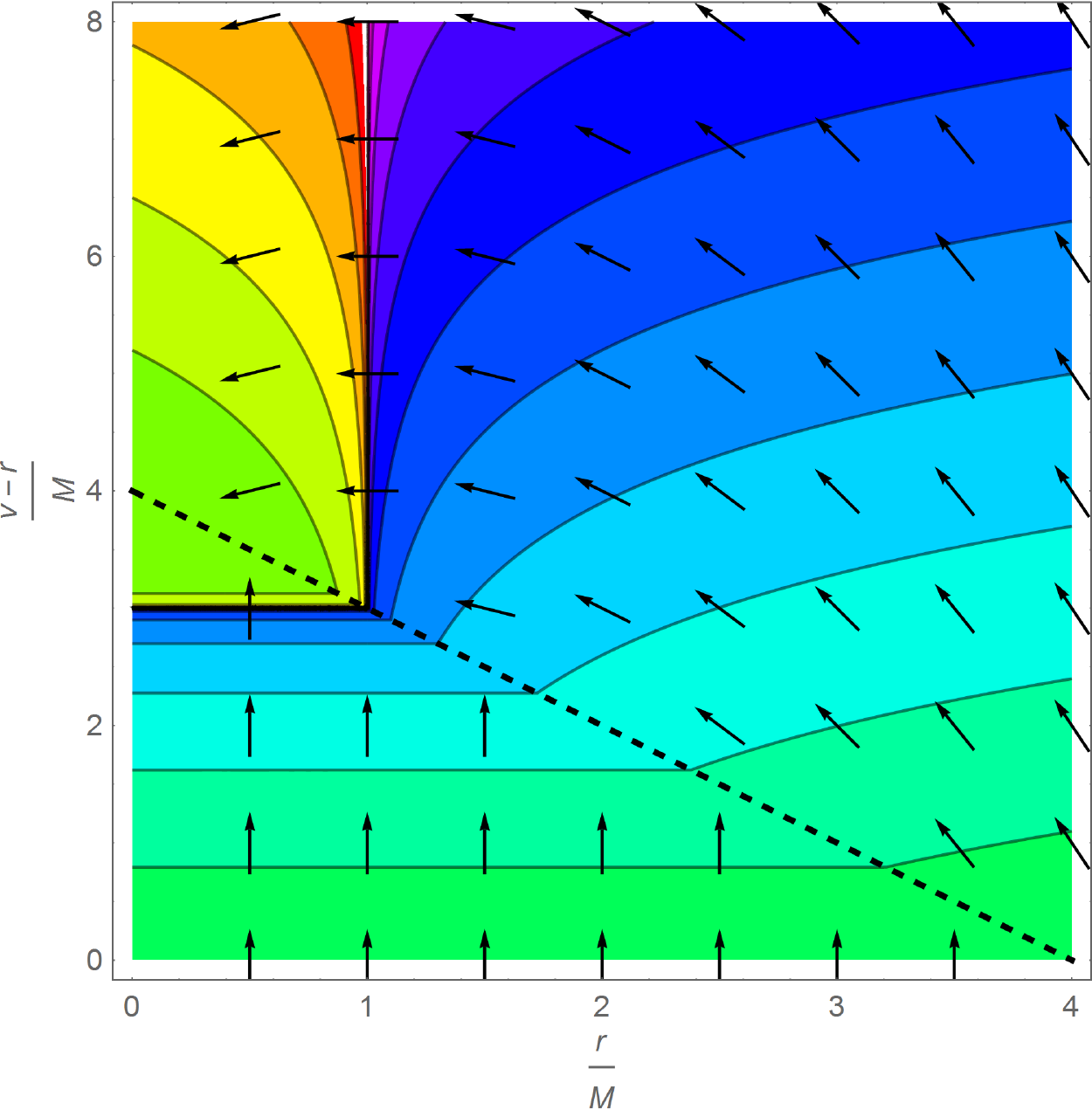} 
\caption{In this figure we show the lines of constant preferred time for the collapsing geometry in the plane $(v-r)/M,r/M$. The dashed line represents the trajectory of the null shell $v=4M$, and arrows show the direction of the \ae ther field $u_\mu$. Notice that the external preferred time $t$ diverges on the universal horizon $r=M$, $v>4M$, whereas the internal time $T$, which is equal to $v - r$ inside the shell, covers the entire space-time. 
} \label{fig:pref}
\end{figure}

In \fig{fig:pref} we show the lines of constant preferred time and the direction of the \ae ther field $u_\mu$ in the $v,r$ plane. The coordinate $t$ is discontinuous across the shell trajectory, as was the null coordinate $u$ in the former section. As in the relativistic case, outside the shell we must use two coordinates $t$ and $t_L$, now on either side of the universal horizon. The inside coordinate $T$ evaluated along the shell, at $v=4M^-$, is a monotonically increasing function of both $t(v=4M^+,r)$ for $r > M$ and of $t_L(v=4M^+,r)$ for $r< M$. So, the foliation of the entire space-time by the inside coordinate $T$ is globally defined and monotonic. 

We consider a real, massless, dispersive field $\Phi$ with a superluminal dispersion relation. Its action is given by \eq{eq:relativisticaction} supplemented by a quartic term:
\begin{equation}\label{eq:action}
S = \int \dd^4x \sqrt{-g} \left[ \lp \pd_\mu \Phi \rp \lp \pd^\mu \Phi \rp - \frac{1}{\Lambda^2} \lp \nabla_\mu \lp h^{\mu \nu} \nabla_\nu \Phi \rp \rp \lp \nabla_\rho \lp h^{\rho \sigma} \nabla_\sigma \Phi \rp \rp\right], 
\end{equation}
where $\nabla_\mu$ is the covariant derivative and $h^{\mu \nu} \equiv g^{\mu \nu} - u^\mu u^\nu$ is the projector on the hyperplane orthogonal to $u^\mu$. $\Lambda$ is the dispersive momentum scale. The field equation reads
\begin{equation}\label{fieldeq}
\nabla_\mu \nabla^\mu \Phi + \frac{1}{\Lambda^2} \lp \nabla_\mu  h^{\mu \nu}  \nabla_\nu \rp^2 \Phi = 0.
\end{equation}
Assuming $\Phi$ is independent on $\theta$ and $\varphi$, \eq{fieldeq} reduces to 
\begin{equation}\label{eq:2DFE}
\left[
\left[ \pd_t + \pd_X V \right] \frac{1}{c} \left[ \pd_t + V \pd_X \right] - \pd_X c \pd_X + \frac{1}{\Lambda^2} \pd_X c \pd_X \frac{1}{c} \pd_X c \pd_X
\right]
\psi = 0
\end{equation}
in the preferred coordinate system. 
As in the relativistic case, $\psi$ is defined by $\psi(t,r) \equiv r \Phi(t,r)$. 
Since \eqref{eq:2DFE} is a self-adjoint equation of order $2$ in $\pd_t$, the Hamiltonian structure of the theory is preserved. In particular, the conserved inner product has the standard form 
\begin{equation}
\lp \psi_1 | \psi_2 \rp
= \ii \int \dd X \lp \psi_1^* \Pi_2 -  \Pi_1^* \psi_2 \rp, 
\label{scal}
\end{equation}
where $\Pi \equiv u^\mu \pd_\mu \psi = \lp \pd_t \psi + V \pd_X \psi \rp /c$ is the momentum conjugated to $\psi$. For more details, see Section~\ref{app:sc}. 

We introduce the Killing frequency $\lambda$, the preferred frequency $\Omega$, and the preferred momentum $P$: 
\begin{align}
& \lambda = - K^\mu \partial_\mu S = - \pd_t S, \label{Kilf} \\ 
& \Omega =  - 
c(X) \, u^\mu \pd_\mu S = \lambda - V(X) P, \\
& P  =  s^\mu \pd_\mu S = \partial_X S. 
\end{align}
In these equations, $S$ may be conceived as the action of a point particle, see~\cite{Brout:1995wp,Balbinot:2006ua,Coutant:2011in}. As explained in these works, $S$ governs the WKB approximation of the solutions of \eq{eq:2DFE}. Notice that \eq{Kilf} only applies outside the shell, whereas all the other equations are valid on both sides. 
The Hamilton-Jacobi equation associated with \eq{eq:2DFE} is
\begin{equation}\label{eq:DR}
{\Omega^2} = {c(X)^2}\left[ P^2 + \frac{P^4}{\Lambda^2}\right]. 
\end{equation}

\subsection{The modes and their characteristics}

To compute the late-time radiation one should identify the solutions of \eq{eq:2DFE} and understand their behavior. In the presence of dispersion, one loses the neat separation of null geodesics into the outgoing $u$ ones and the infalling $v$ ones. In what follows, we call $P^u$ ($P^v$) the roots of the dispersion relation which have a positive (negative) group velocity in the frame at rest with respect to the ``fluid'' of velocity $V$, see~\cite{Macher:2009tw}. Similarly, the corresponding modes will also carry the upper index $u$ or $v$. 

\subsubsection{The in and out asymptotic modes} 

In the internal region $v<4M$, the situation is particularly simple. Since the velocity field $V$ vanishes, the preferred frequency is $\omega = -\pd_T S$ and the dispersion relation \eq{eq:DR} becomes 
\begin{equation}
\omega^2 = P^2 + \frac{P^4}{\Lambda^2}.
\end{equation}
This relation is shown in the left panel of \fig{fig:DRU}. At fixed $\omega$, the positive-frequency modes with wave vectors $P^{u}(\omega) > 0$ and $P^{v}(\omega) < 0$ give the two \textit{in} modes $\phi^{u, \, \rm in}_\omega$ and $\phi^{v, \, \rm in}_\omega$. They both have a positive norm $(\phi, \phi)$, which can easily be set to unity through a normalization factor. The mode $\phi^{u, \, \rm in}_\omega$ is the dispersive version of the relativistic in mode \eq{relat-in-mode}. 

Outside the shell, for $v>4M$, at fixed Killing frequency $\lambda > 0$, the situation is more complicated as the number of real roots depends on $r$. Outside the null horizon, for $r>2M$, one has $c > \left\lvert V \right\rvert$. So, \eq{eq:DR} possesses two real roots $P^{u}(\lambda) > 0$ and $P^v(\lambda)< 0$, which describe outgoing and infalling particles, respectively. The WKB expression for the corresponding stationary modes (the solutions of \eq{eq:2DFE}) is 
\begin{equation}\label{eq:WKBmodes}
\psi_\lambda^{(i)} \lp t, X \rp \approx \frac{\exp \lp -\ii \lp \lambda t  - \int^X P^{(i)}(\lambda, X') \dd X' \rp \rp}{4 \pi\sqrt{\left\lvert \Omega(\lambda,P^{(i)}) /( c(X) \, \pd_\lambda P^{(i)}) \right\rvert}},
\end{equation}
where $P^{(i)}(\lambda, X)$ is a real solution of \eq{eq:DR} at fixed $\lambda$, and $\Omega(\lambda,P^{(i)})$ is the corresponding preferred frequency. These WKB modes generalize the expressions of \cite{Coutant:2009cu,Coutant:2011in} in that $c$ is no longer a constant. Using \eq{scal}, one easily verifies that they have a unit norm. The group velocity along the \textit{i}th characteristic is $\dd X^{(i)}/\dd t = 1/\pd_\lambda P^{(i)}$. When considered far away from the black hole, i.e. $r/(2M) \gg 1$, the $u$-WKB mode is the dispersive version of the relativistic out-mode of \eq{relat-out-mode}. 

From this analysis, we see that there is no ambiguity to define the asymptotic behavior of the in and out modes, solutions of \eq{eq:2DFE}. As before, these two sets encode the black hole radiation through the overlaps \eq{BogC}. To be able to compute these overlaps, we need to construct the modes defined globally. To this end, we must study both the behavior of $\psi_\lambda^{(i)}$ near the horizon and the third kind of stationary modes which propagate in this region.

\subsubsection{Near-horizon modes}

\begin{figure}
\centering
\includegraphics[width = 0.48 \linewidth]{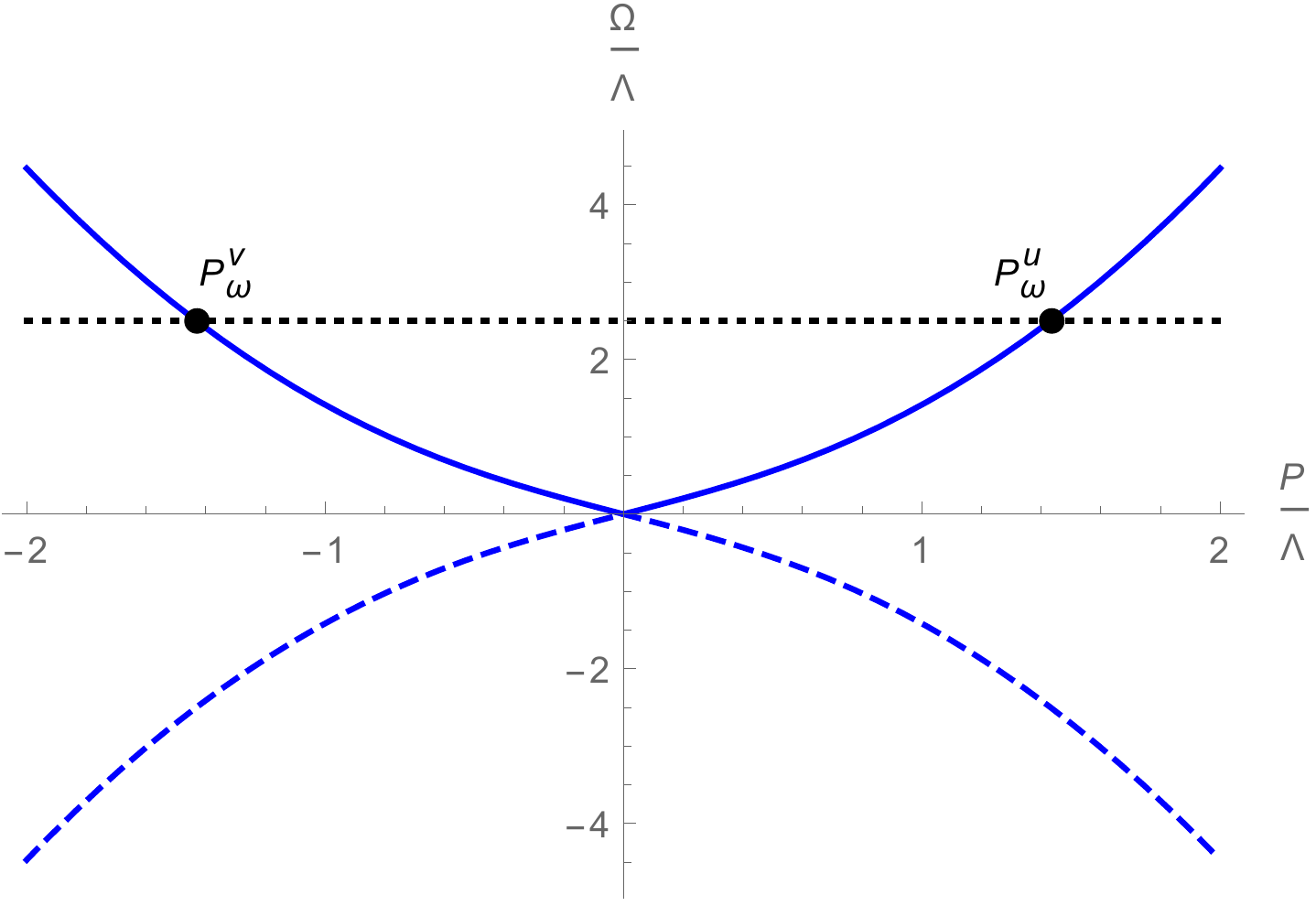} \,
\includegraphics[width = 0.48 \linewidth]{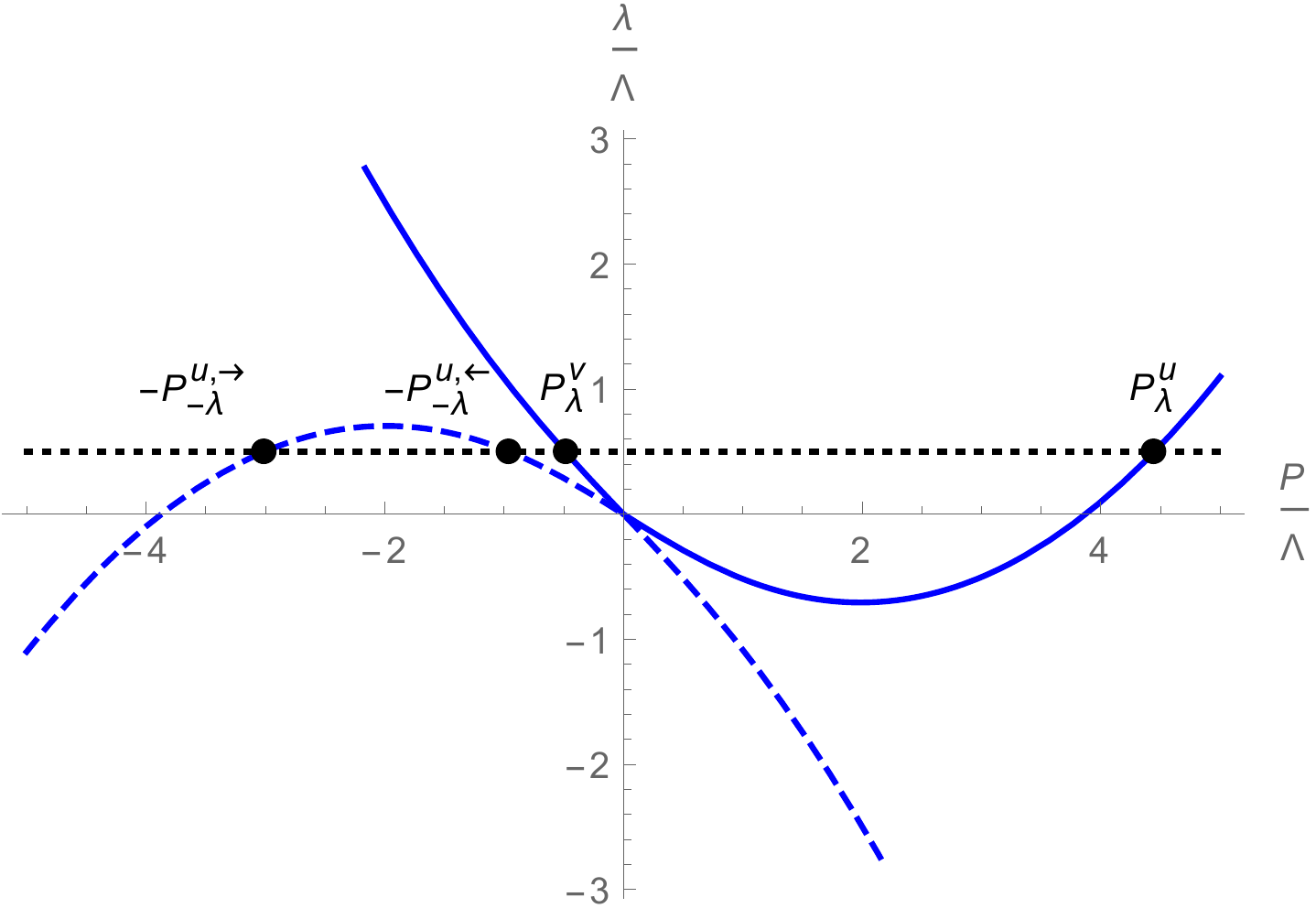} 
\caption{Left panel: Dispersion relation in the internal region, where the preferred frame is at rest, in the $\Omega,P$ plane. The solid line shows $\Omega$ versus $P$ for the positive-norm modes. The dashed line corresponds to negative values of $\Omega$, i.e., to negative-norm modes. The intersections with a line of fixed $\omega >0$ (dotted line) give the two solutions $P^{u}_\om$ and $P^{v}_\om$. Right panel: Dispersion relation in the ``superluminal'' region for $M < r < 2M$ in the $\lambda , P$ plane. The two additional roots on the $u$ branch with negative $\Omega$ are clearly visible. 
} \label{fig:DRU}
\end{figure}

Inside the null horizon but outside the universal horizon, for $ M < r<2M$, one has $c < \left\lvert V \right\rvert$. As can be seen from the right panel of \fig{fig:DRU}, one recovers the two roots $P^{u}(\lambda) > 0$ and $P^{v}(\lambda) < 0$ we just described. One notices that the $u$ root $P^{u}(\lambda)$ has been significantly blueshifted. Locally, in the WKB approximation, the corresponding modes are again given by \eq{eq:WKBmodes}. 

In addition, below a critical frequency $\lambda_c$ which depends on $c$ and $V$, we have two new real roots that we call $-P^{(u, \rightarrow)}_{-\lambda}$ and $-P^{(u, \leftarrow)}_{-\lambda}$, where the arrow indicates the orientation of the group velocity $1/\partial_\lambda P$. (The minus signs in front of these roots and $\lambda$ come from the fact that they have a negative preferred frequency $\Omega$ for $\lambda > 0$. Hence, for $\lambda = - \left\lvert \lambda \right\rvert$, the symmetric roots, $P^{(u, \rightarrow)}_{-|\lambda|}$ and $P^{(u, \leftarrow)}_{-|\lambda|}$, have a positive $\Omega$.) Since $\Omega < 0$, the WKB modes associated with these roots have a negative norm~\cite{Coutant:2011in}. We call the right-moving one $\psir{-\lambda}$ and the left-moving one $\psil{-\lambda}$, so that the modes without complex conjugation have a positive norm. Both of them carry a negative Killing energy $- \lambda$. Using \eq{eq:WKBmodes} and \eq{scal}, one easily verifies that these two modes also have a negative unit norm within the WKB approximation. As we shall see, they describe the negative-energy partners trapped inside the null horizon before and after their turning point, respectively. 
To summarize the situation, it is appropriate to represent the characteristics of the three types of modes. We proceed as in~\cite{Brout:1995wp,Coutant:2011in}.

\subsubsection{The characteristics} 

As said above, the characteristics are solutions of the equation $\frac{\dd X}{\dd T} = \frac{1}{\partial_\lambda P}$. Since the frequency is a constant of motion on each side of the mass shell, they can be computed straightforwardly. In \fig{fig:char}, they are shown in the external region $v>4M$ for a small value of $|\lambda |/ \Lambda = 0.01$ (left panel) and a moderate one $|\lambda| / \Lambda = 1$ (right panel). The solid lines correspond to positive-energy solutions while the dashed ones correspond to negative-energy ones.
 
The infalling $v$-like characteristics corresponding to $\psiv{\lambda}$ (in blue) approach the universal horizon from infinity and cross it at a finite value of $v$. (When sending $\lambda$ to $0$ they asymptote to null infalling geodesics with constant $v$.) Since their wave vectors are finite for $r \to M^+$, these characteristics will play no role in the sequel. As in the relativistic case, the $v$ modes act as spectators in the Hawking effect.~\footnote{Interestingly, $v$-like characteristics have a turning point inside the universal horizon $r<M$. (The presence of the turning point may be understood from the fact that, close to $r=0$, $|V|$ and $c$ go to infinity but $|V|/c$ goes to $1$. So, at fixed $\lambda > 0$ two roots merge at a point $r = r_{\rm tp}(\lambda)>0$. The turning point approaches $r=0$ in the limit $\lambda \to 0$.) For later preferred times, they return towards the universal horizon, which they approach asymptotically for $v \to -\infty$, $t_L \rightarrow + \infty$, and are highly blueshifted. In addition, for $r < M$, there is a new $v$ mode with negative norm for $\lambda > 0$. It is indicated by a dashed green line in \fig{fig:char}. It emerges from the singularity and approaches the universal horizon while closely following the positive Killing frequency characteristic after its turning point. (In fact this new $v$ mode is directly related to the $u$ modes emerging from the singularity in~\cite{Jacobson:2001kz}: inside a universal horizon, $u$ and $v$ modes are swapped because of the vanishing of $c$ at $r=M$.) Since some of the $v$ modes originate from the singularity, and since the blueshift they experience is unbounded for $r \to M^-$, it seems that the $v$ part of the state will not obey Hadamard regularity conditions. This strongly indicates that the {\it inner} side of the universal horizon should be singular. This interesting question goes beyond the scope of the present chapter.\label{vfoot}}

The $u$-like characteristics with positive energy (in red), corresponding to the WKB modes $\psiu{\lambda}$, emerge from the universal horizon from its right ($r>M$) at early times. When $t$ increases, the momentum $P^u_\lambda$ is redshifted while $r$ increases. At a finite time, the characteristics cross the null horizon, and go to infinity as $t \to \infty$ (almost along null outgoing geodesics when $\lambda/\Lambda \ll 1$). 

The third characteristics (orange, dashed line) describe the trajectories followed by the negative-energy partners. For $t \to -\infty$, they also emerge from $r = M^+$. However, when increasing $t$ they have a turning point inside the null horizon, after which they move towards the universal horizon, smoothly cross it, and hit the singularity at $r=0$ at finite values of $v$ and $t_L$. Before the turning point, they are described by the WKB mode $\psir{-\lambda}$, and after the turning point by $\psil{-\lambda}$. 
\begin{figure}
\centering
\includegraphics[width = 0.49\linewidth]{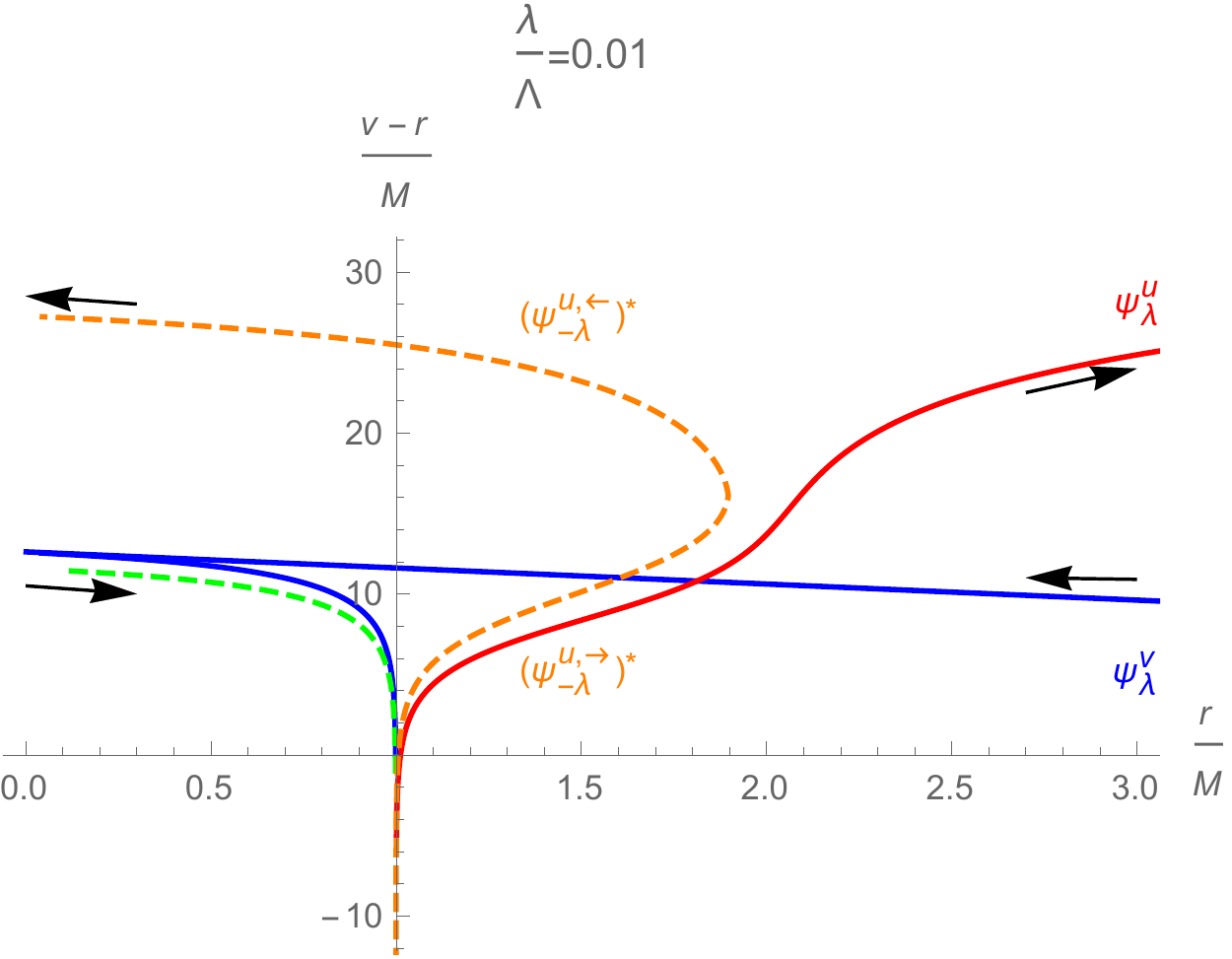} \, 
\includegraphics[width = 0.49 \linewidth]{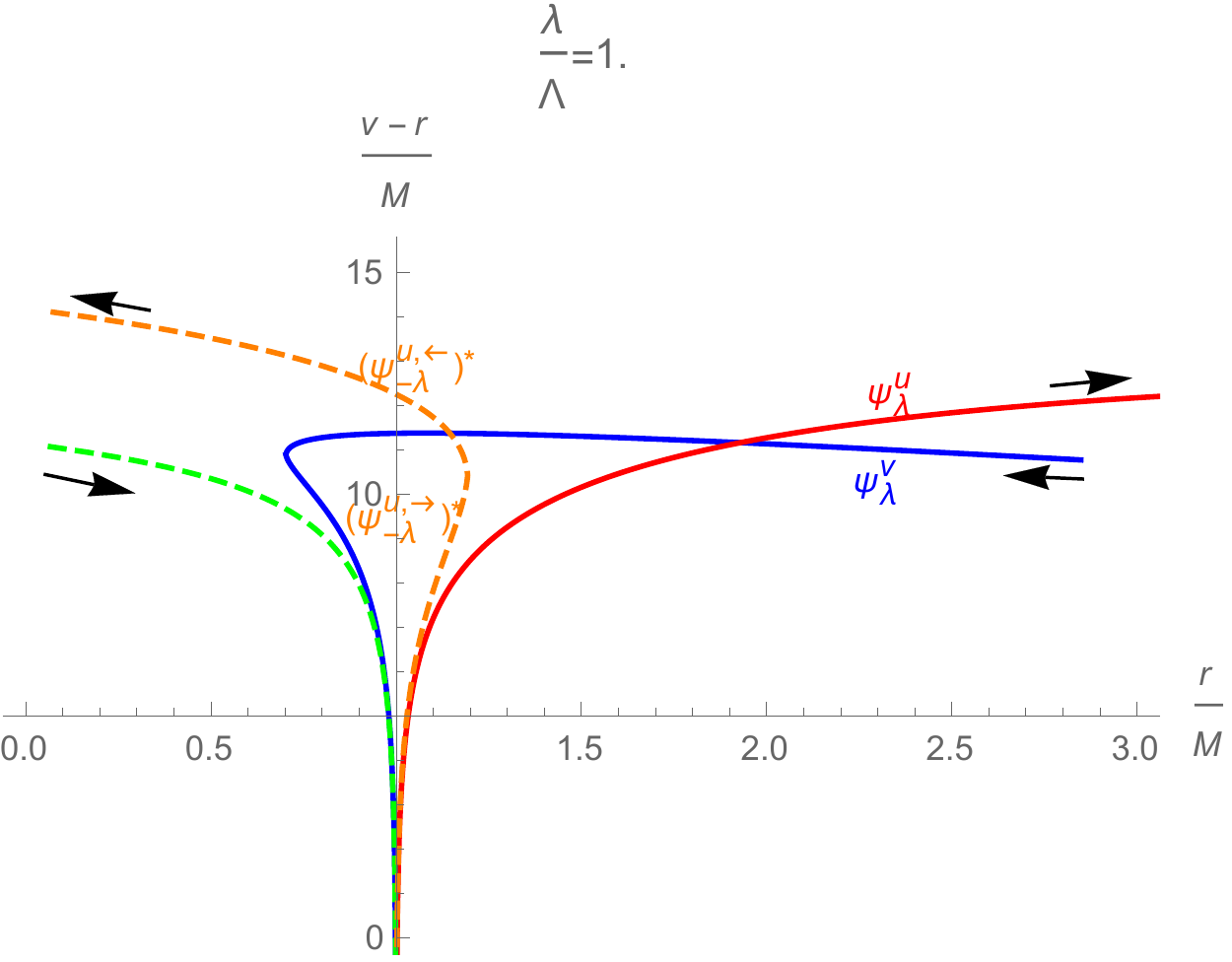}
\caption{Characteristics in a Schwarzschild stationary geometry for $\lambda = 10^{-2} \Lambda$ (left) and $\lambda = \Lambda$ (right). The arrows indicate the direction of increasing preferred time along each characteristic. Solid lines correspond to positive-norm modes and dashed ones to negative-norm modes. For $r > M$, each characteristic is labeled by the corresponding mode. The green dashed line corresponds to an extra $v$ mode confined in the region $r <M$, as discussed in footnote~\ref{vfoot}. In this footnote, we also explain that the infalling $v$ mode (blue line) possesses a turning point inside the universal horizon. The mode corresponding to the orange dashed line is the high-momentum WKB mode $\psir{-\lambda}$ before the turning point, and the low-momentum mode $\psil{-\lambda}$  after it.  
} \label{fig:char}
\end{figure}

It is important to notice that the only novel aspect with respect to the standard dispersive case (treated in full detail in~\cite{Coutant:2011in}) concerns the behavior near the universal horizon. To clarify these new aspects, we represent in \fig{fig:Gchar} the global structure of the characteristics in the collapsing mass shell geometry. 
 
\subsubsection{The characteristics in the collapsing geometry}

In the internal region $v<4M$, the characteristics are straight lines. They are fixed by the value of the inside frequency which is determined (as in the relativistic case), by continuity of the field $\psi$ across the mass shell; see Section~\ref{app:match} for details. As a result, the derivative $\pd_r \psi$ must be continuous across $v = 4M$. At the level of the characteristics (i.e., in the geometrical optic approximation), this implies that $k_v$, the radial momentum at fixed $v$, is continuous along the shell. In terms of the inside and outside preferred momenta $P^u(\omega)$ and $P^u(\lambda,r)$ evaluated at $v=4M^-$ and $v=4M^+$, respectively, the continuity condition gives
\begin{equation}\label{eq:matching_char}
\left\lvert \frac{r}{r-M} \right\rvert \lp \lambda + P^u(\lambda,r) \rp = \omega+P^u(\omega).
\end{equation}
This equation has two solutions, but only one of them has a monotonic preferred time and thus yields a physical, causal trajectory. 
A straightforward calculation using the dispersion relation \eq{eq:DR} also shows that the sign of $\Omega$ is preserved when crossing the mass shell.~\footnote{On the one hand, inside the shell, $\Omega = \om$ and $\abs{\om} > \abs{P}$. So, $\om + P$ and $\Omega$ have the same sign. On the other hand, outside the shell, $\lambda + p = \Omega + (1+V) P = \Omega + c P$. Since $\abs{\Omega} > c \abs{P}$, $\lambda + P$ has the same sign as $\Omega$. \eq{eq:matching_char} thus implies that the sign of $\Omega$ is preserved.}
It should be noted that \eq{eq:matching_char} is the dispersive version of the relativistic equation $|r/(r - 2M)| \lambda = \omega$, which gives back \eq{CarterL} for $r> 2M$, $\omega \gg \lambda$, and when using $u$ rather than $r$. 

\begin{figure}
\centering
\includegraphics[width=0.5\linewidth]{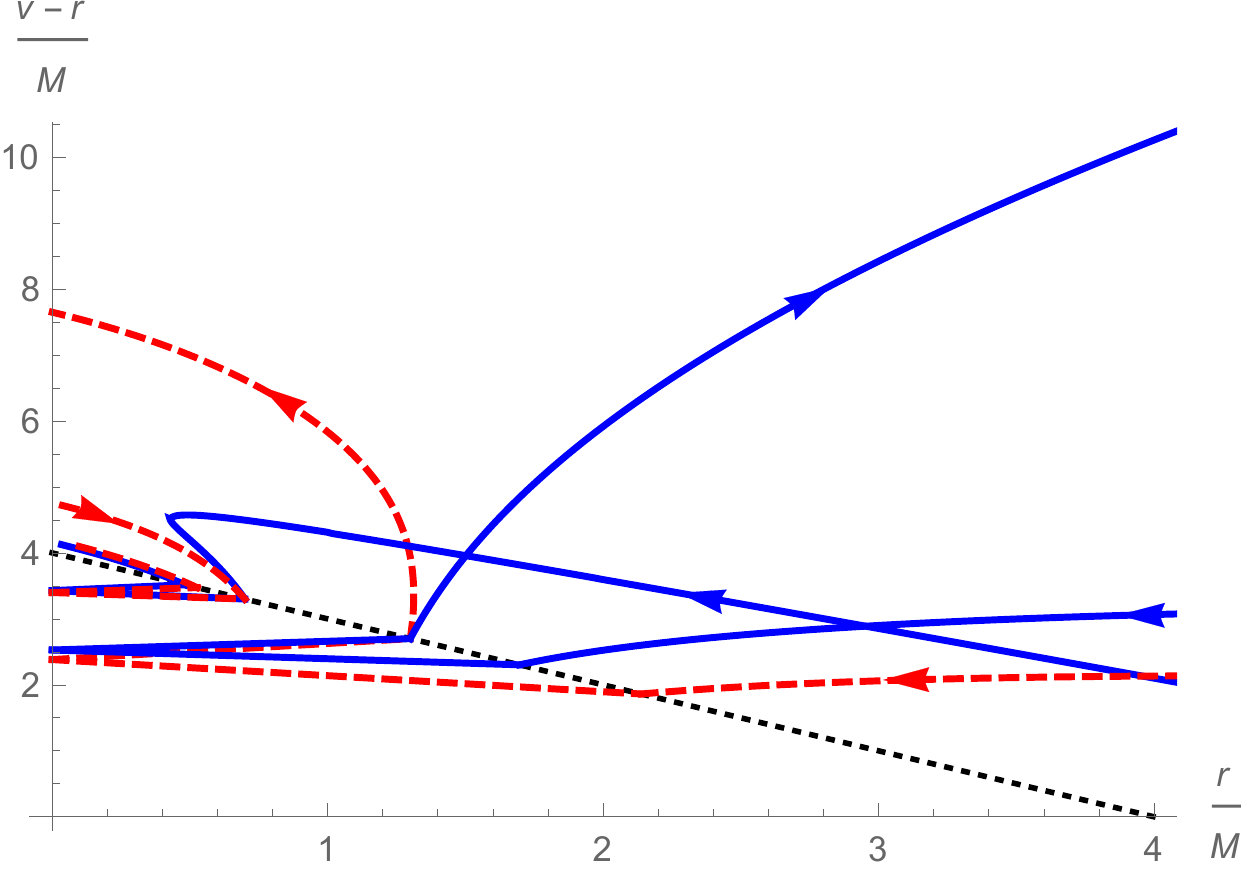} 
\caption{Characteristics crossing the infalling shell in the $v-r,r$ plane. The Killing frequencies of the outgoing $u$ modes and the incoming $v$ modes is $\lambda = \pm 0.5 \Lambda$. The solid (dashed) lines represent characteristics for which the value of the Killing frequency $\lambda$ of the out-going $u$ mode is positive (negative). The arrows indicate the future direction associated with the \ae ther field. When tracing backwards the $u$-like characteristics associated with the Hawking quanta ($\lambda > 0$) and their inside negative energy partners ($\lambda < 0$),  we see that they both originate from infalling $v$-like superluminal characteristics with a high and positive Killing frequency $\lambda_{\rm in}$. The $v$ mode which emanates from the singularity (the dashed line) returns to it after having bounced at the center of the shell. } \label{fig:Gchar}
\end{figure}

It should be also emphasized that all outgoing $u$-like characteristics originate from inside the shell, as in the relativistic case (this is shown in \fig{fig:Gchar}). Therefore, the state of the field inside the shell fully determines the state of the $u$ modes. In this we avoid the problem discussed in~\cite{Jacobson:2001kz}, namely that in the absence of a universal horizon, the $u$ modes of a superluminal field originate from the singularity at $r = 0$. As discussed in  footnote~\ref{vfoot}, these modes still exist, but they are now trapped inside the universal horizon. 

Finally, we notice that the Killing frequency $\lambda_{\rm in}$ of the incoming $v$ modes which generate the outgoing $u$ modes exiting the shell at $r_c \approx M$ is very large. More precisely, when dealing with $u$ characteristics with positive $\Omega$ (i.e., modes with positive norm), irrespective of the sign of their Killing frequency $\lambda$, the Killing frequency $\lambda_{\rm in}$ is positive. A straightforward calculation (based on the continuity of $k_v$ applied to the $v$ modes) shows that it scales as $\lambda_{\rm in} \approx 3 \Lambda M / \lp r_c - M \rp$.

For completeness, we have also represented in \fig{fig:Gchar} a couple of infalling $v$ characteristics which enter the shell for $ 0 < r < M$. One of them comes from $r = 0$ (the dashed line), and one from $r = \infty$ (the solid line). They both reach the singularity after having bounced at $r = 0$ inside the shell. These characteristics, although interesting, play no role in the Hawking process. 

\subsection{Behavior of the WKB modes near the universal horizon}

To be able to compute the late-time behavior of the Bogoliubov coefficients, we need to further characterize the properties of the stationary modes in the immediate vicinity of the universal horizon at $r=M$. For $r>M$, the two roots $P_\lambda^v$ and $P_{-\lambda}^{u,\leftarrow}$ remain finite as $r \to M$. As can be seen in \fig{fig:char}, the associated trajectories smoothly cross the horizon. They thus play no role in the large $\omega$ limit. 

The two other roots $P_\lambda^u$ and $P_{-\lambda}^{u,\rightarrow}$ both diverge as $r\to M$. Importantly, they both satisfy 
\begin{equation}
 P^{\rm in}_{\pm \lambda} = \frac{\Lambda M}{r - M} \pm \frac{r}{M} \lambda  + O \lp 1 - \frac{M}{r} \rp,
\label{Pomo}
\end{equation}
where the + sign applies to $P_\lambda^u$ and the - sign to $P_{-\lambda}^{u,\rightarrow}$. We have added a superscript ``$\textrm{in}$'' to emphasize that this behavior is relevant at early time $t$, just after having crossed the shell. The simple relation between $P_\lambda^u$ and $P_{-\lambda}^{u,\rightarrow}$ implies that, for $r \to M$, the two WKB modes $\psi^u_{\lambda}$ and $\psi_{-\lambda}^{u, \rightarrow}$ are also related to each other by flipping the sign of $\lambda$. In the forthcoming discussion, to underline these points, we shall replace $\psi_{-\lambda}^{u, \rightarrow}$ by $\psi^u_{-\lambda}$, and add a superscript ``$\textrm{in}$'' to the WKB modes $\psi^u_{\pm \lambda}$. 

Although the divergence in $1/ (r-M)$ in \eq{Pomo} resembles what is found in the relativistic case, it has a very different nature due to the different relationship between $r$ and the preferred coordinate $X$. This can be seen by looking at the validity of the WKB approximation for $\psi^{u, \rm in}_{\lambda}$ close to the universal horizon. Deviations from this approximation come from terms in $(\pd_X r) / (r P)$, $(\pd_X (r-M)) / ((r-M) P)$, and $(\pd_X P) / P^2$. 
Using 
\begin{equation}
\pd_X = \frac{r-M}{r} \pd_r, 
\end{equation}
we find that these three terms go to zero as $r \to M$. Therefore, close to the universal horizon, the WKB approximation of~\eq{eq:WKBmodes} becomes exact for $\psi^{u, \rm in}_{\pm\lambda}$. In fact, these modes behave as the dispersive $\textrm{in}$ modes near a null horizon~\cite{Brout:1995wp,Coutant:2011in}. Namely, they have a positive norm for all values of $\lambda$ and, moreover, contain only positive values of $P^u$. We recall that this is the key property which also characterizes the so-called Unruh modes~\cite{Unruh:1976db, Brout:1995rd} for a relativistic field.

These are strong indications that no stationary emission should occur close to the universal horizon, as the pair production mechanism rests on deviations from the WKB approximation. This is confirmed in the next subsection. 

\subsection{Bogoliubov coefficients from the scattering on the shell}

We now have all the elements to determine the scattering coefficients which govern the propagation across the mass shell. Inside the shell, one has the $\textrm{in}$ mode $\phi^{u,\, {\rm in}}_{\omega}$. Along the shell, for $v = 4M^-$, it is a plane wave which behaves as $\phi^{u,\, {\rm in}}_{\omega} \propto \exp{[\ii (\omega + P^u(\omega)) r]}$. After having crossed the shell, for $r/M - 1 \ll 1$, it may be expanded in terms of the four WKB modes (which form a complete basis)
\begin{equation}
\phi^{u,\, {\rm in}}_{\omega} = \int^\infty_{-\infty} \dd \lambda \lp \gamma_{\omega,\lambda} \psi^{u, \rm in}_{\lambda} + \delta_{\omega,\lambda} (\psi^{u, \rm in}_{-\lambda})^* + A_{\omega,\lambda} \psiv{\lambda} + B_{\omega,\lambda} \psil{\lambda} \rp.
\label{modem}
\end{equation}
We are interested in the coefficients $\gamma_{\omega,\lambda}$ and $\delta_{\omega,\lambda}$ which multiply the two modes with divergent wave vectors and opposite norms. The other two coefficients, $A_{\omega,\lambda}$ and $B_{\omega,\lambda}$, multiply the two modes which remain regular across the universal horizon in the $(v,r)$ coordinates. They vanish faster than polynomially in the limit $\omega \to \infty$, and thus can not generate a stationary spectrum.

The calculation of $\gamma_{\omega,\lambda}$ and $\delta_{\omega,\lambda}$ is straightforward in the $(v,r)$ coordinates, see subsections~\ref{app:alpha} and~\ref{app:beta}. For $|\lambda|\lesssim \Lambda$, we find that their ratio decays as
\begin{equation}\label{eq:bovera}
\left\lvert \frac{\delta_{\omega,\lambda}}{\gamma_{\omega,\lambda}} \right\rvert \mathop{=}_{\omega \to \infty} O \lp  \frac{\sqrt{M \Lambda}}{\omega} \exp \lp -2 M P^u(\omega)\rp \rp  ,
\end{equation}
where $P^u(\omega) \sim \sqrt{\omega \Lambda}$ in the high-frequency regime we are considering. Equation \eq{eq:bovera} is the main result of the present chapter. 
It means that, at late time, corresponding to the emission close to the universal horizon and thus to very large values of $P^u_\lambda \sim \Lambda/(r/M - 1)$ (see \eq{Pomo}), the propagation across the shell induces no mode mixing between the inside in mode $\phi^{u, \, {\rm in}}_{\omega}$ and the high-momentum WKB mode with negative norm $(\psi^{u, \rm in}_{-\lambda})^*$, irrespective of the value of $\lambda$. As a result, the state of the field outside the shell at late times is the (stationary) vacuum with respect to the annihilation operators associated with $\psi^{u, \rm in}_{\lambda}$ for $\lambda \in (-\infty, \infty)$.~\footnote{This conclusion differs from that reported in~\cite{Berglund:2012fk}. 
The procedure adopted there apparently implies that the leading term in \eq{Pomo} does not contribute to the ratio of \eq{eq:bovera}, thereby giving rise to a steady thermal radiation governed by the surface gravity of the universal horizon. Instead, the saddle point evaluation of $\delta_{\omega,\lambda}$ detailed in subsection~\ref{app:beta} establishes that in the present model the leading term of \eq{Pomo} gives the exponential damping in $e^{- 2MP}$ of \eq{eq:bovera}.} It thus correspond to the $\textrm{in}$ vacuum as described in~\cite{Brout:1995wp,Coutant:2011in}.~\footnote{To be complete, one should propagate backwards in time the inside field configurations, and verify that they correspond to vacuum $v$-like configurations for $r \to \infty, t \to -\infty$. To verify this, we computed the scattering coefficients encoding a change of the norm of the $v$ modes when crossing the shell. We found that they also decrease exponentially in $\sqrt{\omega \Lambda}$ for $\omega \to \infty$. We also recall here that the Killing frequency of the $v$ modes engendering a stationary $u$ mode diverges as $\lambda_{\rm in} \approx 3 \Lambda M / \lp r_c - M \rp$, where $r_c$ is the radius when the $u$ mode exits the shell.}

\subsection{Genericness of \texorpdfstring{\eq{eq:bovera}}{(5.32)}}
\label{sub:gen}

In this subsection, we distance ourselves from the model we considered and discuss how the above results may be affected. We first consider a modification of the mass shell trajectory close to the universal horizon. From the calculation done in subsections~\ref{app:alpha} and~\ref{app:beta}, the factor $\exp \lp -2 M \sqrt{\Lambda \, \omega} \rp$ in \eq{eq:bovera} comes from the fact that the phase of the mode inside the mass shell is $\theta_{\rm int} \approx -  \omega T \approx \omega \lp r - v \rp$, while that of the mode outside the mass shell is $\theta_{\rm out} \approx \Lambda M / x$, where $x \equiv (r/M) -1$. 
At fixed $v$, we find that the stationary phase condition applied to $\theta_{\rm in} \pm \theta_{\rm out}$ (the upper sign applies to $\gamma$ while the lower sign applies to $\delta$) gives back the large-frequency limit of \eq{eq:matching_char} with $x$ real for $\gamma$, while $x$ is purely imaginary for $\delta$, with a modulus equal to $\sqrt{\Lambda / \omega}$. Let us now consider an arbitrary shell trajectory close to the universal horizon. We define an affine parameter $y$ along this trajectory. The possible saddle points are located where
\begin{equation}
\frac{\dd}{\dd  y} \lp \omega T \mp \frac{\Lambda M }{x} \rp = 0,
\end{equation}
i.e.,
\begin{equation}
\omega \frac{\dd T}{\dd y} \pm \frac{\Lambda M}{x^2} \frac{\dd x}{\dd y} = 0.
\end{equation}
So, the location of the saddle is
\begin{equation}
x^* = \sqrt{\mp M \frac{\dd x}{\dd T} \frac{\Lambda}{\omega}}. 
\end{equation}
We get the same result as before, up to the factor $-M \frac{\dd x}{\dd T}$. Therefore, the ratio $\left\lvert {\delta_{\omega,\lambda}}/{\gamma_{\omega,\lambda}} \right\rvert$ is still suppressed by an exponential factor in $M \sqrt{\Lambda \, \omega}$, with a coefficient depending on the velocity of the mass shell when it crosses $r=M$. 

We now consider a generalization of the dispersion relation \eq{eq:DR} with higher-order terms. Specifically, we consider the dispersion relation
\begin{equation}\label{eq:generalDR}
\frac{\Omega^2}{c^2} = \sum_{j=0}^N \frac{P^{2j}}{\Lambda_j^{2 (j-1)}}.
\end{equation} 
for some $N \in \mathbb{N}$, $N \geq 2$. 
Close to the universal horizon, the divergent wave vectors follow
\begin{equation}\label{eqP}
P \approx \pm \Lambda_N \, x^{\frac{-1}{N-1}}.
\end{equation}
As before, the coefficient $\gamma$ corresponds to $\pm = +$ in \eq{eqP}. The value of the saddle point is then real, and the exponential factor appearing in $\gamma_{\omega, \lambda}$ has a unit modulus. Instead, for the coefficient $\delta$, corresponding to the minus sign in \eq{eqP}, the solutions of the saddle point equation are
\begin{equation}
x^* = \lp \frac{\omega}{\Lambda} \rp^{\frac{1-N}{N}} \e^{\ii \pi \frac{1+2l}{N}}, \; l \in \mathbb{Z}.
\end{equation}
Taking only the saddle points with negative imaginary parts for consistency with the saddle-point approximation, we find that $\delta_{\omega, \lambda}$ is suppressed by a factor which is exponentially small in $\omega^{1/N}$. Interestingly, when using the inside spatial wave number $P^u(\om)$ rather than the inside frequency $\omega$, the modulus of the coefficient $\delta_{\omega, \lambda}$ always decreases as $\exp \lp -M A P^u(\om) \rp$ with $A > 0$.

Similarly, the exponential factor suppressing $\delta_{\omega,\lambda}$ is mildly affected by a change in the metric and/or the form of the \ae ther field, provided the inside wave vector remains smooth, whereas the outside one diverges as 
a power law for $r \to r_{U H}$, where $r_{U H}$ is the radius of the universal horizon. This should remain valid as long as there is no divergence (or cancellation) preventing us from defining preferred coordinates in which the dispersion relation takes the form of~\eq{eq:generalDR} close to the universal horizon. Indeed, the construction of subsection~\ref{app:PC} can be easily extended to a generic space-time with a Killing vector $\chi$, endowed with a generic timelike, normalized \ae ther field $u^\mu$. 

\section{Discussion}
\label{uni:disc}

We have determined the late-time behavior of Hawking radiation in a Lorentz-violating model of a black hole with a universal horizon. To identify the appropriate boundary conditions for the stationary modes of our dispersive field, we worked with a geometry describing the collapse of a thin mass shell and assumed that the inside state of the field is vacuum at (ultra) high inside frequencies $\omega \gg \Lambda$. We then computed the overlap along the shell of the outwards-propagating inside positive-norm modes and the outside stationary modes. In the limit where the shell is close to the universal horizon, we show that the overlap between modes of opposite norms decreases exponentially in the radial momentum $P$. This result is directly related to the peculiar behavior of the momentum when approaching the universal horizon with a fixed Killing frequency, see \eq{Pomo}. Although this behavior was derived in a specific model, we then argued that it will be found for generic (spherically symmetric) regular collapses and superluminal dispersion relations. 
Let us stress that, although the system we consider is not expected to accurately describe the collapse of an initially regular matter distribution because of the discontinuity of the \ae ther field across the mass shell, we think that regularized ones will follow the same qualitative behavior. 
Indeed, the opposite result, namely a nonvanishing late-time Hawking radiation from the universal horizon, would require that deviations from the WKB approximation be stronger in the regular case than in the thin-shell limit. 
In other words, the nonadiabaticities required to have Hawking radiation would have to precisely match those introduced by taking the thin-shell limit for them to cancel each others when the shell's width is sent to zero. 
We thus conjecture that, assuming that deviations from the thin-shell limit considered above can be made small by continuously varying a parameter, the universal horizon of a black hole born by gravitational collapse will not radiate at late times.

As a result, irrespective of the model, at late time, the state of the outgoing field configurations is accurately described, for both positive and negative Killing frequencies, by the WKB modes with large positive momenta $P$ (and a positive norm). In this we recover the standard characterization of outgoing configurations in their vacuum state in the near-horizon geometry. Indeed, the condition to contain only positive momenta $P$ prevails for both relativistic and dispersive fields in the vicinity of the null horizon. The present work, therefore, shows that this simple characterization still applies in the presence of a universal horizon. 

Once this is accepted, the calculation of the asymptotic flux is also standard. It shows that for large black holes the thermality and the stationarity of the Hawking radiation are, to a good approximation, both recovered. This suggests that the laws of black hole thermodynamics should also be robust against introducing high-frequency dispersion.

As a corollary of the divergence of the radial momentum on both sides of the universal horizon, noticing that the inside configurations are blueshifted (towards the future), and that they have no common past with the outside configurations, it seems that the field state cannot satisfy any regularity condition across the universal horizon. It would be interesting to study the space of the field states, and determine whether some dispersive extension of the Hadamard condition can be imposed on the universal horizon. 
More generally, since particles originating from the singularity can fill the region inside the universal horizon, any prediction about the physics inside this region should be considered as speculative.

\section{Additional remarks}
\label{sec:app_uni}

\subsection{Wave equation and Bogoliubov coefficients}
\label{app:details}

In this subsection, we give the general formulas and main steps of the derivation of the results presented in Section~\ref{sec:Horava}.

\subsubsection{Preferred coordinates}
\label{app:PC}

The preferred coordinates $(t,X)$ are defined by the followng four conditions
\begin{itemize}
\item $s^\mu \pd_\mu = \pm \pd_X$ at fixed $t$;
\item $\pd_v = \pm \pd_t$ at fixed $X$;
\item $\pd_r T < 0$ along the shell trajectory;
\item $\pd_r X > 0$ along the shell trajectory.
\end{itemize}
These four conditions uniquely define $t$ and $X$. We find
\begin{equation}\label{eq:pref_t}
t = \left\lbrace
\begin{array}{cc}
v-r, & v<4M ,\\
v- r^*_U, 
 & v>4M \wedge r>M ,\\
- \lp v- r^*_U \rp,
& v>4M \wedge r<M,
\end{array}
\right.
\end{equation}
and
\begin{equation}
X = \left\lbrace
\begin{array}{cc}
r, & v<4M, \\
r^*_U, 
& v>4M \wedge r>M ,\\
- r^*_U, 
& v>4M \wedge r<M .
\end{array}
\right. 
\end{equation}
In these expressions, $r^*_U = r + M \ln | \frac{r}{M}-1 |$ is the tortoise coordinate defined with respect to the universal horizon. 

\subsubsection{Wave equation and scalar product}
\label{app:sc}

The action \eq{eq:action}, extended to complex values of $\Phi$ by replacing every other $\Phi$ by $\Phi^*$, has a $U(1)$ invariance under $\Phi \to \e^{\ii \theta} \Phi$, from which we derive the conserved current density
\begin{equation}\label{eq:current}
\mathcal{J}^\mu \equiv - \ii \sqrt{-g} \lp \Phi \nabla^\mu \Phi^* - \frac{1}{\Lambda^2} h^{\mu \nu} \lp \nabla_\nu \Phi \rp \lp \nabla_\rho h^{\rho \sigma} \nabla_\sigma \Phi^* \rp + \frac{1}{\Lambda^2} \Phi h^{\mu \nu} \nabla_\nu \nabla_\rho h^{\rho \sigma} \nabla_\sigma \Phi^* \rp + c.c.,
\end{equation}
where ``$c.c.$'' stands for the complex conjugate, satisfying
\begin{equation}
\pd_\mu \mathcal{J}^\mu = 0.
\end{equation}
As the wave equation \eq{fieldeq} is linear, one easily shows that $\mathcal{J}^\mu$ defines a conserved (indefinite) inner product in the following way. Considering two solutions $\Phi_1$ and $\Phi_2$ of \eq{fieldeq}, we first define $\mathcal{J}^\mu \lp \Phi_1, \Phi_2 \rp$ by replacing $\Phi^*$ by $\Phi_1^*$ and $\Phi$ by $\Phi_2$ in \eq{eq:current}. The inner product of these two solutions is then defined by
\begin{equation}\label{eq:inner}
\lp \Phi_1, \Phi_2 \rp_\tau \equiv \int \dd^3x \, n_\mu \mathcal{J}^\mu \lp \Phi_1, \Phi_2 \rp,
\end{equation}
where $n_\mu$ is a unit vector orthogonal to the 3-surface of constant $\tau$,
and $\tau$ is a time coordinate. When considering the 3-surfaces defined by $\tau = t$, the above overlap simplifies and gives the standard (Hamiltonian) conserved scalar product of \eq{scal}. 

\subsubsection{Matching conditions on the mass shell}
\label{app:match}

In order to compute the overlap of two modes defined on either side of the mass shell, we need the matching conditions to propagate the modes from the internal region to the external one and vice versa. As we now show, they appear naturally when considering the behavior of $\mathcal{J}^v \equiv \mathcal{J}^\mu \partial_\mu v$ across the shell. To see this, we first rewrite $\mathcal{J}^v \lp \Phi_1, \Phi_2 \rp$ as
\begin{align}\label{eq:innerbis} 
\mathcal{J}^v \lp \Phi_1, \Phi_2 \rp = & - \ii \lp \Phi_2 \sqrt{-g} \lp \nabla^v + \frac{1}{\Lambda^2} h^{v \mu} \nabla_\mu \nabla_\rho h^{\rho \sigma} \nabla_\sigma \rp \Phi_1^* 
- \frac{1}{\Lambda^2} \sqrt{-g} \lp h^{v \mu} 
\nabla_\mu \Phi_2 \rp \lp \nabla_\rho h^{\rho \sigma} \nabla_\sigma \Phi_1^* \rp \rp
\nn
&- \lp \Phi_1^* \leftrightarrow \Phi_2 \rp. 
\end{align}
Inspecting \eq{fieldeq} and requiring that the second term has no singularity which cannot be canceled by the first one, we find that the quantities $\Phi$, $\sqrt{-g} h^{0 \nu} \nabla_\nu \Phi$, $\nabla_\rho h^{\rho \sigma} \nabla_\sigma \Phi$, and \\ $\sqrt{-g} \lp \nabla^0 + \frac{1}{\Lambda^2} h^{0 \nu} \nabla_\nu \nabla_\rho h^{\rho \sigma} \nabla_\sigma \rp \Phi$ are continuous across $v=4M$. 
Since the complex conjugate of a solution of \eq{fieldeq} is still a solution, this applies to $\Phi = \Phi_1^*$ as well as $\Phi = \Phi_2$. Therefore, in evaluating \eq{eq:innerbis} one can evaluate $\Phi_1^*$ and the operators acting on it on one side of the shell,  $v = 4M - \epsilon$, $\epsilon \to 0$, while $\Phi_2$ and the operators acting on it are evaluated on the other side $v = 4M + \epsilon$.

\subsubsection{Calculation of \texorpdfstring{$\gamma_{\omega,\lambda}$}{}} 
\label{app:alpha}

Let us consider two radial modes known on different sides of the mass shell: $\Phi_1$ is known for $v<4M$ and $\Phi_2$ for $v>4M$. The complete expression of the scalar product in the $v,r$ coordinates is somewhat cumbersome, but it greatly 
simplifies in the relevant limit where
\begin{itemize}
\item $\Phi_1$ has a large frequency $\left\lvert \omega \right\rvert \gg \Lambda$;
\item $\Phi_2$ has a large wave vector $\left\lvert k_{v,2} \right\rvert \gg \lambda, \Lambda$.
\end{itemize}
We have introduced the wave vector $k_v \equiv \pd_r S$ at a fixed $v$. For the modes we are interested in, $k_{v,2} = \lp \lambda_2 + P_2 \rp / x$ and $\omega$ are of the order $\Lambda/x^2$, where $x = (r/M)-1$. Keeping only the leading terms in the inner product then gives
\begin{align}
\lp \Phi_1, \Phi_2 \rp_v \approx \frac{4 \ii \pi}{\Lambda^2} \int \dd r \lp 
-\psi_2 (\pd_v+\pd_r)^3 \psi_1^* + \lp 1- \frac{M}{r} \rp \lp \pd_r \psi_2 \rp \lp \pd_v+\pd_r \rp^2 \psi_1^* 
\right. \nn \left.
+ \psi_1^* \lp 1- \frac{M}{r} \rp^3 \pd_r^3 \psi_2 - \lp \lp \pd_v + \pd_r \rp \psi_1^* \rp \lp \frac{M}{r} -1 \rp^2 \pd_r^2 \psi_2
\rp,
\end{align}
with relative corrections of order $x$. When choosing for $\psi_1$ the $\textrm{in}$ mode of frequency $\omega$, and for $\psi_2$ the stationary WKB mode of \eq{eq:WKBmodes} with the large momentum \eq{Pomo}, we get
\begin{align}
\lp \Phi_1, \Phi_2 \rp_v & \approx 4 \pi M \int_{x>0} \dd x \lp \frac{P_{\omega}^3}{\Lambda^2} \pm \frac{P_{\omega}^2}{\Lambda x} \pm \frac{\Lambda}{x^3} + \frac{P_\om}{x^2} \rp \psi_1^* \psi_2 \nn
& \approx \frac{M \e^{4 i M \lp \omega - \lambda \rp} \e^{-\ii M (\omega + P_{\omega})}}{4 \pi \sqrt{\Lambda \left\lvert \omega \lp \frac{\dd  \om}{\dd P} \rp_1 \right\rvert}} \int_{x>0} \dd x \lp \frac{P_{\omega}^3}{\Lambda^2} \pm \frac{P_{\omega}^2}{\Lambda x} \pm \frac{\Lambda}{x^3} + \frac{P_\om}{x^2} \rp 
\nn
& \exp 
\lp \ii \lp \mp \frac{\Lambda M}{x} + \lp 2 \lambda \pm \Lambda \rp M \ln \left\lvert x \right\rvert - M \lp \omega + P_{\omega} \rp x \rp
\rp .
\end{align}
In this equation, as well as in the remainder of this subsection, the sign $\pm$ discriminates between $\gamma$ and $\delta$; see below. In the large-frequency limit, we evaluate this integral through a saddle point approximation. The possible saddle points are the values of $x$ where 
\begin{equation}
\frac{\dd}{\dd x} \lp \mp \frac{\Lambda M}{x} -M \lp \omega + 
P_{\omega} \rp x \rp \approx \frac{\dd}{\dd x} \lp \mp \frac{\Lambda M}{x} -M \omega x \rp = 0,
\end{equation}
i.e.,
\begin{equation}
x^2 \approx \pm \frac{\Lambda}{\omega}.
\end{equation}
This is very similar to the saddle point condition applied to the Bogoliubov coefficients describing the scattering of plane waves on a uniformly accelerated mirror~\cite{Obadia:2002ch, Obadia:2002qe}.

The coefficient $\gamma_{\omega, \lambda}$ is defined for $\pm \omega > 0$. Since the integral runs over $x>0$, we must choose the saddle point $x^*$ at
\begin{equation}
x^*_\gamma \approx \sqrt{\frac{\Lambda}{\left\lvert \omega \right\rvert}}.
\end{equation}
We get
\begin{equation}
\gamma_{\omega, \lambda} \approx \pm \sqrt{\frac{\mp \ii M}{2 \pi \left\lvert \omega \right\rvert}} \exp \lp \ii M \lp 3 \omega - 4 \lambda -P_{\omega} \mp 2 \sqrt{\Lambda \left\lvert \omega \right\rvert} +\frac{1}{2} \lp 2 \lambda \pm \Lambda \rp \ln \lp \frac{\Lambda}{\left\lvert \omega \right\rvert} \rp \mp \Lambda \rp \rp .
\end{equation}
It is easily shown that, under these approximations, the following unitarity relation is satisfied:
\begin{equation}
\int_0^\infty \dd\omega \gamma_{\omega,\lambda}^* \gamma_{\omega,\lambda'} \approx \delta (\lambda - \lambda').
\end{equation}
This indicates that the $\delta_{\omega,\lambda}$ coefficients are suppressed in the limit $\omega \to \infty$, as they would otherwise contribute to the unitarity relation. 

\subsubsection{Calculation of \texorpdfstring{$\delta_{\omega,\lambda}$}{}} 
\label{app:beta}

The calculation of $\delta_{\omega,\lambda}$ follows the same steps. The saddle point equation now is
\begin{equation}
x^{*2}_\delta  = -\frac{\Lambda}{\left\lvert \omega \right\rvert}. 
\end{equation}
To be able to deform the integration contour to include the saddle point, we must choose the solution in the half-plane where the exponential decreases, i.e.,
\begin{equation}
x^*_\delta = - {\rm sgn} \lp \omega \rp \ii \sqrt{\frac{\Lambda}{\left\lvert \omega \right\rvert}}.
\end{equation}
The exponential factor in the integral then gives a suppression factor 
\begin{equation}
\exp \lp -M \lp 2 \sqrt{\Lambda \left\lvert \omega \right\rvert} + \pi \lp -{\rm sgn} \lp \omega \rp \lambda + \frac{\Lambda}{2} \rp + \Lambda \rp \rp.
\end{equation}

In addition, to the order to which the calculation was performed, the prefactor vanishes. As the first relative corrections from neglected terms are of order $O(x^*) = O(\sqrt{\Lambda/\left\lvert \omega \right\rvert})$, we get 
\begin{equation}
\delta_{\omega,\lambda} = O \lp \frac{\sqrt{M \Lambda}}{\left\lvert \omega \right\rvert} \rp \exp \lp -2 M \sqrt{\Lambda \left\lvert \omega \right\rvert} \rp.
\end{equation}

\subsection{Acceleration of the \ae ther field}
\label{app:acc}

The acceleration of the \ae ther field is
\begin{equation}
\gamma^\mu = u^\nu \nabla_\nu u^\mu.
\end{equation}
Using \eq{eq:uands}, this gives for $v \neq 4M$ 
\begin{equation}
\gamma^\mu \gamma_\mu = -\frac{M^2}{r^4} \Theta (v - 4 M).
\end{equation}

For completeness, we now show that, in effectively $(1+1)$-dimensional setups (e.g. under the assumption of spherical symmetry), a stationary universal horizon requires that the \ae ther field has a nonvanishing acceleration, thereby generalizing what was found in de Sitter space in~\cite{Busch:2012ne}. We consider a stationary space-time with Killing vector $K^\mu$, endowed with a timelike \ae ther field $u^\mu$. The universal horizon is defined as the locus where $K^\mu u_\mu = 0$. (Notice that the Killing field must thus be spacelike on the universal horizon.) In particular, $K^\mu$ cannot be aligned with $u^\mu$. Using the Killing equation, the variation of $K^\mu u_\mu$ along the flow of $u^\mu$ is
\begin{equation}
u^\mu \nabla_\mu \lp K^\nu u_\nu \rp = K_\mu \gamma^\mu.
\end{equation}
If $u^\mu$ is freely falling, $\gamma^\mu = 0$ and $u^\mu$ is tangent to the hypersurfaces of constant $K^\mu u_\mu$. In particular, it is tangent to the universal horizon. In $1+1$ dimensions, since $K^\mu$ and $u^\mu$ cannot be aligned, $K^\mu$ is not a tangent vector to the universal horizon, which is thus not stationary. Models with a stationary universal horizon are thus in a different class than those studied in~\cite{Coutant:2011in}.

\begin{figure}
\centering
\includegraphics[width=0.48 \linewidth]{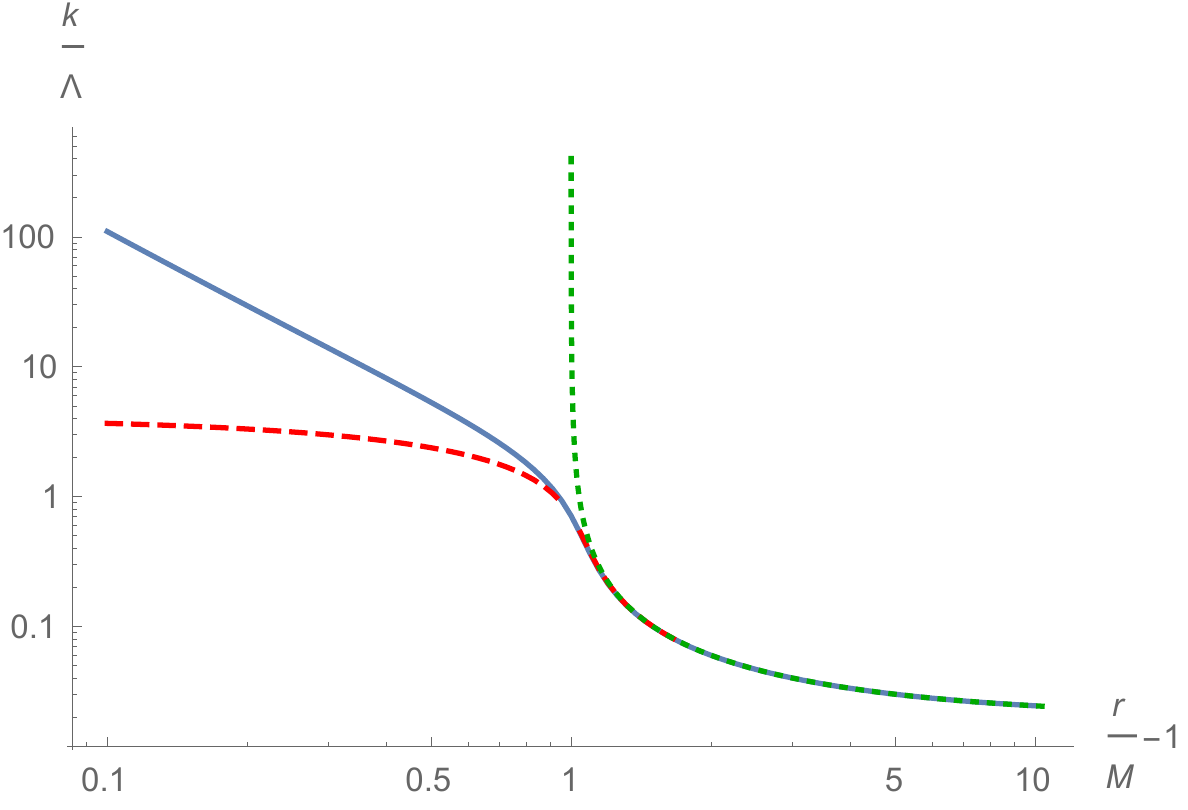}
\caption{Comparison of the wave vectors $k(r,\lambda)/\Lambda$ as a function of $x = r/M - 1$ (both in logarithmic scales) in the relativistic case (green, dotted line), for a freely falling preferred frame (red, dashed line), and in the model of Section~\ref{sec:Horava} (blue, solid line). The Killing frequency is $\lambda = 10^{-2} \Lambda$, and the null horizon is located at $x = 1$. One clearly sees the unbounded growth of the relativistic wave number. More importantly, one also sees that the two dispersive wave vectors behave in the same manner across the null horizon. Hence the acceleration of $u^\mu$ has a significant effect on $k$ only when approaching the universal horizon. } \label{fig:HvsC}
\end{figure}

To see the combined effects of the dispersion and acceleration, we show in \fig{fig:HvsC} the local value of the wave vector in the $v,r$ coordinates, $k_v$, for the outgoing $u$ mode, as a function of $r$. We compare three models with the same parameters, and for $\lambda = 10^{-2} \Lambda$. The blue, solid curve shows the result for the model of Section~\ref{sec:Horava}. The green, dotted curve shows the relativistic case. The red, dashed one shows the result for a dispersive model with a nonaccelerated preferred frame chosen to coincide with the \ae ther frame of Section~\ref{sec:Horava} at $r=2M$.~\footnote{It must be noted that this model is not well defined for $r \to \infty$. The reason is that at $r=2M$, we have $u \cdot \pd_v = 1/2$. A nonaccelerated vector field $w$ which coincides with $u$ at $r=2M$ must thus satisfy the two conditions $w \cdot \pd_t = 1/2$ and $w \cdot w = 1$ at $r=2M$. From the free-fall condition, these two properties extend in the whole domain where the preferred frame is defined. Since they are incompatible in Minkowski space, we deduce that the domain in which the preferred frame can be defined does not extend to $r \to \infty$. A straightforward calculation shows that it extends up to $r = 8 M /3$. However, as this model is well defined close to and inside the null horizon, it can be used to see the qualitative differences between the nonaccelerated and accelerated cases.} We see in \fig{fig:HvsC} that the three models give very similar results for $r$ larger than $2M$. Close to $r=2M$, the relativistic wave vector diverges, while the nonaccelerated dispersive model still closely follows the accelerated one. When $r$ is further decreased, the predictions of the two models separate: the nonaccelerated one gives a finite wave vector at $r=M$ while the accelerated one gives $k \propto \lp r-M \rp^{-2}$. 

\subsection{Hawking radiation from the null horizon}
\label{app:Hawking}

In Section~\ref{sec:Horava}, we showed that the late-time emission from the universal horizon is governed by Bogoliubov coefficients which are exponentially suppressed when the inside frequency $\omega$ is much larger than the outside one $\lambda$. This result was obtained using the WKB approximation of the stationary modes just outside the universal horizon. This approximation is trustworthy as we verified that the deviations from the WKB treatment go to zero when approaching the universal horizon. This implies that, at late time, the inside vacuum is adiabatically transferred across the shell. Hence, the $u$ part of the field state can be accurately described by the WKB mode $\psi^{u, \rm in}_{\lambda}$ with high (preferred) momentum for both signs of $\lambda$. In this we recover the situation described in~\cite{Brout:1995wp,Balbinot:2006ua,Coutant:2011in}. Therefore, the nonadiabaticity that will be responsible for the asymptotic radiation lies in the propagation from the universal horizon to spatial infinity and the late-time emission should essentially come from the stationary scattering near the null horizon. Hence, we expect to get a nearly thermal spectrum governed by the surface gravity of the  null horizon, and with deviations in agreement with those numerically computed in~\cite{Finazzi:2012iu,Robertson:2012ku}.

To verify this conjecture, we numerically propagate the outgoing mode $\phiu{\lambda}$ from a large value of $r/2M$ down inside the trapped region to $r \to M^+$. This mode can be written in the limits $r \to M$ and $r \to \infty$ as 
\begin{align}
& \phiu{\lambda}(r) \mathop{\sim}_{r \to \infty} \psiu{\lambda} + A_\lambda \psiv{\lambda}, \nn
& \phiu{\lambda}(r) \mathop{\sim}_{r \to M} \alpha_\lambda \psi^{u, \rm in}_{\lambda} + \beta_\lambda (\psi^{u, \rm in}_{-\lambda})^*, 
\end{align}
where the WKB modes are as described in Section~\ref{sec:Horava}. The coefficient $A_\lambda$ governs the grey body factor. In our (1+1)-dimensional model, we have verified that it plays no significant role. Hence, as usual, the Hawking effect is essentially encoded in the mode mixing of $u$ modes of opposite norms.  

\begin{figure}
\includegraphics[width=0.48 \linewidth]{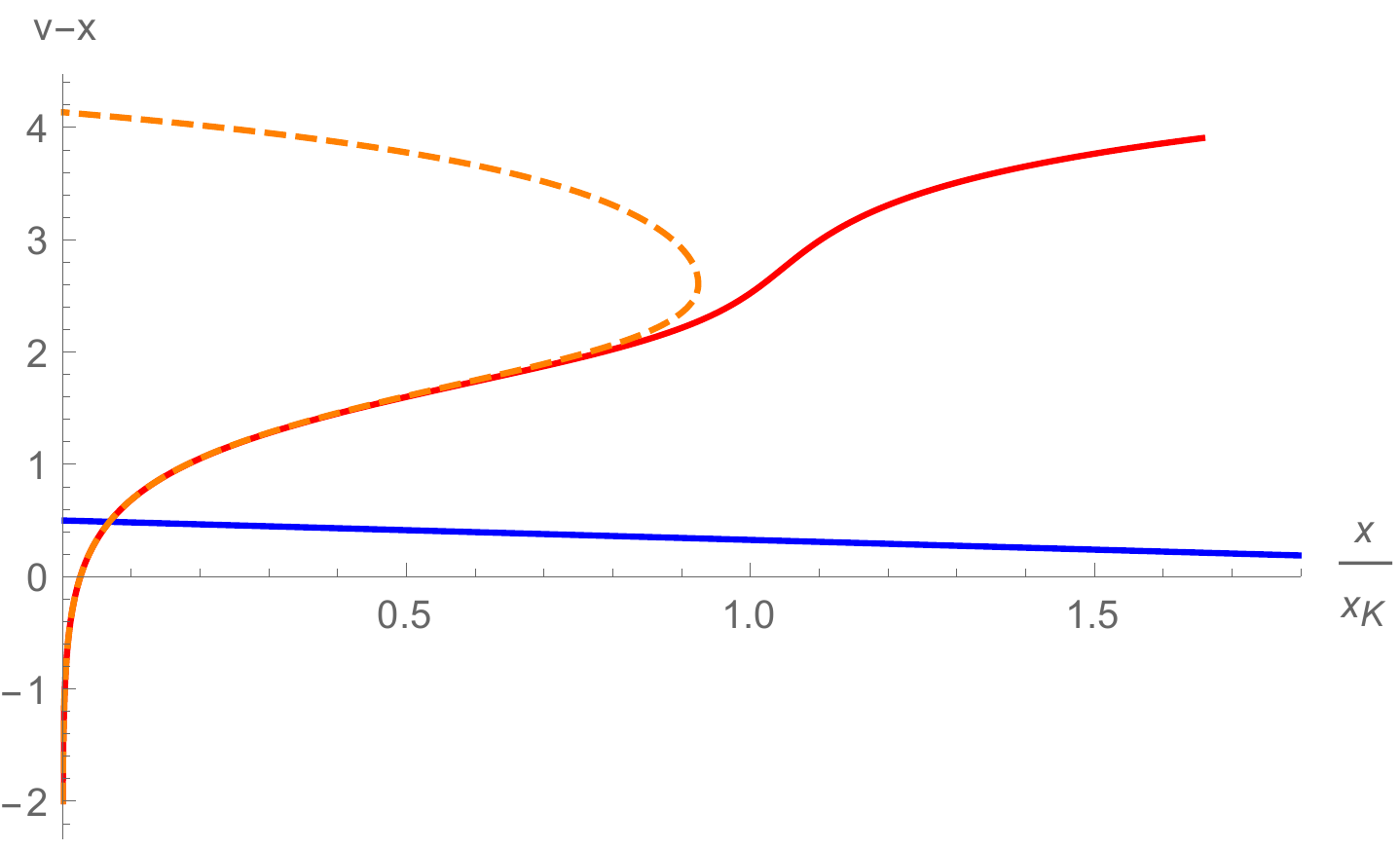} 
\includegraphics[width=0.48 \linewidth]{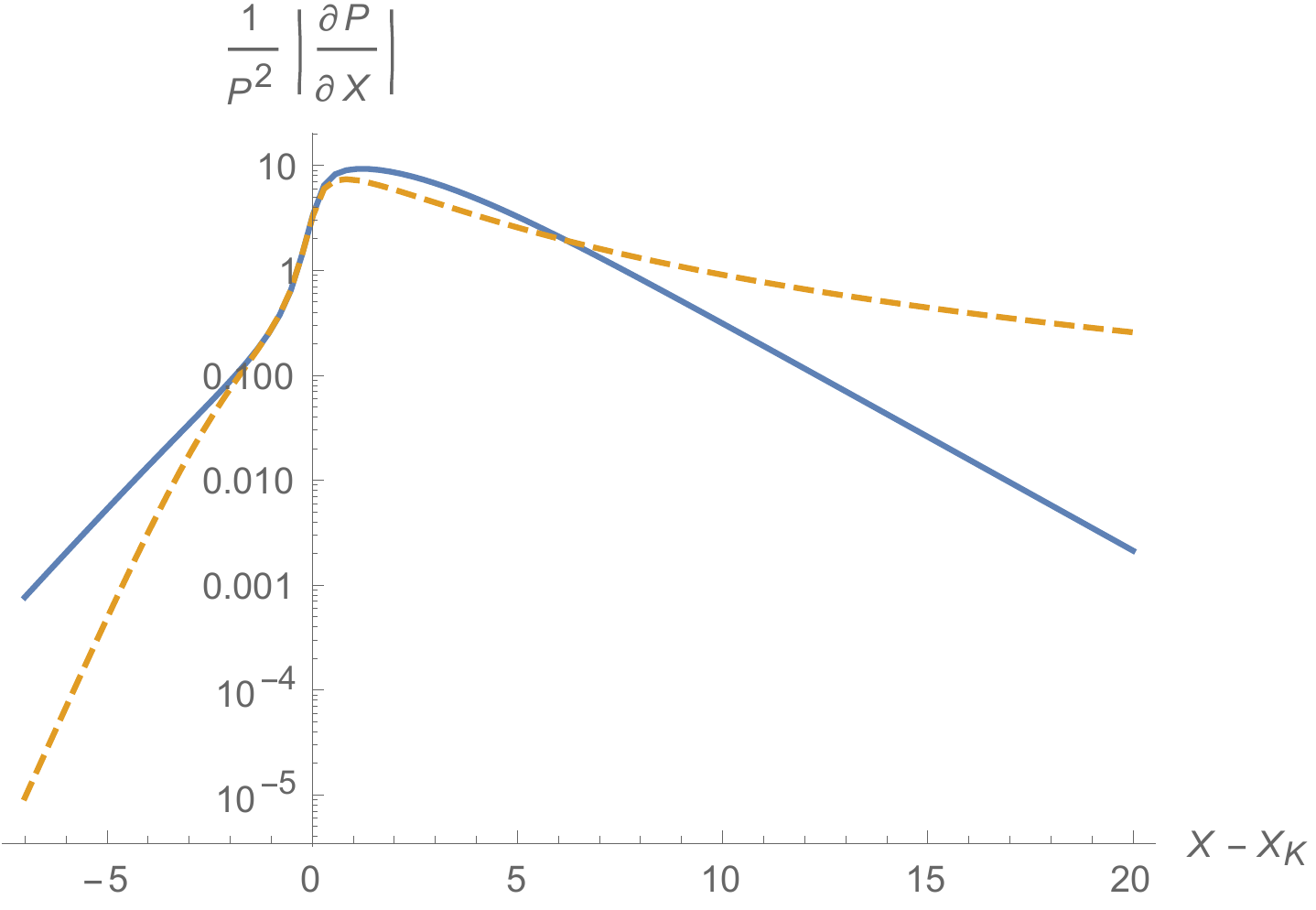}
\caption{Left panel: Characteristics in the $v-x,x/x_K$ coordinates, where $x \equiv r-r_U$ and $x_K$ gives the location of the null horizon. Only the region outside the mass shell and universal horizon is represented. The parameters are $\Lambda/\kappa \sim 1.1$, $X_0 = 0.5$, and $\lambda/\Lambda =0.01$. 
Right panel: Amplitude of the nonadiabatic corrections in the present model (blue, solid line) and the one from Section~\ref{sec:Horava} (orange, dashed line) for $\lambda / \Lambda=0.01$, as a function of the preferred coordinate $X -X_K$, where $X_K$ denotes the position of the null horizon. In this example, the surface gravity is close to $\Lambda$. We verify that the norm of the coefficient $\beta_\lambda$ is of the order of the maximal value of the nonadiabatic parameter, as expected from the analysis of nonadiabaticity~\cite{Massar:1997en}.
} \label{fig:characbis}
\end{figure}

To efficiently perform the numerical analysis, we regularized the metric and \ae ther field. In practice we worked with a metric of the form
\begin{equation}
ds^2 = \lp 1 - 2 f(r) \rp dv^2 - 2 dv dr,
\end{equation}
and a unit-norm \ae ther field
\begin{equation}
u^\mu \pd_\mu = \pd_v - f(r) \pd_r.
\end{equation}
These expressions generalize the model of Section~\ref{sec:Horava} which is recovered for $f(r) =M/r$. The null horizon corresponds to $f(r)=1/2$, and the universal horizon to $f(r)=1$. We can then define the preferred coordinate $X$ along the lines of subsection~\ref{app:PC}. For the numerical integration of \eq{eq:2DFE}, it is appropriate to work with $f$ expressed as a known function of $X$. A convenient choice is 
\begin{equation}\label{eq:frX} 
f(r(X)) = \frac{1}{2} \lp 1- \eta \tanh \lp \frac{X}{X_0} \rp \rp.
\end{equation}
$\eta$ is a positive parameter which must be equal to 1 to have an asymptotically flat space at $r \to \infty$ and a universal horizon at $r \to M$, i.e., $X \to -\infty$. In our numerical simulations, we worked with $\eta < 1$ to avoid large numerical errors due to the divergence of the dispersive roots $P^{\rm in}_{\pm \lambda}$ close to the universal horizon. We then checked that the scattering coefficients become independent of $\eta$ in the limit $\eta \to 1$, as should be the case since the WKB approximations become exact on both sides. 
The advantage of this model is that the metric coefficients and the \ae ther field converge exponentially to their asymptotic values so that the asymptotic modes become exact solutions for $r \to \infty$, provided the decay rate of the exponentially decaying mode is small enough.~\footnote{Notice that an exponential convergence for $X \to -\infty$ is required to have a universal horizon at a finite value of $r = r_U$ with $\dd f/\dd r (r=r_U) \neq 0$.} The characteristics outside the universal horizon $r=r_U$ are shown in \fig{fig:characbis}. 
They exhibit the main properties of \fig{fig:char}. In the right panel we show the parameter governing nonadiabatic corrections, $\lp \pd_X P_\lambda \rp / P_\lambda^2$ as a function of the preferred coordinate $X$ in the present model and the one of Section~\ref{sec:Horava}. In the present model, they go to zero exponentially both for $X \to \infty$ and $X \to -\infty$. 
In the model of Section~\ref{sec:Horava}, they decay exponentially at $X \to -\infty$ but only polynomially at $X \to \infty$. The two models become equivalent, in the sense that the value of $P_\lambda^u(r)$ follows the same law, close to the null horizon when working with the same surface gravity.

\begin{figure}
\centering
\includegraphics[width=0.48 \linewidth]{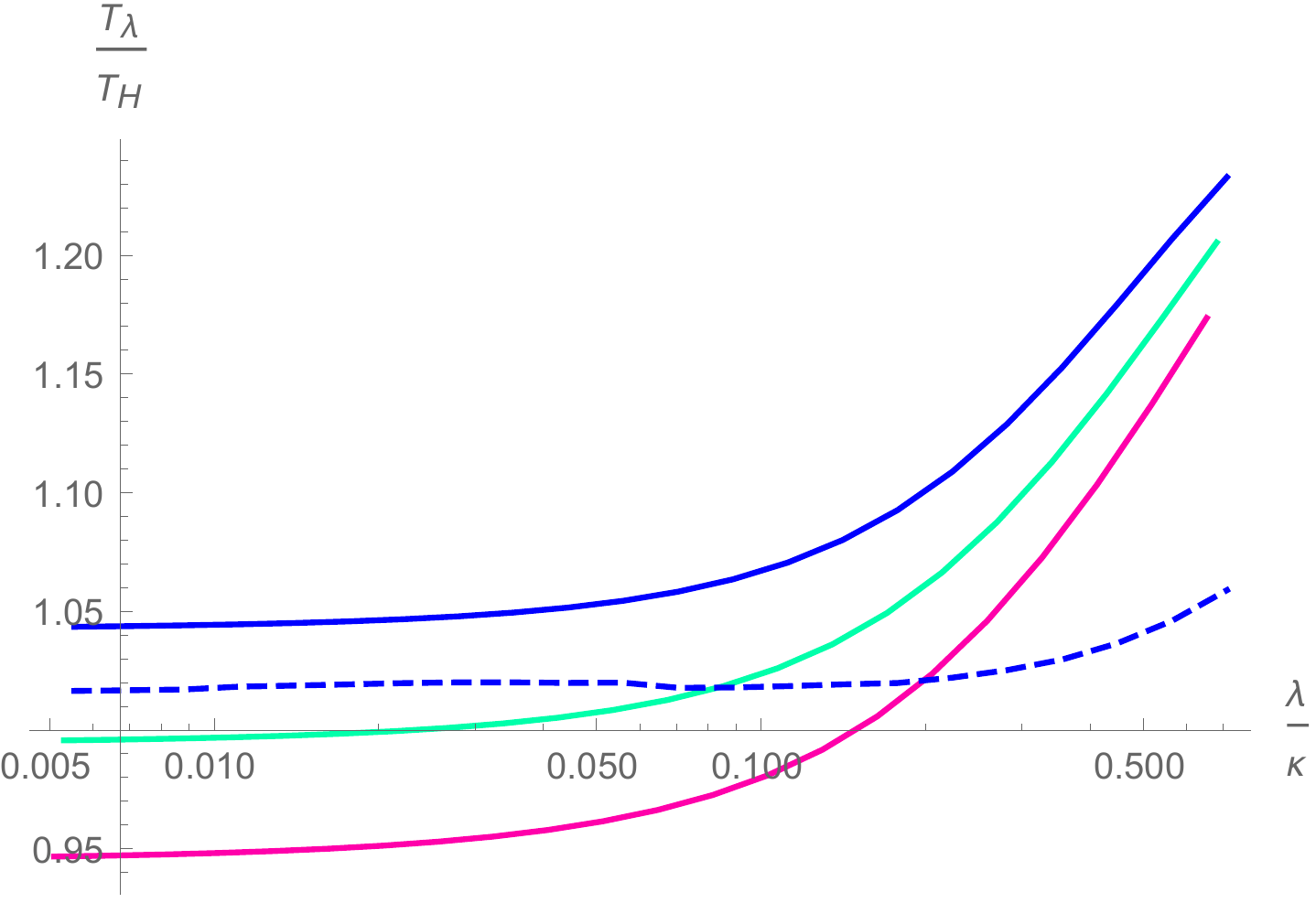} 
\caption{Plot of the effective temperature $T_\lambda$ of \eq{eq:Teff} divided by the Hawking temperature as a function of $\lambda / \kappa$, where $\kappa$ is the surface gravity. The values of $\Lambda/\kappa$ are $1/2$ (solid line) and $3/2$ (dashed line). For the smallest value of $\Lambda$, we show the results for $\eta = 0.9$ (blue line), $0.94$ (cyan line), and $0.98$ (magenta line). For the largest value of $\Lambda$, these three curves are undistinguishable up to numerical errors. This indicates that the limit $\eta \to 1^-$ is well defined, which we checked using a larger range of values for $\eta \in (0.8, 0.99)$. Moreover, when increasing the dispersive scale $\Lambda$, we see that $T_\lambda$ closely agrees with the Hawking value $\kappa/2\pi$ for a larger domain of Killing frequencies.
} \label{fig:KillingScattering}
\end{figure}

The field equation was integrated numerically using~\cite{Mathematica7} and the same techniques as in~\cite{Finazzi:2012iu,Michel:2014zsa}. The results are shown in \fig{fig:KillingScattering}. We obtain two important results. First, at fixed $\Lambda$ and $\kappa$, the effective temperature $T_\la$, defined by
\begin{equation}\label{eq:Teff} 
\left\lvert \beta_\lambda \right\rvert^2 = \frac{1}{e^{\lambda/T_\lambda} - 1}, 
\end{equation}
becomes independent of the regulator $\eta$ as $\eta \to 1^{-}$. Second, at low frequencies $\lambda \ll \Lambda$, we get a Planckian spectrum, i.e., $T_\lambda$ is close to a constant, with deviations from the Hawking temperature compatible with the results of~\cite{Macher:2009tw,Finazzi:2012iu}. This establishes that the propagation between the two horizons does not alter the thermal character of the outgoing spectrum.

\begin{figure}
\centering
\includegraphics[width=0.5 \linewidth]{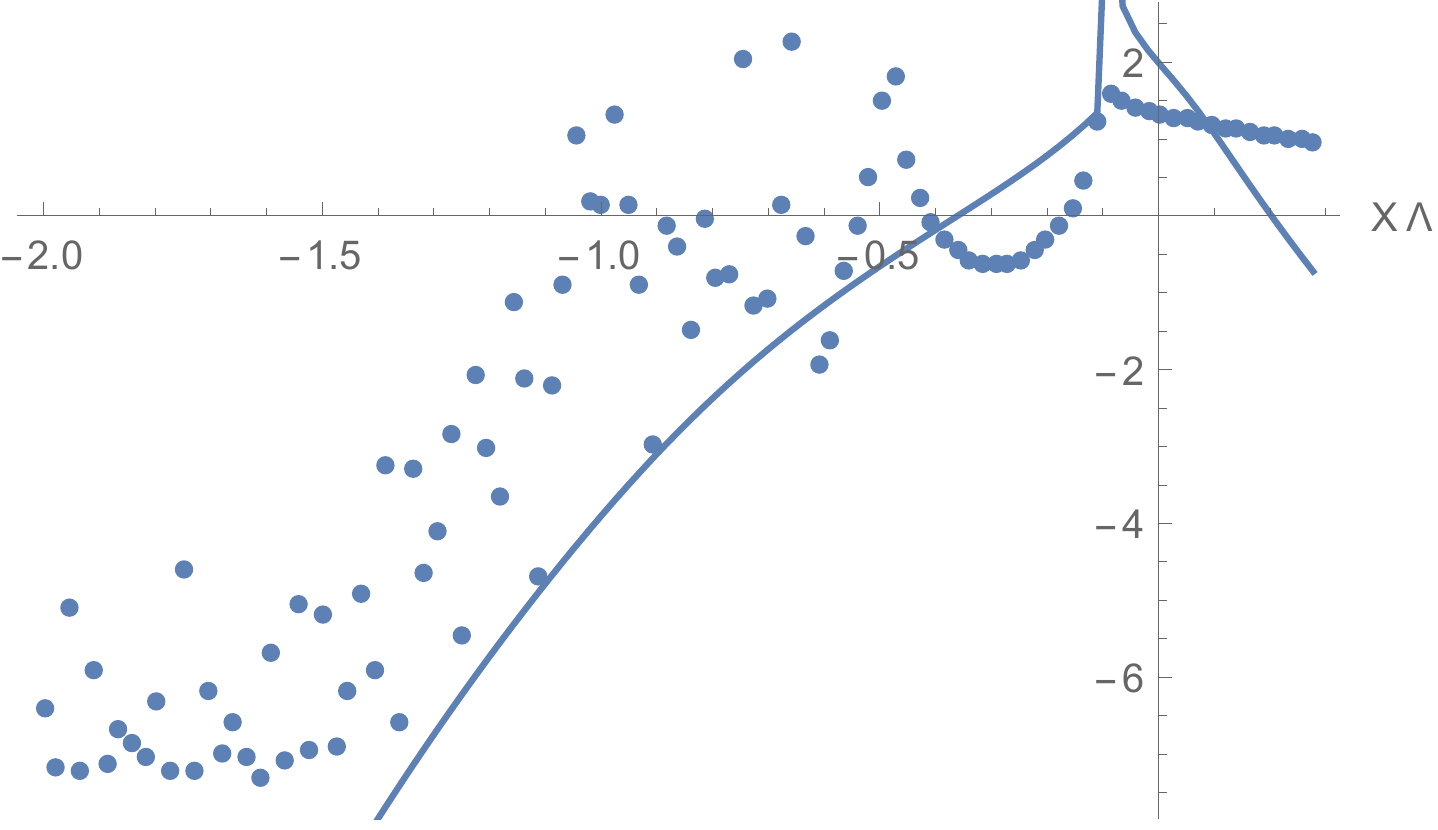} 
\caption{Deviations from the WKB approximation. The solid line shows the natural logarithm of $\left\lvert \lp \pd_X P_\lambda^u \rp / \lp P_\lambda^u \rp^2 \right\rvert$ as a function of the preferred coordinate $X$ in units of $1/\Lambda$, and the points show the logarithm of the relative difference between the solution computed numerically and the corresponding sum of the WKB modes, for $\Lambda=1$, $X_0 = 0.5$, $\eta = 0.99$, and  $\lambda=0.1$. The important spread seems to be due to numerical errors resulting from the increase of the momentum as $X \to -\infty$. The null horizon is located at $X=0$. 
} \label{devWKB}
\end{figure}

To conclude this numerical analysis, we numerically verify that the WKB approximation becomes exact when approaching the universal horizon. To this end, we show in \fig{devWKB} the logarithm of the relative deviation between the numerical solution and the weighted sum of the WKB waves $\alpha_\lambda \psi^{u, \rm in}_{\lambda} + \beta_\lambda (\psi^{u, \rm in}_{-\lambda})^*$. As $X \to -\infty$, we clearly see that the numerical values of the deviations decay following the rough estimation of the WKB corrections given by $\left\lvert \lp \pd_X P_\lambda^u \rp / \lp P_\lambda^u \rp^2 \right\rvert$. (The relatively important spread and the plateau for $X < -1.8$ seem to be due to numerical errors. Indeed, as the wave vector becomes very large, typically of order $10^2$, even a relatively small error in its value gives important and rapidly oscillating errors.) 

\newchapter{Other works and conclusion}
\label{ch:concl}
\begin{tikzpicture}[overlay]
\newcommand*{\xA}{-0.2}
\newcommand*{\xB}{12.2}
\newcommand*{\yA}{4.5}
\newcommand*{\yB}{1.5}
\newcommand*{\epsil}{0.75}
\draw[overlay] (\xA-\epsil,\yA) -- (\xB-\epsil,\yA);
\draw[overlay] (\xA,\yA+\epsil) -- (\xA,\yB);
\draw[overlay] (\xB,\yB-\epsil) -- (\xB,\yA);
\draw[overlay] (\xB+\epsil,\yB) -- (\xA+\epsil,\yB);
\end{tikzpicture}
\begin{flushright}
``Ego plane meis adici posse multa confiteor, nec his solis, sed et omnibus quos edidi'' \\
Gaius Plinius Secundus (23 -- 79), \textit{Naturalis Historia, I}
\end{flushright}
\vspace*{0.5 cm}
\begin{small}
Before concluding, I briefly describe the other works done during the last years. 
My aim is to give the main ideas and results while skipping the unessential technicalities. 
These studies were, in my opinion, not as important as those presented in the previous chapters for the main objectives of this thesis, either because their results were not used in later works, because their completion built on techniques already developed rather than introducing new ones, or because my personal contribution to them was less fundamental. 
However, taken together they underline both the links between the various concepts I used and possible extensions, within and beyond analogue gravity.
I then make concluding remarks to put the achievements of this thesis in a more general framework and, maybe more importantly, on the path which lies ahead.
\end{small}
\newpage


\section[Suppression of infrared instability in trans-sonic flows by condensation of zero-frequency short-wavelength phonons]{Suppression of infrared instability in trans-sonic flows by condensation of zero-frequency short-wavelength phonons \cite{Busch:2014hla}}

It was shown in~\cite{Coutant:2012mf} that analogue white hole flows, which are supercritical in the upstream region and subcritical in the downstream one, have an infra-red instability due to the divergence of the coefficient $\beta_\om$ relating incident, counter-propagating, positive-energy waves to negative-energy outgoing ones in the small-frequency limit $\om \to 0$.  
Although this is only an energetic instability (i.e., there exists negative-energy waves) and not a dynamical instability (there is no  exponentially-growing mode in time), it was predicted to generate a macroscopic undulation with an amplitude growing either linearly or logarithmically in time, depending on the nature of the fluctuations. 
This undulation would start close to the analogue horizon and propagate in the subcritical region (for water waves) or the supercritical one (for density perturbations in Bose-Einstein condensates) depending on the sign of the dispersive term in the dispersion relation. Notice that this mechanism is directly related to the Hawking effect. Indeed, the divergence of $\beta_\om$ as $\om^{-1/2}$ is a consequence of the approximately thermal nature of the spectrum of Hawking radiation, which implies that
\begin{align}
\abs{\beta_\om}^2 \approx \frac{1}{\e^{\hbar \om / (k_B T_H)} - 1}
\mathop{\sim}_{\om \to 0} \frac{k_B T_H}{\hbar \, \om},
\end{align}
where $T_H$ denotes the Hawking temperature. 

Besides this mechanism, undulations can also be produced by flows over an obstacle, or in a potential which depends on $x$, even in the absence of fluctuations. 
Indeed, in general the fluid equations have no solution which is asymptotically homogeneous on both sides for given values of the asymptotic current and density or water depth. 
A flow which is asymptotically homogeneous on one side will thus generally have a finite-amplitude undulation on the other side. 
Undulations are thus a very generic feature of analogue white hole flows. 
Moreover, as mentioned in Chapter~\ref{ch:probing}, they can induce a resonance as their wave vector precisely corresponds to the dispersive zero-frequency root of the dispersion relation. 
One can thus expect that they will have an important effect on the scattering of incoming waves if they extend over many wavelengths.  

\begin{figure}
\includegraphics[width = 0.49 \linewidth]{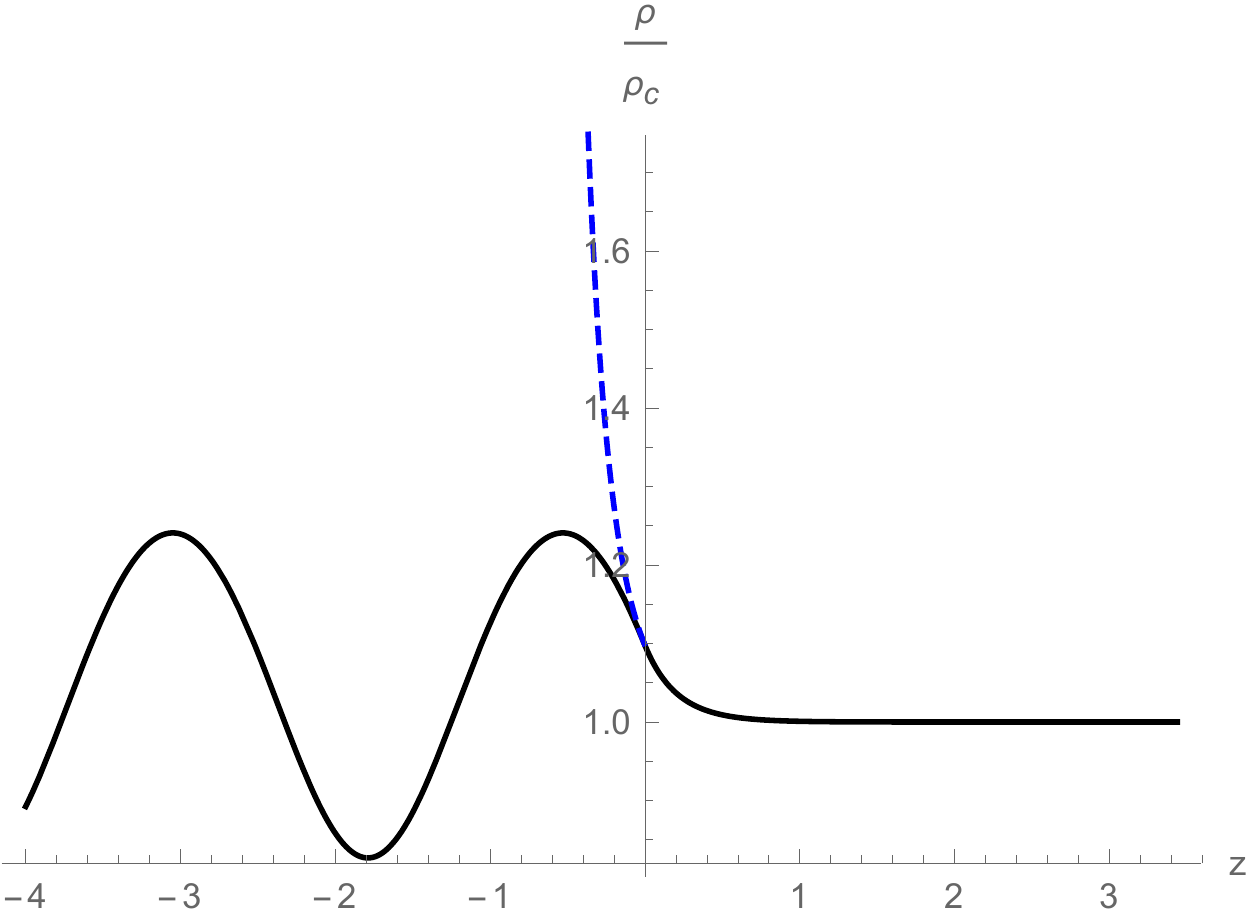}
\includegraphics[width = 0.49 \linewidth]{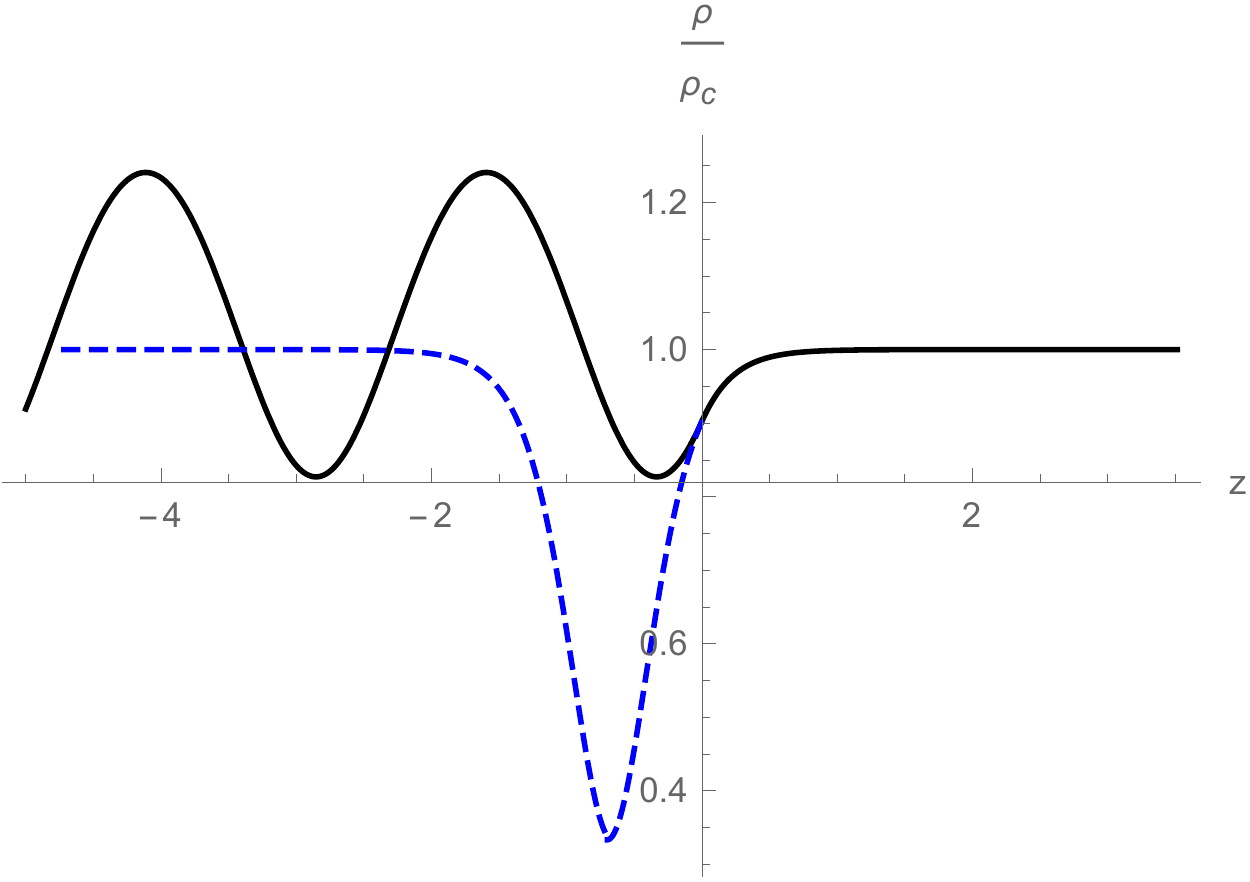}
\caption{We show two solutions of the Gross-Pitaevskii equation in a step-like potential which are asymptotically homogeneous in the downstream subsonic (right) region and contain an undulation in the upstream supersonic (left) one. 
The blue, dashed lines represent a shadow soliton (left) and a soliton (right) in a homogeneous potential. 
The parameters are $g(z>0)=8$, $g(z<0)=1$, $V(z>0)=-28/3$, $V(z<0)=-7/3$, and $J=\sqrt{8/3}$. 
}\label{fig:shsolund}
\end{figure}

In~\cite{Busch:2014hla}, we addressed this issue for white hole flows in one-dimensional Bose-Einstein condensates. 
We first characterized the set of solutions in asymptotically uniform potentials, imposing that
\begin{itemize}
\item the flow is subsonic in the downstream region and supersonic in the upstream one (it then has an analogue white hole horizon);
\item the density $\rho$ is asymptotically homogeneous in the subsonic region.\footnote{In the model we considered, all solutions which do not satisfy this condition have a divergence at a finite distance from the horizon, and are thus not physical.}
\end{itemize} 
We found that, for given values of the potential shape, coupling constant $g$, and asymptotic current $J$, there exists a one-dimensional family of such solutions, which generally have an undulation in the supersonic region. 
Moreover, its minimum amplitude does not vanish unless the potential is fine-tuned. 
In general, the phase of the undulation is fixed only up to $\pi$. 
The two possible phases may be seen as the limits of the ``shadow soliton'' and ``soliton'' solutions obtained in a black hole laser configuration (see Chapter~\ref{ch:saturation}) in the limit where the length of the supersonic region is sent to infinity. 
They are represented in \fig{fig:shsolund}. 

We then solved the Bogoliubov-de Gennes equation numerically over these solutions to determine the effect of the undulation on the spectrum. 
To be able to define the scattering coefficients in the usual way, we worked with localized undulations, whose amplitudes go to $0$ at large values of the space coordinate $-z$. 
While here introduced artificially, this damping is not unphysical: in experiments (see for instance in~\cite{Euve:2014aga, Euve:2015vml}), the undulation amplitude also goes to zero when moving away from the horizon due to dissipation. 
We studied how $\beta_\om$ is modified by the presence of the undulation, in dependence of its amplitude, length, and shape of the damping function. 
We first found that a ``detuned'' potential shape for which all solutions have a nonvanishing undulation does not introduce any significant difference with respect to the ``tuned'' case where an asymptotically homogeneous solution exists.  
We then studied the latter more specifically, and found that the phase of the undulation plays a crucial role. 
For the phase corresponding to the soliton in \fig{fig:shsolund}, the undulation \textit{increases} the divergence in $\beta_\om$. 
This effect is linear in the amplitude and length of the undulation for sufficiently small ones. 
When increasing the amplitude, the infra-red instability even turns to a dynamical instability, i.e., a mode which grows exponentially in time. 

\begin{figure}
\centering
\includegraphics[width = 0.49\linewidth]{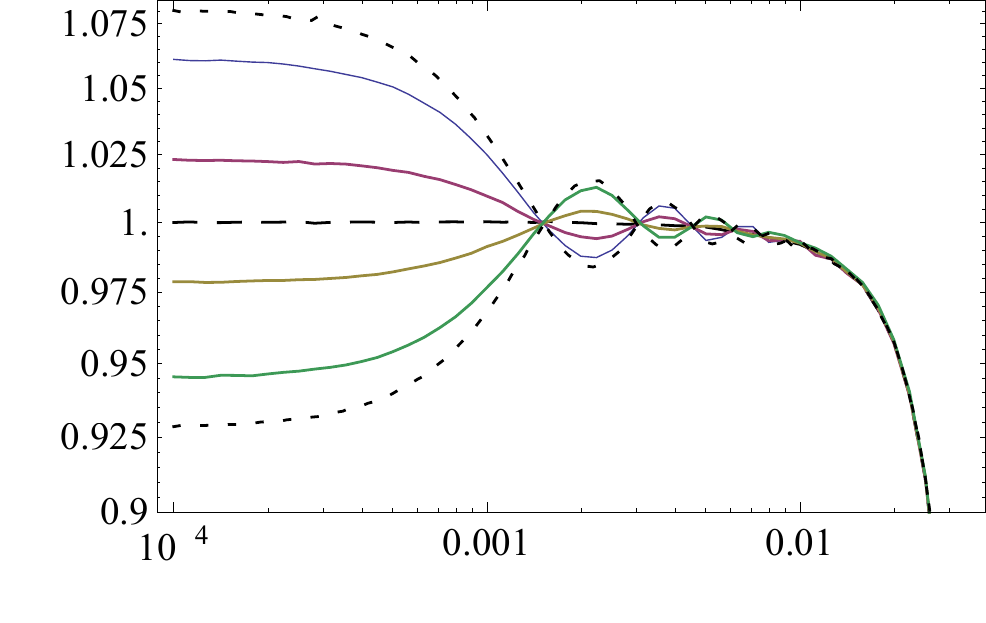} \,
\includegraphics[width = 0.49\linewidth]{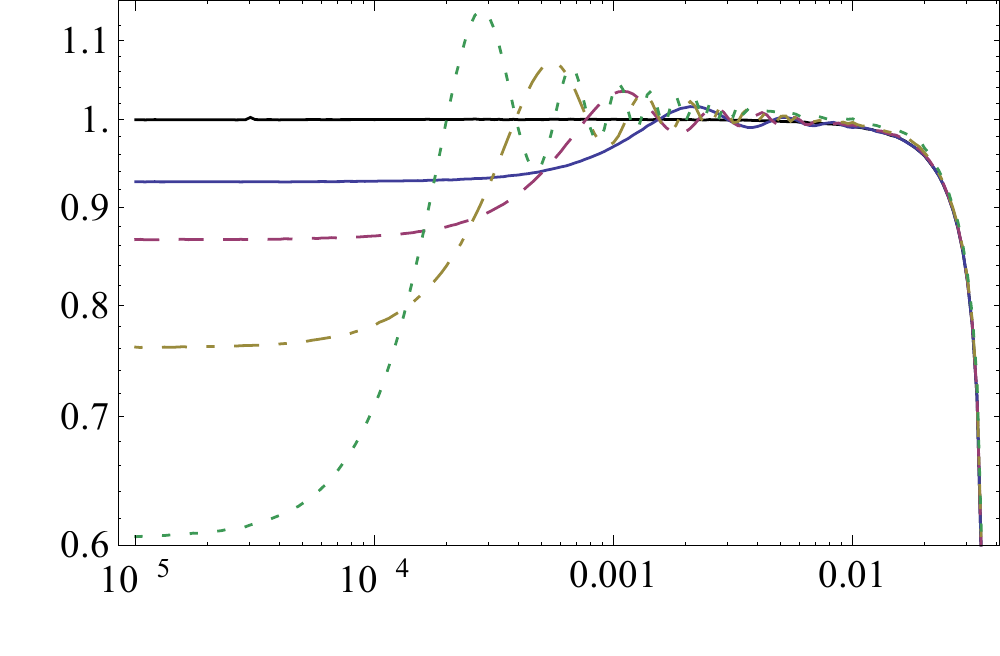}
\begin{tikzpicture}[overlay]
\newcommand*{\vl}{5}
\newcommand*{\hl}{210}
\draw (120pt-\hl,22pt-\vl pt) node{$\om$};
\draw (0pt-\hl,90pt-\vl pt) node{\rotatebox{90}{$T_\om / T_0$}};
\draw (330pt-\hl,22pt-\vl pt) node{$\om$};
\draw (215pt-\hl,90pt-\vl pt) node{\rotatebox{90}{$T_\om / T_0$}};
\end{tikzpicture}
\caption{Effective temperature of \eq{eq:concl_Teff} as a function of $\om$, for various undulations. 
$T_0$ denotes the value of the low-frequency temperature in the absence of undulation. 
The left panel shows the effect of the phase: all the undulations have the same amplitude and length but the position of the small obstacle producing them is varied to change their phases. 
The right panel shows the effect of the length of the undulation for a phase corresponding to the ``shadow soliton'' in \fig{fig:shsolund}, with number of undamped oscillations equal to $20$ (blue, continuous), $40$ (purple, dashed), $80$ (yellow, dash-dotted), and $160$ (green, dotted). 
The continuous black line shows the results without undulation. 
The potentials and other parameters used are given in~\cite{Busch:2014hla}. 
}\label{fig:concl_Teff}
\end{figure}
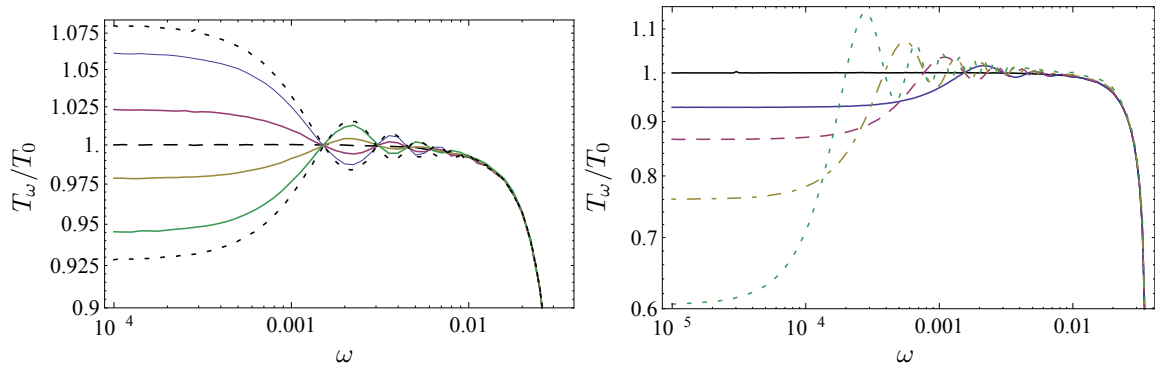

On the contrary, undulations corresponding to the ``shadow soliton'' solution tend to \textit{reduce} the divergence in $\beta_\om$ provided the damping at large values of $-z$ is smooth enough. 
The infra-red divergence, and the low-frequency emission through Hawking effect, are thus weaker. 
Our numerical results indicate that they are completely suppressed in the limit where the length of the undulation is sent to infinity and the damping becomes very smooth. 
This means that, in this limit, the undulation exactly cancels the Hawking effect for $\om \to 0$. 
Results from a few numerical simulations are shown in \fig{fig:concl_Teff}, where we plot the effective temperature, defined by
\begin{align} \label{eq:concl_Teff}
T_\om \equiv \frac{\om}{\ln \lp 1 + \abs{\beta_\om}^{-2} \rp},
\end{align}
as a function of $\om$ for undulations with different lengths and a phase corresponding to the ``shadow soliton'' solution. 

We have also performed numerical simulations for water waves and found similar results. 
For the moment we have no precise explanation for this behavior beyond the fact that the equality between the wave vector of the undulation to linear order and the dispersive, low-frequency root of the dispersion relation can trigger a resonance and thus an important modification of the scattering coefficients even for small amplitudes.  
The analysis sketched in section~\ref{sec:scat_und} of Chapter~\ref{ch:probing} may give some elements towards its understanding. 

Finally, we note that the suppression of the infra-red instability reported in~\cite{Busch:2014hla} bears many similarities with that of the roton-maxon instability studied in~\cite{Pitaevskii1984}. 
There is an important difference, however: In the roton-maxon case the system is stabilized because the negative-energy modes are removed from the spectrum by nonlinear effects. 
In the present case instead, negative-energy waves are still present. 
But their coupling to positive-energy ones seems sufficiently reduced to restore stability.

\clearpage

\section[Wave blocking and partial transmission in subcritical flows over an obstacle]{Wave blocking and partial transmission in subcritical flows over an obstacle \cite{Euve:2014aga}}
\label{sub:exp_paper_1}

The work presented in this article reports, and analyses the data from, observations made in the Pprime institute in Poitiers about water waves propagating on a counter-current with a localized obstacle. 
The experimental setup consisted in a flow of water (maintained by a pump) in a flume. 
In the downstream region, a guillotine was moved up and down periodically using an electric motor, producing waves which propagated against the flow. 
An obstacle with shape similar to that of used in~\cite{Weinfurtner:2010nu} put at the bottom of the flume made the flow accelerate in a localized region. 
The incident wave was partially transmitted and partially reflected by the flow inhomogeneity, see \fig{fig:concl_exp_schema1}. 
Contrary to~\cite{Weinfurtner:2010nu} we focused on transmission, which was measured using two acoustic sensors in the upstream region. 
(For more details, see the subsection 2.A of~\cite{Euve:2014aga}.)
 
We worked with five different values of the current $q$ and asymptotic water depth $h_d$ so as to distinguish the generic features from the peculiarities of each realization.
One important point is that these flows remained globally subcritical: the Froude number $v / c$ was smaller than unity even at the top of the obstacle. 
This property, which is apparently also true for the experiments reported in~\cite{Rousseaux:2007is,Weinfurtner:2010nu}, implies that there was no analogue horizon in the strict sense. 
This choice was made for two reasons. 
The first, technical, one is that it seems difficult to obtain a stable, transcritical flow with obstacles and parameters close to those we used, while subcritical ones can more easily be generated for a wide range of values for $q$ and $h_d$.  
The second reason is that working in conditions similar to those of~\cite{Rousseaux:2007is,Weinfurtner:2010nu} allows a precise comparison between the various results, which helps understand the physics at play in the three experiments. 

The two acoustic sensors monitored the height of the free surface as a function of time, allowing for the determination of the amplitude of its oscillations in the upstream region and thus of the transmission coefficient $|\tilde{A}_\om|$ (defined in \eq{eq:Bsub}). 
Our main objective was to see whether $|\tilde{A}_\om|$ goes to $1$ in the limit $\om \to 0$, as predicted by the analytical and numerical results of~\cite{Michel:2014zsa}(see Chapter~\ref{ch:probing}). 
Results for two different flows are shown in \fig{fig:concl_exp_res1}. 
Although the experimental data (blue curves) are less smooth than the numerical predictions (green and red curves), the agreement is relatively good. 
In particular, the transition from $|\tilde{A}_\om| \approx 0$ to $|\tilde{A}_\om| \approx 1$ in an angular frequency interval of extension close to $2 {\rm Hz}$ when decreasing $\om$ below  $\om_{\rm min}$ is well reproduced. 
This justifies \textit{a posteriori} the approximations done in the numerical analysis, in particular the use of a dispersion relation truncated to quartic order. 

Numerical results were obtained using two different codes. 
The first one, giving the green lines in \fig{fig:concl_exp_res1}, was already used in the study reported in Chapter~\ref{ch:probing}. 
As explained there, it relies on a quartic approximation of the dispersion relation. 
One problem with this approach is that the value of $\om_{\rm min}$, which is the main parameter determining the behavior of the transmission coefficient, is then different than the one from the full dispersion relation. 
Moreover, the quartic approximation is generally not very good close to the turning point of modes with $\om < \om_{\rm min}$.
To circumvent these difficulties, we also used a slightly different code where the coefficients of the quadratic and quartic terms in $\pd_x$ are adjusted in a $\om$-dependent way so that the approximate dispersion relation is always tangent to the exact one at the turning point when there is one, or at the top of the obstacle (where most of the scattering takes place for a sufficiently smooth obstacle\footnote{It is not clear whether this is correct for the obstacle we used because of the sharp upstream slope. However, given the smallness of the differences between results obtained with the two codes, refining the position where the dispersion relations match does not seem crucial at this stage.}) when there is none.  
Results obtained with this code are shown by red, dashed lines in \fig{fig:concl_exp_res1}. 
One can see that the differences between the two numerical estimations, in green and red, are relatively small, indicating that the quartic approximation of Chapter~\ref{ch:probing} is a fairly good one in the present context.

Another feature of this figure is the presence of sharp peaks at small frequencies. 
For the time being we do not have any precise explanation for their presence. 
They may be due to reflections at one or both ends of the flume, involving interferences between incident and reflected waves, or even resonances if the reflection on both ends is important. 

The main addition of~\cite{Euve:2014aga} with respect to the similar experiment~\cite{Weinfurtner:2010nu} was to measure the transmission coefficient, which was not reported in the latter and which we believe to be important for the interpretation. 
As the figure shows, it seems to be in good agreement with the theoretical and numerical predictions, showing in particular that at low frequencies it is this effect which dominates, the production of dispersive waves (analogous to the Hawking effect) being a subdominant process. 
Although we were not able to precisely measure with our setup the quantity $\abs{\beta_\om / \alpha_\om}$, whose behavior is the main result of~\cite{Weinfurtner:2010nu}, we made a few preliminary measurements which are compatible with the values reported in this reference. 

Besides measuring the transmission coefficient in subcritical flows over an obstacle, this work is also a first step towards the transcritical case where the analogy with black-hole physics is clearer. 
In completing it, several hurdles were cleared both on the experimental and numerical side, which paved the way for the next experiment~\cite{Euve:2015vml} and, hopefully, other works to come with higher values of the Froude number and/or a way to measure more accurately the values of $\abs{\alpha_\om}$ and $\abs{\beta_\om}$.   

\begin{myfig}
\includegraphics[width = \linewidth]{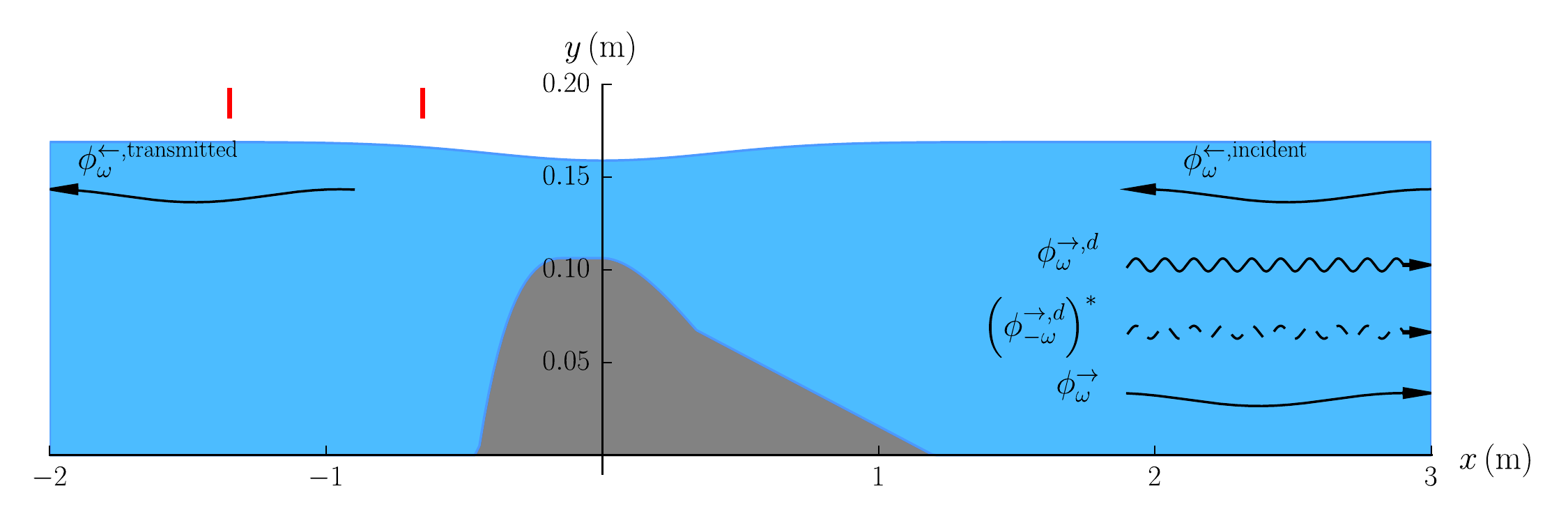}
\capf{Schematic representation of the experimental setup, with water flowing from left to right. 
The grey area shows the obstacle. 
Wiggly arrows correspond to the 5 waves involved in the scattering, the hydrodynamic ones with long wavelength and the dispersive ones with short wavelength. 
They are oriented in the direction of the group velocity. 
The dashed arrow represents the negative-energy wave. 
The two red rectangles show the horizontal position of the upstream acoustic sensors. 
Notice that the horizontal and vertical scales differ by a factor $7.5$, making the obstacle look steeper than it actually is. 
}\label{fig:concl_exp_schema1}
\end{myfig}

\begin{myfig}
\includegraphics[width = 0.49 \linewidth]{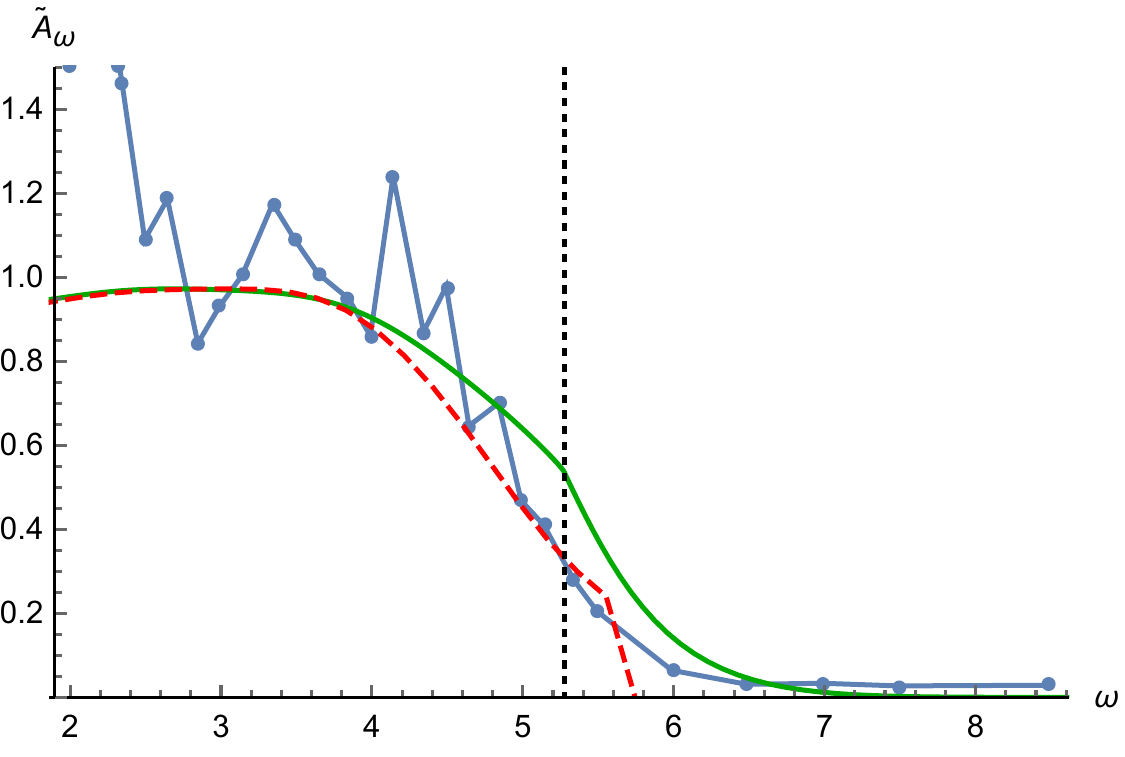}
\includegraphics[width = 0.49 \linewidth]{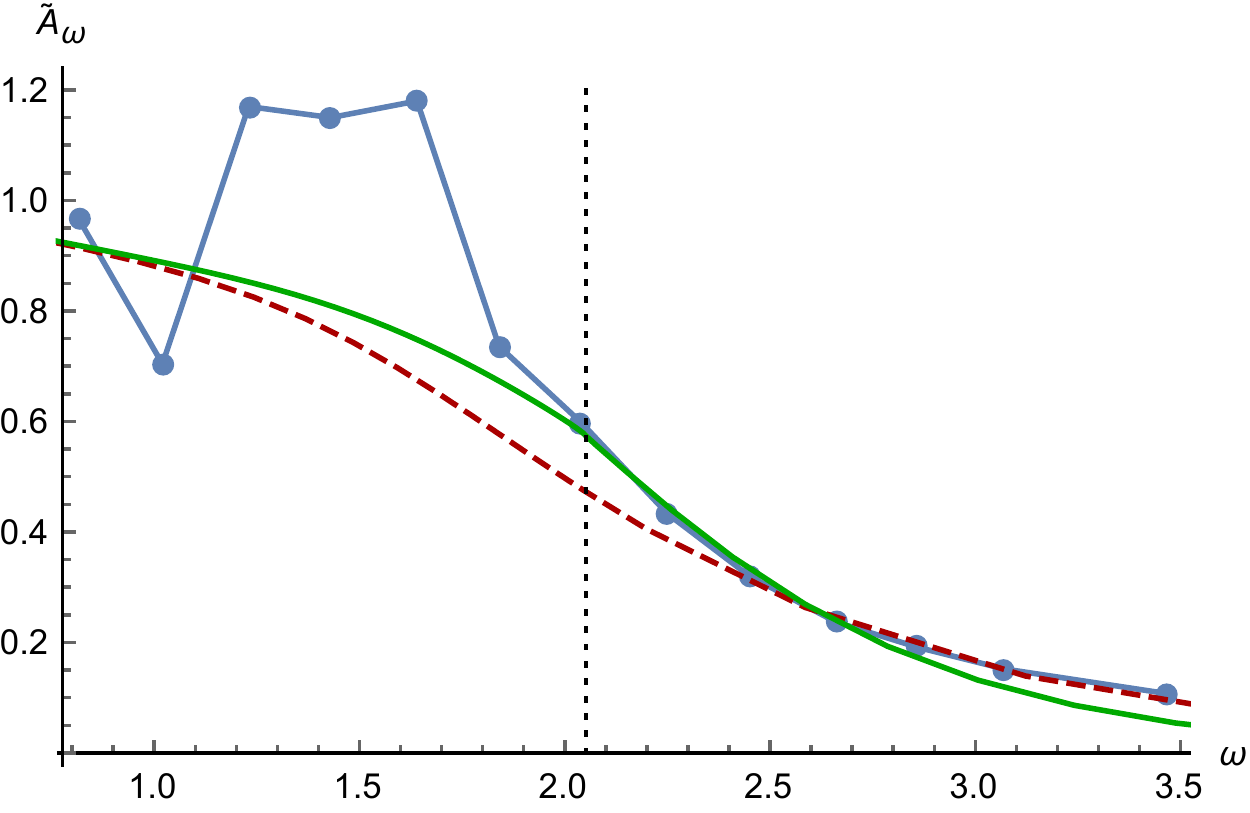}
\capf{Comparison between the numerical predictions and experimental observations for the transmission coefficient of counter-propagating surface water waves in a counterflow with a localized obstacle similar to the one used in~\cite{Weinfurtner:2010nu}. 
Green, continuous lines show the predictions obtained with the code already used in Chapter~\ref{ch:probing}. 
Red, dashed lines show the results using a slightly different code with parameters tuned to match the value of $\om_{\rm min}$ of the full dispersion relation. 
Blue dots are the experimental results. 
The unit of $\om$ is the Hz. 
The left panel shows results for a downstream asymptotic water depth $h_d \approx 0.181 {\rm m}$ and a current $q \approx 0.0276 {\rm m^2 . s^{-1}}$. 
The right panel corresponds to $h_d \approx 0.194 {\rm m}$ and $q \approx 0.045 {\rm m^2 . s^{-1}}$, close to the values used in~\cite{Weinfurtner:2010nu}. 
In each panel, the vertical dotted line indicates the angular frequency $\om_{\rm min}$ above which the characteristics have a turning point. 
}\label{fig:concl_exp_res1}
\end{myfig}

\clearpage

\section[Nonlinear effects in time-dependent transonic flows: An analysis of analog black hole stability]{Nonlinear effects in time-dependent transonic flows: An analysis of analog black hole stability \cite{Michel:2015pra}}

This work was in many ways an intermediate step between~\cite{Michel:2013wpa} and~\cite{Nohair}.  
In the first one, we focused on two problems in the context of black hole lasers: the stationary, nonlinear solutions on one side and the linear, time-dependent perturbations of the homogeneous solution on the other side. 
Deep links between them were conjectured using the properties of an energy functional which seems to accurately govern their behavior under the approximation of small nonlinearities. 
But a precise understanding of the interplay between nonlinear and time-dependent effects was lacking. 
The main aim of~\cite{Michel:2015pra} was to clarify these links by analyzing the nonlinear effects in time-dependent, transonic Bose-Einstein condensates, using the same mean-field approximation. 
It consists of two parts: one dealing with flows whose velocities cross the speed of sound only once, i.e., black and white hole flows, and one on asymptotically subsonic flows with a finite supersonic region, i.e., black hole lasers. 

Let us first focus on the evolution of perturbed black and white hole solutions. 
As done in Chapter~\ref{ch:saturation} for black hole lasers, we focused on the case of step-like potentials, now with only one discontinuity. 
After a brief review of the set of stationary solutions, we showed results from numerical resolutions of the one-dimensional Gross-Pitaevskii equation. 
We developed a code to solve a discretized one-dimensional Gross-Pitaevskii equation on a torus with periodic boundary conditions. 
There are thus two horizons;  
the torus was chosen long enough so that the perturbations going from one horizon to the other during the time of the simulation have a negligible amplitude, so that the evolution of the configurations close to each horizon can be decoupled from each other. 
The initial conditions consisted in homogeneous configurations with densities slightly different from those of known stationary solutions. 

\begin{figure}
\includegraphics[width=0.49\linewidth]{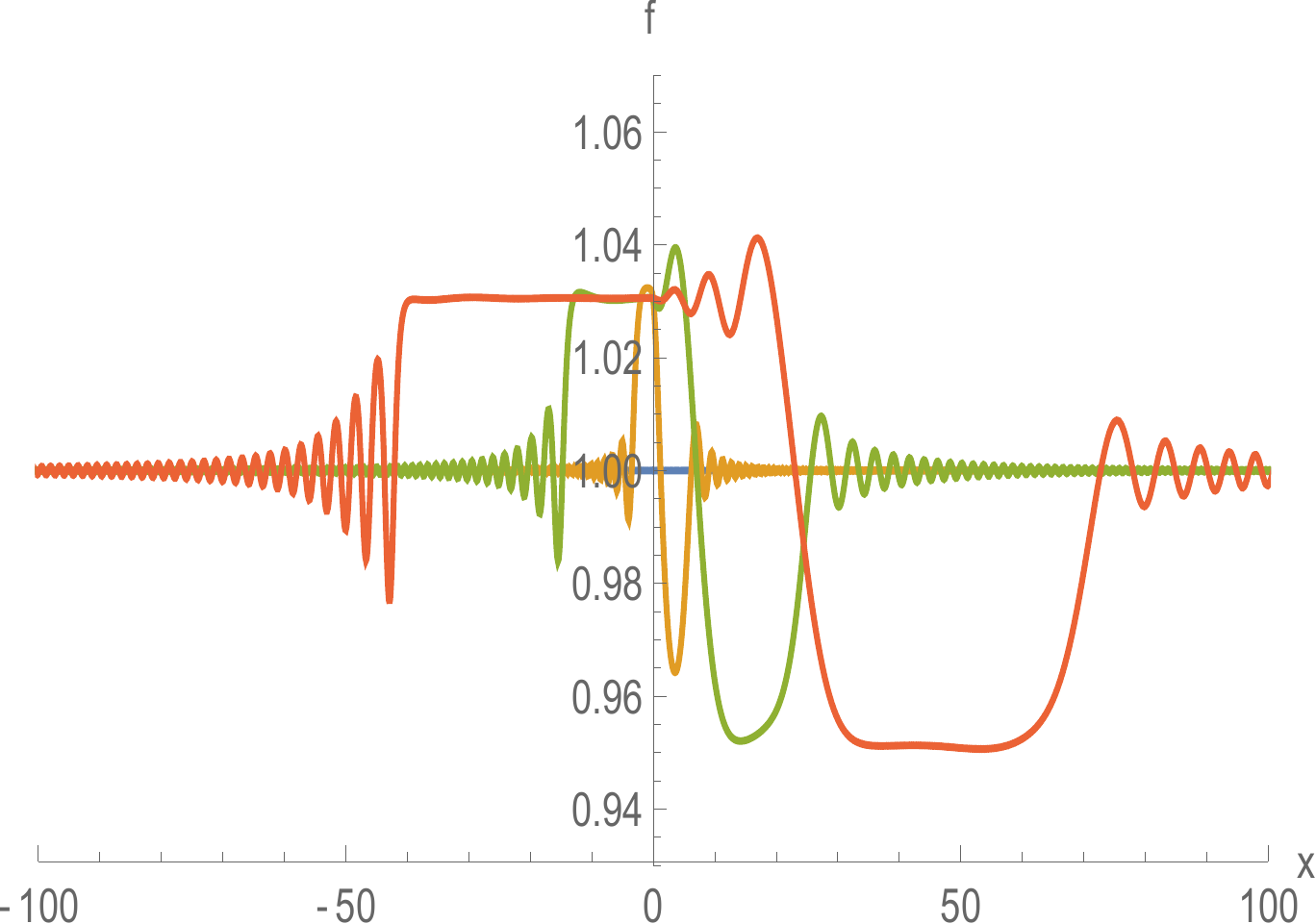}
\includegraphics[width=0.49\linewidth]{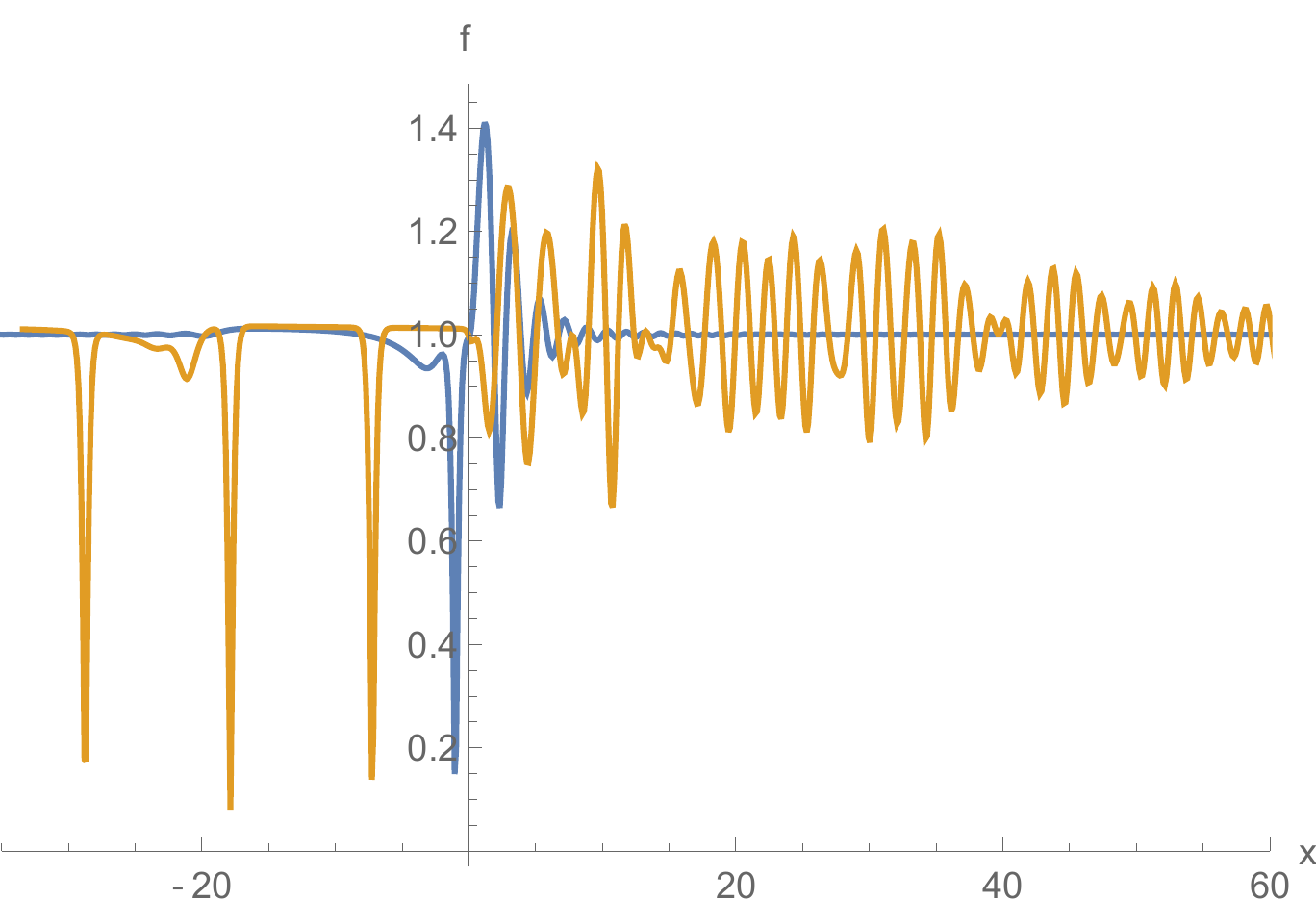}
\caption{Time evolution of $f \equiv \sqrt{\rho}$ in a black hole (left) and a white hole (right) flows subject to a homogeneous perturbation at $t = 0$. 
The parameters are chosen such that a homogeneous stationary solution exists with $f \approx 1.03$ (they are given in the captions of Figs.~3 and~5 of~\cite{Michel:2015pra}). 
The analogue horizon is located at $x = 0$. 
In the left panel, the solution is shown at the non-dimensional times $t = 0$ (blue), $t = 2$ (orange), $t = 10$ (green), and $t = 30$ (red). 
Three nonlinear waves are emitted from the near-horizon region, leaving behind them a homogeneous solution with $f \approx 1.03$, in good agreement with the square root of the density of the homogeneous black hole. 
In the right panel, the solution is shown at $t = 4$ (blue) and $t = 30$ (orange). 
While the parameters and initial conditions are similar to those in the left panel (we slightly changed the former to obtain more visual plots), its evolution is radically different: a perturbed undulation is emitted in the supersonic region and large-amplitude solitons are sent in the subsonic one in an apparently periodic way. 
}\label{fig:concl_BHWHrho}
\end{figure}

We observed very different features between black and white holes, see \fig{fig:concl_BHWHrho}. 
Around a black hole, the initial perturbation is expelled through the emission of three nonlinear waves. This can be motivated by looking at the possible scenarii under the assumptions of weak nonlinearity and that the solution becomes stationary at late times.~\footnote{Similar arguments also tell that these solutions can not exist around white hole horizons as the group velocities of the possible waves have the wrong sign to linear order, i.e., they would move towards the horizon, and not away from it.} 
In the Appendix~C, we derived a few properties of the emitted waves, assuming that the mass and momentum they carry grows slowly in time. 
This laid the foundation for an analytical understanding of this emission process in~\cite{Nohair}, after realizing that these waves are scale-invariant, and to turn these numerical observations into a more precise conjecture, see Chapter~\ref{ch:nohair}. 

The evolution close to the white hole shows a richer phenomenology, which to the best of our knowledge is still not understood analytically. 
It also seems closely related to the aforementioned breakdown of the $\mathbb{Z}_2$ symmetry $\delta \rho \to - \delta \rho$, where $\delta \rho$ is the density perturbation. 
While this symmetry is exact at the level of the (linear) Bogoliubov-de Gennes equation, even-order terms in $\delta \rho$ from the Gross-Pitaevskii equation break it, leading to qualitatively different evolutions:
\begin{itemize}
\item For positive initial perturbations $\delta \rho > 0$, an undulation is generated close to the horizon and propagates in the supersonic region. 
If the perturbation is sufficiently small, the solution becomes stationary at late times (the phase velocity vanishes although the group velocity does not) but not asymptotically homogeneous, containing a periodic density modulation in the supersonic region. 
\item For negative perturbations, this is accompanied by the emission of superposed soliton trains, which seem periodic in time in the domain of parameter space that we probed. 
The solution is thus not stationary nor asymptotically homogeneous at late times. 
\end{itemize}
These properties of white hole flows are also confirmed by new simulations shown in Chapter~\ref{ch:nohair}, although for the time being we have no clear analytical explanation for them.

The second part of~\cite{Michel:2015pra} was partially presented in Chapter~\ref{ch:saturation}. 
It also contains a discussion of the density-density correlation function and the behavior of the averaged density in the presence of lasing modes, motivated by the experiment~\cite{BHLaser-Jeff}. 
In this reference, the observation of density modulations is reported in a black hole laser configuration. 
This modulation grows exponentially in time, as expected for a perturbation due to a lasing mode. 
However, J.~Steinhauer observed this growth both at the level of the correlation function and on the average of the density $\rho$ over approximately $80$ realizations. 
This is at first sight surprising because, assuming the exponentially-growing modes are sourced only by (quantum or classical) random fluctuations with a vanishing mean value, their averaged amplitude should vanish to linear order because of the $\mathbb{Z}_2$ symmetry relating a perturbation $\delta \rho$ and its opposite. 
One possible explanation is that this modulation may be generated by a large-amplitude classical wave with a definite phase and amplitude, which either serves as a seed for the laser effect~\cite{Tettamanti:2016ntx} or directly produces the density modulation through Cerenkov radiation~\cite{Wang:2016jaj}. 
In~\cite{Michel:2015pra} we put forward another possible scenario, namely that nonlinear effects breaking the $\mathbb{Z}_2$ symmetry may make the growth of the perturbations observable even in the absence of deterministic wave. 
Indeed, while two perturbations with opposite signs grow with the same rate at early times when the dynamics is well approximated by the linear approximation, their subsequent evolution strongly differ. 
Moreover, this symmetry breaking is fundamental to understand the evolution of perturbed black hole lasers solutions, see Chapter~\ref{ch:saturation}. 
To the best of our knowledge, it is currently unclear what are the relative importance of these three effects in J.~Steinhauer's experiment. 
Answering this question would be an important step forward in the interpretation of the data, as it would determine to what extent the observed growth was due to random (and thus possibly quantum) perturbations.

\clearpage

\section[Slow sound in a duct, effective transonic flows, and analog black holes]{Slow sound in a duct, effective transonic flows, and analog black holes \cite{Auregan:2015uva}}
\label{sub:slowsound}

In~\cite{Auregan:2015uva}, we considered a new system for studying the analogue Hawking radiation: a flow of gas in a duct with a reduced effective one-dimensional sound velocity due to the use of a compliant wall. 
The latter is made of an array of tubes with a typical diameter of the order of the millimeter, orthogonal to the flow velocity. 
Its effect is to change the boundary condition, reducing the effective speed of sound after integration over the vertical direction, by an amount which depends on the length $b$ of the tubes. 
Therefore, if $b$ changes with the coordinate $x$ in the direction of the flow, so does the effective sound speed, which may then cross the flow velocity and thus produce an analogue black or white hole horizon. 
Notice that this system is very close to the original one of~\cite{Unruh:1980cg}: the main difference is that the Mach number does not have to go above $1$ to have a horizon, thanks to the reduction of the effective sound velocity. 

\begin{myfig}
\includegraphics[width=0.49 \linewidth]{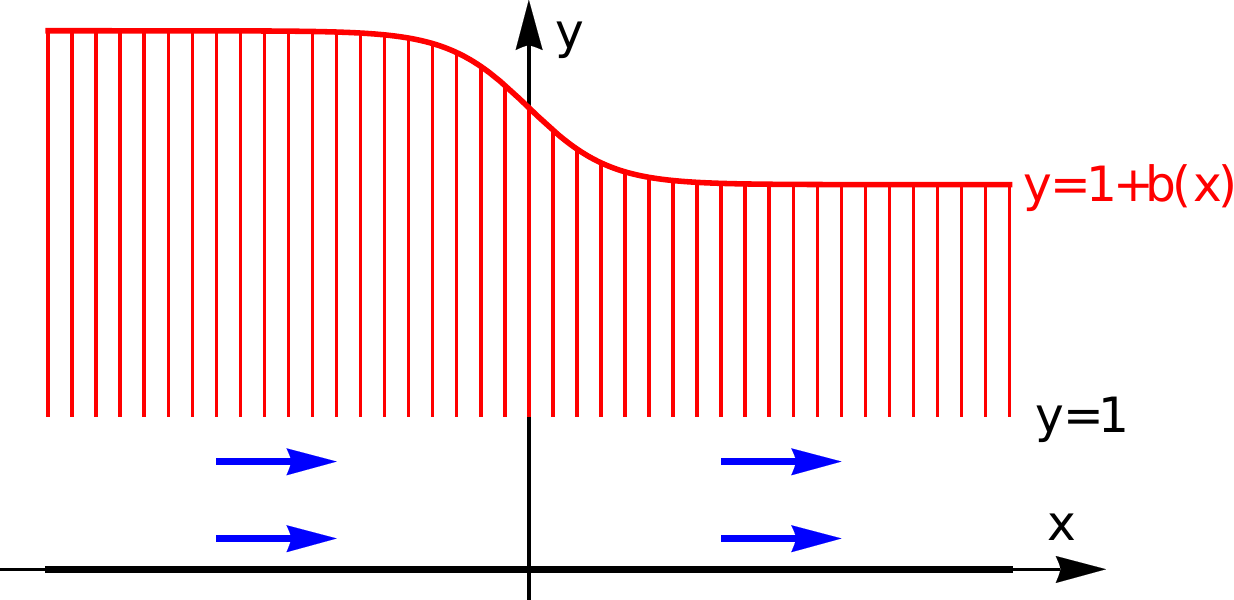}
\includegraphics[width=0.49 \linewidth]{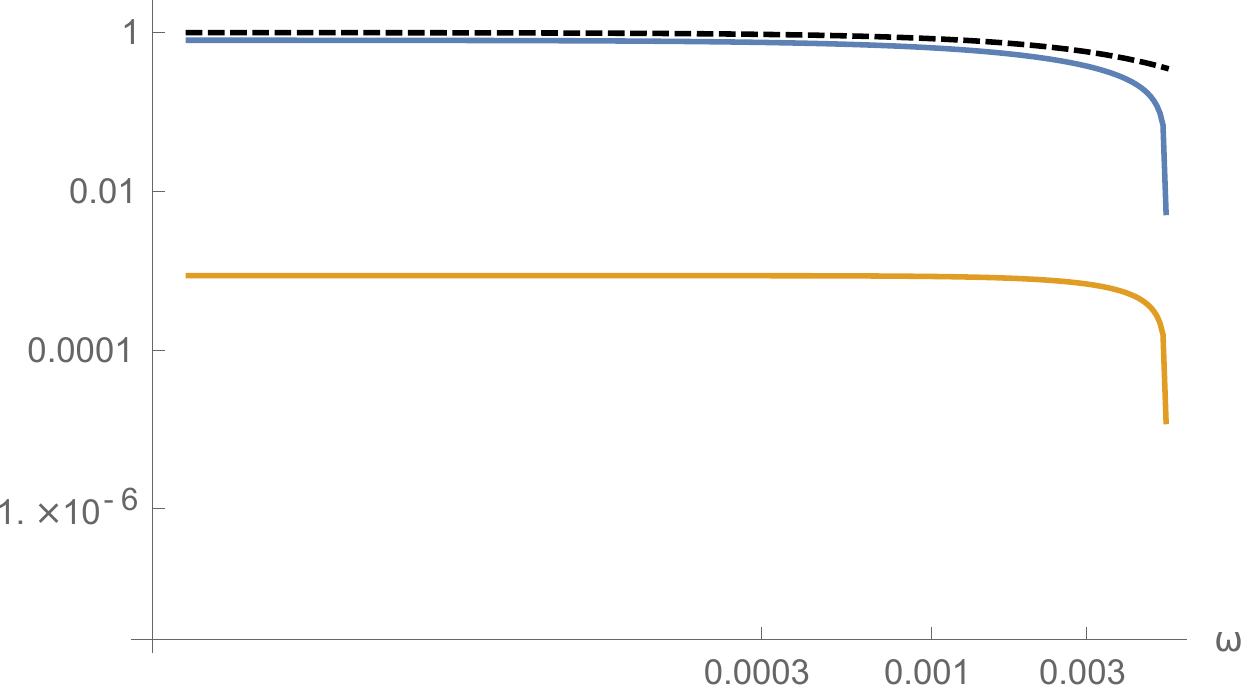}
\capf{Left panel: Schematic drawing of the experimental setup. 
At $y=0$ is a rigid wall, while the compliant wall covers the region $1 < y < 1+b(x)$. 
It is made of a succession of tiny tubes, represented by vertical red lines (not to scale) of height $b(x)$. 
A wiremesh (not represented) is put at $y=1$ to reduce the turbulence, and microphones are added along the line $y = 1+b(x)$ to measure variations of the pressure. 
Blue arrows show the direction of the flow velocity for a white hole configuration. 
Right panel: Numerical prediction for $\om \abs{\beta_\om^2} / T_H$ (blue) and $\om \abs{A_\om^2} / T_H$ (orange), where $T_H$ denotes the Hawking temperature, as functions of $\om$ for realistic values of the parameters. 
We use units where the height of the tube and (three-dimensional) sound speed are equal to unity.
The flow velocity is one third of the speed of sound, and $b: x \mapsto b_1 + b_2 \tanh(x / d_b)$, where $b_1 = 8$, $b_2 = 1$, and $d_b = 1$. 
The black, dashed line shows $\om \abs{\beta_\om^2} / T_H$ for a thermal spectrum with temperature $T_H$.
}\label{fig:concl_slowsound}
\end{myfig}

Let us briefly explain how the dispersion relation and effective metric can be derived. 
Working in a system of units where the three-dimensional sound velocity, the gas mass density, and the height of the duct are equal to $1$, the velocity perturbation $\delta \vec{v}$ and pressure perturbation $\delta p$ are related by
\begin{align}\label{eq:concl_vandP}
D_t p = - \vec{\nabla} \cdot \vec{v}, \, D_t \vec{v} = - \vec{\nabla} p,
\end{align}
where $D_t \equiv \pd_t + v_0 \pd_x$ is the convective derivative and $v_0$ is the mean flow velocity. 
Defining the velocity potential $\phi$ by $\vec{v} = \vec{\nabla} \phi$ and $P = - D_t \phi$, the second equation in \eqref{eq:concl_vandP} ensures $\dd \phi$ is a total differential and the first one gives
\begin{align} \label{eq:phifield}
D_t^2 \phi - \nabla^2 \phi = 0. 
\end{align}
we work in the setup shown in the left panel of \fig{fig:concl_slowsound} and assume $\phi$ is homogeneous in the transverse direction orthogonal to the plane of the figure. 
At $y = 0$ is an unpenetrable wall, giving the boundary condition $\pd_y \phi = 0$. 
The boundary condition at $y = 1$, where lies the compliant wall, is more complicate and involves nonlocal terms~\cite{2015ASAJ..138..605A}.  
However, under the approximation that the flow is near-critical and for small frequencies, it simplifies as
\begin{align}
\pd_y \phi + D_t \lp b(x) D_t \phi \rp = 0. 
\end{align}

In a region where $b$ is homogeneous, one can look for a basis of solutions of the form
\begin{align}
\phi_k \propto \cosh(\alpha_k \s y) \e^{\ii \lp k \s x - \om_k \s t \rp}.
\end{align}
One then obtains the dispersion relation
\begin{align}
\lp \om - v_0 k \rp^2 = c_S^2(b) \, k^2 - \frac{k^4}{\Lambda_b^2} + O \lp k^6 \rp,
\end{align}
where 
\begin{align}
c_S^2(b) \equiv \frac{1}{1+b} \; \text{and} \; \Lambda_b^2 \equiv 3 \frac{\lp 1+b \rp^2}{b^2}. 
\end{align}
We thus see that the effective one-dimensional sound speed $c_S$ is reduced by a factor $1/\sqrt{1+b}$. 

To obtain solutions in the inhomogeneous case where $b$ depends on $x$, it is useful to adopt a Lagrangian formulation. 
To do so, we notice that the field equation \eqref{eq:phifield} and the boundary conditions at $y=0$ and $y=1$ can be derived from a least-action principle with the action
\begin{align}\label{eq:Sslowsound}
S = \frac{1}{2} \int \dd t \int \dd x \int_0^1 \dd y \, \left[ \abs{D_t \phi}^2 - \abs{\vec{\nabla} \phi}^2 + \delta (y-1) b(x) \abs{D_t \phi}^2 \right].
\end{align}
As explained in Chapter~\ref{chap:intro}, the $U(1)$ invariance of \eqref{eq:Sslowsound} implies the conservation of the inner product of two solutions, defined by
\begin{align}
\lp \phi_1 \vert \phi_2 \rp \equiv \ii \int_{- \infty}^{+ \infty} \dd x \int_0^1  \dd y \, \lp \pi_2 \phi_1^* - \pi_1 \phi_2^* \rp, 
\end{align}
where $\pi_i$ denotes the momentum conjugate to $\phi_i$. 
For long-wavelength modes, one can neglect the variations of $\pd_y^2 \phi$ with $y$. 
Using the boundary condition at $y=0$, the field is thus written as
\begin{align}
\phi(x,y,t) \approx \Phi(x,t) + y^2 \Psi(x,t). 
\end{align}
$\Phi$ and $\Psi$ then satisfy a system of two coupled partial differential equations. 
Retaining only the first nontrivial order in $\pd_x$ and $\pd_t$, they may be combined as
\begin{align}
\pd_\mu \lp F^{\mu \nu} \pd_\nu \Phi \rp = 0, 
\end{align}
where 
\begin{align}
F^{\mu \nu}(x) = 
\begin{pmatrix}
v_0^2 - c_S^2(b(x)) & v_0 \\ 
v_0 & 1
\end{pmatrix} . 
\end{align}
Up to multiplication of $F$ by a smooth function, which as mentioned in Chapter~\ref{chap:intro} does not affect the Hawking mechanism\footnote{This factor is the square root of minus the determinant of the matrix $F^{\mu \nu}$, i.e. $c_S$. Since it remains finite and nonvanishing (in fact, equal to $v_0$ by definition) at the horizon, it may  be treated as a constant when considering the near-horizon physics.}, this is the d'Alembert equation in a curved, (1+1)-dimensional space-time with metric $g^{\mu \nu} \propto F^{\mu \nu}$. 
As is the case for water waves or density fluctuations in Bose-Einstein condensates, higher-order terms in the derivatives must be included for shorter-wavelength modes, see the Supplemental Material of~\cite{Auregan:2015uva}.~\footnote{Notice that the equation on $\Phi$ alone involves higher-order derivatives in $t$, which might lead to think that there may be causality issues in our model. However, the set of equations on $\Phi$ and $\Psi$ is second-order and has a canonical Hamiltonian structure. This model is thus well defined from the point of view of Hamiltonian dynamics.} 
To our knowledge, a generic method to solve the resulting equations analytically does not exist. 
But solutions can be found numerically using techniques similar to those of Chapter~\ref{ch:probing}. 
Results for the coefficients $\abs{\beta_\om^2}$ and $\abs{A_\om^2}$, defined as in \eqref{eq:Btrans}, are shown in the right panel of \fig{fig:concl_slowsound} for ``realistic'' parameters, in the sense that we hope they can be realized experimentally. 

The group of Y.~Aurégan is curently trying to realize an analogue white hole flow using this concept in the Laboratoire d'Acoustique de l'Université du Maine, in Le Mans. 
Waves will be sent by a loudspeaker at the downstream end of the duct, and an array of microphones will be added on top of the compliant wall to measure $\pd_t \delta \phi$, from which the scattering coefficients can be extracted. 
Compared with the original proposal of W.~Unruh, it has the important advantage of not requiring the flow velocity to cross the three-dimensional sound speed, which in practice is very difficult to realize while keeping the flow sufficiently stable. 
Indeed, if $b$ is large the effective one-dimensional sound velocity is strongly reduced, allowing for the formation of analogue horizons even for much slower flows. 
In practice, it seems plausible to work with values of $b$ close to $10$, so that reaching a Mach number of $0.3$ could be enough. 
One potential problem is that even for this relatively low value the interface between the duct and compliant wall may trigger important turbulence and whistling, hiding the Hawking effect behind larger-amplitude perturbations. 
To reduce this noise, a wire gauze with low flow resistance is added at the interface. 
We hope that in such a system the classical analogue of the Hawking affect could be measured in the near future. 

\clearpage

\section[Mode mixing in sub- and trans-critical flows over an obstacle: When should Hawking's predictions be recovered?]{Mode mixing in sub- and trans-critical flows over an obstacle: When should Hawking's predictions be recovered? \cite{Michel:2015aga}}

Motivated by the experiments~\cite{Rousseaux:2007is,Weinfurtner:2010nu,Euve:2014aga} in which the scattering of water waves was observed in apparently subcritical flows, we re-examined in~\cite{Michel:2015aga} the problem studied in Chapter~\ref{ch:probing}. 
Using new numerical simulations for various flows with maximum Froude number $F_{\rm max}$ ranging between $0.75$ and $1.25$, we aimed at confirming the results of~\cite{Michel:2014zsa} and obtain a better understanding of the relevent parameters to describe the spectrum. 

Our numerical results first confirm that there are two qualitatively different regimes depending on the value of $F_{\rm max} - 1$, which seems to be the most relevant parameter. 
When $F_{\rm max}$ is significantly larger than $1$ (typically, for $F_{\rm max} > 1.1$ -- although the precise lower bound depends on the shape of the obstacle, in particular its length and slopes), one recovers the Hawking spectrum in the large domain of frequencies satisfying $\om_c \ll \om \ll \om_{\rm max}$ (see Chapter~\ref{ch:probing}, \eq{eq:omc} and the text below it, for the definition of $\om_c$). 
The spectrum is thus essentially fixed by the analogue surface gravity, given (for the incoming counter-propagating mode) by $\pd_x \lp c - v \rp$ at the point where $v-c=0$ on the downstream slope. 
In particular, the upstream slope and length of the obstacle have little effect on the spectrum apart from fixing the small critical frequency $\om_c$. 
On the other hand, for subcritical flows, i.e., $F_{\rm max} < 1$, there seems to be no simple law describing the spectrum. 
As there is no analogue Killing horizon anymore for $x \in \mathbb{R}$~\footnote{However, A.~Coutant and S.~Weinfurtner have shown in~\cite{Coutant:2016vsf} that analogue Killing horizons exist for complex values of $x$, with a strongly suppressed spectrum.}, the low-frequency scattering (for $\om < \om_{\rm min}$) is entirely due to effects which were subdominant in the transcritical case. 
We found numerically that, for instance, the length and upstream slope play an important role. 
For $\om > \om_{\rm min}$, the scattering is still due to the presence of a turning point, as for transcritical flows. 
However, at these frequencies dispersive effects become important. 
They determine in particular the position of the turning point along the downstream slope, and thus the gradients of $c$ and $v$ at this point, which now depends on the frequency of the incoming wave. 

Fig.~\ref{fig:concl:handF} shows the water depth $h$ and Froude number $F$ for six flows with the same shape of $h$ up to a global constant. 
Three of them are subcritical as $F_{\rm max} < 1$. 
One is mildly transcritical, with $1 < F_{\rm max} < 1.1$. 
The two others are transcritical with $F_{\rm max} > 1.1$. 
The behavior of the 4 scattering coefficients of \eqref{eq:Btrans} in these flows is shown in \fig{fig:scoeffs}. 
Focusing first on the transcritical flows, one sees that $\ln \abs{\beta_\om}$ (top, right panel) is linear in $\om$ in a large domain of frequencies. 
For the two highest values of $F_{\rm max}$, it accurately follows \eqref{eq:effT} up to values of $\om$ smaller than those shown in the plot, with a temperature close to the Hawking one $T_H$. 
For the mildly transcritical flow, \eqref{eq:effT} is less good an approximation as frequencies near $\om_c$ are already close to $\om_{\rm max}$,  hence the downward curvature seen in the plot for the smallest values of $\om$. 
However, $\abs{\beta_\om}$ becomes larger than $1$ for moderately small frequencies. 
For the three subcritical flows instead, it never goes above $\e^{-2}$. 
Related to this are the behaviors of the reflexion and transmission coefficients $\abs{\alpha_\om}$ and $|\tilde{A}_\om|$. 
For the transcritical flows, $\abs{\alpha_\om}$ is always larger than $1$ while $|\tilde{A}_\om|$ is very small except at very low frequency.  
This indicates that the incident wave is mostly reflected, with very small residual transmission. 
For the three subcritical flows, we have the opposite when $\om < \om_{\rm min}$: $|\tilde{A}_\om|$ is close to $1$ while $\abs{\alpha_\om}$ becomes very small, indicating that the wave is essentially transmitted. 
Notice that in both cases $\abs{A_\om}$ remains smaller than $\e^{-1}$, so that the mixing between co- and counter-propagating waves in the rest frame of the fluid is always weak. 

\begin{figure}
\centering
\includegraphics[width = 0.49 \linewidth]{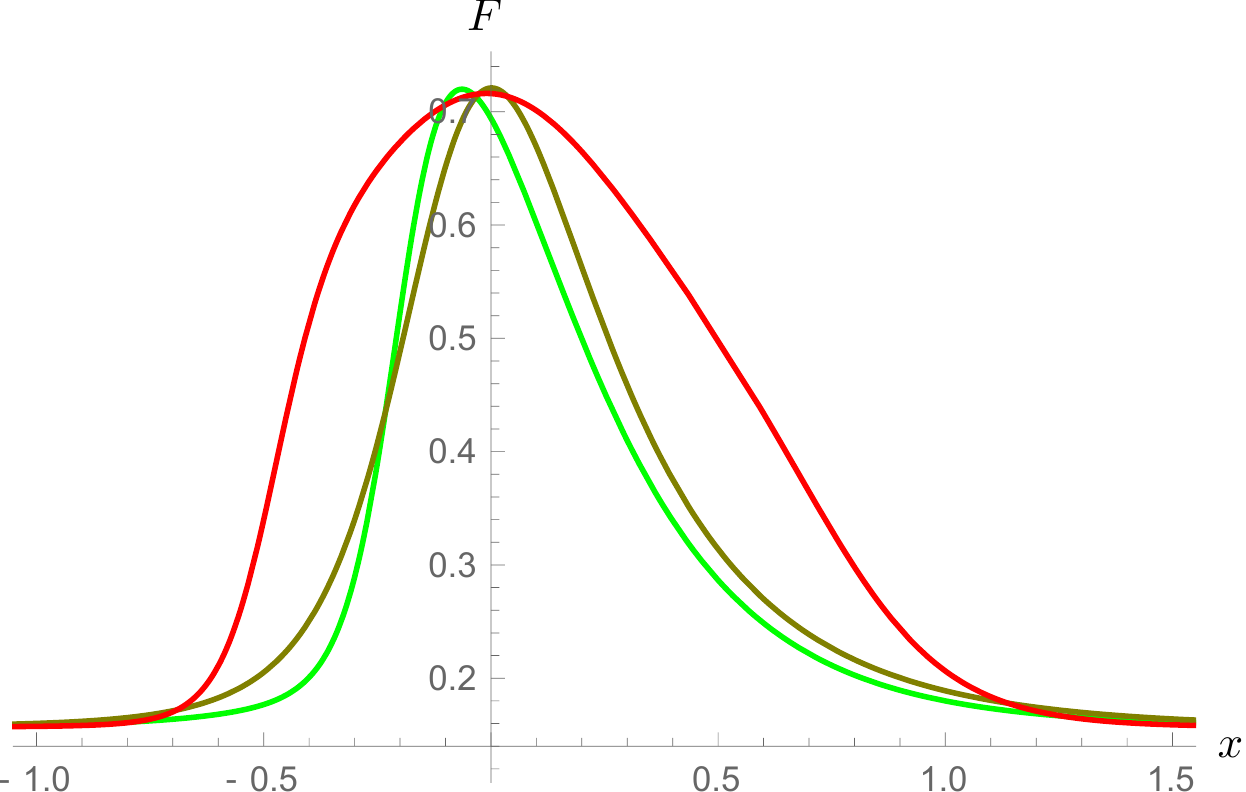}
\includegraphics[width = 0.49 \linewidth]{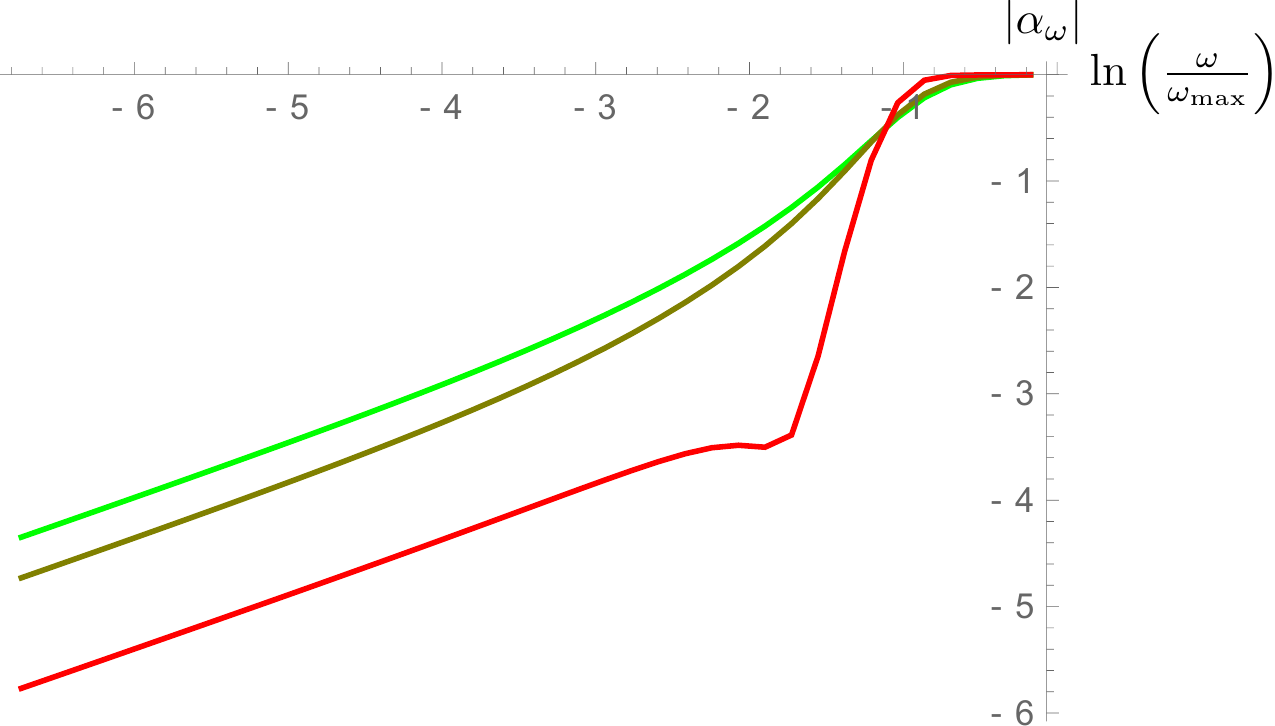}
\includegraphics[width = 0.49 \linewidth]{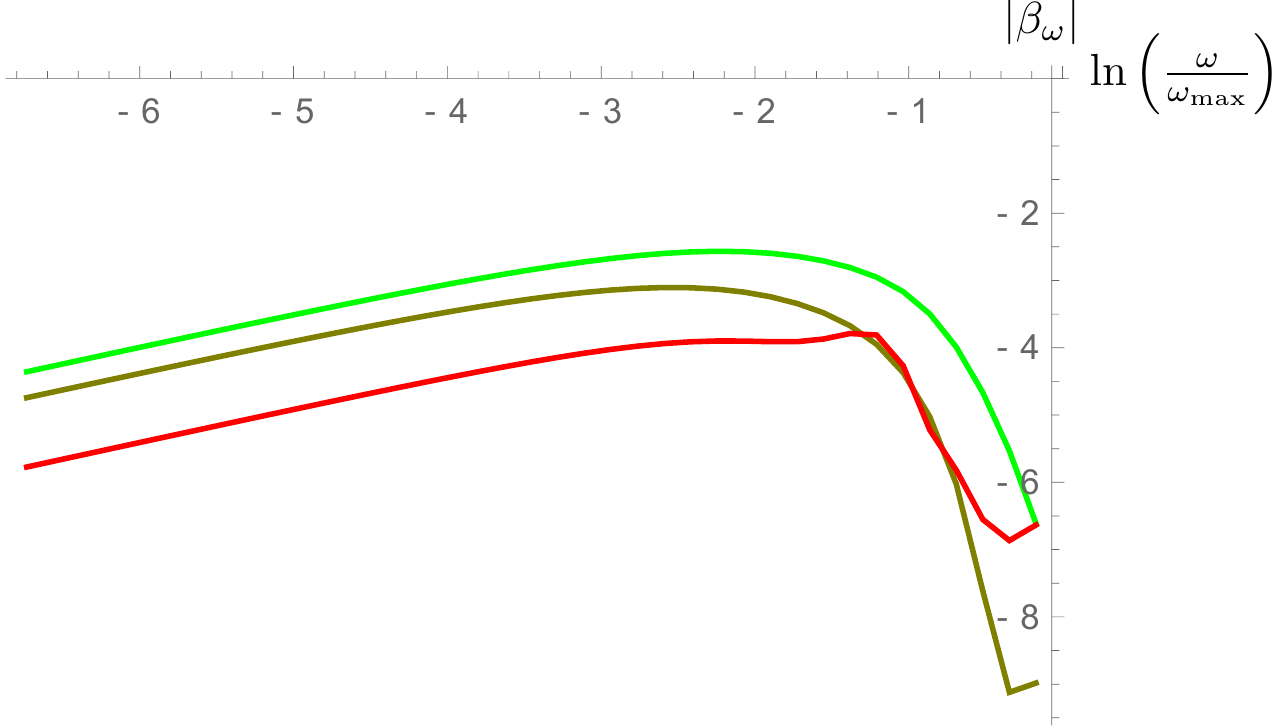}
\includegraphics[width = 0.49 \linewidth]{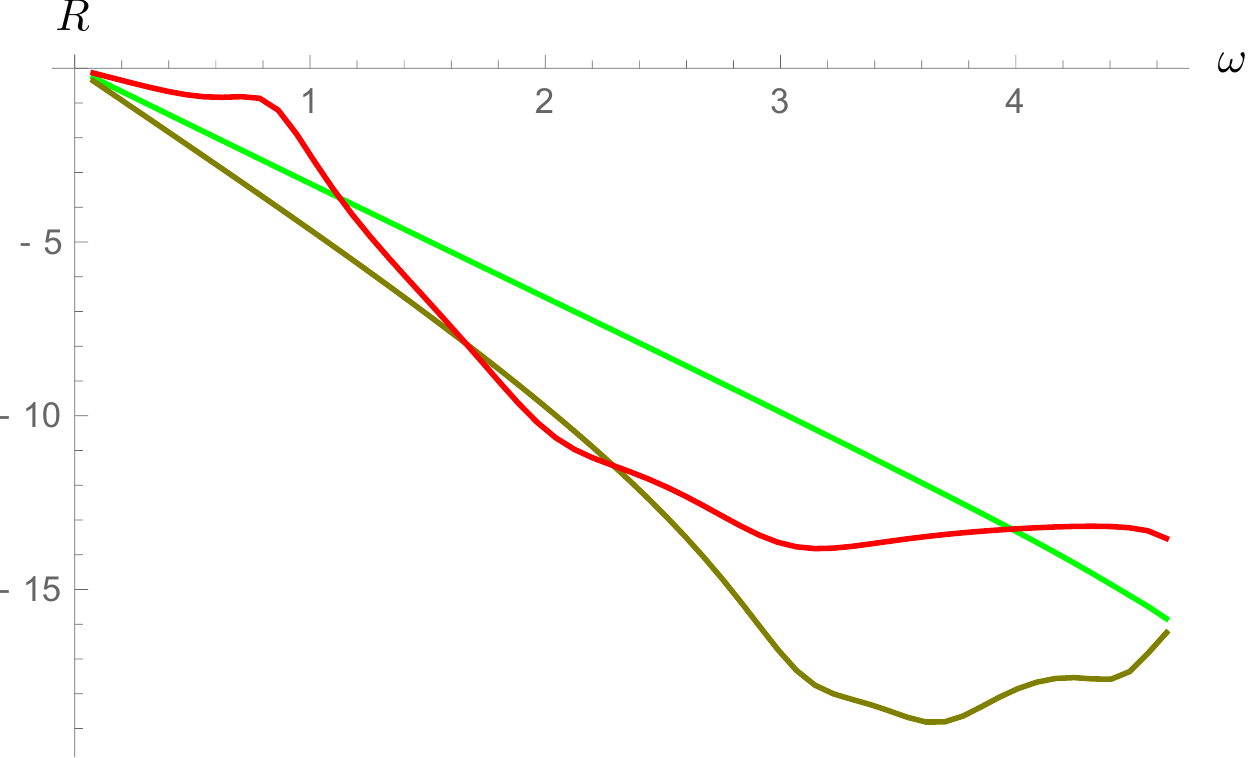}
\caption{We work with three flows with nearby values of $F_{\rm max} \approx 0.72$  but different slopes and widths of the obstacle. 
We show the Froude number (top, left), $\abs{\alpha_\omega}$ (top, right), $\abs{\beta_\omega}$ (bottom, left), and $R \equiv 2 \ln \left( \abs{ \beta_\om / \alpha_\om} \right)$ (bottom, right). 
Green curves correspond to a ``thin'' obstacle with ``large'' upstream slope, see~\cite{Michel:2015aga}. 
Brown curves correspond to a smaller upstream slope, and red curves to a longer obstacle. 
} \label{fig:scoeffsbis}
\end{figure}

To see the effect of the length and upstream slope of the obstacle, we show in \fig{fig:scoeffsbis} the dependance in $\om$ of $\abs{\alpha_\om}$, $\abs{\beta_\om}$, and $R \equiv 2 \ln \left( \abs{ \beta_\om / \alpha_\om} \right)$ for three different flows with nearby values of $F_{\rm max}$ and of the maximum descending slopes. 
Green curves show results for a width and maximum upstream and downstream slopes similar to those of the flow realized in~\cite{Weinfurtner:2010nu}. 
Brown curves are the results for a flow with a smaller upstream slope, and red curves are obtained with a longer obstacle. 
We first notice that, contrary to the case of a transcritical flow where the analogue surface gravity essentially fixes the spectrum, changing either the upstream slope or the length of the obstacle significantly affects $\abs{\alpha_\om}$ and $\abs{\beta_\om}$. 
Second, we see that $R$ is approximately linear in $\om$ provided the upstream slope is large and the obstacle sufficiently short. 
Using this, we could refine the claim made in~\cite{Michel:2014zsa} and conjecture that the linearity of $R$ observed in~\cite{Weinfurtner:2010nu} is, at least partially, due to the important ascending slope of the obstacle used in this work.

\begin{myfig}
\includegraphics[width = 0.49 \linewidth]{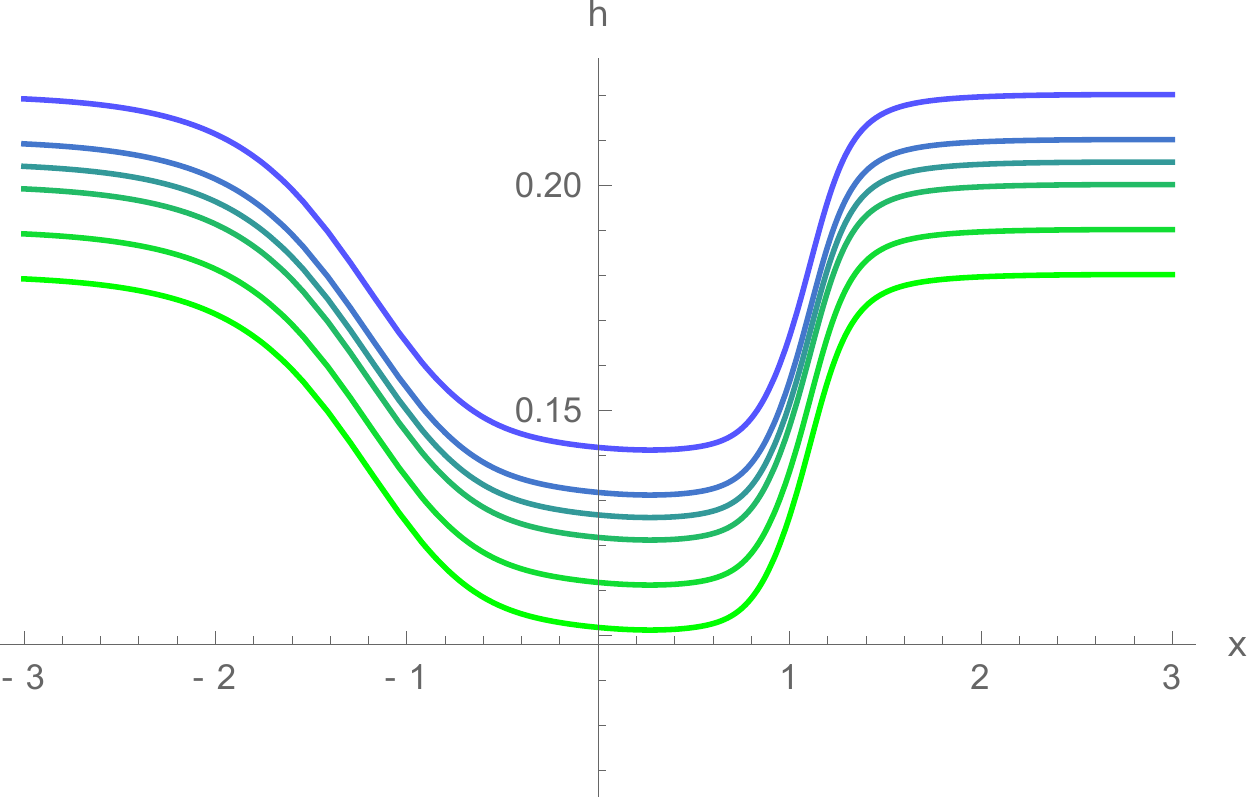}
\includegraphics[width = 0.49 \linewidth]{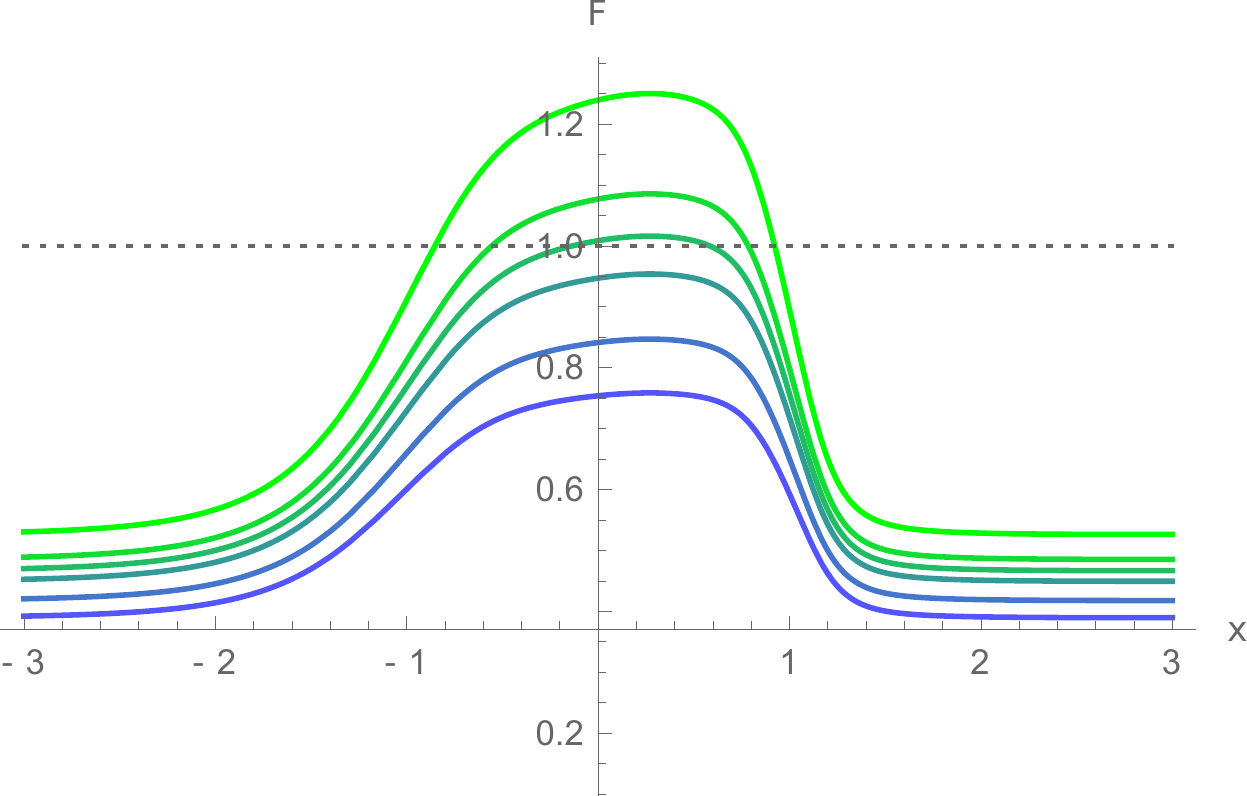}
\capf{Water height $h(x)$ (left) and Froude number $F(x) = v(x) / c(x)$ (right) as functions of the coordinate $x$ along the flume for six flows used in~\cite{Michel:2015aga}. 
The unit of $x$ and $h$ is the meter. 
The horizontal dashed line on the right panel shows $F = 1$. 
One can see that three flows are subcritial and three are transcritical (one of them only barely so).
} \label{fig:concl:handF}
\end{myfig}

\begin{myfig}
\includegraphics[width = 0.49 \linewidth]{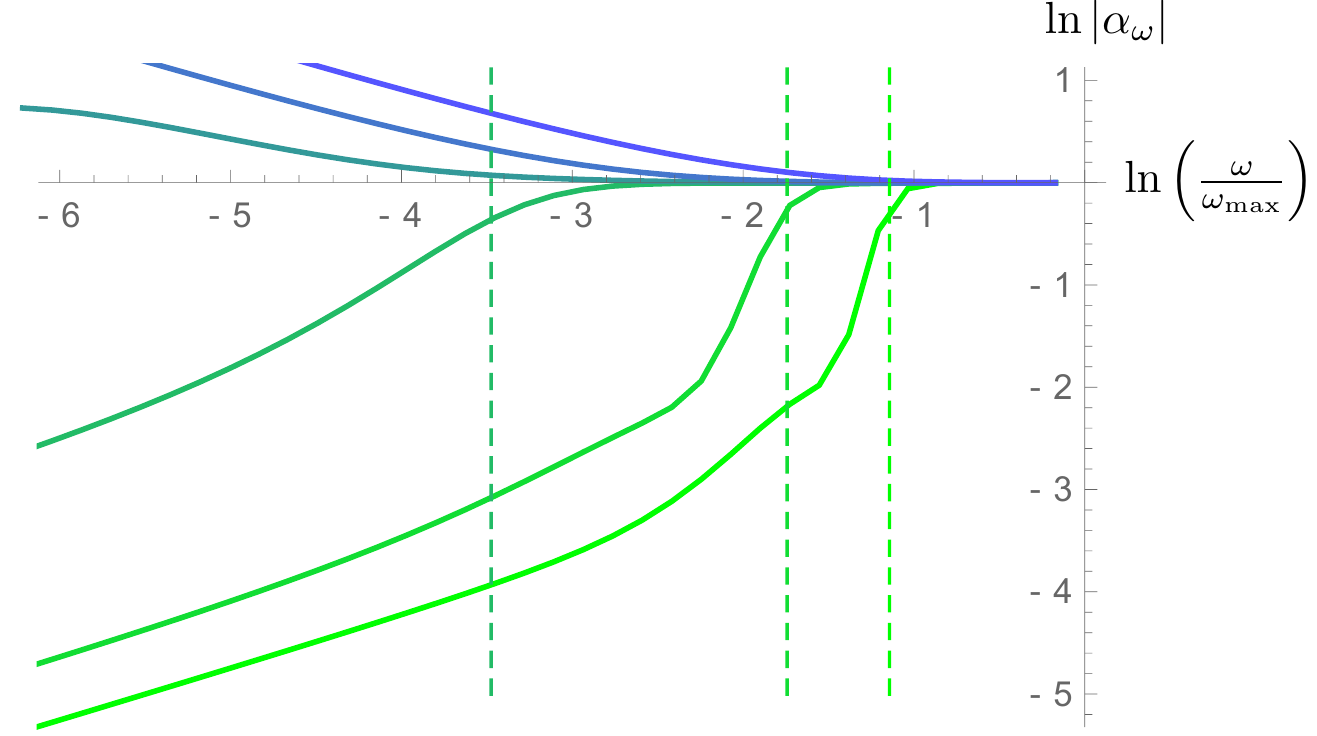}
\includegraphics[width = 0.49 \linewidth]{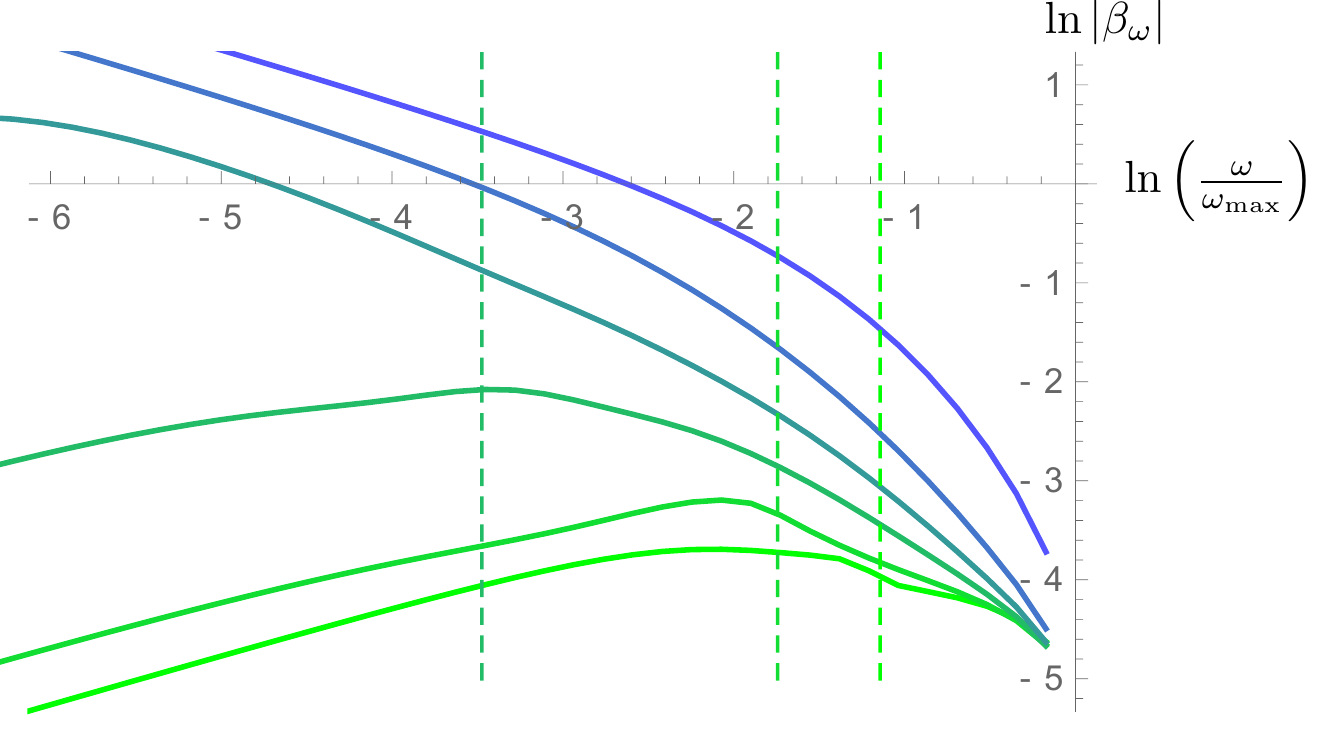}
\includegraphics[width = 0.49 \linewidth]{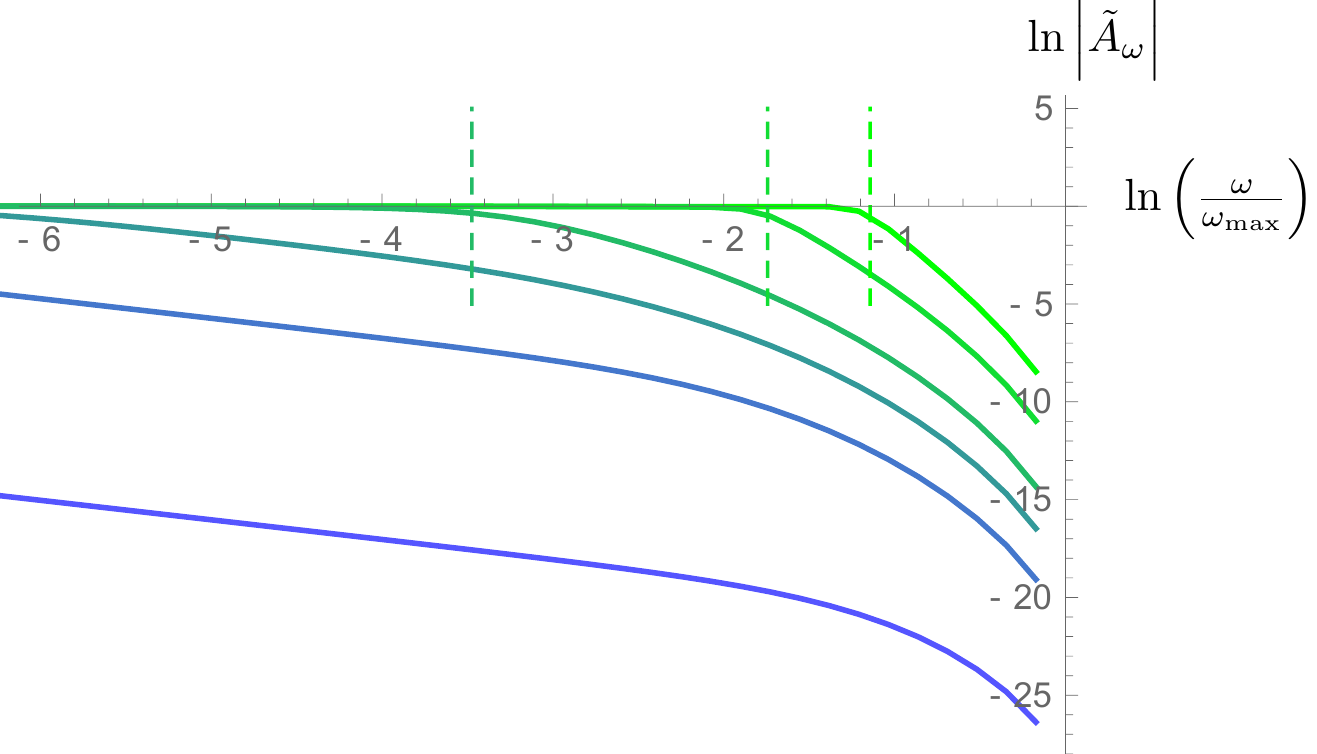}
\includegraphics[width = 0.49 \linewidth]{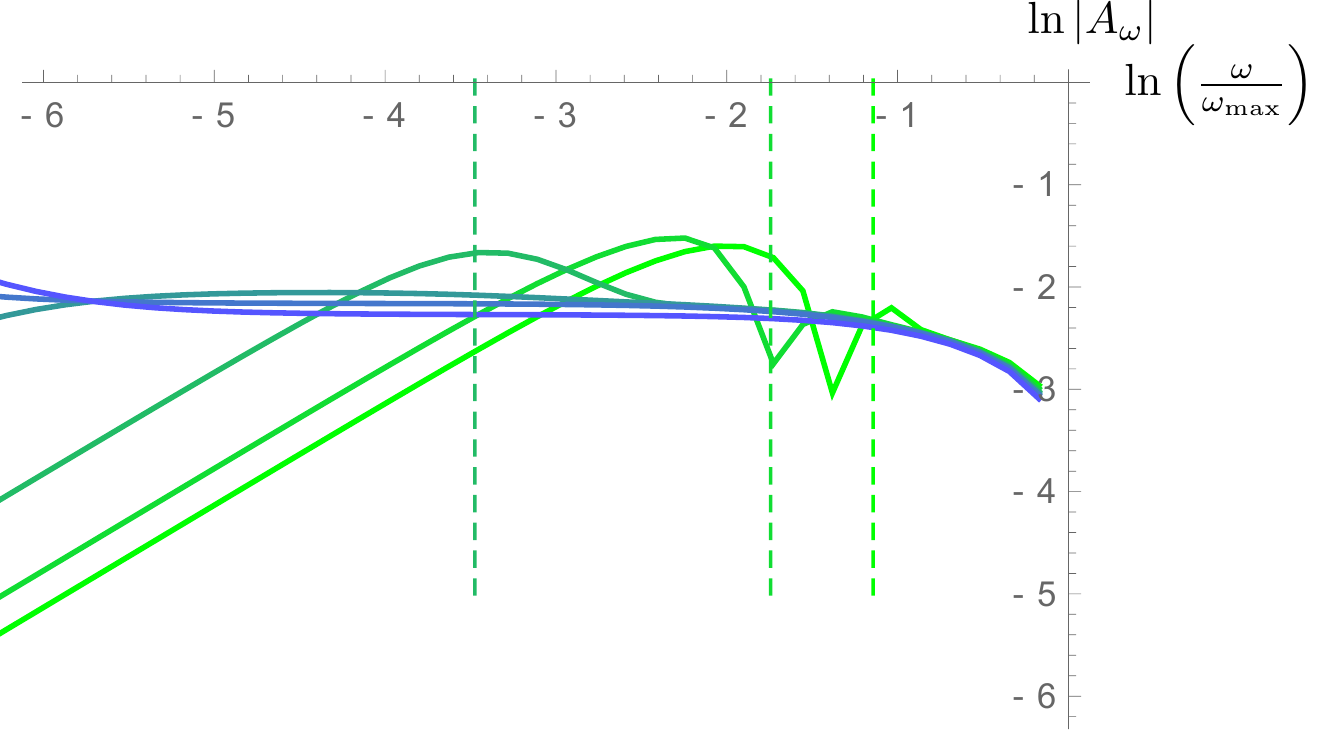}
\capf{We show the 4 scattering coefficients, defined in \eqref{eq:Btrans}, for the 6 flows of \fig{fig:concl:handF} in logarithmic scale. 
The dashed lines show the values of $\om_{\rm min}$ corresponding to each of the subcritical flows. 
} \label{fig:scoeffs}
\end{myfig} 

\clearpage

\section[Dynamical instabilities and quasinormal modes, a spectral analysis with applications to black-hole physics]{Dynamical instabilities and quasinormal modes, a spectral analysis with applications to black-hole physics \cite{Coutant:2016bgk}}
\label{sec:DIQNM}

Three peculiar types of modes play an important role in the study of black hole stability. 
First we have negative-energy modes, leading to energetic instabilities: in the presence of such a mode, and if it can interact with a positive-energy one, energy conservation does not give any bound on its amplitude which, if sourced by fluctuations with a stationary spectrum, can grow without bound -- or until nonlinear effects shift its energy to positive values. 
Second, we have dynamical instability modes (DIM). 
Contrary to an energetic instability, a dynamical instability implies that the system will be unstable even if isolated, as the DIM will grow exponentially (neglecting nonlinear effects) whenever it is present at $t=0$. 
Finally, we have quasinormal modes (QNM), corresponding to resonances of the system. 
They characterize the return of the system to its equilibrium configuration after a perturbation which does not destabilize it. 
The main objective of the work~\cite{Coutant:2016bgk} was too exhibit the links between these three types of modes through the analytic properties of the retarded Green function. 

To this end, we first considered the simple model of a charged scalar field with charge $e$ and mass $m$ in a step-like electrostatic potential $A$ in (1+1) dimensions, following~\cite{Fullingbook}. 
We considered setups where the potential has one or two step-like discontinuities, i.e., the electric field consists in one or two delta functions. 
In the first case, there is no DIM but there are already negative-energy modes allowing for an amplification of incoming waves on the discontinuity of $A$. 
The underlying mechanism can be seen in \fig{fig:yesnoKlein}. 
The mass gap in each region, corresponding to frequencies for which the mode is exponentially decreasing in space, is shown in gray. 
Frequencies above the mass gap correspond to particles, for which the energy and frequency have the same sign. 
Frequencies below the mass gap correspond to antiparticles, whose energy and frequency have opposite signs. 
This can be seen through the conserved inner product of the field equation: the inner product of a mode with itself is always positive above the gap and negative below it. 
If $\abs{e \s A} < 2m$, a frequency corresponding to particles (respectively antiparticles) on one side also corresponds to particles (respectively antiparticles) on the other side. 
On the other hand, if $\abs{e A} > 2 m$, there is an interval of frequencies, called the “Klein region”, corresponding to particles on one side and antiparticles on the other side. 
There is thus a mixing between positive- and negative-energy waves, leading to reflexion coefficients larger than unity. 
In the case of two discontinuities, and if $\abs{e \s A} > 2 m$, DIM are triggered by waves bouncing back and forth between the two discontinuities and being amplified at each reflection. 

\begin{myfig}
\includegraphics[width = 0.45 \linewidth]{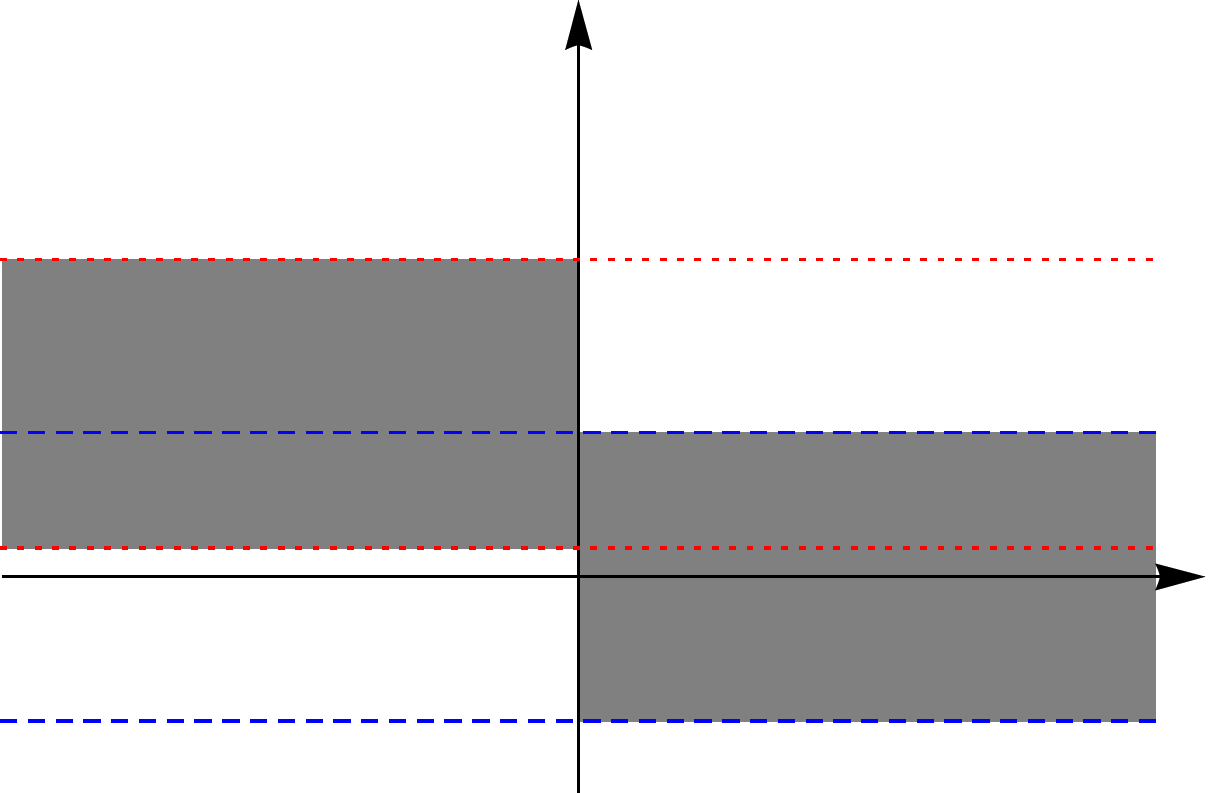}
\hspace{0.5 cm}
\includegraphics[width = 0.45 \linewidth]{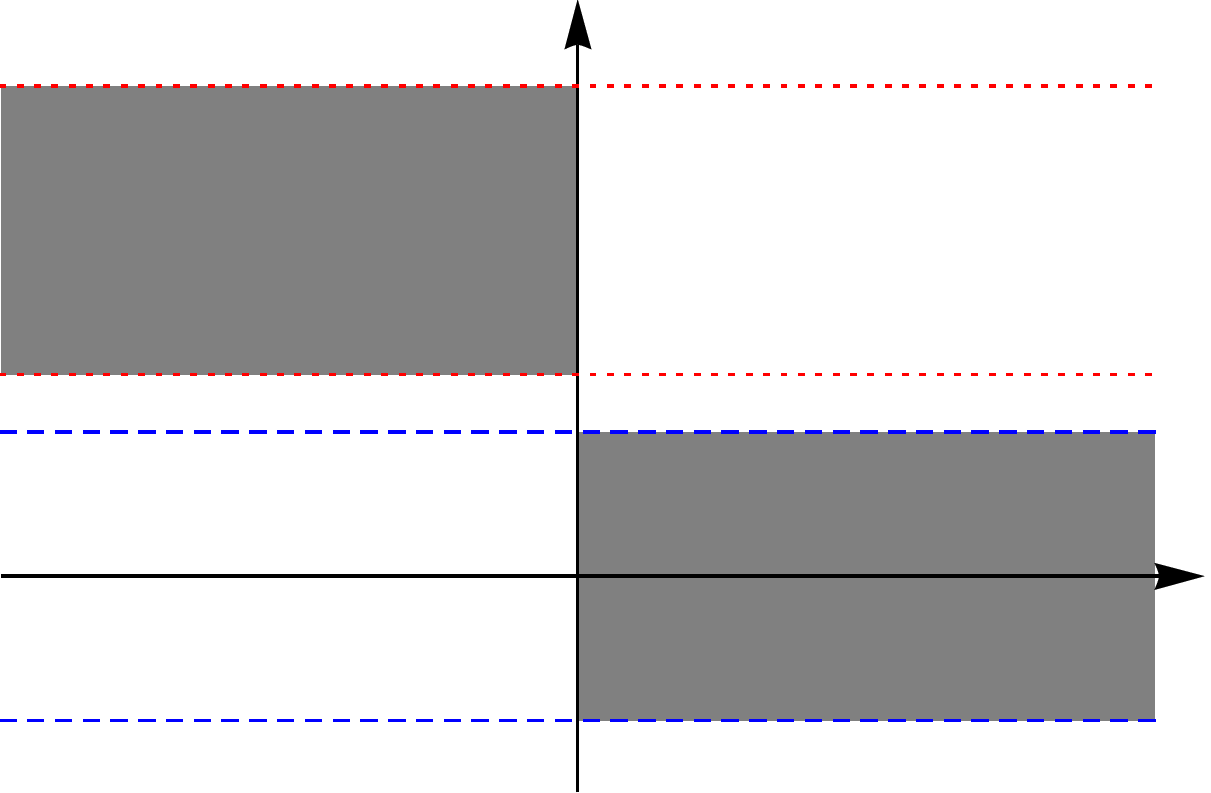} 
\begin{tikzpicture}[overlay]
\put(-223,39){$x$}
\put(-319,128){$\la$}
\put(-10,39){$x$}
\put(-106,128){$\la$}
\end{tikzpicture}
\capf{Forbidden domain of the frequency $\la$ as a function of $x$ for a massive scalar field (mass $m \neq 0$, charge $e$) in a step-like electrostatic potential localized in the left region $x < 0$. 
The blue, dashed lines show the locus $\la = \pm m$, which bounds the forbidden region in the absence of potential. 
The red, dotted ones show $\la = e A \pm m$, where $A$ is the uniform potential in the left region. 
On the left panel, $e A = 1.2 m < 2 m$. 
There is thus no Klein region: a particle (respectively antiparticle) in one region is either totally reflected by the potential or corresponds to a particle (respectively a antiparticle) on the other side. 
On the right panel, $e A = 2.4 m > 2 m$. 
In that case, when $\la$ is in the Klein region $\left] m, e A -m \right[$, the frequency $\la$ corresponds to a particle for $x > 0$ and to an antiparticle for $x < 0$. 
}\label{fig:yesnoKlein}
\end{myfig}

\begin{figure}
\includegraphics[width = 0.49 \linewidth]{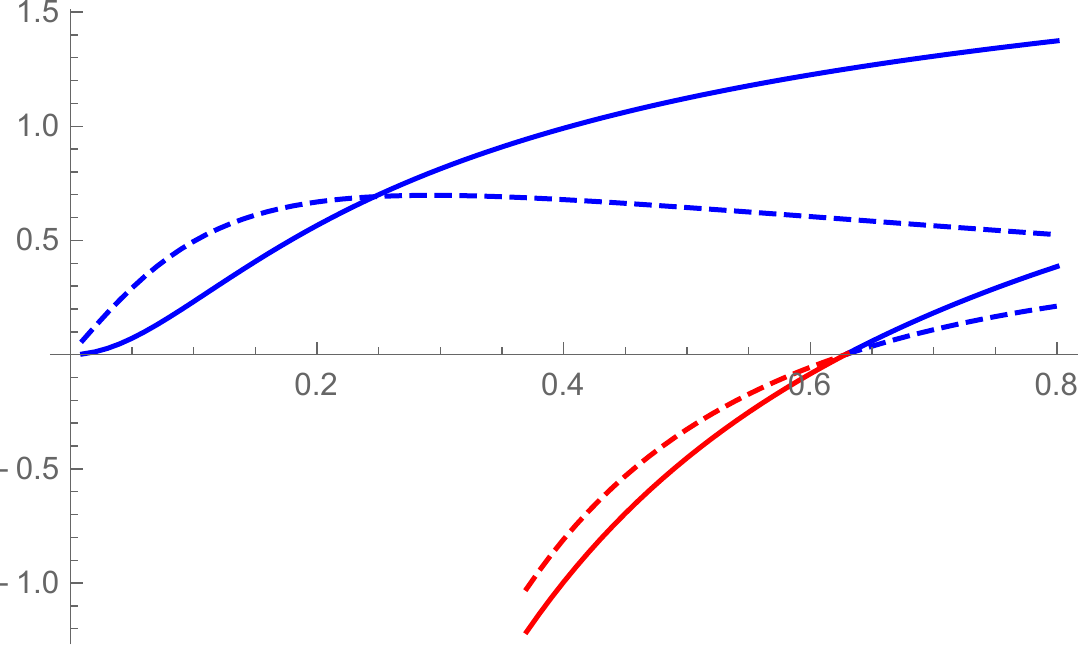}
\includegraphics[width = 0.49 \linewidth]{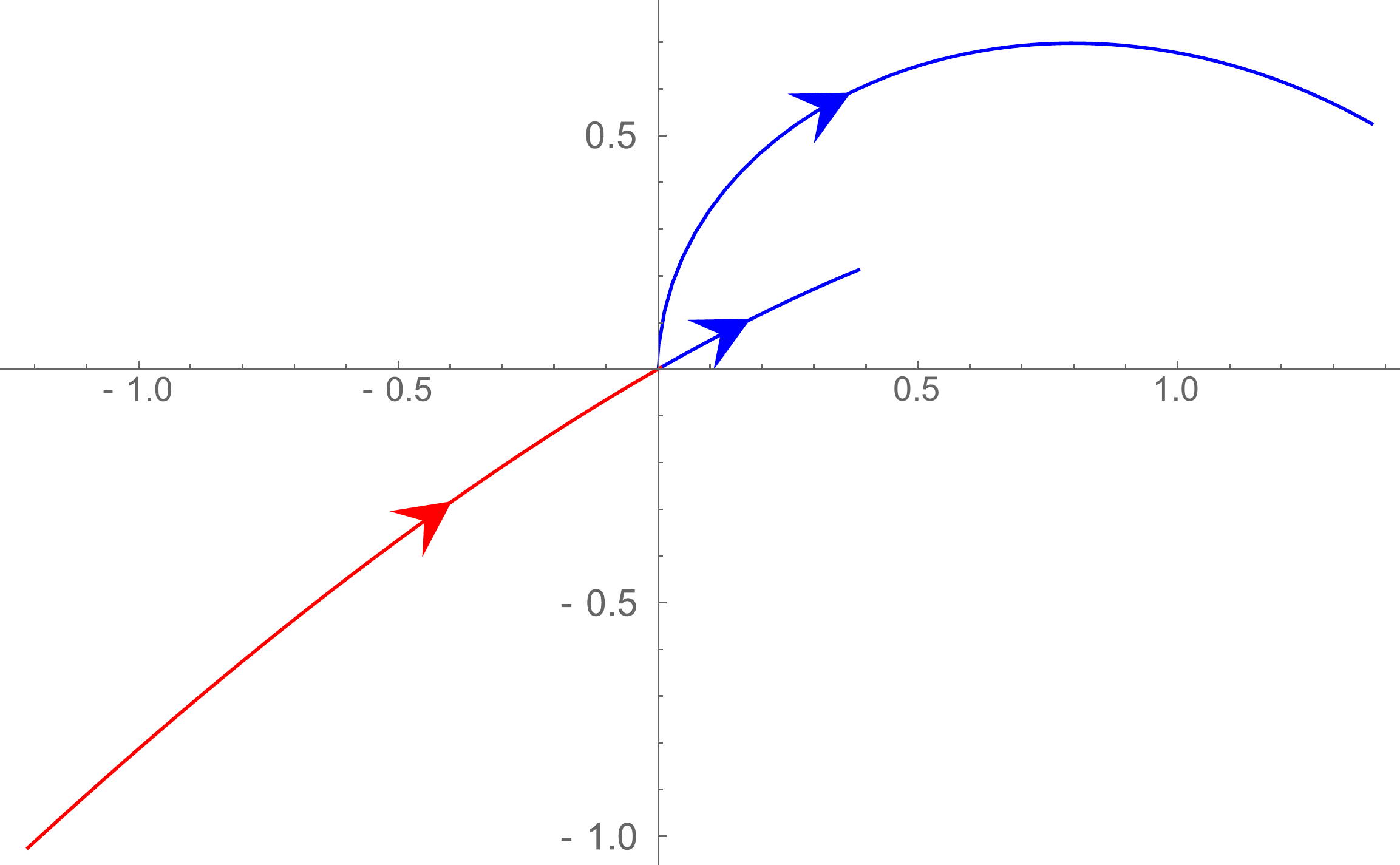}
\begin{tikzpicture}[overlay]
\put(-218,58){\scalebox{0.75}{$L$}}
\put(-419,128){\scalebox{0.75}{$\la, \, \Gamma$}}
\put(-11,75){\scalebox{0.75}{$\la$}}
\put(-116,130){\scalebox{0.75}{$\Gamma$}}
\end{tikzpicture}
\capf{Left: First two complex eigenfrequencies for a massless charged scalar field in a localized electrostatic potential of strength $e A = 2.5$ for $-L < x < + L$, as functions of $L$. 
On the left panel, the real part $\la$ of each frequency is represented by a continuous line, and its imaginary part $\Gamma$ by a dashed line. 
Blue lines correspond to dynamical instability modes (DIM) and red ones to quasinormal modes (QNM). 
One can see that the first eigenfrequency corresponds to a DIM with $\Gamma > 0$ as soon as $L > 0$. 
The second one instead corresponds to a QNM below a critical value of $L$, which smoothly turns to a DIM when increasing this length. 
The right  plot shows the trajectories of these two frequencies in the plane $(\la, \Gamma)$. 
Arrows indicate the direction of increasing $L$, from $0$ to $0.8$. 
}\label{fig:elec_massless}
\end{figure}

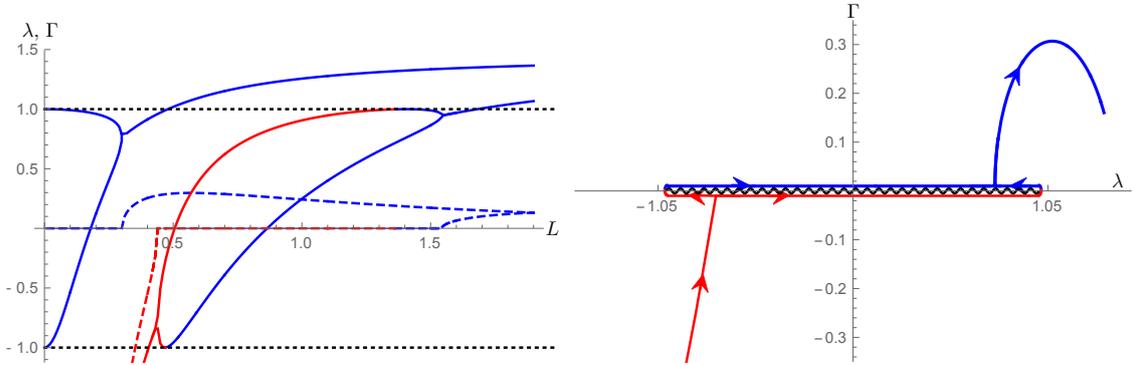
\begin{figure}
\includegraphics[width = 0.49 \linewidth]{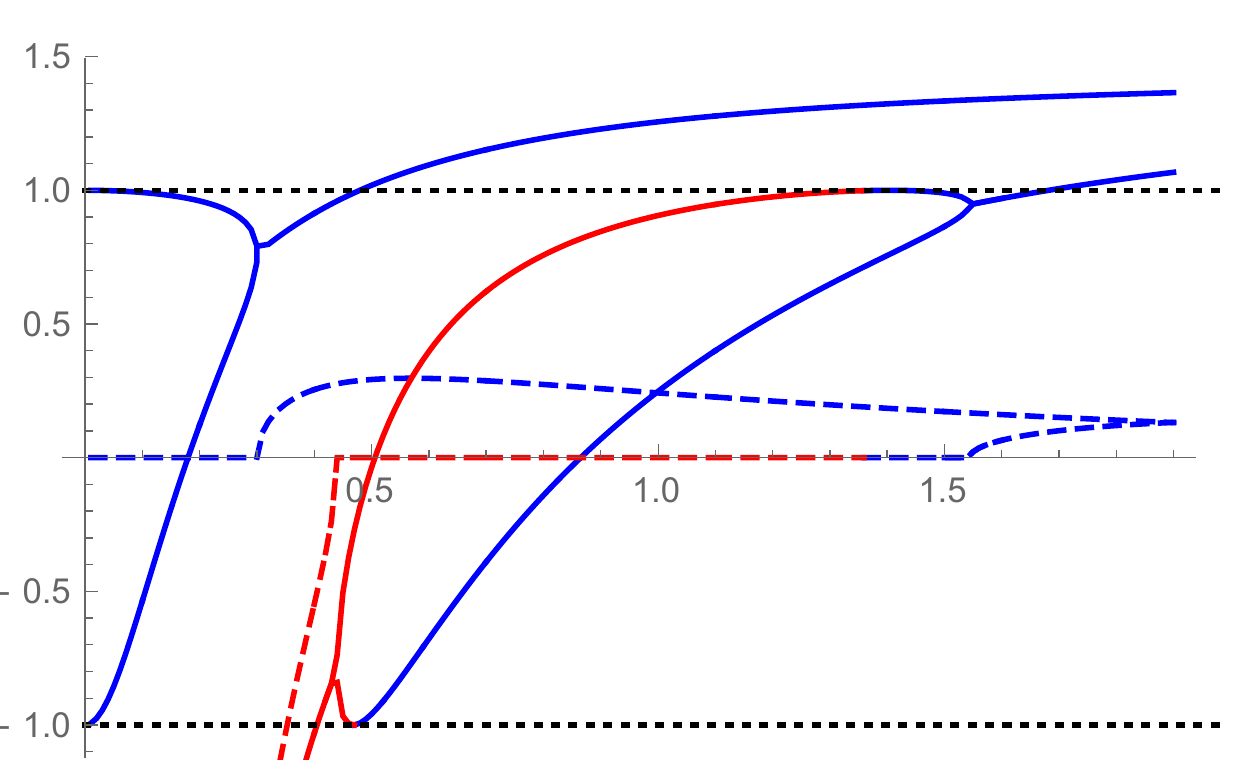}
\includegraphics[width = 0.49 \linewidth]{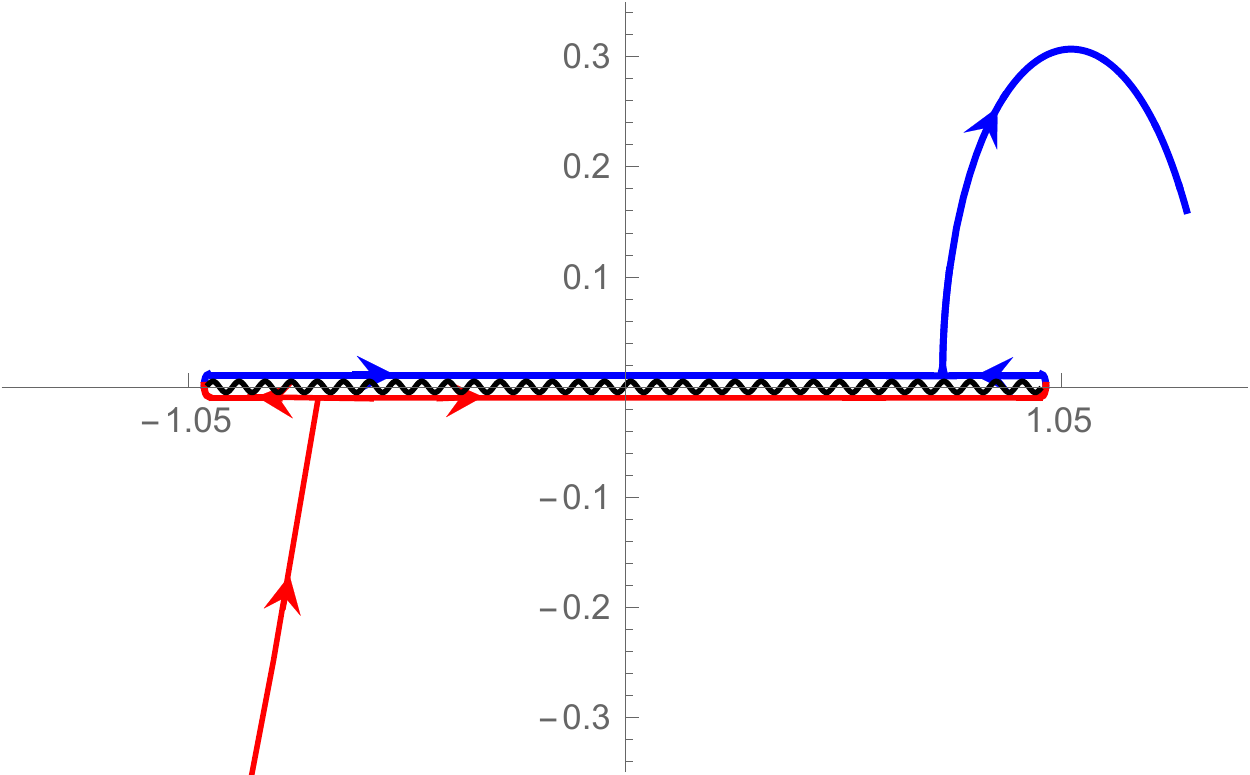}
\begin{tikzpicture}[overlay]
\put(-223,48){\scalebox{0.75}{$L$}}
\put(-419,123){\scalebox{0.75}{$\la, \, \Gamma$}}
\put(-11,66){\scalebox{0.75}{$\la$}}
\put(-110,130){\scalebox{0.75}{$\Gamma$}}
\end{tikzpicture}
\caption{First two complex eigenfrequencies for a massive (mass $m = 1$) charged scalar field in a localized electrostatic potential of strength $e A = 2.5$ for $-L < x < + L$, as functions of $L$. 
The color and style code is the same as in \fig{fig:elec_massless}. 
Contrary to the massless case, for $L = 0^+$ the two eigenfrequencies are real. 
The corresponding modes are exponentially decreasing in the limits $x \to \pm \infty$. 
These two frequencies merge at some critical value of $L$, forming a DIM. 
Similarly, the first QNM does not directly turns to a DIM. 
Instead, it first turns into two real-frequency modes, which separate and then merge again, forming the DIM. 
Right: Trajectory of the second complex eigenfrequency in the plane $(\la, \Gamma)$. The wiggly line shows the branch cut $\la \in \left[ -m, +m \right]$. 
}\label{fig:elec_massive}
\end{figure}

One interesting feature of this model is that its complex eigenfrequencies, as well as those of QNM, can be obtained by solving a transcendental equation instead of a differential one, see~\cite{Coutant:2016bgk} for details. 
One can then explicitly follow the birth of new unstable modes when continuously changing the value of a control parameter, for instance the distance $2L$ between the two discontinuities. 
The massless case $m=0$ is represented in \fig{fig:elec_massless}, where the upper half plane shows DIM frequencies, and the lower half plane shows QNM ones. 
One can prove that at least one DIM exists as soon as $L > 0$. 
As seen in the Figure, it emerges from the zero-frequency modes for $L = 0$. 
All other DIM appear as descendants of QNM when their frequencies cross the real axis (one such transition is shown in the Figure). 
The birth of DIM is slightly more complicate in the massive case, see~\fig{fig:elec_massive}. 
This is due to the branch cut, shown in the left panel, of the relation between the wave vector $k$ and angular frequency $\om$. 
As can be seen in the left panel, when $L$ is below a critical length $L_0$ (close to $0.3$ for the parameters of the figure), there is no DIM. 
Instead, we have two bound state modes (BSM), i.e., real-frequency, square-integrable solutions. 
Their angular frequencies go to $\pm m$ for $L \to 0$. 
When increasing $L$, the modes and their frequencies become closer. 
They merge at $L = L_0$, giving rise to the first DIM. 
Similarly, the transition between a QNM and a DIM involves two BSM. 
A QNM frequency first lands on the branch cut $\om \in \left[ -m, +m \right]$. 
It then turns into two real frequencies, the associated modes still being QNM as they grow exponentially at spatial infinity. 
These frequencies move apart from each other, remaining “just below” the cut, and eventually reach $\pm m$ and move “just above” the cut. 
More precisely, the corresponding modes become asymptotically bounded. 
Further increasing $L$, the two frequencies eventually merge and give rise to a DIM. 

To understand the structure of the solutions of the field equation on more general grounds, we performed a mode decomposition of the field $\phi$ in a generic, asymptotically homogeneous potential. 
Defining the Laplace transform $\bar{\phi}$ of $\phi$, the general solution may be written as 
\begin{align}
\bar{\phi}(x;\la) = \int G_\la (x,x') \, J_\la^0(x') \s \dd x',
\end{align} 
where $G_\la$ is the Green function, satisfying the stationary field equation sourced by a Dirac delta function, and $J_\la^0$ depends on the initial data. 
Performing the inverse Laplace transform, we arrive at the mode decomposition
\begin{align}\label{eq:modedec}
\phi(x,t) = \int_{\sigma_c} \lp a_\om^u \s \e^{-\ii \om t} \s \varphi_\om^u(x)+a^v_\om \s \e^{-\ii \om t} \s \varphi_\om^v(x) \rp \, \dd\om 
+ \sum_{\sigma_d \cap \mathbb R} d_a \s \e^{-\ii \om_a t} \s \chi_a(x) \nn
+ \sum_{\sigma_d \setminus \mathbb R} \lp b_a \s \e^{-\ii \la_a t} \s \varphi_a(x) + c_a \s \e^{-\ii \la_a^* t} \s \psi_a(x)\rp, 
\end{align}
where $\sigma_c$ denotes the continuous spectrum and $\sigma_d$ the discrete spectrum. 
The former is given by the branch cut(s) of $G_\la$ and the latter by its poles. 
Importantly, to arrive at \eqref{eq:modedec} one must choose asymptotically bounded asymptotic conditions. 
The modes $\varphi_\om^u$ and $\varphi_\om^v$ then are linear combinations of plane waves in the asymptotic regions, i.e., the usual scattering modes, while $\chi_a$, $\varphi_a$, and $\psi_a$ are exponentially decreasing. 
$\chi_a$ is a BSM, with a real frequency. 
$\varphi_a$ is a DIM, exponentially decreasing in space in each asymptotic region but exponentially growing in time, while $\psi_a$ decays exponentially in both space (in the two directions) and time (for increasing $t$). 
To make link with the QNM, one can replace the asymptotically bounded boundary conditions in the definition of $G_\la$ by outgoing ones. 
The branch cut along $\sigma_c$ is then erased and replaced by one along the segment $\left[ -m, +m \right]$.  
This does not change the value of the Green function in the upper complex plane, so that its poles there still correspond to DIM. 
However, its poles in the lower half-plane now give the QNM frequencies. 
In particular, performing the inverse Laplace transform with this outgoing Green function, one can show that QNM contribute to the late-time behavior of the solution, and become dominant if the DIM are absent or not excited.  

To study the transition between QNM and DIM when varying $L$ in a general context, we worked with a modified version of the Friedrichs model~\cite{CPA:CPA3160010404}. 
In its original form, it describes the generation of a QNM by the interaction between a BSM and a continuum of modes with norms of positive sign. 
We generalized it to include another continuum with a negative norm. 
The Hamiltonian is 
\begin{align}
H_S =& \int \om' \ket{\phi_{\om'}^{(+)}} \bra{\phi_{\om'}^{(+)}} \dd \om' 
+ \int \om' \ket{\phi_{\om'}^{(-)}} \bra{\phi_{\om'}^{(-)}} \dd \om'
+ \om_0 \ket{\varphi_{\om_0}} \bra{\varphi_{\om_0}} \nn
&+ \int V^{(+)}(\om') \ket{\phi_{\om'}^{(+)}} \bra{\varphi_{\om_0}} \dd \om' + \int V^{(+)*}(\om') \ket{\varphi_{\om_0}} \bra{\phi_{\om'}^{(+)}} \dd \om' \nn 
&+ \int V^{(-)}(\om') \ket{\phi_{\om'}^{(-)}} \bra{\varphi_{\om_0}} \dd \om' - \int V^{(-)*}(\om') \ket{\varphi_{\om_0}} \bra{\phi_{\om'}^{(-)}} \dd \om',
\end{align}
where $\ket{\varphi_{\om_0}}$ is the BSM, $\lb \phi_\om^{(+)} \rb$ the positive-norm continuum, and $\lb \phi_\om^{(-)} \rb$ the negative-norm continuum. 
In the limit of small coupling, assuming $V^{(+)}$ and $V^{(-)}$ are analytic at $\om = \om_0$, we obtain that the imaginary part of the pole of the resolvant is
\begin{align}
\Gamma \approx \pi \lp \abs{V^{(-)}(\om_0)}^2 - \abs{V^{(+)}(\om_0)}^2 \rp
\end{align}
if the norm of the BSM is positive. 
We thus obtain a smooth transition between a QNM ($\Gamma < 0$) and a DIM ($\Gamma > 0$) when varying a control parameter so that $\abs{V^{(-)}(\om_0)}^2 - \abs{V^{(+)}(\om_0)}^2$ goes from negative to positive values, as in the massless case in our toy-model. 
We also showed that the modified Friedrichs model can recover the different steps in the transition between a QNM and a DIM in the massive case when adding a branch cut along $\left[ -m , +m \right]$ to the potentials $V^{(+)}$ and $V^{(-)}$. 

Finally, we also considered what happens when putting the field on a torus with periodic boundary conditions. 
This case is conceptually simpler, although technically mode involved, as the continuous spectrum is replaced by a discrete set of real-frequency modes. 
There is no QNM anymore, and DIM appear through the merging of two real-frequency modes with opposite norms. 
At the level of the Green function, the limit of an infinite torus gives the $G_\la$ obtained with asymptotically bounded modes. 
DIM are thus recovered in this limit, but QNM require an analytic continuation through the branch cut in the lower complex half-plane. 

In the Appendix.~C of~\cite{Coutant:2016bgk}, we explain why this simple model is relevent for black hole physics. 
The idea is that, in both cases (and, in fact, on much more general grounds), DIM come from the interactions between positive- and negative-norm modes with the same frequency in the presence of a trapping potential. 

\clearpage

\section[Scattering of gravity waves in subcritical flows over an obstacle]{Scattering of gravity waves in subcritical flows over an obstacle \cite{Robertson:2016ocv}}

This work continued and clarified~\cite{Michel:2014zsa} (reported in Chapter~\ref{ch:probing}) and~\cite{Michel:2015aga}, with the aim to provide a more systematic analysis able to guide new experiments. 
It consists in a numerical study of the scattering coefficients of linear water waves on a stationary, inviscid, irrotational flow of an ideal fluid over a localized obstacle. 
To simplify the analysis and reduce the number of parameters, we assume that the free surface of the unperturbed flow is asymptotically flat, i.e., that there is no undulation.~\footnote{We further assumed there is no finite undulation.} 
Furthermore, we assume that the unperturbed flow and perturbations are uniform along one direction, so that the problem becomes effectively (2+1)-dimensional, and asymptotically subcritical, as is the case in analogue gravity experiments using water waves. 

\begin{figure}
\centering
\includegraphics[width=0.49 \linewidth]{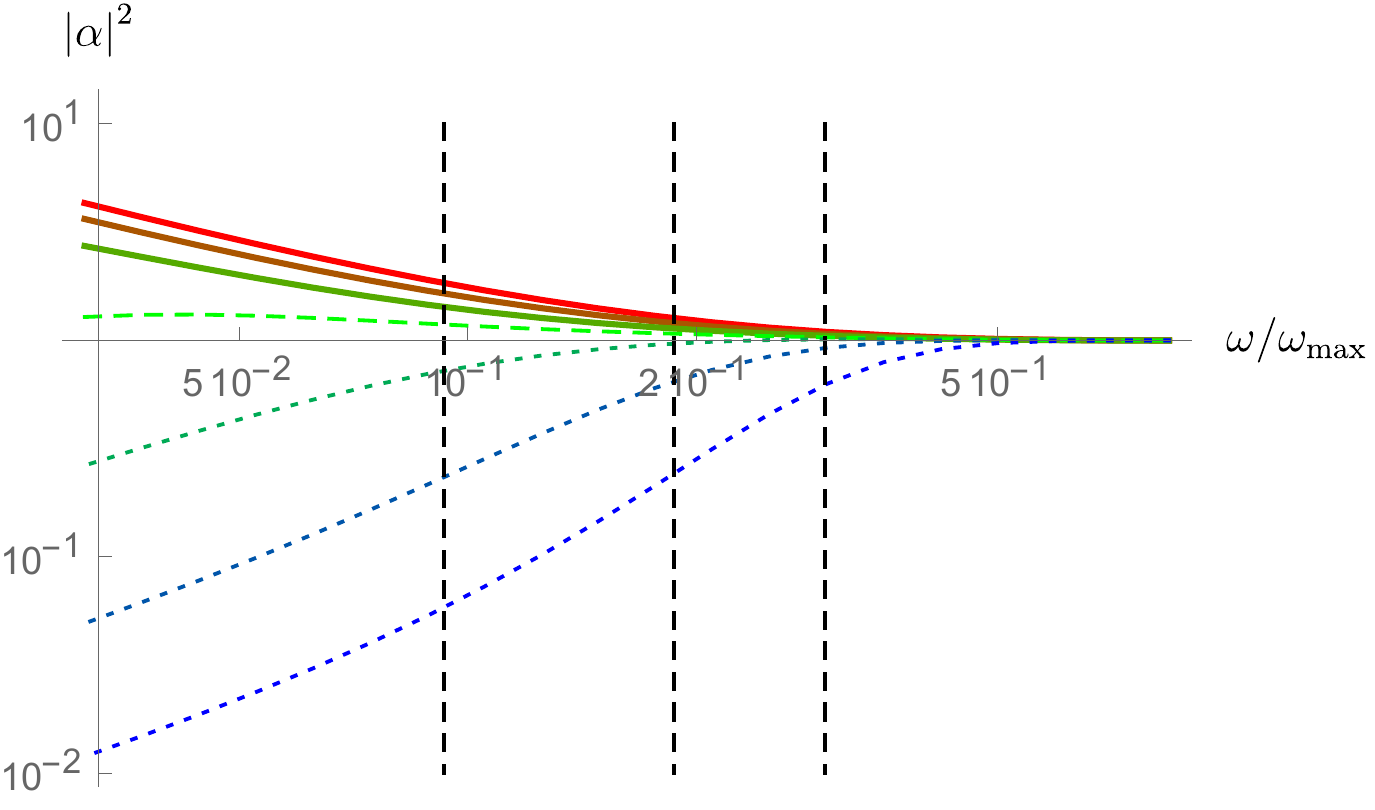}
\includegraphics[width=0.49 \linewidth]{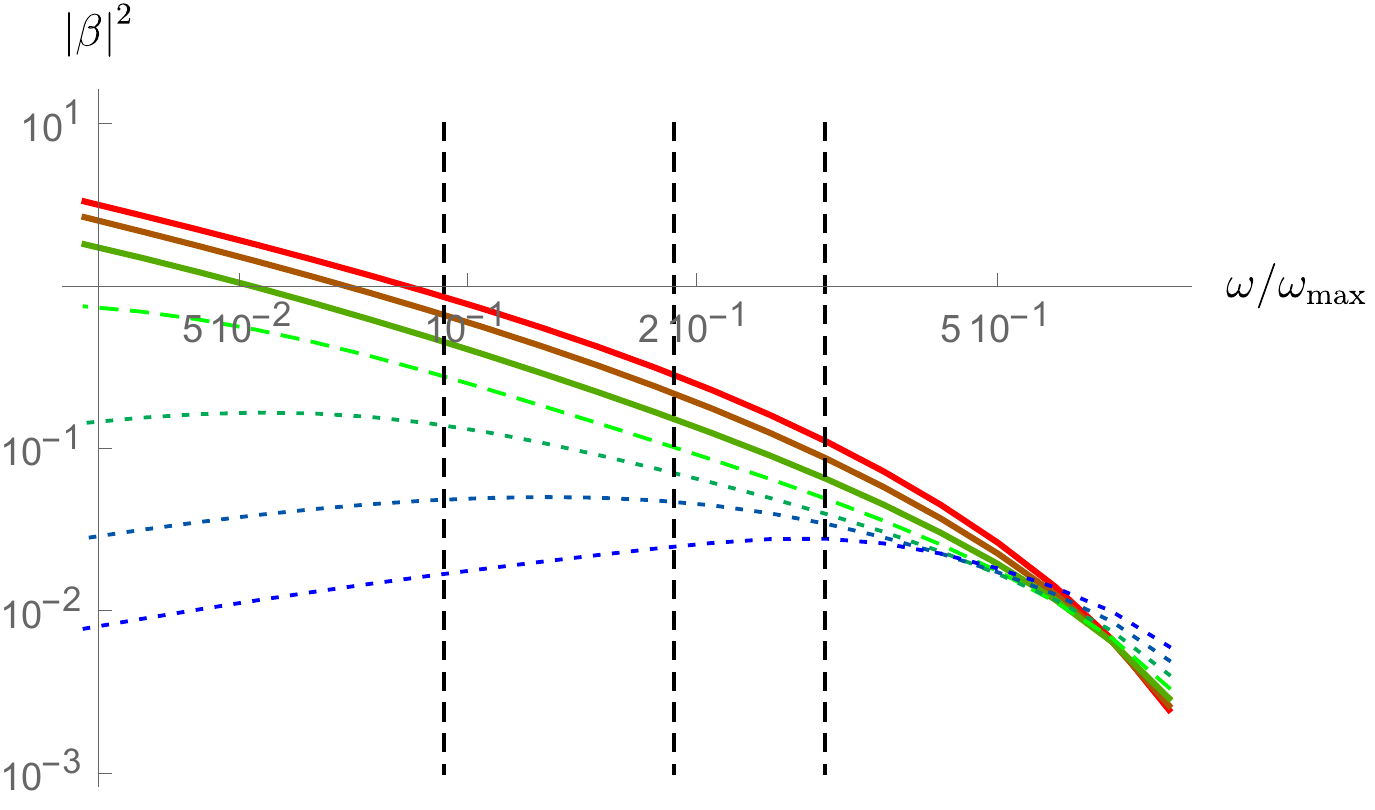}
\includegraphics[width=0.49 \linewidth]{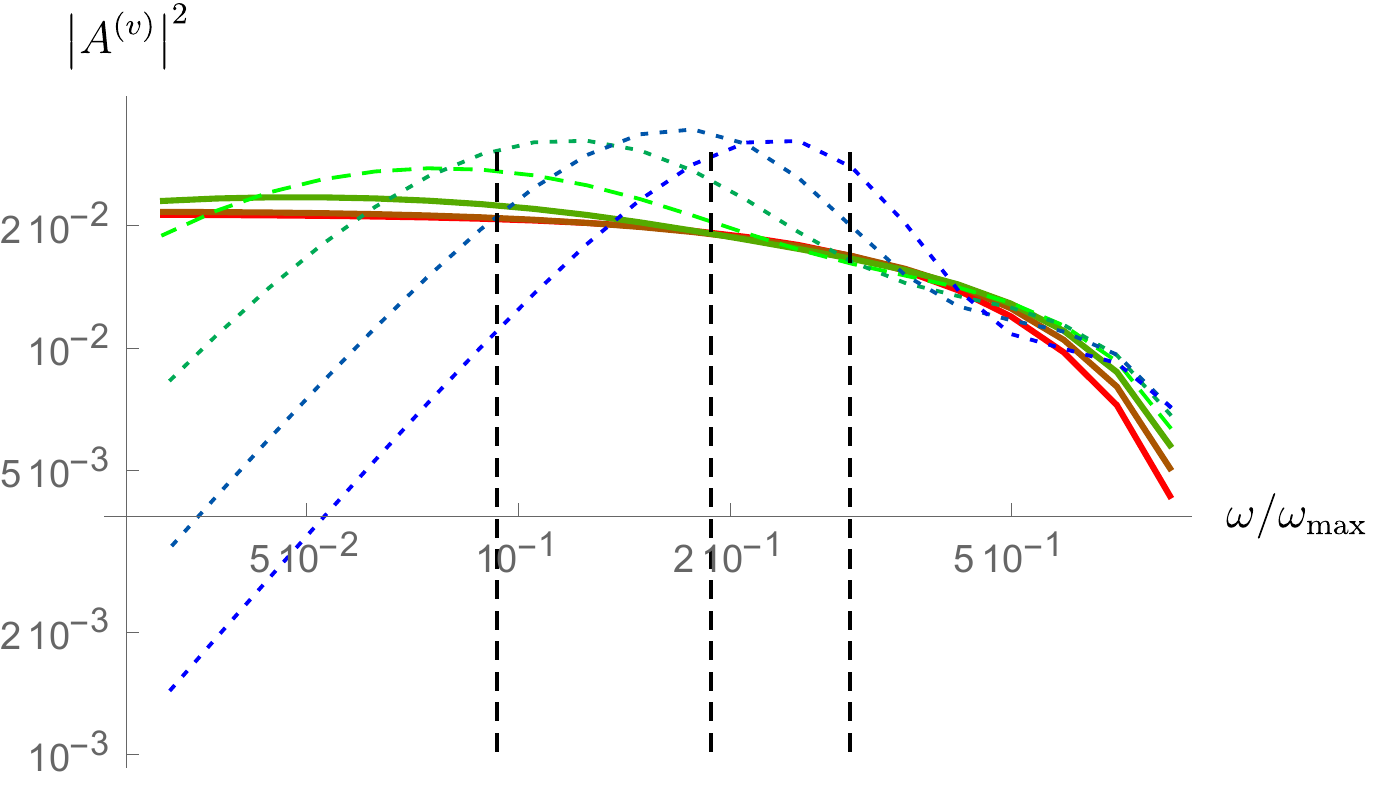}
\includegraphics[width=0.49 \linewidth]{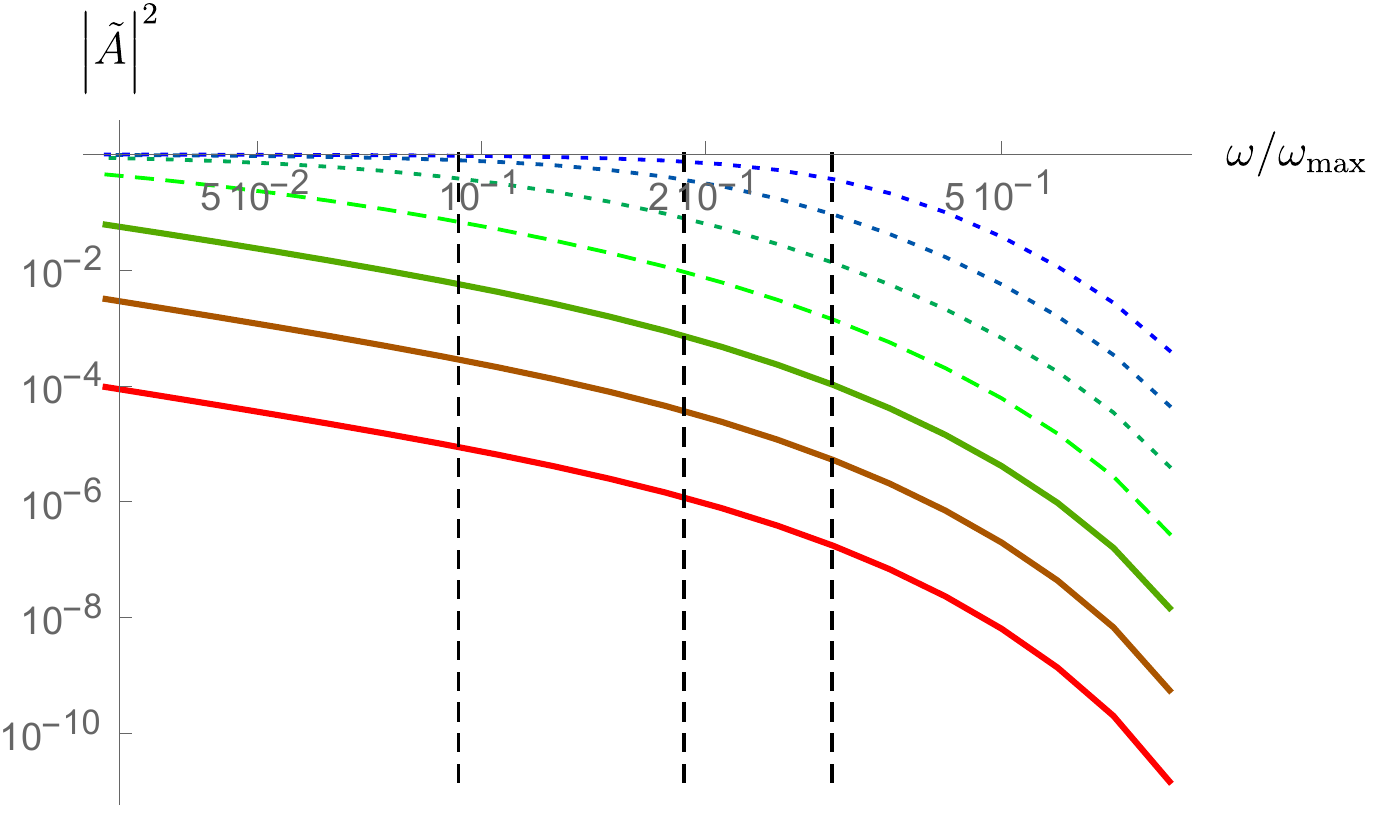}
\caption{Behaviour of the absolute values of the 4 scattering coefficients involving the incoming counter-propagating mode for different values of the maximum Froude number. 
The latter takes equally-spaced values between $1.2$ (red curves) and $0.8$ (blue curve). 
The continuous curves correspond to transcritical flows with $F_{\rm max} > 1$, the dashed one to $F_{\rm max} = 1$, and the dotted ones to subcritical flows with $F_{\rm max} < 1$. 
The vertical dashed lines show the values of $\om_{\rm min}$ for the three subcritical flows.} \label{fig:concl_4coeffs}
\end{figure}

While~\cite{Michel:2014zsa} was mainly focused on subcritical flows and the low-frequency behavior of the coefficients $\beta_\om$ used to define an effective temperature, in~\cite{Robertson:2016ocv} we studied both the transition from a transcritical flow to a subcritical one (see~\fig{fig:concl_4coeffs}) and the main parameters needed to describe the spectrum. 
In particular, building on preliminary results reported in~\cite{Michel:2015aga}, we estimated the influence of the shape of the obstacle.~\footnote{Another effect of the shape of the obstacle, not considered in that work, is the formation of an undulation. In particular, it can be tuned so that the latter's amplitude vanishes.} 
To be specific, we worked with background flows parametrized by
\begin{align} \label{eq:concl_Fx}
F(x) = F_{\rm as} + \lp F_{\rm max} - F_{\rm as} \rp f(x),
\end{align}
where $F(x)$ is the local Froude number, $F_{\rm max}$ its maximum value, $F_{\rm as}$ its asymptotic value, and the function $f$ has the form
\begin{align}
f(x) = \mathcal{N} \, \left[
1 - \tanh\lp a_L \, (x + L / 2) \rp \tanh \lp a_R \, (x - L / 2) \rp
\right],
\end{align}
where $\lp a_L, a_R, L \rp \in \mathbb{R}_+^3$ and $\mathcal{N}$ is chosen such that ${\rm max}_{x \in \mathbb{R}}f(x) = 1$. 
The wave equation \eqref{eq:waveeq} was integrated numerically using a slightly modified version of the code used in Chapter~\ref{ch:probing} to obtain the 16 scattering coefficients defined by
\begin{align} \label{eq:16coeffs} 
\begin{pmatrix}
\symbAtin \\ 
\symbain \\ 
\symbbin \\
\symbAin \\ 
\end{pmatrix} =
\begin{pmatrix}
\tilde{A}_\om & \alpha_\om & \beta_\om &  A_\om^{(v)}  \\ 
\bar{\alpha}_\om &  A_\om & B_\om & \alpha_\om^{(v)} \\  
\bar{\beta}_\om & \bar{B}_\om & \bar{A}_\om 
& \beta_\om^{(v)}  \\
 \bar{A}_\om^{(v)} & \bar{\alpha}_\om^{(v)} & \bar{\beta}_\om^{(v)} & 
A_\om^{(vv)} 
\end{pmatrix}
\begin{pmatrix}
\symbAtout \\ 
\symbaout \\  
\symbbout\\
\symbAout 
\end{pmatrix}.
\end{align} 

\begin{figure}
\centering
\includegraphics[width=0.49 \linewidth]{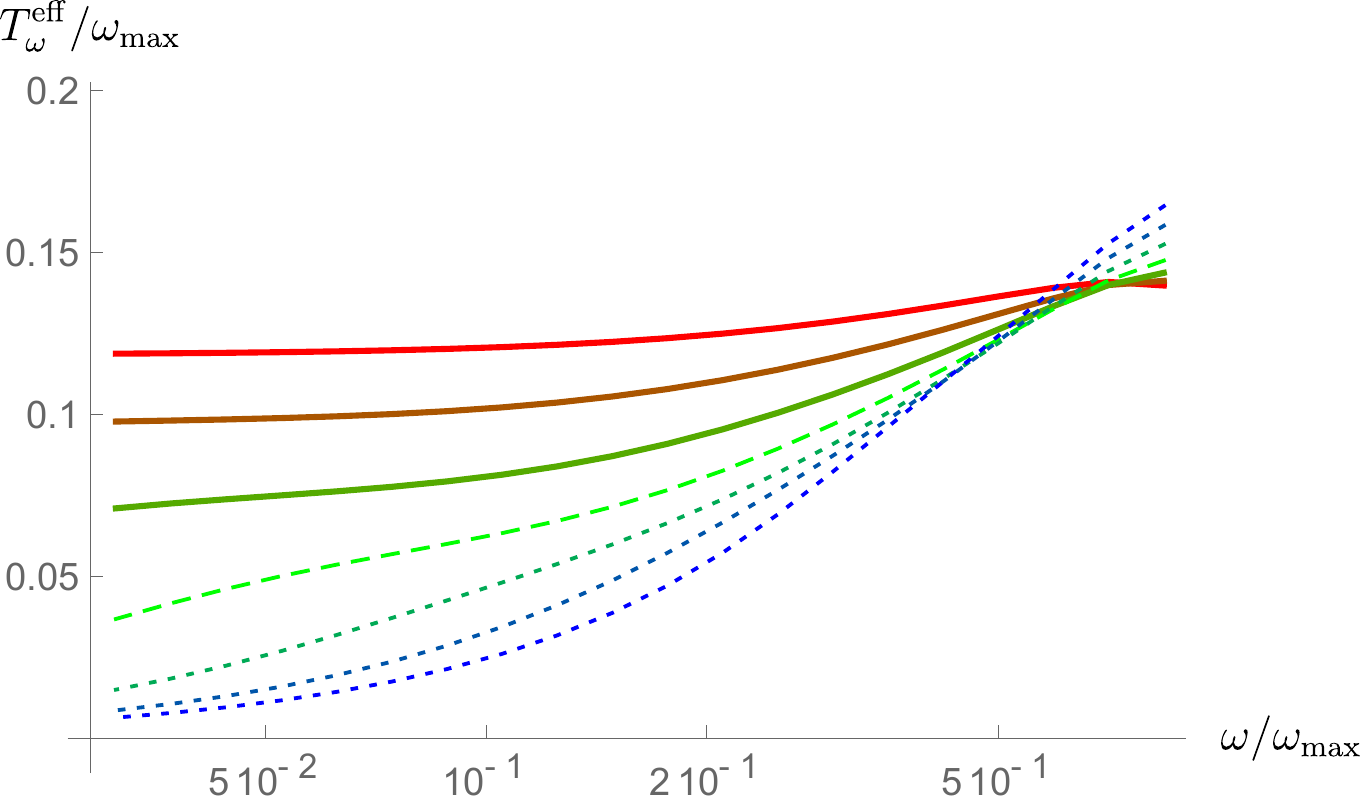}
\includegraphics[width=0.49 \linewidth]{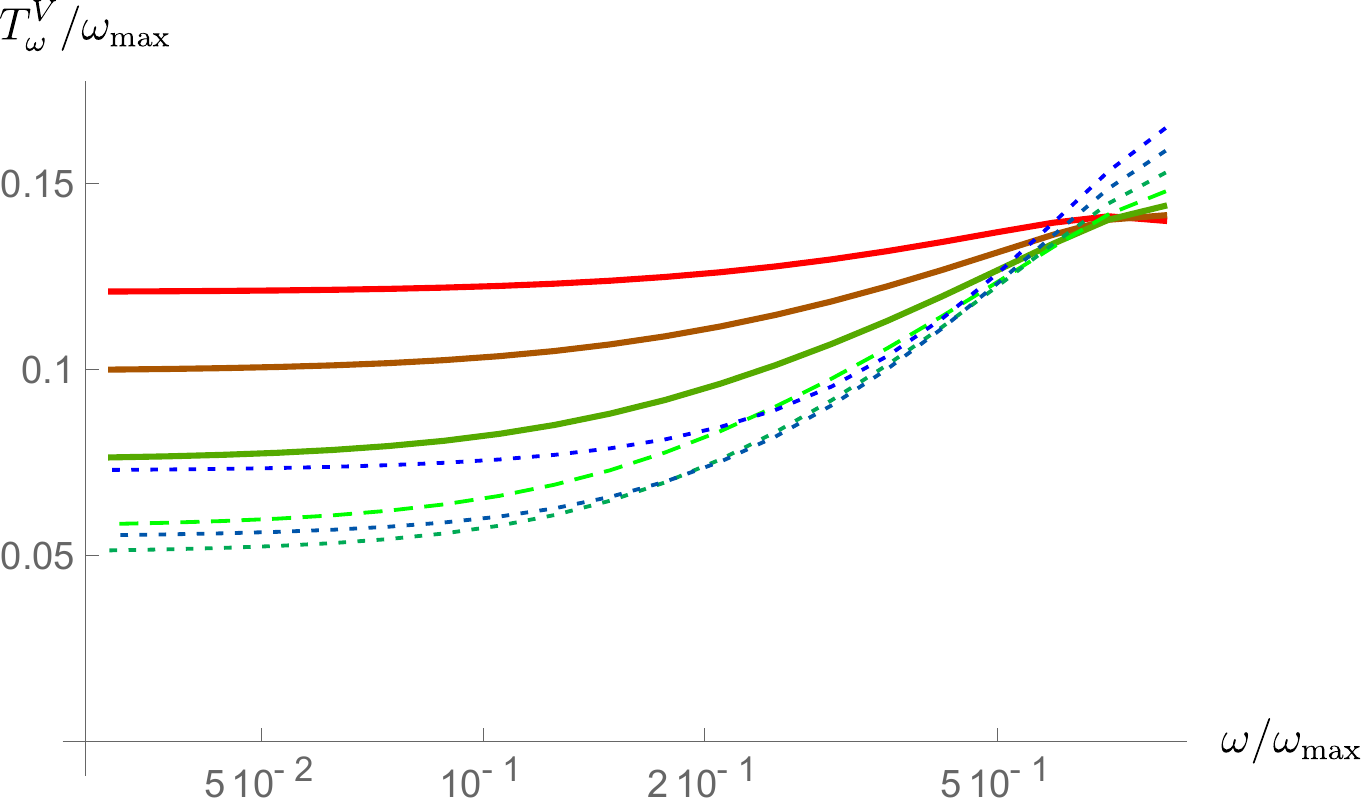}
\caption{Comparison of the two temperatures $T_\om^{\rm eff}$ and $T_\om^V$ for the 7 flows of \fig{fig:concl_4coeffs}.}
\label{fig:concl_temp}
\end{figure}

Their typical behavior in the transcritical case is shown in \fig{fig:concl_coeffs_trans} in dependence of the angular frequency $\om$ of the wave. 
One first notices than the absolute values of some of these coefficients can become significantly larger than unity, growing like $1/\om$ or $1/\om^2$ for decreasing $\om$ above a small cutoff frequency $\om_c$, the latter being a consequence of the finite length of the obstacle. 
(In the limit where its extension goes to infinity, $\om_c$ goes to 0 and these scattering coefficients diverge for $\om \to 0$.) 
This indicates a large over-reflection: the energy of a reflected wave can be much larger than that of the incident one, provided a negative-energy wave carries the energy difference. 
Moreover, when $F_{\rm max}$ is significantly larger than unity (typically, for $F_{\rm max} > 1.2$), the spectrum is nearly thermal in the frequency domain $\om_c \ll \om \ll \om_{\rm max}$, with a temperature close to the Hawking one $T_H$ when $T_H \ll \om_{\rm max}$, see~\fig{fig:concl_temp}. 

In a subcritical flow instead, the absolute values of all scattering coefficients remain smaller than or close to $1$, see~\fig{fig:concl_coeffs_sub}. 
Our numerical results indicate that the spectrum strongly depends on the precise shape of the function $F$ and do not seem to be describable by a single quantity in a large frequency range. 
To further analyze the differences with respect to the transcritical case, it is useful to define two temperatures. 
The first one, $T_\om^{\rm eff}$, is defined by \eqref{eq:effT} (we now add the superscript “$\rm eff$” to avoid possible confusion.) 
The second one, $T_\om^V$, is defined by
\begin{align*}
\left\lvert 
\frac{\beta_\om}{\alpha_\om}
\right\rvert^2
= \e^{- \om / T_\om^V}.
\end{align*}
(We work in units where $\hbar = k_B = 1$.) 
Their dependence in $\om$ are shown in \fig{fig:concl_temp} for several flows with different values of $F_{\rm max}$. 
For transcritical flows, these two temperatures are very close to each other.~\footnote{This can be proved analytically in the frequency range $\om_c \ll \om \ll T_H$ using the unitarity relation and the fact that $\abs{\alpha_\om}$ and $\abs{\beta_\om}$ grow like $\om^{-1/2}$ when decreasing $\om$ while $|A_\om^{(u)}|$ and $|\tilde{A}_\om|$ remain smaller than or of the order of $1$.} 
In subcritical flows, however, they behave very differently at small frequencies: $T_\om^{\rm eff}$ goes to zero like $\om^2$ for $\om \to 0$ while $T_\om^V$ goes to a finite, non-vanishing value in the same limit.  

%

\begin{myfig}
\includegraphics[width=0.49 \linewidth]{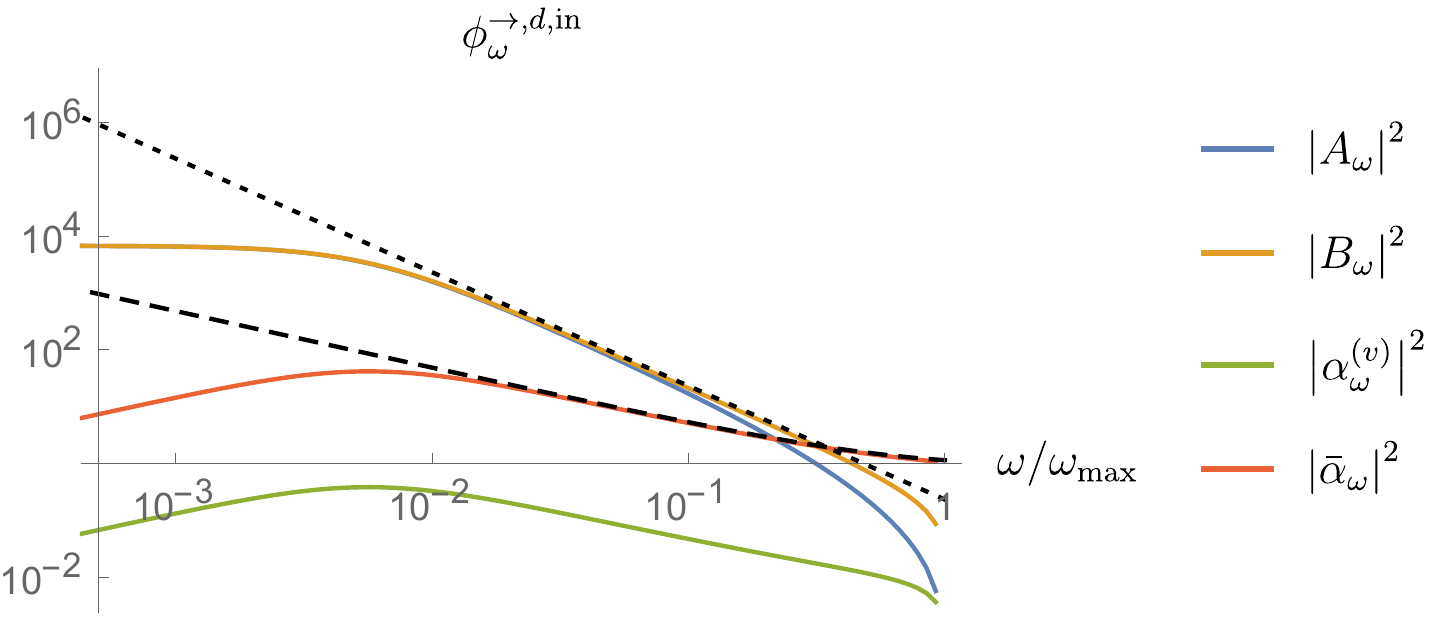}
\includegraphics[width=0.49 \linewidth]{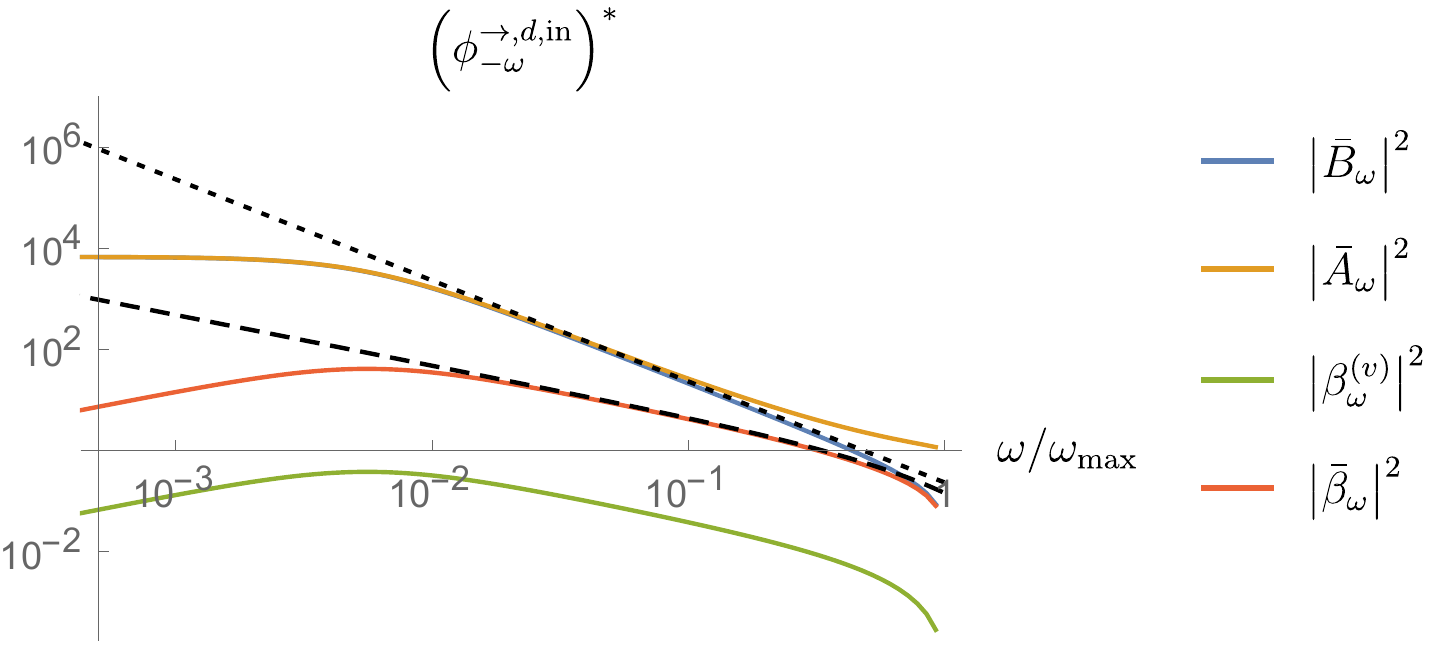}
\includegraphics[width=0.49 \linewidth]{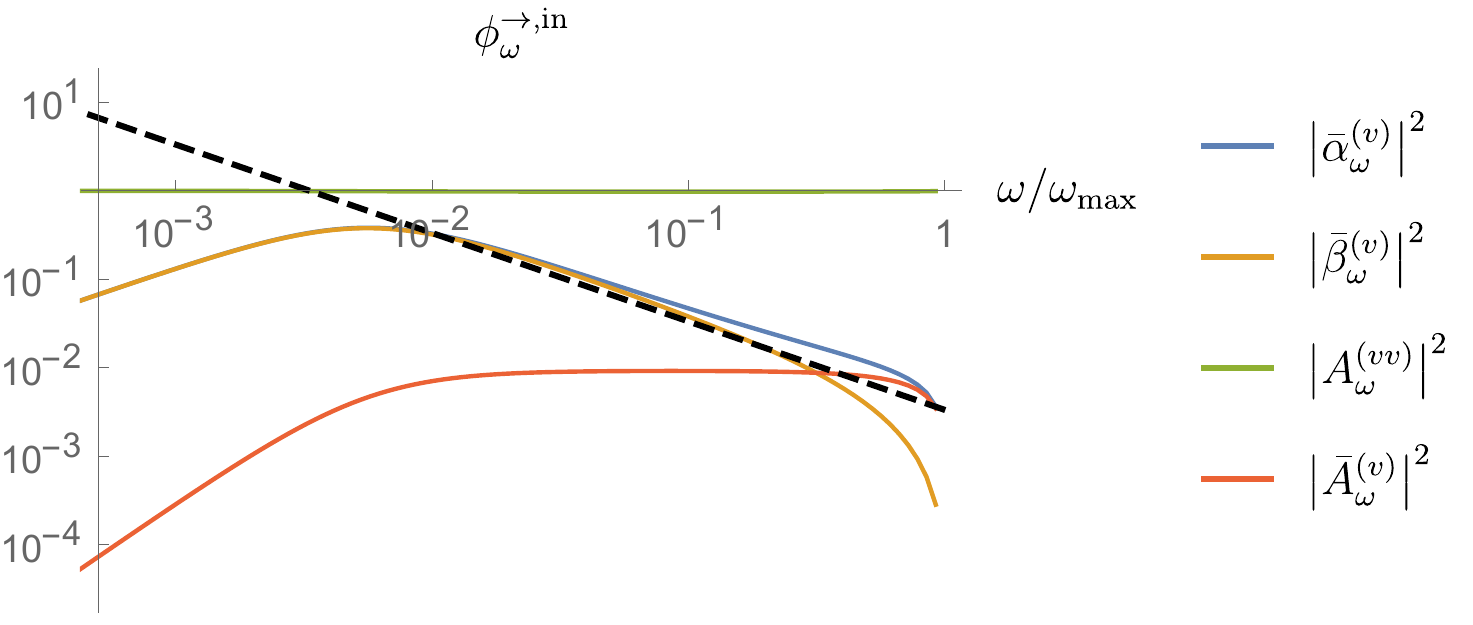}
\includegraphics[width=0.49 \linewidth]{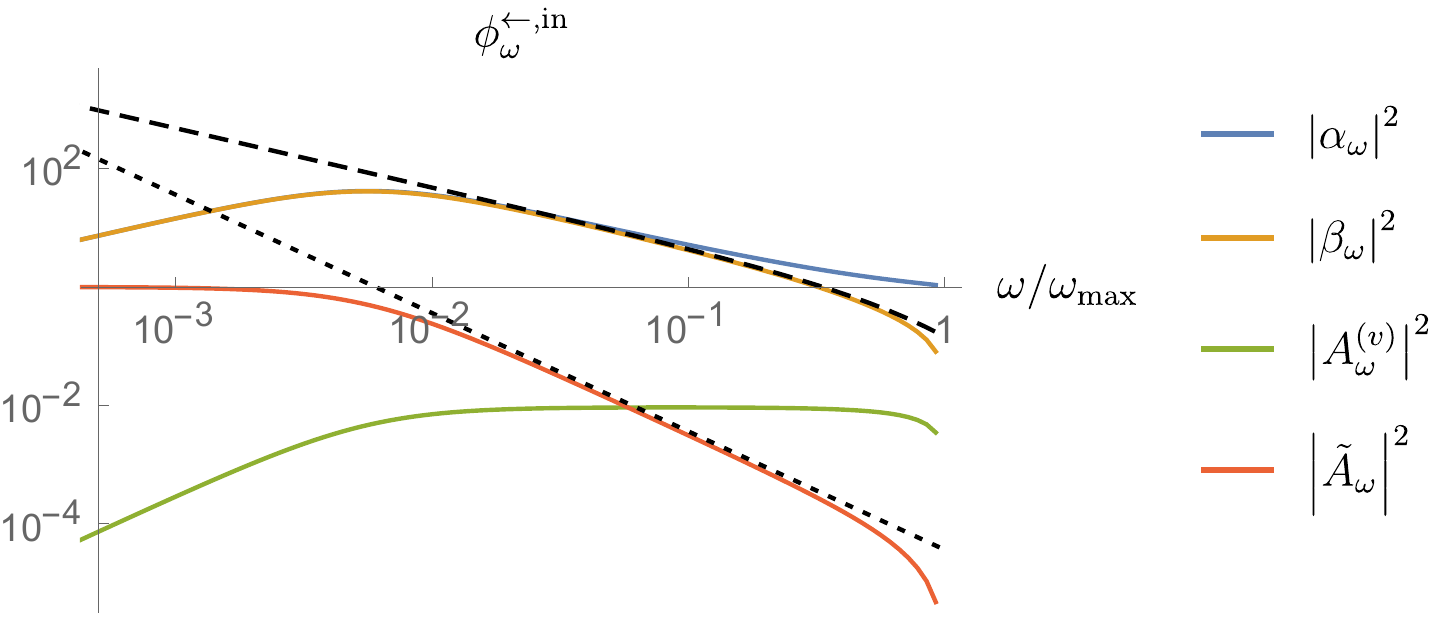}
\capf{Behaviour of the absolute values of the 16 scattering coefficients in a transcritical flow of water over an obstacle. 
The Froude number follows~\eq{eq:concl_Fx} with the parameters $F_{\rm as} = 0.6$, $F_{\rm max} = 1.4$, $a_R = a_L = 2 \, h_{\rm as}^{-1}$, and $L = 2 \, h_{\rm as}$, where $h_{\rm as}$ is the asymptotic water depth. 
Dotted and dashed straight lines are guides for the eye. 
The former are parallel to the line $\om \mapsto \om^{-2}$ and the latter to $\om \mapsto \om^{-1}$.} \label{fig:concl_coeffs_trans}
\end{myfig}

\begin{myfig}
\includegraphics[width=0.49 \linewidth]{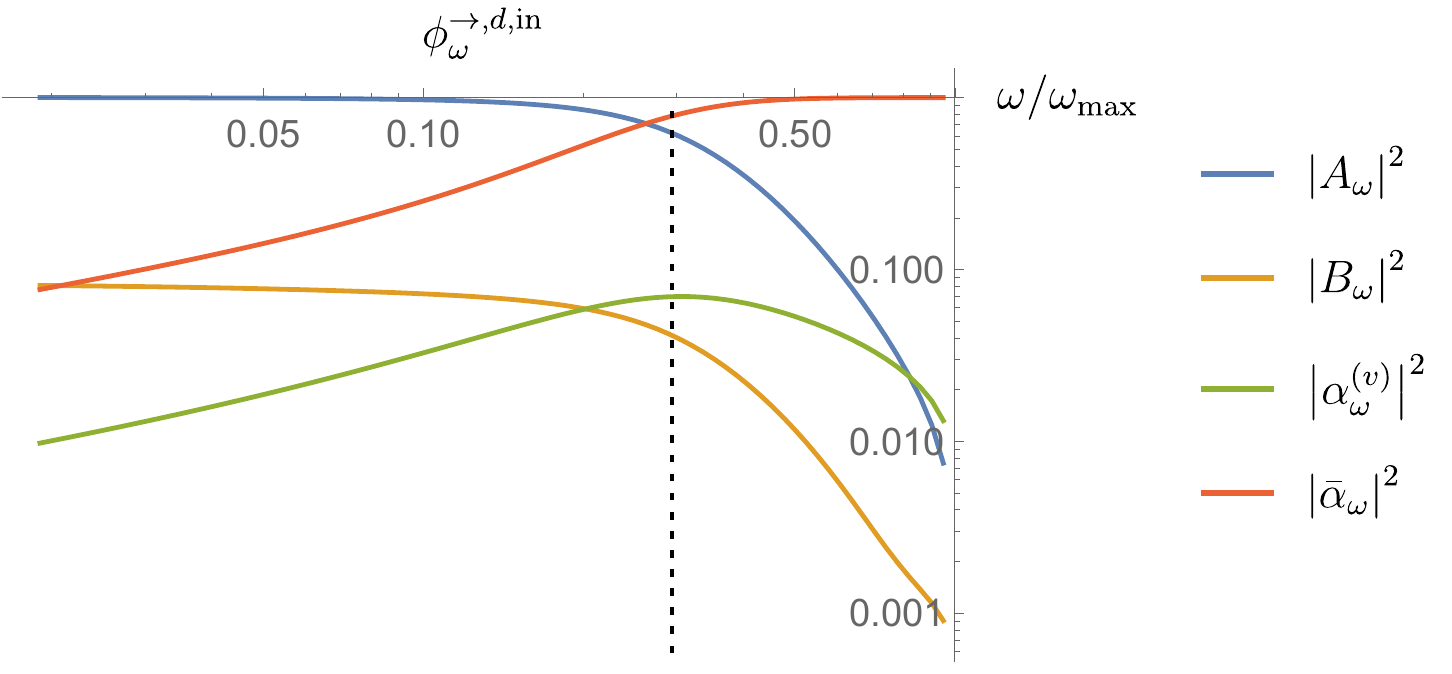}
\includegraphics[width=0.49 \linewidth]{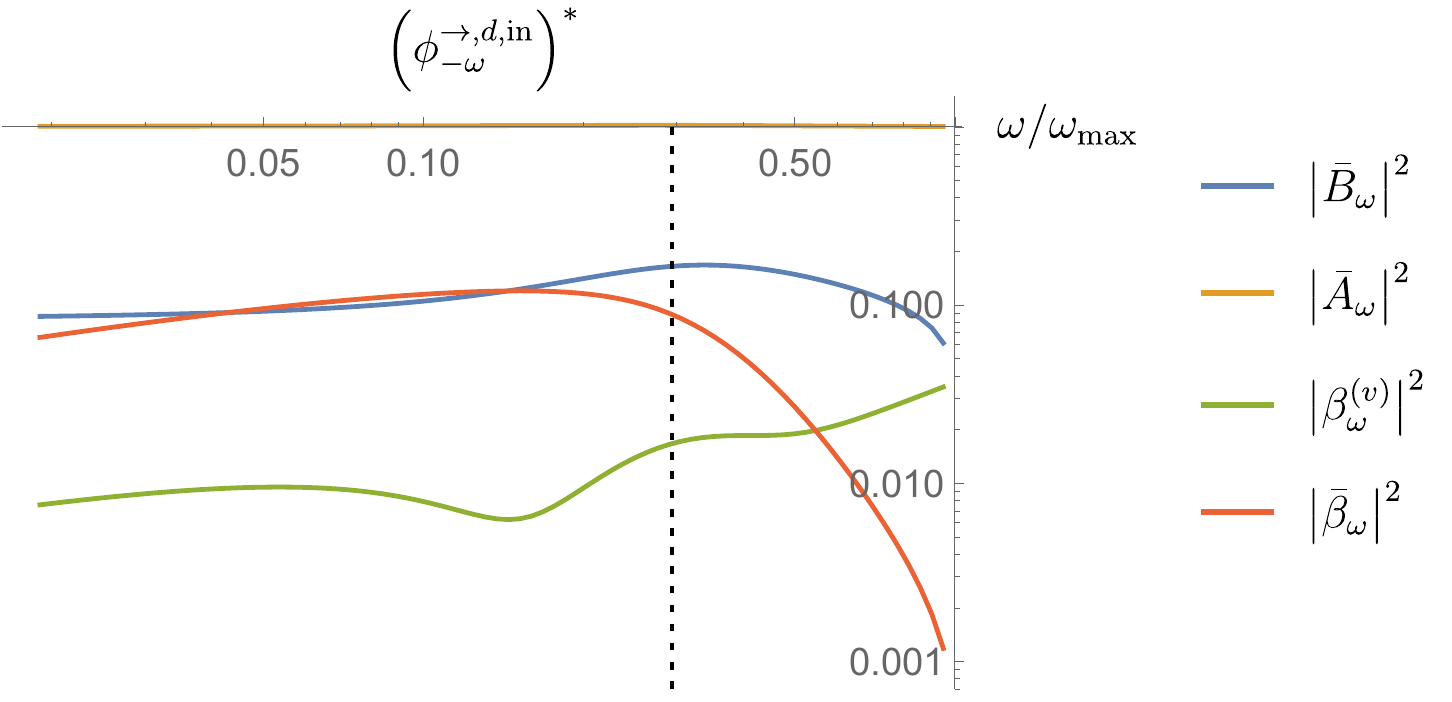}
\includegraphics[width=0.49 \linewidth]{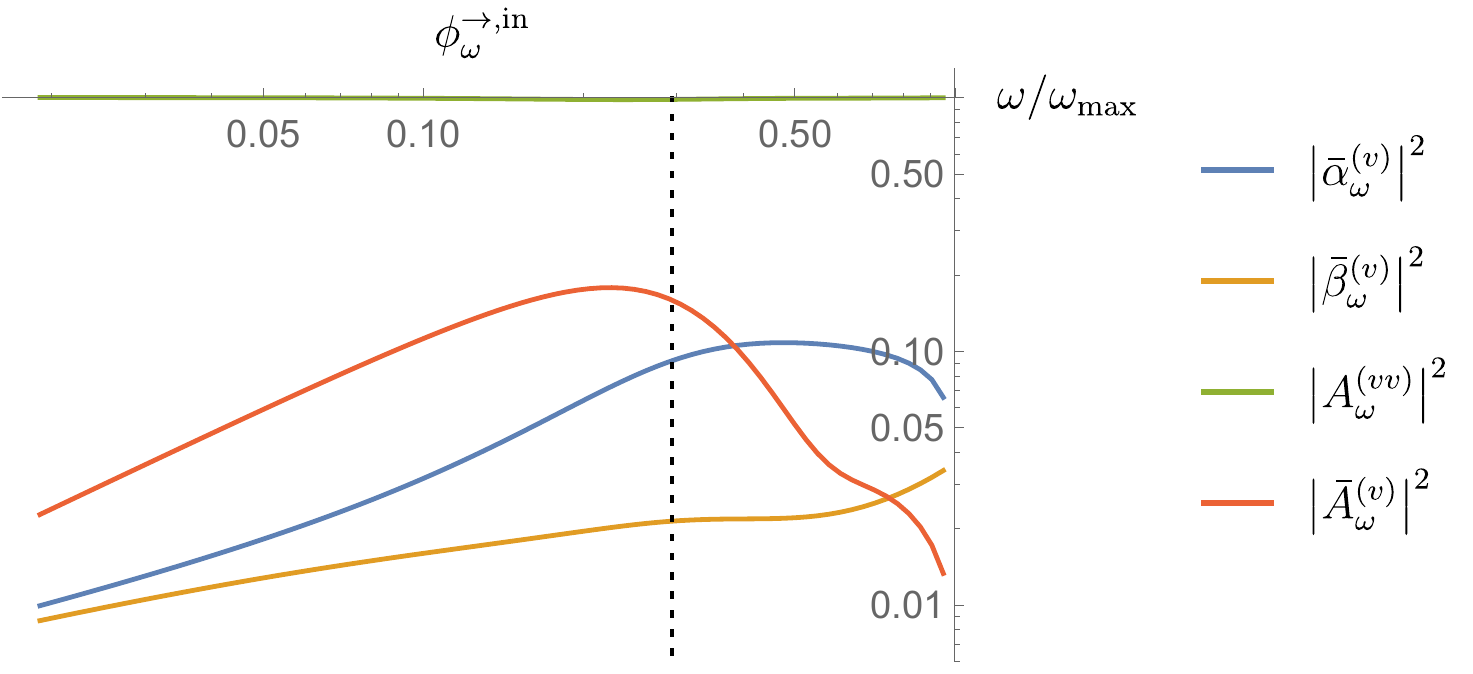}
\includegraphics[width=0.49 \linewidth]{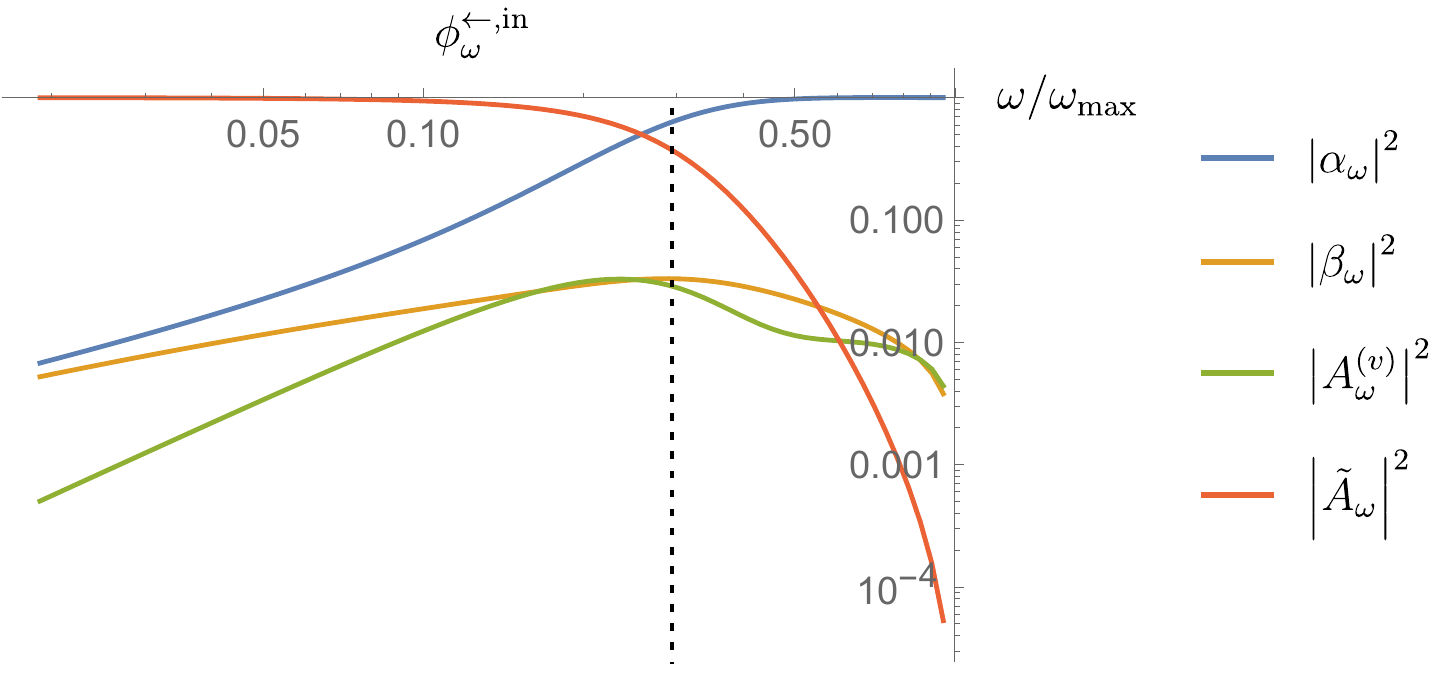}
\capf{Behaviour of the absolute values of the 16 scattering coefficients in a subcritical water flow over an obstacle. The parameters are: $F_{\rm as} = 0.6$, $F_{\rm max} = 0.8$, $a_R = 4 \, a_L = 2 \, h_{\rm as}^{-1}$, and $L = 4 \, h_{\rm as}$. 
The vertical dotted line shows $\om_{\rm min}$.} \label{fig:concl_coeffs_sub}
\end{myfig}

\clearpage

\clearpage

\section[Observation of noise correlated by the Hawking effect in a water tank]{Observation of noise correlated by the Hawking effect in a water tank \cite{Euve:2015vml}}
\label{sub:exp_paper_2}

Experiments designed to detect the analogue Hawking radiation work at several different levels. 
In Vancouver, the group of S.~Weinfurtner~\cite{Weinfurtner:2010nu} used a classical hydrodynamic system to demonstrate the mode conversion at the heart of the Hawking effect. 
Sending an incident wave in a counterflow made inhomogneous by the presence of an obstacle at the bottom of a flume, they observed and measured the amplitudes of the two dispersive waves generated at the blocking point, which were {\it a posteriori} found to agree with numerical calculations~\cite{Michel:2014zsa} and analytical results~\cite{Coutant:2016vsf}. 
J.~Steinhauer, on the other hand, used a gas of cold atoms accelerated by a laser producing a sharp potential change to observe the quantum analogue Hawking effect~\cite{Steinhauer:2015saa}. 
In particular, he used the two-point correlation function of the density perturbations to demonstrate the intrication between phonons on two sides of the horizon. 
In the work~\cite{Euve:2015vml}, we partially bridged the gap between these configurations by measuring the two-point correlation function for classical water waves in a water flume, with the aim to detect the correlations generated by the scattering in a classical system. 

\begin{figure}[ht]
\includegraphics[width=\linewidth]{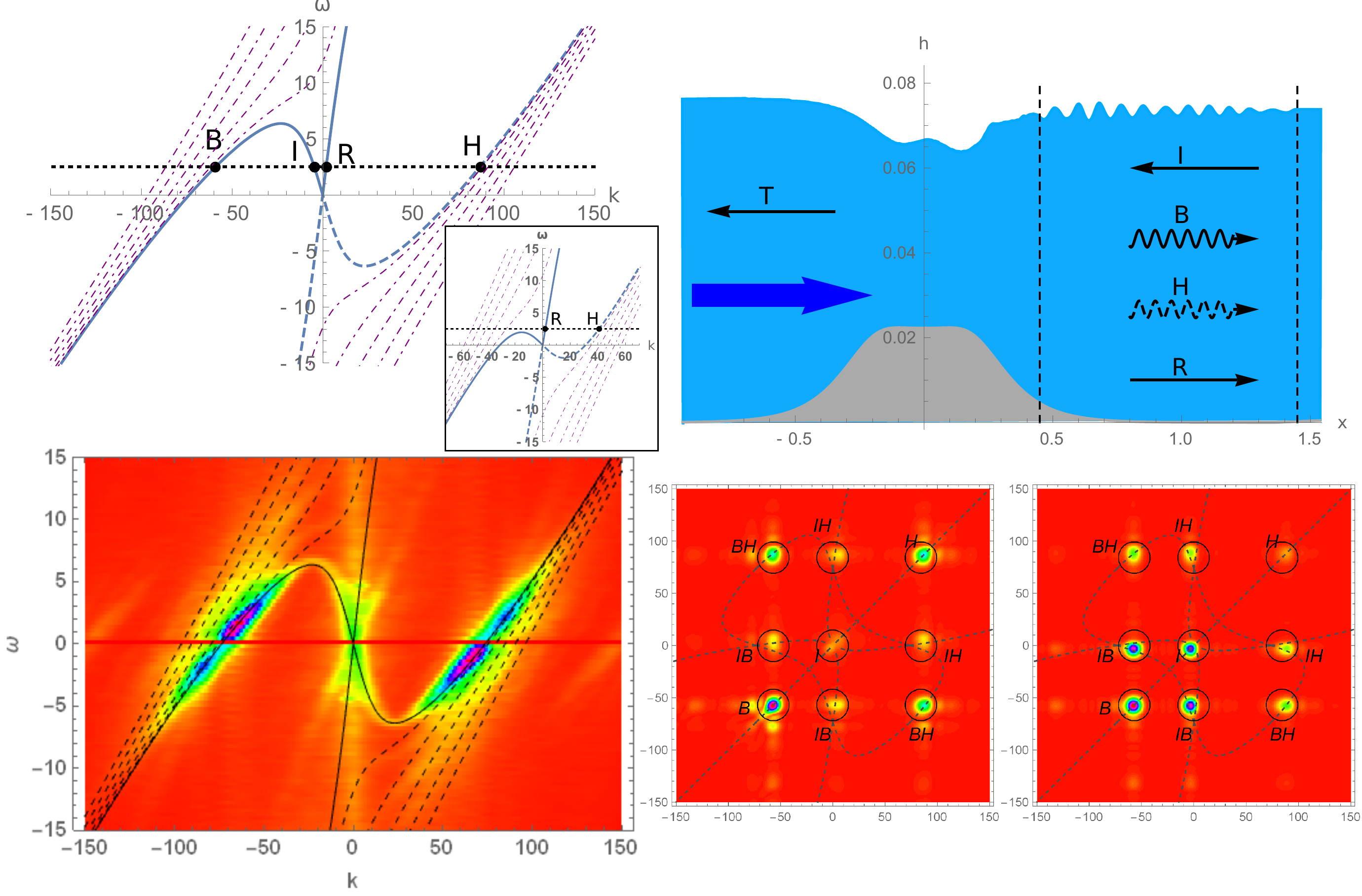} 
\caption{Dispersion relation, schematic drawing of the experimental setup, and main observations of~\cite{Euve:2015vml}. 
{\it Top, left panel:} Dispersion relation in the asymptotic region far from the obstacle. 
Blue lines correspond to “longitudinal” modes, homogeneous in the transverse direction of the flume. 
The continuous and dashed ones show modes with a positive and negative “norm” from~\eqref{eq:intro_KGnorm}, respectively.  
Purple, dot-dashed lines correspond to modes with a nonvanishing, even number of nodes $n_n$ in the transverse direction (we show them for $n_n \leq 10$). 
The four dots show the wave vectors of longitudinal modes with a fixed angular frequency $\omega$, materialized by a horizontal dashed line. 
The inset shows the dispersion relation beyond the turning point for this value of $\omega$.
{\it Top, right panel:} Schematic drawing of the obstacle (grey) and water flow (light blue). 
The thick, dark blue arrow shows the direction of propagation of the flow. 
Thin arrows show the direction of propagation of the waves involved in the scattering of the incident (I) one. 
Wiggly arrows indicate that the wave is dispersive, i.e., has a nonvanishing wave number in the small-frequency limit. 
The dashing indicates the mode with negative “norm”, i.e., negative energy for $\om > 0$. 
The two vertical dashed lines demarcate the region used for the data analysis. 
{\it Bottom, left panel:} Power spectrum of the fluctuations of the free surface due to the noise. 
{\it Bottom, right panel:} Two-point correlations of the component of the signal with angular frequency $\om = 2.5 {\rm Hz}$ for the noise (left) and when sending an I wave (right). 
The $9$ circles are centred on the points where correlations are expected theoretically. 
Dashed lines show their locus when varying $\omega$. 
}
\label{fig:ONCHEWT}
\end{figure}

The experimental setup was very close to the one used in~\cite{Weinfurtner:2010nu,Euve:2014aga},~\footnote{In particular, as was the case in~\cite{Euve:2014aga} and, to the best of our knowledge, in~\cite{Weinfurtner:2010nu}, the maximum Froude number was smaller than unity.} with two important differences:
\begin{itemize}
\item We used an obstacle shape designed using the method sketched in subsection~\ref{sub:NL} of Chapter~\ref{ch:probing} to reduce the amplitude of the undulation of the downstream free surface for a relatively large maximum Froude number $F_{\rm max} \approx 0.86$. 
\item A laser sheet was used to monitor the elevation of the free surface as a function of time and the longitudinal space coordinate. 
\end{itemize}
The second point allowed us to measure not only the amplitudes of the waves emitted by the analogue Hawking mechanism, but also their correlations in space. 
The latter are interesting {\it per se} as one of the signatures of the (analogue) Hawking effect.
They are also useful to determine the scattering coefficients with a better accuracy. 
Indeed, other possible sources of fluctuations, like vibrations of the flume due to the pump, will generally produce uncorrelated waves. 
Measuring the scattering coefficients {\it via} the correlations thus automatically suppresses the noise - or at least its uncorrelated part.~\footnote{The noise can of course indirectly produce correlated waves if it generates incident waves on the obstacles. But it will thus contribute both to the coefficients of the incoming and scattered waves, with ratios close to the scattering coefficients if the noise is not too large inside the window used for the data analysis, see the top, right panel of \fig{fig:ONCHEWT}.}

For each observation, the free surface was recorded by three cameras during $1000 {\rm s}$. 
The resulting data was then divided into $80$ subsets. 
For each of them, we removed the zero-frequency component and performed a Hamming-Fourier transform in time and space using for the latter an integration window of length $1 {\rm m}$, see the top, right panel of \fig{fig:ONCHEWT}. 
For each value of the angular frequency $\omega$, the amplitudes of the different waves were then extracted by convolution with the Hamming function. 
As a first step, we computed the root mean square Fourier amplitude as a function of $\omega$ and the wave vector $k$ for the noise only, i.e., without sending any wave from the wave maker. 
The corresponding power spectrum is shown in the bottom, left panel of \fig{fig:ONCHEWT} (see~\cite{Euve:2015vml} for the precise definition) along with the dispersion relation
\begin{align} \label{eq:DR2}
\lp \om - \vec{v} \cdot \vec{k} \rp^2 = g \, k \tanh \lp k \, h \rp,
\end{align}
also shown in the top, left panel. 
We notice that most of the power lies close to the lines corresponding to “longitudinal” modes, homogeneous in the transverse direction of the flume. 
This confirms that the dispersion relation~\eqref{eq:DR2} accurately describes the system under consideration.~\footnote{It should be noticed, however, that to obtain a good agreement we had to use effective values of the water depth $h$ and the current $J$ which slightly differ from their measured values. These differences may be due to boundary-layer effects on the bottom of the flume.} 
We then computed the two-point function in Fourier space, see the bottom, right panel of the figure. 
As expected, we observed correlations along the antidiagonal, corresponding to the autocorrelation of each wave with itself. 
As also expected for waves produced simultaneously by the scattering process, we found correlations for all pairs $(k,k')$ where $k$ and $k'$ are two solutions of the wave equation for the same angular frequency. 
The results on the right plot, made when sending a wave from the wave-maker at a frequency for which blocking occurs, are quite clear: the strongest correlations are between the incident wave and the two reflected ones, and are close to unity once properly normalized (see~\cite{Euve:2015vml}). 
This shows that the dispersive waves mostly come from the scattering of the incident one. 
On the left plot, done without sending a wave from the wave-maker, one sees correlations essentially between the two dispersive waves. 
This suggests that the latter are already partially correlated in the noise, but less so with the incident wave. 

Finally, we used these results to estimate the scattering coefficients $\alpha_\omega$ and $\beta_\omega$. 
While the former was in relatively good agreement with numerical simulations done with a simplified flow without undulation and a dispersion relation truncated to quartic order, the latter was significantly larger at large frequencies. 
Although we were so far not able to consistently include the effects of the undulations in our numerical simulations, preliminary calculations indicate that it could explain this discrepancy. 
Other possible effects which may affect the results, apart from next orders in the dispersion relation, are the boundary-layer effects on the bottom and walls of the flume and the vorticity which we expect to be present in the downstream region.

\clearpage

\section[Phonon spectrum and correlations in a transonic flow of an atomic Bose gas]{Phonon spectrum and correlations in a transonic flow of an atomic Bose gas \cite{Michel:2016tog}}
\label{sub:JE}

\begin{figure}  
\centering
\includegraphics[width=0.49 \linewidth]{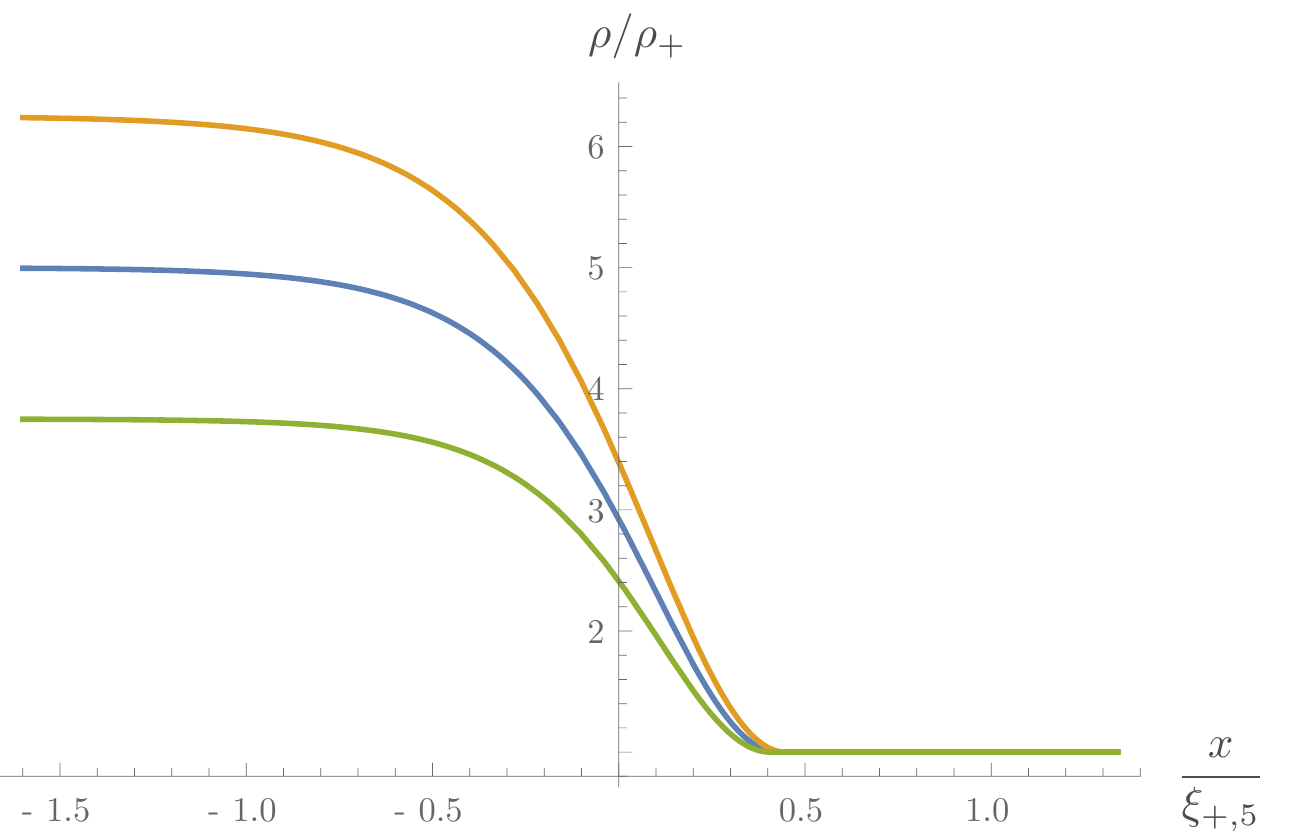} 
\includegraphics[width=0.49 \linewidth]{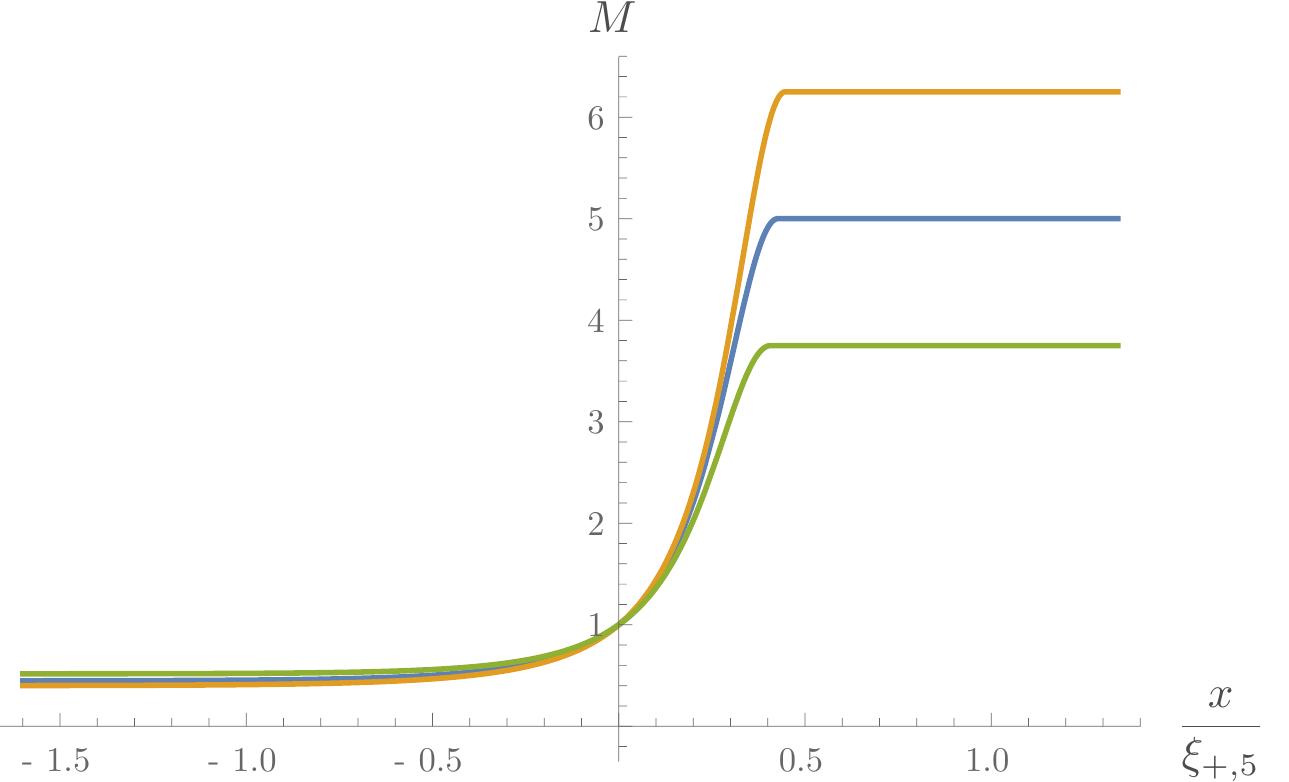} 
\includegraphics[width=0.49 \linewidth]{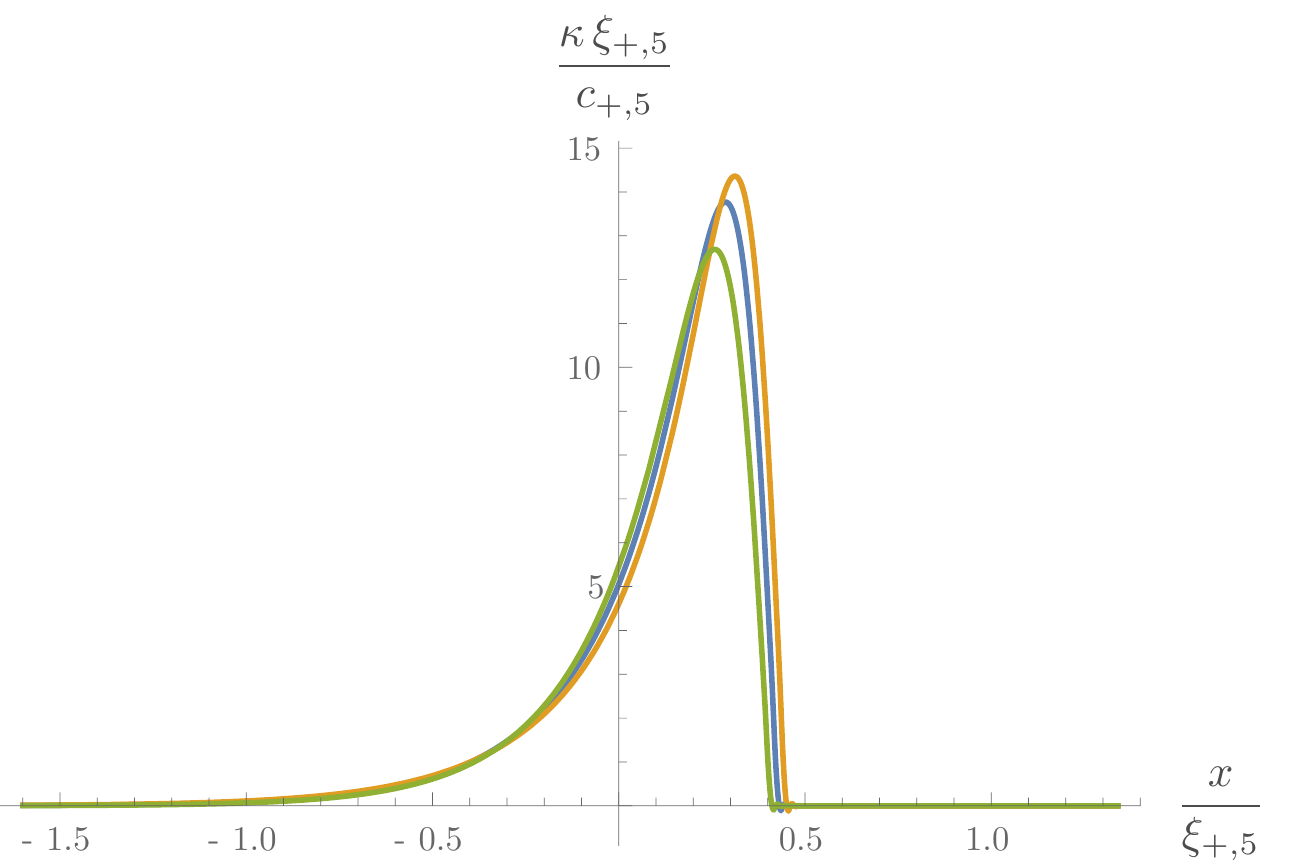} 
\includegraphics[width=0.49 \linewidth]{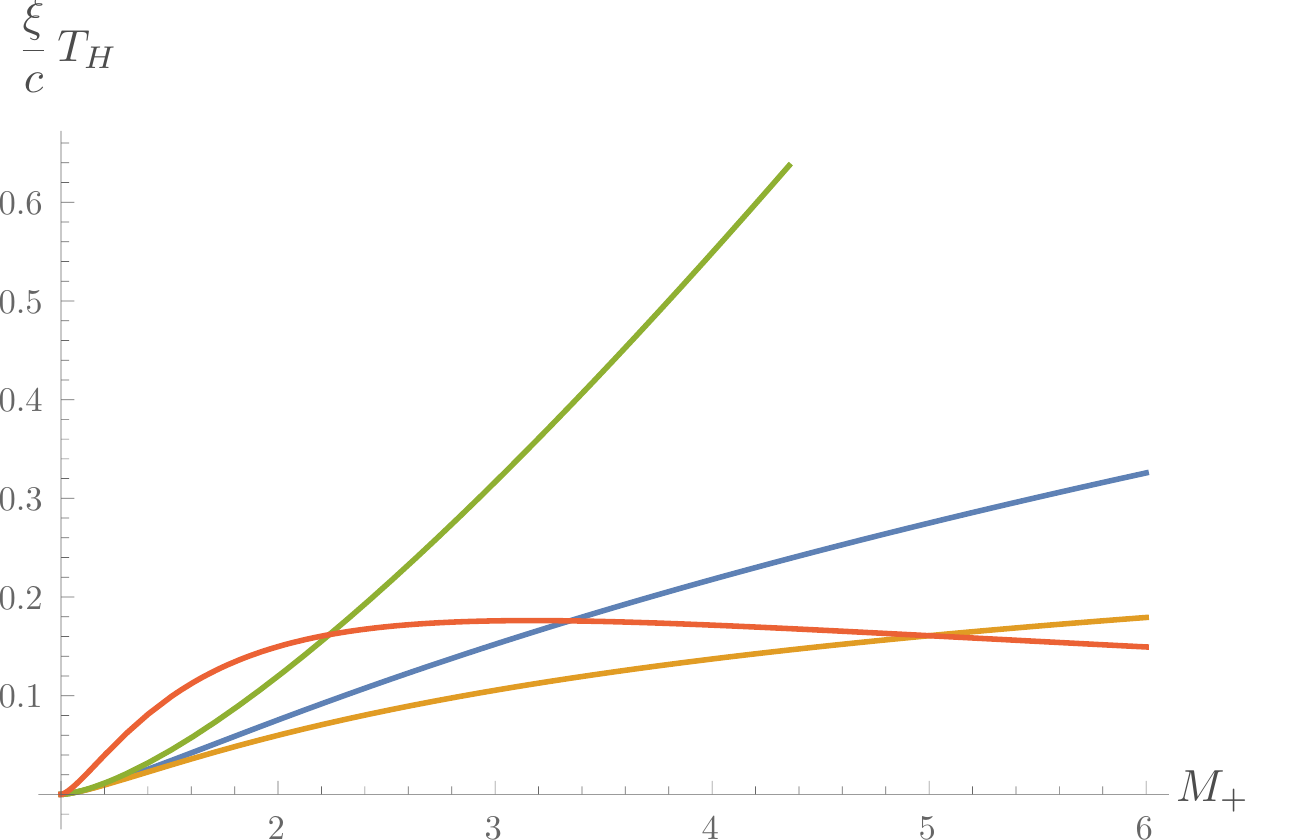} 
\caption{Plots of the rescaled atomic density $\rho/\rho_+$ (top, left) and the Mach number $M$ (top, right) for three waterfall solutions similar to the flow realized in~\cite{Steinhauer:2015saa}. 
The condensate has a positive velocity, is subsonic for $x < 0$, and supersonic for $x > 0$.  
The asymptotic values of the Mach numbers for the green, blue, and orange curves are, respectively, $M_- \equiv M(x \to -\infty) \approx 0.52, 0.45, 0.4$ and $M_+ \equiv M(x \to + \infty) = 3.75, 5, 6.25$. 
The coordinate $x$ is expressed in units of $\xi_{+,5}$, the healing length in the supersonic region for the flow with $M_+ = 5$ (blue curves). Its origin is chosen at the sonic horizon, where $M=1$. 
The bottom left plot shows the adimensionalized gradient $\kappa(x) \xi_{+,5} / c_{+,5}$, where $c_{+,5}$ is the asymptotic downstream sound velocity for the flow with $M_+ = 5$. 
By construction, $\kappa(x)$ identically vanishes on the right of the potential barrier, located near $x/\xi_{+,5} = 0.4$. 
The bottom right plot shows the Hawking temperature $T_H = \kappa(0)/2\pi$ adimensionalized by $\xi(0)/c(0)$ (blue), $\xi_+/c_+$ (green), $\xi_-/c_-$ (orange), and $\xi_{-,5}/c_{-,5}$ (red), for $\rho_+ = 1$, as a function of $M_+$.
Notice that the last adimensionalization gives a nonmonotonic dependence of $T_H$ in $M_+$.} 
\label{fig:bckgd}
\end{figure}

Motivated by the experiment of Jeff Steinhauer~\cite{Steinhauer:2015saa} where the observation of the quantum Hawking radiation in a nearly stationary Bose-Einstein condensate was reported, we studied in~\cite{Michel:2016tog} the spectrum and correlations of phonons at small temperatures in a stationary, one-dimensional condensate.~\footnote{Although there is no true condensation in one dimension~\cite{pitaevskii2003bose}, the behaviour of perturbations in a quasi-condensate are close to those obtained in a true condensate provided the healing length and Compton wavelength  are much larger than the inverse linear number density of atoms~\cite{PhysRevA.67.053615}. From the numbers given in~\cite{Steinhauer:2015saa}, these two conditions seem to be satisfied in the experiment.} 
Assuming the condensate is well described by the Gross-Pitaevskii equation with a uniform, positive two-body coupling and a step-like potential (modeling the sharp potential change realized in~\cite{Steinhauer:2015saa}), all transonic, stationary solutions with asymptotically homogeneous density are of the “waterfall” type~\cite{PhysRevA.85.013621}. 
Up to a few rescalings, they are described by one single parameter, for instance the Mach number $M_+$ in the supersonic region. 
If the flow is oriented to the right, the density has the form 
\begin{align*}
\frac{\rho(x)}{\rho_{1,+}} = \left\lbrace
\begin{array}{cc}
M_+ + \lp 1 - M_+  \rp \lp \cosh \lp \sigma \, \lp x - x_s \rp \rp \rp^{-2} & x \leq x_s \\
1 & x \geq x_s
\end{array}
\right.,
\end{align*}
where $\sigma \equiv \sqrt{M_+ - 1}/ \xi_+$, $\rho_{1,+}$ is the asymptotic density at $x \to + \infty$, $\xi_+$ and $M_+$ are the healing length and Mach number in this region, and $x_s$ is the position of the potential step. 
Three of these solutions are represented in \fig{fig:bckgd}, along with the corresponding Mach number $M \equiv v / c$ and the gradient $\kappa \equiv \pd_x \lp v - c \rp$, where $v$ is the local velocity of the condensate and $c$ is the sound velocity. 
The origin of $x$ is chosen at the analogue horizon. 
The bottom, right panel shows the Hawking temperature $T_H = \kappa(0) / (2 \, \pi)$ as a function of $M_+$. 
Notice that the profile of $\kappa$ becomes very asymmetrical for large values of $M_+$: as can be seen on the bottom left panel, $\kappa$ becomes significantly larger on the right of the horizon than on its left. 

To determine the stability of these solutions, we first solved the time-dependent Gross-Pitaevskii equation with initial data corresponding to perturbed waterfall solutions and found that the initial perturbations are expelled at infinity during the evolution. 
This indicates that these solution seem to be (nonlinearly) stable, acting as local attractor in the sense of Chapter~\ref{ch:nohair}, and thus supports their relevance for describing experiments. 

We then solved the Bogoliubov-de Gennes equation over the solutions represented in \fig{fig:bckgd} to obtain the phonon spectrum. 
More precisely, we computed the scattering coefficients, from which the occupation number of each phonon mode can be extracted once the quantum state is known. 
Our main result is that, if the temperature of the condensate is much smaller than $T_H$, then the spectrum is closely Planckian in the relevant frequency domain, i.e., where most phonons are emitted. 
Strong deviations are only seen at large frequencies, where the mean number of phonons is exponentially small. 
Moreover, the effective temperature is close to the Hawking one, with differences of the order of $10 \%$. 
These solutions are thus suitable for a precise experimental verification of the thermal properties of the analogue Hawking radiation. 

\begin{figure} 
\includegraphics[width=0.49 \linewidth]{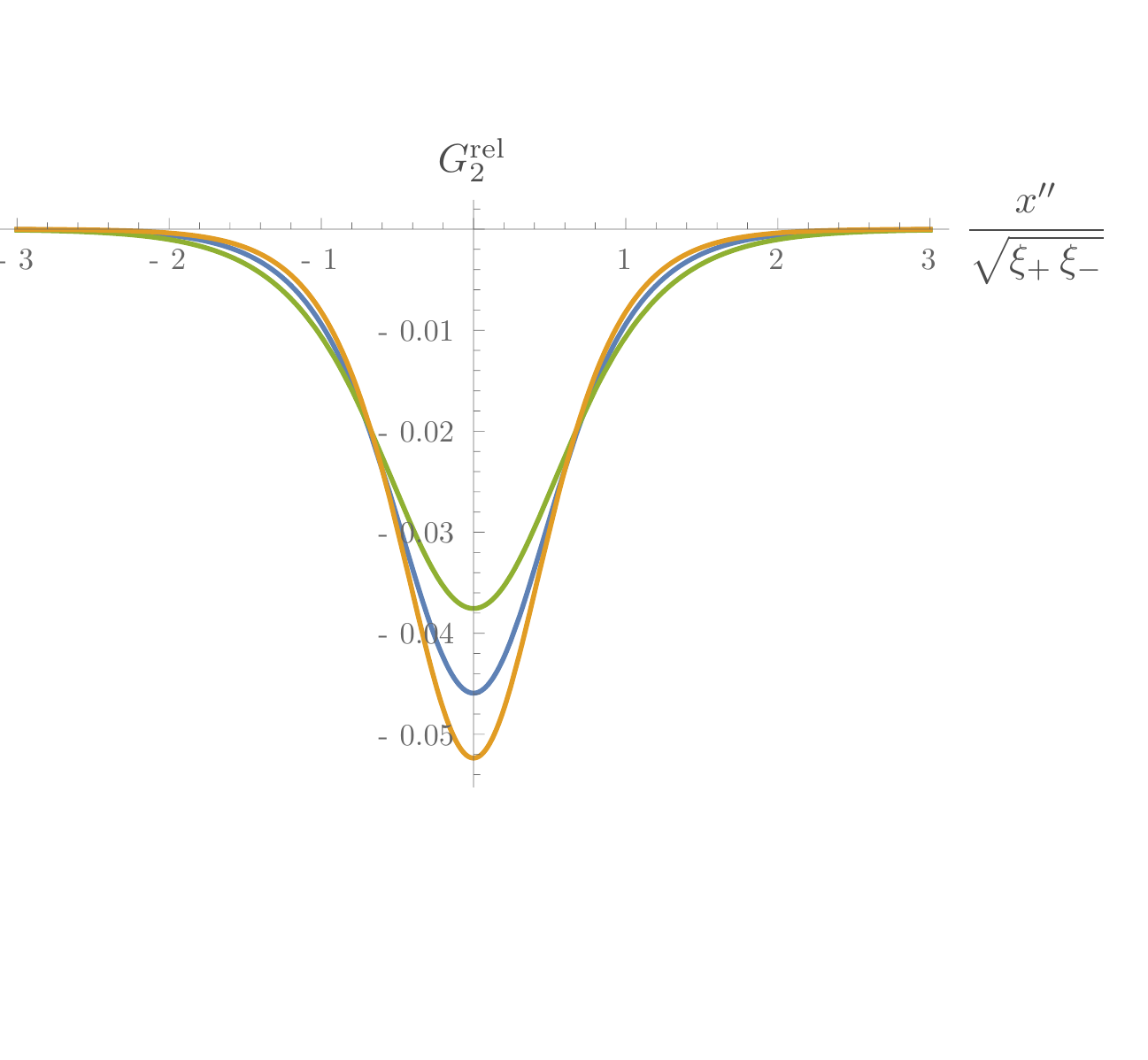}
\includegraphics[width=0.49 \linewidth]{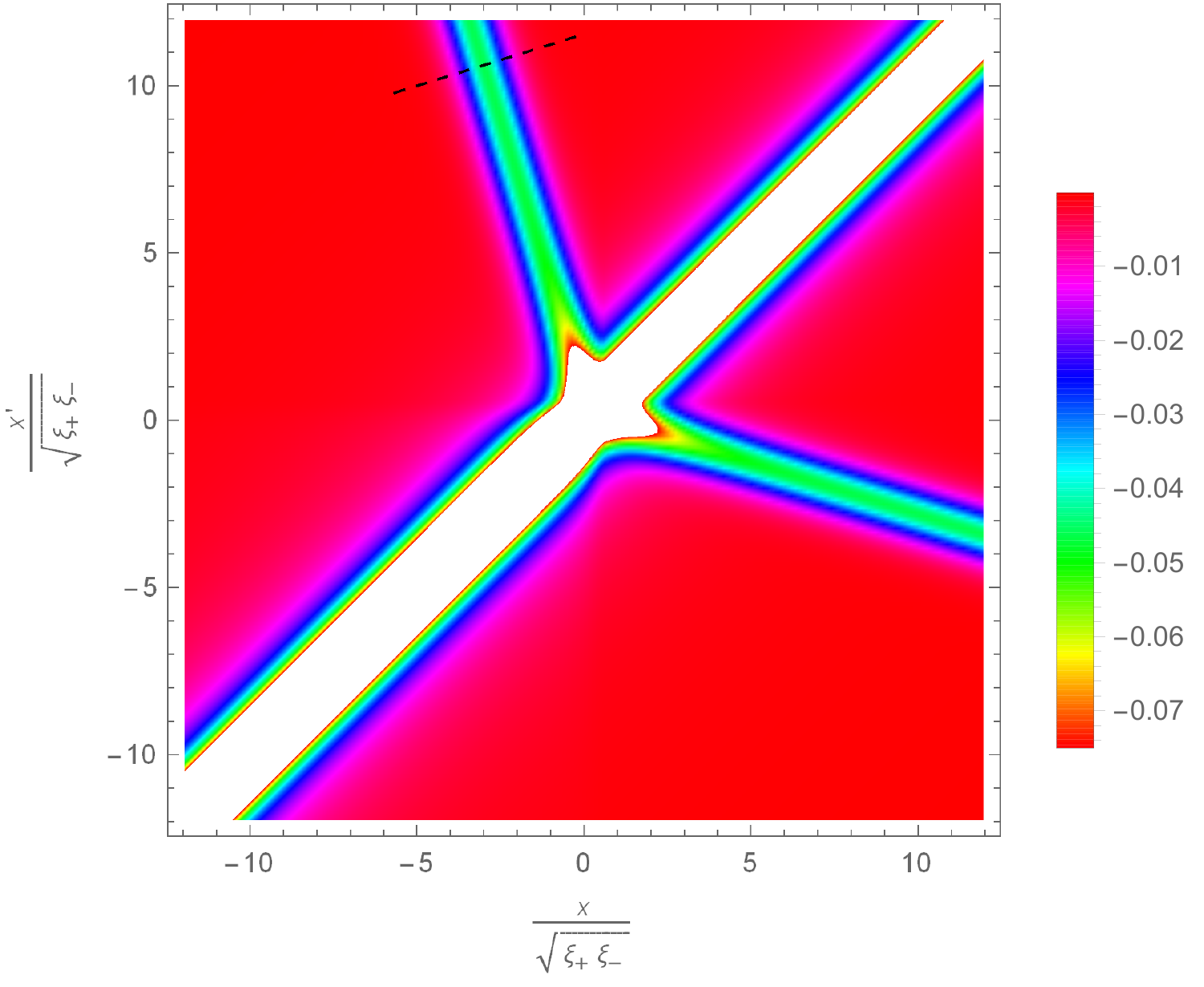}
\caption{Left: We show the profile of the two-point correlation function $G_2^{\rm rel}$, evaluated along a line orthogonal to the locus of its minima and located far from the horizon. 
$x''$ is a coordinate along this line, defined by $x'' = 0$ when $G_2$ reaches its minimum and $\abs{\dd x''} = \sqrt{\dd x^2 + \dd x'^2}$. 
The three curves show the correlation profile for the waterfall solutions with $M_+ = 3.75$ (green), $5$ (blue), and $6.25$ (orange), shown in \fig{fig:bckgd}. 
Right: We show $G_2^{\rm rel}(x,x')$ as a function of $x$ and $x'$ for the waterfall solution with $M_+ = 5$. 
The dashed segment indicates the domain used to represent the correlations on the left panel. The white domain corresponds to values of $G_2^{\rm rel}$ outside the range represented in colors.}
\label{fig:G2rel}
\end{figure} 

Finally, we studied the two-point correlation function $G_2$.  
We first motivated that it accurately follows the relativistic prediction, represented in \fig{fig:G2rel}, obtained by neglecting dispersion. 
Although the qualitative shape agrees with the findings of~\cite{Steinhauer:2015saa}, the width of the correlation band between phonons inside and outside the horizon (see the left panel) is about twice larger than the observed value, while its depth is about twice smaller. 
For the moment we have no clear explanation for these differences. 
It is possible that the waterfall profile we used in this work, while providing a first approximation of the configuration realized in~\cite{Steinhauer:2015saa}, does not match it precisely enough to obtain a quantitative agreement, or misses some physical effects.
One can also notice that the asymptotes of the lines of maxima of $G_2$ do not cross exactly at the horizon, even in the relativistic case. 
This shift, which can be related to the phase of the scattering coefficients, could be observable in future experiments. 
Interestingly, we find that, neglecting dispersion this shift is absent if $c - v$ has the form
\begin{align*}
c(x) - v(x) = a - b \, \tanh(\sigma \, x)
\end{align*}
with $(a,b) \in \mathbb{R}^2$, even in the asymmetric case $a \neq 0$. 

\clearpage

\section[Radial excitations of superconducting strings]{Radial excitations of superconducting strings \cite{Hartmann:2016axn}}

Cosmic strings are one-dimensional topological defects arising in relativistic fields theories with a spontaneously broken $\textrm{U}(1)$ symmetry~\cite{Hindmarsh:1994re,Vilenkin:2000jqa,Vachaspati:2015cma}.~\footnote{More generally, if $G_0$ is the group of symmetries of the theory (endowed with some topology) and $G_1$ its residual, unbroken subgroup, cosmic strings can appear if the first homotopy group of $G_0 / G_1$ is nontrivial, i.e., if $G_0 / G_1$ contains loops which can not be continuously deformed into a single point.} 
As such, they bear many similarities with vortex lines in cold atoms. 
(They are actually vortex lines of a relativistic Bose-Einstein condensate.) 
The qualitative idea is the following: Consider a scalar field $\phi$ with a potential $V \lp \abs{\phi} \rp$ having a global minimum for some strictly positive value $f_0$ of $\abs{\phi}$, and let it evolve while decreasing the temperature of the environment from an initially very high value (such as occured in the early universe). 
For temperatures much larger than the depth of the minimum of $V$, thermal fluctuations are larger than $f_0$ and the drop of the potential between $\abs{\phi} = 0$ and $\abs{\phi} = f_0$ plays no significant role. 
When decreasing the temperature below the depth of the potential, however, $\phi(t,\vec{x})$ will tend to settle down to a configuration of the form $f_0 \, \e^{\ii \varphi(t, \vec{x})}$, for some real-valued function $\varphi$. 
Since this process is local, $\varphi$ initially takes independent values between far-away points. 
Minimization of the kinetic energy will then tend to suppress the spatial variations of $\varphi$. 
However, they can not be completely erased while maintaining continuity of $\phi$ if $\varphi$ varies by a nonvanishing integer multiple of $2 \pi$ along a closed loop. 
Moreover, in that case the phase of $\phi$ can not be defined everywhere in a surface admitting this loop as boundary.~\footnote{One way to see this is to continuously deform the loop to a single point along the surface. The circulation of $\textrm{arg}(\phi)$, which is nonvanishing for the original loop and can change only through discrete steps of $2 \pi$, would be conserved during the continuous deformation if $\textrm{arg}(\phi)$ was defined everywhere, leading to the absurd result that the circulation along a single point is nonvanishing.} 
The field $\phi$ must thus vanish at (at least) one point on the surface.
Applying this argument to a one-parameter family of such surfaces, one finds that the locus (at fixed $t$) where $\phi$ vanishes is of dimension 1. 
One thus obtains a one-dimensional topological defect: a cosmic string.~\footnote{More generally, in $d$ space dimensions, and if the $n^\textrm{th}$ homotopy group of $G_0 / G_1$ is nontrivial, one expects to find topological defects of dimension $d - n - 1$. If $G_0 / G_1$ is not simply connected (corresponding to $n = 0$), one expects topological defects of dimension $d - 1$.}
They generally arise in grand-unified models during one or several phases of the breaking of the grand-unified group to the standard model's $\textrm{SU}(3) \times \textrm{SU}(2) \times \textrm{U}(1)$~\cite{Sakellariadou:2009ev,Allys:2015kge}. 
When including additional fields, a condensate may form in the core of the string, leading to persistent currents~\cite{Witten:1984eb,Davis:1996xs,Peter:1993ww,Peter:1993tm,Davis:1995kk,Hartmann:2008yr,Allys:2015yda}. 

In a previous work~\cite{Hartmann:2012mh}, we studied numerically cosmic-string solutions in a $\textrm{U}(1)_\textrm{local} \times \textrm{U}(1)_\textrm{global}$ model involving two complex scalar fields minimally coupled to gravity. 
The first one, $\phi$, has a gauged $\textrm{U}(1)$ symmetry and a nonvanishing vacuum expectation value in the “true vacuum” minimizing the energy. 
We considered cylindrically-symmetric cosmic string configurations where $\phi$ vanishes at $r = 0$.
The second field, $\sigma$, has a global $\textrm{U}(1)$ symmetry. 
Its vacuum expectation value vanishes in the true vacuum, but not in the region $r \approx 0$ where it condenses and forms a supercurrent. 
We focused on the “fundamental” solutions where the spatial profile of $\sigma$ along $r$ is monotonic, extending the results of~\cite{Peter:1992dw} by taking the coupling to gravity into account. 

In~\cite{Hartmann:2016axn}, we found new solutions of the same model where $\sigma$ has one or several nodes along $r$. 
We first obtained an analytical description of these “excited” solutions in the limit where $\sigma$ is small enough to neglect its self-interactions and back-reaction on $\phi$, approximating $\abs{\phi}$ by a piecewise-linear function, linear for $r$ smaller than a critical value $r_c$ and uniform for $r > r_c$.  
Interestingly, in the limit of large $r_c$ one recovers the solutions of the nonrelativistic Gross-Piaevskii equation in a cylindrically-symmetric harmonic trap. 
We then solved the full set of field equations numerically in Minkowski space to determine how the “excited” condensate affects the properties of the string. 

\begin{figure}
\centering
\includegraphics[width=0.49 \linewidth]{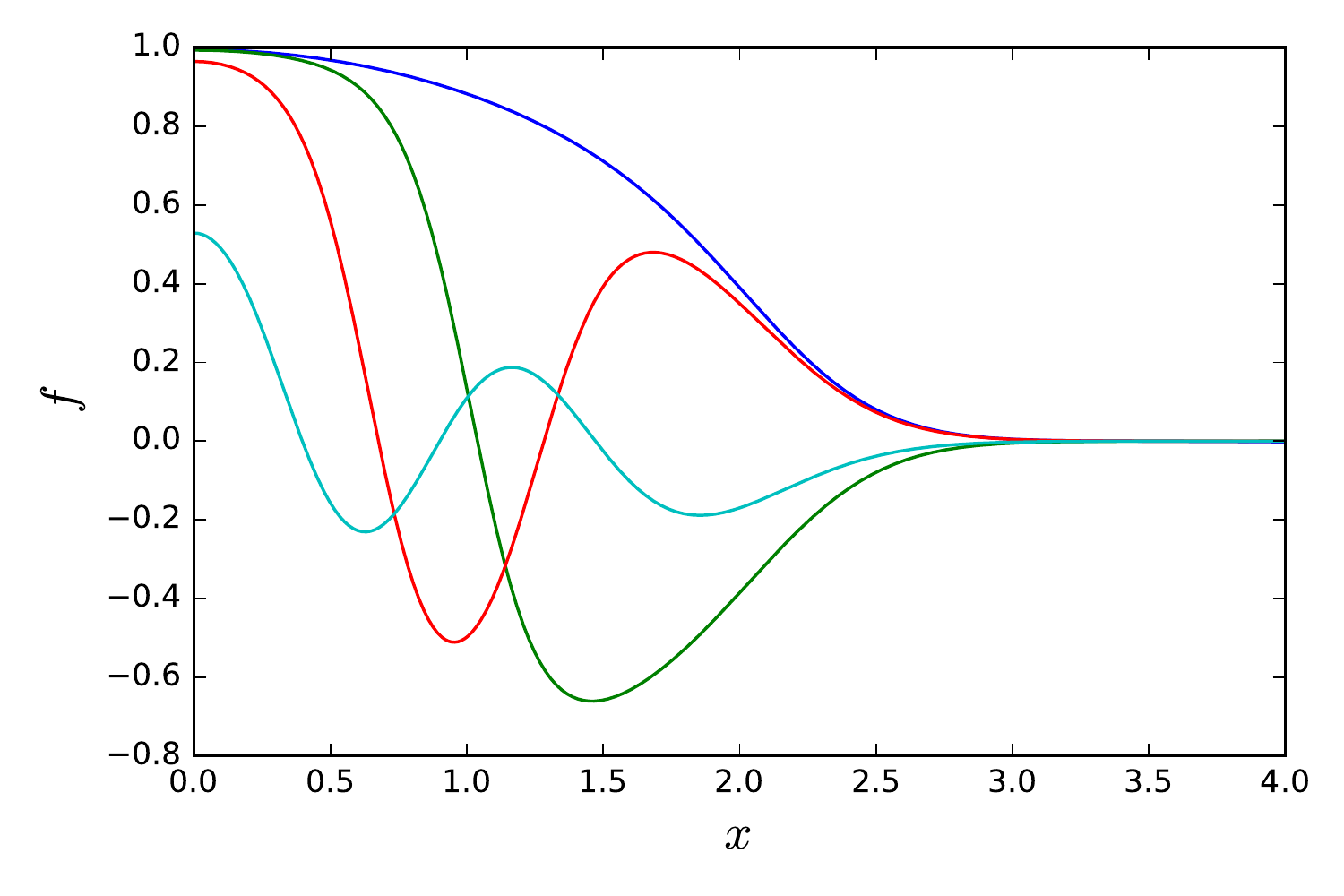}
\includegraphics[width=0.49 \linewidth]{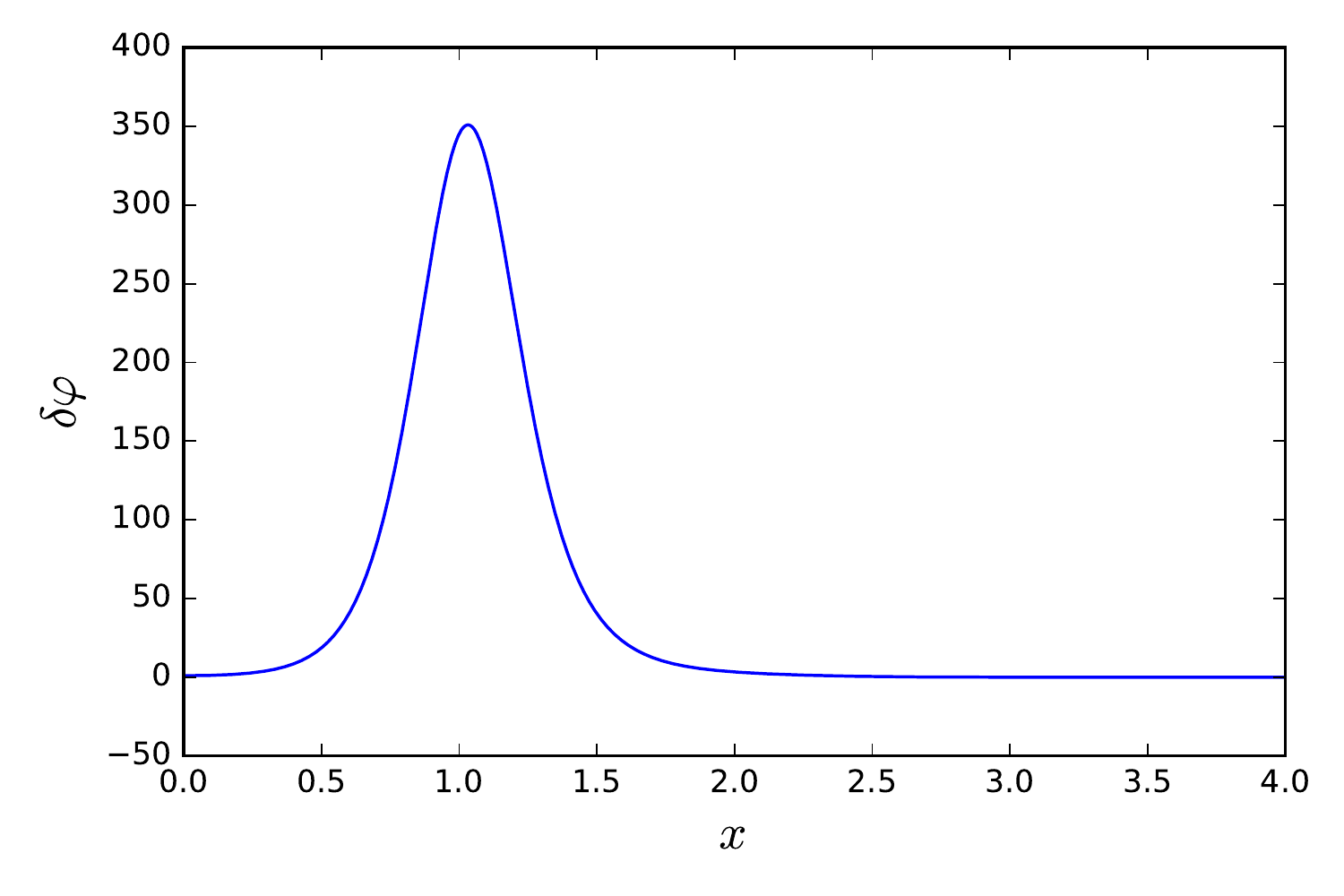}
\includegraphics[width=0.49 \linewidth]{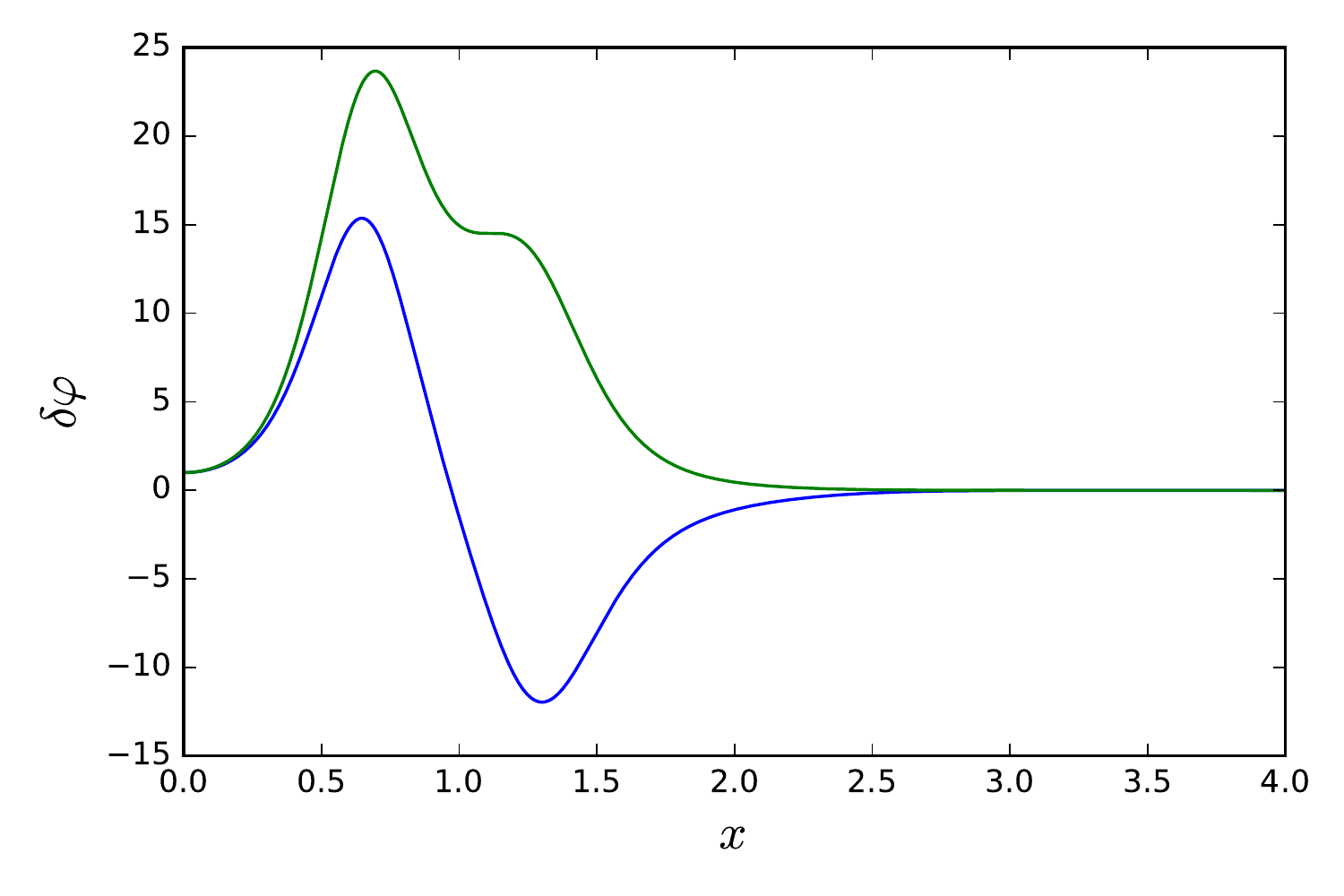}
\includegraphics[width=0.49 \linewidth]{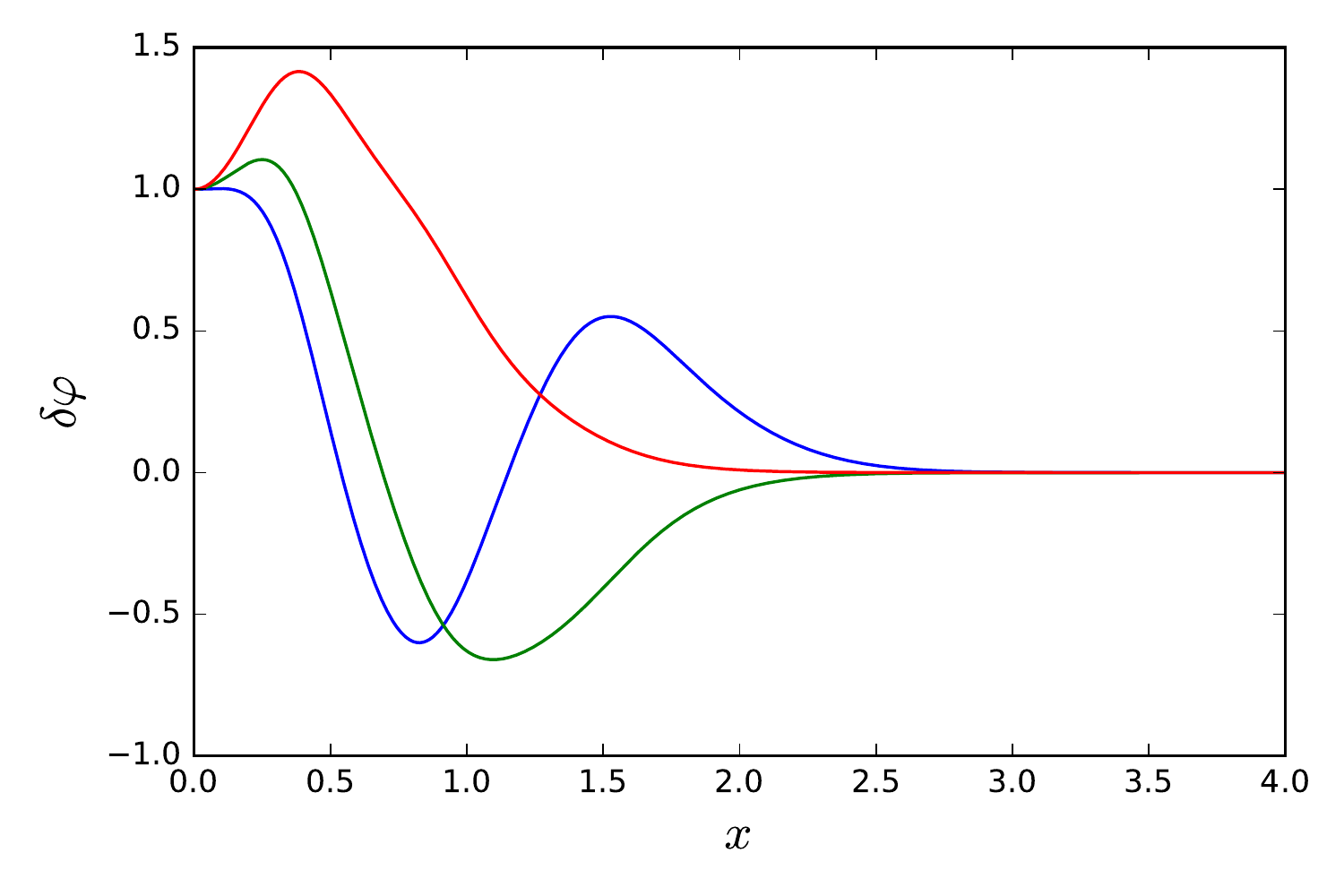}
\caption{The top, left panel shows the fundamental (blue line) and first three excited solutions for the function $f = \abs{\sigma}$ in a harmonic effective potential given by $\abs{\phi(x)} = \sigma_h \, x$. 
We work with $\om^2 - k^2 = 0$, where $\om$ is the angular frequency of the condensate of $\sigma$ and $k$ its wave vector in the direction of the string axis. 
The coupling constants are (see~\cite{Hartmann:2016axn} for their precise definition): $q = 1$, $\gamma_2 = 44$, and $\gamma_3 \, \sigma_h^2 = 9$. 
The three other panels show the spatial profiles of the unstable modes found with $\omega = k = 0$, for the solutions with one (top, right), two (bottom, left), and three (bottom, right) nodes. 
(The normalization is arbitrary.) 
The fundamental solution being dynamically stable, it supports no unstable mode. 
} \label{fig:insta_w0}
\end{figure}

To determine the possible implications of these new solutions for cosmology or their observational signatures, it is crucial to know whether they are stable -- and, if not, what are the growth rates and spatial profiles of the unstable modes. 
As a first step in this direction, we solved the linearized equation on $\sigma$ (the equivalent of the Bogoliubov-de Gennes equation for a nonrelativistic condensate) in a fixed cosmic string background, focusing on modes with complex frequencies. 
Our main results are illustrated in \fig{fig:insta_w0}, where we show 4 solutions for $f \equiv \Re \, \sigma$ (top, left panel) and the unstable modes found over the three ``excited'' solutions. 
We found no complex-frequency mode over the fundamental solution (blue curve in the top, left panel), which indicates that this field configuration is stable -- as expected as it minimizes the energy. 
Turning to the excited solutions, we found that they are all dynamically unstable, the solution with $n$ nodes having $n$ degenerate dynamical instability modes. 
This is actually not unexpected: close to a node, the local value of the condensate density $\abs{\sigma}^2$ is smaller than the one required to minimize the energy. 
An instability can thus arise through condensation of particles in this region. 
The reason why this leads to a dynamical instability, instead of an energetical instability usually found in nonrelativistic Bose-Einstein condensates, is the possibility to fill the hollows in the local density $\abs{\sigma}^2$ by producing pairs of particles and antiparticles.~\footnote{From a more mathematical point of view, the reason is that the conserved quantity associated with the $\textrm{U}(1)$ invariance is the positive number of atoms in the nonrelativistic case but the nonpositive Klein-Gordon inner product in relativistic settings.} 

Finally, we numerically solved the field equations including successively two additional effects. 
First, we coupled the fields minimally to gravity, assuming cylindrical symmetry is preserved. 
As already noticed in~\cite{Hartmann:2012mh} for the fundamental solutions, we found that gravitational effects only mildly affect the structure of the solutions. 
Then we turned the second $\textrm{U}(1)$ symmetry to a local one by adding a new gauge field. 
We found that, for realistic values of the coupling parameters, the structure of the string and condensate in its core are hardly affected, as expected from the results of~\cite{Peter:1992ta}.  

\clearpage

\section[Controlling and observing nonseparability of phonons created in time-dependent 1D atomic Bose condensates]{Controlling and observing nonseparability of phonons created in time-dependent 1D atomic Bose condensates \cite{Robertson:2016evj}}

In~\cite{Robertson:2016evj}, we studied the dynamical Casimir effect (pair production of (quasi)particles due to variations in time of a background field) in cylindrically-symmetric Bose-Einstein condensates in harmonic traps. 
We focused on the entanglement properties of the produced phonons, how to measure them, and in which conditions their quantum state is nonseparable.~\footnote{We remind the reader that a bipartite system is called separable if its density matrix can be written as the sum of tensor products of density matrices of its two constituents times positive prefactors, and nonseparable otherwise. In the present case, the system under consideration is bipartite at fixed wave vector.}  

We first considered the response of the condensed atoms to a time-dependent change in the potential strength, in two regimes: one where this strength is monotonic in time, and one where it is periodically modulated. 
This is a necessary preliminary step because the background field experienced by phonons, i.e, the condensate itself, is not directly accessible in experiments: one generally has control only over the external potential and temperature (and possibly the two-body coupling between the atoms), and not directly to the local density of the condensate. 
To study the phonon pair production in experimental setups, we must thus first know the behavior of the condensate under a change of the potential. 
This is made easier by a scale invariance of the Gross-Pitaevskii equation in $2$ space dimensions for harmonic potentials: assuming the solution is stationary for $t < 0$ where the potential does not change, the solution for $t > 0$ is fully determined by a function of time only, for instance the density at the origin, which obeys a second-order ordinary differential equation. 
To get explicit results, we used a Gaussian approximation for the condensate density at $t < 0$. 
The results thus obtained were then compared with numerical solutions of the Gross-Pitaevskii equation and found to be accurate for experimentally relevent parameters for which the condensate is sufficiently thin. 

We then determined the quantum state of the phonons after changing the potential strength, assuming a thermal state at $t = 0$. 
There are two channels producing correlated pairs: the (purely quantum) scattering of vacuum fluctuations and that of perturbations initially present in the system, which can be described classically. 
To estimate the ``quantumness'' of the state, one thus needs observables able to disentangle these effects. 
In this work we used two of them: the two-point density correlation function and the two-body distribution of atom momenta after releasing the trap, both giving a criterion for nonseparability. 
An important point to bear in mind is that computing each of these quantities requires only the simultaneous measurement of mutually commuting observables, in spite of the fact that their outcome, in the case the state is nonseparable, can not be accounted for by the classical description of the condensate and phonons. 
Indeed, one can write down an inequality, called ``Cauchy-Schwarz'', which is always satisfied in the classical field theory but can break down at the quantum level if the state is nonseparable.~\footnote{More precisely, this is a Cauchy-Schwarz inequality in the usual sense, and thus always satisfied by construction, if assuming all quantities entering it are complex numbers. It can however break down in a quantum gaz where they are noncommuting operators. For the inequality we used, one can show that it is always satisfied if the state is separable. Its breakdown is thus a sufficient condition for the state to be nonseparable.}

We showed that the two-point density correlation function can be used as a probe of the nonseparability of the phonon state. 
Considering first a monotonic variation of the low-frequency phonon speed $c$, we found the state is indeed nonseparable provided the initial temperature is not too large and the variation of $c$ is fast enough. 
When considering a periodic modulation instead, we found a resonance leading to an exponential growth of the number of produced pairs around a specific wave vector, making the state nonseparable. 
This exponential growth should have an important back-reaction on the condensate when the number of phonons becomes large, which seems to explain the results of the experiment reported in~\cite{Jaskula-et-al}. 
Finally, we studied the effects of the release of the trap. 
We found that although the expansion of the gas produced an additional dynamical Casimir effect, the later is small at high wave vectors. 
Moreover, it can be reduced by slowing the decrease of the potential strength. 
Assuming it is sufficiently slow, the correlations between the momenta of the atoms after the release can thus also be used as a reliable probe of the nonseparability of the state before it took place.

\clearpage

\section{Concluding remarks}

This thesis was devoted to a few aspects of classical and quantum field theories in nonuniform and/or time-dependent backgrounds, which are ubiquitous in condensed matter, fluid dynamics, and curved space-times. 
Particular emphasis was put on models with an effective, low-energy description as a relativistic massless scalar field in a metric with an event horizon and the resulting (analogue) Hawking effect. 
Three aspects of these models were studied: their nonlinear classical behavior, linear (classical or quantum) perturbations, and links with experimental realizations. 
I now briefly conclude on each of them by summarizing my collaborators and I's main contributions, before pointing out the main open questions they raise. \\

Since they are, by essence, beyond the mathematical correspondence between ``analogue'' condensed-matter models and fields around black holes, nonlinear effects in analogue gravity had not been extensively studied in spite of their importance for the dynamics and for experimental realizations. 
We partially bridged this gap by considering three problems where nonlinear terms and the analogue Hawking effect both play a prominent role. 

Focusing on configurations with only one horizon, we motivated that black hole flows are stable under both linear and nonlinear perturbations, so that they can in principle be realized experimentally without fine-tuning the external potential nor the flow parameters. 
Moreover, the corresponding solutions of the field equations are uniquely determined by a few macroscopic conserved quantities, and the emission of nonlinear waves during the transition from an initially perturbed configuration to a late-time stationary one follows general rules which can be obtained analytically. 
This is very reminiscent of the ``uniqueness'' and ``no-hair'' results in general relativity plus Maxwell theory. 

White hole flows, on the other hand, are generically unstable to long-wavelength perturbations. 
Numerical simulations indicate that this instability generates a periodic modulation with a growing amplitude, as predicted by the linear theory, as well as periodic soliton trains. 
The presence of solitons, whose properties seem to strongly depend on the flow and initial perturbation, as well as the evolution of the amplitude of the modulation, show that nonlinear terms play a crucial role in the dynamics. 
To understand the effects of this instability on Hawking radiation, we solved numerically the linearized field equations in a periodically-modulated background solution of the stationary nonlinear equation. 
We found that the latter can dramatically modify the low-frequency behavior, and even suppress the distinctive divergence found for thermal spectra. 
This may be seen as a saturation mechanism: since the infra-red divergence of the scattering coefficients relating positive-energy waves to negative-energy ones is at the source of the instability, its regularization by nonlinear effects should stabilize the system. 

We also studied configurations with two horizons, extending to nonlinear models the interplay between the behaviors of black and white hole flows which had previously been studied only to linear order. 
Considering ``black hole laser'' configurations where the two horizons form a resonant cavity in which Hawking radiation is self-amplified, we explained how nonlinearities can make the latter observable in spite of its vanishing ensemble average at linear order. 
We also discussed the late-time behavior of the system, showing that it can either saturate on a stable stationary solution, through a process akin to black hole evaporation, or enter a regime of periodic emission of superposed soliton trains. 
This surprizing result indicates that the time-evolution of Hawking radiation is -- at least in this context -- more intricate than could have been expected, a small change in the initial conditions leading to qualitative differences at late times in a seemingly chaotic way. 
This seems to be a manifestation of the instability of white hole flows, regularized by the black hole horizon when the distance between them is small enough.

Related ideas were used in effective $(2+1)$-dimensional problems with cylindrical symmetry in cold atoms and in a relativistic field theory. 
In the first case, the nonlinear field equation determines the evolution of the background condensate under a time-dependent variation of the external potential. 
This allowed us to follow the evolutions of phonons and the quantum entanglement of the final state in configurations analogous to the reheating phase of the early universe. 
In the second case, these techniques shed light on the behavior of cosmic strings with an ``excited'' condensate in their core. 
We used them to find new nonlinear solutions, describe their structure and main properties, and discuss their relevance for future observations. 

In all the systems we considered, we have thus shown that nonlinear effects can strongly affect the stationary and dynamical properties of analogue models. 
Including them is therefore required to obtain a precise description, both at the conceptual level -- in relation with extensions of the gravitational analogy or back-reaction effects -- and for experiments as they modify the behavior of observables. 
We have studied them in several models with the aim to exhibit their main general features, which can be used to understand qualitatively the behavior of analogue models and to refine the design of experiments. \\

At the level of the linear theory, one important question concerns the structure of the instabilities which can arise. 
Indeed, from an experimental point of view, the presence or absence of unstable modes and, in the former case, their growth rates and spatial profiles, are crucial elements to determine if their realization can be realistically envisaged. 
But instabilities are also interesting from a purely theoretical perspective: being a common denominator to widely different analogue systems and strongly related to superradiance, Hawking radiation, and quasi-normal modes, they offer an angle complementary to the study of nonlinear solutions, showing more clearly the necessary and sufficient conditions to obtain a given behavior. 
We first studied dynamical instabilities (DI) in a black hole laser and found they can be of two different types. 
Besides the already-known modes with complex frequencies, which have a clear interpretation as self-amplified Hawking radiation, we found modes with a purely imaginary frequency not directly associated with any radiation. 
These modes are in a sense intermediate between quasinormal modes (QNM) and the standard unstable modes, as revealed by the transitions between them occurring when varying the parameters of the model. 
Generalizing these results led us to a classification of DI in (1+1)-dimensional systems and their relations with superradiance. 
We showed, in particular, that DI and QNM frequencies are poles of the retarded Green function, and that a QNM can continuously turn to a DI (or conversely) when varying external parameters. 
This offers a general framework allowing for a precise understanding of the black hole laser as well as other systems which can be described effectively by a (1+1)-dimensional field equation, such as the Klein paradox, Gregory-Laflamme instability, and the aforementioned cosmic string solutions. 

Returning to stable flows of ideal fluids, we aimed at determining which properties of the spectrum depend or not on the presence of a horizon. 
To this end, we studied the scattering of water waves on flows over a localized obstacle with a view to characterizing the residual Hawking-like effect when there is no analogue event horizon. 
Indeed, while the scattering in the presence of a horizon was already well understood, the horizonless case had attracted much less attention -- although experiments done with water waves seem to operate in this regime. 
As expected, we found that the scattering coefficients involved in the Hawking effect do not independently follow a thermal spectrum, even at low frequencies (although their ratio still does). 
Moreover, their behavior strongly depends on the details of the flow, in contrast to the case with horizon whose surface gravity is the only relevant parameter at low frequency. 
To be complete and describe systems operating in an intermediate regime, we also studied the transition occurring when lowering the maximum speed of the flow which goes smoothly from a transcritical one (with horizon) to a subcritical one (without horizon). 
The thermal nature of the spectrum is then progressively suppressed and restricted to lower frequencies, until it completely disappears when the flow becomes subcritical.

On a different note, we considered a collapsing mass shell forming a black hole in gravity theories with high-energy dispersion. 
In our model, the additional ``universal'' horizon does not radiate at late times, so that the radiation spectrum is still given by the surface gravity of the null Killing horizon -- as is the case in general relativity (without dispersion) and in analogue models (with dispersion). 
The fact that the khronon field we used is discontinuous across the shell has since then led to questioning the validity of this result, as in more realistic models it is supposed to be regular. 
I think that this discontinuity of the khronon field should not be a problem. 
Indeed, the theories we worked with are reparametrization invariant, meaning that the khronon $\tau$ can be replaced by any function of itself without affecting the physical results. 
Moreover, since we did not use $\tau$ for the mode matching, our calculation is unchanged when performing two different reparametrizations inside and outside the shell, which can be used to make the khronon field locally continuous across it. 
I thus expect that its discontinuity plays no role. 
However, the \ae ther field also has a discontinuity, which can not be removed by reparametrization and thus may affect our results. 
To the best of my knowledge, whether it is the case or not remains an open question. \\

Another side of our work was to make link between the theoretical description of analogue models and their practical realizations. 
In particular, we used our analytical and numerical results to make proposals aiming at improving existing experimental setups and obtaining a closer link with the Hawking effect. 
We also discussed how they may affect the interpretation of previous observations. 
We worked with general, simplified models which aim at capturing the essential ingredients of the physics at play, and where the relations between observables and theoretical concepts can be drawn unequivocally.  
This is complementary to approaches, followed by other groups, relying on a precise description of specific systems, including additional properties of the setup but lacking a clear distinction between generic and specific results. 
I believe that both points of view should be used in synergy to obtain a clear interpretation of analytical and experimental results: the later gives a more precise description accounting for effects that general models tend to miss, while the former sheds light on the genericness of the observations and their link with gravity.

Considering water waves experiments, we showed that two different notions of ``thermality'' can be naturally defined, giving similar results in transcritical flows (with an analogue horizon) but differing qualitatively in subcritical ones. 
This difference clarifies in which sense the emission spectrum can be said ``thermal'' when, as was apparently the case in experiments done so far, there is no analogue event horizon. 
It also underlines that the presence of a horizon is necessary if one is to have a clear link with the Hawking effect. 
On a similar note, we proposed a method to reduce the amplitude of the downstream undulation which is generally present in experiments and hampers a precise estimation  of the scattering coefficients.  
It was then used in collaboration with the group of Germain Rousseaux to measure (to our knowledge, for the first time) independently the scattering coefficients of surface waves in a flow resembling a white hole one -- although \textit{stricto sensu} no horizon was present. 
We hope this important step will open the way for observing the thermal behavior of the spectrum in a water analogue of a white hole in the near future. 

We also contributed to the theoretical modeling and analysis of experiments in Bose-Einstein condensates. 
In particular, the transition from linear to nonlinear behaviors in black hole lasers and the behavior of the two-point correlation function in Bose-Einstein condensates mentioned in the previous chapters and sections may help in explaining J.~Steinhauer's recent observations and clarify which features are specifically due to Hawking radiation. 
While we were so far not able to make a precise link, closer analysis of the experimental data could allow for an estimation of the effects we studied.

Finally, in collaboration with Yves Aurégan and Vincent Pagneux, we described a new analogue model using sound waves in a pipe with a compliant wall. 
While closer to the original proposal of W.~Unruh than previous realizations, the reduction of the effective sound velocity could make it easier to realize a transcritical flow in this new setup. \\

These works leave open several questions which will hopefully be addressed in the near future. 
A first one concerns the mechanism generating superposed soliton trains in black hole laser and white hole flows. 
The soliton trains themselves are well known: since they propagate at a constant speed, they simply correspond to stationary, periodic solutions of the Gross-Pitaevskii equation in a Galilean frame, and can be written in terms of Jacobi elliptic functions. 
Their superpositions seem less easy to describe analytically, but this should be possible using inverse scattering methods. 
However, to the best of my knowledge the mechanism producing them remains elusive. 
Similarly, the precise relations between the properties of the initial data and those of the soliton trains are still poorly understood. 
As a first step, a more systematic numerical analysis than those we performed so far would be welcome to try and exhibit general features which could shed light on the path to an analytical description. 
A second open issue regards the effects of an undulation on Hawking radiation. 
We have already performed numerical simulations which seem to indicate an important effect at low frequencies, to the point that the spectrum would lose its thermal character. 
We are currently working on an analytical description of the propagation of modes over the undulation. 
The next step will be to combine this analysis with the scattering on analogue white holes to try and understand the numerical observations -- and determine whether they are generic or model- and/or parameter-dependent. 
While this question seems to be within closer reach than the previous one, obtaining a good description of actual experimental setups would also require taking into account additional effects like dissipation into account, opening a new and challenging path ahead. 
A third problem requiring further investigation concerns, precisely, the effects not included in current descriptions of analogue models. 
One can think for instance of vorticity, surface tension, and viscosity for water waves or dissipation due to interactions between the condensate and noncondensed atoms in cold gases. 
Finally, it would be interesting to extend our analysis of scalar fields in a collapsing-shell geometry producing a black hole with a universal horizon to models where the \ae ther field is regular, as this could settle the question whether its discontinuous character strongly affects our results. 
On a similar note, the study of (quantum) fields in the recently found rotating black hole solutions without universal horizon in Ho\v{r}ava gravity~\cite{Sotiriou:2014gna} could shed light on the causal structure of the theory. \\

From a broader point of view, these works offer a glimpse at the variety of questions and issues raised by quantum fields in inhomogeneous backgrounds or curved spacetimes, of which analogue gravity is a prominent example. 
At the crossroads between general relativity and water waves, nonlinear physics and quantum field theory, cold atoms and black holes, it exhibits the generic features under phenomena which seem at first sight unrelated, allowing one to put them in a consistent framework where the similarities and differences between the various models are clearly seen. 
Doing so, it sheds new light on counter-intuitive aspects like negative-energy waves, their mathematical origin, and their observable consequences. 

Analogue gravity has seen rapid and important developments during the last few years. 
Using cold atoms, J.~Steinhauer~\cite{Steinhauer:2015saa} has reported the first observation of the quantum Hawking emission from an analogue black hole. 
This is the very first direct observation requiring both (analogue) curved space and quantum mechanics to be explained -- which was the original aim of analogue gravity~\cite{Unruh:1980cg}. 
On the hydrodynamic side, the group of S.~Weinfurtner~\cite{Torres:2016iee} recently observed the superradiant emission from a vortex, which could serve as a first analogue of a Kerr-like metric and open up the still uncharted realm of analogue gravity in higher-dimensional systems. 
In parallel, progress has been made on models involving polaritons~\cite{PhysRevLett.114.036402}, sound waves in a gas~\cite{2015ASAJ..138..605A}, and optical fibers~\cite{Roger:2016slp}. 
In each case, a precise study of the interplay between the generic features and specificities of the model yields a deeper understanding both of the physical system \textit{per se} and of the Hawking effect in more general frameworks than was originally envisioned. 

These developments, as well as the works reported here, also bring new questions regarding, for instance, higher-dimensional effects, the role of boundary conditions, and the possibility of a fully quantum and nonlinear treatment. 
These issues, closely (although indirectly) related to those raised by tentative quantum theories of gravity such as the unitarity of black hole evaporation, could serve as a long-term program for analogue gravity, while its initial aspirations, namely the observation of the analogue Hawking radiation and a characterization of the regularizing effects in models with high-energy dispersion, seem to have been attained or to be within reach. 
While solutions to these problems would not directly extend to their counterparts in the gravity realm, which would require mastering the quantum aspects of gravity itself, they can provide both a better understanding of the analogue models and inspiration for devising possible scenarios, which might then be realized in actual quantum gravity models. 
For this reason, I strongly believe that this field of research still has a lot to offer to whomever will take up the challenge, the variety of the involved topics and concepts providing a unique opportunity for progress in our understanding of Nature. 

\printbibliography[heading=bibintoc, env=bibliographyNUM, title=List of publications, keyword=primary]
\printbibliography[heading=bibintoc, title=Bibliography, keyword=secondary]

\cleardoublepage

\appendix
\selectlanguage{french}
\backgroundsetup{
scale=1,
angle=0,
opacity=1,
color=black,
contents={
	\ifodd\value{page}
		\begin{tikzpicture}[remember picture,overlay]
		\fill [blue!50!black] (-10.2,-15) rectangle (-9.95,15); 
		\end{tikzpicture}
	\else
		\begin{tikzpicture}[remember picture,overlay]
		\fill [blue!50!black] (10.2,-15) rectangle (9.95,15); 
		\end{tikzpicture}
	\fi
}}

\pgfdeclareimage{parts1}{soutenance/figures/parts1}
\pgfdeclareimage{parts2}{soutenance/figures/parts2}
\pgfdeclareimage{parts3}{soutenance/figures/parts3}
\pgfdeclareimage{parts4}{soutenance/figures/parts4}

\fontsize{12}{14}\selectfont

\pagestyle{plain}


\chapter*{Résumé de la démarche et des principaux résultats}
\addcontentsline{toc}{chapter}{\appendixname: Résumé de la démarche et des principaux résultats}
\addtocounter{chapter}{1}

\begin{small}
\begin{center}
\begin{minipage}{0.8 \linewidth}
Ce document, annexe au manuscrit \textit{Nonlinear and quantum effects in analogue gravity}, présente un résumé des travaux réalisés pendant ma thèse, effectuée sous la direction de Renaud Parentani, ainsi que leurs principaux résultats. 
Mon objectif est d'en donner une description débarrassée autant que possible des détails techniques, que le lecteur intéressé trouvera dans le manuscrit, en me concentrant sur les éléments ayant motivé leur analyse.
\end{minipage}
\end{center}
\end{small}

\vspace*{0. cm}


\clearpage 

\section{Introduction}

\subsection{Gravité analogue: motivation}

La relativité générale et la mécanique quantique sont deux des piliers de la physique moderne, permettant à la fois d'expliquer des observations antérieures telles que l'avance du périhélie de Mercure ou l'effet photoélectrique et prédisant de nouveaux phénomènes physiques, notamment les ondes gravitationnelles ou la structure atomique, ayant été depuis vérifiés avec une grande précision. 
Ces deux théories, cependant, semblent mutuellement incompatibles, ce qui pose problème pour obtenir une description unifiée de monde qui nous entoure ou, plus prosaïquement, pour étudier des systèmes où la gravité et les effets quantiques jouent tous deux un rôle important. 
Les premiers instants de l'univers ou le comportement de la matière au voisinage d'un trou noir en sont deux exemples flagrants, l'absence de théorie unifiée limitant considérablement notre capacité à en obtenir un degré de compréhension satisfaisant. 
Bien que des tentatives pour aboutir à une théorie de \og gravité quantique \fg{}, à même de décrire la gravitation dans le cadre de la mécanique quantique, existent (citons notamment les théories des cordes, la gravité quantique à boucle et les modèles tensoriels), celles-ci ne fournissent pas encore de prédictions suffisamment précises pour pouvoir les vérifier. 
Il est donc important d'avoir des modèles combinant certains aspects gravitationnels et quantiques, pouvant servir de guide ou démontrer des résultats généraux qui seront éventuellement appliqués à une théorie de gravité quantique. 

C'est exactement ce que fournit la gravité analogue. 
Ce domaine d'étude, initié par William Unruh en 1981, repose sur l'existence d'une correspondance mathématique précise entre le comportement de la matière au voisinage d'un trou noir et celui des ondes sonores dans un flot transsonique. 
De tels modèles, qualifiés d'\og analogues \fg{}, ont un lien avec la relativité générale via la présence d'un horizon aux propriétés similaires à la frontière d'un trou noir. 
En outre, certains d'entre eux tels les condensats de Bose-Einstein sont intrinsèquement quantiques. 
D'autres, par exemple les ondes à la surface de l'eau, tout en restant classiques (au sens de \og non quantiques \fg{}), présentent le phénomène d'amplification d'ondes à l'origine de l'effet Hawking (voir ci-dessous). 
Cela permet, du moins en principe, d'étudier les phénomènes quantiques en présence d'un champ gravitationnel dans un cadre plus général, voire de réaliser l'analogue d'un trou noir en laboratoire. 
À titre d'illustration, je représente sur la figure~\ref{fig:intro_char} les caractéristiques des ondes sonores dans deux flots, analogues à un trou noir et à un trou blanc (renversé temporel d'un trou noir).

\subsection{Effet Hawking}

Nous avons tous en tête l'image d'un trou noir comme un corps au champ gravitationnel si intense qu'il ne peut rayonner: la vitesse de libération à l'intérieur de celui-ci étant supérieure à la vitesse de la lumière, rien ne peut s'en échapper. 
Un trou noir ne peut ainsi pas émettre de radiation et devrait rester invisible; d'où le qualificatif \og noir \fg{}. 
Cette description naïve doit cependant être revue dans un cadre plus réaliste: si le trou noir n'est pas isolé mais entouré d'un nuage de matière, ce dernier, accéléré et déformé par les forces gravitationnelles, voit ses particules s'entrechoquer à grande vitesse. 
L'énergie dissipée par les collisions les chauffe à de hautes températures, menant à l'émission d'énergie sous forme lumineuse. 
Ainsi, bien que nulle onde ne s'échappe véritablement de l'intérieur du trou noir, celui-ci devient indirectement visible grâce à la matière située juste à l'extérieur qui, elle, rayonne.

\begin{myfig}
\centering
\includegraphics[width = 0.49 \linewidth]{figures/intro/char_an_BH.pdf}
\includegraphics[width = 0.49 \linewidth]{figures/intro/char_an_WH.pdf}
\includegraphics[width = 0.49 \linewidth]{figures/intro/char_an_BH_b.pdf}
\includegraphics[width = 0.49 \linewidth]{figures/intro/char_an_WH_b.pdf}
\capf{Caracteristiques des ondes sonores dans le plan $(t,x)$, où $t$ désigne le temps et où la vitesse du fluide est orientée selon $x$. 
Cette dernière est donnée par $v_0 > 0$ et la vitesse du son par $c_0$. 
Le panneau de gauche représente un flot de type \og trou noir \fg{} avec $v_0(x) - c_0(x) = \tanh(x) + 0.3 \tanh(x)^2$. 
Le panneau de droite représente un flot de type \og trou blanc \fg{} avec $v_0(x) - c_0(x) = -\tanh(x) + 0.3 \tanh(x)^2$. 
Dans les deux cas, l'horizon analogue se trouve en $x=0$. }\label{fig:intro_char}
\end{myfig}

Jusque là, notre raisonnement est totalement classique. 
Lorsque les effets quantiques sont inclus, on sait que le \og vide \fg{} est siège de création perpétuelle de paires de particules et d'antiparticules, qui s'annihilent rapidement entre elles et sont donc en général inobservables. 
Ce qui nous mène à la question suivante: ces particules pourraient-elles agir comme notre nuage de matière classique et rayonner en présence du champ gravitationnel d'un trou noir? 
Celui-ci pourrait-il être ainsi visible, même isolé? 
Stephen Hawking a répondu à ces questions dans un article paru en 1975: oui, un trou noir peut émettre un rayonnement! 
L'image intuitive est la suivante: l'attraction gravitationnelle peut séparer les paires de particules et antiparticules produites juste au-dessus de l'horizon, là où les forces de marée sont très importantes, les secondes tombant dans le trou noir alors que les premières s'échappent, donnant lieu au \og rayonnement de Hawking \fg{}. 
En outre, ce rayonnement est thermique, avec une température inversement proportionnelle à la masse du trou noir. 

Ces résultats, rares exemples de cas connus où la relativité générale et la mécanique quantique jouent tous deux un rôle fondamental, sont cruciaux pour notre compréhension actuelle de la physique des trous noirs. 
Mais ils souffrent de deux problèmes, à ce jour non résolus:
\begin{itemize}
\item Le problème dit \og transplanckien \fg{}: le calcul de Hawking repose sur une hypothèse de basse énergie, mais fait indirectement intervenir des particules très énergétiques, ce qui pose la question de la validité des approximations effectuées. 
\item Le problème dit \og de l'unitarité \fg{}: il semble, au vu de ces résultats, que l'état final du système, à savoir un ensemble thermique de particules, soit indépendant de ce qui a pu former le trou noir, par exemple l'effondrement d'une étoile. 
L'information sur cet état initial serait donc perdue, ce qui briserait la loi d'\og unitarité \fg{} en mécanique quantique.
\end{itemize}
Plusieurs explications ont été avancées pour résoudre ces deux problèmes, mais aucune, à ce jour, n'apporte de solution qui soit pleinement satisfaisante et puisse être vérifiée par des observations. 
En outre, la température du rayonnement de Hawking serait extrêmement faible pour toues les trous noirs connus à ce jour, ce qui rendrait sa détection très difficile. 

\begin{myfig}
\begin{tikzpicture}
\def\xo{4.};
\def\yo{-2.5};
\newcommand{\rr}{1. };
\newcommand{\lel}{1. };
\def\l{4.};
\newcommand{\lax}{1.2};
\newcommand{\lay}{2.7};
\newcommand{\lat}{5.};
\filldraw[fill=black!25] (\xo-\rr,\yo) arc (180:0:\rr and \rr/2) -- (\xo+\rr,\yo+\l) arc (360:540:\rr and \rr/2) -- cycle;
\filldraw[fill=black!35] (\xo-\rr,\yo) arc (180:360:\rr and \rr/2) -- (\xo+\rr,\yo+\l) arc (360:180:\rr and \rr/2) -- cycle;
\draw[dashed] (\xo-\rr,\yo) arc (180:0:\rr and \rr/2);
\draw[dashed] (\xo-\rr,\yo) -- (\xo-\rr,\yo-\lel);
\draw[dashed] (\xo+\rr,\yo) -- (\xo+\rr,\yo-\lel);
\draw[dashed] (\xo-\rr,\yo+\l) -- (\xo-\rr,\yo+\lel+\l);
\draw[dashed] (\xo+\rr,\yo+\l) -- (\xo+\rr,\yo+\lel+\l);
\draw[black,-stealth] (\xo,\yo) -- (\xo-\lax,\yo-\lax) node[above]{$x$};
\draw[black,-stealth] (\xo,\yo) -- (\xo+\lay,\yo) node[above]{$y$};
\draw[black,-stealth] (\xo,\yo) -- (\xo,\yo+\lat) node[right]{$t$};
\draw (\xo+\rr+0.1,\yo+1) node{\scalebox{0.5}{\pgfbox[center,bottom]{\pgfuseimage{parts1}}}};
\draw (\xo+\rr+4.1,\yo+1) node{\scalebox{0.5}{\pgfbox[center,bottom]{\pgfuseimage{parts2}}}};
\draw (\xo+\rr+3.1,\yo+3) node{\scalebox{0.5}{\pgfbox[center,bottom]{\pgfuseimage{parts3}}}};
\draw (\xo-\rr-1.1,\yo+2.5) node{\scalebox{0.5}{\pgfbox[center,bottom]{\pgfuseimage{parts4}}}};
\end{tikzpicture}
\capf{Illustration qualitative de l'effet Hawking. Ici, $x$ et $y$ sont deux coordonnées spatiales et $t$ désigne le temps, la troisième dimension spatiale n'étant pas représentée. Le cylindre gris représente la trajectoire de l'horizon d'un trou noir. Les lignes bleues montrent les trajectoires de paires de particules/antiparticules générées par des fluctuations quantiques. Si elles apparaissent loin de l'horizon, elles s'annihilent mutuellement après un temps très court et ne peuvent donc être observées. Mais si elles apparaissent près de l'horizon, les importantes forces de marée peuvent les séparer avant qu'elles s'annihilent, l'une tombant dans le trou noir alors que l'autre s'échappe -- et devient alors observable.}
\end{myfig}

\subsection{Que peut-on apprendre de la gravité analogue?}

Notons que les deux ingrédients intervenant dans le calcul de Stephen Hawking, à savoir la présence d'un horizon pouvant séparer les paires de particules/antiparticules et la génération de ces dernières par des fluctuations quantiques, sont aussi présents dans les modèles \og analogues \fg{}. 
L'effet Hawking doit donc également avoir lieu dans ces systèmes, ce qui constitue le principal intérêt de ce domaine. 
Mais s'il y a bien un lien entre eux et les trous noirs astrophysique, toute analogie a des limites et il convient avant toute chose de savoir jusqu'où elle peut être poussée, c'est-à-dire ce que l'on peut véritablement en apprendre.

Pour répondre à cette question, commençons par voir ce que nous ne pouvons \textit{pas} en apprendre. 
Quoi qu'il en soit, la gravité analogue ne pourra résoudre les problème transplanckien et de l'unitarité. 
En effet, ceux-ci font intervenir des effets, comme la réaction d'une particule émise sur le trou noir lui-même, qui sont au-delà de l'analogie entre les trous noirs astrophysiques et les modèles analogues. 
Les résoudre dans le cadre analogues ne prouvera donc rien pour ce qui est du domaine gravitationnel. 

Cela étant dit, l'étude des trous noirs analogues peut se révéler intéressante pour au moins trois raisons. 
Tout d'abord, ces modèles permettent d'illustrer le mécanisme de Hawking (prouvant au passage qu'il n'y a pas d'incompatibilité logique entre la présence d'horizons au sens de la relativité et les principes quantiques) dans un cadre plus général, incluant par exemple des phénomènes dispersifs ou dissipatifs. Cela montre l'aspect très générique de ce mécanisme. 
Deuxièmement, malgré les objections formulées ci-dessus, il est possible que la résolution des problèmes susmentionnés dans le cadre analogue puisse s'appliquer à certains modèles de gravité quantique. 
Cela peut en outre servir de guide lors de l'élaboration de ces théories, en montrant quels sont les ingrédients nécessaires pour obtenir telle ou telle propriété physique. 
Enfin, et de manière peut-être plus terre-à-terre, cette étude apporte un nouvel éclairage sur les modèles analogues eux-mêmes, permettant de clarifier la description de certains phénomènes ou d'en découvrir de nouveaux. 

Ainsi, s'il convient de ne pas étendre les résultats obtenus en gravité analogue plus loin que leur domaine d'application, ceux-ci ont beaucoup à nous enseigner tant sur la physique de basse énergie, donc non directement concernée par la gravité quantique, que sur de possibles voies vers une description unifiée du monde qui nous entoure. 

Au cours de cette thèse, mes collaborateurs et moi nous sommes intéressés à trois aspects de la gravité analogue, à savoir les effets non-linéaires, la diffusion linéaire et les liens entre modèles théoriques et expériences. 
Je vais à présent les présenter successivement, tâchant de souligner leur pertinence et leurs implications. 

\section{Aspects non-linéaires}


La correspondance mentionnée ci-dessus est, \textit{stricto sensu}, valable seulement à l'ordre linéaire.  Les effets non-linéaires, au-delà de l'analogie, ont ainsi été relativement peu étudiés. 
En outre, les travaux les concernant se sont en général concentrés sur quelques modèles spécifiques, à même de décrire plus ou moins fidèlement les expériences mais ne permettant pas de distinguer clairement les aspects généraux et les particularités des cas étudiés. 
Il est cependant important d'obtenir une bonne compréhension de ces effets de manière plus globale, et ce pour au moins deux raisons. Tout d'abord, il est nécessaire, afin d'élaborer des expériences, de savoir quel domaine de l'espace des paramètres permet de réaliser des flots propices à l'observation du rayonnement de Hawking (ou d'autres effets) sans que le signal soit noyé sous un bruit trop important. 
Par ailleurs, d'un point de vue plus conceptuel, il peut être intéressant de voir si l'analogie avec les trous noirs astrophysiques mentionnée ci-dessus peut être étendue et, si oui, dans quelle mesure.

L'approche que mes collaborateurs et moi avons suivie au cours de cette thèse se fonde sur l'étude de modèles simplifiés mais plus généraux, permettant de voir quels aspects sont propres à la physique d'un horizon et lesquels sont plus particuliers. 
L'utiliser en synergie avec celle d'autre groupes, souvent plus spécifique, me semble nécessaire pour aboutir à une bonne compréhension des systèmes étudiés.

\subsection{Dynamique des trous noirs}

Une question importante, tant pour les expériences que pour obtenir une compréhension satisfaisante des trous noirs analogues, concerne la nature des flots stationnaires de type \og trou noir \fg{} (c'est à dire présentant l'analogue de l'horizon d'un trou noir) et leur stabilité. 
Afin d'y répondre, nous nous sommes concentrés sur deux classes de modèles: un condensat de Bose-Einstein transsonique en (1+1) dimensions, décrit par l'équation de Gross-Pitaevskii, et un fluide parfait dont les ondes de surface sont décrites par l'équation de Korteweg-de Vries. 
Nous nous sommes d'abord intéressés à la structure des solutions stationnaires et avons montré que celles-ci dépendent seulement de quelques paramètres macroscopiques. 
En outre, elles peuvent dans des cas simples être décrites analytiquement. 

Concernant la stabilité de ces solutions, une question préliminaire concerne leur stabilité linaire, \textit{i.e.}, sous des perturbations de faibles amplitudes correctement décrites par les équations linéarisées. 
Nous avons montré que les flots de type \og trou noir \fg{} sont bien linéairement stables, les perturbations décroissant, localement, polynomialement en temps. 
Le cas des perturbations non-linéaires, requérant une résolution des équations de Gross-Pitaevskii et Korteweg-de Vries, est plus complexe. 
Mais un calcul analytique est possible sur un modèle simple et des simulations numériques permettent d'étudier les cas plus complexes (et plus réalistes). 
Notre résultat principal est le suivant: l'ensemble des solutions de type \og trou noir \fg{} est un attracteur local, au sens où toute perturbation initiale (pourvu que son amplitude ne soit pas trop grande) est expulsée à l'infini, la solution convergeant localement vers un de ses élément. 
En outre, la transition entre la configuration initiale perturbée et le trou noir obtenu aux temps longs s'effectue \textit{via} l'émission d'ondes non-linéaires dont le nombre et les propriétés peuvent être déterminés analytiquement. 
Celles-ci sont de deux types: onde de choc dispersive ou onde simple, représentées sur la figure~\ref{fig:solsNLS}.

Notons que ces résultats sont très similaires aux théorèmes d'unicité et de stabilité des trous noirs astrophysiques. 
En particulier, dans les deux cas (analogue et astrophysique), les solutions stationnaires sont entièrement décrites par quelques quantités macroscopiques reliées à des courants conservés et le retour à l'équilibre après une perturbation s'effectue \textit{via} des ondes non-linéaires puis des modes quasi-normaux (définis plus loin). 
Il semble donc que l'analogie persiste, dans une certaine mesure, pour les effets non-linéaires. 

\begin{myfig}
\includegraphics[width=0.49 \linewidth]{figures/analogue_no-hair/DSWKdV}
\includegraphics[width=0.49 \linewidth]{figures/analogue_no-hair/SWKdV}
\capf{Exemples d'une onde de choc dispersive (gauche) et d'une onde simple (droite) pour l'équation de Korteweg-de Vries.}\label{fig:solsNLS}
\end{myfig}

\subsection{Dynamique des trous blancs}

Un trou blanc est, en quelque sorte, l'\og inverse \fg{} d'un trou noir, ce mot devant être entendu au sens temporel. 
Alors qu'il est possible d'entrer \og dans \fg{} un trou noir (au sens où une particule peut traverser l'horizon de l'extérieur vers l'intérieur) mais pas d'en sortir, un objet initialement à l'intérieur d'un trou blanc peut en sortir, mais pas y entrer à nouveau.  
Du point de vue des modèles analogues, un flot de type \og trou blanc \fg{} s'obtient à partir d'un \og trou noir \fg{} en inversant simplement la vitesse du fluide, ce qui revient à un retournement temporel. 

À l'ordre linéaire, ces flots sont légèrement instables: des perturbations de grande longueur d'onde génèrent une ondulation de la surface libre (pour les ondes de surface) ou de la densité (pour un condensat de Bose-Einstein) croissant, selon les propriétés statistiques du bruit, logarithmiquement (pour une température nulle) ou linéairement (à température non nulle) en temps. 
Son amplitude finit donc par atteindre des valeurs pour lesquelles les effets non-linéaires deviennent importants. 
Nous nous sommes intéressés à l'évolution aux temps longs de cette instabilité et avons observé numériquement deux comportements très différents selon le signe de la perturbation initiale:
\begin{itemize}
\item pour certaines valeurs des paramètres, l'ondulation sature à une amplitude finie et la solution devient stationnaire;
\item pour d'autres valeurs, elle s'accompagne de l'émission périodique de solitons (perturbations localisées et de grande amplitude) à partir de l'horizon. 
\end{itemize}
Il semble qu'une solution stationnaire ne soit jamais atteinte dans le second cas, l'émission de solitons continuant indéfiniment. 

\subsection{Dynamique des «~amplificateurs de Hawking-Corley-Jacobson~»}

Il est intéressant de combiner ces aspects en considérant une configuration avec deux horizons: un de type \og trou blanc \fg{} et un de type \og trou noir \fg{}. 
La région séparant les deux horizons peut alors agir comme une cavité résonnante, le rayonnement de Hawking étant réfléchie et amplifiée par chacun d'eux. 
Une telle configuration est appelée \og black hole laser \fg{} dans les publications en langue anglaise. Une traduction littérale pouvant se révéler ici trompeuse, j'utiliserai la dénomination d'\og amplificateurs de Hawking-Corley-Jacobson \fg{}, ou \og amplificateur HCJ \fg{}. 
(Steven Corley et Theodore A. Jacobson furent à ma connaissance les premiers, en 1998, à s'intéresser à un tel système.) 

À l'ordre linéaire, sous certaines conditions, il existe une instabilité dynamique: des perturbations croissent exponentiellement en temps. 
Nous avons montré que cette instabilité peut être de deux types distincts, voir la figure~\ref{fig:QNM}. 
Le premier, directement relié au rayonnement de Hawking, fait intervenir des modes de fréquences complexes dont les parties réelles et imaginaires sont toutes deux non nulles. 
Le second fait intervenir des modes de fréquences imaginaires pures. 
Il n'est pas directement relié à l'effet Hawking, bien qu'il puisse être vu comme un progéniteur du précédent, et plus intrinsèque à la cavité formée par les deux horizons. 

En incluant les effets non-linéaires, nous avons observé comme pour les trous blancs deux types de comportement, selon la distance entre les horizons et la configuration initiale. 
Dans certains cas, le système sature sur une solution stationnaire, dont le profil et les propriétés peuvent être obtenus analytiquement dans un modèle simple. 
Essentiellement, la densité entre les deux horizons croît suffisamment pour supprimer l'instabilité en diminuant le nombre de Mach. 
Il est intéressant de noter que la structure des solutions stationnaires est intimement reliée à celle des modes instables, voir la figure~\ref{fig:ene-9-eq}. 
Dans d'autres cas, une solution stationnaire n'est jamais atteinte et l'horizon de type \og trou blanc \fg{} émet des groupes de solitons de manière périodique. 
À l'image des trous blancs, les amplificateurs HCJ ont donc un comportement non-linéaire complexe, qui semble même dans certains cas devenir chaotique. 
Il serait très intéressant d'explorer ces questions, en particulier le mécanisme de formation des solitons, plus avant. 

\begin{myfig}
\includegraphics[scale=0.6]{figures/saturation/QNM.pdf} \hspace*{0. cm}
\includegraphics[scale=0.4]{figures/saturation/trajC.pdf}
\capf{Gauche: Évolution des fréquences angulaires des deux premiers modes instables d'un amplificateur HCJ en fonction de la demi-distance $L$ entre les deux horizons, dans les unités d'une longueur critique $L_0$. Les courbes bleue et verte représentent leurs parties réelles et les courbes orange et rouge leurs parties imaginaires.
Droite: trajectoire de la fréquence angulaire du premier mode instable dans le plan complexe lorsque $L$ varie.} \label{fig:QNM}
\end{myfig}

\begin{myfig}
\includegraphics[width=0.49 \linewidth]{figures/saturation/ene-4-eq.pdf} 
\includegraphics[width=0.49 \linewidth]{figures/saturation/ene-5-eq.pdf} 
\capf{Énergies des solutions non-linéaires moins celle de la configuration de HCJ initiale, en fonction de la distance $L$ entre les deux horizons. Le panneau de gauche montre les énergies des solutions pouvant être continument déformées en la solution initiale en modifiant $L$, et celui de droite celles qui lui restent déconnectées.}\label{fig:ene-9-eq}
\end{myfig} 

\section{Diffusion et effet Hawking}

Un des principaux objectifs et intérêts de la gravité analogue concerne l'observation et l'étude du rayonnement de Hawking. 
Comme pour les effets non-linéaires décrits ci-dessus, nous nous sommes efforcés d'exhiber certaines de ses propriétés générales plus que de décrire avec précision un système particulier. 
Nous nous sommes principalement intéressés à trois aspects intimement liés:
\begin{itemize}
\item l'importance de l'horizon pour les propriétés statistiques du rayonnement;
\item les différents types d'instabilités à même de survenir en (1+1) dimensions;
\item l'effet d'un horizon \og universel \fg{} (défini ci-dessous). 
\end{itemize}
Les trois sous-parties suivantes détaillent chacune un de ces points. 

\subsection{Ondes de surface dans un fluide parfait: horizons et thermicité}

Il s'agit de l'un des systèmes analogues les plus simples: le flot d'un fluide parfait sur un obstacle l'accélérant localement et générant, lorsqu'il est transcritique, un horizon analogue. 
Or, dans les expériences réalisées à ce jour, le flot reste apparemment souscritique, ne présentant donc pas d'horizon au sens strict. 
Une équipe expérimentale dirigée par Silke Weinfurtner a cependant, en 2010, rapporté l'observation d'un spectre thermique, comparable à ce que donnerait l'effet Hawking. 
Ceci pose la question de l'importance de l'horizon: un phénomène similaire au rayonnement de Hawking pourrait-il avoir lieu en son absence? 
Si oui, comment expliquer que ses propriétés statistiques restent thermiques alors que la température est fixée, en présence d'un horizon, par la gravité de surface de ce dernier? 
Pour répondre à ces questions, nous avons calculé analytiquement (sur un modèle simplifié) et numériquement les coefficients de diffusion d'une onde de surface sur un flot inhomogène souscritique. 
La relation de dispersion est donnée figure~\ref{fig:DR} et le processus de diffusion est représenté schématiquement sur la figure~\ref{fig:modes}.  
Nos résultats montrent l'existence de deux notions de \og température \fg{}: une définition \og statistique \fg{} reliée à la dépendance en énergie du flux émis et une définition plus effective. 
Pour un flot transcritique (avec un horizon), ces deux notions sont essentiellement équivalentes et le spectre d'émission est bien thermique. 
Par contre, pour un flot souscritique, elles deviennent très différentes. 
Le spectre reste thermique, au sens où la température effective ne dépend pas de la fréquence des ondes, en utilisant la seconde définition, mais pas pour la première. 
Nous pouvons tirer deux conclusions de ces observations. 
D'une part, le caractère thermique du spectre observé expérimentalement semble dû à l'usage d'une définition effective de la température, insensible à certaines déviations. 
D'autre part, un horizon semble crucial pour avoir un lien univoque avec l'effet Hawking, dont le spectre a une température bien définie au sens statistique. 
Ce second point est renforcé par le fait que, dans le cas souscritique, les ondes émises dépendent des détails de l'obstacle, alors que dans le cas transcritique elles sont essentiellement déterminées par la gravité de surface du trou noir analogue, comme dans le cas relativiste.
\textcolor{white}{Une lueur ambrée \\ Tombait du crépuscule \\ Et sa main majuscule \\ Disait l'eau colorée \\}
\begin{myfig}
\vspace*{0.1cm}
\includegraphics[width = 0.5 \linewidth]{figures/probing/disprel.pdf}
\capf{Relation de dispersion entre le vecteur d'onde $k$ et la fréquence angulaire $\om$ dans le modèle d'ondes de surface utilisé pour les calculs numériques. 
Les points colorés montrent les vecteurs d'onde possibles pour deux valeurs de $\om$. 
} 
\label{fig:DR}
\end{myfig}
\vfill

\begin{myfig}
\includegraphics[width = 0.49 \linewidth]{figures/probing/figmodes1.pdf}
\includegraphics[width = 0.49 \linewidth]{figures/probing/figmodes2.pdf}
\capf{Diffusion en présence (gauche) et absence (droite) de point de rebroussement. 
Les couleurs correspondent à celles des points colorés sur la figure~\ref{fig:DR}. 
L'onde incidente est représentée en trait plein et les modes générés par la diffusion en tirets. }
\label{fig:modes}
\end{myfig}

\subsection{Instabilités en (1+1) dimensions}

Motivés par les deux types de solutions exponentiellement croissantes en temps observées dans un amplificateur HCJ, nous avons étudié les instabilités à même de survenir en théorie des champs en (1+1) dimensions sous un angle plus général.
Cette étude montre des liens profonds entre trois types de modes (solutions d'une équation linéaire) appartenant au spectre discret: les modes d'instabilité dynamique (MID), quasi-normaux (MQN) et liés (ML). 
Les MID sont exponentiellement croissants en temps et décroissants en espace. 
Ils sont ainsi de carré sommables et donc appartiennent au spectre de la théorie au sens usuel. 
les MQN sont décroissants en temps mais croissants en espace. 
Ils ne font donc pas partie de spectre proprement dit, mais restent cruciaux pour comprendre la dynamique des champs aux temps longs. 
Enfin, les ML sont stationnaires et spatialement décroissants. 
Comme les MID, ils appartiennent au spectre de la théorie.  
Nous avons montré que les fréquences de ces trois types de mode peuvent être obtenues comme pôles de la fonction de Green retardée, \textit{i.e.}, calculée en imposant des conditions limites sortantes. 
Par ailleurs, une transformation de Laplace permet de décomposer la solution générale sur une base de modes incluant, en plus des précédents, ceux du spectre continu, et montre la pertinence des MQN pour la décrire aux temps longs lorsqu'il n'existe aucun DIM ou ML.  
Nous avons brièvement discuté de la pertinence de ces observations pour la physique des trous noirs, soulignant les conditions sous lesquelles des DIM sont présents. 
Cette étude fournit également une explication à l'émergence des deux types de MID dans un amplificateur de HCJ en utilisant les symétries du problème et la structure Hamiltonienne de la théorie.

Nous avons également introduit la notion de \og mode quasi-quasi-normal \fg{} (MQQN) pour décrire des configuration avec plusieurs régions de diffusion nettement séparées et/ou des conditions limites différentes. 
Les MQQN sont définis de la même manière que les MQN mais en imposant des conditions limites sortantes autour d'une région donnée. 
Ils peuvent ainsi être calculés dans des cas où il n'y a pas de MQN pour le problème complet, par exemple lorsque celui-ci est défini sur un tore, et décrivent la dynamique à des échelles de temps intermédiaires.  

\subsection{Effet Hawking et horizon universel}

Cet aspect de mon travail de thèse concerne, plus que la gravité analogue proprement dite, deux classes de théories de la gravitation dont l'objectif est d'inclure une partie des effets quantiques dans une description effective: les théories dites de Ho\v{r}ava et d'Einstein-\AE ther. 
À l'image des modèles analogues, ces théories font intervenir des termes dispersifs qui régularisent certaines divergences apparaissant en relativité générale, ce qui, du moins pour les premières, fait naître l'espoir qu'elles puissent être utilisées comme théories de gravité quantique effectives à basse énergie.  
Le parallèle avec les modèles analogues est en fait très profond, les idées et concepts développés du côté gravitationnel pouvant dans une large mesure être appliqués à ces modèles et réciproquement. 
En particulier, dans les deux cas, les termes dispersifs brisant l'invariance de Lorentz permettent à des particules de dépasser la vitesse de la lumière et donc de traverser l'horizon d'un trou noir dans les deux sens -- horizon qui n'en est ainsi plus vraiment un au sens premier du terme.   

Les trous noirs stationnaires des théories de Ho\v{r}ava et d'Einstein-\AE ther ont cependant un ingrédient supplémentaire par rapport à la majorité des modèles analogues: un second horizon, dit \og universel \fg{}, qu'une particule ne peut traverser que dans un sens quelle que soit sa vitesse, même supérieure à celle de la lumière. 
(Des trajectoires typiques de particules sans masse sont représentées sur la figure~\ref{fig:char}.)
La présence de deux horizons pose la question de la nature du rayonnement de Hawking ainsi que son origine, qui pourrait en principe avoir trait à chacun d'eux. 
Pour voir ce qu'il en est, nous avons d'abord calculé le spectre émis en ne prenant en compte que la région proche de l'horizon universel. 
Nous avons montré que, dans le cas où le trou noir est formé par l'effondrement d'une fine couche de matière, cette région ne rayonne pas dans la limite des temps longs.
Un calcul numérique plus global indique que mécanisme de Hawking a toujours lieu autour de l'horizon \og standard \fg{}, exactement comme en gravité analogue. 
Le spectre est ainsi approximativement thermique avec une température donnée par sa gravité de surface,  l'horizon universel ne semblant pas jouer de rôle significatif. 

Nous avons également donné des arguments, reposant sur la continuité des observables et l'analyse des ingrédients menant à ces résultats, suggérant (sans cependant constituer une preuve formelle) que ces derniers devraient s'étendre à des modèles plus réalistes d'effondrement stellaire formant un trou noir. 
Il serait intéressant de  pouvoir faire le calcul explicitement sur un tel modèle afin de vérifier si tel est bien le cas -- et, si non, quelle est alors l'origine du rayonnement de Hawking.

\section{Applications à des réalisations expérimentales}

Les résultats présentés ci-dessus sont étroitement liés à des considérations expérimentales. 
En effet, toute expérience de gravité analogue requiert à la fois un bon contrôle des aspects non-linéaires du système étudié, afin de déterminer le profil et les propriétés d'un éventuel horizon analogue, ainsi qu'une étude de la diffusion permettant d'estimer l'amplitude de l'effet attendu puis d'interpréter les observations. 
Au cours de cette thèse, mes collaborateurs et moi nous sommes intéressés à trois systèmes: les ondes de surface dans un canal à houle en présence d'un obstacle, les perturbations de densité dans un condensat de Bose-Einstein et les ondes sonores dans un tube avec une paroi \og molle \fg{} (en un sens donné ci-dessous). 
Le premier a mené à deux expériences, effectuées par le groupe de Germain Rousseaux à l'institut Pprime de l'université de Poitiers, dont nous avons contribué à analyser les résultats. 
Le second a été utilisé par Jeff Steinhauer pour observer -- à ma connaissance, pour la première fois -- le rayonnement de Hawking analogue. 
Le troisième est actuellement réalisé au Laboratoire d'Acoustique de l'Université du Maine, au Mans, par le groupe d'Yves Aurégan et Vincent Pagneux. 

\begin{myfig}
\includegraphics[width = 0.49\linewidth]{figures/universal/char.pdf} 
\includegraphics[width = 0.49 \linewidth]{figures/universal/char2.pdf}
\capf{Trajectoires en présence d'un horizon universel pour des particules sans masse de basse énergie devant l'échelle dispersive (gauche) et de haute énergie comparable à cette échelle (droite). 
L'horizon universel se trouve en $r = M$ et l'horizon \og standard \fg{} en $r = 2M$, où $M$ désigne la masse du trou noir. 
} \label{fig:char}
\end{myfig}

Il est important de noter que, bien que deux de ces expériences soient classiques (au sens de \og non quantiques \fg{}), elles présentent un lien étroit avec le rayonnement de Hawking. 
En effet, ce dernier peut être vu comme l'amplification de fluctuations quantiques les changeant en particules \og réelles \fg{}, observables. 
Mais le phénomène d'amplification lui-même est déjà présent dans la théorie classique. 
Il est ainsi possible d'observer le mécanisme à l'origine de l'effet Hawking dans un cadre purement classique. 

\subsection{Ondes de surface: choix de l'obstacle}

Une composante importante des expériences utilisant des ondes de surface est l'obstacle placé au fond du canal. 
En effet, c'est lui qui accélère le courant, le nombre de Froude maximal atteint et le profil du courant dépendant donc de sa forme. 
En outre, il convient pour avoir des résultats univoques de minimiser l'amplitude de l'ondulation de la surface libre en aval, ce qui requiert une forme d'obstacle précise. 
En effet, pour ne pas que les ondes réfléchies aient une trop grande longueur d'onde qui les rendrait difficiles à observer et poserait des problèmes liés à la réflexion sur les bords du canal, les ondes sont dans la plupart des cas envoyés de l'aval vers l'obstacle, et passent donc sur l'ondulation. 
Celle-ci peut donc induire des effets parasites importants à moins que son amplitude ne soit très faible. 

Nous nous sommes intéressés à cette question et écrit un code \textit{Mathematica} utilisant les propriétés d'holomorphie du potentiel de vitesse pour obtenir une forme d'obstacle à partir des données de la surface libre et du courant asymptotique. 
Ces derniers peuvent être choisis arbitrairement, chaque possibilité donnant un profil d'obstacle différent et pouvant par la suite être affinée pour obtenir les propriétés désirées. 
Ce code, reposant sur les hypothèses d'un fluide parfait, irrotationnel et sans tension de surface, a été utilisé pour élaborer l'obstacle de la seconde expérience du groupe de Germain Rousseaux. 

\subsection{Ondes de surface: coefficients de transmission et corrélations} 

La première expérience réalisée à l'institut Pprime, dont le schéma est donné figure~\ref{fig:concl_exp_schema1}, utilise le même obstacle que celui du groupe de Silke Weinfurtner, à Vancouver. 
Notre objectif était de mesurer le coefficient de transmission et vérifier qu'il tend vers $1$ dans la limite des basses fréquences, comme le prédit notre analyse théorique. 
Les mesures montrent un relativement bon accord avec les résultats numériques, en particulier dans le domaine de fréquences intermédiaire entre les hautes valeurs où la transmission est quasiment absente et les basses valeurs où celle-ci domine. 
L'écart le plus significatif concerne des pics de transmission à basses fréquences, qui pourraient être dus à des réflexions sur les bords du canal. 
Ces résultats semblent valider les approximations utilisées dans notre étude numérique. \\

\begin{myfig}
\includegraphics[width = \linewidth]{figures/concl/schema2.pdf}
\capf{Schéma du dispositif expérimental utilisé dans la première expérience, l'écoulement se faisant de la gauche vers la droite. 
La surface grise représente l'obstacle et les flèches les différentes ondes impliquées dans le processus de diffusion. 
Les deux rectangles rouges donnent les positions de senseurs acoustiques utilisés pour mesurer la hauteur de la surface libre. 
}\label{fig:concl_exp_schema1}
\end{myfig}

Pour la seconde expérience, un obstacle a été élaboré en utilisant la méthode décrite ci-dessus. 
Celui-ci permet d'obtenir un nombre de Froude au-dessus de l'obstacle un peu plus important avec une ondulation de plus faible amplitude. 
L'idée de cette expérience était de mesurer les corrélations entre les ondes en aval de l'obstacle (incidente et réfléchies) afin de réduire le bruit dans l'estimation des coefficients de diffusion. 
Nous avons observé que certains coefficients se comportent comme prédits par la théorie, alors que d'autres sont plus importants, ce qui peut être dû à l'ondulation résiduelle et sa résonance avec les ondes réfléchies. 
Cette étude est, à mes yeux, un premier pas vers l'élaboration d'un système permettant d'avoir un flot transcritique stable et avec une ondulation de très faible amplitude, ce qui permettrait d'observer de manière univoque le mécanisme sous-jacent à l'effet Hawking pour des ondes de surfaces.

\subsection{Étude de l'effet Hawking dans un condensat de Bose-Einstein}

Comme je l'ai déjà mentionné, Jeff Steinhauer a récemment réalisé une expérience lui permettant d'observer l'effet Hawking dans un condensat de Bose-Einstein contenant un trou noir analogue. 
Nous avons peu après décrit théoriquement un modèle proche de son système, bien que simplifié, afin de voir dans quelle mesure ses résultats s'inscrivent dans un cadre général. 
Ce travail donne d'abord une description des solutions stationnaires et asymptotiquement homogènes en présence d'une marche de potentiel, qui semble modéliser de manière approchée mais correcte celui utilisé par Jeff Steinhauer. 
Nous avons ensuite résolu l'équation linéarisée dans un cas proche de celui de l'expérience et calculé la fonction de corrélation à deux points, sur laquelle les conclusions expérimentales sont fondés. 
Nos résultats sont assez proches des observations, mais présentent des différences non négligeables, qui
pourraient être dues au fait que le condensat utilisé dans l'expérience n'est pas exactement stationnaire.  

\subsection{Ralentir le son pour \og écouter \fg{} l'effet Hawking}

Une autre expérience, dont nous avons participé à la description théorique, utilise de l'air s'écoulant dans un tube muni d'une paroi \og molle \fg{}. 
Ici, \og molle \fg{} n'est pas à prendre au sens strict, mais comme qualifiant les conditions limites qu'elle impose. 
Celle-ci est en fait constituée de tubes d'épaisseur millimétrique, imposant une condition plus faible qu'une paroi lisse, voir schéma figure~\ref{fig:concl_slowsound}. 
Son principal intérêt est une réduction de la vitesse du son effective à l'intérieur du tube: 
il est ainsi possible de réaliser un horizon analogue avec un nombre de Mach restant inférieur à $1$, ce qui devrait beaucoup simplifier l'obtention d'un flot stable. 
Des microphones peuvent être placés au-dessus des tubes pour mesurer les variations de pression et ainsi d'estimer les coefficients de diffusion. 
Nous avons réalisé une étude théorique qui semble indiquer que, pourvu que les effets négligés ne deviennent pas trop importants, l'observation de l'effet Hawking analogue dans ce système est accessible.
La réalisation expérimentale est en cours au Laboratoire d'Acoustique de l'Université du Maine.
\begin{myfig}
\includegraphics[width=0.49 \linewidth]{figures/concl/configuration.pdf}
\capf{Schéma de l'expérience. 
La paroi en $y = 0$ est \og dure \fg{} et celle en $y = 1+b(x)$ est \og molle \fg{}, au sens défini dans le texte. 
Elle se compose d'une succession de petits tubes, représentés par des traits verticaux, de longeur $b(x)$. 
Les flèches bleues donnent la direction de l'écoulement pour un flot de type \og trou blanc \fg{}.
}\label{fig:concl_slowsound}
\end{myfig}

\section{Ouverture: Cordes cosmiques et autres lignes de vortex}

Une corde cosmique est une solutions en théorie des champs où l'un d'entre eux prend une valeur différente de celle minimisant localement l'énergie à cause d'un défaut topologique de dimension 1. 
Par exemple, pour un champ complexe $\phi$ avec un potentiel $V$ ne dépendant que de sa valeur absolue $\abs{\phi}$, si $V$ a un minimum absolu pour une valeur $\phi_0 > 0$ de $\abs{\phi}$, le champ va évoluer dynamiquement pour que $\abs{\phi}$ tende vers $\phi_0$ en tout point, afin de minimiser son énergie. 
Mais si cette évolution a lieu de manière indépendante entre des points séparés par une longue distance, alors la phase de $\phi$ prendra des valeurs différentes en des régions de l'espace éloignées les unes des autres. 
Celle-ci peut ainsi varier d'un nombre entier non nul de fois $2 \s \pi$ sur un cercle de grand rayon. 
Dans ce cas, $\phi$ doit s'annuler en au moins un point sur toute surface admettant ce cercle comme frontière, formant une ligne le long de laquelle $\phi = 0$ et la densité d'énergie est donc importante. 
Les régions autour de telles lignes sont appelées cordes cosmiques. 
De nombreux modèles de grande unification (unifiant l'électromagnétisme et les interactions fortes et faibles) prédisent leur formation lors des transitions de phase qui ont eu lieu dans l'univers primordial. 
Leur épaisseur typique est de l'ordre de l'échelle de grande unification, soit $10^{-32}\mathrm{m}$, alors que leur longueur peut atteindre des valeurs cosmologiques.\footnote{Étant donnée l'absence de structure additionnelle, du moins dans les modèles les plus simples, le terme de \og fil \fg{} serait peut-être plus approprié que \og corde \fg{} pour les décrire. L'usage a cependant retenu ce dernier.}

Sous certaines conditions, ces cordes peuvent porter des charges et courants, dûs à la condensation de champs de matière en leur sein.
Ces derniers sont en général monotones, leur norme décroissant rapidement depuis un maximum sur l'axe de la corde pour s'annuler asymptotiquement.  
Nous avons cependant montré en collaboration avec Betti Hartmann et Patrick Peter que, dans un cadre relativement général, il existe aussi des solutions \og excitées \fg{} pour lesquelles le champ n'est pas monotone et passe une ou plusieurs fois par $0$ avant d'y tendre asymptotiquement. 
Celles-ci sont révélées par une étude analytique sous une approximation linéaire, qui permet d'obtenir certaines de leurs propriétés, et numériquement par calcul explicite. 
Il est intéressant de noter que les équations les régissant sont, dans une certaine limite, identiques à celles obtenues pour un condensat de Bose-Einstein non relativiste dans un piège harmonique dans deux directions et homogène dans la troisième. 
Plus généralement, il existe un lien étroit avec les lignes de vortex dans des condensats à plusieurs composantes, que nous espérons étudier dans un avenir proche. 
Trois exemples de solutions \og excitées \fg{} sont représentés figure~\ref{fig:insta_w0}.

Nous avons également débuté l'étude de la stabilité des solutions \og excitées \fg{}. 
Il semble que celles-si soient instables, présentant autant de MID que de surfaces où le champ s'annule. 
Cela peut être motivé par le fait qu'il est alors énergétiquement favorable de créer des paires de particules et d'antiparticules, mais une preuve rigoureuse nous échappe encore. 
En outre, il n'est pas exclu que des solutions \og excitées \fg{} puissent se révéler stables dans des domaines de l'espace des paramètres que nous n'avons pas encore explorés, avec des conséquences potentiellement importantes pour l'évolution du réseau de cordes, prédit par de nombreux modèles de grande unifications, au cours de l'histoire de l'univers. \\
\begin{myfig}
\includegraphics[width=0.49 \linewidth]{figures/concl/CS/4sols0.pdf}
\caption{Amplitude $f$ du condensat en fonction de la distance $x$ à l'axe de la corde cosmique pour quatre solutions: une solution \og fondamentale \fg{} (en bleu) pour laquelle $f$ ne s'annule pas et trois solutions \og excitées \fg{} (en vert, rouge et cyan).} \label{fig:insta_w0}
\end{myfig}

\section{Conclusion}

Mes collaborateurs et moi avons étudié durant cette thèse quelques aspects des théories des champs en espaces courbes. 
Je me suis surtout intéressé à des modèles dits de \og gravité analogue \fg{}, non relativistes mais présentant une symétrie de Lorentz effective à basse énergie menant à une correspondance mathématique précise (bien que partielle) avec des champs relativistes dans un espace-temps courbe. 
Celle-ci permet d'étudier des phénomènes impliquant à la fois des concepts issus de la relativité générale et de la mécanique quantique, en particulier le rayonnement des trous noirs prédit par Stephen Hawking. 

Les travaux que nous avons réalisés se concentrent sur trois angles d'approche.  
Le premier concerne les effets non-linéaires. 
Nous avons montré d'une part que ceux-ci sont souvent cruciaux pour la dynamique des systèmes analogues et d'autre part qu'ils révèlent une extension qualitative de l'analogie avec la gravitation. 
Le second a trait à l'ordre linéaire. Nos résultats le concernant soulignent et clarifient l'importance d'un horizon pour que le lien entre l'émission d'ondes dans un flot inhomogène et le rayonnement de Hawking soit univoque. 
Enfin, les divers systèmes expérimentaux, leurs similarités et différences, montrent que cette analogie nous permet d'en apprendre autant sur les systèmes \og analogues \fg{} eux-mêmes que sur l'aspect générique de l'effet Hawking. 

Plus généralement, la variété des modèles analogues et le lien étroit avec la théorie des champs en espace courbe comme avec certaines des questions qui s'imposent lorsque l'on tente de quantifier la gravitation font de la gravité analogue un formidable point de rencontre de multiples domaines de la physique moderne, de la relativité générale aux ondes à la surface de l'eau en passant par les atomes froids et les principes généraux des théories quantiques des champs. 
Pour les échanges d'idées qu'il permet entre différentes communautés scientifiques, pour l'étendue des questions qu'il propose et les possibles applications de ses résultats à d'autres problèmes, je pense et j'espère que ce sujet transversal de la physique moderne verra encore de nombreux développements au cours des années à venir. 

\nocite{Michel:2013wpa}
\nocite{Michel:2014zsa}
\nocite{Busch:2014hla}
\nocite{Euve:2014aga}
\nocite{Michel:2015pra}
\nocite{Auregan:2015uva}
\nocite{Michel:2015rsa}
\nocite{Michel:2015aga}
\nocite{Michel:2015mlr}
\nocite{Euve:2015vml}
\nocite{Coutant:2016bgk}
\nocite{Robertson:2016ocv}
\nocite{Michel:2016tog}
\nocite{Hartmann:2016axn}
\nocite{Robertson:2016evj}
\printbibliography[heading=bibintoc, env=bibliographyNUM, title=Liste des publications, keyword=primary]

\cleardoublepage

\stepcounter{chapter}

\clearpage
\ifodd\thepage\hbox{}\newpage\else\fi
\thispagestyle{empty}\parindent=0pt

\backgroundsetup{contents={}}
\begin{textblock}{10}(2,.3)
\logoED
\end{textblock}

\begin{textblock}{13.9}(1.3,1.6)
\vfill
\setlength{\fboxsep}{5mm}
\noindent\fbox{\parbox{\linewidth-2\fboxrule-2\fboxsep}{
\large{\textbf{Titre :~}\PhDTitleFR}\\

\small{\textbf{Mots clefs :~} \PhDkeywordsFR 
\begin{multicols}{2}
\fontsize{10}{12}\selectfont
\textbf{R\'esum\'e :~}\PhDsumFR
\end{multicols}}
}}

\vfill
\bigskip

\vfill
\setlength{\fboxsep}{5mm}
\noindent\fbox{\parbox{\linewidth-2\fboxrule-2\fboxsep}{
\large{\textbf{Title :~}\PhDTitleEN}\\

\small{\textbf{Keywords :~} \PhDkeywordsEN 
\begin{multicols}{2}
\fontsize{10}{12}\selectfont
\textbf{Abstract :~}\PhDsumEN
\end{multicols}}
}}

\vfill
\end{textblock}

\begin{textblock}{13}(13,14.6)
\includegraphics[height=2cm]{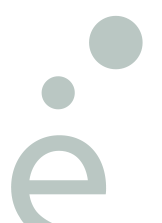}
\end{textblock}

\parindent=0pt 
\begin{textblock}{10}(2.2,14.8)\color{blue!20!red!45!black}{\footnotesize{\textbf{Universit\'e Paris-Saclay}\\Espace Technologique / Immeuble Discovery \\
Route de l'Orme aux Merisiers RD 128 / 91190 Saint-Aubin, France}}
\end{textblock}

\vspace*{\stretch{1}}

\addtocounter{chapter}{5}

\end{document}